\documentclass[a4paper,11pt,french,twoside,openright]{memoir} 
\usepackage[T1]{fontenc}
\usepackage[latin1]{inputenc}
\usepackage{textcomp}
\usepackage{graphicx}
\usepackage{ae,aeguill}
\usepackage{float}
\usepackage{amsfonts}
\usepackage{array}
\usepackage[french]{babel}
\usepackage{geometry}

\usepackage[svgnames]{xcolor}
\usepackage{color}
\usepackage{colortbl}
\usepackage{empheq}
\newcommand*\mybluebox[1]{\colorbox{AntiqueWhite}{\hspace{1em}#1\hspace{1em}}}

\usepackage{fancyhdr}
\fancypagestyle{plain}{ %
  \fancyhf{} % remove everything

	\fancyfoot[RE]{\textit{Pol Grasland-Mongrain}}
	\fancyfoot[LO]{\textit{Applications de la force de Lorentz en acoustique médicale}}
	\fancyfoot[RO]{\thepage}
	\fancyfoot[LE]{\thepage}
}

\fancypagestyle{vide}{ %
  \fancyhf{} % remove everything

}

\fancypagestyle{titre}{ %
  \fancyhf{} % remove everything
	\fancyhead{}
	\fancyfoot{}
	\renewcommand{\headheight}{0pt}

}

\makechapterstyle{mathese}{%

}
\chapterstyle{mathese}

\DisemulatePackage{setspace}
\usepackage{setspace}
\usepackage{sistyle}

\usepackage{amsmath}
\DeclareMathOperator{\grad}{\mathbf{grad}}
\DeclareMathOperator{\Div}{div}
\DeclareMathOperator{\rot}{\mathbf{rot}}
\newcommand{\sinc}{{\textrm{sinc}}}

\begin{document}

%\begin{titlepage}
\thispagestyle{titre}
\enlargethispage*{5cm}
	
\begin{minipage}[c]{.5\linewidth}
	\begin{flushleft}
			\includegraphics[width=.5\linewidth]{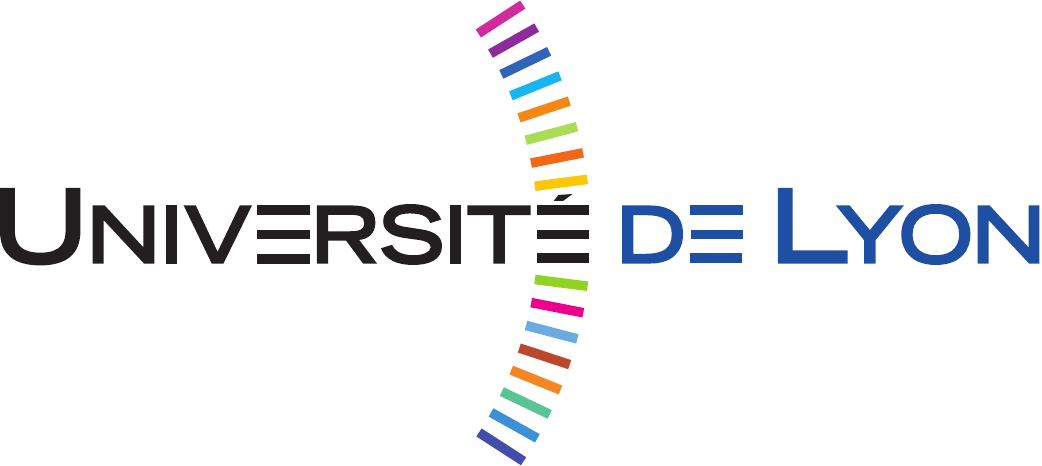}
	\end{flushleft}
\end{minipage} \hfill
\begin{minipage}[c]{.5\linewidth}
	\begin{flushright}
			\includegraphics[width=.5\linewidth]{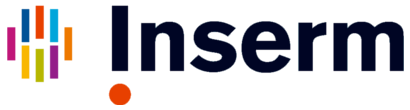}
	\end{flushright}
\end{minipage}

\begin{center}
	\LARGE{\textbf{\textsc{Thèse de Doctorat}\\}}
	\large{Sp\'ecialit\'e Ingénierie biologique et médicale\\}
	\large{École Doctorale Interdisciplinaire Science-Santé\\}
	\vspace*{1cm}
	\large{Présentée pour obtenir le grade de\\}%\LARGE{}
	\Large{\textbf{\textsc{Docteur de l'Université de Lyon}}\\}
	\vspace*{1cm}
	\large{Par\\}%\vspace*{.4cm}
	\LARGE{\textbf{Pol \textsc{Grasland-Mongrain}}\\}
	\vspace*{1cm}
	\huge{\textbf{Applications de la force de Lorentz\\en acoustique médicale}}
	\vspace*{.4cm}
\end{center}

	\vspace*{1cm}

\begin{figure}[!ht]
	\begin{center}
		\includegraphics[width=.4\linewidth]{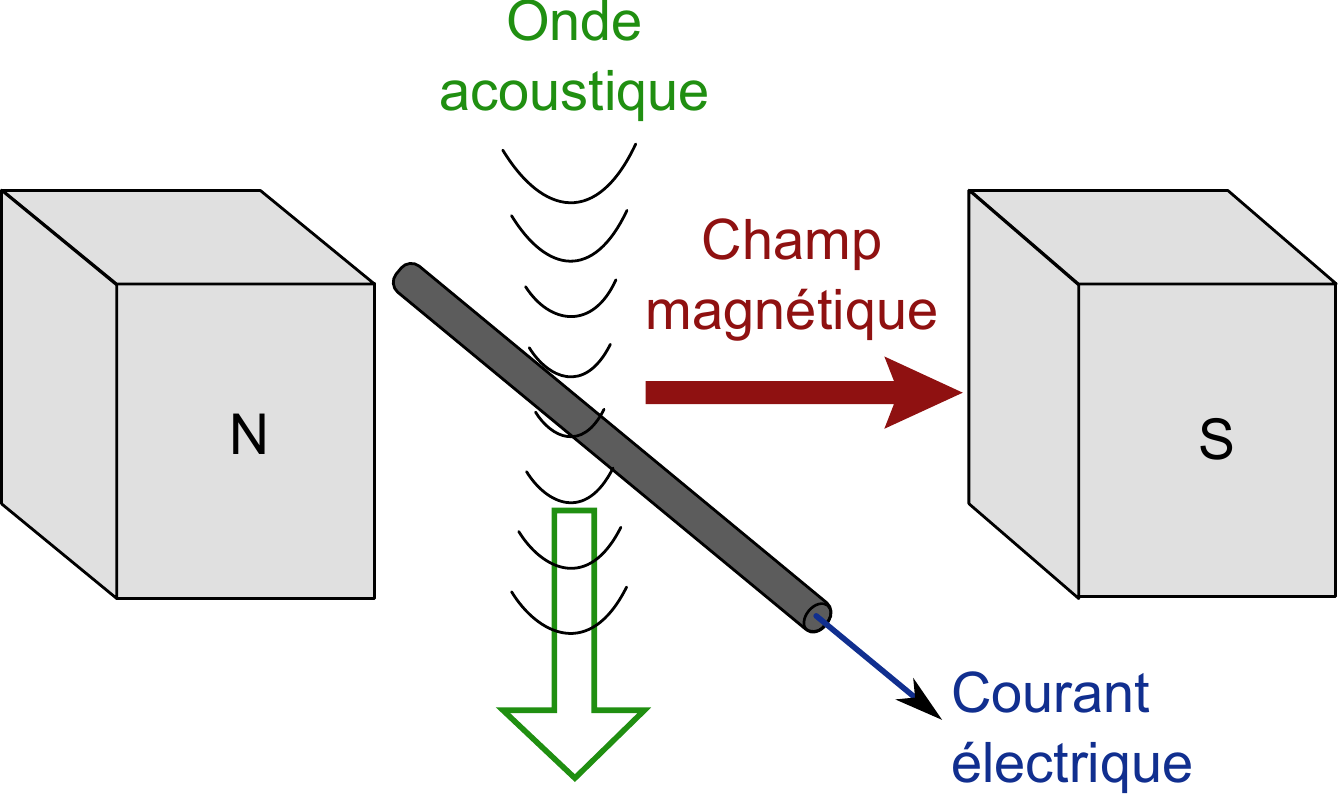}
	\end{center}
\end{figure}

	\vspace*{1cm}

{Thèse soutenue le 12 décembre 2013 devant le jury composé de :\vspace*{.4cm}\\
%\large{Soutenance de thèse prévue le 12 décembre 2013 devant le jury composé de :\\
\begin{tabular}{l l l l}
& & M. Cyril \textsc{Lafon} & \textit{Directeur de thèse}\\
& & M. Mickaël \textsc{Tanter} & \textit{Rapporteur}  \\
& & Mme Gail \textsc{ter Haar} & \textit{Rapporteur} \\
& & M. Stefan \textsc{Catheline} & \\
&	& M. Pierre \textsc{Celsis} & \\
& & M. Bruno \textsc{Gilles} & \\
& & M. Eric \textsc{McAdams} & \\
& & M. Serge \textsc{Mensah} & \\
\end{tabular}}

\begin{flushright}
\small{\textit{Numéro d'ordre : 280-2013}}
\end{flushright}
%\end{titlepage}

\cleardoublepage
\thispagestyle{plain}

\hrule
\vspace*{.1cm}
\textsc{Encadrement\\}
	Directeur de thèse : Cyril \textsc{Lafon}\\
	Co-directeur de thèse : Jean-Yves \textsc{Chapelon}\\
\vspace*{.2cm}

\hrule
\vspace*{.1cm}
\textsc{Titre complet de la thèse (français)\\}
	{Applications de la force de Lorentz en acoustique médicale :\\}
	{Hydrophone à force de Lorentz\\}
	{Tomographie d'Impédance Electrique par Force de Lorentz\\}
	{Imagerie d'ondes de cisaillement induites par force de Lorentz\\}
\vspace*{.2cm}

\hrule
\vspace*{.1cm}
\textsc{Titre complet de la thèse (anglais)\\}
	{Applications of Lorentz force in medical acoustics:\\}
	{Lorentz force hydrophone\\}
	{Lorentz Force Electrical Impedance Tomography\\}
	{Imaging of shear waves induced by Lorentz force\\}
\vspace*{.2cm}

\hrule
\vspace*{.1cm}
\textsc{Mots-clés (français)\\}
force de Lorentz ; acoustique médicale ; hydrophone ; Ultrasons Focalisés à Haute Intensité ; Tomographie Magnéto-Acousto-Electrique ; Tomographie d'Impédance Electrique par Force de Lorentz ; impédance électrique ; élastographie\\
\vspace*{.2cm}

\hrule
\vspace*{.1cm}
\textsc{Mots-clés (anglais)\\}
Lorentz force ; medical acoustics ; hydrophone ; High Intensity Focused Ultrasound ; Magneto-Acoustic-Electrical Tomography ; Lorentz Force Electrical Impedance Tomography ; electrical impedance ; elastography\\
\vspace*{.2cm}

\hrule
\vspace*{.1cm}
\textsc{Etablissement délivrant le grade de docteur\\}
Université Claude Bernard Lyon 1\\
\vspace*{.2cm}

\hrule
\vspace*{.1cm}
\textsc{Intitulé et adresse du laboratoire d'accueil\\}
Laboratoire de Thérapie Appliquée par Ultrasons,\\
INSERM U1032,\\
151 Cours Albert Thomas,\\
69424 Lyon Cedex 03 France\\
\cleardoublepage 

\chapter*{Remerciements}
Ecrire des remerciements est un exercice délicat, surtout que cette partie sera probablement la plus lue de ce manuscrit. Je souhaite donc adresser ma sincère gratitude à ceux qui m'ont aidé dans ma thèse et plus largement à tous ceux que mon travail a intéressés, en implorant par avance de l'indulgence pour ma mémoire imparfaite si j'en oublie certains.

Je commence donc par Cyril Lafon, mon directeur de thèse, qui m'a encadré tout au long de mes trois années de thèse et m'a prodigué moult conseils pour ma future (j'espère) carrière de chercheur. Cyril, merci pour tes encouragements, ta confiance et ton soutien sans faille ! Je remercie également Jean-Yves Chapelon, directeur du LabTAU et co-directeur de thèse, pour m'avoir accueilli au sein du laboratoire et pour toutes nos discussions, scientifiques ou non.

Je tiens à remercier Stefan Catheline, qui m'a encadré notamment sur la partie onde de cisaillement, pour son expertise, son intérêt, sa bonne humeur. Stefan, tu m'as montré qu'on peut être à la fois un chercheur de grande renommée et simplement quelqu'un de cool ! Je voudrais aussi remercier Bruno Gilles, avec qui j'ai travaillé sur l'hydrophone électromagnétique, qui m'a ouvert les yeux sur ce que je sais et surtout sur tout ce que je ne sais pas en mécanique des fluides...

Je remercie sincèrement les deux rapporteurs de ce manuscrit : Gail ter Haar, que j'ai pu rencontrer avec John Civale lors de mon séjour à l'ICR, et Mickaël Tanter, qui a été mon maître de stage officiel en master 2 et avec qui les échanges ont toujours été très cordiaux. Je suis flatté de pouvoir vous présenter ces travaux de recherche aujourd'hui !

Un grand merci aux autres membres du jury : Eric McAdams, que j'ai rencontré lors d'un entretien très sympathique sur les problèmes d'impédance électrique, ainsi que Serge Mensah et Pierre Celsis, qui m'ont tous les deux été chaudement recommandés par mon directeur de thèse. J'espère que ce travail vous intéressera !

J'adresse aussi mes remerciements les plus chaleureux à Jean-Martial Mari, sans qui je ne serai sans doute pas ici en ce moment à rédiger ces lignes. Jean-Martial, je loue ici tes conseils sur mon travail, les fins de journée à essayer de faire fonctionner tant bien que mal une expérience, la dix-huitième relecture de cet article qu'on n'arrive plus à lire... Mais aussi les blagues, les gadgets, les sorties, le cat-sitting, le déménagement, et bien d'autres... L'une de tes citations favorites \footnote{``Hâtez-vous lentement, et sans perdre courage,\\Vingt fois sur le métier remettez votre ouvrage,\\Polissez-le sans cesse, et le repolissez,\\Ajoutez quelquefois, et souvent effacez.''} est également devenue mienne ! Je remercie aussi Rémi Souchon pour ses conseils précieux de méthodologie, pour son intérêt pour mon travail et sa bonne humeur quotidienne. Je voudrais aussi citer Amalric Montalibet et Jacques Jossinet pour leurs travaux sur la Tomographie d'Impédance Electrique par Force de Lorentz au laboratoire.

Je salue ici les stagiaires pas si anonymes que j'ai eu la chance d'encadrer directement ou non, qui ont tous contribué à une part de cette thèse : Sandra Montalescot, la première, qui a fait preuve d'un enthousiasme et d'un investissement qui forcent le respect, Benjamin Roussel et Alexandre Petit, qui ont réalisé un travail très sérieux et ce toujours dans une ambiance agréable, Florian Cartellier, qui a fait avancer le projet d'élastographie avec une grande maitrise tout en préparant un concours et en organisant son mariage, ainsi que Mickaël Parisi, qui a réussi à dépasser ses objectifs de stage.

Ces trois années de thèse se sont passées dans un contexte agréable au sein du laboratoire, et je souhaite donc remercier tout particulièrement mes camarades de râlerie, pardon, de bureau, à savoir Elodie et Apoutou. Je remercie aussi toutes ces personnes avec qui j'ai eu des échanges, qu'ils soient professionnels ou amicaux, Adrien, Adrien et Adrien, Pauline, Charlène, Jacqueline, Cyril, Jérémy et Jérémy, Arash, Andrew, Alain, Maxime, Giovanna, Ali, Mike, Françoise, Jean-Louis, Fabrice, Frédéric et Frédéric, Wilfried, Nicolas, David, Heldmuth El Colombiano, Au, Abbas, Thomas, Cécilia, Mathieu, Isabelle, Sandrine et Sandrine, Anthony, Lucie et Lucie, Irvin, Alexandre, Jean-Christophe, Claude, Manuela, Yves, Bernard, Benjamin...

Je remercie les personnes du Département de Mathématiques Appliquées à l'ENS de Paris qui m'ont toujours accueilli très chaleureusement, donc un grand merci à Habib Ammari, Laurent Seppecher, Pierre Millien et Thomas Boulier, dont le manuscrit et la soutenance ont été une précieuse source d'inspiration.

Lors de mon monitorat, j'ai pu rencontrer des personnes formidables, notamment Valérie Martinez qui fait preuve de qualités humaines exceptionnelles, mais aussi Julien De Bonfils, Sylvain Mousset, Laurent Ducroux, ainsi que mes collègues de monitorat. Je pense aussi à Cédric Ray, le directeur de mon master 2, que j'apprécie énormément. Et j'adresse un remerciement tout particulier à Valérian Reithinger, quelqu'un ayant de grandes qualités humaines et professionnelles. Valérian, dans un an, tu seras dans la même situation que moi, bon courage !

Je voudrais citer mes colocataires qui pour certains m'ont supporté pendant presque deux années : Mathieu, un ami plus qu'un coloc, Mathilde, Lucie, et plus fugacement, Sandra et Pauline. Cette coloc a été bien plus qu'une simple cohabitation, merci à vous ! J'ai également une pensée pour Aquarelle, qui au milieu de ma thèse a voulu vérifier expérimentalement la loi de la chute des corps de Newton. Heureusement, elle ne s'est pas intéressée aux travaux de Schrödinger. Et je remercie ici Florence pour avoir rendu la fin de ma thèse bien plus plaisante.

J'adresse un remerciement un peu spécial à Solène, qui m'a supporté tout au long de ma thèse, dans les moments faciles et surtout dans les moments difficiles, même si la fin de la thèse n'a pas été comme nous l'avions prévue. Solène, si mes heurs et malheurs ne t'ont pas découragée d'entreprendre un doctorat, je te souhaite bonne chance !

Enfin, \textit{last but not least} comme le dirait un certain docteur, je souhaite dédier cette thèse à mon grand-père Eugène, qui n'a malheureusement pas pu en voir la fin, et plus largement à ma famille. Sans elle, je n'en serai pas là aujourd'hui, et je remercie ici tout particulièrement Loïc, Johanne, Erwan, Yiming ainsi que Tissia qui nous a rejoints il y a plus de deux ans déjà !

\cleardoublepage
\thispagestyle{vide}
\vspace{\fill}
\begin{flushright}
	\textit{Le pessimiste se plaint du vent,\\l'optimiste espère qu'il va changer,\\le réaliste ajuste ses voiles.}\\William Arthur Ward,\\écrivain américain
\end{flushright}
\vspace{\fill}
\cleardoublepage

\tableofcontents

\chapter{Introduction}
La force de Lorentz présente une propriété remarquable, celle de relier un courant électrique à un mouvement mécanique. Ainsi, le déplacement d'un conducteur dans un champ magnétique induit par force de Lorentz un courant électrique, et vice-versa. La force de Lorentz, dans une acception large qui sera retenu dans cette thèse, inclut les phénomènes d'induction électromagnétique et de force électromotrice. Elle est d'ailleurs l'une des principales manifestations de l'interaction électromagnétique, qui fait partie des quatre interactions élémentaires de la physique.

Cette force a eu de nombreuses applications depuis le XIXe siècle, en étant la base d'objets aussi variés que les générateurs de courant électrique, les accélérateurs de particules ou les spectromètres de masse. Si elle a déjà été utilisée en acoustique, étant la base de nombreux microphones, elle a été rarement employée en acoustique médicale. Ce dernier est un domaine large d'utilisation des ondes acoustiques pour notamment l'imagerie et le diagnostic de pathologies mais aussi les traitements médicaux.

\section{La force de Lorentz}
\label{secForceLorentz}

\subsection{Historique}
Si les premières manifestations du magnétisme ont été observées dès l'antiquité grecque, il a fallu attendre le XVIIIe siècle pour voir les premières descriptions quantitatives d'une ``force électromagnétique'' \cite{grant1991electromagnetism}: une première relation a été proposée par Johann Tobias Mayer en 1760 pour des objets magnétiques et par Henry Cavendish en 1762 pour les particules chargées électriquement, avec des vérifications expérimentales par Charles-Augustin de Coulomb en 1784.

De nombreuses manifestations de cette ``force électromagnétique'' ont été ensuite observées au cours du XIXe siècle. Ainsi, Hans Oersted découvrit en 1820 qu'un fil parcouru par un courant électrique dévie l'aiguille d'une boussole à proximité -- donc qu'un champ électrique induit un champ magnétique \cite{vershuur1993hidden} ; Michel Faraday constate en 1831 qu'un champ magnétique variable induit un courant électrique. Les différentes théories de l'électrostatique, de l'électrocinétique et du magnétisme ont finalement été synthétisées par James Maxwell en 1865 par quatre équations \cite{jackson1998classical}, \cite{darrigol2000electrodynamics} :

\begin{empheq}[box=\mybluebox]{align}
	\begin{array}{r c l r c l}
		\Div \mathbf{E} & = &  \frac{\rho}{\epsilon_0} ; & \rot \mathbf{E} & = & -\frac{d\mathbf{B}}{dt} ;\\
		\Div \mathbf{B} & = & 0                        ; & \rot \mathbf{B} & = & \mu_0 \mathbf{j} + \mu_0 \epsilon_0 d\mathbf{E}{dt}\\
	\end{array}
	\label{eqIntroEqMaxwell}
\end{empheq}

On donne souvent les noms respectifs de Maxwell-Gauss, Maxwell-Faraday, Maxwell-flux et Maxwell-Ampère à ces quatre équations pour marquer la contribution de chaque chercheur.

Ces équations, bien qu'elles aient été données au milieu du XIXe siècle, ont ``résisté'' à l'introduction de théories du XXe siècle comme la relativité restreinte et sont toujours valables aujourd'hui.
Ce domaine de la physique reste néanmoins l'objet de nombreuses recherches. Pour n'en citer qu'une, afin de rendre les quatre équations beaucoup plus symétriques avec des expressions de $E$ et $B$ similaires, des physiciens cherchent des ``monopoles magnétiques'', ou ``charges magnétiques'', à l'instar des charges électriques. Bien que Paul Dirac ait prédit une existence théorique possible dans le cadre de la physique quantique \cite{dirac1931quantised}, l'absence de résultat malgré plus d'un siècle de recherche et la description du champ magnétique comme une manifestation relativiste du champ électrique laisse douter de l'existence de tels monopoles.

\subsection{Expressions de la force de Lorentz} 

La force de Lorentz peut s'exprimer de différentes manières selon que l'on s'intéresse à une particule chargée simple ou une distribution de charges dans un volume.

\paragraph{Pour une particule simple}
Soit une particule de charge $q$ se déplaçant à une vitesse $\mathbf{v}$ dans un champ électromagnétique $(\mathbf{E},\mathbf{B})$

\begin{figure}[!ht]
		\begin{center}
	   		\includegraphics[width=.5\linewidth]{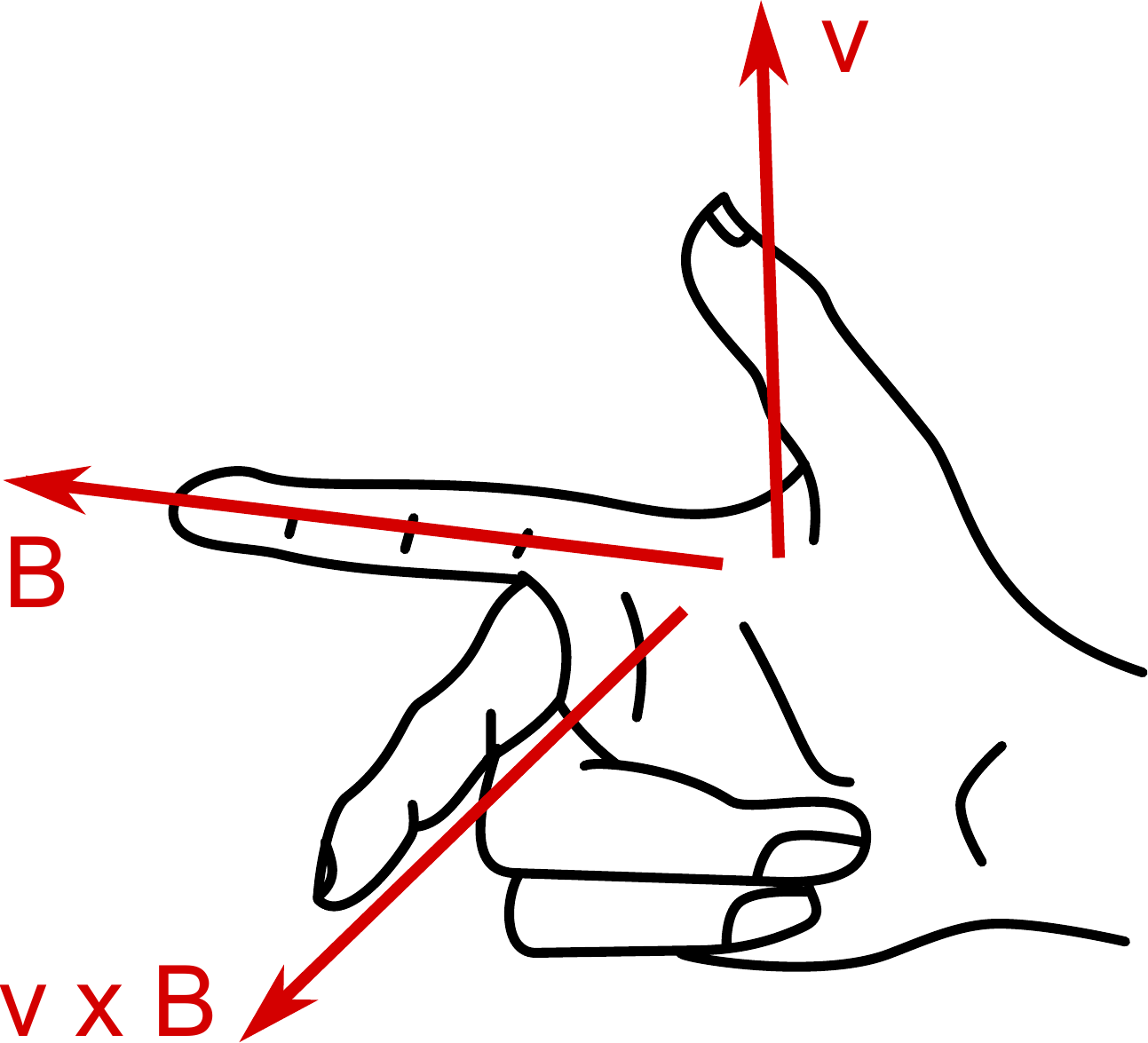}
	   		\caption{\textbf{Règle de la main droite}. En l'absence de champ électrique $\mathbf{E}$, une particule de charge positive se déplaçant avec une vitesse $\mathbf{v}$ dans la direction et le sens du pouce et un champ magnétique $\mathbf{B}$ dans la direction et le sens de l'index subira une force dans la direction et le sens du majeur.}
	   		\label{figIntroForceLorentzMain} 
		\end{center}
\end{figure}

La force de Lorentz que subit la particule vaut alors :
\begin{empheq}[box=\mybluebox]{align}
	\mathbf{F} = q\mathbf{E} + q\mathbf{v} \wedge \mathbf{B}
	\label{eqIntroForceLorentz1}
\end{empheq}
avec $\wedge$ le produit vectoriel, qui permet de définir la direction et le sens du terme $q\mathbf{v}\wedge \mathbf{B}$ comme rappelé dans le figure \ref{figIntroForceLorentzMain}.

Concrètement, cela signifie qu'une particule sera localement accélérée dans la direction du champ électrique $\mathbf{E}$ (dans un sens ou l'autre selon sa charge $q$) et qu'elle changera de direction perpendiculairement à sa vitesse $\mathbf{v}$ et au champ magnétique $\mathbf{B}$ (dans un sens ou l'autre selon sa charge $q$).%, comme représenté dans la figure \ref{figIntroForceLorentzParticule}).
%
%\begin{figure}[!ht]
		%\begin{center}
	   		%\includegraphics[width=.3\linewidth]{figIntroForceLorentzParticule.pdf}
	   		%\caption{Une particule se déplaçant dans un champ magnétique $\mathbf{B}$ à une vitesse $\mathbf{v}$ va être déviée selon sa charge.}
	   		%\label{figIntroForceLorentzParticule} 
		%\end{center}
%\end{figure}

On peut alors parler de \textit{force électrique} pour le terme $q\mathbf{E}$ et de \textit{force magnétique} pour le terme $q\mathbf{v}\wedge \mathbf{B}$, la force de Lorentz étant la somme de ces deux termes. Cependant, certains auteurs désignent improprement la force de Lorentz comme la force magnétique uniquement.

%\paragraph{Pour une distribution de charges volumique}
%Soit une distribution de charge de densité $\rho$ dans un volume et une densité de courant $\mathbf{j}$ égal à $\rho \mathbf{v}$, dans un champ électromagnétique $(\mathbf{E},\mathbf{B})$
%
%La force de Lorentz volumique $\mathbf{f}=\frac{d\mathbf{F}}{dV}$ s'écrit alors :
%\begin{empheq}[box=\mybluebox]{align}
	%\mathbf{f} = \rho \mathbf{E} + \mathbf{j} \wedge \mathbf{B}
	%\label{eqIntroForceLorentz2}
%\end{empheq}

\paragraph{Pour un conducteur parcouru par un courant}
Lorsqu'un conducteur fin placé dans un champ magnétique $\mathbf{B}$ est parcouru par un courant $I$ (pouvant être lui-même issu d'un champ électrique $\mathbf{E}$), la force subie par ce conducteur est égale à :
\begin{empheq}[box=\mybluebox]{align}
	\mathbf{F} = \int_L {I \mathbf{dl} \wedge \mathbf{B}}
	\label{eqIntroForceLorentz2}
\end{empheq}
avec $\mathbf{dl}$ un petit élément du conducteur de longueur totale $L$.

On appelle parfois cette force la \textit{force de Laplace}, plus fréquemment dans les ouvrages francophones qu'anglophones\footnote{La nationalité française de Pierre-Simon de Laplace n'y est peut-être pas étrangère.}, mais on réserve généralement ce terme pour les conducteurs filiformes.

On peut réécrire cette équation sous une forme volumique, avec un petit élément de volume $d\tau$ :
\begin{empheq}[box=\mybluebox]{align}
	\frac{d\mathbf{F}}{d\tau} = \mathbf{j} \wedge \mathbf{B}
	\label{eqIntroForceLorentz3}
\end{empheq}
avec $\mathbf{j}$ la densité de courant électrique dans l'élément de volume $d\tau$ du conducteur.

\begin{figure}[!ht]
		\begin{center}
	   		\includegraphics[width=.8\linewidth]{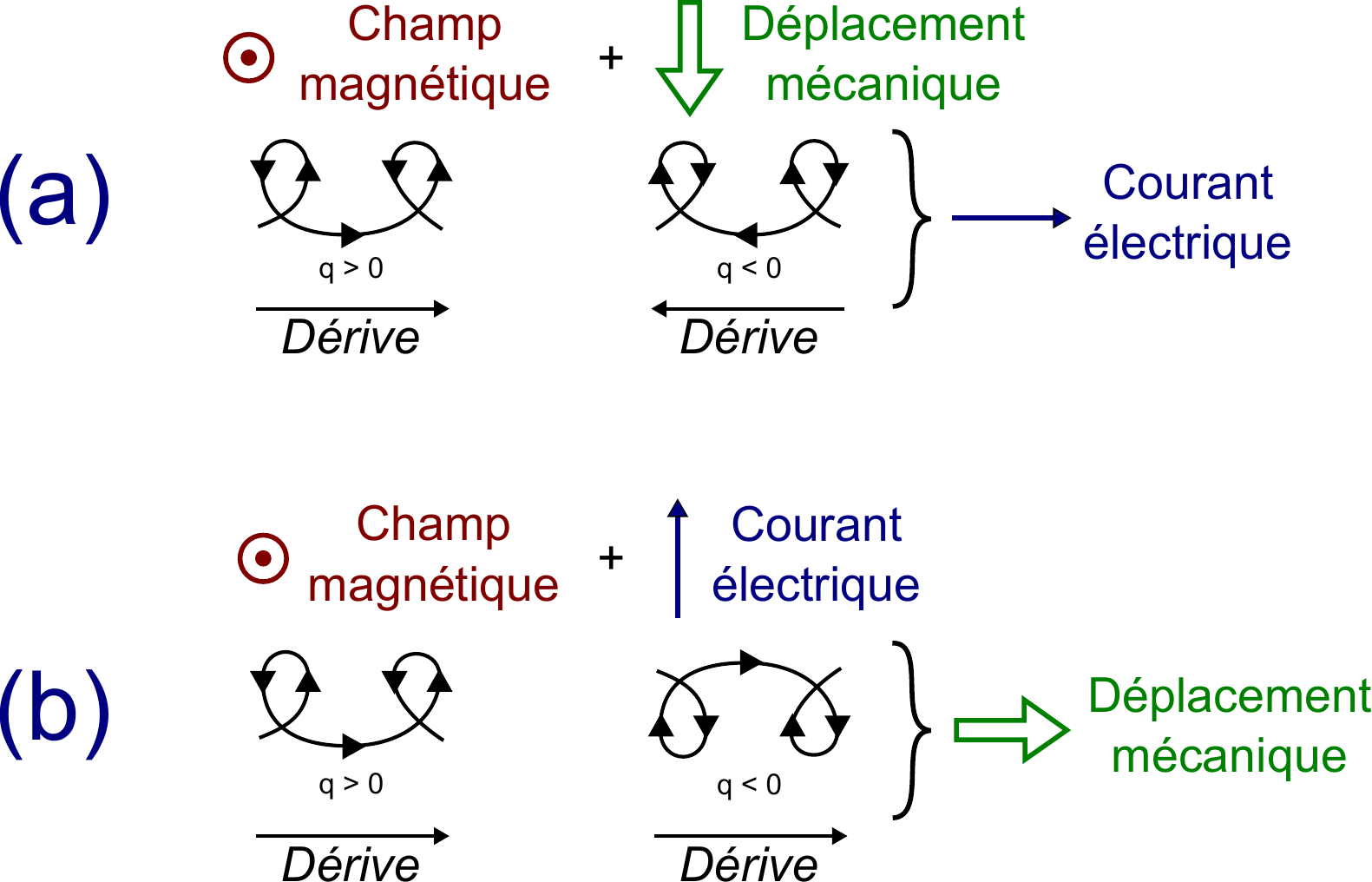}
	   		\caption{(a) Sous l'influence d'un champ magnétique et d'un déplacement mécanique, les particules conductrices dérivent dans la même direction mais dans des sens opposés selon leur signe : cela induit un courant électrique. (b) Sous l'influence d'un champ magnétique et d'un courant électrique, toutes les particules conductrices dérivent dans la même direction et le même sens : cela induit un déplacement mécanique global. }
	   		\label{figIntroForceLorentzSens} 
		\end{center}
\end{figure}

C'est sous cette forme que la force de Lorentz sera le plus souvent utilisée dans cette thèse : un courant électrique dans un champ magnétique induit un déplacement mécanique, et vice-versa. Cela peut s'expliquer simplement en observant le déplacement de particules conductrices, notamment les ions dans les tissus biologiques, qui est résumé sur la figure \ref{figIntroForceLorentzSens}. Sous l'influence d'un champ magnétique et d'un déplacement mécanique, les particules conductrices dérivent dans la même direction mais dans des sens opposés selon leur signe, ce qui induit un courant électrique. La force de Lorentz sera utilisée sous cette forme dans le chapitre \ref{secHydroEM} consacré à un hydrophone et dans le chapitre \ref{secLFEIT} qui étudie une méthode d'imagerie d'impédance électrique. Sous l'influence d'un champ magnétique et d'un courant électrique, toutes les particules conductrices dérivent dans la même direction et le même sens, ce qui induit un déplacement mécanique global. La force de Lorentz sera utilisée ainsi dans le chapitre \ref{secElastoEM} consacré à une technique d'élastographie.

\paragraph{Induction de Lorentz, induction de Neumann}
Une formulation particulière de la force de Lorentz est donnée par la force électromotrice due à un mouvement d'un conducteur dans un champ magnétique \cite{jackson1998classical}. Cette grandeur est égale à :
\begin{empheq}[box=\mybluebox]{align}
	e=\int_L{(\mathbf{v}\wedge \mathbf{B}).\mathbf{dl}}
	\label{eqIntroForceLorentz4}
\end{empheq}
où l'intégration se fait sur la portion considérée du conducteur. Cette force électromotrice, malgré son nom, n'est pas une force : elle est homogène à une tension, et ce nom n'est encore utilisé que pour des raisons historiques. Ainsi, le mouvement d'un conducteur à une vitesse $v$ dans un champ magnétique $B$ créé une tension $e$ à ses bornes. On parle alors d'\textit{induction de Lorentz}.

Par ailleurs, si le champ magnétique est variable au cours du temps, il induit un champ électrique, comme décrit par l'équation de Maxwell-Faraday. Ce champ électrique peut alors produire une force électromotrice dans le conducteur même si celui-ci est immobile : on parle dans ce cas d'\textit{induction de Neumann}.

Ces deux approches ont été rassemblées en une seule à travers l'équation de Faraday $e=-\frac{d\Phi}{dt}$ avec $\Phi$ le flux de champ magnétique à travers la surface $S$ formée par le circuit du conducteur (une boucle pour un circuit fermé). Cette équation remarquable, qui résume deux phénomènes a priori différents, a été l'une des motivations qui ont mené à la relativité restreinte \cite{einstein1905elektrodynamik}. Mais ceci dépasse le cadre de cette thèse, où l'on utilisera uniquement des champs magnétiques fixes.

\section{L'acoustique médicale}
L'acoustique médicale est un domaine utilisant les ondes acoustiques pour deux grands types d'applications : pour le traitement des pathologies d'une part et pour l'imagerie et le diagnostic d'autre part.

\subsection{La thérapie par ultrasons}
Lors de ses recherches sur le sonar pendant la première Guerre Mondiale, Paul Langevin a constaté que les ondes acoustiques avaient un potentiel de destruction important \cite{newman1998history}. Cela a conduit naturellement à l'utilisation des ondes acoustiques pour la thérapie. Ainsi, les frères Fry ont évalué les effets des ultrasons sur les tissus biologiques dès les années 50 \cite{fry1979biological}. La thérapie par ultrasons utilise deux effets principaux : les effets mécaniques et les effets thermiques. Les effets mécaniques regroupent les phénomènes autres que thermiques dus au déplacement des tissus par l'onde acoustique. En particulier, les fortes pressions acoustiques peuvent créer des bulles de cavitation d'énergie considérable. Les effets thermiques, eux, sont créés par l'absorption de l'onde ultrasonore par les tissus.

\begin{figure}[!ht]
		\begin{center}
	   		\includegraphics[width=.5\linewidth]{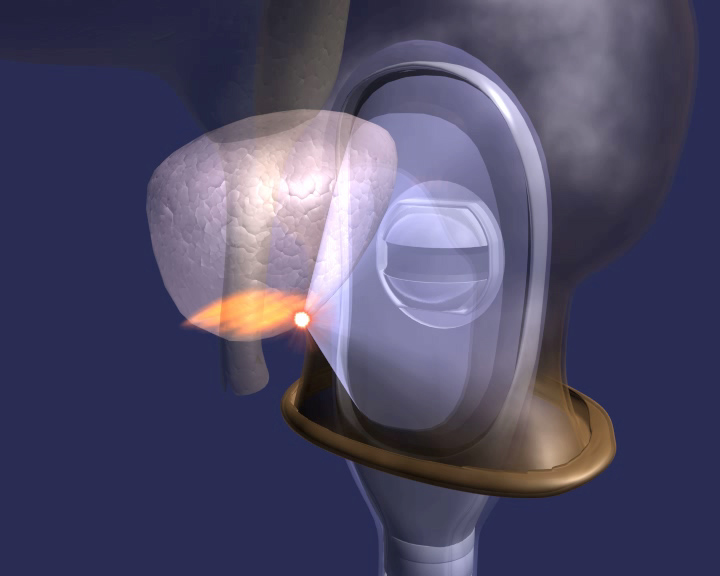}
	   		\caption{Principe d'une sonde de traitement du cancer de la prostate par onde ultrasonore focalisée à haute intensité. La sonde endorectale brûle au point focal le tissu prostatique sans endommager le rectum (source page \pageref{source11}).}
	   		\label{figIntroAcoustiqueMedicaleHIFU} 
		\end{center}
\end{figure}

Ces effets ont mené à l'élaboration de thérapies par ultrasons, et notamment les traitements par HIFU (High Intensity Focused Ultrasound) qui, comme leur nom l'indique, utilisent une onde ultrasonore focalisée à haute puissance. Ainsi, on peut traiter des tissus pathologiques en les brûlant, comme des tumeurs cancéreuses par exemple, comme représenté sur la figure \ref{figIntroAcoustiqueMedicaleHIFU}. Une zone entière peut alors être traitée précisément point par point en déplaçant la zone focale des ultrasons, sans traiter les couches intermédiaires entre le transducteur et la zone focale. Cette technique a été appliquée entre autres pour le traitement du foie \cite{kennedy2004high}, \cite{melodelima2009thermal}, du sein \cite{kennedy2005high}, \cite{gianfelice2003}, de la prostate \cite{rewcastle2006high}, \cite{gelet1996treatment} (notamment avec la commercialisation d'appareils cliniques comme l'Ablatherm et le Sonablate \cite{ahmed2009minimally}), etc.

Cependant, ce type de traitement est encore jeune et moins connu que la radiothérapie ou la chimiothérapie, et les normes internationales pour assurer un traitement fiable et précis évoluent encore aujourd'hui \cite{shaw2008calibration}. En particulier, la caractérisation précise du champ ultrasonore de thérapie à pleine puissance reste aujourd'hui un challenge, par le développement de capteurs à la fois précis et résistants -- dont cette thèse proposera un prototype.

Parmi les autres applications des ondes acoustiques à des fins de thérapie, on peut citer la lithotripsie qui cherche à détruire des calculs rénaux grâce à des ``ondes de choc'' (des ondes acoustiques puissantes et brèves) \cite{martin1986ultrasound} ; la sonochimiothérapie et la sonoporation, qui améliorent grâce à des ondes ultrasonores la délivrance de drogues, soit en ``activant'' ces dernières, soit en améliorant leur pénétration dans les cellules \cite{somaglino2011delivrance} ; la régénération des os, qui est accélérée par l'application d'ondes acoustiques à faible puissance \cite{lavandier2009experimental}.

\subsection{L'imagerie acoustique}
La forme la plus répandue d'imagerie acoustique est l'échographie, technique que l'on retrouve dans le domaine médical, vétérinaire ou industriel, qui a été développée dès la fin des années 40 par Georges Ludwig puis John Wild \cite{wild1951use}.

\begin{figure}[!ht]
		\begin{center}
	   		\includegraphics[width=.5\linewidth]{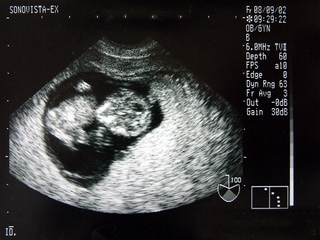}
	   		\caption{Echographie ``mode B'' d'un f\oe tus de 9 semaines (source page \pageref{source10}).}
	   		\label{figIntroAcoustiqueMedicaleEchographie} 
		\end{center}
\end{figure}

Dans cette technique, une sonde émet une onde ultrasonore dans une direction. Chaque interface acoustique va donner lieu à une onde réfléchie, ou écho, captée par la sonde initiale. En mesurant le temps d'aller-retour des différents échos, et en connaissant la vitesse de propagation de l'onde, on peut en déduire la distance des interfaces. Cela constitue l'échographie ``mode A''. Si l'on convertit l'amplitude des échos en niveaux de gris et que l'on observe plusieurs directions successives grâce à l'onde ultrasonore, on peut former ligne par ligne une image, ce qui est nommé échographie ``mode B'', dont on peut voir un exemple sur la figure \ref{figIntroAcoustiqueMedicaleEchographie}. Cette technique présente de nombreux avantages qui expliquent son succès : résolution spatiale de l'ordre du millimètre, affichage des images en temps réel, innocuité, facilité d'utilisation, coût faible de mise en \oe uvre. Cependant, la qualité d'une échographie dépend beaucoup de la position de la sonde qui n'est pas connue a priori : l'échographie est ``opérateur-dépendant'' où souvent une même personne fait l'image et l'interprétation à la fois, contrairement à des radiographies ou des images IRM.

L'échographie Doppler, développée à partir des années 1960-70 notamment avec L. Pourcelot \cite{pourcelot1974applications}, est issue de cette technique. Dans cette dernière, on observe les variations de fréquence de l'onde reçue par rapport à l'onde émise. Cela donne une information sur le déplacement dans le tissu biologique étudié, notamment les vaisseaux sanguins.

Mais on peut utiliser les ondes acoustiques d'autres manières pour l'imagerie médicale. Par exemple, on peut utiliser le déplacement local du tissu dû aux ultrasons pour en déduire des paramètres physiques -- cela est utilisé entre autres dans la tomographie d'impédance électrique par force de Lorentz, présentée dans cette thèse. Une onde lumineuse ou un courant électrique peuvent également être perturbés par la présence d'une onde acoustique, ce qui a été mis à profit dans les techniques d'imagerie acousto-optique \cite{bossy2005} et acousto-électrique \cite{ammari2012b}.

Enfin, une voie prometteuse d'imagerie médicale a été ouverte avec l'élastographie ultrasonore dynamique qui permet de réaliser des images d'élasticité des tissus \cite{ophir1991}. Cette technique utilise des ondes acoustiques pour faire des images échographiques et pour provoquer un déplacement dans un tissu, observable sur les images échographiques. Cette technique sera décrite plus en détails dans le chapitre de cette thèse consacré à une utilisation particulière de la force de Lorentz en élastographie.

\section{Structure de la thèse}
L'objectif de cette thèse a donc été d'étudier différentes façons d'utiliser la force de Lorentz dans le domaine de l'acoustique médicale. La force de Lorentz présente en effet la propriété de relier, via un champ magnétique, un courant électrique à un mouvement mécanique.

La première partie de cette thèse concerne le développement d'un hydrophone pour effectuer des champs de vitesse acoustique. Cet hydrophone était constitué d'un fil de cuivre vibrant dans un champ magnétique, ce qui induit par force de Lorentz un courant électrique. Un modèle hydrodynamique a été élaboré pour trouver une relation entre le courant électrique mesuré et la pression ultrasonore. Un prototype a ensuite été conçu et sa résolution spatiale, réponse fréquentielle, sensibilité, résistance et réponse directionnelle ont été examinées. Cette partie est décrite dans le chapitre \ref{secHydroEM}.

Une méthode d'imagerie appelée Tomographie d'Impédance Electrique par Force de Lorentz a aussi été étudiée. Dans cette méthode, un tissu biologique conducteur est déplacé par ultrasons dans un champ magnétique, ce qui induit un courant électrique par force de Lorentz. L'impédance électrique du tissu peut ensuite être déduite de la mesure du courant. Cette technique a été appliquée pour réaliser des images de fantômes de gélatine et de muscle de b\oe uf, qui ont donné un contraste supplémentaire aux images ultrasonores. Nous avons aussi pu détecter une lésion thermique dans un échantillon de muscle de poulet avec cette méthode. Cela constitue le chapitre \ref{secLFEIT}.

Enfin, cette thèse a démontré la possibilité de générer des ondes de cisaillement dans des tissus mous en utilisant la force de Lorentz. Cela a été réalisé en appliquant un courant électrique par deux électrodes dans un solide mou placé dans un champ magnétique. Des ondes de cisaillement ont été observées dans des échantillons de gélatine et de foie. Les ondes de cisaillement ont été utilisées pour trouver l'élasticité des échantillons et ainsi que leur conductivité électrique. Ces techniques sont présentée dans le chapitre \ref{secElastoEM}.

Enfin, le chapitre \ref{secConclusion} dresse un bilan des avancées scientifiques obtenues dans le cadre de cette thèse, avec une liste des communications scientifiques.

\chapter{Hydrophone à force de Lorentz}
\label{secHydroEM}

Avec le développement des transducteurs ultrasonores destinés à l'imagerie ou la thérapie, la connaissance du champ acoustique émis est devenue cruciale \cite{harris1988hydrophone}. En effet, une incertitude sur le champ acoustique lors d'une thérapie par ultrasons focalisés à haute intensité peut avoir des conséquences dramatiques \cite{ter2000intervention}, \cite{shaw2008buoyancy}, \cite{barnett2000international} : les zones pathologiques peuvent être sous-traitées, des tissus vitaux endommagés. Il est d'ailleurs obligatoire aujourd'hui de préciser les paramètres acoustiques des équipements médicaux ultrasonores \cite{medical1993}. Ces techniques mesurent l'intensité acoustique, la pression acoustique, le déplacement ou la vitesse particulaire \cite{preston1991}. Cependant, il y a encore peu de méthodes bien adaptées aux champs acoustiques à haute puissance \cite{fda1985}. Ce chapitre va donc présenter l'ensemble des techniques disponibles et introduira un concept d'hydrophone basé sur la force de Lorentz, dont l'un des avantages serait de pouvoir faire des mesures de champs acoustiques à haute puissance.

\section{La mesure de champs ultrasonores}

\subsection{Les ultrasons à haute puissance}
Le terme de ``haute puissance'' en acoustique n'est pas trivial \cite{hodnett1997strategy}. A haute puissance, deux phénomènes peuvent endommager les appareils de mesure : le chauffage et les dommages créés par la cavitation. La cavitation, dépendant de plusieurs paramètres, peut apparaitre dans des gammes de 100 kPa à 100 MPa, comme indiqué dans le tableau \ref{tableHydroEMIntroCavitation} et peut facilement endommager des capteurs fragiles. Pour un milieu donné, le risque de cavitation augmente avec la pression et le nombre de cycles par pulse et diminue avec la fréquence.

D'autre part, une onde acoustique se modifie à cause d'effets non-linéaires lors de sa propagation. Cet effet augmente avec la puissance de l'onde. Ainsi, avec une onde ultrasonore à haute puissance, on peut obtenir une ``onde choquée'', qui présente une forme en dents de scie. Le spectre de cette onde possède de nombreuses harmoniques, donc un capteur d'onde ultrasonore à haute puissance devra avoir une bande passante fréquentielle large pour correctement représenter le signal.

\begin{table}
	\begin{tabular}{|p{.20\linewidth}|p{.18\linewidth}|p{.14\linewidth}|p{.16\linewidth}|p{.16\linewidth}|}
		\hline
		\rowcolor{SandyBrown} Application & Fréquence d'utilisation & Nombre de cycles par tir & Pression acoustique typique & Seuil de \mbox{cavitation} \\
		\hline
		Imagerie \mbox{ultrasonore} & 1 à 20 MHz & 1 & 1 MPa & 10 MPa \\
		\hline
		Lithotripsie & 0,2 MHz & 1 & 100 MPa & 10 MPa \\
		\hline
		Imagerie \mbox{Doppler} & 1 à 30 MHz & 1 à 100 & 1 à 10 MPa & 1 à 10 MPa \\
		\hline
		Thérapie \mbox{ultrasonore} & 0,5 à 10 MHz & Continu & 1 à 10 MPa & 100 kPa \\
		\hline
	\end{tabular}
	\caption{Revue des principales applications acoustiques médicales, avec les fréquences typiques d'utilisation, le nombre de cycles acoustiques par tir, la pression acoustique généralement utilisée (plus exactement, l'amplitude du pic positif de la pression) et la pression acoustique au-dessus de laquelle des nuages de cavitations risquent d'apparaitre. Certaines applications évitent la cavitation, tandis que d'autres l'exploitent (source : \cite{hodnett1997strategy}).}
	\label{tableHydroEMIntroCavitation}
\end{table}

Un paramètre majeur est constitué par l'intensité acoustique $I$, qui donne le taux d'énergie par unité de suface perpendiculaire à la direction de propagation :

\begin{empheq}[box=\mybluebox]{align}
	I=\frac{1}{T}\int{p\mathbf{u}dt}
	\label{eqHydroEMIntroIntensite}
\end{empheq}
avec $T$ la durée considérée, $p$ la pression acoustique et $\mathbf{u}$ la vitesse particulaire.

\subsection{Les mesures de pression acoustique}

\subsubsection{Les hydrophones piézoélectriques}
Les premiers hydrophones utilisaient, à l'instar des transducteurs ultrasonores classiques, l'effet piézoélectrique pour mesurer des pressions ultrasonores. Par cet effet, la pression acoustique sur une membrane piézoélectrique crée une tension aux bornes de ce matériau, tension que l'on peut mesurer et donc relier à la pression. On distingue parfois les hydrophones piézoélectriques selon le matériau de la partie sensible, qui peut être piézocomposite ou pyroélectrique (comme le polyfluorure de vinylidène, plus connu sous le nom de PVDF) \cite{harris2000}.

Les hydrophones à membrane sont composés d'une membrane piézoélectrique de diamètre typique entre 0,2 mm et 75 mm \cite{retat2011} acoustiquement transparente. Ils présentent de nombreuses caractéristiques intéressantes pour le mesure de pression acoustique \cite{dereggi1977} : leur réponse est plate sur une large plage de fréquences \cite{bacon1982characteristics}, elle est linéaire sur une gamme importante de pressions, elle est stable au cours du temps et la membrane est quasiment invisible aux ultrasons \cite{smith1989hydrophones}.

\begin{figure}[!ht]
   \begin{minipage}[c]{.5\linewidth}
		\begin{center}
	   		\includegraphics[width=1\linewidth]{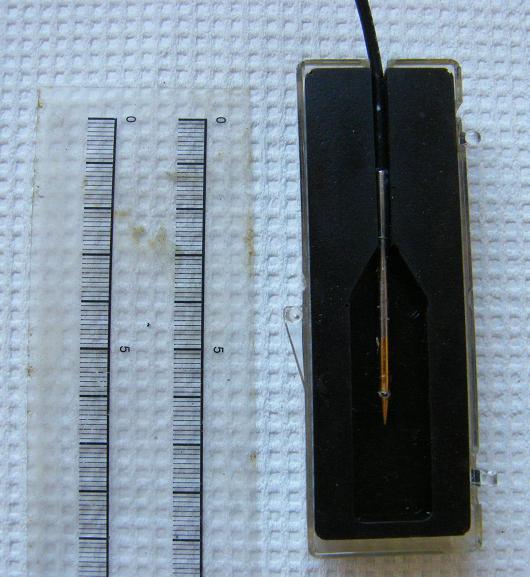}
		\end{center}
   \end{minipage} \hfill
   \begin{minipage}[c]{.5\linewidth}
		\begin{center}
	   		\includegraphics[width=1\linewidth]{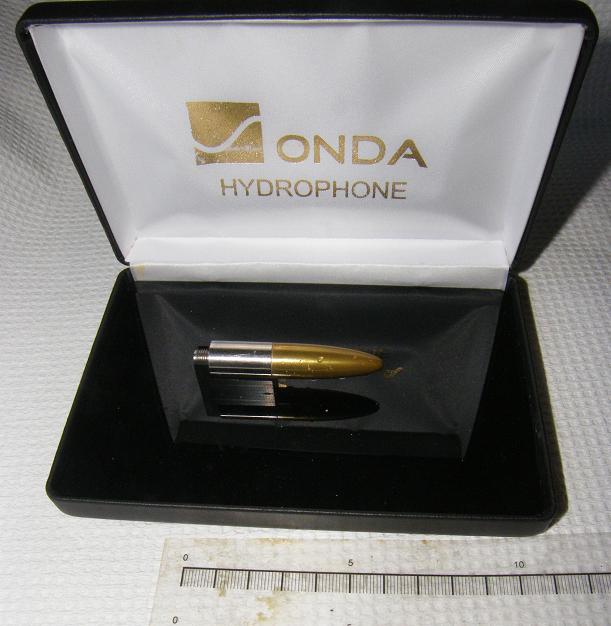}
		\end{center}
   \end{minipage}
 	\caption{(a) Photographie d'un hydrophone piézoélectrique aiguille (PZT-Z44-0400-H816, Speciality  Engineering  Associates, Sunnyvale, CA, USA). (b) Photographie d'un hydrophone piézoélectrique de type ``Lipstick'' (HGL-200-1113, Onda Corporation, Sunnyvale, CA, USA). }
 	\label{figHydroEMIntroPiezo} 
\end{figure}

Les hydrophones à aiguilles sont généralement plus fins que les hydrophones à membrane, leur diamètre pouvant atteindre 0,04 mm \cite{retat2011}. Cela permet d'avoir une meilleure résolution spatiale au prix d'une sensibilité moindre. La partie sensible est placée à la pointe de l'aiguille, ce qui permet de peu perturber l'onde acoustique. Un hydrophone à aiguille de type ``Lipstick'' est représenté sur la figure \ref{figHydroEMIntroPiezo}.

\subsubsection{Les hydrophones à fibre optique}
Plusieurs designs d'hydrophones utilisant des fibres optiques existent.

Un premier type soumet une portion d'une fibre optique à l'onde acoustique. Celle-ci modifie la propagation d'une onde lumineuse dans la fibre, comme l'indice de réfraction, un changement de polarisation... La résolution spatiale de cet hydrophone est cependant limitée, car les changements sont intégrés le long du fil soumis à l'onde acoustique \cite{flax1982}.

\begin{figure}[!ht]
		\begin{center}
    \includegraphics[width=.6\linewidth]{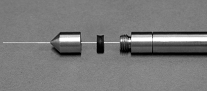}
 		\caption{Vue éclatée d'un hydrophone à fibre optique avec son support (source page \pageref{source9}).}
		\label{figHydroEMIntroFibreOptique}
 		\end{center}
\end{figure}

Un deuxième type d'hydrophone, plus fréquent, utilise une pointe de fibre optique comme élément de mesure de la pression. En effet, l'indice de réfraction d'un fluide dépend de la pression locale, on peut donc observer des changements dans la réflexion de l'onde lumineuse à la pointe de la fibre, à cause du changement d'indice à l'interface fibre/milieu \cite{klann2005}, \cite{staudenraus1993fibre}. On peut voir une vue éclatée de ce type d'hydrophone sur la figure \ref{figHydroEMIntroFibreOptique} (FOPH 2000, RP acoustics, Leutenbach, Allemagne). Ces hydrophones sont réputés extrêmement résistants, jusqu'à plusieurs centaines de mégapascals, mais manquent de sensibilité, avec une limite basse autour de 1 MPa \cite{bucaro1977}. Cela restreint l'utilisation de ces hydrophones à des signaux intenses comme celui produit par des lithotripteurs ou stables dans le temps pour effectuer des moyennes.

\subsubsection{L'imagerie Schlieren}
\begin{figure}[!ht]
		\begin{center}
    \includegraphics[width=.5\linewidth]{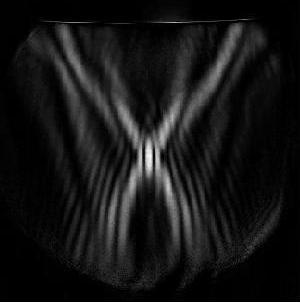}
 		\caption{Exemple de champ de pression reconstruit par un système d'imagerie Schlieren. Les zones plus claires correspondent à des pressions plus élevées (source page \pageref{source8}).}
		\label{figHydroEMIntroSchlieren}
 		\end{center}
\end{figure}

L'imagerie Schlieren est une méthode de visualisation des variations d'indice, dépendant de la pression acoustique, dans un milieu transparent en utilisant un rayon lumineux, généralement un laser \cite{hanafy1991}, \cite{korpel1987}, \cite{newman1973observations}. Les rayons lumineux émis traversent le milieu étudié, puis sont focalisés par une lentille convergente sur une fine lame opaque. Les rayons lumineux n'ayant pas été déviés par les changements d'indice du milieu sont absorbés sur la lame opaque, tandis que les autres sont détectés par un capteur optique. Cette technique, qui donne des images telles que celle sur la figure \ref{figHydroEMIntroSchlieren}, permet de visualiser des champs de pression rapidement, mais nécessite un milieu transparent. Cette technique permet en théorie d'avoir des mesures quantitatives \cite{schneider1996quantitative}, mais ne donne des résultats fiables que pour des basses pressions et des difficultés subsistent dans la fabrication du système \cite{pitts2000three}, \cite{pitts2001optical}, \cite{pitts2002optical}.

\subsection{Les réflecteurs ponctuels}
Il est également possible de mesurer des propriétés acoustiques d'un champ ultrasonore avec la diffusion des ultrasons par un obstacle. Cet obstacle est généralement une sphère métallique placée à la pointe d'un support \cite{bernier1995practical}. Les ultrasons diffusés sont alors observés avec un transducteur externe, plus rarement le transducteur d'émission. La mesure est cependant relative, car le transducteur ne capte qu'une faible partie de l'onde diffusée. L'un des facteurs limitants de cette technique est le support de la boule, qui ne doit pas perturber le champ acoustique, et la taille de la boule. Un fil long et fin a donc été proposé comme obstacle \cite{raum1997}, où ce fil est déplacé en translation et rotation pour reconstruire le champ complet par tomographie.

\subsection{Les mesures d'intensité acoustique}

\subsubsection{Les sondes thermiques}
L'un des premiers systèmes de mesure acoustique était simplement constitué d'un thermocouple, avec des premières évaluations autour de 1950 \cite{fry1954determination1}, \cite{fry1954determination2}. La température augmente en effet jusqu'à ce qu'il y ait équilibre entre les pertes de chaleur et l'absorption d'énergie par le thermocouple. Cet équilibre est généralement atteint en moins d'une seconde \cite{fry1954determination2}. Cette température d'équilibre permet alors d'obtenir l'intensité ultrasonore -- les pertes de chaleur sont proportionnelles à la différence entre la température de la sonde et celle du milieu \cite{martin1980use}. Cette technique est utilisée à la fois pour des mesures ponctuelles ou pour des mesures volumiques.

\subsubsection{Les balances acoustiques}
Lorsqu'un obstacle est placé dans la direction de propagation des ultrasons, il va subir une force de radiation acoustique. L'évaluation de cette force permet de calculer l'intensité acoustique. Ainsi, il existe un système nommé ``balance acoustique'' composé d'un absorbant ultrasonore suspendu à une balance ; lorsque des ultrasons sont émis sur l'absorbant, ce dernier va être soumis à la fois à son poids et à la force de radiation acoustique : la balance indiquera alors une masse différente. Cela permet d'en déduire la force sur l'absorbant, et donc l'intensité ultrasonore \cite{preston1991}, \cite{davidson1991ultrasonic}. Cependant, cette technique ne permet d'obtenir qu'une intensité moyenne au cours du temps et sur la surface de la masselotte : la résolution spatiale et temporelle est donc en pratique faible \cite{hodnett1997strategy}. Dans le même esprit, un type de balance acoustique mesure le changement de poids d'un liquide absorbant, lié à son chauffage par ultrasons \cite{shaw2008buoyancy}.

\begin{figure}[!ht]
		\begin{center}
    \includegraphics[width=.5\linewidth]{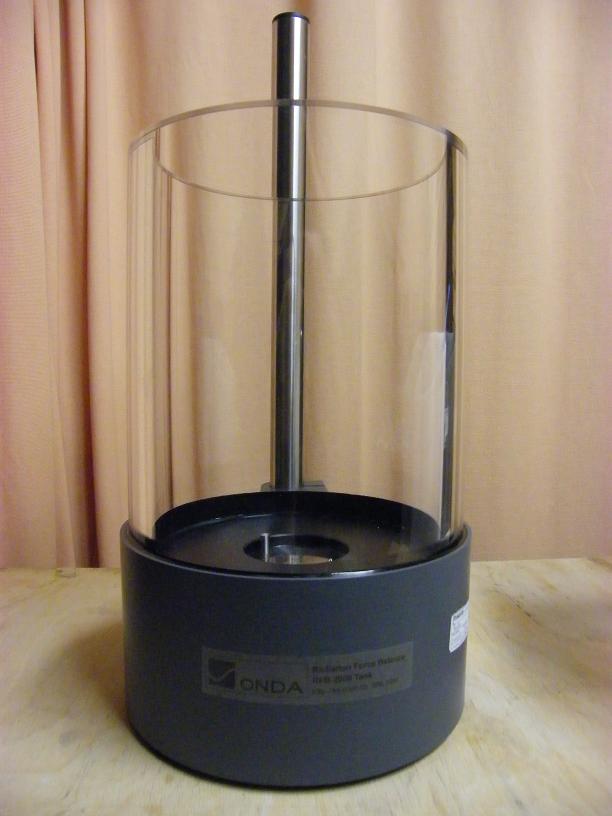}
 		\caption{Photographie d'une balance acoustique. Un transducteur est placé dans le tube et émet des ultrasons sur l'absorbant placé en bas. La force de radiation acoustique est calculée grâce au changement de poids apparent de l'absorbant (RFB-2000 Radiation Force Balance, Onda Corporation, Sunnyvale, CA, USA)}
		\label{figHydroEMIntroBalance}
 		\end{center}
\end{figure}

\subsection{Les mesures de déplacement ou de vitesse particulaire}

\subsubsection{Les interféromètres}
Un interféromètre de Fabry-Pérot peut être utilisé comme hydrophone. En effet, l'onde acoustique peut modifier le chemin optique d'une cavité de Fabry-Pérot, ce qui module la phase des signaux. Ce changement de phase est détectable par interférométrie, et on peut donc en déduire le déplacement particulaire à la surface de la cavité \cite{cox2007}.

Il est également possible d'utiliser un interféromètre de Michelson. Dans cet interféromètre, une membrane servant de miroir acoustique semi-réfléchissant est placée dans le champ acoustique. Une technique d'interféromètrie mesure alors les déplacements à la surface de la membrane, ce qui permet d'en déduire la vitesse des particules \cite{preston1991}.

\subsubsection{Les hydrophones à force de Lorentz}
\begin{figure}[!ht]
   \begin{minipage}[c]{.5\linewidth}
		\begin{center}
	   		\includegraphics[width=1\linewidth]{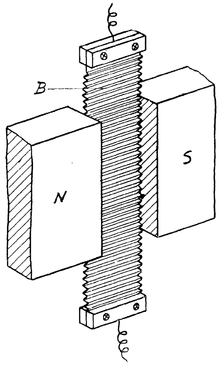}
		\end{center}
   \end{minipage} \hfill
   \begin{minipage}[c]{.5\linewidth}
		\begin{center}
	   		\includegraphics[width=1\linewidth]{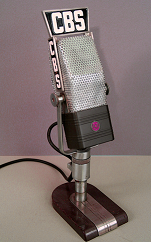}
		\end{center}
   \end{minipage}
 	\caption{(a) Principe d'un microphone à ruban : un ruban plissé est placé au milieu d'un aimant. Le son fait vibrer le ruban dans le champ magnétique, ce qui induit par force de Lorentz un courant électrique mesurable (source page \pageref{source4}). (b) Photographie d'un microphone à ruban (source page \pageref{source3}).}
 	\label{figHydroEMIntroHydroEM} 
\end{figure}

L'utilisation de la force de Lorentz pour détecter des ondes acoustiques par mesure de courant électrique date du début du XXe siècle, avec par exemple le microphone à ruban présenté en 1929 par H. Olson \cite{olson1931} représenté sur la figure \ref{figHydroEMIntroHydroEM}. Ce microphone est constitué d'un ruban conducteur placé dans l'entrefer d'un aimant. Lorsqu'un son arrive sur le ruban, celui-ci vibre dans un champ magnétique, ce qui induit par force de Lorentz un courant électrique. Ce type de microphone est réputé pour la qualité du son rendu, grâce à une large bande passante, mais des bruits secs et intenses peuvent déchirer la membrane. Cet appareil a donc laissé la place à d'autres microphones, même si l'on constate un retour depuis quelques années notamment grâce à l'amélioration de la solidité de la membrane. Ce microphone n'est cependant pas adapté à la cartographie de champs de pression acoustiques, notamment car la résolution spatiale serait faible.

Plus tard, en 1969, L. Filipczynski a présenté le principe d'un hydrophone électromagnétique \cite{filipczynski1969}. Celui-ci est constitué d'un absorbant ultrasonore sur lequel est fixé un fil électrique, le tout placé dans le champ magnétique d'un aimant. Lorsque des ultrasons pénètrent dans l'absorbant, l'interface entre le milieu et l'absorbant vibre, et notamment le fil fixé dessus. Comme le fil vibre dans un champ magnétique, cela induit un courant électrique. Contrairement au microphone à ruban, cet hydrophone est très résistant -- des tests avec pressions de 30 MPa ne l'ont pas endommagé \cite{etienne1997}. La résolution spatiale est cependant limitée par la longueur du fil.

\begin{figure}[!ht]
		\begin{center}
    \includegraphics[width=.8\linewidth]{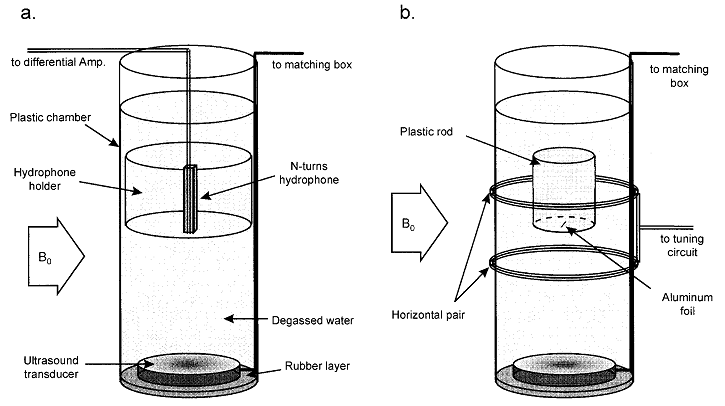}
 		\caption{Deux dispositifs d'hydrophone à force de Lorentz proposés par Sharf et al. (a) Dispositif avec un absorbant et un fil fixé autour. (b) Dispositif avec un morceau de fil fixé à un absorbant et une bobine réceptrice \cite{sharf1999}.}
		\label{figHydroEMIntroHydroEM2}
 		\end{center}
\end{figure}

En 1997, Y. Sharf a amélioré le dispositif en proposant de placer un fil court sur l'absorbant sans être relié à un appareil de mesure \cite{sharf1999}, comme représenté sur le schéma \ref{figHydroEMIntroHydroEM2}. Le courant électrique est récupéré par induction magnétique, grâce à une bobine placée autour du fil. L'intérêt est de pouvoir raccourcir la longueur du fil fixé à l'absorbant, en améliorant par là la résolution spatiale. Mais cela se fait au détriment de la sensibilité de la mesure.

L'hydrophone à force de Lorentz étudié dans cette thèse est un dispositif convertissant le mouvement du fluide dû aux ultrasons en courant électrique par force de Lorentz. A la différence des hydrophones présentés ci-dessus, il utilise un fil long et n'a pas d'absorbant. L'hydrophone est composé d'un ou plusieurs aimants créant un champ magnétique homogène dans lequel est placé un fil électrique. Lorsque les ultrasons mettent en mouvement le milieu, cela déplace par frottement le fil -- si ce dernier est suffisamment fin, il perturbe peu le champ acoustique. Le mouvement du milieu déplace par frottements le fil, ce qui induit un courant électrique dans celui-ci. Une bonne résolution spatiale est alors atteinte en effectuant des rotations et des translations du fil, ce qui permet de réaliser une tomographie du champ acoustique.

\subsection{Résumé}
Les caractéristiques de l'ensemble des techniques vues ci-dessus sont résumées sur le tableau \ref{tableHydroEMIntroRésumé}. On observe que le principal défaut des hydrophones à force de Lorentz par rapport aux autres techniques est sa mauvaise résolution spatiale.

\begin{table}
	\begin{tabular}{|p{.18\linewidth}|p{.15\linewidth}|p{.14\linewidth}|p{.11\linewidth}|p{.22\linewidth}|p{.05\linewidth}|}
		\hline
		\rowcolor{SandyBrown} Type & Paramètre \mbox{mesuré} & Résolution spatiale & Résistance à la \mbox{cavitation} & Commentaires & Réf. \\
		\hline
		Hydrophone piézoélectrique & Pression acoustique & 40 \micro m & - & & \cite{retat2011}, \cite{hodnett1997strategy} \\
		\hline
		Hydrophone à fibre \mbox{optique} & Pression acoustique & 10 \micro m & + & Pression minimale de quelque centaines de kPa & \cite{hodnett1997strategy}, \cite{klann2005}, \cite{staudenraus1993fibre}\\
		\hline
		Imagerie Schlieren & Pression acoustique & Taille du faisceau lumineux & +++  & Utilise une technique de reconstruction tomographique ; quantitatif à basse pression seulement & \cite{schneider1996quantitative}, \cite{pitts2000three}, \cite{pitts2001optical}, \cite{pitts2002optical}\\
		\hline
		Réflecteurs ponctuels & Pression acoustique réfléchie & 500 \micro m & 0 & Mesure relative ; Meilleure résolution spatiale avec une reconstruction tomographique & \cite{anand2004monitoring}, \cite{bernier1995practical}, \cite{raum1997} \\
		\hline
		Sonde \mbox{thermique} & Intensité acoustique & 10 \micro m & - & Peu cher ef facile à \mbox{utiliser} & \cite{fry1954determination1}, \cite{fry1954determination2}, \cite{martin1980use}\\
		\hline
		Balance \mbox{acoustique} & Intensité acoustique & Taille de l'absorbant & 0 & Intensité moyenne au cours du temps & \cite{preston1991}, \cite{davidson1991ultrasonic}\\
		\hline
		Interféromètre & Déplacement ou vitesse particulaire & 50 \micro m & - & & \cite{cox2007}, \cite{preston1991} \\
		\hline
		Hydrophone à force de Lorentz & Déplacement ou vitesse particulaire & 5 mm & + & Résolution spatiale dépendante de la direction & \cite{filipczynski1969}, \cite{etienne1997}, \cite{sharf1999} \\
		\hline
	\end{tabular}
	\caption{Caractéristiques des techniques de mesure des paramètres acoustiques présenté dans la section. Les symboles -, 0 et + indiquent respectivement que la cavitation endommage facilement, modérément ou difficilement l'appareil de mesure. Le système d'imagerie Schlieren y est insensible.}
	\label{tableHydroEMIntroRésumé}
\end{table}

\section{Principe de l'hydrophone à force de Lorentz}
\subsection{Expression de la force de Lorentz}
\begin{figure}[!ht]
		\begin{center}
    \includegraphics[width=.6\linewidth]{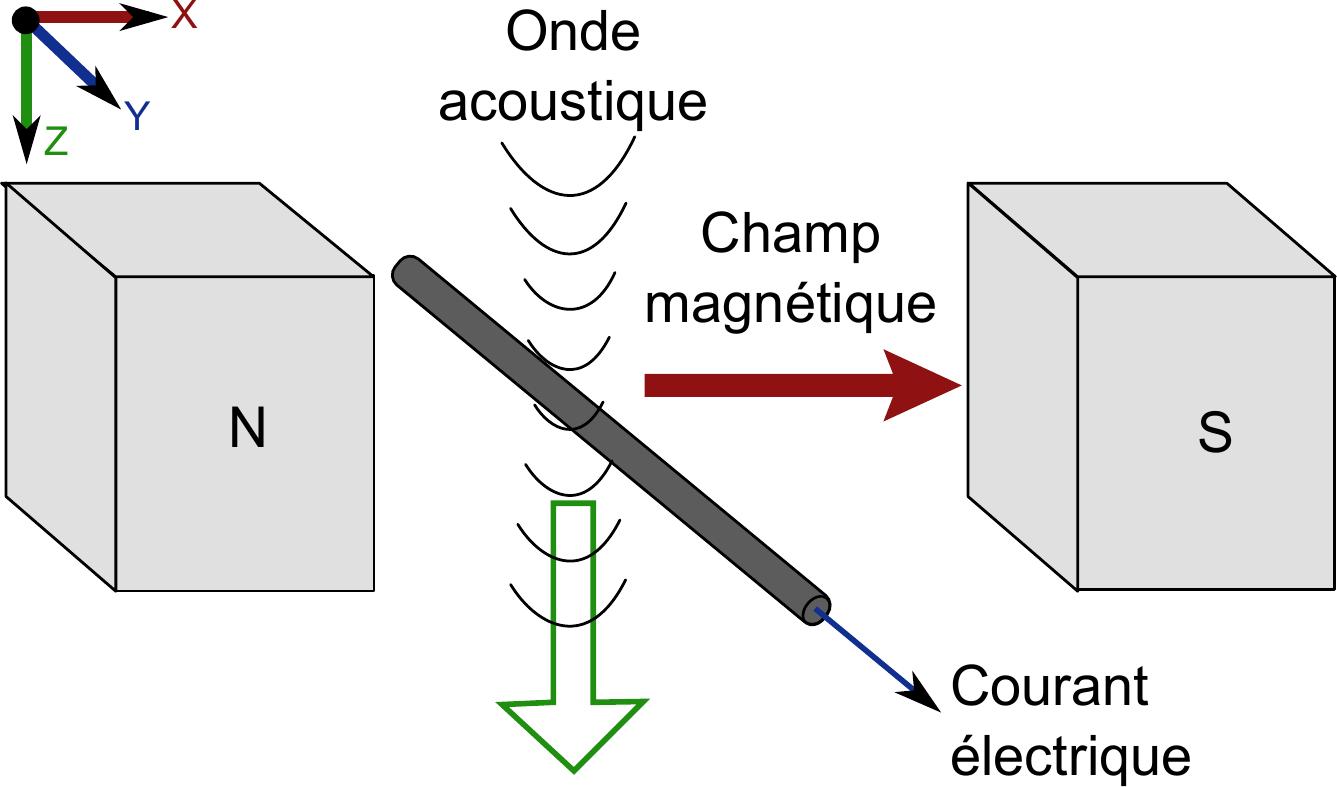}
 		\caption{\label{figHydroEMTheoriePrincipe} Principe d'un hydrophone à force de Lorentz : un fil électrique plongé dans un champ magnétique vibre sous l'effet d'une onde ultrasonore, ce qui induit un courant électrique.}
 		\end{center}
\end{figure}

On suppose dans cette partie que le champ magnétique est dans la direction X, le fil électrique dans la direction Y et l'axe ultrasonore dans la direction Z, comme représenté sur le schéma \ref{figHydroEMTheoriePrincipe}.

Le mouvement d'un conducteur dans un champ magnétique induit un courant électrique perpendiculaire au mouvement et au champ magnétique. Ainsi, pour un conducteur de longueur $L$, la force électromotrice induite $e$ est proportionnelle à la vitesse $v$ du conducteur multipliée par le champ magnétique $B$, intégrée le long du conducteur, ce qui s'exprime :

\begin{empheq}[box=\mybluebox]{align}
			e = \int_L{(\mathbf{v} \wedge \mathbf{B}) . \mathbf{dl}}
	\label{eqHydroEMTheorie1}
\end{empheq}

Si le fil est isolé, ou si la conductivité électrique du milieu est suffisamment faible comparée à celle du fil, les pertes de courant du fil au milieu environnant peuvent être négligées.

Le déplacement induisant un courant électrique provient de (a) la vibration interne des charges dans le fil et (b) de la vibration du fil lui-même à cause des ultrasons. Pour un fil de diamètre $D$ = 200 \micro m, à une longueur d'onde ultrasonore $\lambda$ = 1,5 mm (correspondant à une fréquence de 1 MHz) avec une vitesse du son égale à 1500 m.s$^{-1}$, on considère que le diamètre du fil est négligeable face à la longueur d'onde. La vitesse du son étant de 5000 m.s$^{-1}$ environ dans le fil, la variation de pression à l'intérieur du fil est par conséquence petite et toutes les charges peuvent être supposées se déplaçant ensemble. Dans le cas présenté ici, le phénomène (a) est donc négligeable et le signal provient quasi exclusivement du phénomène (b). La section suivante s'attachera donc à déterminer la vitesse du fil en fonction de la vitesse ultrasonore.

\subsection{Lien entre vitesse du fluide et vitesse du fil}
La relation entre la vitesse du fluide $u_0$ et la celle du fil $v$ peut être évaluée théoriquement \cite{landau1987}. Dans la démonstration suivante, le fil est modélisé comme un cylindre infiniment long. La démonstration se fait dans un plan 2D perpendiculaire au fil et toutes les quantités physiques comme la force et l'énergie doivent être considérées par unité de longueur.
Pour une onde plane, la pression $p$ et la vitesse du fluide $u_0$ sont proportionnels :
\begin{empheq}[box=\mybluebox]{align}
			u_0 = \frac{p}{\rho c}
	\label{eqHydroEMTheorieLien1}
\end{empheq}
avec $\rho$ la densité du fluide, $c$ la vitesse du son.

En prenant p = 1 MPa (pression typique dans la zone focale des ultrasons en échographie), $\rho$ = 1000 kg.m$^{-3}$ (densité de l'eau) et $c$ = 1480 m.s$^{-1}$ (vitesse du son dans de l'eau douce à 20 $^o$C), la vitesse du fluide $u_0$ vaut approximativement 0,7 m.s$^{-1}$.

Pour évaluer l'importance des forces inertielles par rapport aux forces visqueuses, on peut utiliser le nombre de Reynolds $Re$ qui est égal au rapport entre ces deux forces. Dans le cas du cylindre infini, le nombre de Reynolds vaut
\begin{empheq}[box=\mybluebox]{align}
			Re=\frac{u_0 D}{\nu}
	\label{eqHydroEMTheorieLien2}
\end{empheq}
avec $D$ le diamètre du cylindre et $\nu$ la viscosité cinématique du milieu. En prenant $D$ = 200 \micro m, $\nu$ = $1.10^{-6}$ m$^2$.s$^{-1}$ (eau à 20 $^o$C) et $u_0$ = 0,7 m.s$^{-1}$ comme dans le cas précédent, le nombre de Reynolds vaut 180. Cette valeur indique que les forces visqueues sont négligeables, contrairerement à celles des forces de pression, et le fluide peut par conséquent être considéré comme parfait.

D'autre part, si le cylindre est petit comparé à la longueur d'onde ultrasonore et que les turbulences sont faibles au cours du temps (compatible avec la valeur de $Re$ inférieure à 1000), le fluide loin du cylindre est considéré irrotationnel avec une vitesse constante $u_0$. Si les dimensions typiques des variations de pression (la longueur d'onde ultrasonore $\lambda$) sont bien plus grandes que les dimensions de l'obstacle (le diamètre D), le fluide peut de plus être considéré comme incompressible. Sous ces conditions, dans un référentiel où le fluide loin du cylindre est au repos, la vitesse du fluide $\mathbf{\tilde{u}}$ est égale au gradient d'un potentiel $\phi$ solution de l'équation de Laplace $\Delta \phi = 0$.

Les solutions pour $\phi$ satisfaisant les conditions aux limites sont les dérivées de ln($\mathbf{r}$), où $\mathbf{r}$ est le vecteur orthogonal à l'axe du cylindre, $R_0$ le rayon du cylindre et $\mathbf{\tilde{v}}$ la vitesse du cylindre dans le référentiel considéré. On peut donc écrire $\phi$ sous la forme :
\begin{empheq}[box=\mybluebox]{align}
			\phi = -\frac{R_0^2}{r^2} \mathbf{\tilde{v}}.\mathbf{r}
	\label{eqHydroEMTheorieLien4}
\end{empheq}

Le gradient de potentiel $\mathbf{\tilde{u}}$ s'écrit alors :
\begin{empheq}[box=\mybluebox]{align}
			\mathbf{\tilde{u}} = \grad \phi = \frac{R_0^2}{r^2} (\frac{2\mathbf{r}(\mathbf{\tilde{v}}.\mathbf{r})}{r^2} - \mathbf{\tilde{v}})
	\label{eqHydroEMTheorieLien5}
\end{empheq}

Une relation plus pratique à utiliser entre $\mathbf{\tilde{u}}$ et $\mathbf{\tilde{v}}$ peut être trouvée en utilisant l'énergie cinétique totale du fluide $E_c$, simplement égale à :
\begin{empheq}[box=\mybluebox]{align}
			E_c=\frac{\rho}{2}\int{u^2 dV}
	\label{eqHydroEMTheorieLien6}
\end{empheq}

Le domaine d'intégration de $E_c$ est l'ensemble de l'espace considéré cylindrique de rayon $R$ avec $R$ tendant vers l'infini, moins le volume du cylindre, de rayon $R_0$. On peut de plus réécrire $\tilde{u}^2$ en $\tilde{v}^2 + \tilde{u}^2 - \tilde{v}^2$, ce qui donne pour l'énergie cinétique totale :
\begin{empheq}[box=\mybluebox]{align}
			E_c=\frac{\rho}{2}\int{\tilde{v}^2 dV}+\frac{\rho}{2}\int{(\mathbf{\tilde{u}}+\mathbf{\tilde{v}})(\mathbf{\tilde{u}}-\mathbf{\tilde{v}}) dV}
	\label{eqHydroEMTheorieLien7}
\end{empheq}

Dans la seconde intégrale, on peut remplacer $(\mathbf{\tilde{u}}+\mathbf{\tilde{v}})$ par $\grad (\phi + \mathbf{\tilde{v}}.\mathbf{r})$. En utilisant les équations de continuité $\Div\mathbf{\tilde{u}}=0$ et $\Div \mathbf{\tilde{v}}=0$, l'énergie cinétique totale peut être réécrite :
\begin{empheq}[box=\mybluebox]{align}
			E_c=\frac{\rho}{2}\int{\tilde{v}^2 dV}+\frac{\rho}{2}\int{\Div ((\phi + \mathbf{\tilde{v}}.\mathbf{r})(\mathbf{\tilde{u}}-\mathbf{\tilde{v}})) dV}
	\label{eqHydroEMTheorieLien8}
\end{empheq}

La première intégrale est simplement égale à $\frac{\rho}{2}\tilde{v}^2(\pi R^2 - \pi R_0^2$). En utilisant le théorème de Green-Ostrogradsky, la seconde intégrale peut être remplacée par l'intégrale de $(\phi + \mathbf{\tilde{v}}.\mathbf{r})(\mathbf{\tilde{u}}-\mathbf{\tilde{v}}).\mathbf{n}$ sur la surface des cylindres de rayons $R$ et $R_0$. D'après les équations de continuité, $\mathbf{\tilde{v}}.\mathbf{n}=\mathbf{\tilde{u}}.\mathbf{n}$ à la surface du cylindre de rayon $R_0$, ce qui donne une intégrale nulle. Après avoir remplacé $\mathbf{\tilde{u}}$ et $\phi$ par leurs expressions en \ref{eqHydroEMTheorieLien4} et \ref{eqHydroEMTheorieLien5} et en négligeant les termes tendant vers 0 lorsque $R$ tend vers l'infini, on trouve finalement pour l'énergie cinétique
\begin{empheq}[box=\mybluebox]{align}
			E_c=\frac{\rho}{2}\pi R_0^2 \tilde{v}^2
	\label{eqHydroEMTheorieLien9}
\end{empheq}

En appliquant le théorème de la puissance cinétique, on peut calculer une force de trainée $\mathbf{F}$ définie à partir de l'énergie cinétique :
\begin{empheq}[box=\mybluebox]{align}
			E_c=-\mathbf{F}.\mathbf{\tilde{v}} = -(\rho \pi R_0^2 \frac{d\mathbf{\tilde{v}}}{dt}).\mathbf{\tilde{v}}
	\label{eqHydroEMTheorieLien10}
\end{empheq}

Dans le référentiel du laboratoire, on peut remplacer $\mathbf{\tilde{v}}$ par $\mathbf{\tilde{v}}-\mathbf{\tilde{u_0}}$ avec $\mathbf{\tilde{v}}$ la vitesse du cylindre et $\mathbf{\tilde{u}_0}$ la vitesse du fluide loin du cylindre. $\mathbf{F}$ est alors égal à :
\begin{empheq}[box=\mybluebox]{align}
	\label{eqHydroEMTheorieLien11}
			\mathbf{F} = \rho \pi R_0^2 (\frac{d\mathbf{\tilde{u}_0}}{dt} - \frac{d\mathbf{\tilde{v}}}{dt})
\end{empheq}

La deuxième loi de Newton énonce que l'accélération du cylindre est égale à la somme de la force de trainée et de la force reliée au déplacement du volume de fluide égal à celui occupé par le cylindre, ce qui peut s'exprimer sous la forme :
\begin{empheq}[box=\mybluebox]{align}
	\begin{array}{rcl}
			\rho_0 \pi R_0^2 \frac{d\mathbf{\tilde{v}}}{dt} &=& \rho \pi R_0^2 (\frac{d\mathbf{\tilde{u}_0}}{dt}-\frac{d\mathbf{\tilde{v}}}{dt}) + \rho \pi R_0^2 \frac{d\mathbf{\tilde{u}_0}}{dt} \\
			\Rightarrow (\rho_0 + \rho) \pi R_0^2 \frac{d\mathbf{\tilde{v}}}{dt} &=& 2 \rho \pi R_0^2 \frac{d\mathbf{\tilde{u}_0}}{dt}
	\end{array}
	\label{eqHydroEMTheorieLien12}
\end{empheq}

Après intégration par rapport au temps, on trouve finalement la relation \ref{eqHydroEMTheorieLien13} entre la vitesse du fluide $\mathbf{\tilde{u}_0}$ et celle du cylindre $\mathbf{\tilde{v}}$ :
\begin{empheq}[box=\mybluebox]{align}
			\mathbf{\tilde{v}}= K(\rho,\rho_0) \mathbf{\tilde{u}_0}
	\label{eqHydroEMTheorieLien13}
\end{empheq}
avec $K(\rho,\rho_0) = \frac{2\rho}{\rho_0 + \rho}$. Cette relation, remarquablement simple, ne dépend que de la densité du fluide et du cylindre. Remarquons que si le cylindre et le fluide ont la même densité, ils se déplacent à la même vitesse.

\subsection{Reconstruction tomographique}
Dans le principe de l'hydrophone à force de Lorentz, le signal induit par force de Lorentz est créé tout le long du fil, donc la mesure n'est pas ponctuelle. Pour réaliser des champs ultrasonores, on peut réaliser une tomographie, grâce à des rotations et translations successives du fil.

La transformée de Fourier de l'équation \ref{eqHydroEMTheorie1} s'écrit :
\begin{empheq}[box=\mybluebox]{align}
			E(k,\theta)=\int{v_z B_{\delta}e^{-j2\pi \delta k} d\delta}
	\label{eqHydroEMTheorieTomo1}
\end{empheq}
avec $\delta$ la coordonnée le long de la direction $k$ d'angle $\theta$ avec l'axe X, $k$ étant défini par $k=x\cos \theta + y\sin \theta$:

La vitesse ultrasonore en chaque point est alors calculée en prenant la transformée de Fourier inverse à deux dimensions de l'équation \ref{eqHydroEMTheorieTomo1}, ce qui permet de donner la vitesse ultrasonore en chaque coordonnée $(x,y$ :
\begin{empheq}[box=\mybluebox]{align}
			v(x,y)=\frac{1}{B}\int_0^{\pi}{\int_{-\infty}^{+\infty}{E(k,\theta) \left|k\right|e^{j2\pi \delta k}d\theta dk}}
	\label{eqHydroEMTheorieTomo2}
\end{empheq}
La mesure du courant le long de directions $k$ pour différents angles $\theta$ donne donc une reconstruction du champ de vitesses : l'hydrophone à force de Lorentz mesure donc la vitesse (et non la pression) ultrasonore. Cet hydrophone pourrait mener à l'intensité ultrasonore s'il est combiné avec un hydrophone mesurant la pression ultrasonore \cite{shaw2008buoyancy}.

Cependant, lorsque l'hypothèse d'onde plane peut être fait, la pression et la vitesse du fluide sont proportionnelles. Cela permet de calculer la pression ultrasonore en chaque coordonnée $(x,y)$:
\begin{empheq}[box=\mybluebox]{align}
			p(x,y)=\frac{\rho c}{B}\int_0^{\pi}{\int_{-\infty}^{+\infty}{E(k,\theta) \left|k\right|e^{j2\pi \delta k}d\theta dk}}
	\label{eqHydroEMTheorieTomo3}
\end{empheq}

En pratique, le nombre de translations donnera la résolution spatiale de l'image tandis que la qualité de la reconstruction sera reliée au nombre de rotations.

\subsection{Résonance des fils}
La partie sensible de l'hydrophone est donc un fil électrique maintenu à ses extrémités. On peut chercher la fréquence de résonance mécanique de ce fil en le modélisant comme une corde de guitare.
Les fréquences de résonance $f_n$ sont alors définies par :
\begin{empheq}[box=\mybluebox]{align}
	\label{eqHydroEMTheorieResonance1}
	f_n = n\frac{c}{2L} = n\frac{1}{2L} \sqrt{\frac{T}{w}}
\end{empheq}
avec $c$ la vitesse du son dans le fil, $T$ la tension, $w$ la masse par unité de longueur, $L$ la longueur totale de fil et $n$ l'harmonique considérée.

En prenant une contrainte de 220 MPa (contrainte maximale avant rupture pour le cuivre) pour une surface de $S$ = 4 $10^{-9}$ m$^{-2}$ pour un fil de 70 \micro m de diamètre, ce qui donne une tension $T$ = 0,9 N, $w$ = 1,3 $10^{-4}$ kg.m$^{-1}$ et $L$ = 15 cm, la fréquence de résonance mécanique $f_0$ est égale à 300 Hz. Les harmoniques non négligeables seront principalement en-dessous de 1000 Hz : c'est deux à trois ordres de grandeurs sous les fréquences de travail, donc la résonance mécanique n'a quasiment pas d'effet sur le signal.

\subsection{Dépendance angulaire}
\begin{figure}[!ht]
   \begin{minipage}[c]{.5\linewidth}
		\begin{center}
	   		\includegraphics[width=1\linewidth]{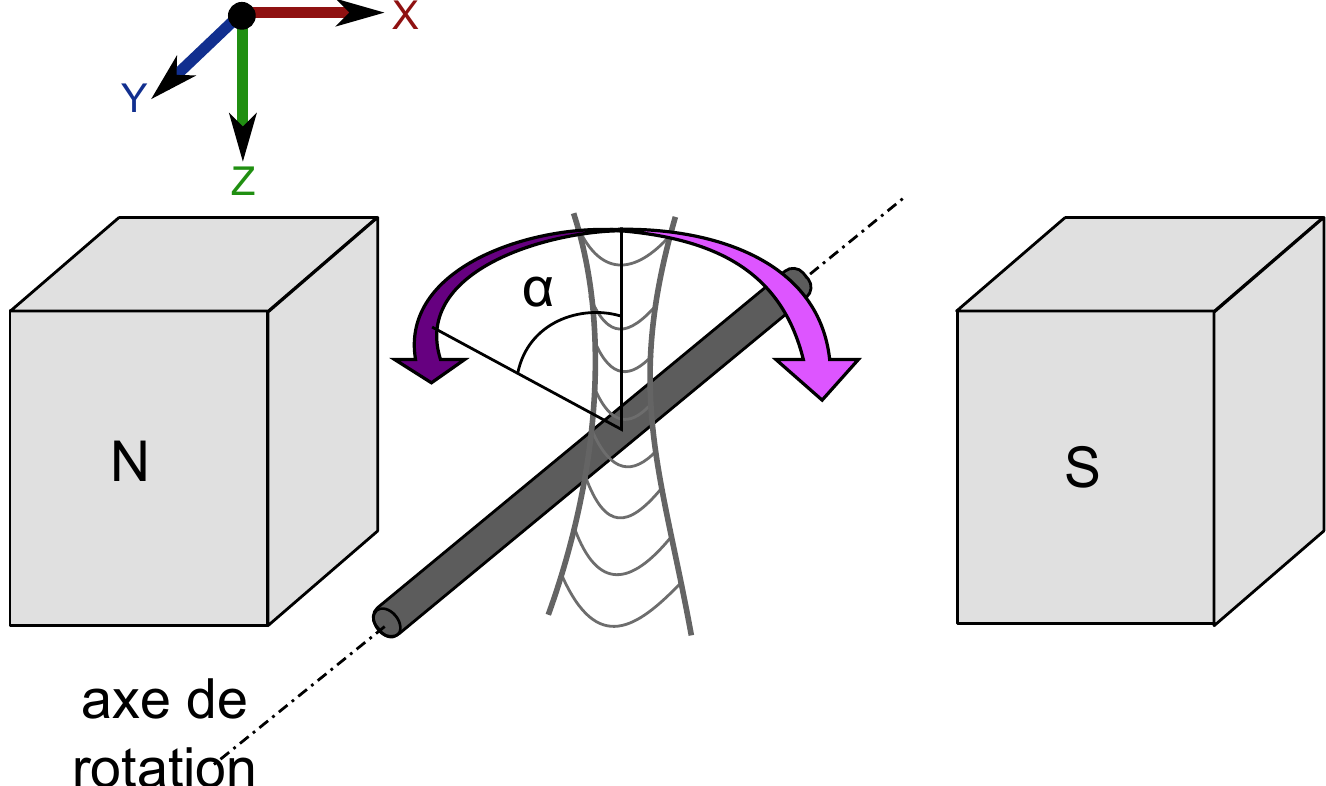}
		\end{center}
   \end{minipage} \hfill
   \begin{minipage}[c]{.5\linewidth}
		\begin{center}
	   		\includegraphics[width=1\linewidth]{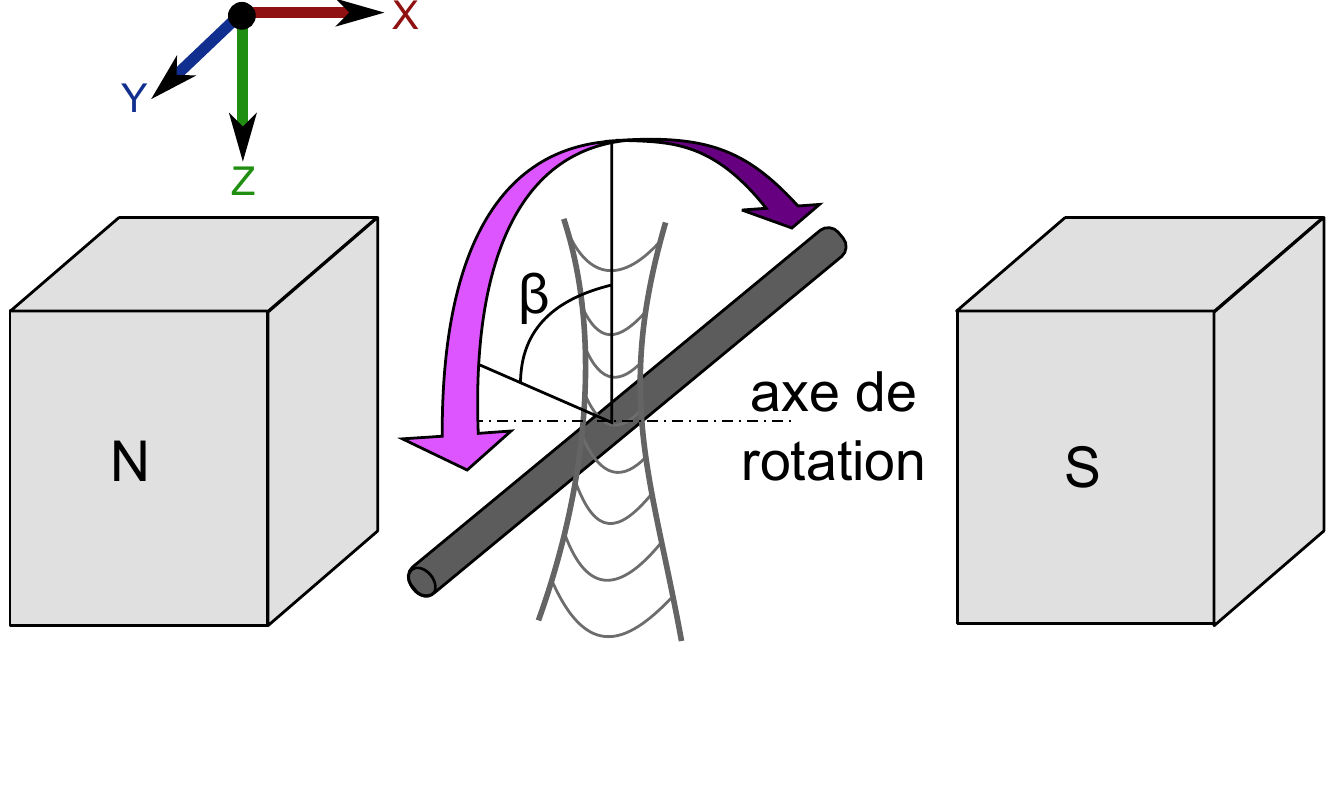}
		\end{center}
   \end{minipage}
 	\caption{(a) Réponse directionnelle avec un axe de rotation dans la direction Y. (b) Réponse directionnelle avec un axe de rotation dans la direction X.}
 	\label{figHydroEMTheorieAxeRot} 
\end{figure}

Le modèle peut être affiné en regardant la réponse directionnelle de l'hydrophone. Cette réponse directionnelle est différente si l'axe de rotation est parallèle ou perpendiculaire au fil électrique.
Nous supposerons donc dans la suite que le champ magnétique est toujours selon X, le fil selon Y, mais que les ultrasons se propagent dans le plan XZ (axe de rotation dans la direction Y) ou YZ (axe de rotation dans la direction X), comme représenté respectivement sur les figures \ref{figHydroEMTheorieAxeRot}-(a) et \ref{figHydroEMTheorieAxeRot}-(b).

Avec un axe de rotation dans la direction Y, on peut s'attendre à une dépendance sinusoidale à cause du produit vectoriel dans l'expression de la force électromotrice induite par force de Lorentz :
	\begin{empheq}[box=\mybluebox]{align}
		e = \int{(\mathbf{v} \wedge \mathbf{B}). \mathbf{dl}} = \int{v_z B \cos(\alpha)dl}
	\label{eqHydroEMTheorieDepAngle1}
	\end{empheq}
avec $\alpha$ l'angle entre l'axe de propagation des ultrasons et la direction du champ magnétique.

De même, la réponse directionnelle avec un axe de rotation dans la direction X s'écrit :
\begin{empheq}[box=\mybluebox]{align}
		e = \int{(\mathbf{v} \wedge \mathbf{B}). \mathbf{dl}} =  \int{v_z B \cos(\beta) dl}
	\label{eqHydroEMTheorieDepAngle2}
\end{empheq}
avec $\beta$ l'angle entre l'axe de propagation des ultrasons et la direction du fil.

Dans ce deuxième cas, l'amplitude mesurée ne présentera pas forcément une dépendance sinusoïdale, car le fil peut être exposé simultanément à des pics et creux de pression. En supposant que l'onde ultrasonore a une forme sinusoïdale, la vitesse du fluide $u(\xi)$ étant égale à $U_0 \cos( \frac{2\pi}{\lambda}\xi )$, avec $U_0$ la vitesse maximum du fluide et $\xi$ = $d \tan(\beta)$, $d$ étant la largeur du faisceau ultrasonore. Le calcul de l'intégrale entre 0 et L de l'équation \ref{eqHydroEMTheorieDepAngle2} donne finalement une force électromotrice :
\begin{empheq}[box=\mybluebox]{align}
	e = K \frac{U_0 B}{d} \sinc(\frac{2 \pi d}{\lambda} \tan(\beta))
	\label{eqHydroEMTheorieDepAngle3}
\end{empheq}

Il faut cependant bien garder en tête que cette dernière relation dépend de la forme du signal ultrasonore émis, et n'a donc pas un caractère général comme pour le cas où l'axe de rotation est dans la direction Y.

\section{Dispositifs testés}
L'hydrophone a été mis en \oe uvre en deux temps, avec d'abord un simple prototype servant de preuve de concept, puis un deuxième plus élaboré qui avait pour but d'être caractérisé quantitativement.

\subsection{Preuve de concept avec un simple aimant et un fil}
\begin{figure}[!ht]
   \begin{minipage}[c]{.4\linewidth}
		\begin{center}
	   		\includegraphics[width=1\linewidth]{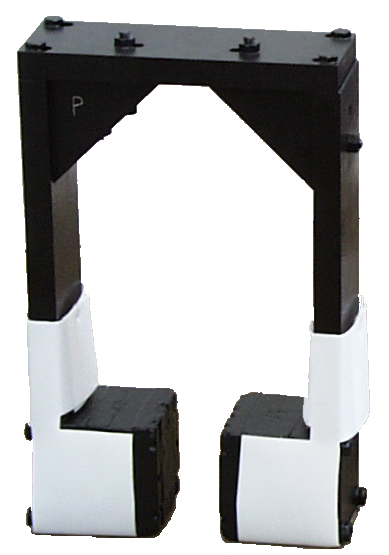}
		\end{center}
   \end{minipage} \hfill
   \begin{minipage}[c]{.6\linewidth}
		\begin{center}
	   		\includegraphics[width=1\linewidth]{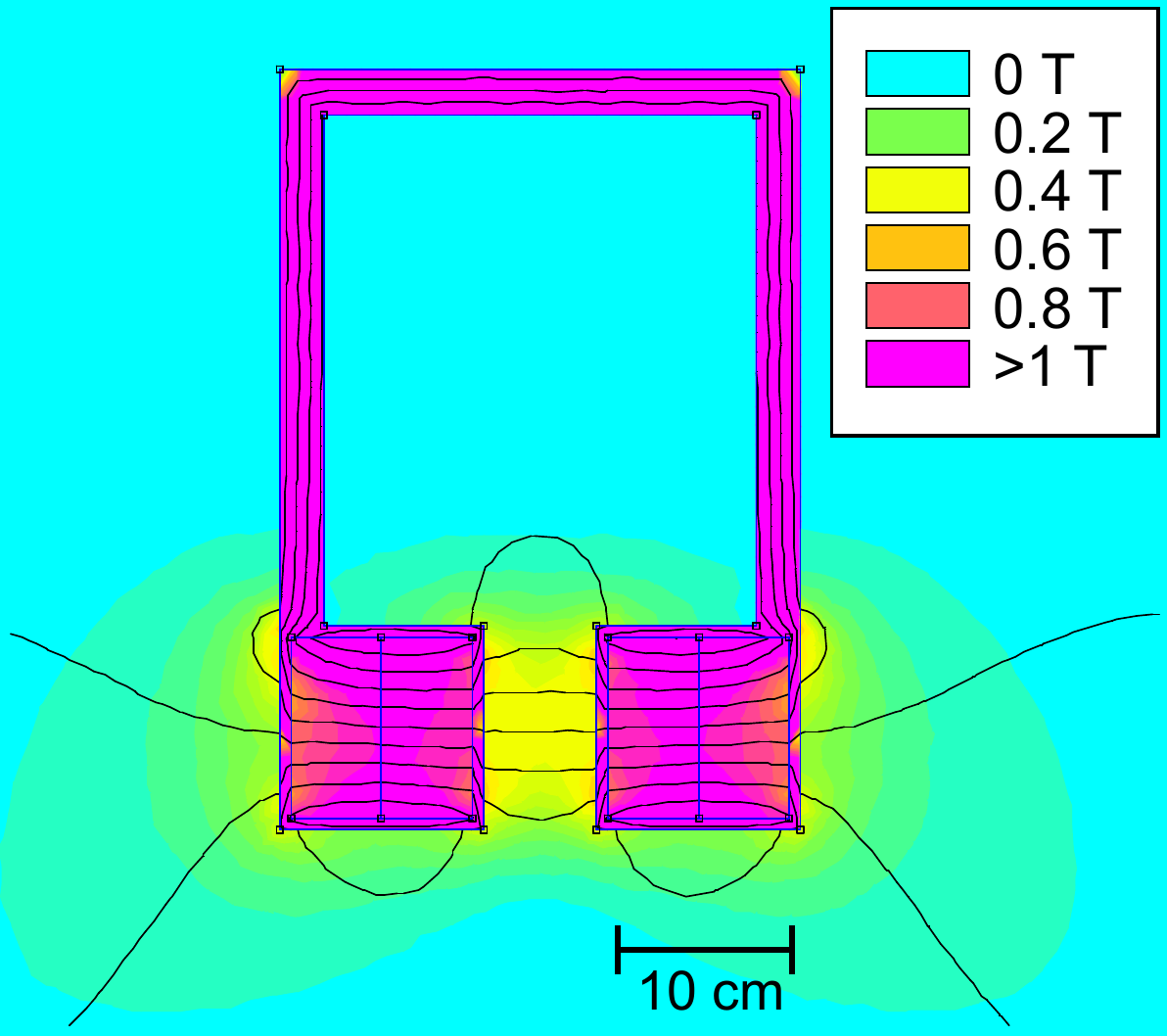}
		\end{center}
   \end{minipage}
 	\caption{(a) Photographie de l'aimant en U, composé d'une structure en acier recouvert d'un vernis protecteur, et de deux paires d'aimants NdFeB. (b) Simulation de l'intensité du champ magnétique créé par l'aimant en U avec les lignes de champ en noir. Le champ magnétique atteint 350 mT au centre des deux pôles du bas.}
 	\label{figHydroEMProto1} 
\end{figure}

\begin{figure}[!ht]
		\begin{center}
    \includegraphics[width=.3\linewidth]{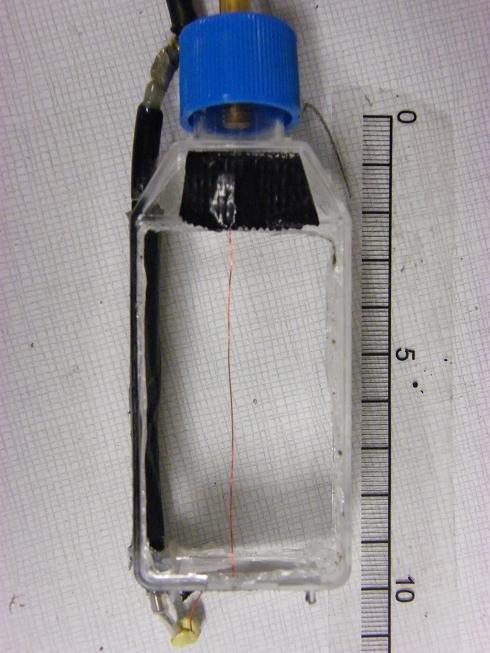}
 		\caption{\label{figHydroEMProto1Fil} Photographie du fil électrique du premier prototype.}
 		\end{center}
\end{figure}

Le premier prototype est composé d'un fil de cuivre verni de 200 \micro m de diamètre placé dans l'entrefer d'un aimant permanent en forme de U. Cet aimant, visible sur la photographie \ref{figHydroEMProto1}-(a) est composé de deux pôles, chacun formé par deux aimants NdFeB de 5x5x3 cm$^3$ (BLS Magnet, Villers la Montagne, France). L'espace entre les pôles était de 4,5 cm. Le champ magnétique était relativement inhomogène, comme le montre une simulation avec le logiciel FEMM 4.2 (figure \ref{figHydroEMProto1}-(b)), de 200 à 350 mT le long du fil. Le fil électrique, visible sur la photographie \ref{figHydroEMProto1Fil} est soudé à un câble BNC puis relié à un amplificateur de courant de 1 MV.A$^{-1}$ (HCA-2M-1M, Laser Components, Olching, Allemagne), pour pouvoir observer le signal avec un oscilloscope. Le fil de cuivre était tenu de manière lâche de sorte qu'il puisse suivre en chaque point le mouvement du fluide tout en étant suffisamment tendu pour rester droit.

Ce prototype a servi de preuve de concept pour montrer qu'il est possible de faire des mesures de champ acoustique à l'aide d'un dispositif simple composé d'un fil électrique et d'un aimant.

\subsection{Deuxième prototype : hydrophone à force de Lorentz}
\begin{figure}[!ht]
   \begin{minipage}[c]{.45\linewidth}
		\begin{center}
	   		\includegraphics[width=1\linewidth]{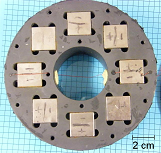}
		\end{center}
   \end{minipage} \hfill
   \begin{minipage}[c]{.55\linewidth}
		\begin{center}
	   		\includegraphics[width=1\linewidth]{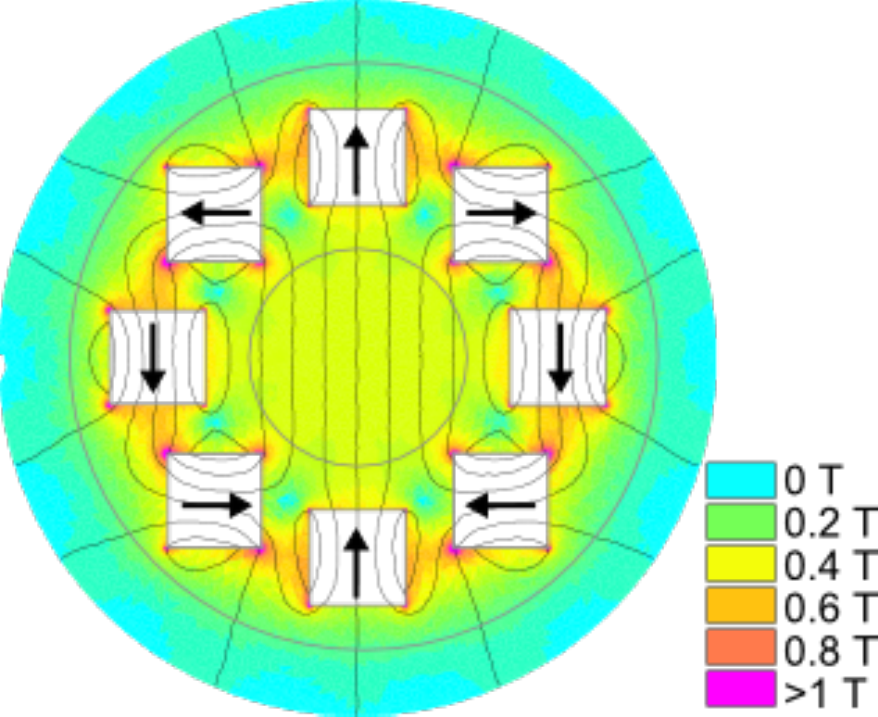}
		\end{center}
   \end{minipage}
 	\caption{(a) Photographie du prototype d'hydrophone à force de Lorentz. Cet hydrophone est composé de huit aimants placés dans un support en PVC. Lors des expériences, le fil était plus tendu, grâce à deux vis non visibles sur la photographie. (b) Simulation du champ magnétique créé par le prototype d'hydrophone à force de Lorentz avec les lignes de champ en noir et la direction des aimants indiqués par une flèche. Le champ magnétique est égal à 0,17 $\pm$ 0,03 T au centre du cylindre.}
 	\label{figHydroEMProto2} 
\end{figure}

Un nouveau prototype a ensuite été conçu avec pour objectif, entre autres, d'obtenir un champ magnétique homogène le long du fil électrique, comme visible sur la figure \ref{figHydroEMProto2}-(a). Nous avons opté pour un arrangement d'aimants dans une configuration de Halbach \cite{halbach1980}. Cette configuration comporte huit aimants permanents créant un champ magnétique de 200 mT le long du fil, comme le montre la simulation du logiciel FEMM 4.2 (figure \ref{figHydroEMProto2}-(b)). Ces aimants étaient supportés par un cylindre de PVC. Le champ magnétique a été vérifié à l'aide d'un teslamètre et s'est avéré être entre 160 et 200 mT le long du fil, le maximum étant au centre.

Un fil a été introduit dans le creux du cylindre, ressortant à travers deux vis trouées. Ces vis servaient notamment à tendre plus ou moins le fil électrique (la tension, comme il sera démontré lors d'une expérience, a une influence négligeable sur le signal). Ce fil était facilement interchangeable pour les besoins de l'expérience. Le fil avec un diamètre de 200 \micro m et était fait en cuivre verni. Il était soudé à un câble BNC puis relié à un amplificateur de courant de 1 MV.A$^{-1}$ (HCA-2M-1M, Laser Components, Olching, Germany), pour pouvoir observer le signal avec un oscilloscope.

\section{Caractérisation}
La caractérisation de l'hydrophone à force de Lorentz s'est basée sur la norme International Electrotechnical Commission 62127-3, afin d'avoir des éléments comparables avec d'autres hydrophones. Certains éléments spécifiques ont été rajoutés, comme l'influence de la tension du fil électrique, d'autre supprimés car non pertinents dans notre cas, comme la taille de la partie sensible.

Le générateur de fonctions était un AFG 3022B (Tektronix, Beaverton, OR, Etats-Unis), l'amplificateur de signal avait une puissance de 200 W (LA200H, Kalmus Engineering, Rock Hill, SC, Etats-Unis). Le transducteur ultrasonore ``principal'' a été monté au laboratoire avec une membrane piézoélectrique (piezoelectric ceramic PZ-28, Kvistgraad, Meggitt, Denmark)  d'une fréquence centrale de 0,5 MHz. Il était sphérique avec un diamètre de 50 mm et un rayon de courbure de 50 mm. La largeur du faisceau à 0,5 MHz était de 8 mm dans la zone focale.

Pour les expériences de résolution spatiale, de réponse fréquentielle et de sensibilité, les caractéristiques ont été comparées avec celles d'un hydrophone lipstick, avec une partie sensible de 200 \micro m de diamètre, calibré avec son propre préamplificateur à 20 dB (HGL-200-1113 avec préamplificateur AH2010, Onda Corporation, Sunnyvale, CA, Etats-Unis).

\subsection{Résolution spatiale}
Dans la théorie, nous avons vu que la pression ultrasonore ne peut être directement mesurée grâce à l'hydrophone à force de Lorentz : il faut réaliser une tomographie par des translations et rotations successives du transducteur. La qualité des champs va être confrontée à l'aide de deux références : une simulation par ordinateur et une mesure par un hydrophone piézoélectrique de dimensions comparables à celles de l'hydrophone à force de Lorentz.

\paragraph{Matériel et méthode}
Le générateur émettait un signal de 1 V pic à pic à 1,1 MHz avec 2 sinusoïdes par salve toutes les 10 ms, amplifié, jusqu'à un transducteur de 1,1 MHz, de diamètre et de distance focale 50 mm. Pour réaliser un champ acoustique, il était nécessaire de réaliser des translations et des rotations de l'hydrophone. Pour des raisons d'encombrement et de poids, seul le transducteur (et non l'hydrophone) était déplacé. Ce déplacement était réalisé grâce à deux platines motorisées dirigées par un contrôleur (Newport Corporation MM4005, Irvine, CA, Etats-Unis). Le premier prototype a été utilisé dans cette expérience et le fil était placé à la zone focale du transducteur. Le champ de pression était ensuite reconstruit à partir des amplitudes pic à pic enregistrées par l'oscilloscope à chaque position par une transformée de Radon, calculée avec la fonction \textit{iradon} de Matlab.

La qualité du champ de pression donné par l'hydrophone à force de Lorentz a été contrôlée de deux manières : en comparant avec une simulation numérique du champ de pression basée sur la résolution linéaire de l'intégrale de Rayleigh du champ acoustique \cite{cathignol1997} et avec le champ de pression acquis par l'hydrophone piézoélectrique lipstick.

\begin{figure}[!ht]
		\begin{center}
    \includegraphics[width=1\linewidth]{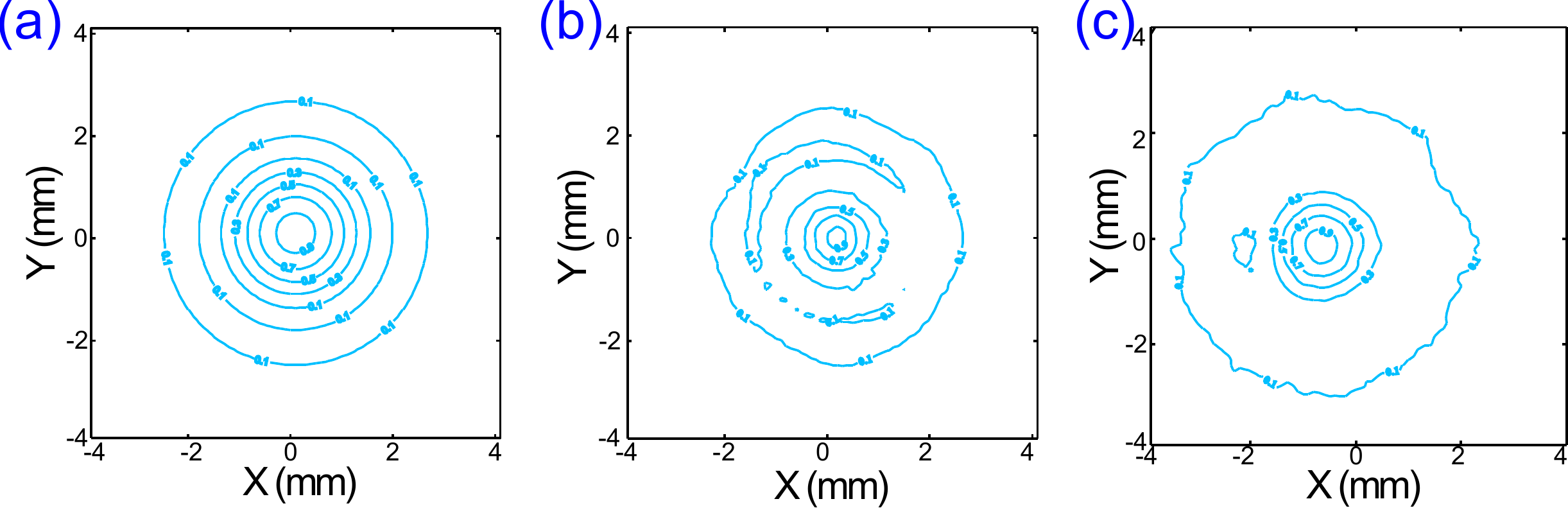}
 		\caption{\label{figHydroEMCaracResSpaResultat1} (a) Champ de pression normalisé dans le plan focal du transducteur simulé par ordinateur. (b) Champ de pression normalisé dans le plan focal du transducteur acquis par l'hydrophone piezoélectrique. (c) Champ de pression normalisé dans le plan focal du transducteur reconstruit par l'hydrophone à force de Lorentz. Les contours extérieurs sont sensiblement identiques.}
 		\end{center}
\end{figure}

\begin{figure}[!ht]
		\begin{center}
    \includegraphics[width=.6\linewidth]{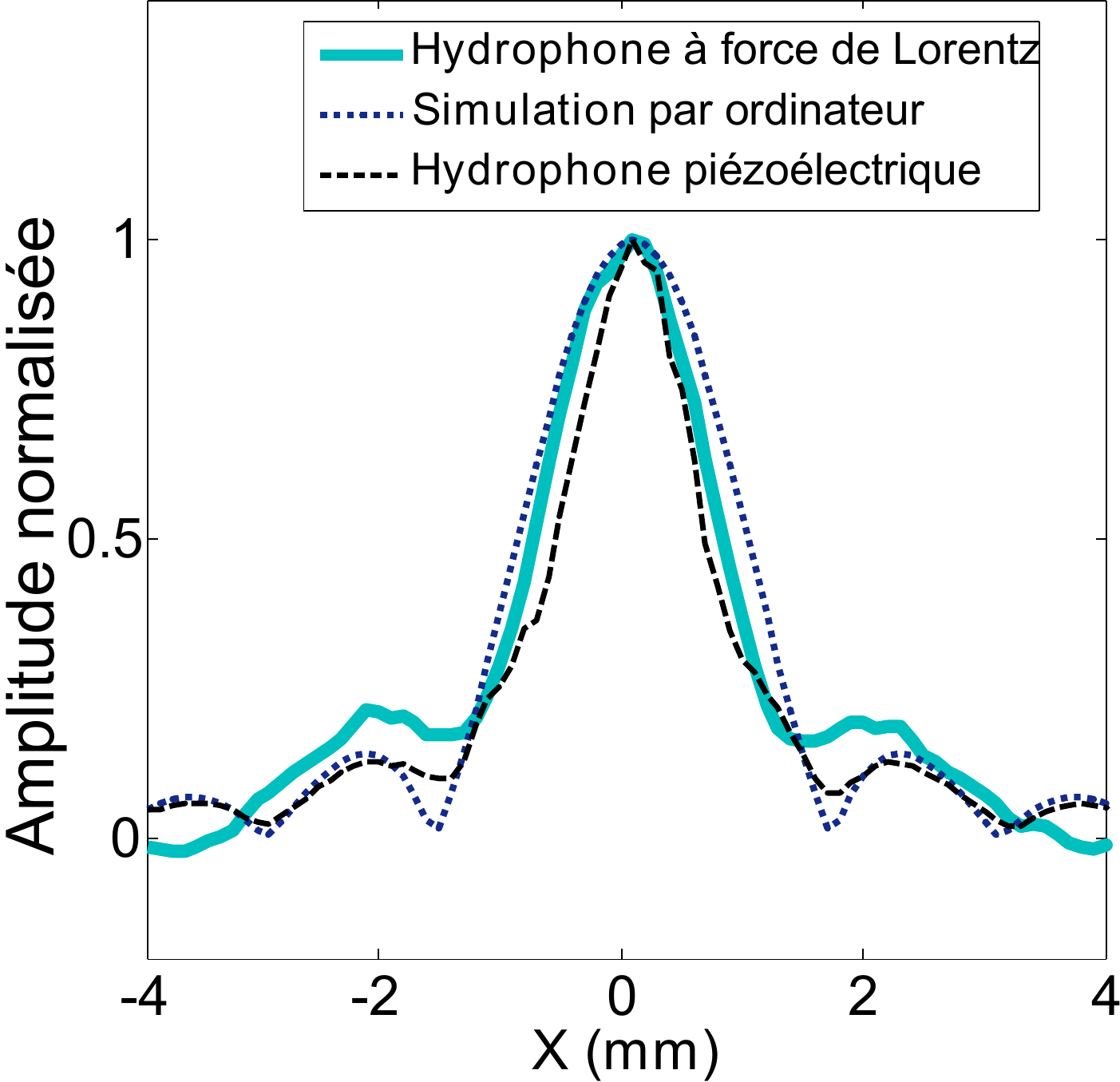}
 		\caption{\label{figHydroEMCaracResSpaResultat2} Champ de pression normalisé dans le plan focal du transducteur le long d'une ligne passant par le maximum d'amplitude, selon la simulation par ordinateur, l'acquisition par l'hydrophone piezoélectrique et la reconstruction par l'hydrophone à force de Lorentz. La largeur à mi-hauteur du pic central vaut 1,62 mm pour la simulation, 1,23 mm pour l'hydrophone piezoélectrique et 1,57 mm l'hydrophone à force de Lorentz. Loin du pic central, l'hydrophone à force de Lorentz est moins bien résolu que les autres méthodes. Sa résolution spatiale est néanmoins millimétrique.}
 		\end{center}
\end{figure}

\paragraph{Résultats et discussions}
Les figures \ref{figHydroEMCaracResSpaResultat1}-(a), -(b) et -(c) montrent respectivement le champ de pression simulé par ordinateur, acquis par l'hydrophone piezoélectrique, et par l'hydrophone à force de Lorentz, dans un plan XY à 50 mm du transducteur (distance focale). La figure \ref{figHydroEMCaracResSpaResultat2} représente la pression normalisée le long d'une ligne passant par le maximum central pour les trois champs de pression, afin de comparer la largeur du pic central.

On peut remarquer une grande similarité entre les différents champs de pression. La largeur de pic à mi-hauteur est de 1,62 mm pour la simulation par ordinateur, de 1,23 mm pour la mesure par hydrophone piezoélectrique et de 1,57 mm pour la mesure par hydrophone à force de Lorentz. L'hydrophone à force de Lorentz donne donc une largeur de pic central intermédiaire entre la simulation et l'hydrophone piezoélectrique, les deux références. A basse pression et loin du pic central, le champ est cependant de moins bonne qualité, avec en particulier des valeur négatives par endroit.

Cette expérience permet cependant de montrer que la résolution de l'hydrophone à force de Lorentz est au moins millimétrique.

\subsection{Réponse fréquentielle}
Le modèle hydrodynamique présenté dans la partie théorique suppose que le fil électrique doit être fin comparé à la longueur d'onde ultrasonore. Un fil de 200 \micro m d'épaisseur est du même ordre de grandeur que la longueur d'onde d'un signal ultrasonore à 7,5 MHz. Nous avons donc évalué la réponse fréquentielle de l'hydrophone avec plusieurs diamètres de fil électrique.

\paragraph{Matériel et méthode}
Le générateur émettait un signal de 1 V pic à pic de fréquence entre 100 kHz à 5 MHz par pas de 20 kHz avec 4 sinusoïdes par salve toutes les 10 ms, amplifié, jusqu'à un transducteur large bande centré à 2 MHz, de diamètre 50 mm et de distance focale 210 mm. L'amplificateur de courant HCA-2M-1M avait une fréquence de coupure à 2 MHz, donc a été remplacé par un amplificateur large bande (HCA-10M-100k, Laser Components, Olching, Allemagne).

Le deuxième prototype a été utilisé dans cette expérience et le fil électrique était placé à la zone focale du transducteur. Quatre diamètres de fil ont été testés : 70 \micro m, 100 \micro m, 200 \micro m et 400\micro m.

D'autre part, afin de mieux comprendre la réponse fréquentielle, l'impédance électrique de l'hydophone a été évaluée. Pour cela, l'hydrophone à force de Lorentz avec un fil à 200 \micro m de diamètre était plongé dans l'eau et branché directement (sans amplificateur) à un impédancemètre (ZVB-4 Vector Network Analyzer, Rohde Schwarz GmbH, Munich, Allemagne). Ce dernier évaluait l'impédance réelle et imaginaire de .15 à 100 MHz.

\begin{figure}[!ht]
   \begin{minipage}[c]{.5\linewidth}
		\begin{center}
	   		\includegraphics[width=1\linewidth]{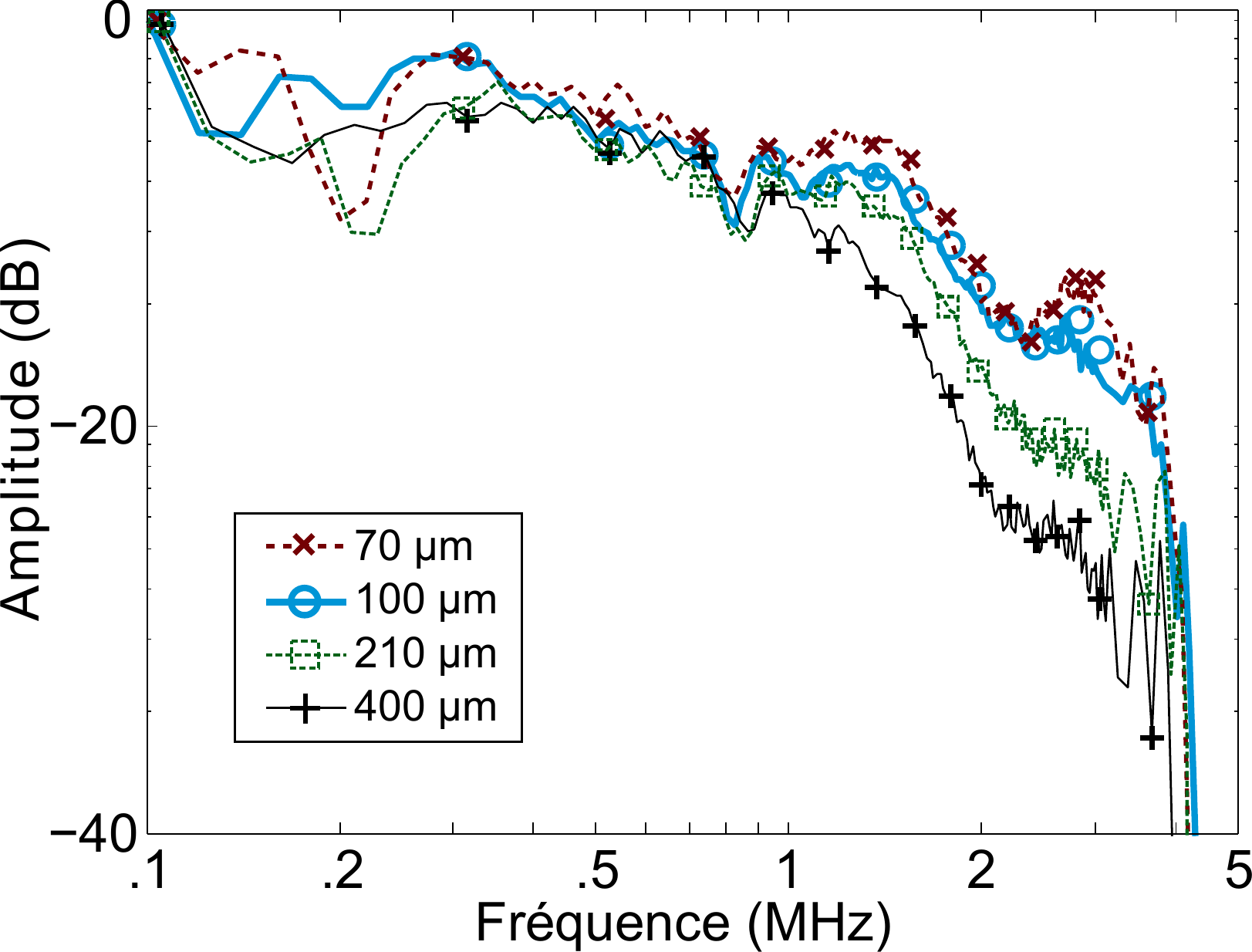}
		\end{center}
   \end{minipage}
   \begin{minipage}[c]{.5\linewidth}
		\begin{center}
	   		\includegraphics[width=1\linewidth]{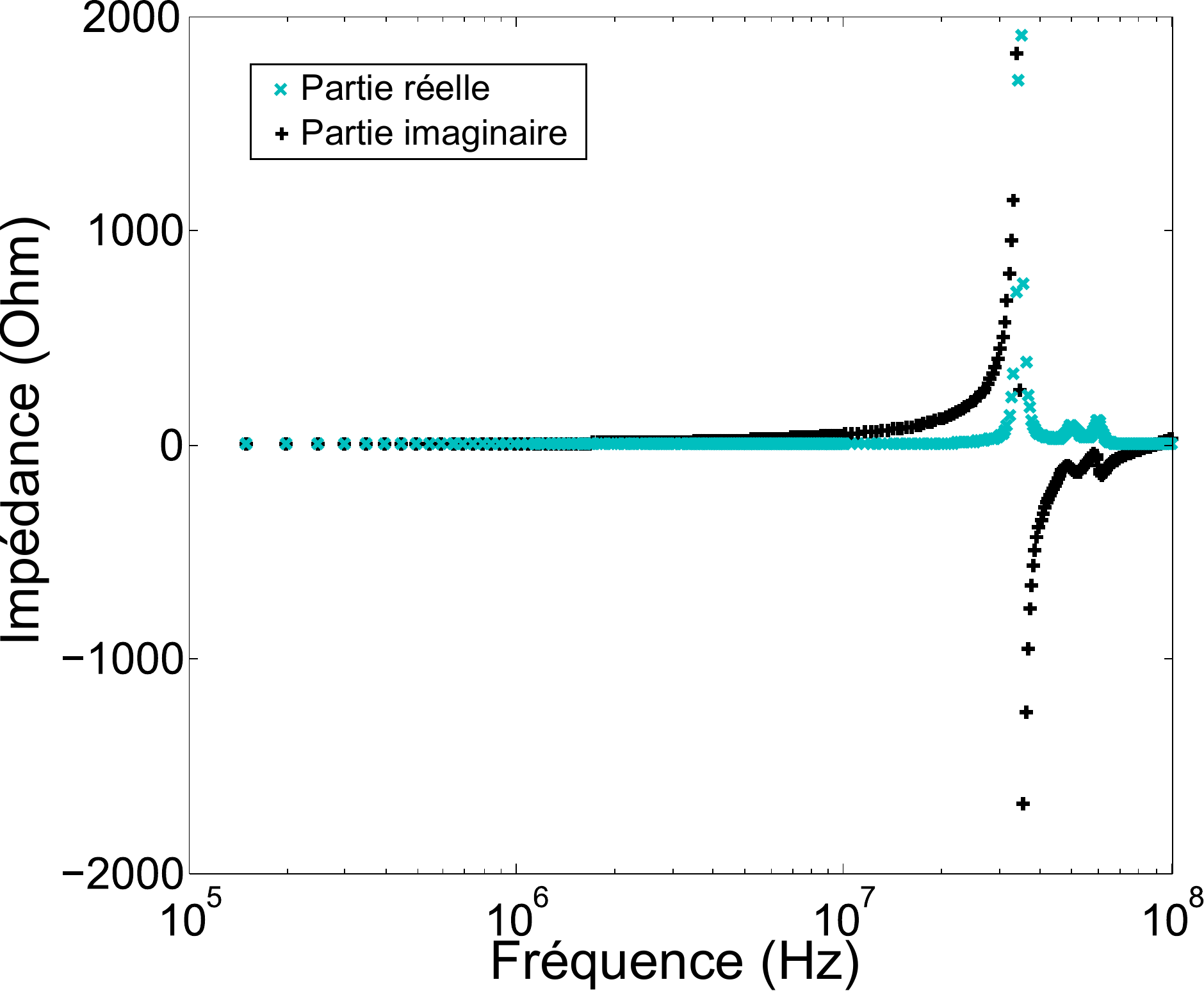}
		\end{center}
   \end{minipage} \hfill
 	\caption{(a) Réponse fréquentielle de l'hydrophone à force de Lorentz entre 0,1 et 5 MHz pour quatre diamètres de fil : 70 \micro m (rouge), 100 \micro m (bleu), 210 \micro m (vert) et 400 \micro m (noir). La réponse décroit avec la fréquence, avec une fréquence de coupure haute entre 1 et 2 MHz. La fréquence de coupure varie faiblement avec la taille du fil. (b) Impédance réelle ($x$ bleus) et imaginaire ($+$ noirs) en fonction de la fréquence de l'hydrophone à force de Lorentz entre 0,15 et 100 MHz. Aucun pic de résonance n'apparait en-dessous de 30 MHz, donc cela n'explique pas la réponse fréquentielle obtenue.}
 	\label{figHydroEMCaracRepFreqResultat} 
\end{figure}

\paragraph{Résultats et discussions}
La réponse fréquentielle pour les quatre diamètres de fil est représentée sur la figure \ref{figHydroEMCaracRepFreqResultat}-(a). On remarque une bande passante plate de 150 kHz à 1 MHz environ, puis une diminution assez rapide. La taille du fil semble avoir une faible influence sur la fréquence de coupure haute. La fréquence de coupure basse n'a pas pu être mesurée.

L'impédance électrique de l'hydrophone entre 0,15 et 10 MHz est représenté sur la figure \ref{figHydroEMCaracRepFreqResultat}-(b). Le pic de résonance électrique le plus bas apparait vers 30 MHz. Aucune résonance électrique n'influence donc la réponse fréquentielle de l'hydrophone.

Cette réponse fréquentielle ne s'explique pas bien par le modèle hydrodynamique retenu, où on suppose seulement que le fil électrique est de taille négligeable devant la longueur d'onde des ultrasons, et devrait faire l'objet de recherches ultérieures.

\subsection{Sensibilité}
La dépendance du signal induit en fonction de la pression a été étudiée afin de comparer les résultats au modèle et surtout de voir à quelle gamme de pressions est adapté l'hydrophone.

\paragraph{Matériel et méthode}
Le générateur émettait un signal de tension variable, de fréquence 0,5 MHz avec 4 sinusoïdes par salve toutes les 10 ms, amplifié, jusqu'au transducteur principal. L'amplification pouvait être nulle, faite par l'amplificateur de puissance de 200 W, ou par un amplificateur de 500 W (1040L, Electronics and Innovation Engineering, Rochester, NY, Etats-Unis), afin d'obtenir une vaste gamme de pression mesurable.

Le deuxième prototype a été utilisé dans cette expérience et le fil était placé à la zone focale du transducteur. Comme la tension de sortie maximale de l'amplificateur de signal HCA-2M-1M était égale à 4V, les plus hautes pressions ne pouvaient pas être évaluées avec celui-ci. Deux expériences ont donc été réalisées, l'une avec amplificateur de signal et l'autre sans. Le modèle hydrodynamique a été utilisé pour comparer les résultats avec la sensibilité théorique de l'hydrophone.

\begin{figure}[!ht]
   \begin{minipage}[c]{.5\linewidth}
		\begin{center}
	   		\includegraphics[width=1\linewidth]{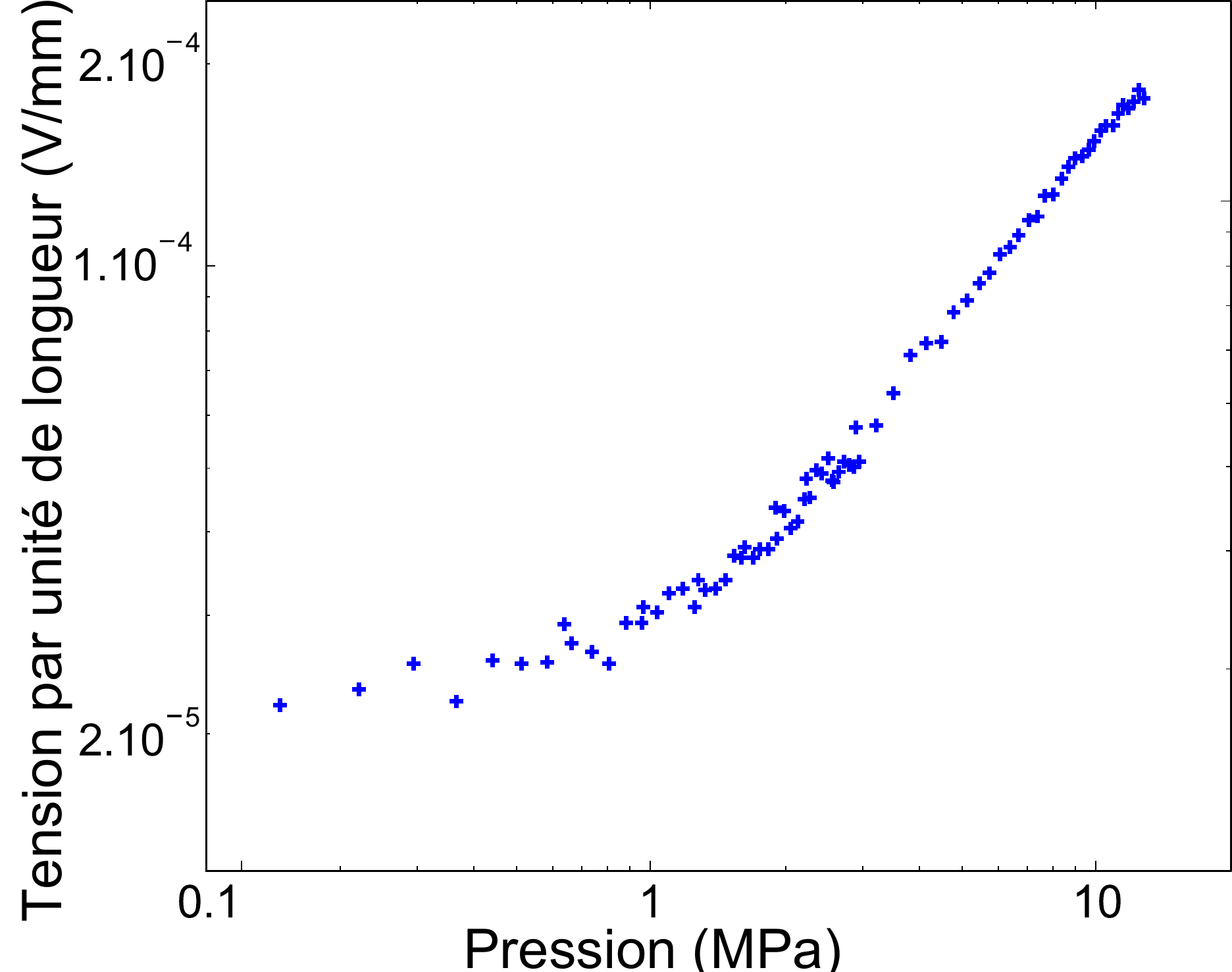}
		\end{center}
   \end{minipage} \hfill
   \begin{minipage}[c]{.5\linewidth}
		\begin{center}
	   		\includegraphics[width=1\linewidth]{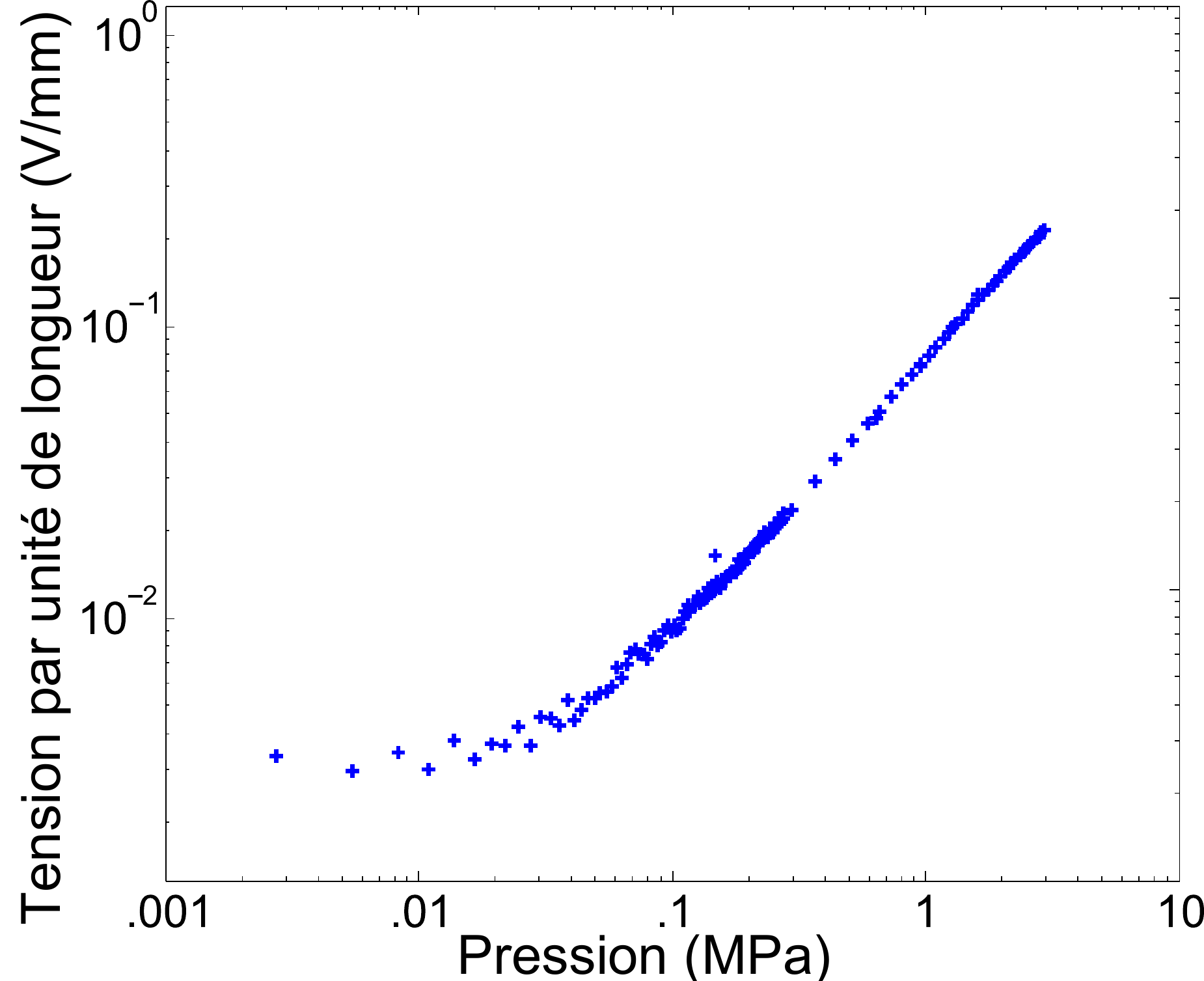}
		\end{center}
   \end{minipage}
 	\caption{(a) Tension acquise par l'hydrophone à force de Lorentz en fonction de la pression sans amplificateur de signal. Une régression linéaire donne une sensibilité de  $1,4 \pm 0,1\ 10^{-11}$ V.mm$^{-1}$.Pa$^{-1}$, soit -217 à -216 dB re \micro V.mm$^{-1}$.Pa$^{-1}$ avec un coefficient de régression linéaire $R^2$ égal à 0,997. (b) Tension acquise par l'hydrophone à force de Lorentz en fonction de la pression avec amplificateur de signal. Une régression linéaire donne une sensibilité de $8,6 \pm 0,1\ 10^{-8}$ V.mm$^{-1}$.Pa$^{-1}$, soit -141,4 à -141,1 dB re \micro V.mm$^{-1}$.Pa$^{-1}$ avec un coefficient $R^2$ égal à 0,9981 avec l'amplificateur de signal.}
 	\label{figHydroEMCaracSensResultat} 
\end{figure}

\paragraph{Résultats et discussions}
Les figures \ref{figHydroEMCaracSensResultat}-(a) et \ref{figHydroEMCaracSensResultat}-(b) montrent le signal acquis par l'hydrophone respectivement sans et avec amplificateur de courant. Une régression linéaire donne une sensibilité de  $1,4 \pm 0,1\ 10^{-11}$ V/mm/Pa = -217 à -216 dB re \micro V/mm/Pa avec un coefficient de régression linéaire $R^2$ égal à 0,997 sans l'amplificateur de signal et de $8,6 \pm 0,1\ 10^{-8}$ V/mm/Pa = -141,4 à -141,1 dB re \micro V/mm/Pa avec un coefficient $R^2$ égal à 0,9981 avec l'amplificateur de signal. Les valeurs élevées de $R^2$ montrent une excellente linéarité en fonction de la pression à la fréquence de 0,5 MHz de 10 kPa à 10 MPa. Aux pressions faibles, les parasites électromagnétiques deviennent prédominants par rapport au signal induit.

Ces valeurs doivent néanmoins être comparées avec celles données par le modèle hydrodynamique. En prenant une densité de l'eau $\rho = 1000 \pm 10$ kg.m$^{-3}$, la vitesse du son dans l'eau $c = 1470 \pm 10$ m.s$^{-1}$, le champ magnétique au niveau du fil électrique $B = 0,17 \pm 0,03$ T, la largeur du faisceau ultrasonore $l = 8 \pm 1$ mm, le facteur $K(\rho,\rho_0) = 5,0 \pm 0,3$ avec $\rho_0 = 9000 \pm 500$ kg.m$^{-3}$, la sensibilité sans amplification est égale à $2,3 \pm 0,7\ 10^{-11}$ V.mm$^{-1}$.Pa$^{-1}$, soit -215 à -210 dB re \micro V/mm/Pa. Cela donne une différence de 40\% avec la sensibilité expérimentale. Avec une amplification $\alpha = 1 10^6$ V.A$^{-1}$ et une résistance du fil plus entrée de l'amplificateur $R = 251 \pm 1$ $\Omega$, la sensibilité est égale à $9,4\pm 2,7\ 10^{-8}$ V.mm$^{-1}$.Pa$^{-1}$, soit -143 to -138 dB re \micro V.mm$^{-1}$.Pa$^{-1}$, ce qui donne une différence de 10\% avec la sensibilité expérimentale. Notons que l'incertitude totale est majoritairement déterminée par l'incertitude sur le champ magnétique, de l'ordre de 20\%.

Ces résultats montrent que le modèle hydrodynamique donne une compréhension correcte du phénomène, mais encore insuffisante : une calibration reste nécessaire pour faire des mesures précises.

\subsection{Résistance à la cavitation}
Comme vu dans l'introduction, à puissance ultrasonore élevée, deux phénomènes peuvent endommager les hydrophones : le chauffage et les dommages créés par la cavitation. L'augmentation de la température à cause de l'atténuation supérieure dans le fil sera faible car (1) le fil est petit par rapport à la longueur d'onde ultrasonore et (2) le fil est refroidi par le fluide environnant. Même si une augmentation globale de la température peut changer la résistance du fil, celle-ci restera faible par rapport à la résistance d'entrée de l'amplificateur. Des nuages de cavitation peuvent cependant endommager le fil électrique et donc avoir une influence sur le signal induit. Une expérience a donc été mise en oeuvre en utilisant un matériel spécifique pour obtenir périodiquement un nuage de cavitation sur le fil pendant une longue période de temps.

\paragraph{Matériel et méthode}
Un ordinateur avec un module d'émission / réception (module NI5781R, National Instruments, Austin, TX, Etats-Unis) produisait un signal électrique, amplifié par un amplificateur de 69 dB (Prana RD, Brive, France), pour un transducteur ultrasonore de fréquence centrale égale à 550 kHz avec un diamètre et une distance focale de 10 cm. Les mesures ont été réalisées en trois étapes :
\begin{enumerate}
	\item Excitation du transducteur avec un signal avant amplification de 0,05 V avec 28 sinusoïdes par émissions toutes les 25 ms. La pression était enregistrée par le deuxième prototype d'hydrophone à force de Lorentz placé à la distance focale.
	\item Excitation du transducteur avec un signal de tension régulé avant amplification d'environ 2 V avec un rapport cyclique de 0,1 toutes les 25 ms au transducteur. L'amplitude était régulée de telle sorte qu'un nuage de cavitation apparaissait chaque seconde pendant 4800 secondes dans la zone focale, avec une pression pic à pic de 15 MPa. L'index de cavitation, défini comme $CI=20\log(\left|E\right|)$, avec $E$ la transformée de Fourier du signal entre 0,1 et 3 MHz \cite{desjouy2013}, est égal à 15 dB.
	\item Même excitation qu'en (1) avec un signal avant amplification de 0,05 V avec 28 sinusoïdes par émissions toutes les 25 ms. La pression était enregistrée par le deuxième prototype d'hydrophone à force de Lorentz placé à la distance focale.
\end{enumerate}

La modification du signal à cause de la cavitation était évaluée entre les deux signaux grâce à leur coefficient de corrélation $R$ défini par :
\begin{empheq}[box=\mybluebox]{align}
	R = \frac{  \sum{\Psi _1 \Psi _2}  }   {\sqrt{   \sum{\Psi _1 ^2}\sum{\Psi _2 ^2}   }}
	\label{eqHydroEMCaracSensMateriel}
\end{empheq}
avec $\Psi1$ et $\Psi2$ les deux signaux considérés. Ce coefficient est égal à 1 quand deux signaux $\Psi_1$ et $\Psi_2$ sont identiques, -1 quand exactement opposés et 0 quand totalement non corrélés. 

\begin{figure}[!ht]
	   \includegraphics[width=.90\linewidth]{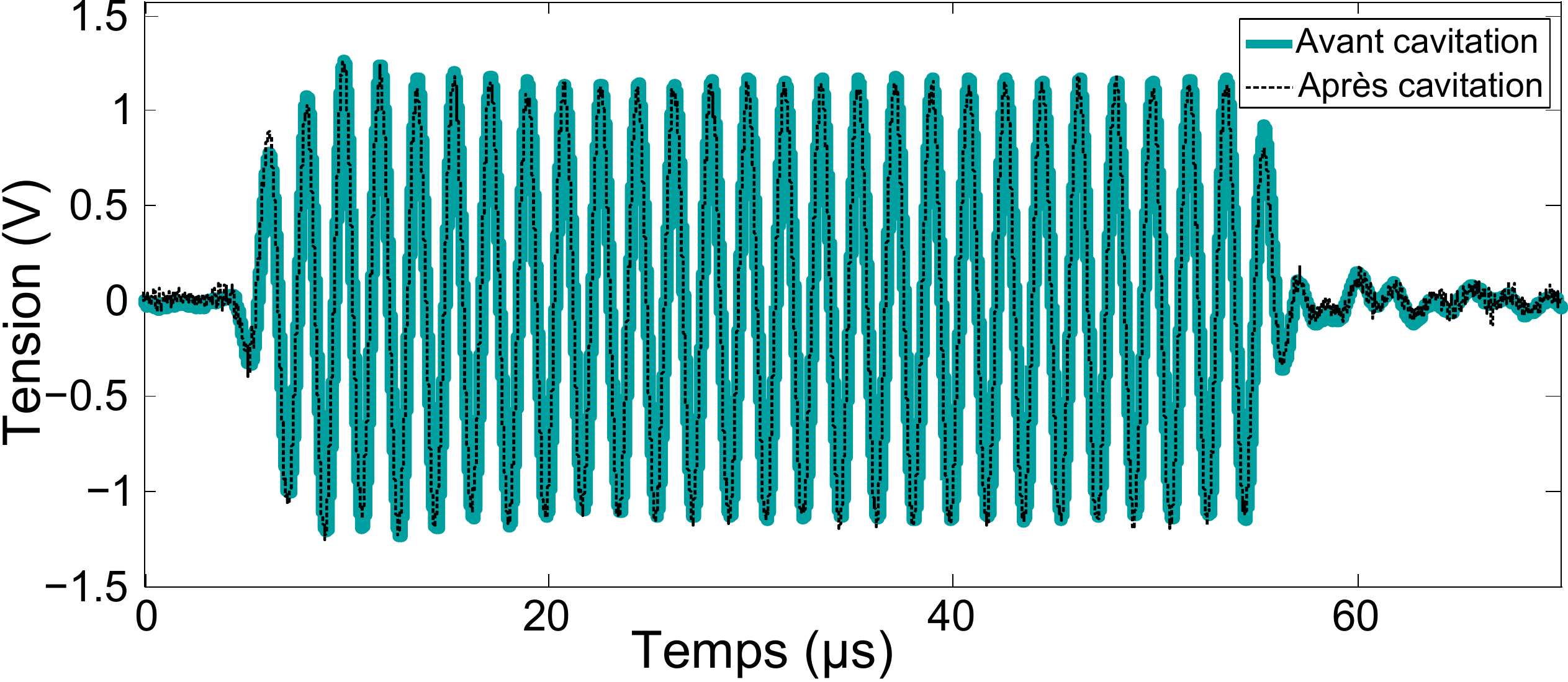}
	   \caption{\label{figHydroEMCaracSensResultat3} Signal avant (bleu continu) et après (pointillés noirs) 4800 s de cavitation avec un nuage de cavitation par seconde et un index de cavitation égale à 15 dB. Les signaux sont quasiment identiques, avec un coefficient de corrélation égal à 0,9963, ce qui montre que la cavitation a eu peu d'impact sur l'hydrophone.}
\end{figure}

\paragraph{Résultats et discussion}
L'amplitude des signaux avant et après 4800 secondes de cavitation est représentée respectivement en bleu continu et en pointillés noirs sur la figure \ref{figHydroEMCaracSensResultat3}). Le coefficient de corrélation est égal à 0,9963, valeur qui montre que les signaux sont quasiment identiques.

La cavitation a donc peu d'impact sur le signal induit et l'hydrophone se révèle robuste aux dommages à haute pression. L'hydrophone peut même faire des mesures pendant la cavitation ; mais la longueur du fil rend la résolution spatiale médiocre, tandis qu'une tomographie ne donnera pas la valeur en temps réel, surtout que le nuage de cavitation a un caractère aléatoire qui ne sera pas rendu par la reconstruction.

\subsection{Réponse directionnelle}
La réponse directionnelle présentée dans la partie théorique a été évaluée grâce à une expérience où le faisceau ultrasonore n'était pas orthogonal au champ magnétique ou à la direction du fil électrique.

\paragraph{Matériel et méthodes}
Le générateur émettait un signal de 1 V pic à pic, de fréquence 0,5 MHz avec 4 sinusoïdes par salve toutes les 10 ms, amplifié, jusqu'au transducteur principal. L'hydrophone à force de Lorentz était placé au point focal avec différents angles par rapport au faisceau ultrasonore.

L'angle entre le transducteur et l'hydrophone variait de -35$^o$ à +35$^o$ avec des pas de 0,5 $^o$ dans deux directions : avec l'axe de rotation selon la direction X ou Y. Des angles plus élevés n'ont pas pu être évalués à cause de l'espace disponible au centre du support de PVC de l'hydrophone.

\begin{figure}[!ht]
   \begin{minipage}[c]{.5\linewidth}
		\begin{center}
	   		\includegraphics[width=1\linewidth]{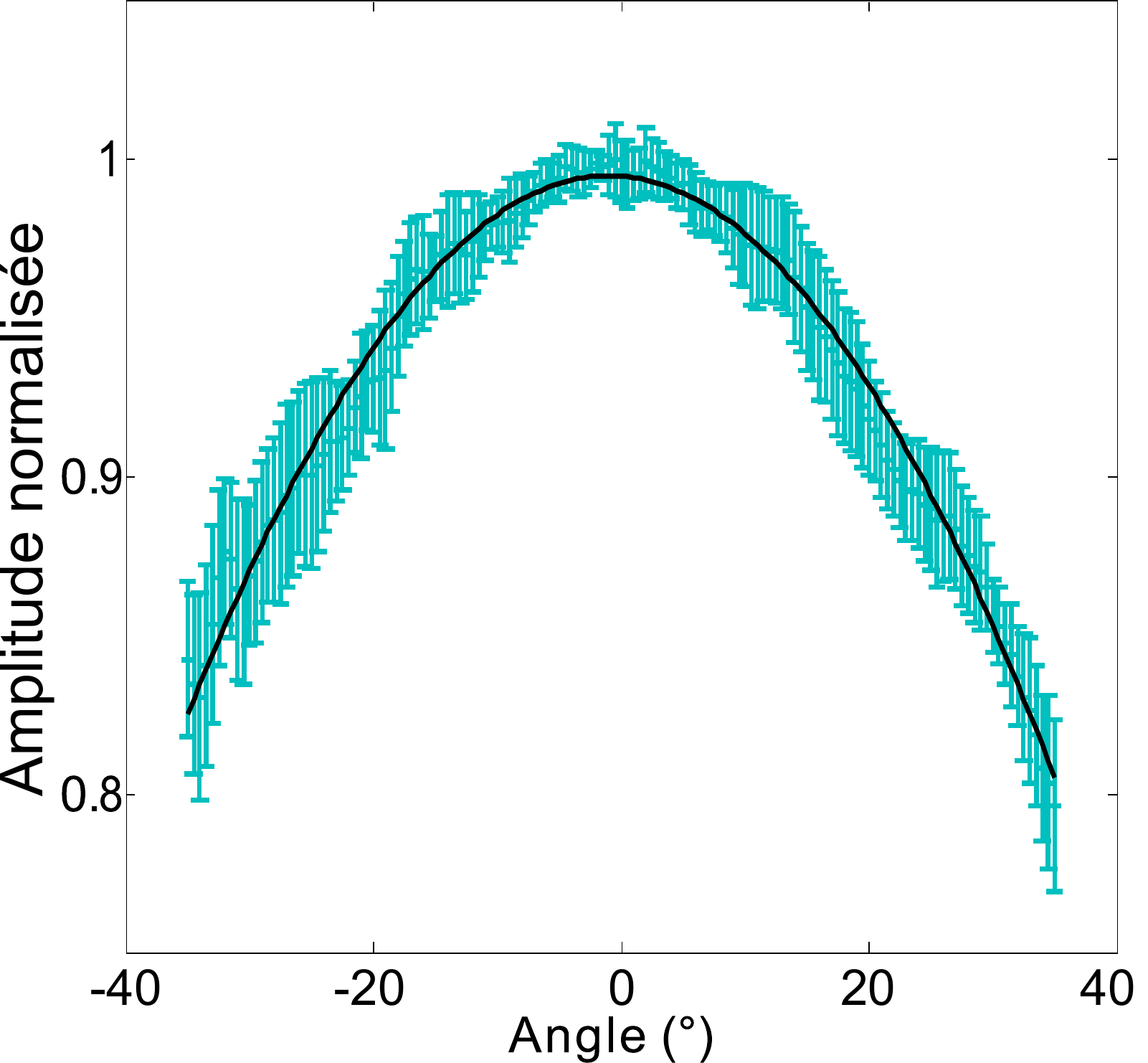}
		\end{center}
   \end{minipage} \hfill
   \begin{minipage}[c]{.5\linewidth}
		\begin{center}
	   		\includegraphics[width=1\linewidth]{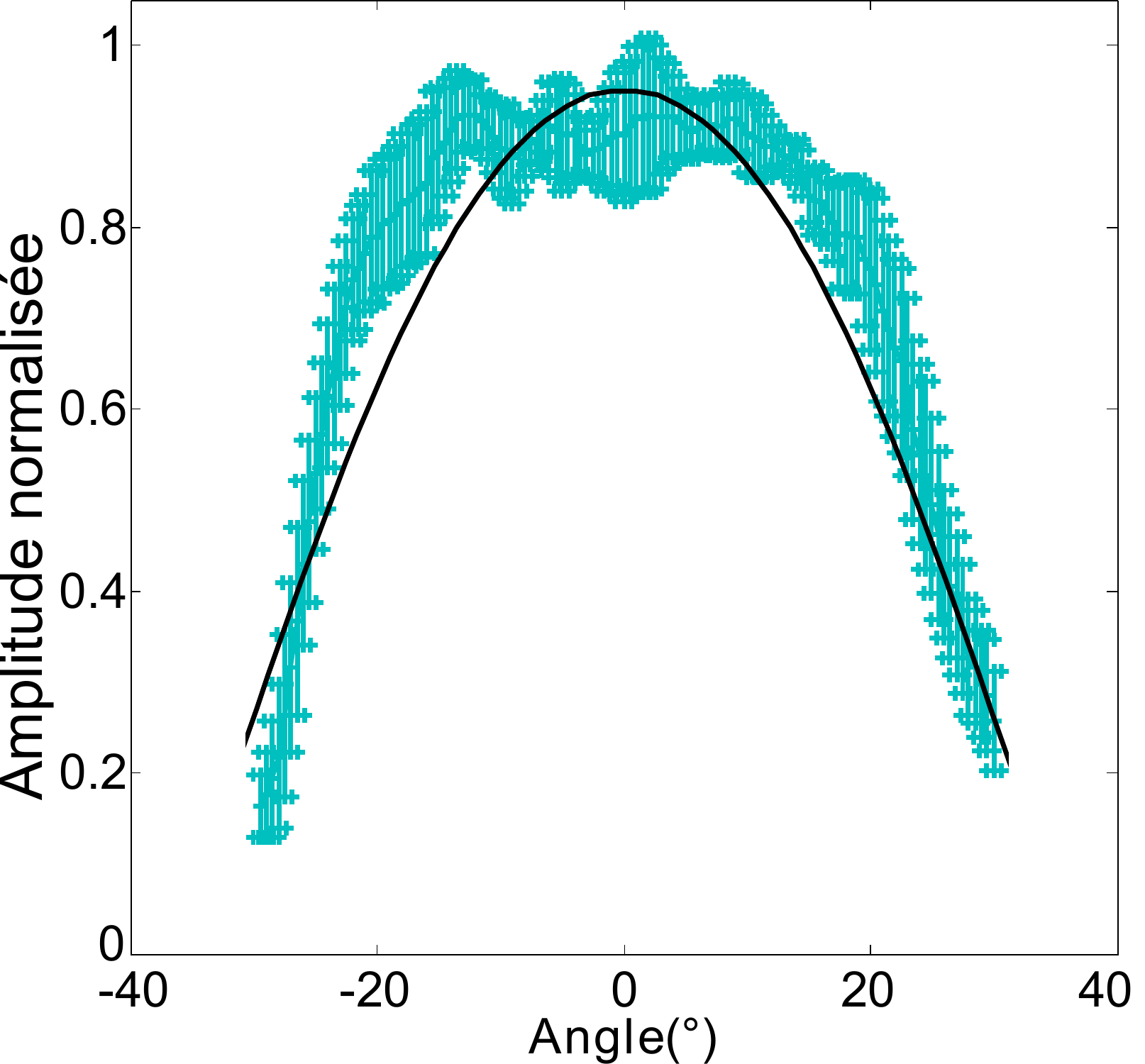}
		\end{center}
   \end{minipage}
 	\caption{(a) Amplitude normalisée du signal en fonction d'un angle $\alpha$ avec l'axe Z pour un axe de rotation dans la direction Y et fonction $\cos(\alpha)$ en fonction de l'angle $\alpha$, avec un écart-type sur huit expériences. La courbe théorique est bien adaptée aux données expérimentales. L'amplitude est supérieure à 80\% du maximum pour des angles de -35 à +35$^o$. (b) Amplitude normalisée du signal en fonction d'un angle $\beta$ avec l'axe Z pour un axe de rotation dans la direction X et fonction $\sinc(\frac{2\pi}{\lambda}\tan(\beta))$ en fonction de l'angle $\beta$, avec un écart-type sur huit expérience. L'amplitude est supérieure à 50\% du maximum pour des angles de -25 à +25$^o$.}
 	\label{figHydroEMCaracRepDirResultat} 
\end{figure}

\paragraph{Résultats et discussions}
L'amplitude normalisée du signal en fonction de l'angle est représenté sur la figure \ref{figHydroEMCaracRepDirResultat}-(a) pour un axe de rotation dans la direction Y et sur la figure \ref{figHydroEMCaracRepDirResultat}-(b) pour un axe de rotation dans la direction X. Les barres d'erreur représentent la déviation standard obtenues par huit acquisitions différentes. Pour vérifier la théorie initiale, la fonction cosinus a été représentée sur la figure \ref{figHydroEMCaracRepDirResultat}-(a) et celle donnée en \ref{eqHydroEMTheorieDepAngle3} est représentée sur la figure \ref{figHydroEMCaracRepDirResultat}-(b). L'amplitude est supérieure à 80\% du maximum pour des angles de -35 à +35$^o$ pour un axe de rotation dans la direction Y et supérieure à 50\% du maximum pour des angles de -25 à +25$^o$ pour un axe de rotation dans la direction X.

La réponse directionnelle avec un axe de rotation dans la direction Y se vérifie plutôt bien tandis que les résultats avec l'axe de rotation dans la direction X sont moins proches, ce qui peut s'expliquer par les hypothèses d'onde sinusoïdale effectuées sur le signal émis.

\subsection{Influence de la tension du fil électrique}
Le modèle hydrodynamique présenté suppose que le fil suit parfaitement le mouvement du fluide en tout point, mais ceci peut être modifié par la tension du fil.

\paragraph{Matériel et méthodes}
Le générateur émettait un signal de 1 V pic à pic, de fréquence 0,5 MHz avec 4 sinusoïdes par salve toutes les 10 ms, amplifié, jusqu'au transducteur principal.

Le deuxième prototype a été utilisé dans cette expérience et le fil était placé à la zone focale du transducteur. Il était posé horizontalement, donc le fil électrique avait une forme de chainette à cause de son poids. Cette forme, décrite il y a plus de 300 ans par Bernoulli, Huyghens et Leibniz \cite{routh1891}, a une hauteur $y$ reliée à la position $x$ :
\begin{empheq}[box=\mybluebox]{align}
	y(x)=a\cosh(x/a)+c
	\label{eqHydroEMExpTensionMateriel}
\end{empheq}
avec $c$ la hauteur aux extrémités du fil et $a = \frac{H}{w g}$ avec $H$ la composante horizontale de la tension, $w$ la masse par unité de longueur et $g$ l'accélération de pesanteur égale à 9,81 m.s$^{-1}$.

Les ultrasons étaient focalisés dans la zone la plus basse du fil et la force électromotrice induite mesurée. Ensuite, lorsque la tension mécanique était augmentée par la rotation de deux vis, le fil se tendait et le point le plus bas se rapprochait du transducteur. Le délai entre l'émission des ultrasons par le transducteur et sa réception par l'hydrophone, modifié, était mesuré. En prenant une vitesse du son de 1470 m/s, le déplacement correspondant était calculé, ce qui permettait d'en déduire la tension. La tension a été ensuite divisée par la surface de la section du fil (un disque de 70 \micro m de diamètre) pour obtenir une contrainte en pascals, afin d'étendre ce résultat plus facilement à d'autres fils. La contrainte initiale a été augmentée graduellement jusqu'à la destruction estimée à 220 MPa \cite{wildi2000}.

Pour quantifier l'influence de la tension sur le signal, nous avons calculé le coefficient de corrélation $R$ entre deux signaux comme défini précédemment par
\begin{empheq}[box=\mybluebox]{align}
	R = \frac{  \sum{\Psi _1 \Psi _2}  }   {\sqrt{   \sum{\Psi _1 ^2}\sum{\Psi _2 ^2}   }}
	\label{eqHydroEMCaracTensionMateriel1}
\end{empheq}
Dans cette expérience, le coefficient de corrélation a été calculé pour chaque signal (le retard éventuel de l'écho étant compensé) avec un signal de référence, le premier signal avec une contrainte de 300 kPa.

\begin{figure}[ht]
		\begin{center}
	   \includegraphics[width=.90\linewidth]{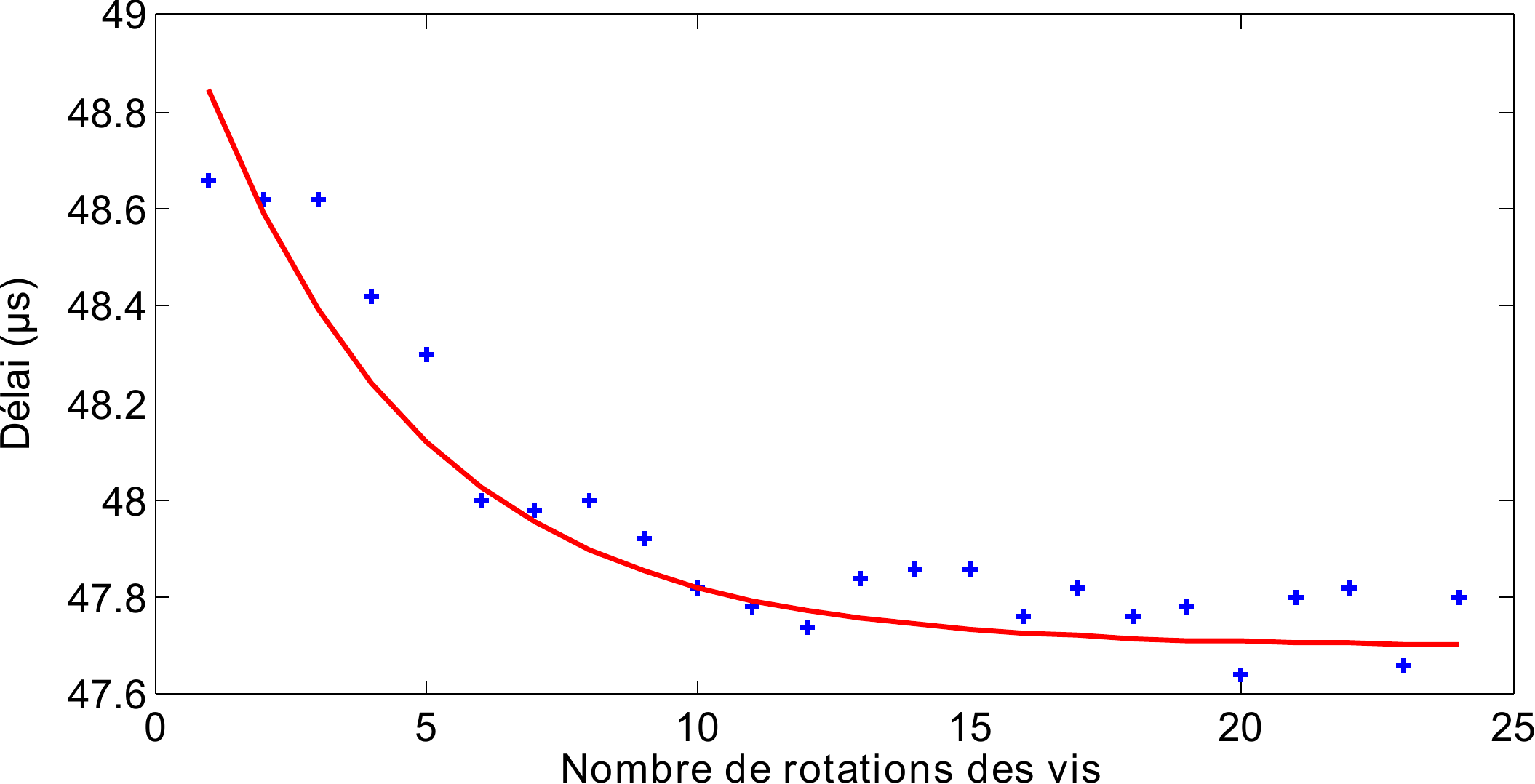}
	   \caption{\label{figHydroEMCaracTensionFilResultat} Délai entre l'émission des ultrasons et leur réception par l'hydrophone à chaque augmentation de tension, plus ajustement des données par une exponentielle décroissante. Lorsque la tension croît à cause des rotations des vis, le fil se tend et se rapproche du transducteur.}
		\end{center}
\end{figure}

\paragraph{Résultats et discussions}
Le délai entre l'émission des ultrasons et leur réception par l'hydrophone à chaque augmentation de tension est représenté sur la figure \ref{figHydroEMCaracTensionFilResultat}. Le délai diminue lorsque la tension augmente, mais les dix derniers déplacements étaient quasiment indétectables : ils ont été extrapolés à partir des premières mesures et de la contrainte de rupture, avec un ajustement d'une exponentielle décroissante. Le déplacement maximal est égal à 1,5 mm (délai de 1 \micro s avec une vitesse de 1470 m/s), ce qui indique que la contrainte initiale était égale à 300 kPa.

Le tableau \ref{tableHydroEMCaracTensionFilResultat1} montre le coefficient de corrélation comme défini en \ref{eqHydroEMCaracTensionMateriel1} entre chaque signal et le signal de référence (contrainte de 0,3 MPa) en fonction de la contrainte du fil estimée (en MPa). Le coefficient de corrélation est toujours supérieur à 0,97, avec des variations indépendantes de l'augmentation de la tension, ce qui signifie que la tension n'a pas d'effet significatif sur le signal.

\begin{table}
	\begin{tabular}{|c|c|}
		\hline
		\rowcolor{SandyBrown} Estimation de la contrainte du fil (MPa) & Coefficient de correlation  R\\
		\hline
		0,3 &1\\
		0,4 &0,9843\\
		0,7 &0,9840\\
		1   &0,9751\\
		2   &0,9765\\
		3   &0,9768\\
		5   &0,9657\\
		7   &0,9760\\
		10  &0,9802\\
		20  &0,9786\\
		40  &0,9775\\
		50  &0,9711\\
		80  &0,9874\\
		100 &0,9833\\
		200 &0,9829\\
		\hline
	\end{tabular}
	\caption{Coefficient de corrélation entre chaque signal et le signal de référence (tension de 0,3 MPa) en fonction de l'estimation de la tension du fil (en MPa).}
	\label{tableHydroEMCaracTensionFilResultat1}
\end{table}

\section{Bilan du chapitre}

\subsection{Principaux résultats}
L'objectif de ce travail était de concevoir un hydrophone basé sur la force de Lorentz. Un modèle hydrodynamique et électromagnétique de l'hydrophone a été proposé, à partir duquel un prototype a été fabriqué puis caractérisé en suivant la norme de l'International Electrotechnical Commission dédiée aux hydrophones.

Le design est basé sur une approche tomographique avec une reconstruction du champ acoustique par une transformée de Radon inverse. Utiliser la tomographie n'est pas rare pour des mesures acoustiques, avec des utilisations en imagerie Schlieren \cite{pitts1994tomographic} ou l'utilisation de diffuseurs à fils pour la caractérisation de transducteurs \cite{raum1997}. Pour diminuer le nombre total d'acquistions, on peut s'affranchir des rotations si l'on sait que le champ acoustique présente une symétrie de rotation. De plus, le nombre de fils peut être multiplié, ce qui permet de faire des acquistions simultanées. L'influence de chaque fil sur ses voisins devrait cependant être proprement évalué dans ce cas. A la fin de cette thèse, l'étude théorique, expérimentale et par simulation de la faisabilité d'un hydrophone multifils était encore en cours.

La première étape de la conception du prototype d'hydrophone a été de contrôler le champ magnétique au niveau du fil électrique. L'arrangement de Halbach a permis d'obtenir un champ magnétique homogène le long du fil, ce qui facilite l'approche tomographique. D'autre part, il a fallu évaluer l'impact de la tension mécanique du fil électrique sur le signal induit. Le coefficient de corrélation entre les signaux acquis à différentes tensions était supérieur à 0,97, ce qui indique que la tension a un impact quasiment négligeable. Le fil doit cependant être légèrement tendu pour éviter des erreurs de localisation des signaux.

Comme suggéré par la norme IEC, la réponse fréquentielle a été étudiée. Les mesures d'impédance électrique n'ont montré aucune résonance dans les fréquences de travail entre 0,15 et 10 MHz, avec un premier pic de résonance à 30 MHz. Les résonances électriques n'ont donc pas d'impact sur les mesures de l'hydrophone à force de Lorentz. Une expérience a montré que la fréquence haute de coupure se trouve entre 1 et 2 MHz, valeur qui semble faiblement dépendre du diamètre du fil. Ceci est assez mal expliqué par le modèle hydrodynamique que nous avons utilisé qui n'est valable que pour un diamètre du fil petit comparé à la longueur d'onde ultrasonore. La fréquence basse de coupure n'a pu être étudiée avec l'équipement disponible mais est inférieure à 100 kHz.

La réponse directionnelle a également été évaluée. Pour la plupart des hydrophones, cela a un impact sur la taille effective de l'élément sensible, mais la forme cylindrique de cet hydrophone à force de Lorentz rend ce paramètre peu pertinent. La réponse directionnelle à -3 dB est supérieure à $\pm 35^o$ avec un axe de rotation dans la direction Y, la limite étant l'espace intérieur du cylindre de PVC, et égale à $\pm 25^o$ avec un axe de rotation dans la direction X. Une modèle de réponse directionnelle a été proposé et est en accord avec les résultats expérimentaux.

Les sensibilités expérimentales et théoriques sont relativement proches, avec des différences inférieures à 40\%. Le modèle hydrodynamique donne donc une bonne compréhension du phénomène, même si une calibration reste nécessaire pour des mesures précises. A basse pression, inférieure à 50 kPa, deux problèmes surviennent : (1) les parasites électromagnétiques et le signal induit par force de Lorentz sont du même ordre de grandeur, donc un blindage serait nécessaire ; et (2) le nombre de Reynolds devient très proche de 1 (pour un fil avec un diamètre de 200 \micro m à 10 kPa, $Re = 1,3$) et la principale hypothèse du modèle hydrodynamique n'est pas satisfaite. Inversement, le signal devrait être linéaire à des pressions supérieures à 12 MPa. De plus, l'hydrophone s'est révélé résistant à la cavitation. Pour des comparaisons avec d'autres hydrophones, il faut cependant garder à l'esprit que l'hydrophone à force de Lorentz mesure surtout la vitesse du fluide, et que la relation utilisée entre la vitesse et la pression n'est valide que dans certaines conditions, notamment d'onde plane. 

\subsection{Avenir du dispositif}
Les prochaines études devraient améliorer le modèle, en particulier à l'aide de simulations informatiques. Cela pourrait aider notamment à trouver un moyen d'augmenter la fréquence de coupure. En effet, la réponse fréquentielle de cet hydrophone est en-deçà des attentes pour des applications médicales. Cet hydrophone fonctionne bien pour des fréquences inférieures au mégahertz, sans que le modèle hydrodynamique puisse expliquer sa forme.

Le principal atout de l'hydrophone à force de Lorentz pour la caractérisation de champs acoustiques est sa robustesse. Le fil électrique est très résistant aux hautes pressions, quasiment insensible à la chaleur, et s'est révélé robuste à la cavitation. A titre anecdotique, on peut remarquer que le principal défaut du microphone à ruban proposé par H. Olson, la fragilité de la partie sensible, se révèle paradoxalement être la principale force de l'hydrophone à force de Lorentz !

Un autre avantage de cet hydrophone est son faible coût - si le fil est abimé, il peut être facilement remplacé à bas prix. La destruction du vernis sur le fil n'aurait qu'un impact faible sur le signal, car la conductivité électrique de la plupart des électrolytes est faible par rapport à la conductivité électrique du fil (conductivité d'une salée inférieure à 5 S.m$^{-1}$, conductivité du cuivre environ égale à 60 MS.m$^{-1}$, et les pertes de courant dans le milieu environnant seraient être négligeables.

La mesure de vitesse plutôt que de pression peut être intéressante pour des mesures précises acoustique. Cet hydrophone pourrait en particulier être combiné avec un hydrophone mesurant la pression ultrasonore afin calculer une intensité acoustique sans approximation d'onde plane, voire pour étudier les non-linéarités acoustiques \cite{shaw2008buoyancy}.

En l'état, une application possible de cet hydrophone est donc la caractérisation de dispositifs médicaux à haute puissance et faible fréquence, comme la lithotripsie.

Enfin, une autre voie de développement pourrait être le domaine de l'acoustique sous-marine, qui utilise des fréquences inférieures au mégahertz et donc évite le principale défaut de ce dispositif.

\chapter{Tomographie d'impédance électrique par force de Lorentz}
\label{secLFEIT}

\section{L'impédance électrique des tissus biologiques}

\subsection{Les types de milieux biologiques}
Du point de vue électrique, les tissus biologiques sont des milieux complexes qui ressemblent peu aux autres matériaux et présentent une grande variabilité de propriétés au sein du corps humain \cite{mcadams1995tissue}. Pour étudier les propriétés électriques des tissus, on peut néanmoins les classer en trois grands groupes : les liquides, les tissus conjonctifs et l'épithélium, ces tissus composant les organes tels que le cerveau, le c\oe ur, le foie...

Les liquides sont composés essentiellement d'eau, de minéraux et d'éléments organiques. Cela rassemble des liquides aussi variés que le liquide céphalo-rachidien, l'urine, l'humeur aqueuse...

Les tissus conjonctifs sont composés de cellules séparées par un milieu intercellulaire. Il peut s'agir par exemple du sang, dont le milieu intercellulaire est liquide, d'os, où le milieu intercellulaire calcifié, de tissus adipeux majoritairement composés de poches de graisse, etc.

Enfin, l'épithélium est composé de cellules jointes. C'est le cas de l'épiderme à la surface de la peau, mais on en trouve également dans le corps humain, comme l'épithélium intestinal ou l'épithélium gastrique.

\subsection{Les propriétés électriques des milieux biologiques}
D'un point de vue électromagnétique, les tissus biologiques peuvent être caractérisés comme amagnétiques, conducteurs et diélectriques à pertes \cite{bernard2007these}.

Mises à part certaines protéines comme la méthémoglobine qui sont paramagnétiques, l'ensemble des tissus biologiques est en effet diamagnétique. Le corps humain est donc globalement considéré comme insensible au champ magnétique, et l'on prend quasi systématiquement une perméabilité magnétique égale à celle du vide.

Les propriétés électriques d'un milieu sont décrites grâce à sa conductivité électrique complexe $\underline{\sigma}(\omega)$, définie comme :
\begin{empheq}[box=\mybluebox]{align}
	\mathbf{j} = \underline{\sigma}(\omega) \mathbf{E}
	\label{eqLFEITIntroConductivite}
\end{empheq}
avec $\mathbf{j}$ la densité de courant en un point, $\mathbf{E}$ le champ électrique en ce point et $\omega$ la pulsation du champ électrique. Cette conductivité électrique complexe peut être écrite comme la somme d'une partie réelle et imaginaire selon la relation : $\underline{\sigma}(\omega) = \sigma'(\omega) + i \sigma''(\omega)$, avec $\sigma'(\omega)$ la conductivité électrique réelle, $i$ le nombre complexe dont le carré vaut -1 et $\sigma''(\omega)$ la conductivité électrique imaginaire. Cette grandeur représente la conductance d'un volume de matière et a pour unité les Siemens par mètre (S.m$^{-1}$). La résistivité $\rho$ est définie comme l'inverse de $\underline{\sigma}(\omega)$, avec pour unité des $\Omega$.m.

La conductivité électrique réelle $\sigma'(\omega)$ représente l'ensemble des pertes dans le milieu dû au déplacement des charges libres et liées. Cette grandeur peut être séparée en la somme de deux composantes, $\sigma_S$ qui représente les pertes ohmiques dans le milieu (résistance au mouvement des charges libres) et $\sigma_d(\omega)$ qui représente les pertes diélectriques, dépendant de la fréquence (résistance au mouvement des charges liées).

La conductivité électrique imaginaire $\sigma''(\omega)$ traduit la polarisabilité du matériau, donc sa capacité à stocker de l'énergie sous forme d'un champ électrique. Plusieurs mécanismes de polarisation existent, comme la polarisation d'orientation de molécules dipolaires rigides, la polarisation ionique, la déformation du nuage électronique des atomes... Chaque type de polarisation présente une réponse maximale à une fréquence du champ électrique définie, ce qui correspond à une résonance conduisant à une absorption d'énergie par le milieu -- c'est ce que l'on appelle le phénomène de relaxation. On peut définir la permittivité relative $\epsilon_r'$ comme la conductivité électrique imaginaire divisée par la pulsation : $\epsilon' = \frac{\sigma''}{\epsilon_0 \omega}$. Cette grandeur est plus souvent utilisée pour les tissus biologiques car elle présente une amplitude de variation moins importante que $\sigma''(\omega)$.

\begin{figure}[!ht]
 \begin{center}
	 \includegraphics[width=0.5\columnwidth]{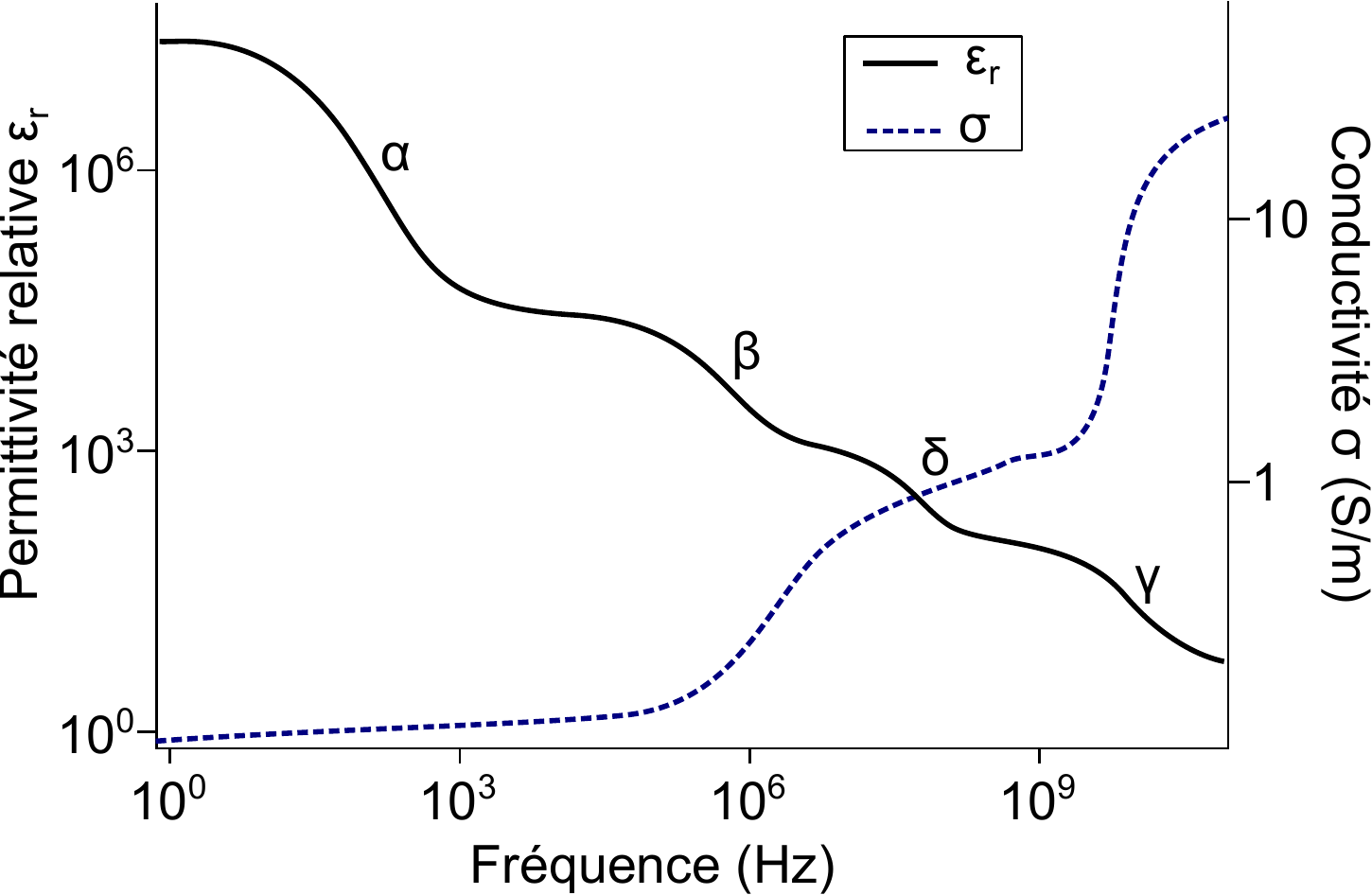}
	 \caption{Variation de la conductivité électrique et de la permittivité relative typique d'un tissu en fonction de la fréquence. On observe plusieurs sauts, correspondants à des phénomènes de relaxation, que l'on nomme $\alpha$, $\beta$, $\delta$ et $\gamma$ \cite{schwan1957conductivity}}
	 \label{figLFEITIntroConducElecTissus1}
 \end{center}
\end{figure}

Comme représenté sur la figure \ref{figLFEITIntroConducElecTissus1} issue d'un modèle donné par Schwan \textit{et al.} \cite{schwan1957conductivity}, on observe expérimentalement quatre grands phénomènes de relaxation dans les tissus biologiques à des fréquences inférieures au GHz, que l'on nomme respectivement $\alpha$, $\beta$, $\delta$ et $\gamma$ \cite{grimnes2002interface}.

La relaxation $\alpha$ apparait entre quelques Hz et quelques kHz. Elle correspond à une diffusion des ions au niveau de la membrane des cellules : la formation d'une double couche d'ions à la surface des membranes fait apparaitre des dipôles électriques locaux. La conductivité électrique réelle varie peu tandis que la permittivité relative $\epsilon_r$ diminue fortement.

La relaxation $\beta$ se situe entre 500 kHz et 20 MHz. A ces fréquences, la membrane n'est plus isolante, et la conductivité électrique réelle devient représentative de la conductivité intracellulaire et extracellulaire. D'autre part, la membrane se polarise moins bien, donc la permittivité relative diminue.

La relaxation $\delta$, autour de 50 MHz, est plutôt faible et est due à la relaxation de molécules d'eau au voisinage de macromolécules.

Enfin, la relaxation $\gamma$ n'intervient qu'à très haute fréquence, autour de 17 GHz, et a pour origine l'orientation dipolaire des molécules d'eau.

\subsection{Les différents modèles de conductivité électrique des tissus biologiques}

Les liquides biologiques sont bien modélisés par des électrolytes. Différents modèles empiriques ont été proposés pour modéliser la dépendance fréquentielle des propriétés électriques des autres tissus, en particulier les tissus conjonctifs \cite{duck1990physical}.

\begin{figure}[!ht]
 \begin{center}
	 \includegraphics[width=0.3\columnwidth]{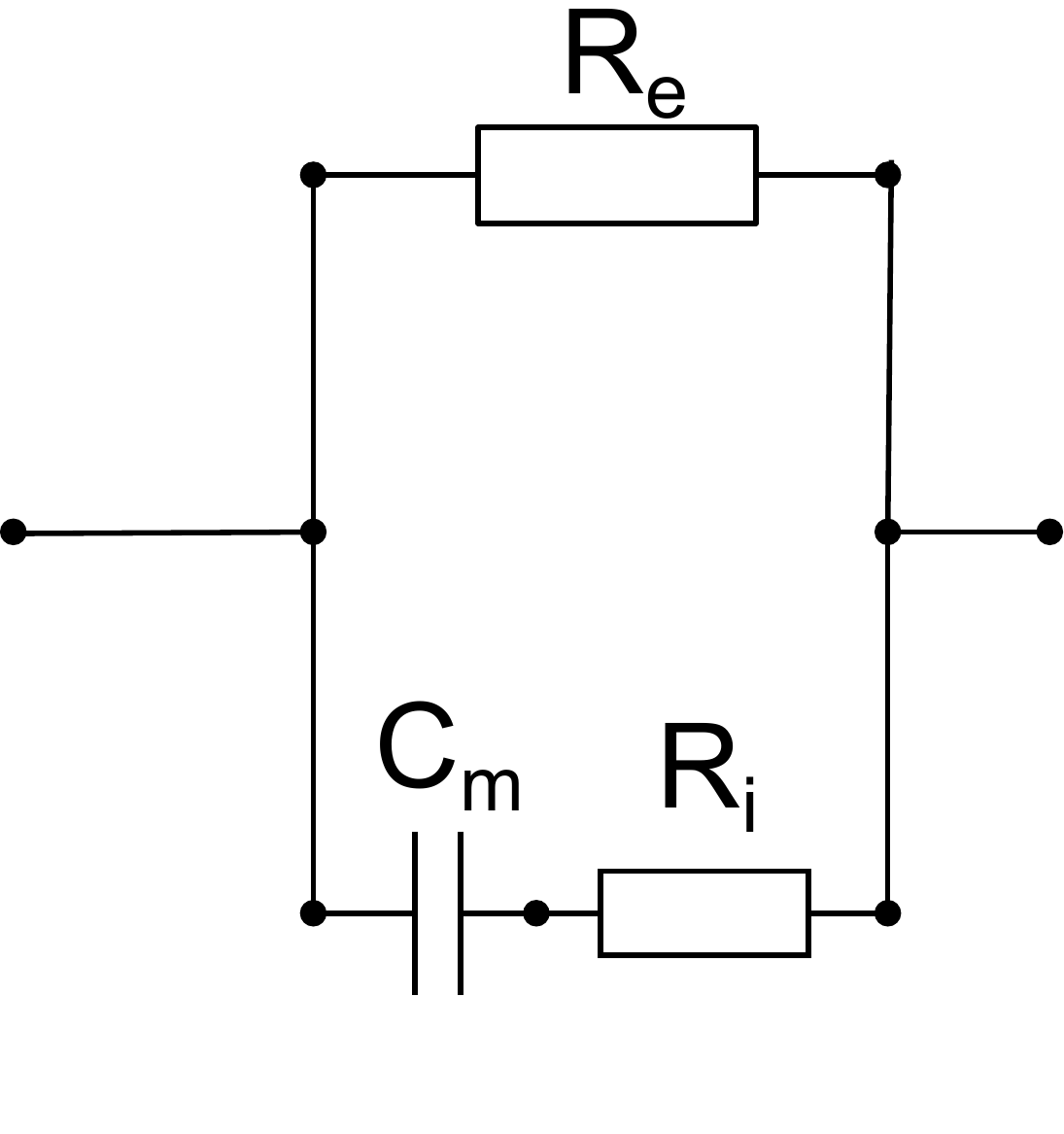}
	 \caption{Modélisation par un circuit électrique d'un tissu biologique simple. Ce milieu est composé de cellules dont la membrane a une capacité $C_i$ et un milieu intracellulaire de résistance $R_i$, et baignent dans un milieu extracellulaire de résistance $R_e$}
	 \label{figLFEITIntroCircuitElec}
 \end{center}
\end{figure}

\paragraph{Le modèle de Fricke}
Pour des milieux biologiques peu complexes, on peut se contenter d'un modèle de conductivité électrique simple comme le modèle de Fricke \cite{fricke1932}. Ce modèle suppose que l'on a un tissu conjonctif isotrope, sans inhomogénéité ni molécules polaires, et essentiellement composé de cellules baignant dans un liquide. On considère que les cellules sont composées d'une membrane de quelques micromètres d'épaisseur, de capacité électrique $C_m$, renfermant un noyau et diverses structures que l'on désignera sous le terme de milieu intracellulaire, avec une résistance $R_i$. Cette membrane, partiellement perméable, isole le milieu intracellulaire de l'extérieur, le milieu extracellulaire, de résistance $R_e$. On peut alors donner un circuit électrique équivalent avec en parallèle une résistance $R_i$ et un dipôle condensateur $C_m$ et résistance $R_e$, comme représenté sur la figure \ref{figLFEITIntroCircuitElec}.

\paragraph{Le modèle de Debye}
La plupart des tissus ne peuvent cependant pas être modélisés aussi simplement. En particulier, les tissus biologiques présentent plusieurs phénomènes de relaxation qui ne sont pas prédits par le modèle ci-dessus. Si on considère plusieurs seuils de relaxation $k$, on peut adopter le modèle de Debye \cite{debye1929polar}. Dans ce modèle, la conductivité électrique $\underline{\sigma}(\omega)$ peut être écrite sous la forme :
\begin{empheq}[box=\mybluebox]{align}
	\underline{\sigma}(\omega) = \sigma_{\infty} + i\omega\epsilon_0\epsilon_{\infty} + \sum_k{\frac{\Delta\sigma_k}{1-i \omega \tau_k}}
	\label{eqLFEITIntroModeleDebye}
\end{empheq}
avec $\sigma_{\infty}$ la conductivité électrique du tissu limite à haute fréquence, $\epsilon_{\infty}$ la permittivité diélectrique à très haute fréquence, $\Delta\sigma_k$ la variation de conductivité électrique pour la relaxation $k$ et $\tau_k$ la constante de temps du phénomène de relaxation $k$ considéré.

\paragraph{Le modèle de Cole-Cole}
Le modèle de Debye peut être amélioré en prenant en compte une distribution de constantes de temps plutôt qu'une série discrète, ce qui donne le modèle de Cole-Cole \cite{cole1941dispersion} défini selon l'équation :
\begin{empheq}[box=\mybluebox]{align}
	\underline{\sigma}(\omega) = \sigma_{\infty} + i\omega\epsilon_0\epsilon_{\infty} + \sum_k{\frac{\Delta\sigma_k}{1+ (-i \omega\tau_k)^{1-\alpha_k}}}
	\label{eqLFEITIntroModeleColeCole}
\end{empheq}
avec $\alpha_k$ un paramètre empirique qui caractérise le temps caractéristique du phénomène de relaxation $k$. A noter que ce modèle est essentiellement empirique mais n'a pas spécialement de justification physique \cite{mcadams1996problems}.

\subsection{Courbes représentatives (muscle, sang, graisse, os)}
\begin{figure}[!ht]
 \begin{center}
	 \includegraphics[width=0.95\columnwidth]{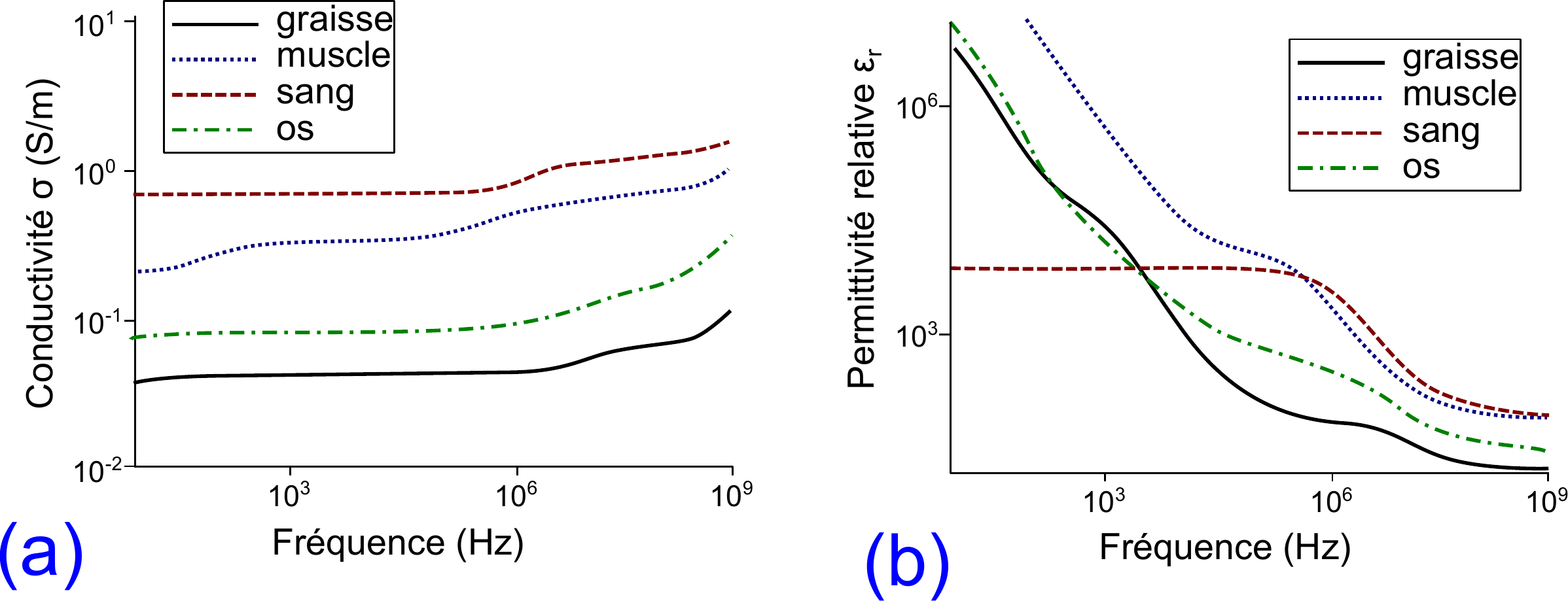}
	 \caption{Variation de la conductivité électrique (a) et de la permittivité relative (b) de la graisse, du sang, d'un muscle et d'un os en fonction de la fréquence. A 500 kHz, fréquence principale étudiée dans cette étude, on observe une conductivité électrique dix fois inférieure pour la graisse que pour le muscle \cite{gabriel1996dielectric1}.}
	 \label{figLFEITIntroConducElecTissus2}
 \end{center}
\end{figure}

La conductivité électrique de plusieurs milieux biologiques est représentée sur la figure \ref{figLFEITIntroConducElecTissus2}, à partir d'une revue de Gabriel \textit{et al.} faisant référence en la matière \cite{gabriel1996dielectric1}, \cite{gabriel1996dielectric2}. Dans les expériences de ce chapitre, on se placera principalement à une fréquence de 500 kHz. A cette fréquence, on observe que la conductivité électrique d'un morceau de tissu adipeux, égale à 0,08 S/m, est environ dix fois inférieure à celle d'un morceau de muscle, égale à 0,5 S/m (on ne tient pas compte des effets d'anisotropie du muscle, qui sont faibles à une fréquence supérieure à une centaine de kilohertz). Des valeurs issues d'un modèle de Cole-Cole sont également indiquées en annexe.

\subsection{Tissus sains et tissus tumoraux}
Plusieurs études montrent que les tissus tumoraux présentent des propriétés électriques différentes des tissus sains. Par exemple, Haemmerich \textit{et al.} \cite{haemmerich2003} montre qu'à 1 MHz, les tumeurs hépatiques présentent une conductivité électrique de (2,69 $\pm$ 0,91) 10$^{-1}$ S.m$^{-1}$ tandis que les tissus sains avaient une conductivité électrique de (4,61 $\pm$ 0,42) 10$^{-1}$ S.m$^{-1}$ : cela montre que les deux types de tissus peuvent être différenciés par leur conductivité électrique. La mesure à différentes fréquences (spectroscopie d'impédance électrique) permet par ailleurs de bien distinguer les tissus normaux des tissus cancéreux, par exemple dans le sein \cite{jossinet1999review}, pour la peau \cite{aberg2004skin} ou pour la prostate \cite{halter2008electrical}.

\subsection{Les lésions thermiques dans les tissus}
\label{secLFEITIntroLesionThermique}
La température affecte notablement la conductivité électrique des tissus. Duck \cite{duck1990physical} montre par exemple qu'un morceau de muscle chauffé entre 20 et 40°C présente à une fréquence de 100 kHz une augmentation de conductivité électrique de 2,1 \% .$^o$C$^{-1}$ (valeur qui se retrouve dans de nombreux tissus). On peut donc faire des suivis de température par imagerie de conductivité électrique. Cela serait particulièrement utile pour suivre la formation de lésions thermiques \cite{jossinet2005impedance}, notamment par ultrasons focalisés à haute intensité.

Un modèle particulièrement intéressant a été proposé par Pop \textit{et al.} \cite{pop2003}, où le changement de conductivité électrique dans les tissus chauffés est séparé en deux parties : une partie réversible, simplement liée à l'augmentation de la température, et une partie irréversible, liée à des modifications structurelles du tissu. De cette étude ressort que les tissus peuvent emmagasiner une certaine quantité de chaleur sans dommage. Lorsque la quantité de chaleur dépasse une certaine valeur, des dommages irréversibles apparaissent. Ainsi, lorsque le tissu est refroidi à sa température initiale, les valeurs de conductivité et de permittivité ne reviennent pas à leur valeur initiale si la quantité de chaleur absorbée a été suffisante.

\section{Les méthodes d'imagerie d'impédance électrique}
Bien que les propriétés électriques des tissus biologiques aient été intensément étudiées comme montré dans la section précédente, peu de méthodes existent pour faire des mesures non invasives de conductivité électrique. En imagerie médicale, la méthode la plus ancienne et la plus développée est la Tomographie d'Impédance Electrique, mais ne reste que rarement utilisée en clinique. Plusieurs autres méthodes tentent de produire des images de l'impédance électrique des tissus, mais elles n'en sont à l'heure actuelle qu'au stade de recherche et développement.

\subsection{La tomographie d'impédance électrique}
\begin{figure}[!ht]
   \begin{minipage}[c]{.5\linewidth}
		\begin{center}
	   		\includegraphics[width=1\linewidth]{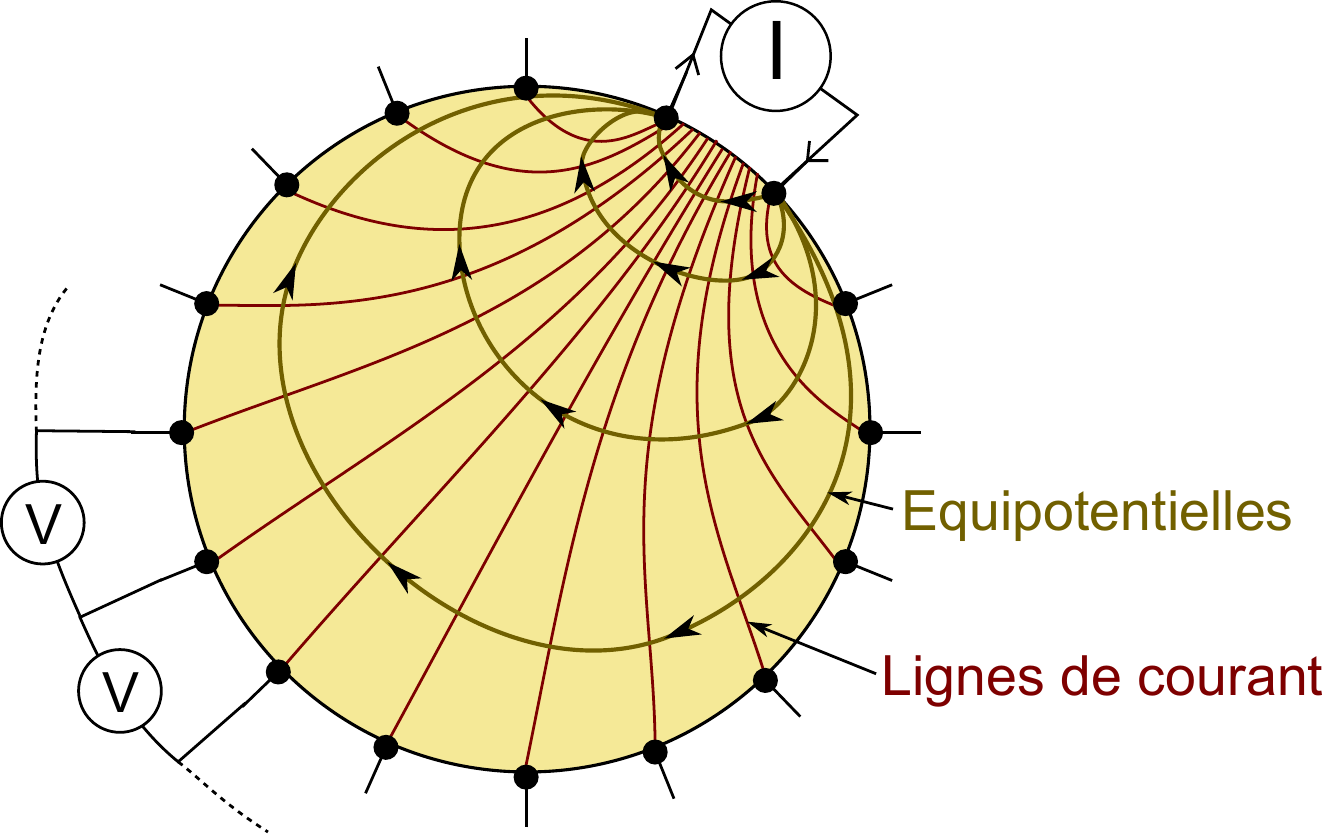}
		\end{center}
   \end{minipage} \hfill
   \begin{minipage}[c]{.5\linewidth}
		\begin{center}
	   		\includegraphics[width=1\linewidth]{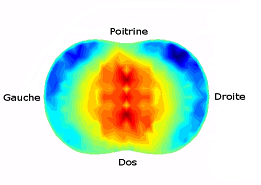}
		\end{center}
   \end{minipage}
 	\caption{(a) Principe de la tomographie d'impédance électrique. Des électrodes sont placées tout autour d'un tissu à étudier. Un courant électrique est injecté dans le tissu par deux électrodes et le potentiel électrique est capté par toutes les autres. Le schéma représente des lignes de courant en rouge et des équipotentielles en beige. (b) Résultat d'une tomographie d'impédance électrique au niveau de la poitrine d'un patient, avec un niveau de couleur relié à la conductivité électrique locale. La résolution spatiale est médiocre, mais permet pour un \oe il expérimenté de savoir si les poumons sont sains ou non \cite{lionheart1997electrical}.}
 	\label{figLFEITMethodesEIT} 
\end{figure}

La technique la plus avancée aujourd'hui se nomme la Tomographie d'Impédance Electrique \cite{cheney1999} (EIT pour ``Electrical Impedance Tomography''). Dans cette technique, schématisée sur la figure \ref{figLFEITMethodesEIT}, des dizaines d'électrodes sont placées sur le patient. Un courant électrique suffisamment faible pour être inoffensif est injecté entre deux électrodes et le potentiel dans le tissu est mesuré par toutes les autres. Puis on change successivement d'électrodes d'injection afin d'avoir un ensemble de mesures. A partir de l'ensemble des mesures, une image tridimensionnelle de la conductivité électrique peut être reconstruite. Il s'agit d'un problème inverse dit ``mal posé'' à cause du faible nombre de mesures comparé au nombre de points à reconstruire. Ainsi, malgré des progrès mathématiques importants et l'utilisation d'hypothèses de reconstruction, la résolution spatiale reste faible par rapport aux autres méthodes d'imagerie médicale \cite{brown2003electrical}. Cette technique est cependant utilisée cliniquement pour certaines pathologies du poumon \cite{frerichs2002detection}, en complément d'une technique d'imagerie classique. D'autres appareils sont en cours de développement, comme pour l'imagerie de la prostate avec une sonde endoscopique \cite{jossinet2002electrical}. On peut également citer un appareil utilisé cliniquement nommé T-scan basé sur ce principe, même s'il ne produit pas d'image à proprement parler. Cet appareil permet de détecter des cancers du sein par l'injection dans le corps d'une patiente d'un courant électrique récupéré au niveau du sein : un courant anormal peut indiquer la présence de nodules cancéreux \cite{ammari2004b}.

\subsection{L'imagerie d'impédance électrique par résonance magnétique}
\begin{figure}[!ht]
 \begin{center}
	 \includegraphics[width=0.9\columnwidth]{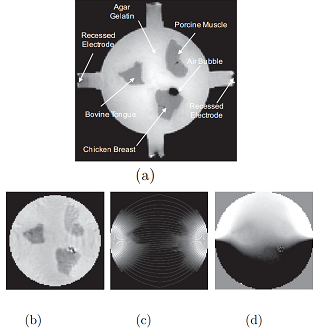}
	 \caption{(a) Image IRM d'un fantôme d'agar cylindrique contenant trois échantillons de tissu biologique. (b) Image de conductivité électrique du même échantillon en utilisant la méthode de tomographie d'impédance électrique par résonance magnétique. (c) Image des lignes de courant et de l'amplitude de la densité de courant J en chaque point. Le courant était injecté de l'électrode de gauche vers celle de droite. (d) Densité de flux de champ magnétique B modifié par la densité de courant montré en (c) \cite{seo2011magnetic}.}
	 \label{figLFEITMethodesMREIT}
 \end{center}
\end{figure}

La Tomographie d'Impédance Electrique par Résonance Magnétique (MREIT, pour ``Magnetic Resonance Electrical Impedance Tomography'') est une technique développée à partir de 2002 d'imagerie d'impédance électrique utilisant, comme son nom l'indique, un appareil d'imagerie par résonance magnétique \cite{kwon2002}, \cite{woo2008}. Le principe, représenté sur le figure \ref{figLFEITMethodesMREIT}, est d'acquérir des images IRM du patient. Ensuite, un courant électrique faible est envoyé dans un organe à l'aide par des électrodes. Le courant se propage dans les tissus, avec une densité de courant supérieure dans les zones d'impédance électrique faible. La propagation de ce courant modifie le champ magnétique principal B$_z$, ce qui perturbe donc le signal de résonance magnétique. Ces signaux perturbés permettent alors de calculer l'impédance électrique locale. La résolution spatiale est identique à celle de l'IRM \cite{ammari2008}. L'une des difficultés actuelle est la détermination la quantité de courant à injecter pour avoir un rapport signal sur bruit suffisant tout en ne présentant pas de risque pour le patient \cite{oh2003}.

\subsection{L'imagerie acousto-électrique}
\begin{figure}[!ht]
 \begin{center}
	 \includegraphics[width=0.8\columnwidth]{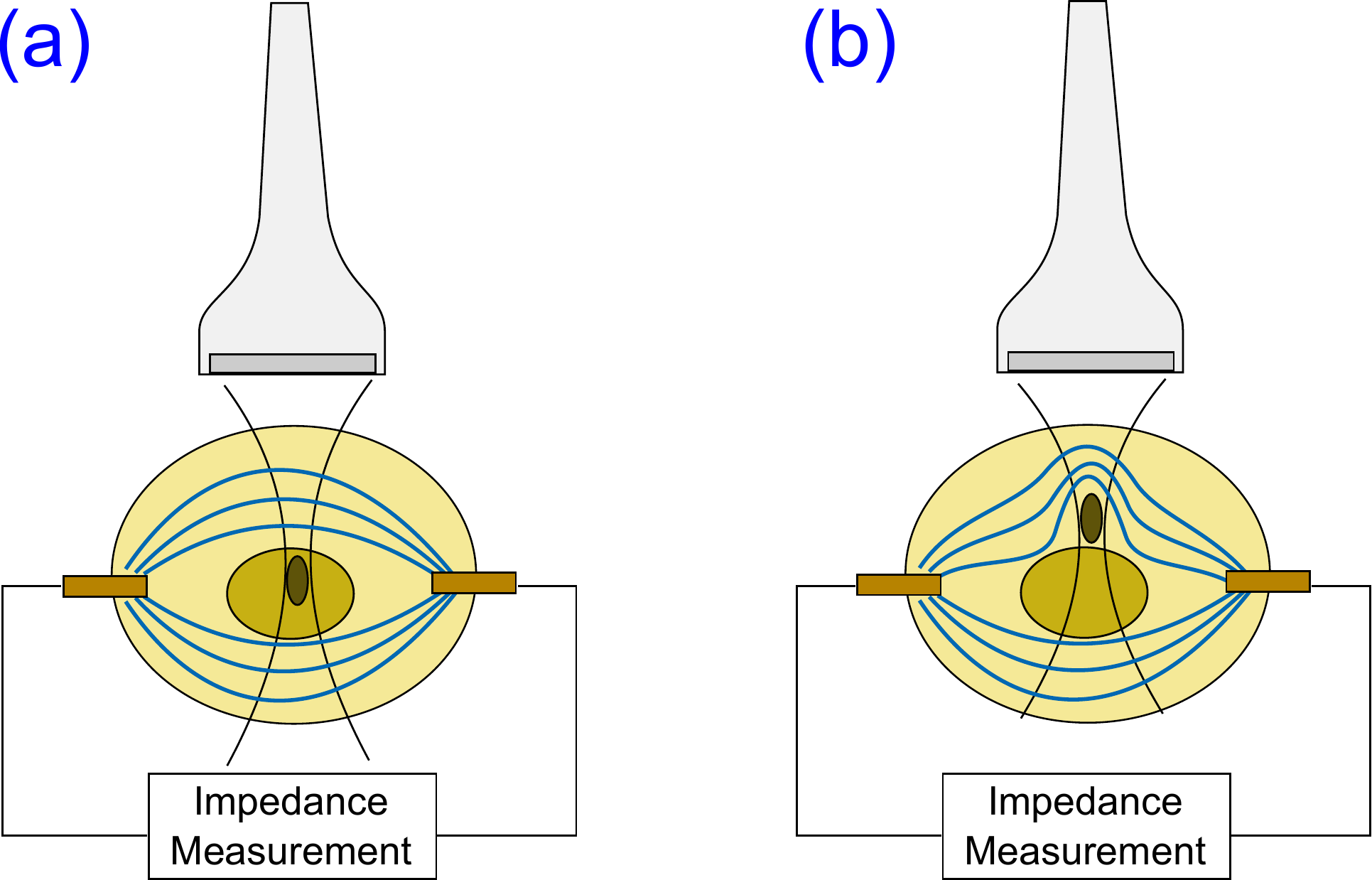}
	 \caption{Principe de l'imagerie acousto-électrique. Un courant électrique est injecté dans un tissu via deux électrodes. Une onde ultrasonore est focalisée en un point du tissu, et les changements mesurés au niveau des électrodes permettent de déduire la conductivité électrique dans la zone focale des ultrasons.}
	 \label{figLFEITMethodesAcoustoElectrique}
 \end{center}
\end{figure}

Si un courant électrique est injecté dans un organe par des électrodes, il se propage dans toutes les directions, avec une intensité qui dépend de la conductivité électrique locale. Mais il est difficile de savoir exactement par où le courant électrique passe : c'est l'une des principales difficultés de la tomographie de conductivité électrique. Cet inconvénient est contourné dans la technique d'imagerie acousto-électrique\cite{ammari2012b} (aussi appelée tomographie acousto-électrique \cite{zhang2004}), proposée au début des années 2000. Le principe de cette technique, schématisée sur la figure \ref{figLFEITMethodesAcoustoElectrique}, est de faire passer un courant électrique dans un tissu, puis de modifier la conductivité électrique en un point en y focalisant une onde acoustique. Cela modifie alors la propagation locale du courant électrique. Et plus la conductivité électrique dans la zone focale de l'onde acoustique est forte, plus la modification du courant électrique sera importante, ce qui est mesurable avec des électrodes \cite{lavandier2000}, \cite{jossinet1998phenomenology}. Ainsi, en focalisant l'onde acoustique en différents endroits du tissu, on peut obtenir une image de conductivité électrique des tissus avec une résolution spatiale de l'ordre du millimètre \cite{olafsson2006}.

\subsection{La tomographie d'impédance électrique par force de Lorentz}
\label{secLFEITMethodesEIT}
\begin{figure}[!ht]
 \begin{center}
	 \includegraphics[width=0.8\columnwidth]{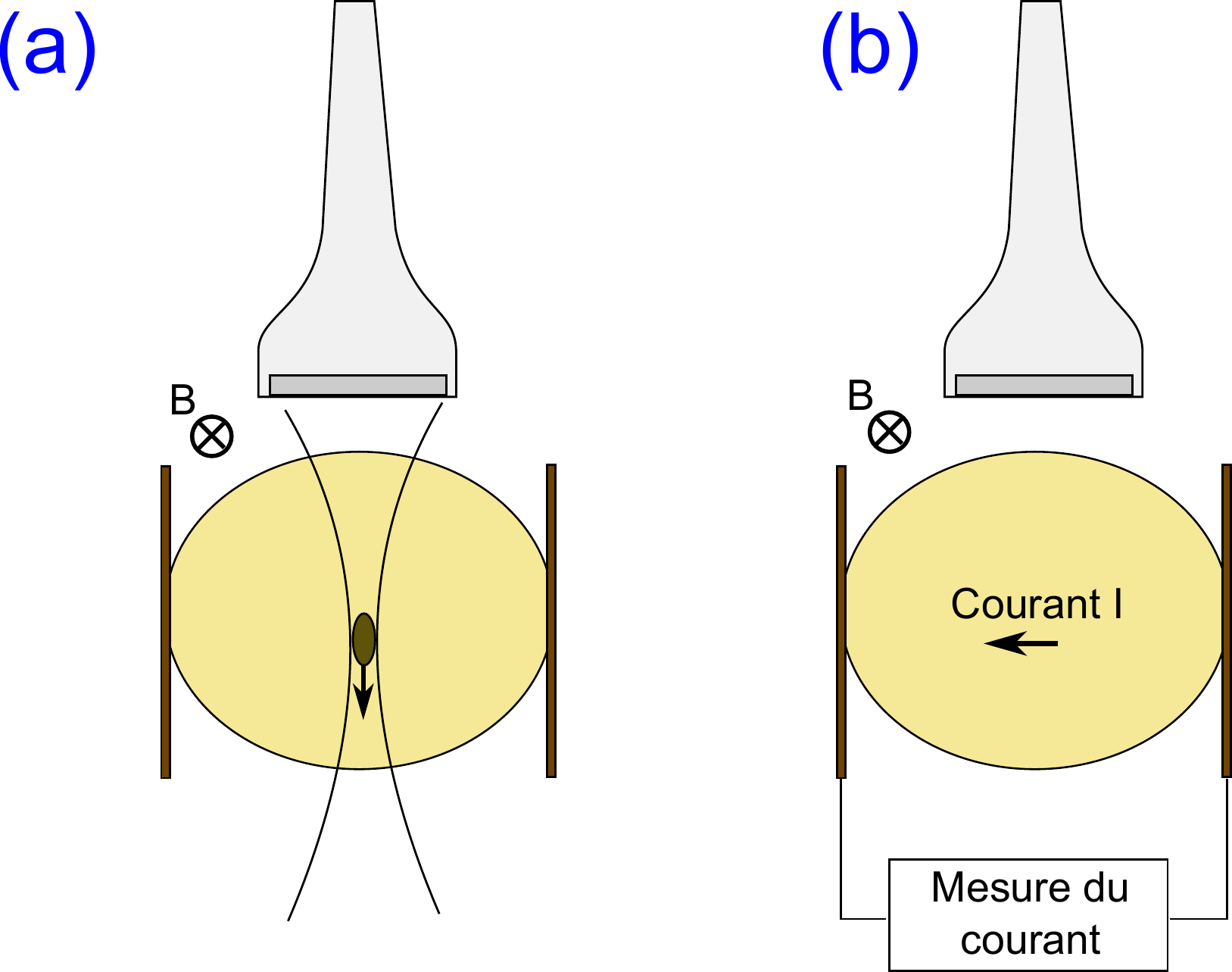}
	 \caption{Principe de la tomographie d'impédance électrique par force de Lorentz. Des ultrasons sont focalisés dans un tissu (a), et celui-ci est mis en mouvement dans la zone focale (ellipse marron). Le déplacement du tissu dans un champ magnétique crée un courant électrique mesurable par des électrodes (b).}
	 \label{figLFEITMethodesLFEIT}
 \end{center}
\end{figure}

Une autre technique d'imagerie d'impédance électrique a été proposée sous le nom d'Imagerie par Effet Hall (HEI pour ``Hall Effect Imaging'') en 1998 \cite{wen1998}. Ce nom a ensuite été critiqué par certains auteurs \cite{roth1998comments}, car le phénomène utilisé n'est selon eux pas l'effet Hall, si bien que d'autres appellations ont été proposées, comme la Tomographie Magnéto-Acousto-Electrique (MAET, pour ``Magneto-Acousto-Electrical Tomography'') \cite{xu2007} ou la Tomographie d'Impédance Electrique par Force de Lorentz (LFEIT pour ``Lorentz Force Electrical Impedance Tomography'') \cite{grasland2013LFEIT}. Ce dernier nom, qui a été proposé dans le cadre de cette thèse, permet à la fois de caractériser le paramètre étudié (l'impédance électrique) et la méthode pour y parvenir (la force de Lorentz). Dans cette méthode, représentée sur la figure \ref{figLFEITMethodesLFEIT} et qui sera décrite plus amplement dans la suite de ce chapitre, un transducteur émet des ultrasons dans un tissu biologique soumis à un champ magnétique. Le tissu, conducteur, se déplace donc dans un champ magnétique, ce qui induit un courant électrique par force de Lorentz. Ce courant peut être mesuré par des électrodes placées autour de l'échantillon. Il a été montré que ce courant est relié à l'impédance électrique, ou plus précisément aux variations d'impédance électrique, dans la zone focale des ultrasons. En focalisant dans différentes zones du tissu, une image des interfaces d'impédance électrique peut donc être reconstruite, avec une résolution spatiale proche de celle des ultrasons \cite{roth2011}.

\subsection{La tomographie magnéto-acoustique}
\label{secLFEITMethodesMATMI}

On peut utiliser le principe de la tomographie d'impédance électrique par force de Lorentz en sens ``inverse'' \cite{towe1988}, \cite{wen1998}, \cite{montalibet2002these}. Cette technique combine un courant électrique et un champ magnétique pour créer des ondes ultrasonores.

\begin{figure}[!ht]
   \begin{minipage}[c]{.6\linewidth}
		\begin{center}
	   		\includegraphics[width=1\linewidth]{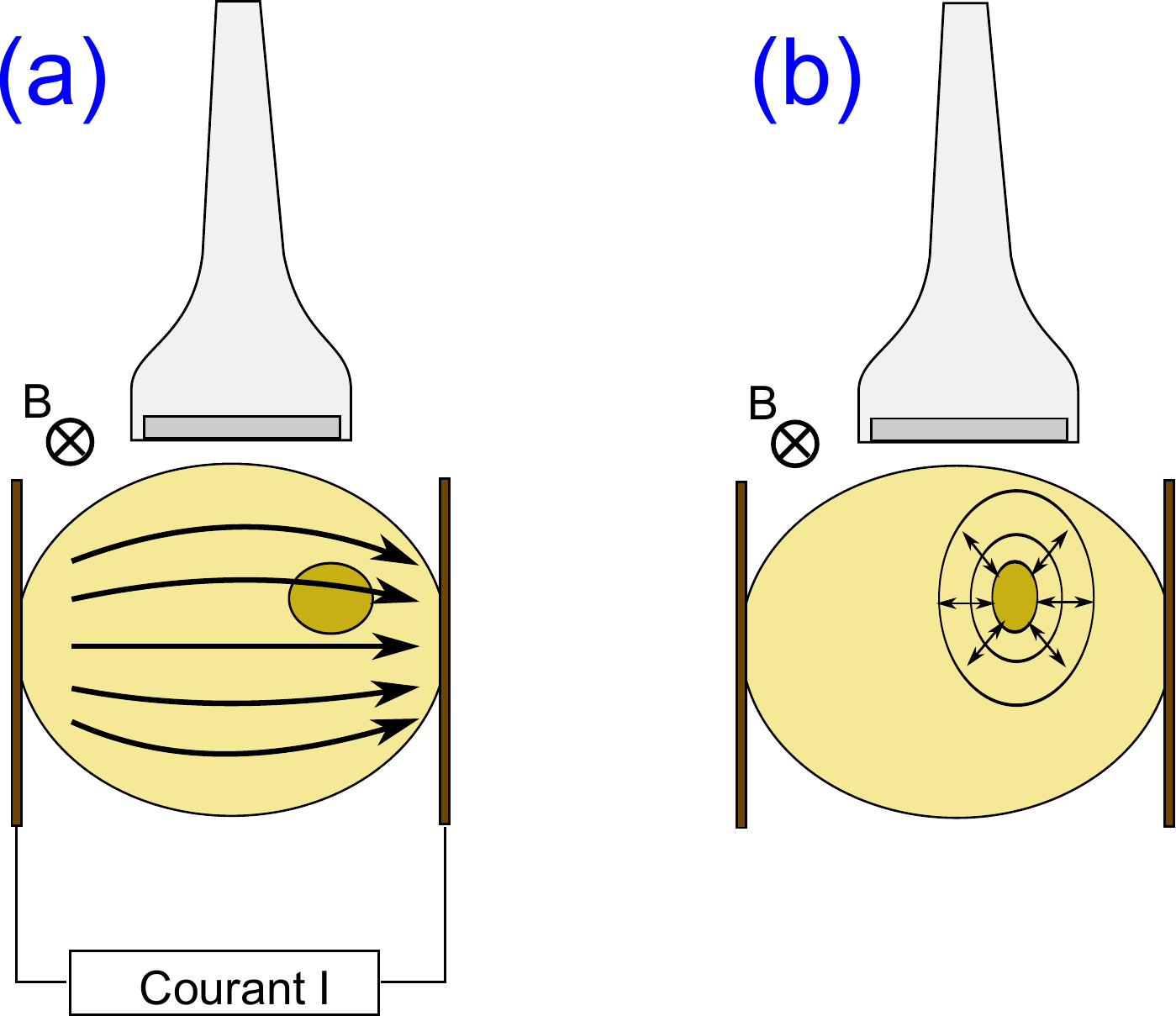}
		\end{center}
   \end{minipage} \hfill
   \begin{minipage}[c]{.4\linewidth}
		\begin{center}
	   		\includegraphics[width=1\linewidth]{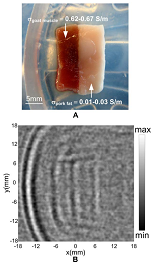}
		\end{center}
   \end{minipage}
 	\caption{Principe de la tomographie magnéto-acoustique. Un courant électrique est appliqué dans un tissu conducteur soumis à un champ magnétique (a), ce qui induit aux interfaces de conductivité électrique (ellipse beige) une onde ultrasonore détectable par des transducteurs placés autour du tissu (b). Dans le cas de la tomographie magnéto-acoustique avec induction magnétique, le courant électrique est induit grâce à un champ magnétique transitoire. (A) Photographie d'un échantillon comprenant un morceau de muscle de chèvre et un morceau de graisse de porc dont une image de tomographie magnéto-acoustique avec induction magnétique. (B) Image de conductivité électrique prise avec la technique MAT-MI de l'échantillon présenté en (A) \cite{hu2011}.}
 	\label{figLFEITMethodesMATMI} 
\end{figure}

Dans cette technique, schématisée sur les figures \ref{figLFEITMethodesMATMI}-(a) et -(b), on émet dans un tissu biologique un courant électrique à l'aide de deux électrodes. Le tissu étant placé dans un champ magnétique, des mouvements seront induits par force de Lorentz. Le mouvement sera d'amplitude différente de part et d'autre de chaque interface de conductivité électrique, ce qui va donner lieu à la formation d'une onde de compression -- aux fréquences utilisées, supérieures à la centaine de kilohertz, on peut négliger l'apparition des ondes de cisaillement car celles-ci sont très rapidement atténuées. Les ondes de compression peuvent être détectées par des transducteurs ultrasonores placés tout autour de l'échantillon, ce qui permet grâce à leurs temps de propagation de retrouver les sources des ondes, donc de réaliser des images des interfaces de conductivité électrique \cite{montalibet2002scanning}, \cite{montalibet2001electric}.

Une amélioration du dispositif a été proposée où le courant électrique est injecté dans l'échantillon non pas à l'aide d'une paire d'électrodes, mais par induction avec un champ magnétique variable créé une bobine : cela a donné lieu à la tomographie magnéto-acoustique avec induction magnétique (MAT-MI pour ``Magneto-Acoustic Tomography with Magnetic Induction'') \cite{xu2005}. Cette méthode présente l'avantage d'être sans contact et donc d'éviter les problèmes de contact avec les électrodes. Les derniers résultats sont plutôt encourageants, avec des essais sur des tissus biologiques ex-vivo \cite{hu2011}, \cite{hu2011magnetoacoustic}, dont un résultat est montré sur les figures \ref{figLFEITMethodesMATMI}-(A) et -(B).

%
%
%\begin{figure}[!ht]
 %\begin{center}
	 %\includegraphics[width=0.7\columnwidth]{figLFEITMethodesMATMI.pdf}
	 %\caption{Principe de la tomographie magnéto-acoustique. Un courant électrique est appliqué dans un tissu conducteur soumis à un champ magnétique (a), ce qui induit aux interfaces de conductivité électrique (ellipse beige) une onde ultrasonore détectable par des transducteurs placés autour du tissu (b). Dans le cas de la tomographie magnéto-acoustique avec induction magnétique, le courant électrique est induit grâce à un champ magnétique transitoire.}
	 %\label{figLFEITMethodesMATMI}
 %\end{center}
%\end{figure}
%
%
%\begin{figure}[!ht]
 %\begin{center}
	 %\includegraphics[width=0.7\columnwidth]{figLFEITMethodesMATMIResultat.png}
	 %\caption{Résultat de tomographie magnéto-acoustique par induction magnétique. (A) photographie d'un échantillon comprenant un morceau de muscle de chèvre et un morceau de graisse de porc ; (B) image de conductivité électrique prise avec la technique MAT-MI de l'échantillon présenté en (A). (source : \ref{source2})}
	 %\label{figLFEITMethodesMATMIResultat}
 %\end{center}
%\end{figure}

\section{Principe de la tomographie d'impédance électrique par force de Lorentz}

On supposera dans cette partie que le champ magnétique est selon l'axe X, la direction de propagation des ultrasons selon l'axe Z et le courant électrique principalement selon l'axe Y. Les calculs suivants se basent sur ceux effectués par A. Montalibet \cite{montalibet2002these}.

\subsection{Mouvement des ions sous l'action des ultrasons dans un champ magnétique}
Lorsqu'un champ ultrasonore est appliqué dans un tissu biologique, modélisé comme un électrolyte, le tissu est mis en mouvement. Les ions présents dans la solution sont supposés solvatés ; on peut alors décrire leur mouvement comme s'il était créé par une force de friction avec le milieu déplacé par les ultrasons. Ce mouvement suit généralement le mouvement global du milieu. Cependant, en présence d'un champ magnétique B, les ions sont en plus déviés par force de Lorentz à cause de leur charge. Cela permet d'écrire l'équation du mouvement pour un ion :
\begin{empheq}[box=\mybluebox]{align}
  m_s \frac{d\mathbf{u}}{dt} = -k (\mathbf{u} - \mathbf{v}) + q\mathbf{u}\wedge \mathbf{B}
  \label{eqLFEITPrincipeMouvement1}
\end{empheq}
avec $m_s$ masse de l'ion solvaté, $\mathbf{u}$ vitesse de l'ion, $\mathbf{v} = v\mathbf{e_z}$ vitesse du fluide mise en mouvement par les ultrasons, $k$ coefficient de frottement, $\mathbf{B}$ champ magnétique et $q$ charge de l'ion.

Le coefficient $k$ peut simplement être évalué avec une expérience où un ion est déplacé par un champ électrique $\mathbf{E}$ dans un fluide où il a atteint sa vitesse limite $v_{lim}$, comme décrit par l'équation \ref{eqLFEITPrincipeMouvement2} :
\begin{empheq}[box=\mybluebox]{align}
\begin{array}{rcl}
  k v_{lim} &=& q E\\
  \Rightarrow \frac{k}{q} &=& \frac{E}{v_{lim}} = \frac{1}{\mu}\\
\end{array}
\label{eqLFEITPrincipeMouvement2}
\end{empheq}
$\mu$ étant défini comme la mobilité de l'ion. Ce coefficient vaut typiquement 5.10$^8$ T$^{-1}$ pour un ion solvaté.
 
En remplaçant $k$ par $q/\mu$ dans l'équation (\ref{eqLFEITPrincipeMouvement1}), on obtient donc simplement
\begin{empheq}[box=\mybluebox]{align}
  \frac{d\mathbf{u}}{dt} + \frac{q}{\mu m_s} (\mathbf{u} - v\mathbf{e_z}) = \frac{q}{m_s} \mathbf{u}\wedge \mathbf{B}
  \label{eqLFEITPrincipeMouvement3}
\end{empheq}

Il n'y a aucun mouvement dans la direction X, donc on peut étudier le mouvement dans le plan YZ seulement. L'équation \ref{eqLFEITPrincipeMouvement3} se décompose selon les directions Y et Z en deux équations différentielles couplées.

On suppose que $v_z$ est de la forme $V \sin(\omega t)$ à une fréquence $f$ = 500 kHz, soit $\omega \approx 3.10^{6}$ rad.s$^{-1}$, et on prend comme ordre de grandeur $q = 1,6.10^{-19}$ C, $m_s = \frac{200}{6,02.10^{23}}$ kg, $B = 1$ T et $\mu = 5.10^{8}$ T$^{-1}$.

La résolution de ces équations donnent pour les expressions de $u_z$ et $u_y$ :
\begin{empheq}[box=\mybluebox]{align}
\left\{
	\begin{array}{rcl}
		u_z & = & \frac{V \omega}{\alpha} ( \cos \omega t - e^{-\alpha t} \cos \beta t ) + \frac{V \omega^2 \beta}{\alpha^2} (-e^{-\alpha t} \sin \beta t - \cos \omega t) + V\sin (\omega t)\\
		u_y & = & \frac{V \omega}{\alpha} ( - e^{-\alpha t} sin \beta t ) + \frac{V \omega^2 \beta}{\alpha^2} (e^{-\alpha t} \cos \beta t - \cos \omega t ) - V \frac{\omega^2 \beta}{\alpha^2} \cos (\omega t)
	\end{array}
\right.
\label{eqLFEITPrincipeMouvement8}
\end{empheq}
avec $\alpha = \frac{q}{\mu m_s}\approx 9.10^{9}$ s$^{-1}$ et $\beta =\frac{qB}{m_s}\approx 500$ s$^{-1}$

En gardant les termes dominants à $t \approx 1/\omega\approx 2$ \micro s, on obtient alors :
\begin{empheq}[box=\mybluebox]{align}
\left\{
	\begin{array}{rcl}
		u_z &=& v_0 \sin \omega t \\
		u_y &=& v_0 \mu B \sin \omega t
	\end{array}
\right.
\label{eqLFEITPrincipeMouvement9}
\end{empheq}
On retrouve ainsi l'équation donnée par Montalibet \cite{montalibet2002these}: $ u_y = \mu B u_z$. Tout se passe comme si le mouvement selon Z était uniquement dû aux ultrasons et le mouvement selon Y à la déviation par le champ magnétique. Dans toute la suite, on supposera donc que la vitesse $u_z$ des ions sera égale à $v_z$ et que la vitesse $u_y$ sera égale à $\mu B u_z$.
Cela donne, pour un déplacement de l'ordre de un micromètre dans la direction Z, une déviation d'environ dix femtomètres dans la direction Y, soit quelques noyaux atomiques !

\subsection{Densité de courant créée}
Le déplacement différent des charges positives et négatives entraine l'apparition d'une densité de courant électrique $\mathbf{j}$, définie par $\sum_k{\rho_k \mathbf{u_k}}$, avec $\rho_k$ la densité de charge $k$ et $\mathbf{u_k}$ la vitesse moyenne des charges $k$.

En décomposant selon les axes Y et Z, on peut donc écrire
\begin{empheq}[box=\mybluebox]{align}
	\mathbf{j}=\sum_k{\rho_k u_{k,z} \mathbf{e_z}} + \sum_k{\rho_k u_{k,y} \mathbf{e_y}}
\label{eqLFEITPrincipeDensité1}
\end{empheq}

Après un temps de transition négligeable, la vitesse des ions $u_{k,z}$ est égale à la vitesse du milieu $v_z$, pour les ions positifs comme pour les ions négatifs. On obtient alors $\sum{\rho_k u_{k,z} \mathbf{e_z}} = v_z \mathbf{e_z} \sum{\rho_k} = \mathbf{0}$, car le milieu est globablement neutre. La composante selon Z de la densité de courant est donc nulle.

Remplaçons $u_{k,y}$ par $\mu B u_{k,z}=\mu B v_z$ : cela donne $\sum{\rho_k u_{k,y} \mathbf{e_y}} = v_z B \mathbf{e_y} \sum{\rho_k \mu_k}$. Cette grandeur n'est pas nulle car la mobilité dépend de la charge. Cette grandeur est définie comme égale à $\underline{\sigma}$, la conductivité électrique du milieu.

La densité de courant locale peut donc être définie par :
\begin{empheq}[box=\mybluebox]{align}
	\mathbf{j}=\sigma v_z B \mathbf{e_y}
\label{eqLFEITPrincipeDensité2}
\end{empheq}

\subsection{Détection du courant}
En mesurant la densité de courant $j$ sur une surface $S$ (par exemple, la surface de l'électrode de mesure, avec une normale selon $\mathbf{e_z}$), on obtient un courant électrique global $I$ :
\begin{empheq}[box=\mybluebox]{align}
	I=\int{\int{\mathbf{j}.\mathbf{dS}}}=\int{\int{\sigma v_z B dx dz}}
\label{eqLFEITPrincipeDétection1}
\end{empheq}

La vitesse particulaire et la pression d'un fluide sont reliées par la relation $\rho \frac{d\mathbf{v}}{dt} = -\mathbf{\nabla p}$, que l'on peut réécrire sous la forme $v_z = -\frac{1}{\rho}\int{\frac{dp}{dz}dt}$. On peut donc écrire :
\begin{empheq}[box=\mybluebox]{align}
	I=-\int{\int{( \frac{\sigma}{\rho}(\int{\frac{dp}{dz}dt}) B dx dy)}}
\label{eqLFEITPrincipeDétection3}
\end{empheq}

Une intégration par parties de l'équation (\ref{eqLFEITPrincipeDétection3}) donne alors :
\begin{empheq}[box=\mybluebox]{align}
	I=[-\frac{\sigma}{\rho}B\int{p dt}]_{z_1}^{z_2} +\int{\frac{d(\sigma\frac{B}{\rho})}{dz}(\int{p dt})dz)}
\label{eqLFEITPrincipeDétection4}
\end{empheq}
Le premier terme est nul si l'onde ultrasonore n'atteint pas les électrodes \cite{montalibet2002these}.

\begin{figure}[!ht]
 \begin{center}
	 \includegraphics[width=0.7\columnwidth]{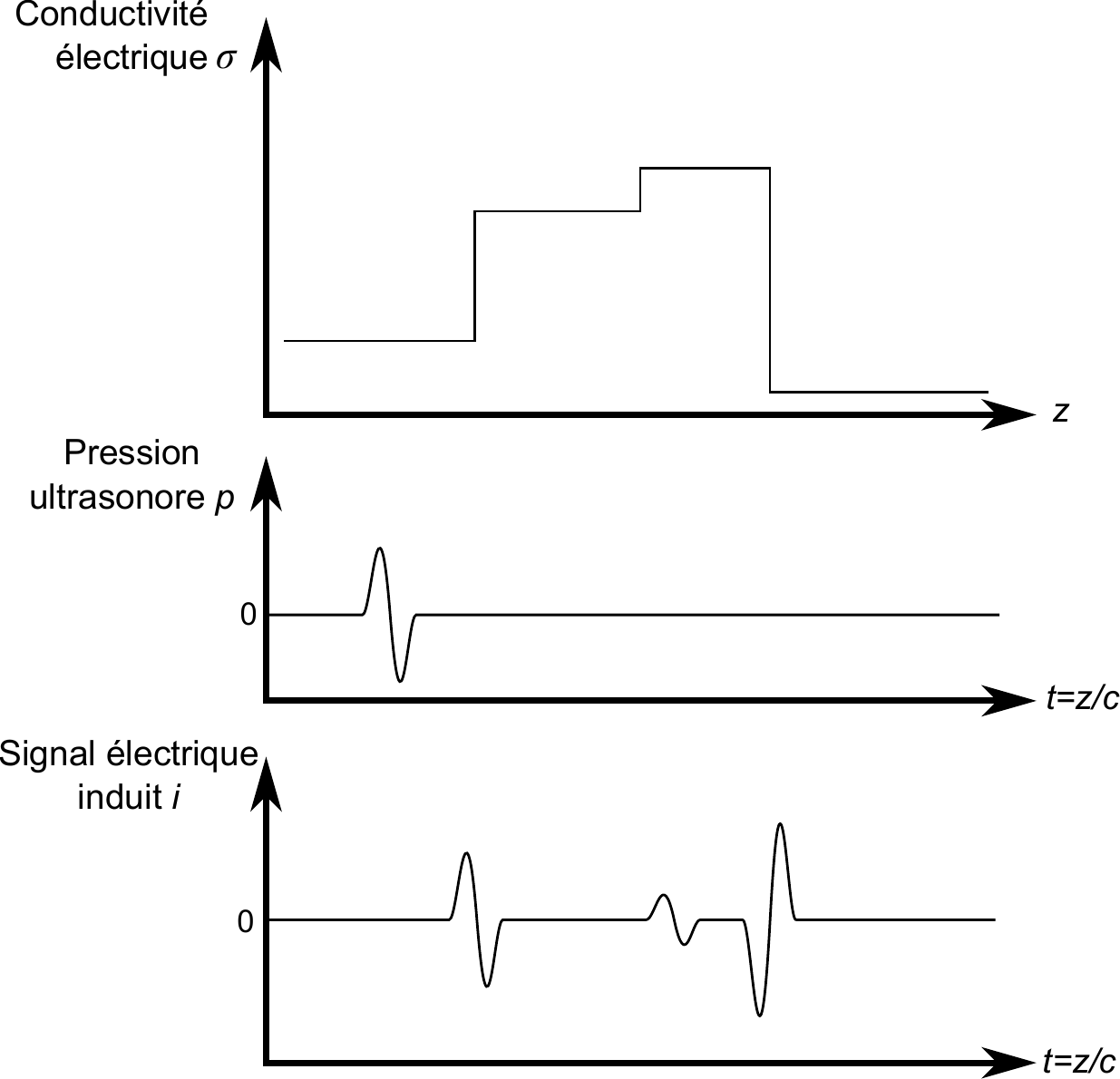}
	 \caption{Supposons que la conductivité électrique présente trois variations dans la direction Z, direction selon laquelle se propage une impulsion ultrasonore composée d'une seul sinusoïde. Le courant électrique mesuré par les électrodes, proportionnel au produit de convolution entre le gradient de conductivité électrique avec la pression ultrasonore, aura temporellement la forme d'une sinusoïde à chaque variation de conductivité électrique.}
	 \label{figLFEITPrincipeDetection}
 \end{center}
\end{figure}

Supposons que l'onde soit progressive et non atténuée dans la zone de mesure : on a donc un lien entre le temps et la distance et $\int{p(t,z)dt} = f(t-z/c)$. En supposant que le champ magnétique $B$ et la densité du milieu $\rho$ sont homogènes dans le milieu (ce dernier terme présentant des variations inférieures à 10 \% dans les tissus mous), on obtient :
\begin{empheq}[box=\mybluebox]{align}
  I(t)=\frac{B\rho}{c}\int_0^{\infty}{H(\tau) f(t-\tau)d\tau} = \frac{B\rho}{c}(H\star P)(t)
\label{eqLFEITPrincipeDétection5}
\end{empheq}
avec $\tau=z/c$, $H(\tau)=\frac{d\sigma}{dz}$ et $\star$ le symbole du produit de convolution. Cette équation est donc le produit de convolution entre la variation de conductivité électrique $H$ et la forme du signal ultrasonore $P$. En d'autre mots, le courant électrique détecté par les électrodes représente temporellement le signal ultrasonore à chaque interface de conductivité électrique, ce qui est représenté sur la figure \ref{figLFEITPrincipeDetection}

\subsection{Existence d'un \textit{speckle} acousto-électrique}
Comparons le signal donné par l'équation \ref{eqLFEITPrincipeDétection5} à un signal d'échographie. Supposons que l'onde acoustique, se propageant toujours selon l'axe Z, ne soit que très peu atténuée par les hétérogénéités acoustiques : celles-ci ne réfléchissent donc qu'une faible partie des ultrasons, et on reste dans un régime de diffusion simple (non multiple).

La pression mesurée au temps t sera égale à \ref{eqLFEITPrincipeSpeckle1} :
\begin{empheq}[box=\mybluebox]{align}
	p(t)=\sum_k{R(z_k) f(t-2z_k/c)}
	\label{eqLFEITPrincipeSpeckle1}
\end{empheq}
avec $z_k$ la position de chaque hétérogénéité acoustique $k$, $R(z_k)$ le coefficient de réflexion de chaque hétérogénéité (petit devant 1), $f$ la forme du signal ultrasonore.

Dans un modèle continu, cela donne avec $\tau=z/c$ :
\begin{empheq}[box=\mybluebox]{align}
  p(t)=\int_0^{\infty}{R(\tau) f(t-2\tau)d\tau}
	\label{eqLFEITPrincipeSpeckle2}
\end{empheq}

Ce qui correspond à un facteur 2 près au produit de convolution entre la distribution de coefficients de réflexion $R$ et la forme du signal ultrasonore $f$. Cela signifie que si les hétérogénéités sont suffisamment proches les unes des autres et que le signal ultrasonore est suffisamment long, on peut observer des interférences entre les ondes réfléchies. Cela donne des zones brillantes ou sombres, selon que les interférences soient respectivement constructives ou destructives : on nomme ce phénomène les \textit{tavelures ultrasonores}, ou plus fréquemment \textit{speckle ultrasonore} \cite{catheline1998interferometrie} par analogie avec le \textit{speckle optique}.

Les équations (\ref{eqLFEITPrincipeSpeckle2}) et (\ref{eqLFEITPrincipeDétection5}) sont finalement tout à fait similaires, en remplaçant le coefficient de réflexion $R$ (relié à la variation d'impédance acoustique) par la variation de conductivité électrique $\frac{d\sigma}{dz}$. On peut donc définir un \textit{speckle électro-acoustique} : il est de nature électrique, car il s'agit d'inhomogénéités de conductivité électrique, et acoustique, car la taille des inhomogénéités est reliée à la longueur d'onde acoustique.

\section{Dispositif expérimental}
\begin{figure}[!ht]
 \begin{center}
	 \includegraphics[width=1\columnwidth]{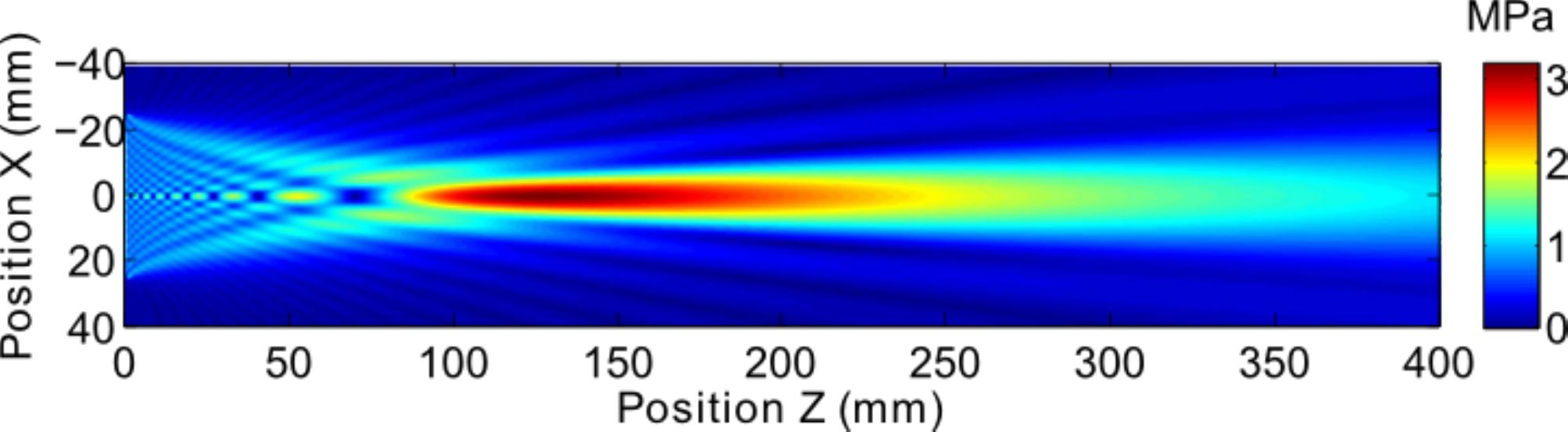}
	 \caption{Simulation du champ de pression du transducteur dans le plan YZ, pour X = 0 mm, Y = -40 à 40 mm, Z = 0 à 400 mm, avec pas de 0,2 mm. Le faisceau est à peu près cylindrique de Z = 100 à 400 mm, avec un diamètre un peu inférieur à 20 mm. La pression diminue de 3,2 MPa à Z = 150 mm à 1 MPa à Z = 400 mm.}
	 \label{figLFEITDispositifChampLong}
 \end{center}
\end{figure}

\begin{figure}[!ht]
 \begin{center}
	 \includegraphics[width=0.5\columnwidth]{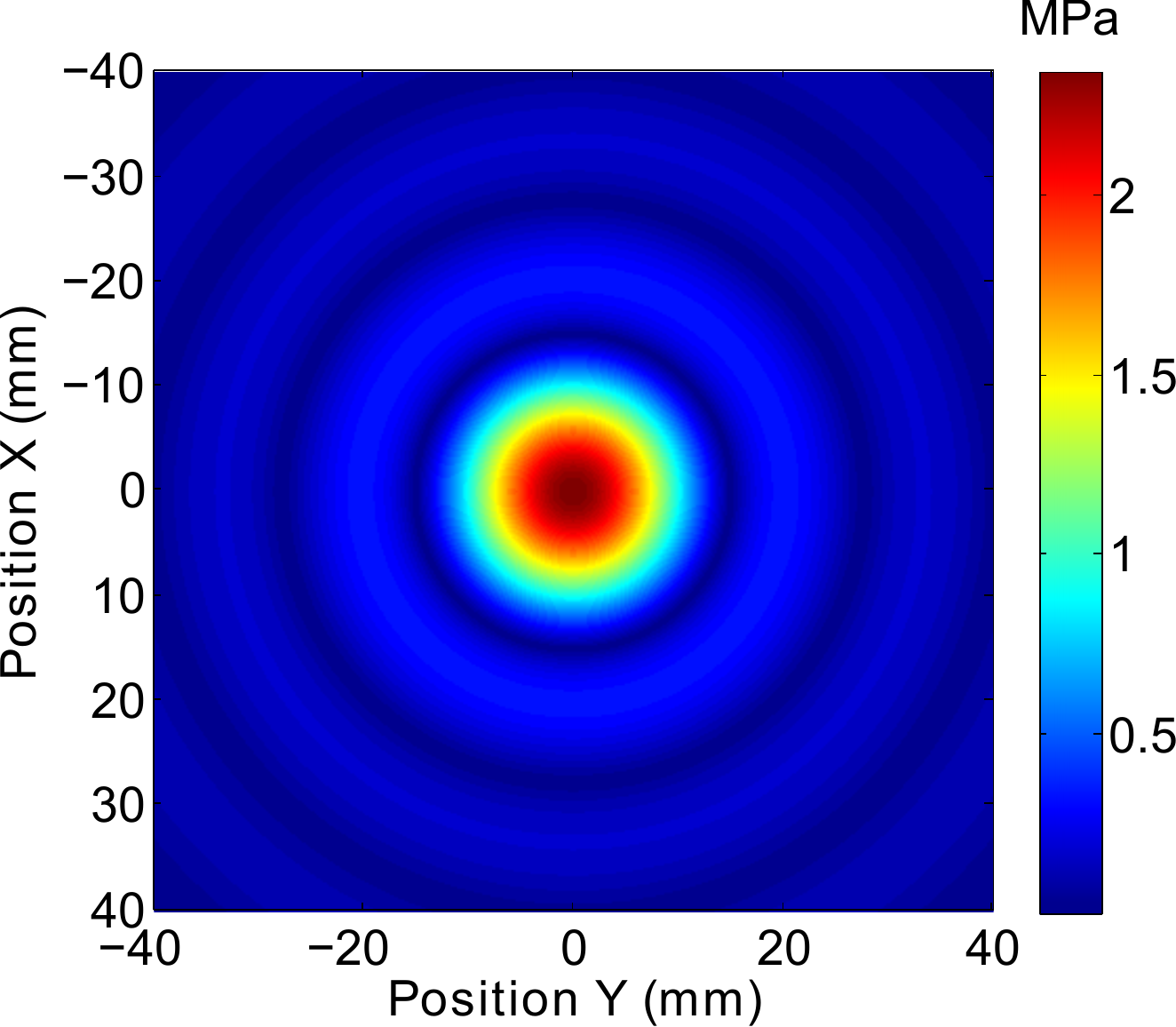}
	 \caption{Simulation du champ de pression du transducteur dans le plan XY, pour X = -40 à 40 mm, Y = -40 à 40 mm, Z = 210 mm, avec pas de 0,2 mm. On observe une décroissance du faisceau autour de l'axe, avec quelques oscillations. La largeur à mi-hauteur du faisceau vaut 18 mm.}
	 \label{figLFEITDispositifChampLat}
 \end{center}
\end{figure}

\begin{figure}[!ht]
 \begin{center}
	 \includegraphics[width=0.9\columnwidth]{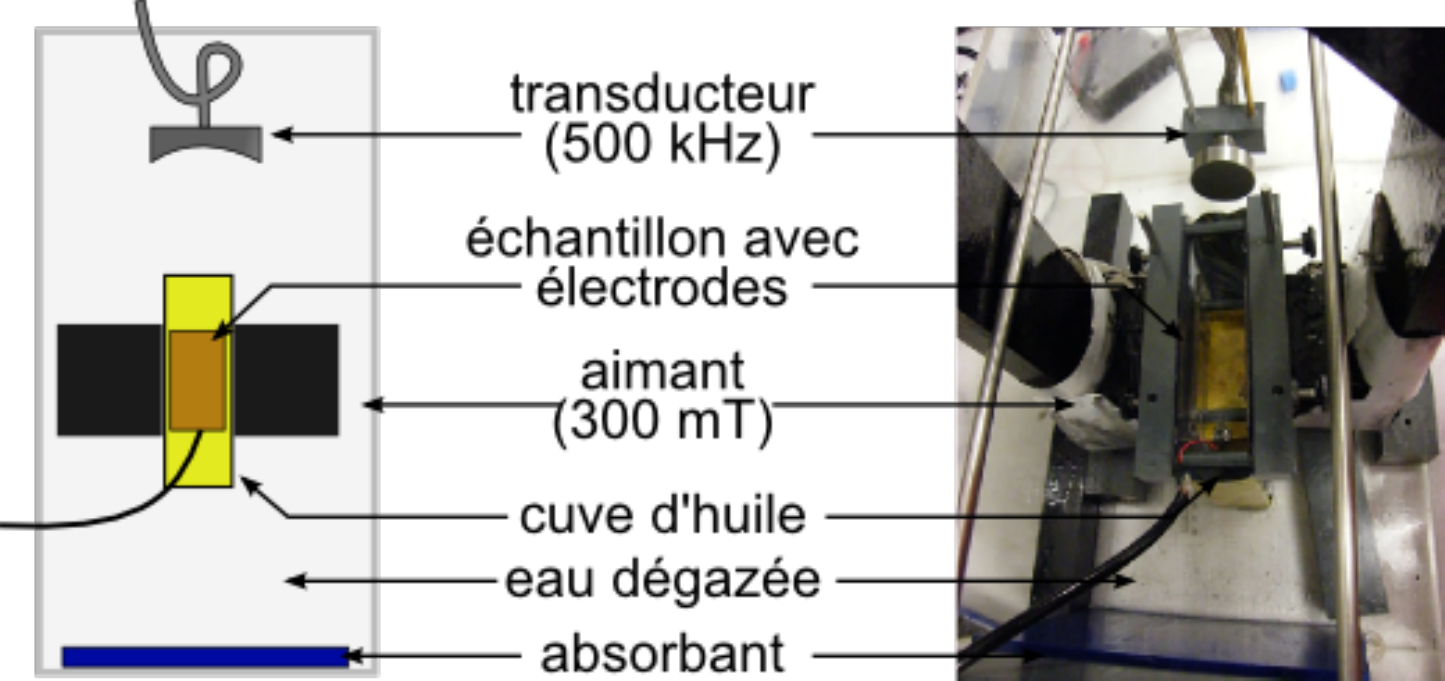}
	 \caption{Dans le dispositif de base, un transducteur à 500 kHz émet des ultrasons sur un échantillon placé entre deux électrodes dans une cuve d'huile au milieu d'un champ magnétique d'environ 300 mT créé par un aimant en U. Le courant électrique induit est reçu par les deux électrodes puis mesuré. Le tout est placé dans une cuve d'eau dégazée.}
	 \label{figLFEITDispositif}
 \end{center}
\end{figure}

L'expérience effectuée est schématisée sur la figure \ref{figLFEITDispositif}. Un courant électrique de 1 V à 0,5 MHz de trois sinusoïdes par salve à une fréquence de répétition de 100 Hz était émis par un générateur (HP33120A, Agilent, Santa Clara, CA, Etats-Unis) relié à un amplificateur de puissance de 200 W (LA200H, Kalmus Engineering, Rock Hill, SC, Etats-Unis). Ce courant était transmis à un transducteur à 0,5 MHz, de diamètre 50 mm et focalisé à 210 mm placé dans une cuve d'eau dégazée de 100x50x50 cm$^3$. Les champs de pression simulés montrent que le faisceau ultrasonore a une largeur à mi-hauteur d'environ 2 cm, avec une pression efficace au point focal de 3 MPa, comme montré par les figures \ref{figLFEITDispositifChampLong} et \ref{figLFEITDispositifChampLat}.

Une cuve d'huile de 20x4x15 cm$^3$ était placé de 15 à 35 cm du transducteur le long de l'axe ultrasonore. Cette cuve d'huile était placée au milieu d'un aimant permanent en forme de U décrit dans le chapitre \ref{secHydroEM}, avec photographie et simulation de champ magnétique sur la figure \ref{figHydroEMProto1}. Cette cuve servait d'isolant pour éviter tout courant parasite entre les électrodes et le transducteur.

Les différents échantillons étaient placés dans cette cuve de 22 à 28 cm du transducteur. Trois types d'échantillons ont été testés : des petites cuves d'eau salée à 30\% en sel (soit 300 g/L) avec le long de l'axe acoustique des membranes perméables aux ultrasons ; des fantômes de gélatine à 10\% en sel et 5\% en gélatine ; ou des morceaux de muscle animal, souvent avec une interface de graisse visible (b\oe uf, porc, poulet).

Une paire d'électrodes en cuivre de taille 10x3x0,1 cm$^3$ était placée au contact de l'échantillon, respectivement dessus et dessous. Les électrodes étaient reliées via un amplificateur de courant de 1 MV/A (HCA-2M-1M, Laser Components, Olching, Germany) et un filtre de 0,1 à 2 MHz (NF Corporation FV-628B, Yokohama, Japon) à un oscilloscope avec une entrée à 50 $\Omega$ (WaveSurfer 422, LeCroy, Chestnut Ridge, NY, Etats-Unis) qui effectuait des moyennes sur 2000 acquisitions. Les signaux étaient ensuite enregistrés et analysés par le logiciel Matlab (The MathWorks, Natick, MA, Etats-Unis). Ce logiciel filtrait les signaux entre 0,1 et 1 MHz et calculait la valeur absolue de la transformée de Hilbert du signal pour trouver l'enveloppe de ce signal. Cette enveloppe était ensuite convertie en niveaux de couleur.

Comme vu dans la partie théorique, le signal donne le gradient de conductivité électrique le long d'une ligne seulement. Afin de réaliser une image complète des gradients de conductivité électrique, le transducteur pouvait être déplacé pas à pas verticalement pour former une image ligne par ligne.

\section{Expériences mises en \oe uvre}

\subsection{Objectivation du signal}
Cette partie avait pour but de créer un courant par force de Lorentz, de détecter celui-ci et de s'assurer que ce courant est bien dû à l'interaction entre les ultrasons et le champ magnétique. Dans ces expériences, l'échantillon était une petite cuve d'eau salée isolée du milieu externe par une membrane perméable aux ultrasons. La salve ultrasonore était composée de dix sinusoïdes, afin que le signal soit bien visible.

\subsubsection{Le signal détecté dépend-il de la présence d'ultrasons ?}
Nous avons d'abord voulu vérifier que la présence d'ultrasons dans le fantôme est bien nécessaire à l'obtention du signal.

\paragraph{Matériel et méthodes}
Pour cela, une expérience a été faite en plaçant un absorbant ultrasonore épais de 2 cm entre le transducteur ultrasonore et l'échantillon.

\begin{figure}[!ht]
 \begin{center}
	 \includegraphics[width=0.9\columnwidth]{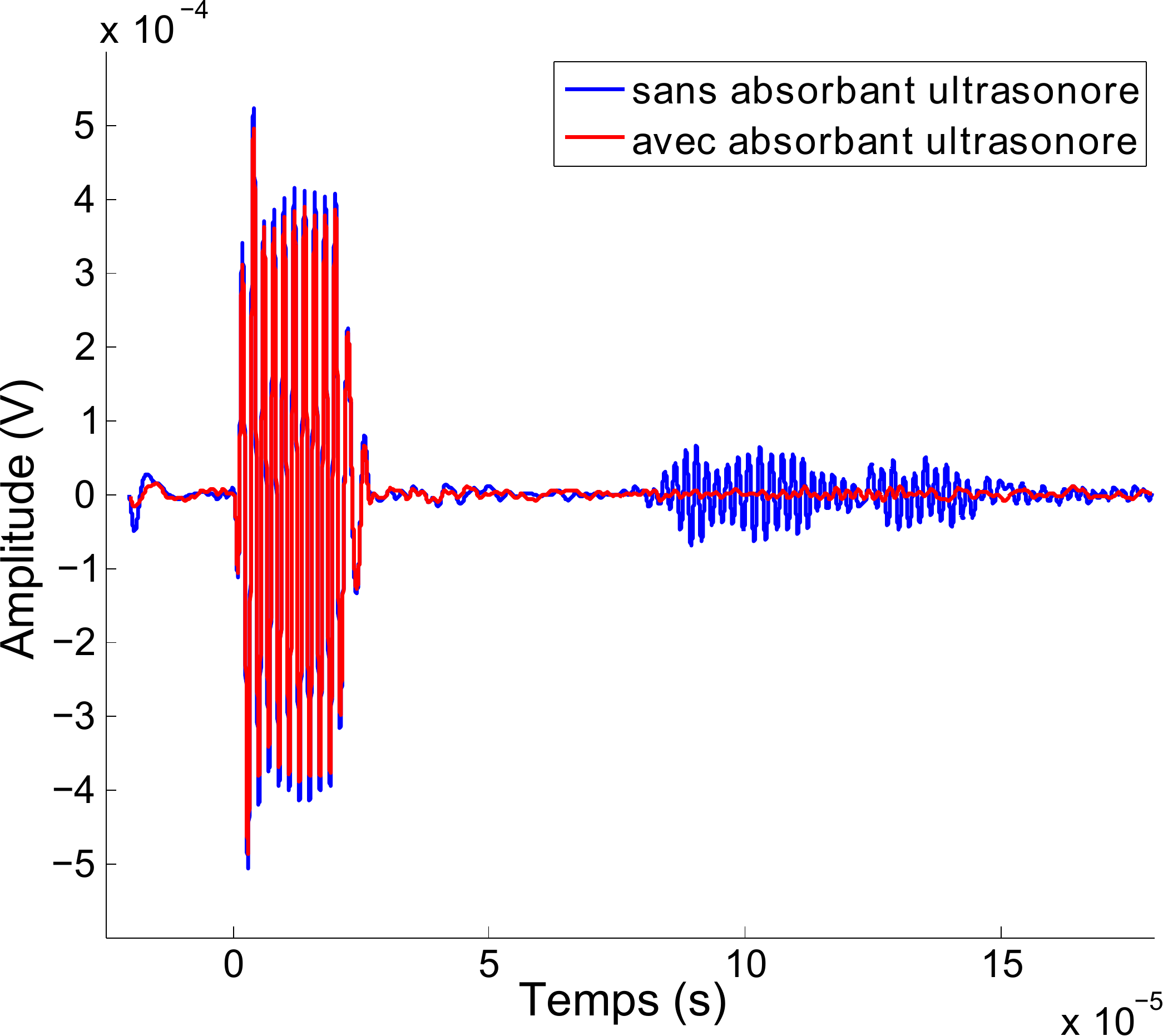}
	 \caption{Le signal issu des électrodes est mesuré sans absorbant entre le transducteur et le fantôme (signal bleu) puis avec un absorbant entre le transducteur et le fantôme (signal rouge). On observe que le premier ensemble de pics est quasiment inchangé tandis que le deuxième disparait après ajout de l'absorbant.}
	 \label{figLFEITExpObjAvecSansChampUS}
 \end{center}
\end{figure}

\paragraph{Résultats et discussions}
On observe sur la figure \ref{figLFEITExpObjAvecSansChampUS} que le premier ensemble de pics reste quasiment inchangé entre les deux expériences, alors que le second disparait. Les ultrasons sont donc nécessaires au deuxième ensemble de pics mais pas au premier.

\subsubsection{Le signal détecté dépend-il de la présence d'un champ magnétique ?}
D'après le modèle théorique, le champ magnétique est nécessaire à l'obtention d'un courant électrique par force de Lorentz.

\paragraph{Matériel et méthodes}
Cette hypothèse a été testée en retirant l'aimant sans autre modification dans le dispositif.

\begin{figure}[!ht]
 \begin{center}
	 \includegraphics[width=0.9\columnwidth]{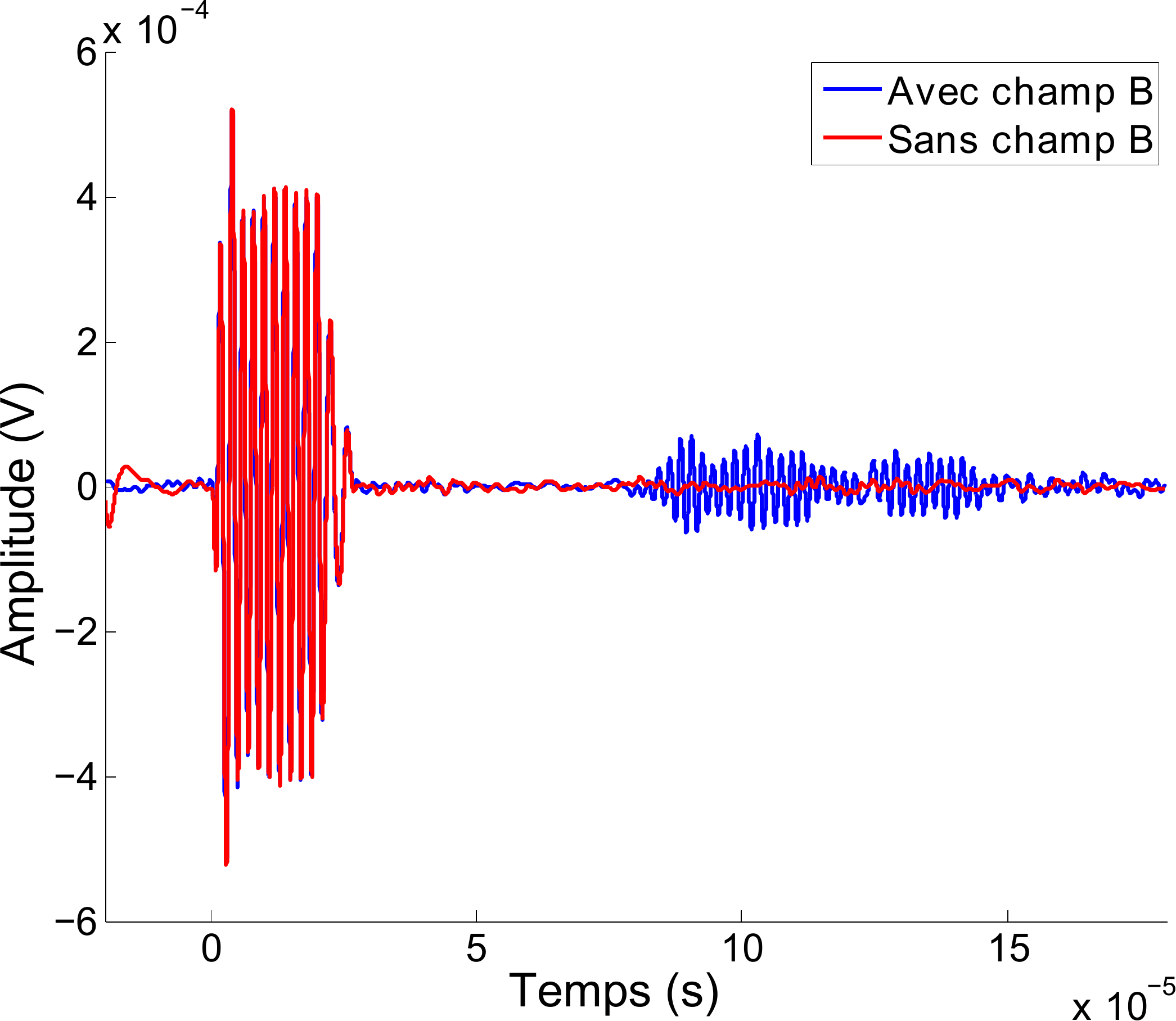}
	 \caption{Le signal issu des électrodes est mesuré avec un aimant autour du fantôme (signal bleu) puis sans aimant (signal rouge). Le premier ensemble de pics est quasiment inchangé tandis que le deuxième disparait après retrait de l'aimant.}
	 \label{figLFEITExpObjAvecSansChampB}
 \end{center}
\end{figure}

\paragraph{Résultats et discussion}
Les résultats, visibles sur la figure \ref{figLFEITExpObjAvecSansChampB}, montrent que le premier ensemble de pics reste quasiment inchangé entre les deux expériences, mais que le deuxième disparait. Le champ magnétique est donc nécessaire au deuxième ensemble de pics mais pas au premier.

\subsubsection{Le signal détecté est-il décalé dans le temps si on déplace l'échantillon par rapport au transducteur ?}
Nous avons regardé si la distance entre le transducteur et l'échantillon a une influence sur le temps de réception du signal.

\paragraph{Matériel et méthodes}
Une expérience a été réalisée avec l'échantillon placé à 16 puis à 10 cm du transducteur.

\begin{figure}[!ht]
 \begin{center}
	 \includegraphics[width=0.9\columnwidth]{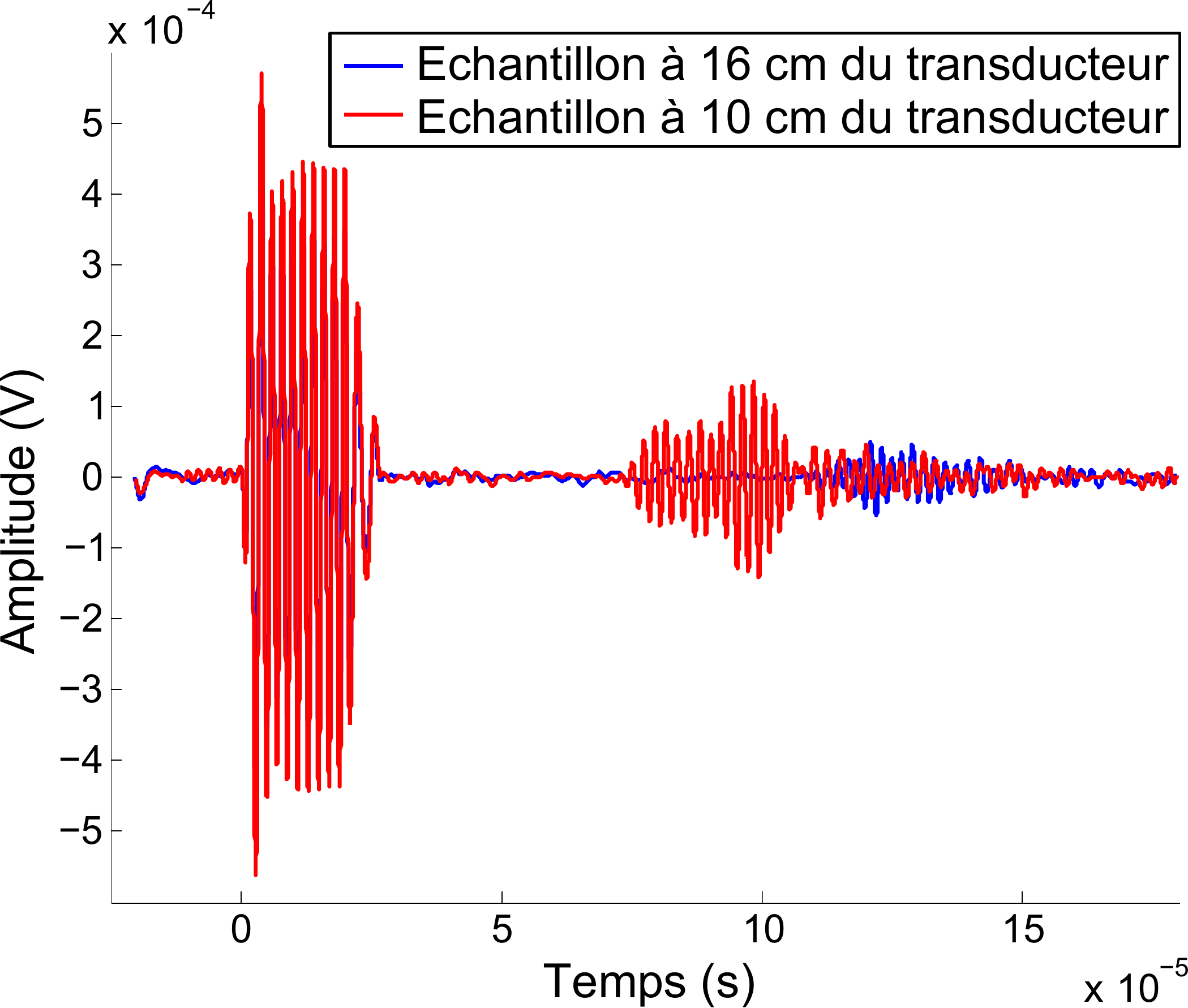}
	 \caption{Le signal issu des électrodes est mesuré avec le fantôme à 16 cm du transducteur (signal bleu) puis à 10 cm du transducteur (signal rouge). On observe que le signal arrive plus tôt, avec un temps compatible avec la durée de propagation des ultrasons.}
	 \label{figLFEITExpObjFantomeDeplace}
 \end{center}
\end{figure}

\paragraph{Résultats et discussions}
On observe sur la figure \ref{figLFEITExpObjFantomeDeplace} que le deuxième ensemble de pics commence vers 70 \micro s après l'émission avec le fantôme à 10 cm du transducteur, et vers 110 \micro s avec le fantôme à 16 cm du transducteur. Avec une vitesse de propagation de 1500 m.s$^{-1}$, cela correspond à des distances entre l'émission et la réception de 10,5 et 15,5 cm respectivement ce qui correspond approximativement à la distance entre le transducteur et le fantôme. Cependant, le premier ensemble de pics n'a pas été modifié et n'est donc pas été influencé par la position du fantôme.

\subsubsection{Le signal détecté dépend-il de la conductivité électrique de l'échantillon ?}
Le signal électrique est supposé dépendre au gradient de conductivité électrique. La relation étant plutôt complexe (produit de convolution), nous avons simplement vérifié qu'un gradient de conductivité électrique plus faible donnait effectivement un signal plus faible.

\paragraph{Matériel et méthodes}
Une expérience a donc été mise en oeuvre avec un échantillon à 30 g.L$^{-1}$ en sel puis un deuxième de taille identique au précédent sans ajout de sel, donc avec une conductivité électrique égale à celle de l'eau du robinet.

\begin{figure}[!ht]
 \begin{center}
	 \includegraphics[width=0.9\columnwidth]{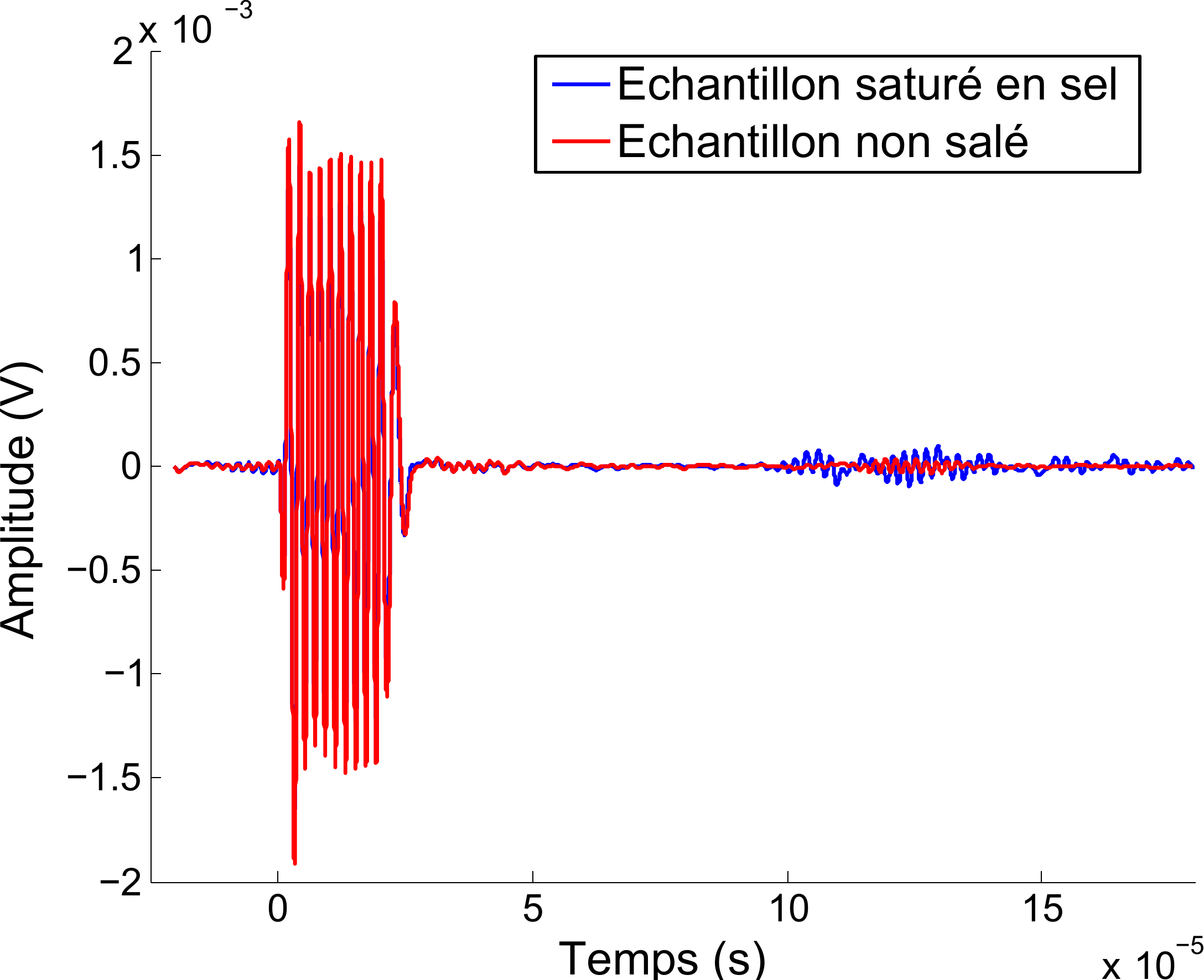}
	 \caption{Le signal issu des électrodes est mesuré avec échantillon salé (signal bleu) puis un échantillon sans sel (signal rouge). Le premier ensemble de pics est quasiment inchangé, et le deuxième diminue notablement lorsque le gradient de conductivité électrique est plus faible.}
	 \label{figLFEITExpObjAvecSansSel}
 \end{center}
\end{figure}

\paragraph{Résultats et discussions}
On constate sur la figure \ref{figLFEITExpObjAvecSansSel} que l'amplitude de l'ensemble de pics apparaissant vers 100 \micro s diminue drastiquement entre l'échantillon salé, qui donne un gradient de conductivité électrique important, qu'avec l'échantillon non salé,  avec un gradient de conductivité électrique plus faible. Cependant, le premier ensemble de pics n'est pas modifié notablement.

\subsubsection{Le signal détecté dépend-il de l'émission/réception d'ultrasons par le transducteur ?}
Les expériences précédentes montrent un signal au moment de l'émission qui ressemble fortement au signal électrique émis par le générateur. Ce signal pourrait être expliqué par la conversion du courant électrique en ultrasons qui émet des parasites électromagnétiques détectés par les électrodes.

\paragraph{Matériel et méthodes}
Pour vérifier cela, un réflecteur de PVC de 3 cm de large a été placé entre le fantôme et le transducteur, à 2 cm de ce dernier. Il était orthogonal à l'axe ultrasonore de façon à ce que les échos ultrasonores reviennent au niveau du transducteur.

\begin{figure}[!ht]
 \begin{center}
	 \includegraphics[width=0.9\columnwidth]{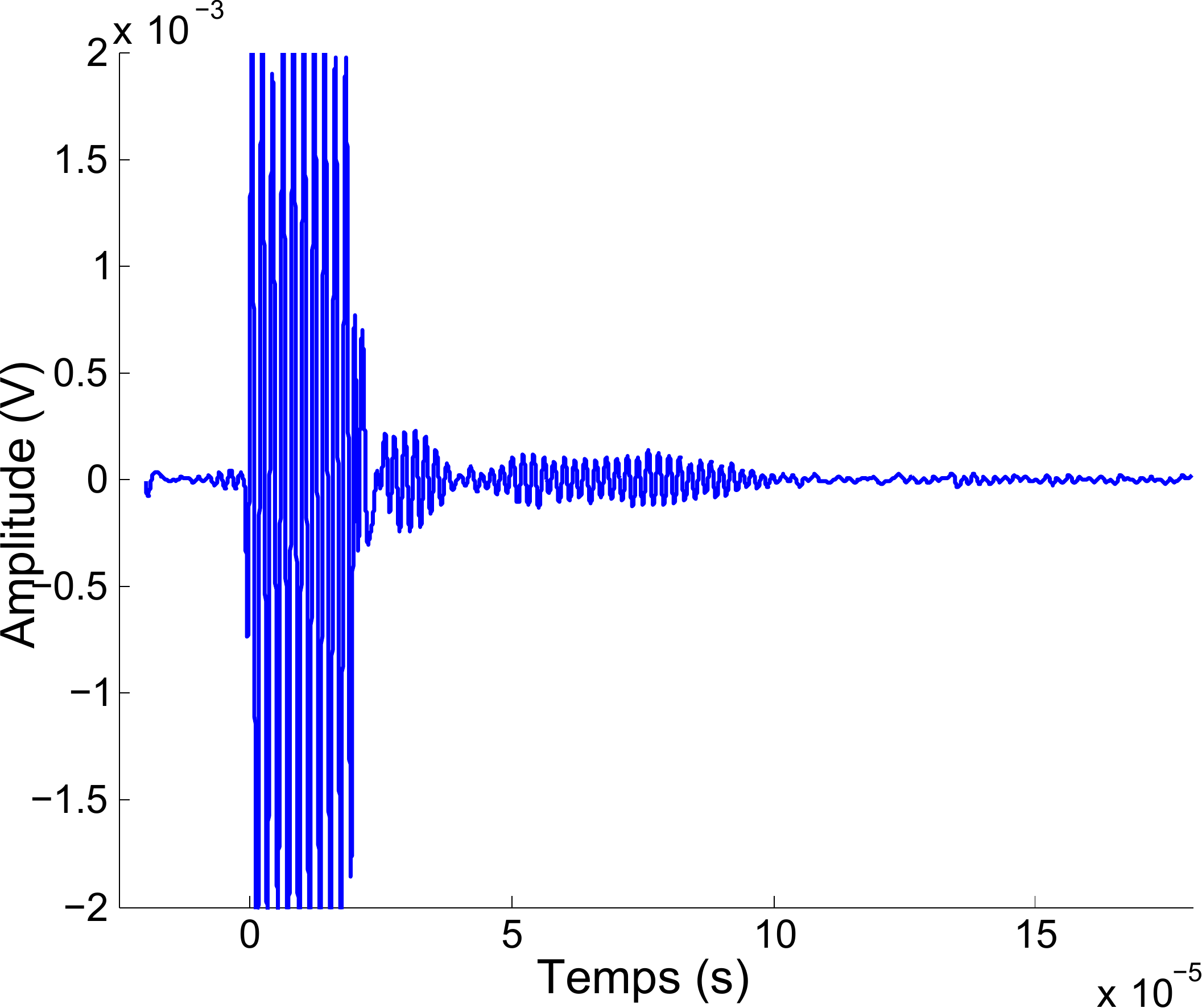}
	 \caption{Le signal issu des électrodes est mesuré avec un réflecteur de PVC face au transducteur, à 2 cm de celui-ci. On observe l'apparition d'un nouvel ensemble de pics à environ 25 \micro s de l'émission.}
	 \label{figLFEITExpObjReflecteur.pdf}
 \end{center}
\end{figure}

\paragraph{Résultats et discussions}
On remarque sur la figure \ref{figLFEITExpObjReflecteur.pdf} qu'un signal est apparu à partir de 25 \micro s environ. Cela correspond à une distance d'environ 4 cm, soit 2 cm aller-retour. On constate également dans les expériences ultérieures la présence systématique de ce signal à une distance deux fois supérieure aux surfaces réfléchissantes (toujours avec une vitesse de propagation de 1500 m.s$^{-1}$. On peut donc en conclure que le mécanisme de conversion électrique/mécanique produit un signal électrique au niveau des électrodes, que ce soit à l'émission ou la réception.

\subsubsection{Conclusion}
D'après les éléments ci-dessus, le signal détecté est composé de deux parties :
\begin{itemize}
	\item des parasites électromagnétiques émis lors de l'émission ou de la réception d'ultrasons par le transducteur. Ces parasites se propagent du transducteur aux électrodes qui font office d'antenne, et ce à une vitesse très élevée (proche de la vitesse de la lumière) face à la vitesse moyenne des ultrasons (1500 m/s). Ces parasites peuvent être réduits grâce à un blindage et une mise à la masse des différents éléments du dispositif.
	\item un signal qui dépend des ultrasons, du champ magnétique, de la conductivité électrique et qui apparait aux interfaces de conductivité électrique. Il s'agit du signal acousto-magnétique que l'on souhaite détecter.
\end{itemize}

Afin de s'affranchir des parasites électromagnétiques qui ont une amplitude souvent proche voire supérieure au signal recherché, on peut blinder les éléments du dispositif, mais cela s'est avéré insuffisant dans notre dispositif. Il est également possible enregistrer le signal acoustique réfléchi, comme en échographie mode A, et soustraire celui-ci au signal acousto-magnétique. Cette idée s'avère néanmoins très sensible au bruit et donc n'a pas pu être utilisée dans nos expériences. Enfin, on peut laisser une distance suffisante entre le transducteur et le fantôme pour que les signaux acousto-magnétiques recherchés apparaissent avant que les échos ultrasonores n'arrivent au niveau du transducteur. Cette méthode, facile à mettre en oeuvre, a été retenue pour la suite.

Enfin, nous avons remarqué que les signaux sont tout à fait reproductibles au cours du temps, aucune variation n'étant détectée d'une expérience à l'autre, toute choses étant égales par ailleurs.

\subsection{Image d'un échantillon de gélatine salée}

\begin{figure}[!ht]
 \begin{center}
	 \includegraphics[width=0.8\columnwidth]{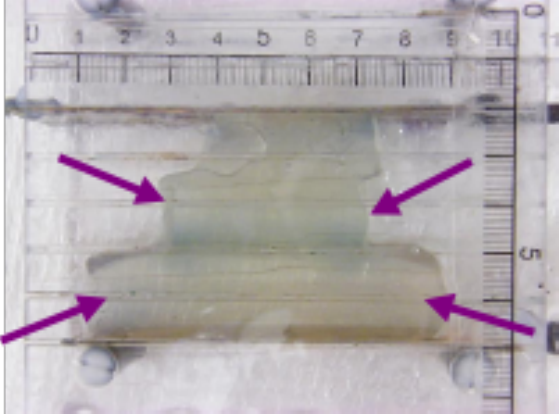}
	 \caption{Photographie du fantôme de gélatine utilisé dans l'expérience. On observe deux interfaces d'entrée et deux interfaces de sortie, comme indiqué par les quatre flèches violettes.}
	 \label{figLFEITExpImageFantomePhoto}
 \end{center}
\end{figure}

\paragraph{Matériel et méthodes}
La technique d'imagerie a été appliquée sur un fantôme que l'on peut voir sur la figure \ref{figLFEITExpImageFantomePhoto}, composé de 10\% en gélatine et 10 g.L$^{-1}$ de sel, donnant une conductivité électrique égale à 2 S/m. Ce fantôme avait une forme particulière, avec deux interfaces d'entrée et deux interfaces de sortie. Des acquisitions ont été faites ligne par ligne, avec 140 lignes séparées chacune de 0,25 mm. Avec 2000 acquisitions par ligne à 2 minutes par ligne, l'acquisition a pris approximativement 5 heures. Simultanément, une image ultrasonore a été acquise et reconstruite ligne par ligne à partir des échos ultrasonores.

\begin{figure}[!ht]	
 \begin{center}
	 \includegraphics[width=0.7\columnwidth]{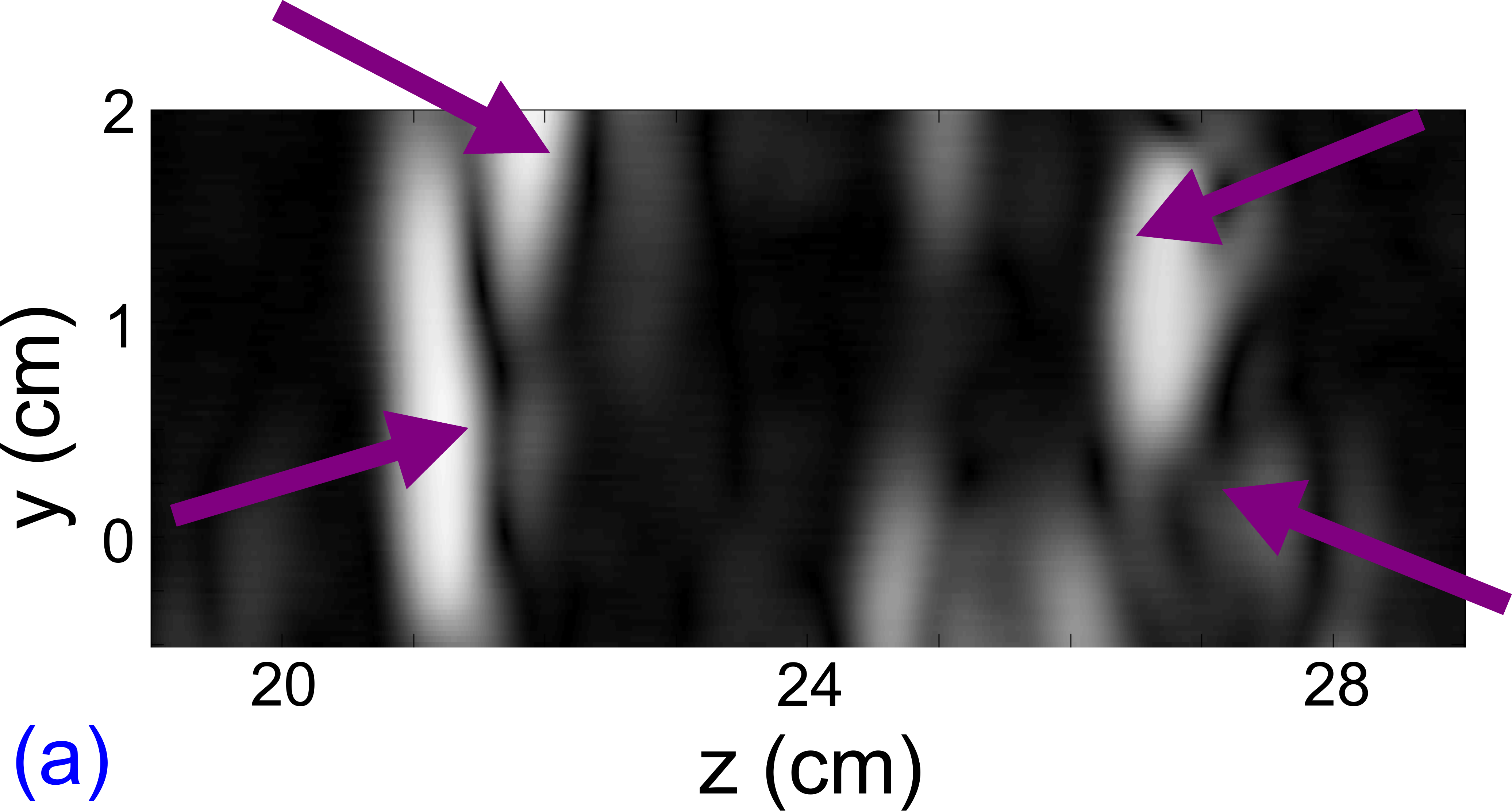}
 \end{center}
 \begin{center}
	 \includegraphics[width=0.7\columnwidth]{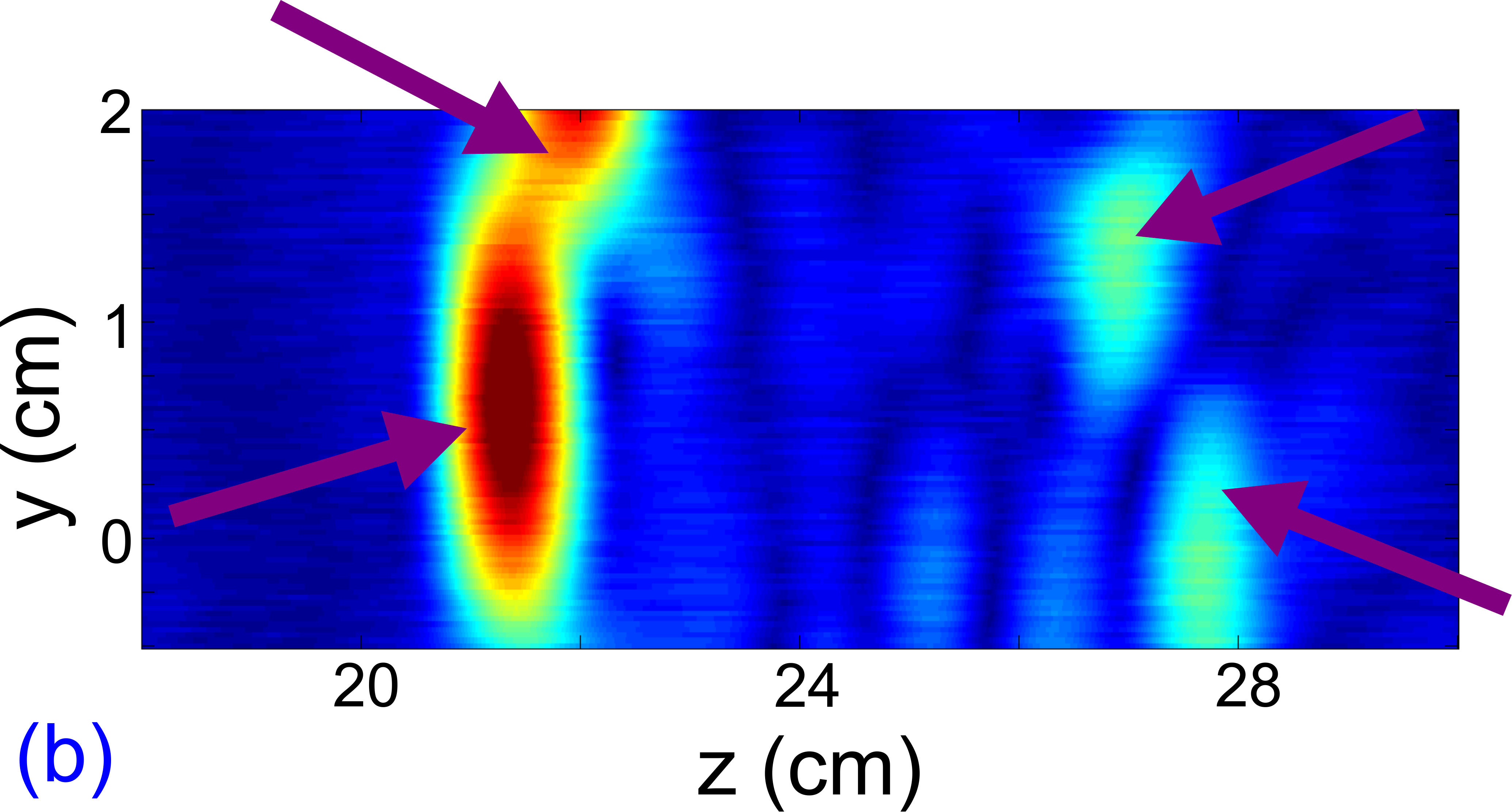}
	 \caption{(a) Image ultrasonore du fantôme de gélatine. (b) Image d'impédance électrique correspondante du fantôme de gélatine. Le faisceau ultrasonore ayant une largeur de presque deux centimètres, on voit plusieurs interfaces sur une même ligne. On distingue néanmoins sur l'image d'impédance électrique comme sur l'image ultrasonore les deux interfaces d'entrée et les deux de sortie comme indiqué par les flèches violettes.}
	 \label{figLFEITExpImageFantome}
 \end{center}
\end{figure}

\paragraph{Résultats et discussions}
L'image ultrasonore du fantôme est représentée sur la figure \ref{figLFEITExpImageFantome}-(a) et l'image d'impédance électrique par force de Lorentz (que l'on nommera simplement ``image d'impédance électrique'' par la suite) est visible sur la figure \ref{figLFEITExpImageFantome}-(b). On peut reconnaitre les interfaces d'entrée et de sortie (flèches violettes) à la fois sur l'image ultrasonore et l'image d'impédance électrique. Le faisceau ultrasonore ayant une largeur supérieure à un centimètre, on voit des interfaces différentes sur une même ligne.

Cette expérience montre que le dispositif est adapté pour observer des interfaces de conductivité électriques.

\subsection{Image de tissus organiques}
\begin{figure}[!ht]
 \begin{center}
	 \includegraphics[width=0.6\columnwidth]{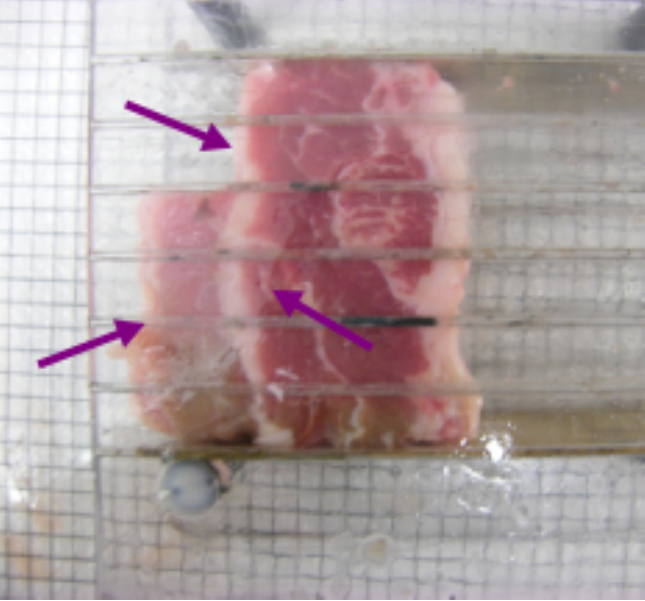}
	 \caption{Photographie de la côte de boeuf utilisée dans l'expérience. Cet échantillon présente deux interfaces d'entrée, une couche de graisse marquée et des couches de muscle peu homogène.}
	 \label{figLFEITExpImageTissuPhoto}
 \end{center}
\end{figure}

\paragraph{Matériel et méthodes}
La technique d'imagerie a été appliquée sur un morceau de côte de b\oe uf issu d'une boucherie, que l'on peut voir sur la figure \ref{figLFEITExpImageTissuPhoto}. Cet échantillon présente deux interfaces d'entrée et une couche de graisse marquée ; le muscle lui-même est peu homogène, avec de nombreuses inclusions de graisse. Une image a été faite ligne par ligne, avec 96 lignes séparées chacune de 0,5 mm. Avec 2000 acquisitions par ligne à 2 minutes par ligne, l'acquisition a pris approximativement 3,5 heures.

\begin{figure}[!ht]
   \begin{minipage}[c]{.5\linewidth}
		\begin{center}
	   		\includegraphics[width=1\linewidth]{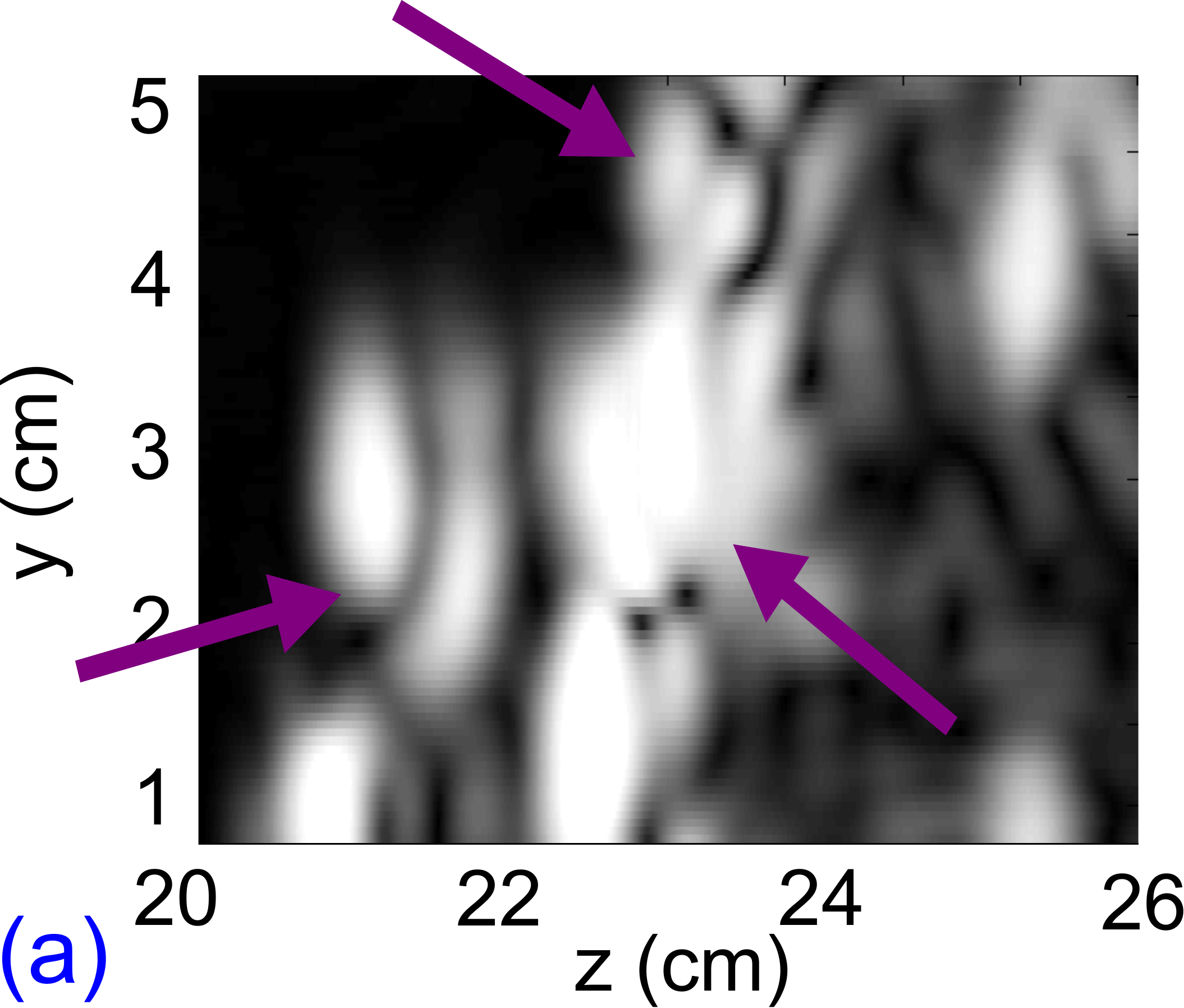}
		\end{center}
   \end{minipage} \hfill
   \begin{minipage}[c]{.5\linewidth}
		\begin{center}
	   		\includegraphics[width=1\linewidth]{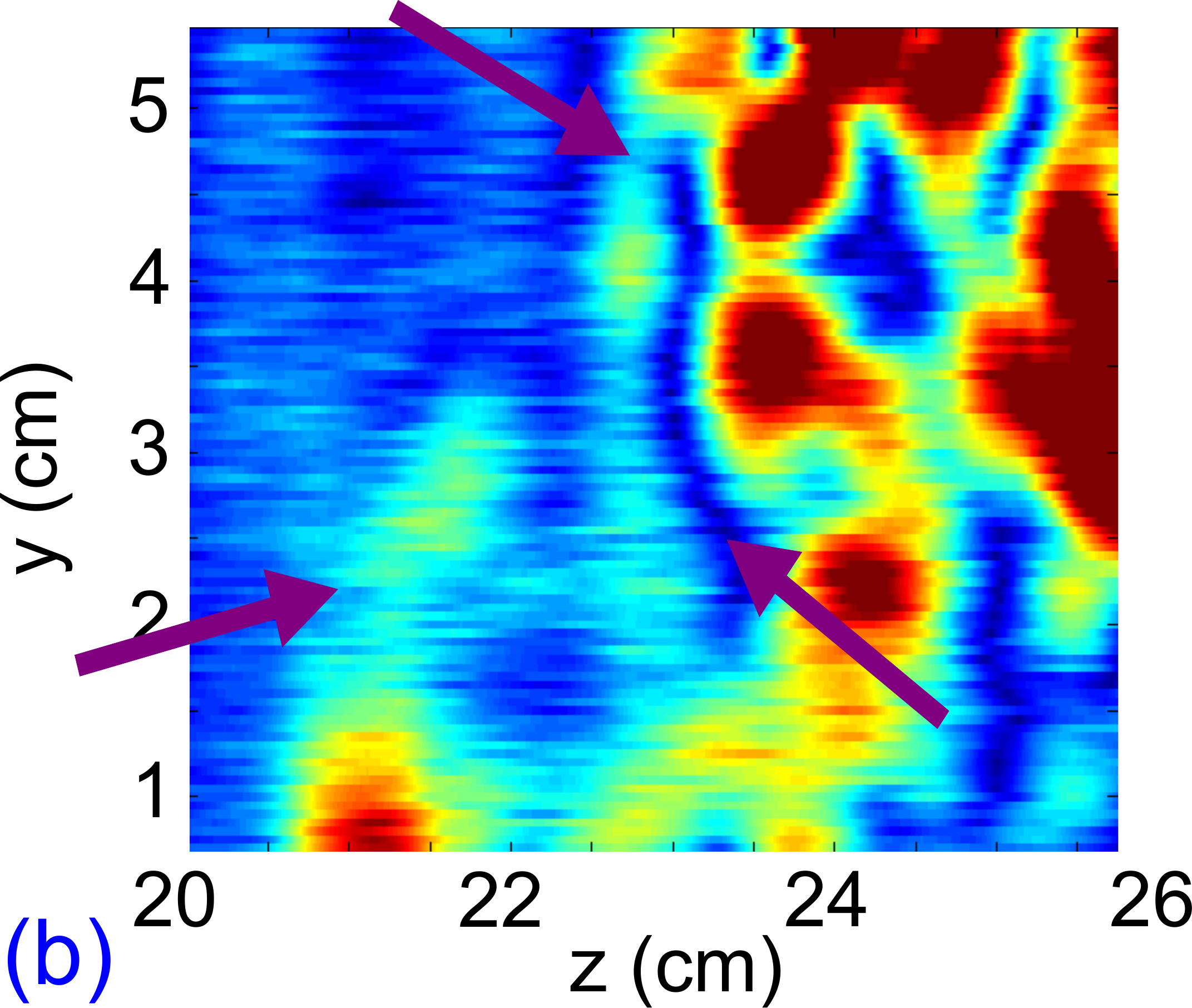}
		\end{center}
   \end{minipage}
	   		\caption{(a) Image ultrasonore de la côte de boeuf ; (b) Image d'impédance électrique de la côte de boeuf.}
	   		\label{figLFEITExpImageTissu} 
\end{figure}

\paragraph{Résultats et discussions}
L'image ultrasonore du fantôme est représentée sur la figure \ref{figLFEITExpImageTissu}-(a) tandis que l'image d'impédance électrique est visible sur la figure \ref{figLFEITExpImageTissu}-(b).

On peut reconnaitre les interfaces d'entrée (flèches violettes) à la fois sur l'image ultrasonore et l'image d'impédance électrique, bien que l'interface du bas soit difficilement visible sur l'image d'impédance électrique. L'interface de graisse semble être visible sur l'image d'impédance électrique sans l'être sur l'image ultrasonore -- ce qui semblerait normal vu la grande différence et de conductivité électrique et la faible différence d'impédance acoustique entre la graisse et le muscle. La résolution spatiale est comparable entre l'image ultrasonore et l'image d'impédance électrique, maise reste cependant médiocre avec notamment un ensemble de taches qui seront discutées dans le paragraphe suivant. La résolution spatiale pourrait être améliorée dans l'axe ultrasonore en augmentant la fréquence ultrasonore pour diminuer la longueur d'onde et dans l'axe transverse en utilisant des faisceaux ultrasonores plus fins.

\subsection{Observation du \textit{speckle} électro-acoustique}
D'après la partie théorique, les images d'impédance électrique de tissus présentant des hétérogénéités devraient présenter une texture granuleuse similaire de \textit{speckle} acousto-électrique similaire au \textit{speckle} ultrasonore. Cependant, le \textit{speckle} électro-acoustique n'est pas forcément identique au \textit{speckle} ultrasonore, car il peut y avoir des hétérogénéités électriques et non acoustiques et vice-versa : par exemple une petite inclusion de graisse aura une conductivité électrique différente du muscle alentours, mais quasiment la même impédance acoustique.

\begin{figure}[!ht]
 \begin{center}
	 \includegraphics[width=0.6\columnwidth]{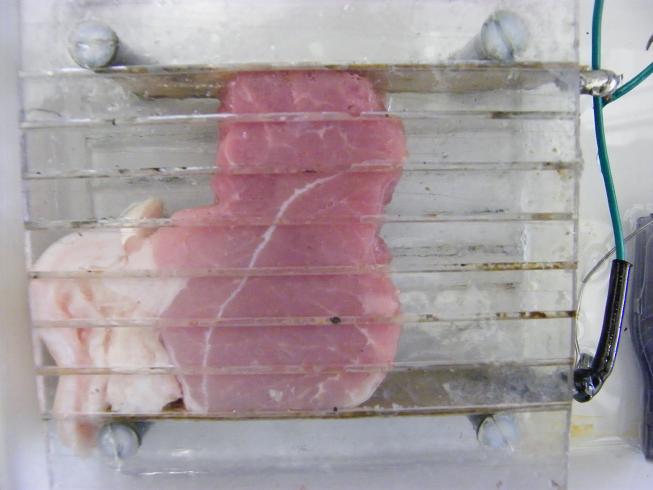}
	 \caption{Photographie de la côte de boeuf utilisée dans l'expérience. On observe une couche de graisse, puis un morceau de muscle avec une forme en ``L''}
	 \label{figLFEITExpSpecklePhoto}
 \end{center}
\end{figure}

\paragraph{Matériel et méthodes}
Afin de caractériser ce \textit{speckle}, une expérience a été faite en modifiant l'angle d'imagerie. Deux images faites à partir d'un angle différent présenteront en effet un \textit{speckle} différent si l'angle est suffisamment important : il s'agit typiquement d'un angle donnant dans l'image un déplacement d'une longueur d'onde \cite{wagner1983}. L'imagerie composée (ou \textit{compound imaging}) en ultrasons est d'ailleurs basée sur ce principe : on réalise des images à partir de plusieurs angles proches, la moyenne des images diminue le \textit{speckle} des images tout en conservant les interfaces d'impédance acoustique \cite{jespersen1998}.
Pour cela, des images d'un morceau de côte de porc visible sur la figure \ref{figLFEITExpSpecklePhoto} ont été faites avec cinq angles différents. Les images ultrasonores et d'impédance électrique ont été acquises simultanément. Les angles choisis étaient de -2$^o$ à 2$^o$ avec un pas de 1 $^o$, l'axe Z étant pris comme angle égal à 0$^o$ et le centre de rotation étant au milieu du transducteur. La distance entre le transducteur et le fantôme étant de 22 cm, cela donne un déplacement approximatif vertical de -8 à +8 mm, la longueur d'onde étant de 3 mm à 500 kHz.

\begin{figure}[!ht]
   \begin{minipage}[c]{.5\linewidth}
		\begin{center}
	   		\includegraphics[width=.58\linewidth]{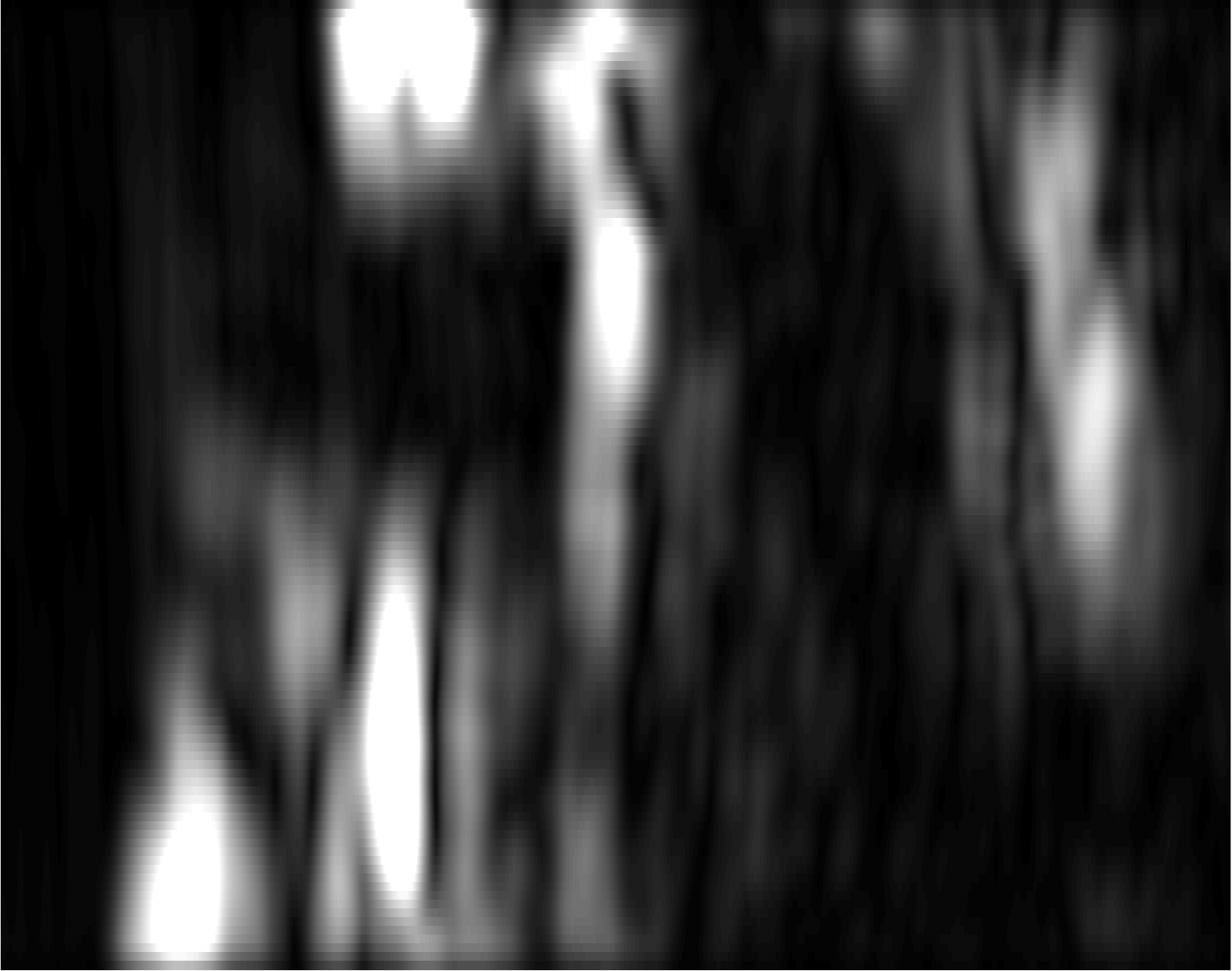}
		\end{center}
   \end{minipage} \hfill
   \begin{minipage}[c]{.5\linewidth}
		\begin{center}
	   		\includegraphics[width=.58\linewidth]{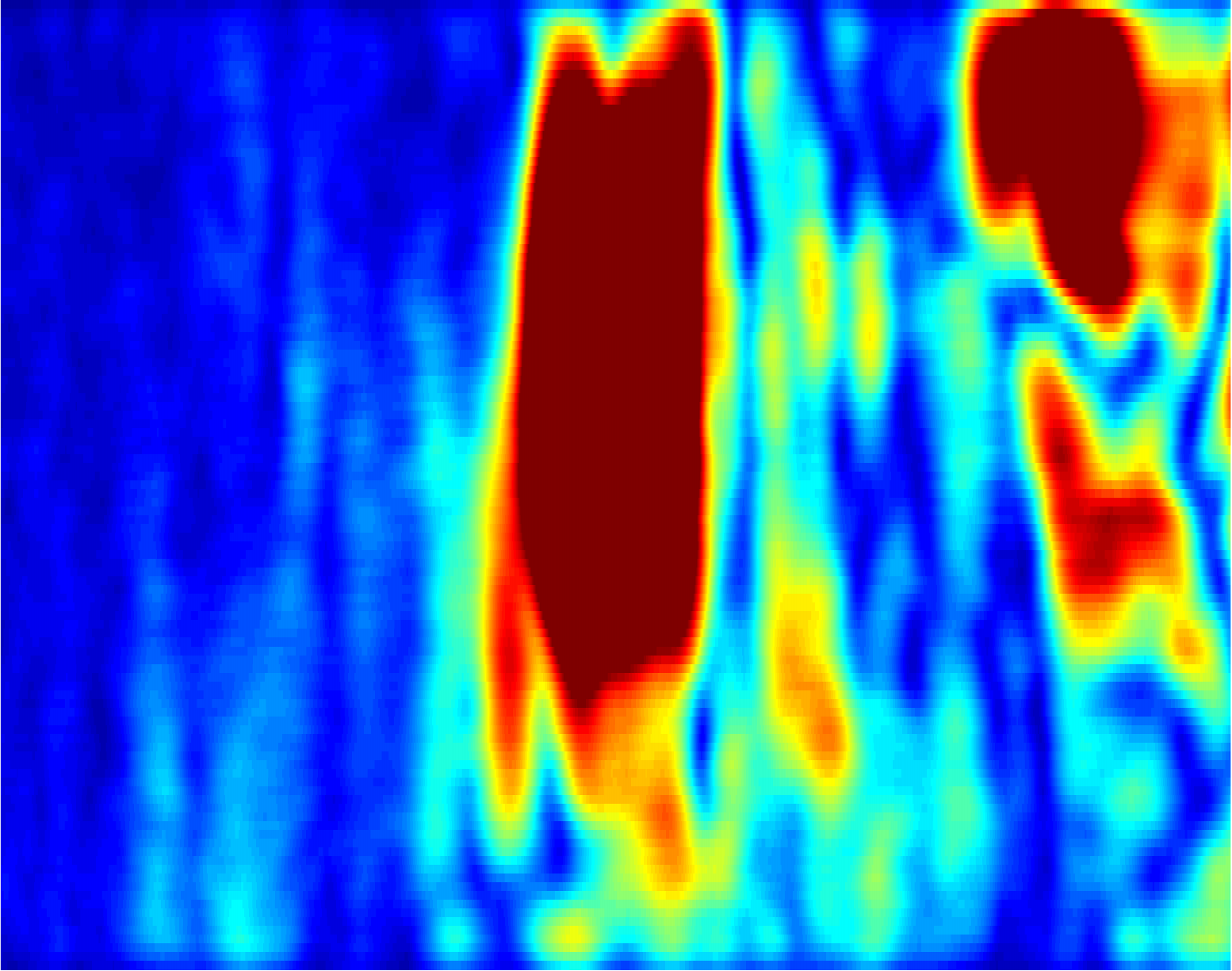}
		\end{center}
   \end{minipage}
   \begin{minipage}[c]{.5\linewidth}
		\begin{center}
	   		\includegraphics[width=.58\linewidth]{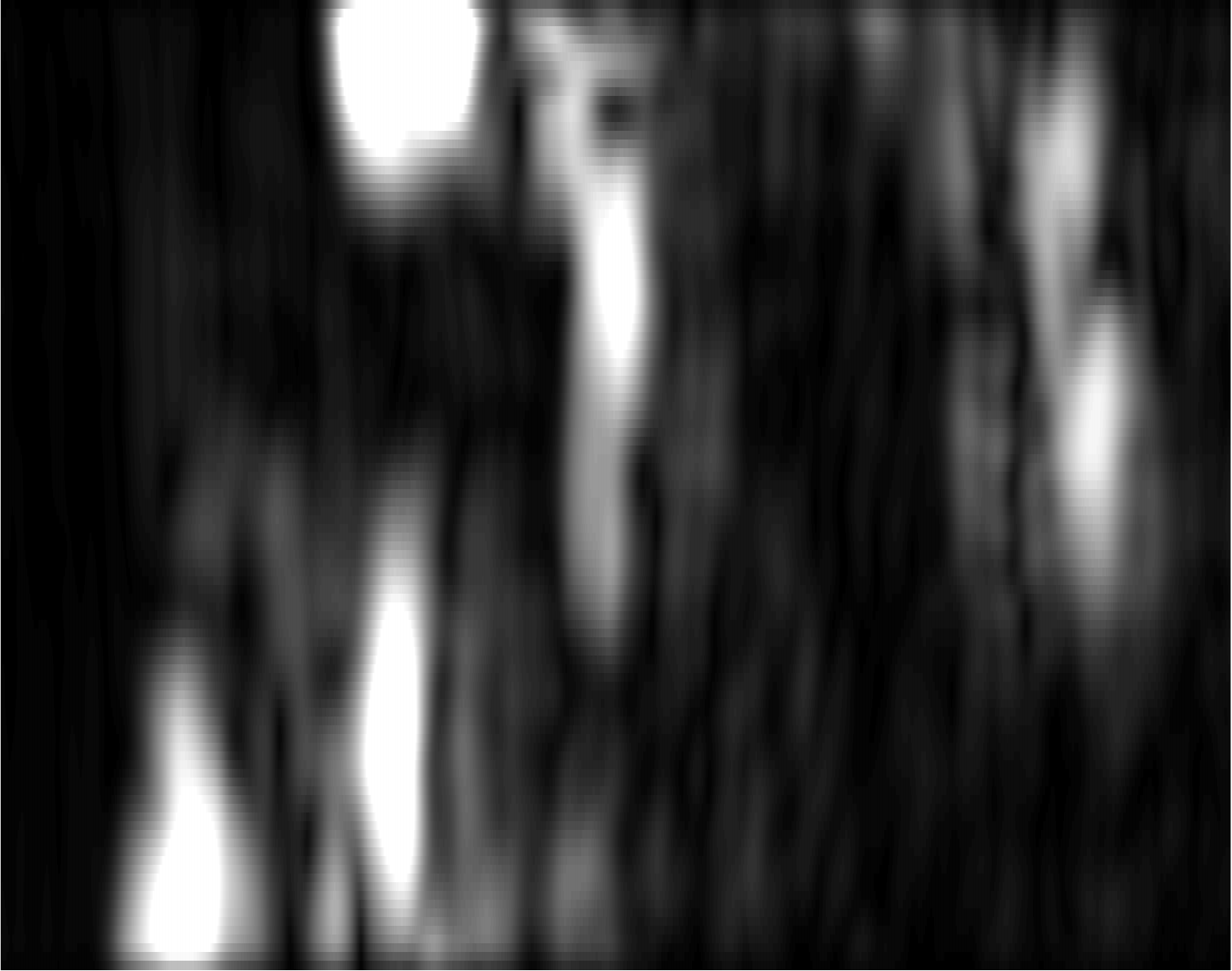}
		\end{center}
   \end{minipage} \hfill
   \begin{minipage}[c]{.5\linewidth}
		\begin{center}
	   		\includegraphics[width=.58\linewidth]{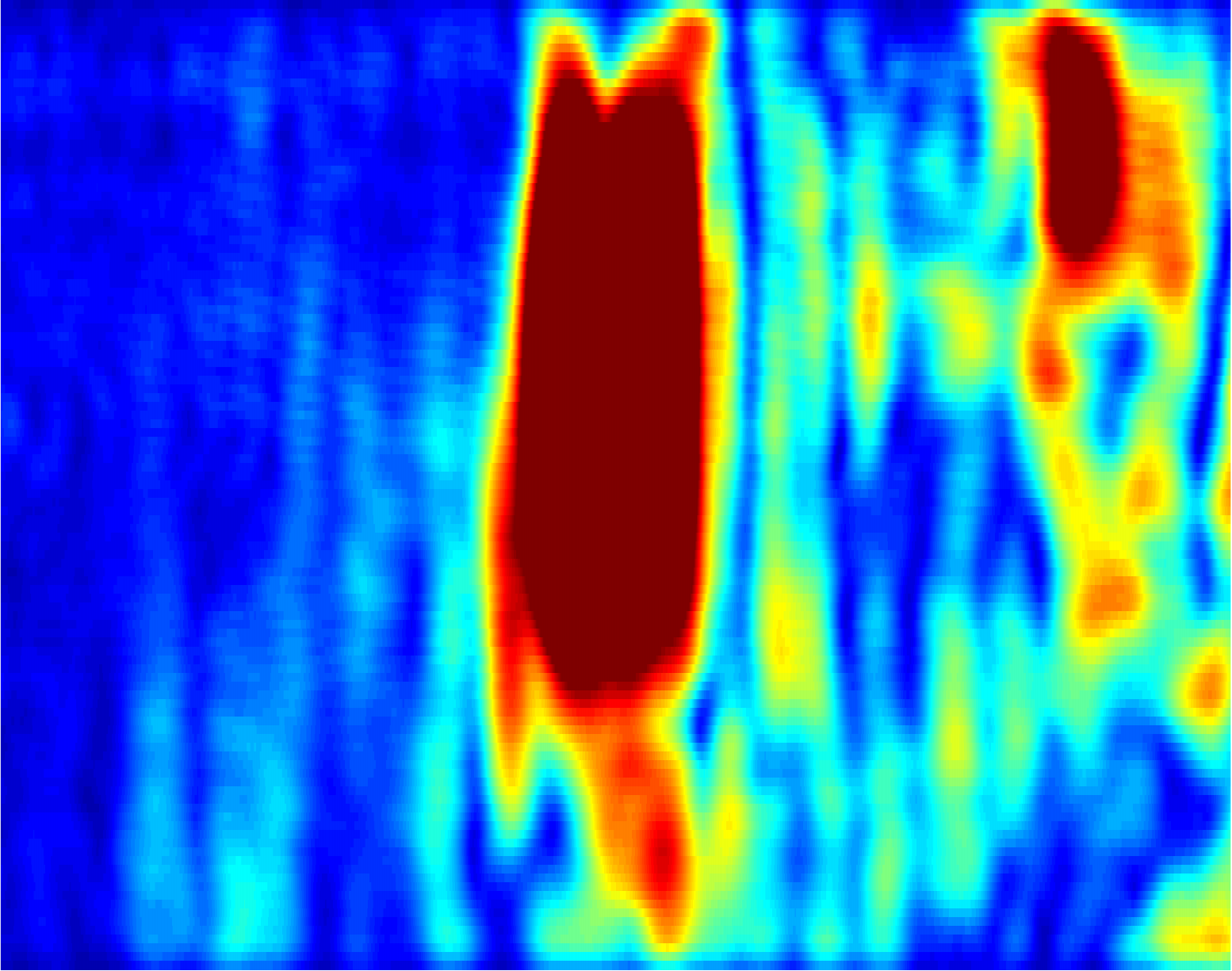}
		\end{center}
   \end{minipage}
   \begin{minipage}[c]{.5\linewidth}
		\begin{center}
	   		\includegraphics[width=.58\linewidth]{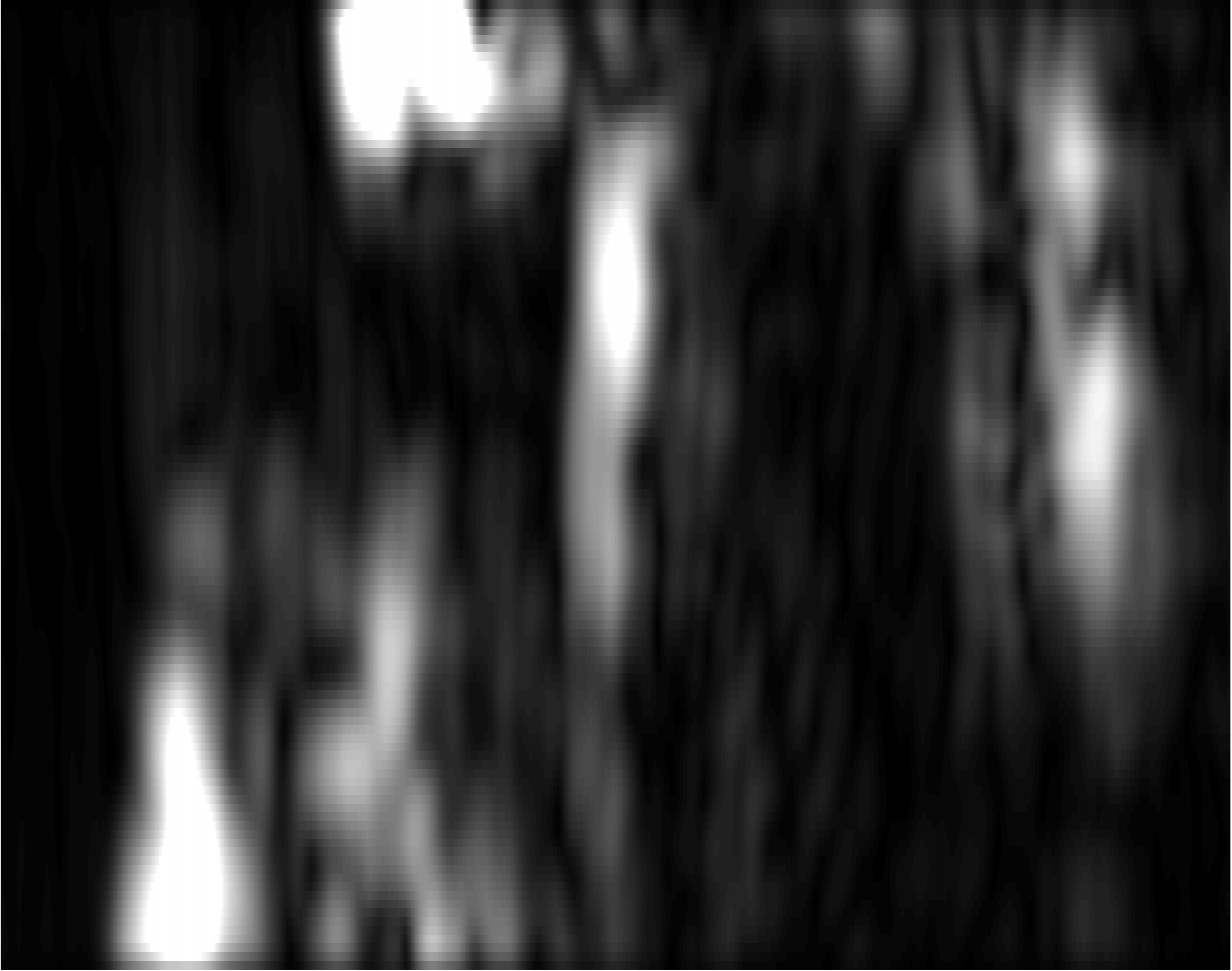}
		\end{center}
   \end{minipage} \hfill
   \begin{minipage}[c]{.5\linewidth}
		\begin{center}
	   		\includegraphics[width=.58\linewidth]{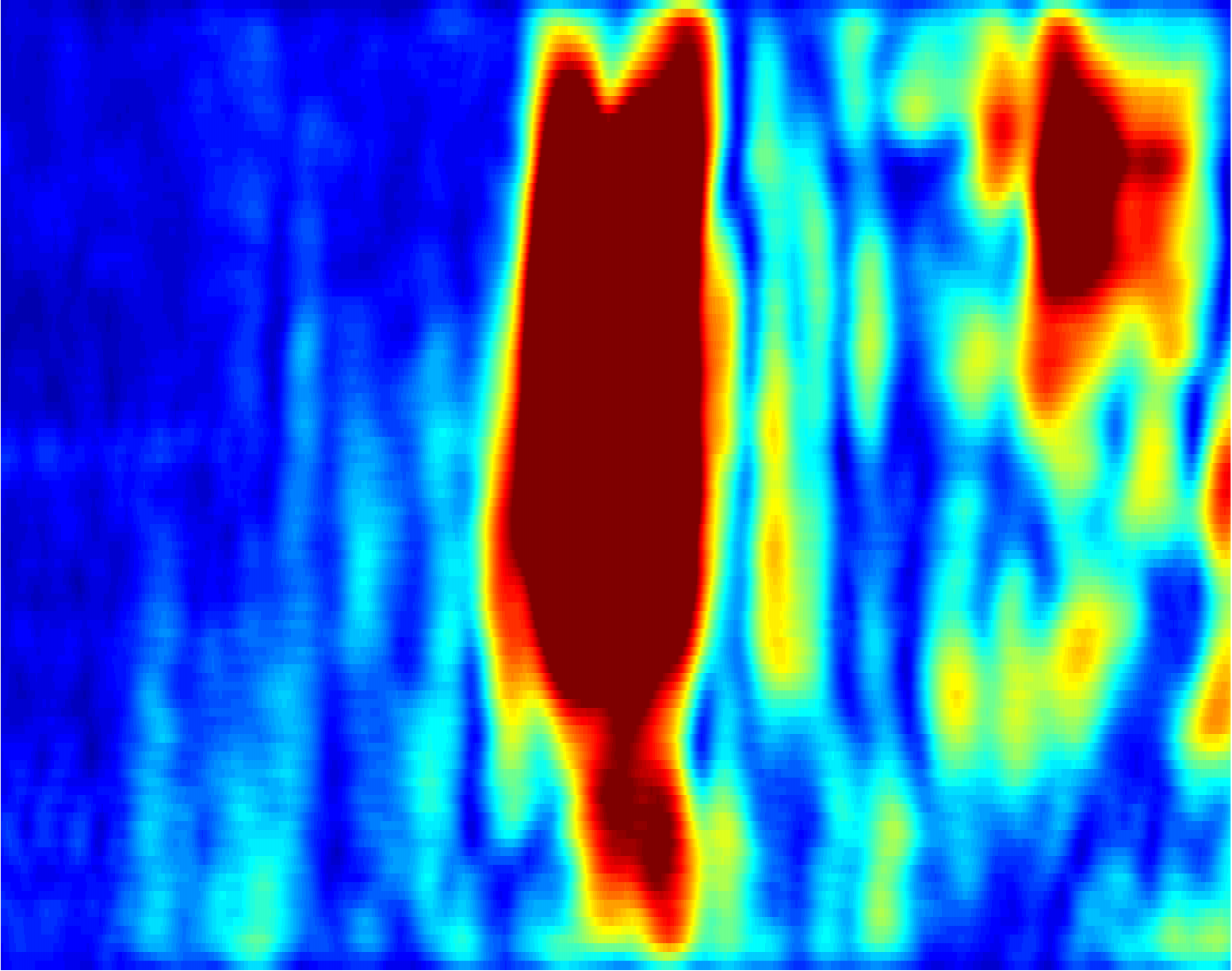}
		\end{center}
   \end{minipage}
   \begin{minipage}[c]{.5\linewidth}
		\begin{center}
	   		\includegraphics[width=.58\linewidth]{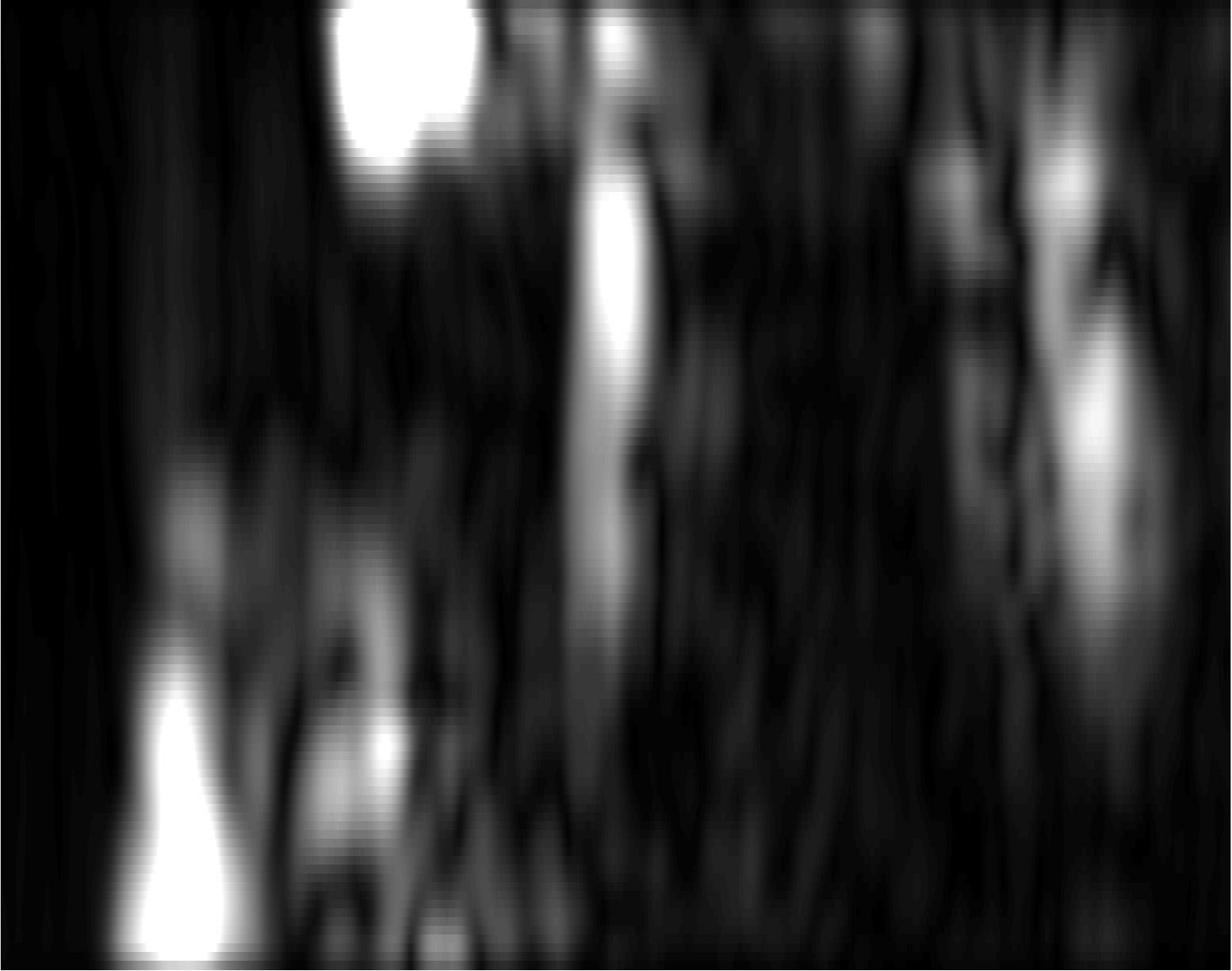}
		\end{center}
   \end{minipage} \hfill
   \begin{minipage}[c]{.5\linewidth}
		\begin{center}
	   		\includegraphics[width=.58\linewidth]{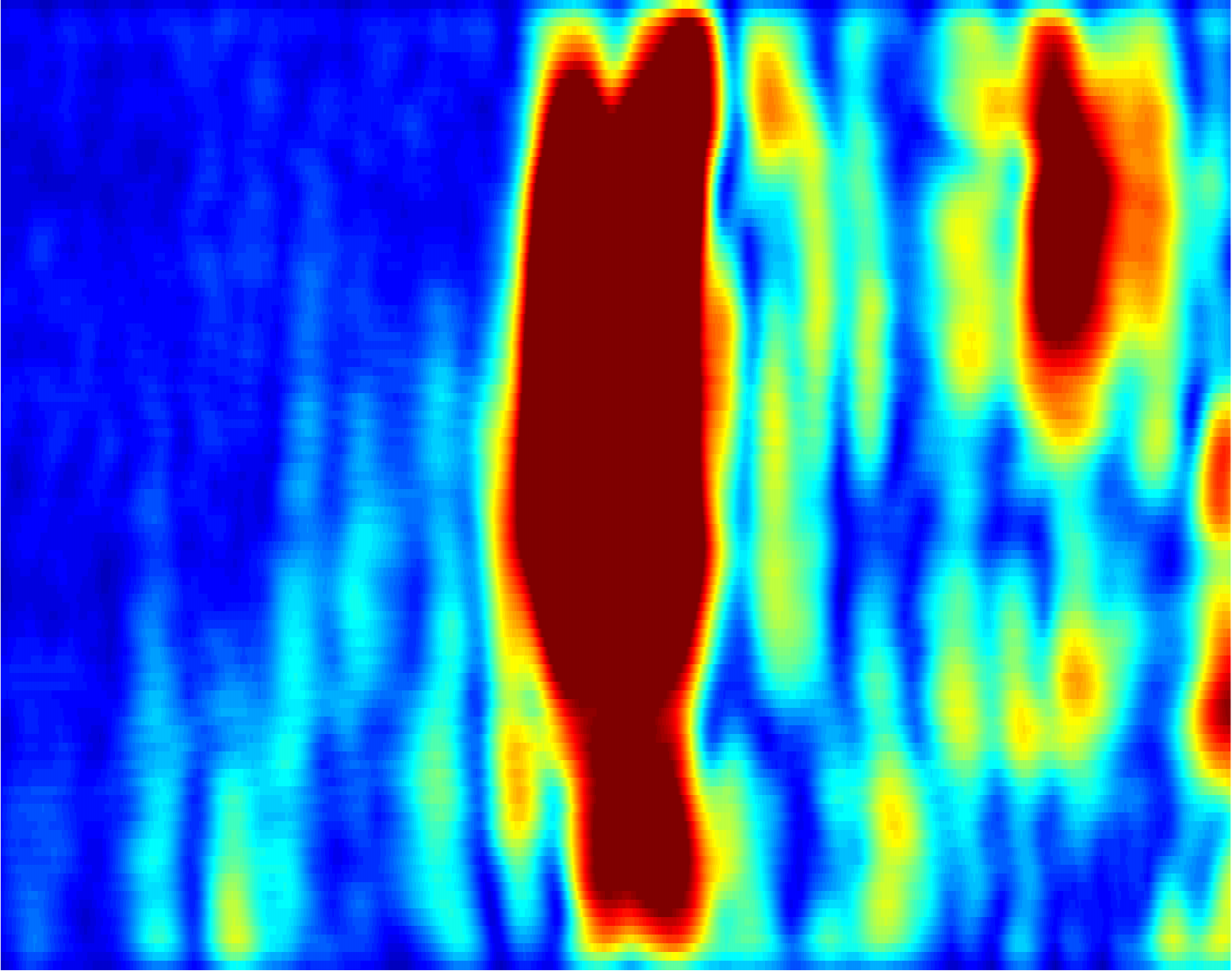}
		\end{center}
   \end{minipage}
   \begin{minipage}[c]{.5\linewidth}
		\begin{center}
	   		\includegraphics[width=.58\linewidth]{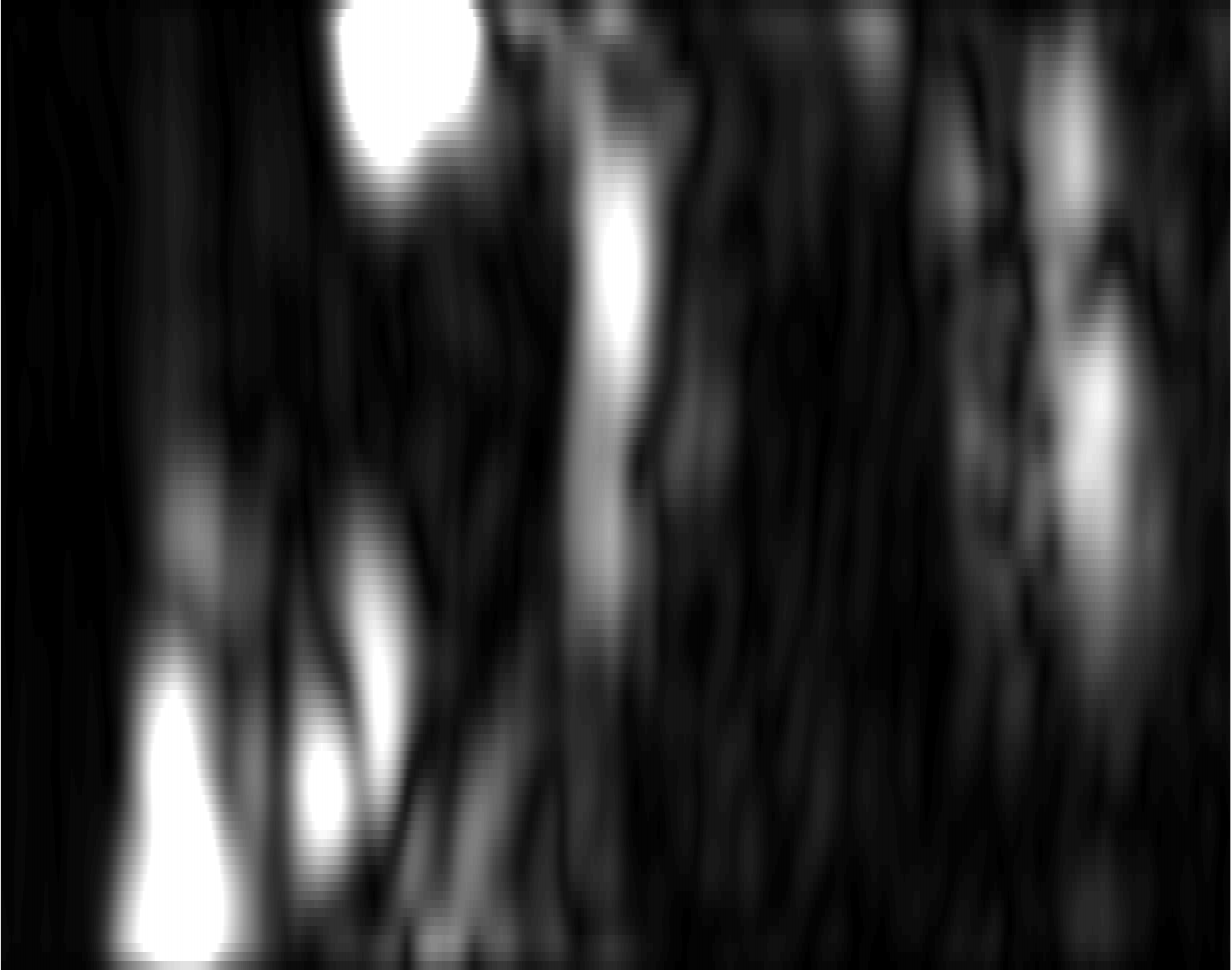}
		\end{center}
   \end{minipage} \hfill
   \begin{minipage}[c]{.5\linewidth}
		\begin{center}
	   		\includegraphics[width=.58\linewidth]{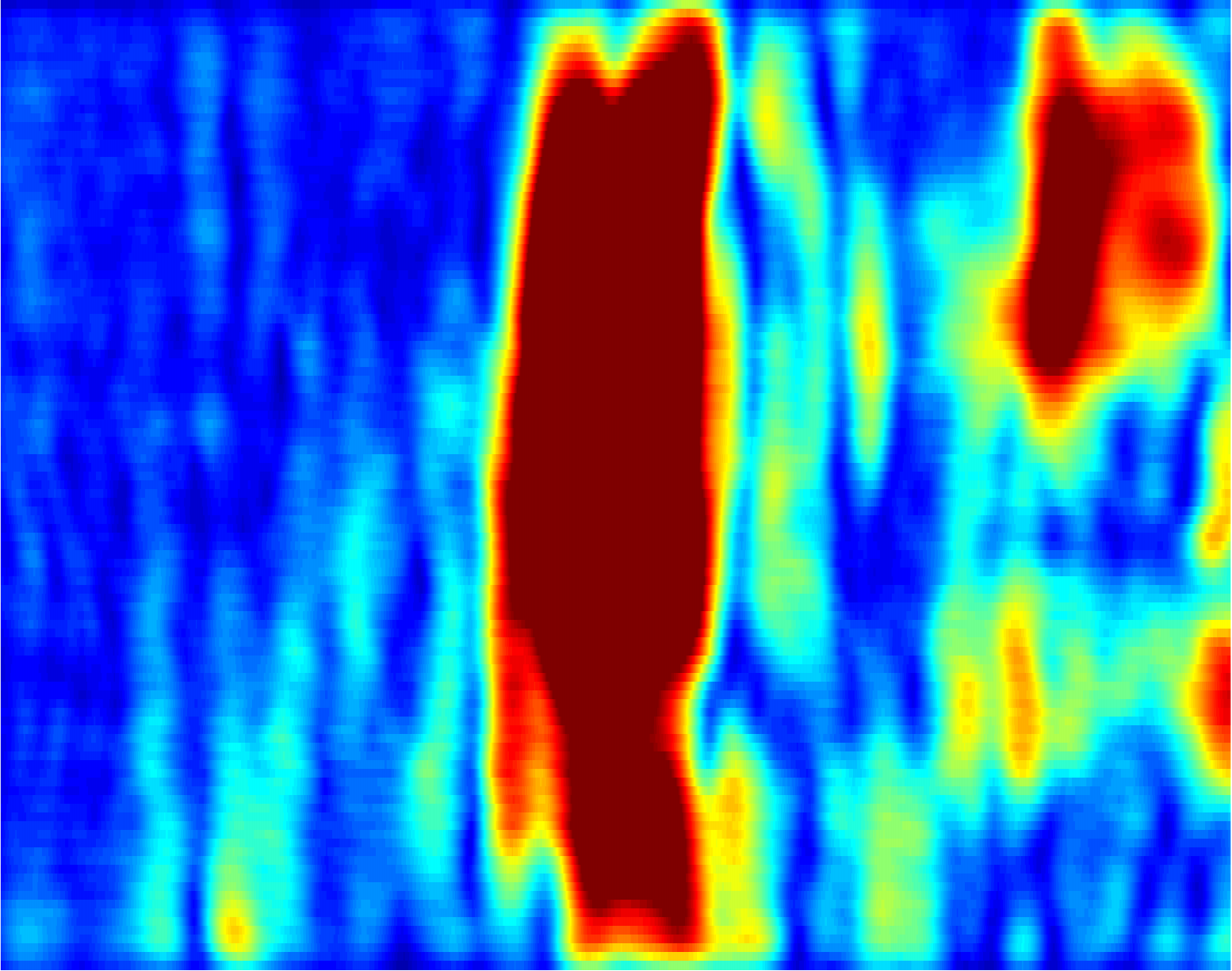}
		\end{center}
   \end{minipage}
	  \caption{(a) Images ultrasonores de la côte de porc, aux angles -2, -1, 0, 1 et 2 $^o$. (b) Image d'impédance électrique correspondantes de la côte de porc, aux angles -2, -1, 0, 1 et 2 $^o$. Les interfaces principales sont nettement marquées sur l'ensemble des images. Les variations plus faibles, interprétées comme du \textit{speckle}, varient entre les différents angles, surtout entre les angles ``extrêmes''.}
	  \label{figLFEITExpSpeckle}
\end{figure}

\paragraph{Résultats et discussions}
La figure \ref{figLFEITExpSpeckle} représente les images ultrasonores et d'impédance électrique à 5 angles différents : -2$^o$, -1$^o$, 0$^o$, +1$^o$ et +2$^o$.

On peut observer sur les images que les interfaces principales sont nettement visibles. On voit néanmoins des changements de texture entre ces interfaces, d'autant plus importants que l'angle est différent, bien que cela reste difficile à quantifier. Ces changements ne sont pas dus à des variations temporelles aléatoires des signaux, étant donné la reproductibilité des expériences au cours du temps.

\begin{figure}[!ht]
   \begin{minipage}[c]{.5\linewidth}
		\begin{center}
	   		\includegraphics[width=.58\linewidth]{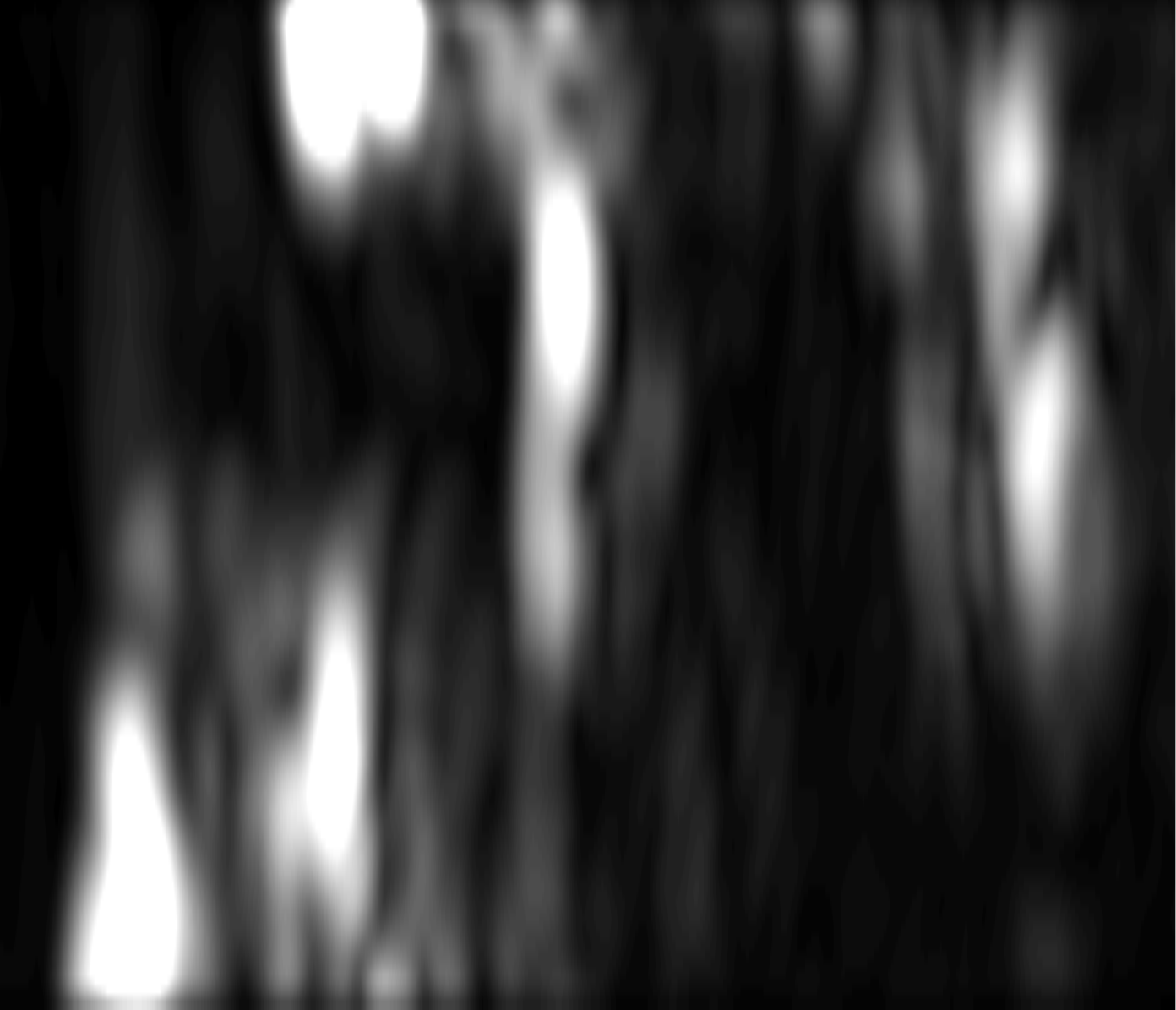}
		\end{center}
   \end{minipage} \hfill
   \begin{minipage}[c]{.5\linewidth}
		\begin{center}
	   		\includegraphics[width=.58\linewidth]{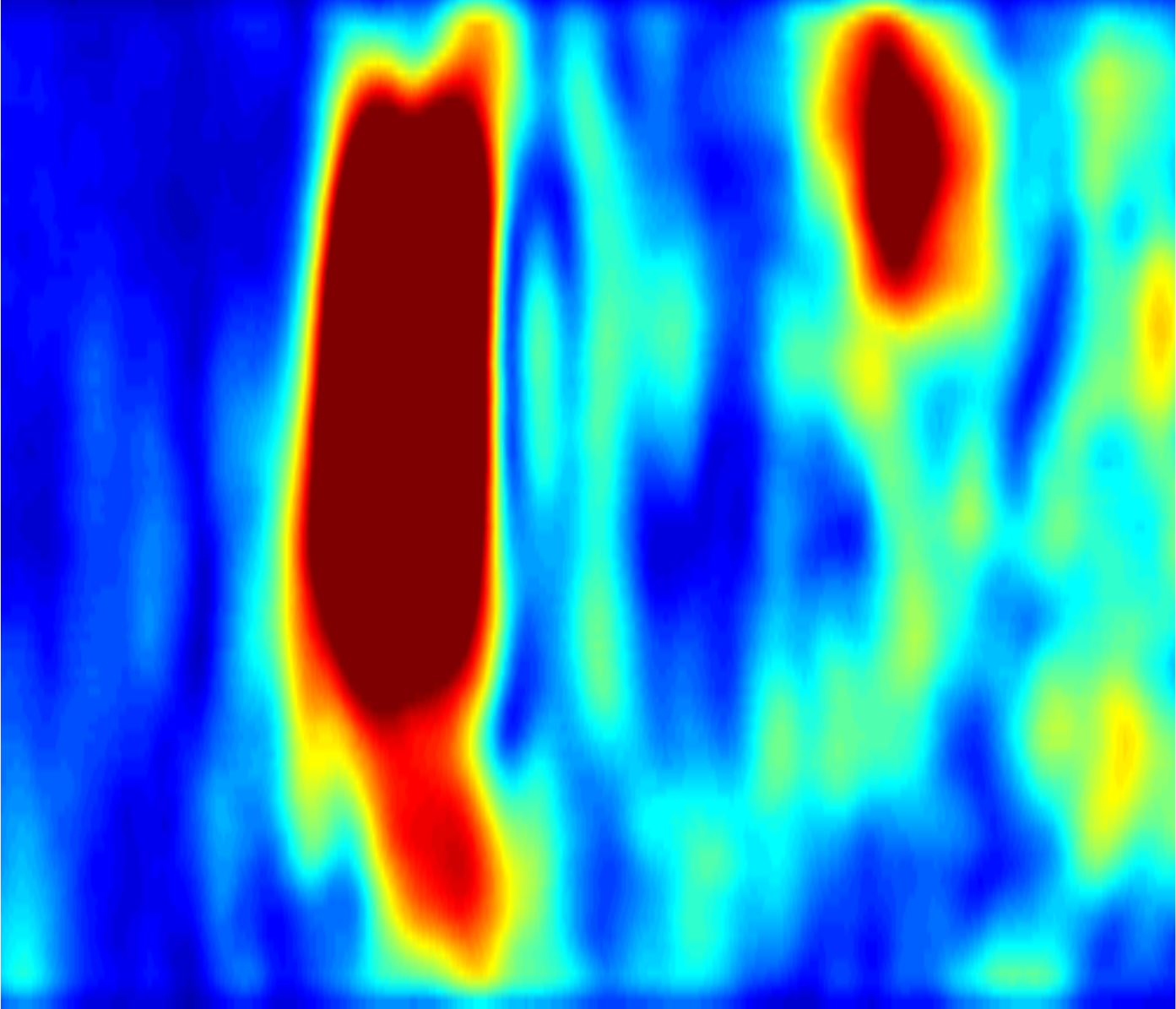}
		\end{center}
   \end{minipage}
	   		\caption{(a) Image ultrasonore composée à partir d'une moyenne des acquisitions cinq angles différents. (b) Image d'impédance électrique dcomposée à partir d'une moyenne des acquisitions cinq angles différents.}
	   		\label{figLFEITExpSpeckleCompound} 
\end{figure}

On peut donc concevoir une image composée en réalisant des moyennes sur les acquisitions à cinq angles différents, comme représenté sur l'image \ref{figLFEITExpSpeckleCompound}. Cela permet de diminuer la présence du \textit{speckle} sur les images, et donc améliorer légèrement la qualité des images - même si cela n'est pas évident sur l'image présentement obtenue. Cependant, l'application la plus intéressante serait non pas de réduire ce \textit{speckle}, mais de l'utiliser pour caractériser les tissus et leur état, comme le font les médecins en échographie.

%\newline

\subsection{Suivi de lésion thermique}
Dans la section \ref{secLFEITIntroLesionThermique} consacré à l'influence de la température sur l'impédance électrique des tissus, nous avons vu qu'une lésion thermique modifiait la conductivité électrique des tissus. Le but de cette expérience était la méthode était suffisamment efficace pour observer un changement suite à une lésion thermique.

\paragraph{Matériel et méthodes}
Pour cela, une image d'un échantillon de blanc de poulet de 4x3x5 cm$^3$ était observé par le transducteur avec 60 lignes séparées de 2 mm. Afin d'obtenir un signal plus important, chaque salve était composée de cinq sinusoïdes au lieu de trois. Cet échantillon était placé de 22 à 27 cm du transducteur. Puis, le transducteur était retiré à l'aide d'un banc motorisé pour laisser place à un transducteur de thérapie ultrasonore. Ce deuxième transducteur, placé à 2 cm de la face avant de l'échantillon, était focalisé à 50 mm et avait une fréquence centrale de 0,5 MHz. Un générateur amplifié par un amplificateur 500 W (1040L, Electronics and Innovation Engineering, Rochester, NY, Etats-Unis) émettait durant 60 secondes avec un cycle de 50 \% un signal électrique de puissance 16W comme mesuré par un wattmètre (392.4017.02 NAP, Rhode \& Schwartz GmbH, Munich, Allemagne). Le transducteur a effectué 30 lésions successives séparées de 1 mm chacune, formant une bande verticale de lésions sur 3 cm de hauteur et une zone non insonifiée de 1 cm dessous. Puis, le transducteur était retiré, et après 10 minutes d'attente pour que le tissu refroidisse, le transducteur initial d'imagerie était replacé à sa position initiale pour acquérir une nouvelle image. Enfin, l'échantillon a été retiré et coupé en deux afin d'évaluer l'étendue de la lésion HIFU. 

\begin{figure}[!ht]
   \begin{minipage}[c]{.5\linewidth}
		\begin{center}
	   		\includegraphics[height=4cm]{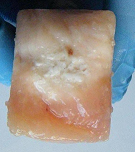}
		\end{center}
   \end{minipage} \hfill
   \begin{minipage}[c]{.5\linewidth}
		\begin{center}
	   		\includegraphics[height=4cm]{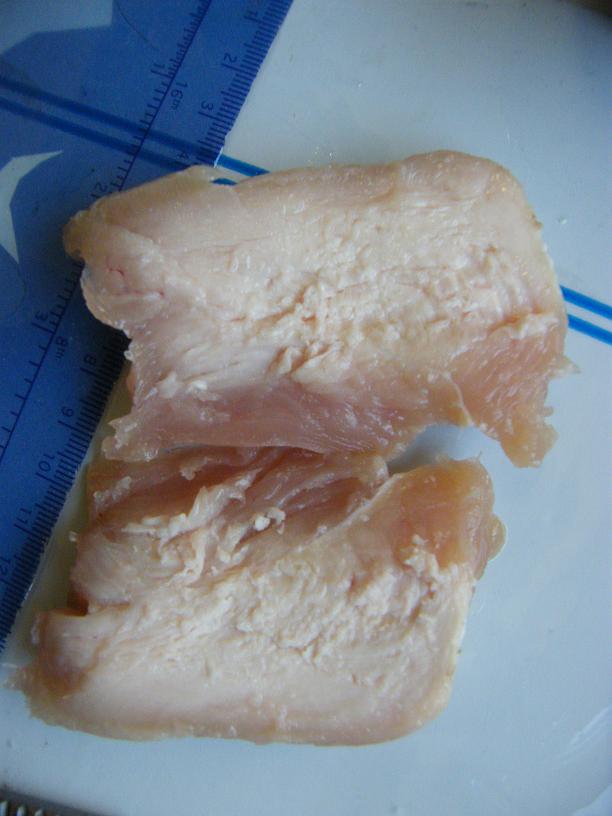}
		\end{center}
   \end{minipage}
	   		\caption{(a) Photographie de la face arrière de l'échantillon de blanc de poulet après lésion thermique. (b) Photographie d'une coupe centrale l'échantillon de blanc de poulet après lésion thermique.}
	   		\label{figLFEITExpHIFU} 
\end{figure}

\paragraph{Résultats et discussions}
La figure \ref{figLFEITExpHIFU}-(a) montre la face arrière de l'échantillon de poulet après lésion thermique. La lésion s'est propagée jusqu'aux bords de l'échantillon, sauf celui du bas. La figure \ref{figLFEITExpHIFU}-(b) montre une coupe centrale du blanc de poulet après lésion thermique. On observe une bande de tissu lésé de 3 cm de haut, de 1 cm de la face avant et à la face arrière. Une bande 1 cm de haut de tissu non lésé est présente sous la lésion thermique.

\begin{figure}[!ht]
   \begin{minipage}[c]{.5\linewidth}
		\begin{center}
	   		\includegraphics[width=1\linewidth]{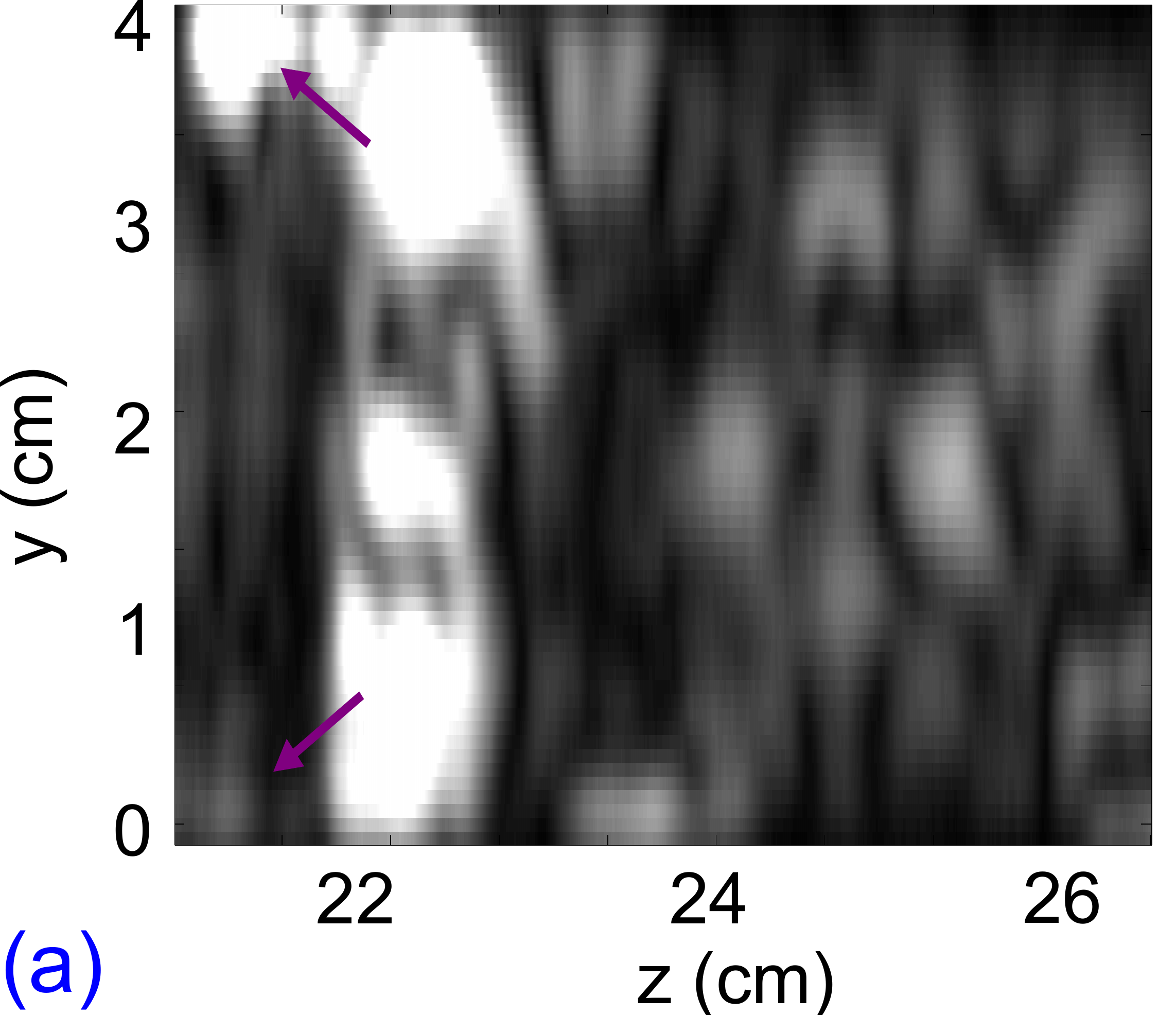}
		\end{center}
   \end{minipage} \hfill
   \begin{minipage}[c]{.5\linewidth}
		\begin{center}
	   		\includegraphics[width=1\linewidth]{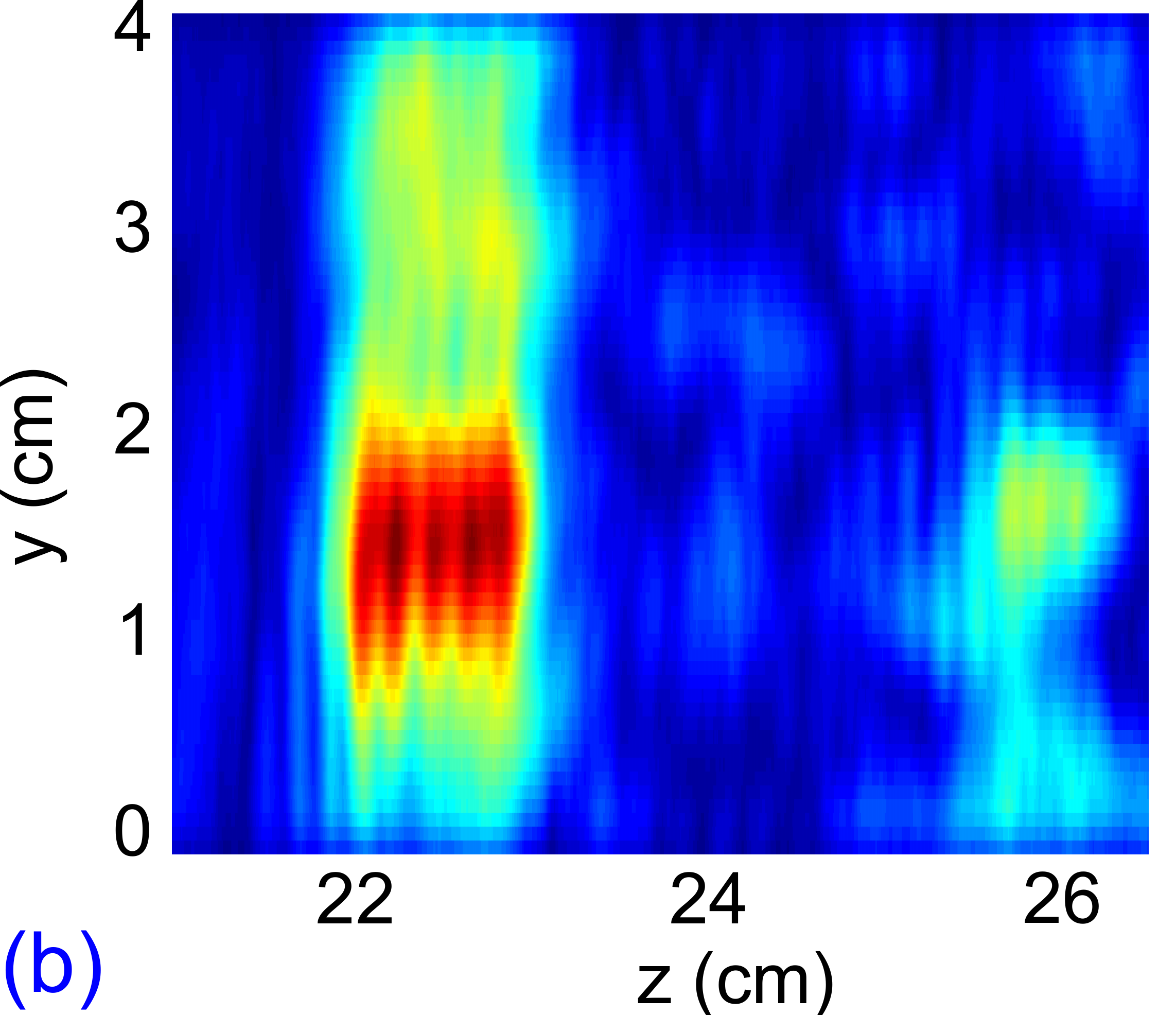}
		\end{center}
   \end{minipage}
 	\caption{(a) Image ultrasonore de l'échantillon avant lésion thermique. (b) Image d'impédance électrique de l'échantillon avant lésion thermique.}
 	\label{figLFEITExpHIFUBefore}
   \begin{minipage}[c]{.5\linewidth}
		\begin{center}
	   		\includegraphics[width=1\linewidth]{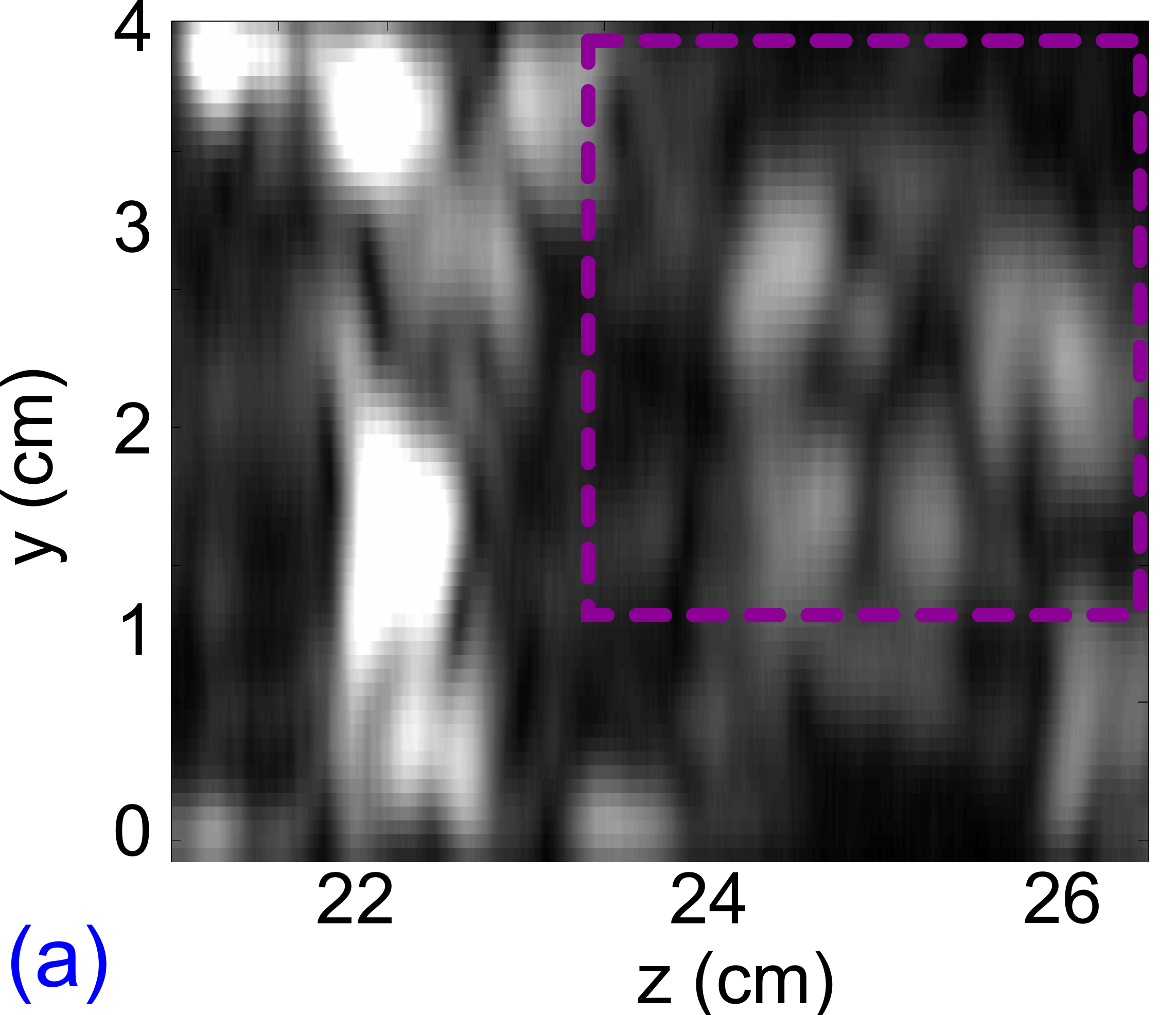}
		\end{center}
   \end{minipage} \hfill
   \begin{minipage}[c]{.5\linewidth}
		\begin{center}
	   		\includegraphics[width=1\linewidth]{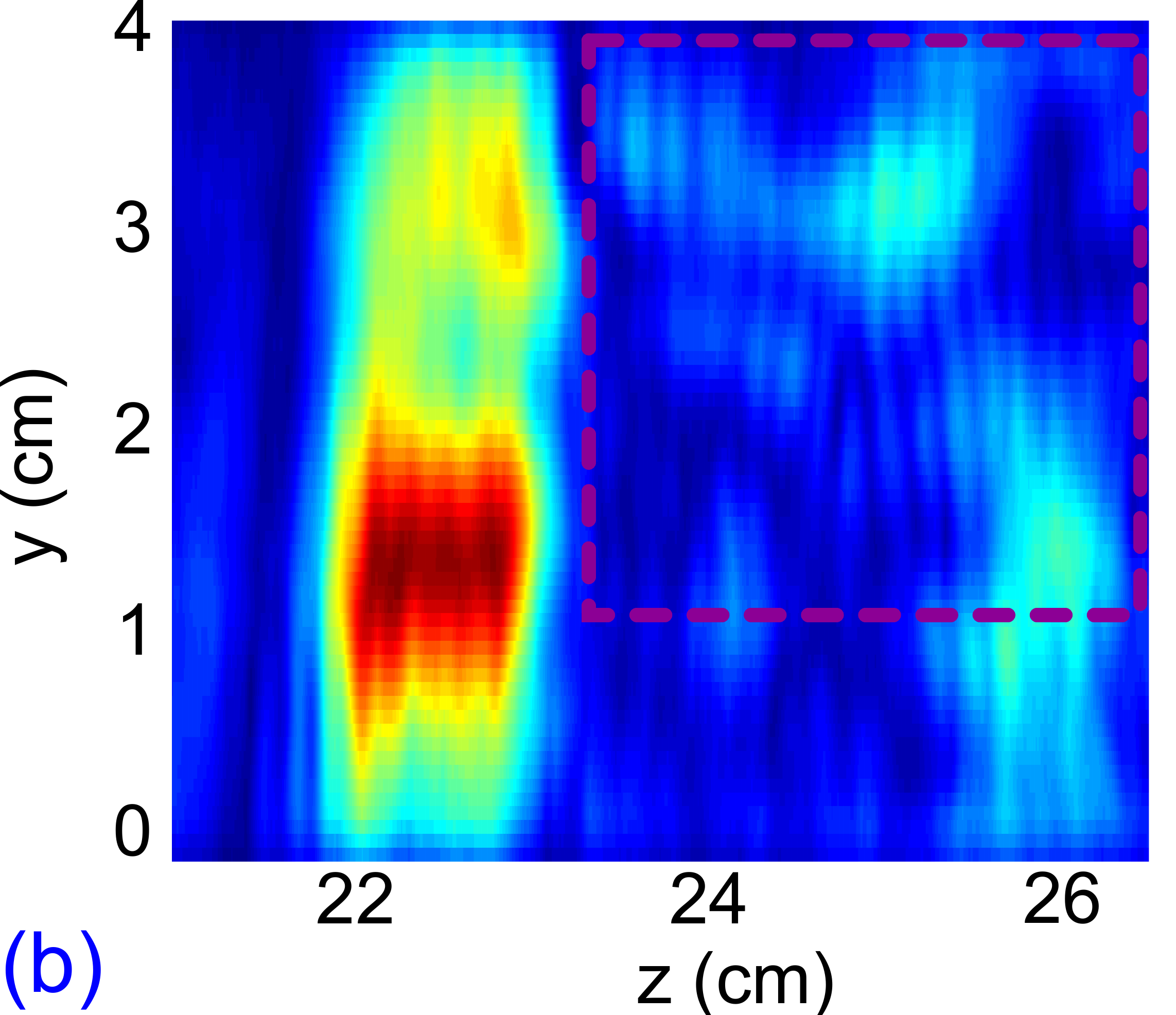}
		\end{center}
   \end{minipage}
 	\caption{(a) Image ultrasonore de l'échantillon après lésion thermique, avec la zone correspondant à la lésion thermique encadrée en pointillés violets. (b) Image d'impédance électrique de l'échantillon avant lésion thermique, avec la zone correspondant à la lésion thermique encadrée en pointillés violets.}
 	\label{figLFEITExpHIFUAfter} 
\end{figure}

Les images ultrasonore et d'impédance électrique avant et après lésion thermique sont représentées sur les figures \ref{figLFEITExpHIFUBefore}-(a) et (b) et \ref{figLFEITExpHIFUAfter}-(a) et (b) respectivement. On observe sur toutes les images l'interface d'entrée et une texture granuleuse à l'intérieur du tissu. Les taches blanches en haut à gauche et l'autre en bas à gauche de l'image ultrasonore, indiquées par des flèches violettes, correspondent à des réflexions sur les entretoises du support ; ces entretoises, plongées dans le bain d'huile non conducteur, sont elles-mêmes non conductrices et donc ne sont pas visibles sur l'image d'impédance électrique. Les interfaces ont une épaisseur assez importante et ont un aspect de bandelettes verticales, surtout sur l'image d'impédance électrique, ce qui est dû à l'utilisation de cinq sinusoïdes lors des séquences d'imagerie.

\begin{figure}[!ht]
		\begin{center}
	   		\includegraphics[width=.5\linewidth]{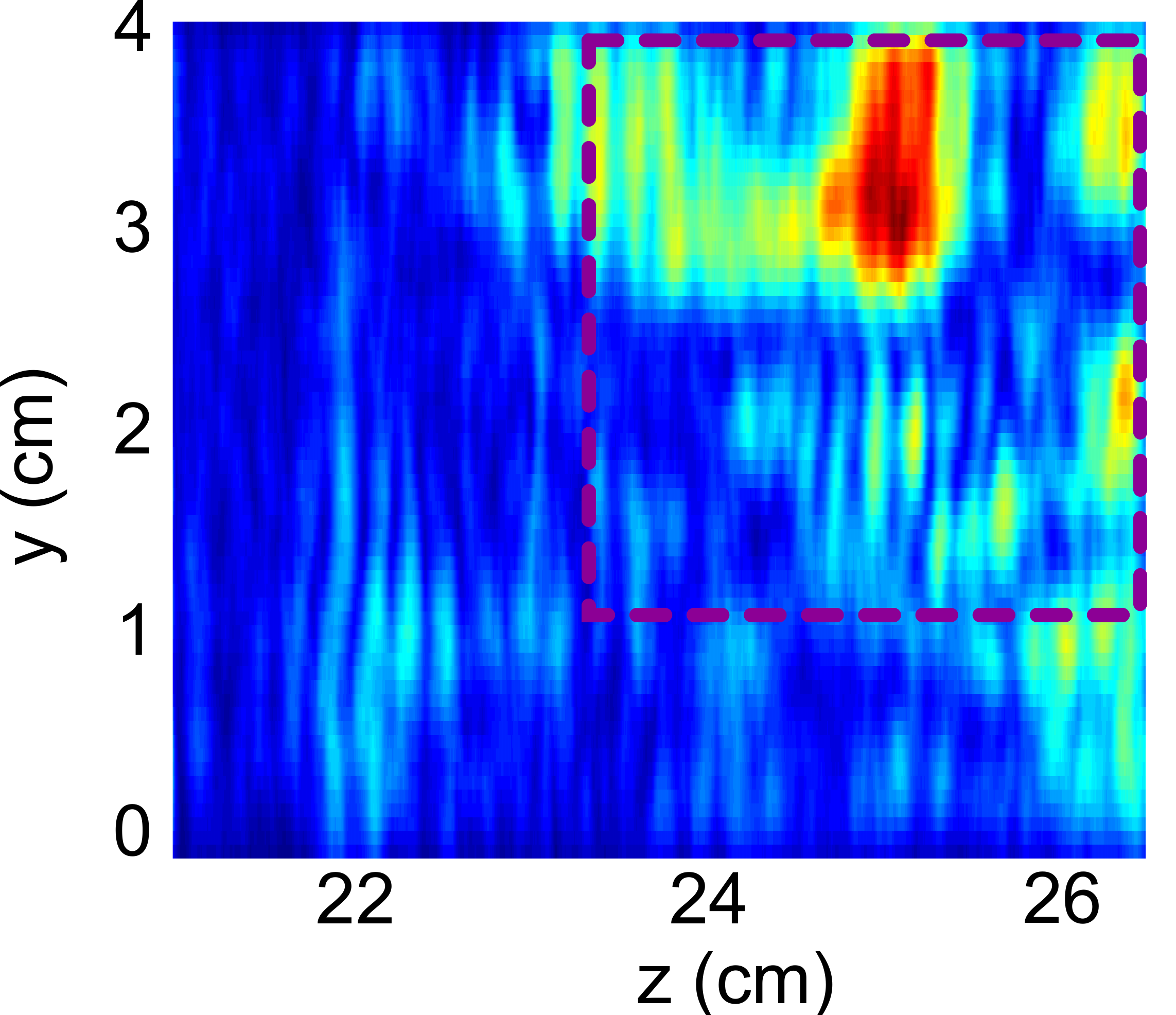}
	   		\caption{Différence entre l'image d'impédance électrique \textit{avant} et \textit{après} lésion thermique. Les principales modifications sont observées dans la zone lésée, comme indiquée par l'encadré violet.}
	   		\label{figLFEITExpHIFUDiffLFEIT} 
		\end{center}
\end{figure}

On voit sur les deux images après lésion thermique les mêmes interfaces d'entrée que sur les images avant lésion thermique. Cependant, plus en profondeur et sur la partie haute de l'image, dans la zone encadrée par un pointillé violet, on remarque une modification de texture. Cette zone se superpose à la zone lésée. En-dehors de cette zone, on observe quelques changements sur l'image ultrasonore, mais quasiment pas sur l'image d'impédance électrique. Ces observations sont plus claires si l'on soustrait les signaux acquis avant à ceux acquis après après lésion thermique, ce qui permet de reconstruire une image des différences comme sur la figure \ref{figLFEITExpHIFUDiffLFEIT}.

Cette expérience a montré que la tomographie d'impédance électrique par force de Lorentz peut être potentiellement utilisée pour suivre des lésions thermiques. Les différences sur les images d'impédance électrique avant et après lésion sont certainement dues aux modifications de conductivité électrique dus à la lésion thermique. Cependant, il n'est pas suffisamment clair sur les images s'il s'agit de modifications d'hétérogénéités changeant le \textit{speckle} acousto-électrique, de changements globaux de conductivité électrique, voire éventuellement d'effets de la température si le tissu n'était pas parfaitement refroidi. Pour investiguer cela, il est nécessaire d'améliorer la qualité des images, et tout particulièrement la résolution spatiale.

\section{Bilan du chapitre}

\subsection{Principaux résultats}
L'objectif de ce travail était d'étudier une méthode d'imagerie de la conductivité électrique que nous avons nommée Tomographie d'Impédance Electrique par Force de Lorentz. Dans cette méthode, on met en mouvement grâce à des ultrasons focalisés un morceau de tissu biologique placé dans un champ magnétique. Cela induit un courant électrique par force de Lorentz, mesurable par des électrodes. Ce courant est relié aux variations d'impédance électrique le long de l'axe ultrasonore. Le principal intérêt de cette méthode réside dans sa résolution spatiale : celle-ci est en théorie proche de la longueur d'onde ultrasonore. 

La première partie de ce travail a consisté à objectiver le signal afin s'assurer que le courant détecté est effectivement celui induit par force de Lorentz. Il s'avère que l'émission et la réception d'ultrasons par le transducteur s'ajoute au signal induit, ce qui a posé plusieurs difficultés expérimentales.

La technique a ensuite été appliquée sur des fantôme de gélatine et de morceaux de tissus biologiques. Les images d'impédance électrique et les images ultrasonores sont cohérentes entre elles, lorsque les interfaces d'impédance électrique et acoustique étaient présentes aux mêmes emplacements. L'interface de graisse d'un tissu biologique a notamment été observée sur une image d'impédance électrique sans être visible sur l'image ultrasonore associée, ce qui a été expliqué par la forte variation d'impédance électrique et la faible variation d'impédance acoustique de cette couche par rapport au reste du muscle.

A l'instar des échographies, les images issues de cette méthode sont censées présenter un \textit{speckle} électroacoustique. Celui-ci est en effet dû à des hétérogénéités de nature électrique mais reliées à la longueur d'onde acoustique. La résolution spatiale des images étant cependant médiocre et le signal détecté faible, ce \textit{speckle} est difficilement observable. Néanmoins, des variations faibles d'angle d'imagerie montrent des interfaces quasiment identiques, mais des motifs à l'intérieur du tissu légèrement différents, ce qui est un indice de la présence du \textit{speckle} recherché. Le principal intérêt ne serait cependant pas de réduire ce \textit{speckle} avec des techniques d'imagerie composée, comme présentée dans ce chapitre, mais de l'utiliser pour caractériser les tissus.

Enfin, nous avons tenté de faire des images de lésion thermique avec cette méthode. En effet, comme les tissus lésés thermiquement présentent une conductivité électrique différente des tissus normaux, la méthode peut potentiellement localiser ces zones. Malgré la faible résolution spatiale de la méthode, on observe quand même une modification dans l'image d'impédance électrique, sans que cette modification ne soit clairement délimitée.

\subsection{Avenir de la méthode}
La tomographie d'impédance électrique par force de Lorentz induite par ultrasons présente de nombreuses similarités avec l'échographie : images des interfaces d'impédance, présence de \textit{speckle}, résolution millimétrique. Plusieurs des méthodes dérivées de l'échographie standard sont donc théoriquement envisageables dans cette technique, comme l'imagerie composée. Afin de caractériser le \textit{speckle} électroacoustique comme présenté dans cette thèse, il serait intéressant de faire des images à différentes fréquences ultrasonores afin de'éavluer l'évolution de la taille typique des grains -- cela n'a pu être fait avec le matériel disponible au laboratoire.

La position des électrodes n'étant pas critique \cite{montalibet2002these}, on peut envisager une sonde spéciale comprenant un transducteur ultrasonore et deux électrodes placées de part et d'autre pour récupérer le courant.

La principale difficulté reste cependant la faiblesse du courant électrique créé. Il est possible d'augmenter celui-ci d'un facteur 10 environ à l'aide d'un champ magnétique plus puissant. En utilisant un appareil IRM, même si l'on perdrait en même temps en légèreté et en coût on gagnerait en puissance et en homogénéité du champ magnétique. L'utilisation de larges électrodes permet de recueillir plus de courant, mais il est probable que le gain potentiel soit inférieur à un ordre de grandeur par rapport aux configurations actuellement utilisées. Le capteur de courant peut par contre être largement optimisé : comme on connait exactement le signal recherché, les méthodes de filtrage peuvent extraire le signal utile sans être obligé d'effectuer des moyennes sur de nombreuses acquisitions.

Un autre axe de développement serait de développer une méthode d'imagerie avec des ondes ultrasonores planes acquises à différents angles, comme en échographie ultrarapide. On gagnerait en vitesse d'acquisition et peut-être en qualité de reconstruction. Ceci se rapproche notamment de techniques d'imagerie acousto-électrique ultrarapide actuellement en développement \cite{provost2013ultrafast}.

Plusieurs applications sont envisageables. Comme évoqué précédemment, cette méthode présente un contraste différent de l'IRM, de l'échographie ou des radiographies : on peut donc observer des choses invisibles avec d'autres méthodes, notamment grâce au \textit{speckle} des images. De plus, comme cela a été fait durant cette thèse, le matériel peut réaliser simultanément une échographie, donnant par là une image multimodale en temps réel.

Une application particulièrement intéressante serait d'utiliser l'influence de la température sur la conductivité électrique des tissus, afin de réaliser suivi en temps réel des traitements par ultrasons focalisés.

\chapter{Imagerie d'ondes de cisaillement induites par force de Lorentz}
\label{secElastoEM}

\section{L'élastographie par ondes de cisaillement}

\subsection{L'élasticité des tissus biologiques}
Comme montré sur le tableau \ref{tableElastoEMElastoOrigine}, certains tissus pathologiques comme les tumeurs du sein ou de la prostate présentent une élasticité différente des tissus sains, élasticité que l'on peut définir par le module d'Young $E$ du tissu.

\begin{table}
	\begin{tabular}{|c|c|c|}
		\hline
		\rowcolor{SandyBrown}\multicolumn{2}{|c|}{Type de tissu biologique} & module d'Young (kPa) \\
		\hline
		Sein 			& Normal adipeux & 18-24 \\
		\cline{2-3} & Normal glandulaire & 28-66 \\
		\cline{2-3} & Tissu fibreux & 96-244 \\
		\cline{2-3} & Carcinome & 22-560 \\
		\hline
		Prostate 		& Normal antérieur & 55-63 \\
		\cline{2-3} & Normal postérieur & 62-71 \\
		\cline{2-3} & Hypertrophie bénigne & 36-41 \\
		\cline{2-3} & Carcinome & 96-241 \\
		\hline
		Foie 			& Normal & 0,4-6 \\
		\cline{2-3} & Cirrhose & 15-100 \\
		\hline
	\end{tabular}
	\caption{Quelques valeurs du module d'Young de tissus biologiques sains et pathologiques.}
	\label{tableElastoEMElastoOrigine}
\end{table}
Le module d'Young d'un tissu est défini à partir de la loi de Hooke. Cette loi, valable pour des faibles déformations en traction ou compression, énonce que l'allongement $\Delta l$ est proportionnel à la force appliquée $F$, avec un coefficient de proportionnalité $k$ : $F=k \Delta l$.

\begin{figure}[!ht]
 \begin{center}
	 \includegraphics[width=0.25\columnwidth]{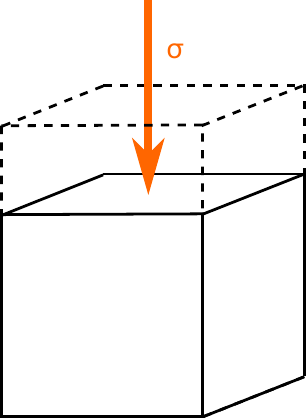}
	 \caption{Le module d'Young est le coefficient de proportionnalité entre la contrainte appliquée $\sigma$ et l'allongement relatif $\epsilon$ du matériau.}
	 \label{figElastoEMIntroElasticiteModuleYoung}
 \end{center}
\end{figure}

En divisant cette équation par la surface d'application de la force et par la longueur totale du matériau, on peut relier la contrainte $\sigma$ en pascals et la déformation relative $\epsilon$ en pourcent, avec un coefficient de proportionnalité $E$ (voir schéma \ref{figElastoEMIntroElasticiteModuleYoung}) :
\begin{empheq}[box=\mybluebox]{align}
	\sigma = E \epsilon
	\label{eqElastoEMElastoPrincipe2}
\end{empheq}
Le coefficient $E$ est le module d'Young, homogène à une pression.

\begin{figure}[!ht]
 \begin{center}
	 \includegraphics[width=0.4\columnwidth]{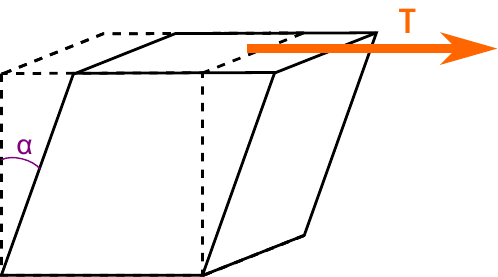}
	 \caption{Le module de cisaillement est le coefficient de proportionnalité entre la contrainte appliquée $\sigma$ et la tangente de l'angle de déformation $\alpha$ du matériau.}
	 \label{figElastoEMIntroElasticiteModuleCisaillement}
 \end{center}
\end{figure}

De la même façon, pour de faibles déformations en cisaillement, la contrainte de cisaillement (ou cission) $\tau$ en pascals est reliée à l'angle de déformation relatif $\alpha$ en radian (voir schéma \ref{figElastoEMIntroElasticiteModuleCisaillement}) :
\begin{empheq}[box=\mybluebox]{align}
	\tau = \mu \tan(\alpha)
	\label{eqElastoEMElastoPrincipe3}
\end{empheq}
avec $\mu$ appelé module de cisaillement (aussi appelé deuxième coefficient de Lamé), homogène à une pression.

\begin{figure}[!ht]
 \begin{center}
	 \includegraphics[width=0.4\columnwidth]{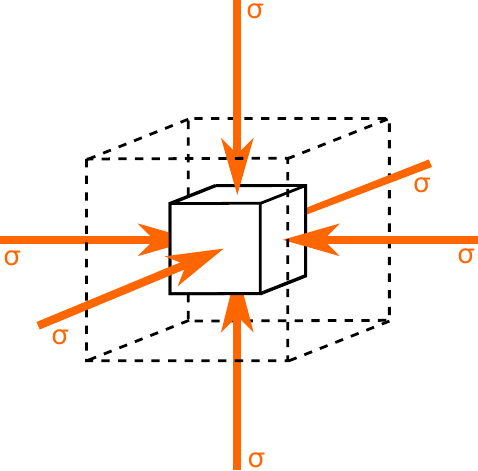}
	 \caption{Le module d'Young est le coefficient de proportionnalité entre la contrainte appliquée isostatique $\sigma_{tot}$ et l'allongement relatif total $\epsilon_tot$ du matériau.}
	 \label{figElastoEMIntroElasticiteModuleCompressionIsostatique}
 \end{center}
\end{figure}

Enfin, la déformation relative totale $\epsilon_{tot}$ en pourcent est proportionnelle à la contrainte isostatique $\sigma_{tot}$ en pascals (voir schéma \ref{figElastoEMIntroElasticiteModuleCompressionIsostatique}) :
\begin{empheq}[box=\mybluebox]{align}
	\sigma_{tot} = K \epsilon_{tot}
	\label{eqElastoEMElastoPrincipe4}
\end{empheq}
avec $K$ le module d'élasticité isostatique en Pa.

Ces trois grandeurs sont liées entre elles selon l'équation :
\begin{empheq}[box=\mybluebox]{align}
	E=\frac{9K\mu}{3K + \mu}
	\label{eqElastoEMElastoPrincipe5}
\end{empheq}

\begin{figure}[!ht]
 \begin{center}
	 \includegraphics[width=0.8\columnwidth]{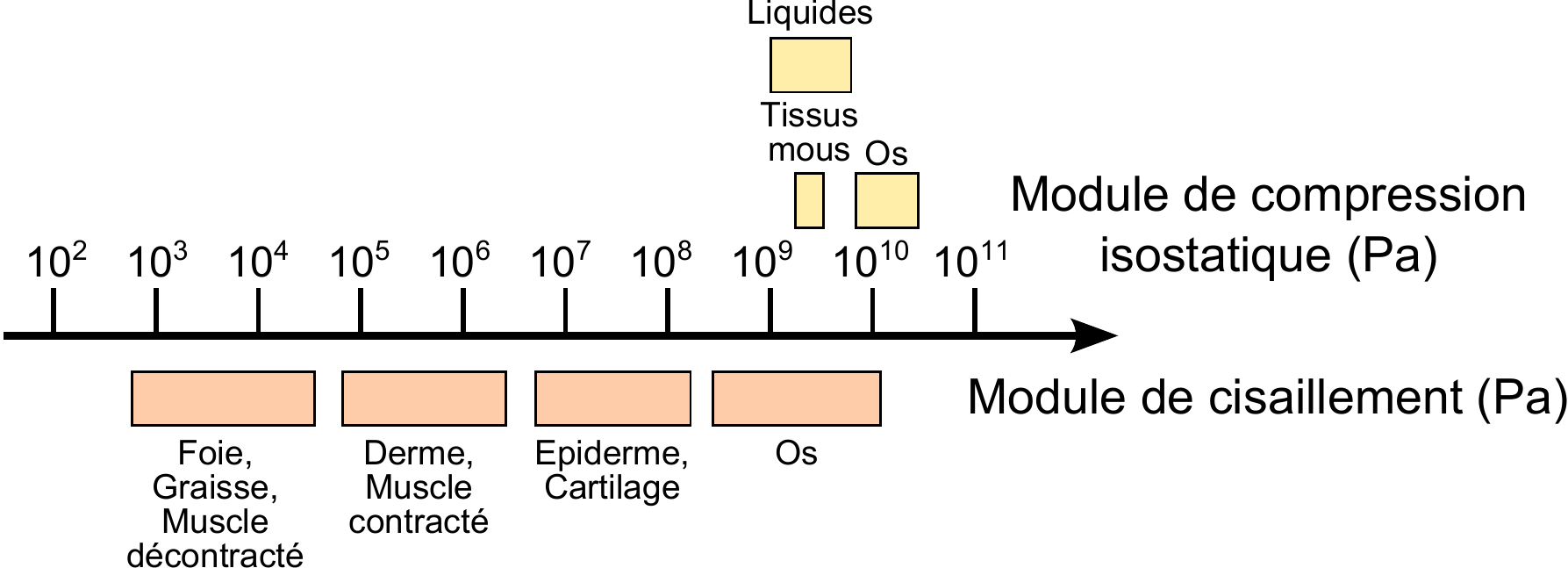}
	 \caption{Le module d'élasticité isostatique $K$ vaut environ 10 GPa pour la plupart des tissus biologiques. Selon les tissus, le module de cisaillement varie de plusieurs ordres de grandeurs, de 100 Pa à 10 GPa \cite{sarvazyan1998}.}
	 \label{figElastoEMIntroElasticiteTissus}
 \end{center}
\end{figure}

Alors que le module d'élasticité isostatique $K$ varie peu selon les tissus, le module de cisaillement montre une grande variabilité au sein du corps humain, comme représenté sur la figure \ref{figElastoEMIntroElasticiteTissus}

\subsection{L'élastographie statique}
La palpation manuelle d'un organe, principalement sensible au module de cisaillement du tissu, permet de détecter des nodules durs. Cette technique assez simple est malheureusement non quantitative, dépend de l'opérateur et nécessite d'avoir des nodules gros et durs pour être détectables \cite{hall1977progress}. Pour améliorer les diagnostics des médecins, des méthodes d'imagerie de l'élasticité des tissus ont été développées à partir des années 1990.

Une première méthode a été introduite par Ophir \textit{et al.} en 1991 \cite{ophir1991}, que l'on nomme aujourd'hui élastographie statique. Une première image ultrasonore est réalisée, puis une contrainte appliquée grâce à une tige extérieure, suivie d'une deuxième image ultrasonore. Cette technique obtient le déplacement à partir de corrélations locales entre les deux images, technique appelée interférométrie des tavelures ultrasonores, ou plus souvent \textit{speckle-tracking}. Le principal problème vient de la loi de Hooke, qui a la formulation d'une matrice 6x6 lorsqu'on étudie des volumes à trois dimensions :
\begin{empheq}[box=\mybluebox]{align}
	\sigma_{ij} = C_{ijkl} \epsilon_{kl}
	\label{eqElastoEMElastoLoiHookeMatricielle}
\end{empheq}
avec la convention de sommation d'Einstein pour $i$, $j$, $k$, $l$, chacune prenant une valeur parmi X, Y et Z. Ainsi, lors d'une déformation appliquée manuellement, de nombreuses valeurs de contraintes sont inconnues, donc il est difficile d'en déduire les $C_{ijkl}$ et le module d'Young du matériau. Pour éviter ce problème, des méthodes dynamiques utilisant la propagation d'ondes mécaniques dans les tissus ont été proposées.

\subsection{La vitesse des ondes mécaniques dans les tissus biologiques}
\label{secElastoEMIntroVitesse}

\begin{figure}[!ht]
 \begin{center}
	 \includegraphics[width=0.8\columnwidth]{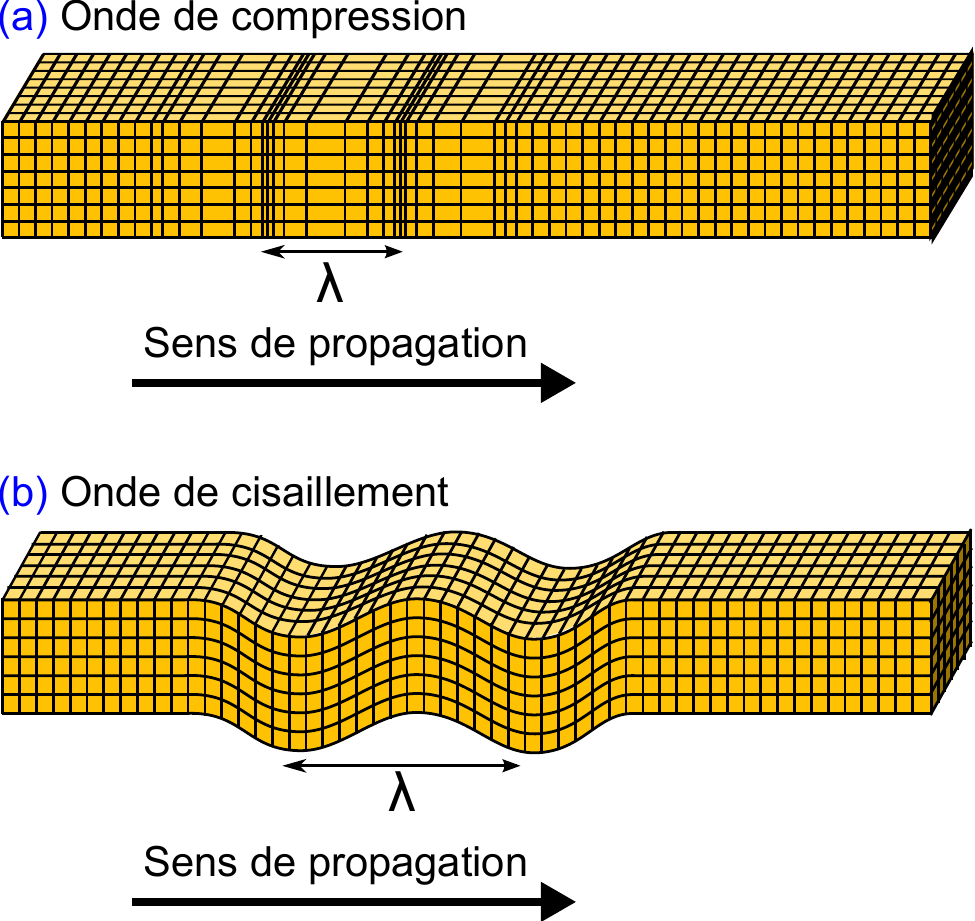}
	 \caption{(a) Onde de compression plane se propageant dans un milieu infini. La propagation se fait dans la même direction que le déplacement local. (b) Onde de cisaillement plane se propageant dans un milieu infini. La propagation se fait dans un sens orthogonal au déplacement local.}
	 \label{figElastoEMIntroPrincipeOndePS}
 \end{center}
\end{figure}

Comme dans tous les solides, deux types d'ondes mécaniques existent dans les tissus biologiques : les ondes de compression et les ondes de cisaillement, dont la propagation dans un milieu infini est illustrée sur la figure \ref{figElastoEMIntroPrincipeOndePS}.

Rappelons que l'onde de compression n'est longitudinale (propagation dans la direction du déplacement) et celle de cisaillement transversale (propagation dans une direction orthogonale au déplacement) que pour des ondes planes. Pour une onde non plane, il est donc tout à fait possible d'avoir des ondes de cisaillement longitudinales.

La démonstration de l'existence et la description de ces deux ondes peut se faire à partir de l'équation de Navier, aussi appelée équation de l'élastodynamique :
\begin{empheq}[box=\mybluebox]{align}
	\rho\frac{d^2\mathbf{u}}{dt^2} = (K + \frac{4}{3}\mu) \grad \Div \mathbf{u} + \mu \rot \rot \mathbf{u} + \frac{d\mathbf{F}}{d\tau}
	\label{eqElastoEMIntroVitesse1}
\end{empheq}
avec $\rho$ la densité du milieu, $u$ le déplacement local du tissu, $K$ le module de compression isostatique, $\mu$ le module de cisaillement et $\frac{d\mathbf{F}}{d\tau}$ une force volumique extérieure.

Supposons qu'aucune force extérieure n'est appliquée sur le tissu. On peut prendre la divergence de l'équation \ref{eqElastoEMIntroVitesse1}, ce qui donne en définissant $p=\Div \mathbf{u}$ :
\begin{empheq}[box=\mybluebox]{align}
		\frac{ d p^2}{dt^2} = v_p^2 \Delta p 
	\label{eqElastoEMIntroVitesse2}
\end{empheq}
avec $v_p = \sqrt{(K + \frac{4}{3}\mu)/\rho}$. Cette équation correspond donc à la propagation de la divergence du déplacement local, ce qui permet de définir $v_p$ comme la vitesse de l'onde de compression.

De même, on peut prendre le rotationnel de l'équation \ref{eqElastoEMIntroVitesse1}, ce qui donne en définissant $\mathbf{s}=\rot \mathbf{u}$ :
\begin{empheq}[box=\mybluebox]{align}
		\frac{ d \mathbf{s}^2}{dt^2} = v_s^2 \Delta \mathbf{s}
	\label{eqElastoEMIntroVitesse3}
\end{empheq}
avec $v_s = \sqrt{\mu/\rho}$. Cette équation correspond à la propagation du rotationnel du déplacement local, ce qui permet de définir $v_s$ comme la vitesse des ondes de cisaillement.

Dans les tissus biologiques, on observe que $v_p$ vaut typiquement environ 1500 m/s et $v_s$ de 1 à 10 m/s, ce qui donne une valeur de $K$ supérieur de quatre à cinq ordres de grandeur à $\mu$. L'expression du module d'Young donnée dans l'équation \ref{eqElastoEMElastoPrincipe5} peut dans ce cas être simplifiée en $E \approx 3 \mu = 3 \rho v_s^2$.

D'après l'expression de $v_s$ dans l'équation \ref{eqElastoEMIntroVitesse3} et la relation approchée entre le module d'Young et le module de cisaillement, la mesure de la vitesse d'une onde de cisaillement en un point d'un milieu permet de calculer son module d'Young en ce point : c'est le principe de l'élastographie par onde de cisaillement. Cette technique est composée de deux parties : un procédé pour créer un déplacement mécanique source d'une onde de cisaillement et un dispositif d'imagerie des déplacements dus à la propagation de l'onde de cisaillement.

Cependant, les ondes de cisaillement sont rapidement atténuées dans les tissus biologiques, avec des coefficients d'atténuation de l'ordre de $10^{5}$ dB/cm pour une fréquence de 10 MHz, alors que les ondes de compression ont un coefficient d'atténuation de l'ordre de 9 dB/cm à cette fréquence. Les ondes de cisaillement ne peuvent donc être observées qu'à basse fréquence, de quelques dizaines à quelques centaines de Hertz.

\subsection{L'élastographie par onde de cisaillement}
Les premières mesures par ondes de cisaillement se sont faites grâce à des ondes de cisaillement monochromatiques, donc avec une excitation continue. Sur ce principe, Greenleaf \textit{et al.} \cite{muthupillai1995magnetic} ont développé une méthode d'élastographie visualisée par IRM. Le principal défaut de cette méthode est de mélanger plusieurs types d'ondes, notamment les ondes de compression et celles de cisaillement. La vitesse mesurée en chaque point est donc une combinaison a priori non connue de $v_p$ et de $v_s$, ce qui induit des erreurs dans les images d'élasticité des tissus. Pour pallier ce problème, Sarvazyan \textit{et al.} \cite{sarvazyan1998} et Catheline \textit{et al.} \cite{catheline1998interferometrie} ont proposé d'utiliser des impulsions plutôt que des signaux continus, le premier avec une visualisation par IRM et le deuxième par ultrasons. Ainsi, grâce à la différence de vitesse de propagation, il est possible de séparer la vitesse de l'onde de compression de celle de l'onde de cisaillement, et donc d'obtenir des images quantitatives d'élasticité. Mais l'une des principales difficultés réside dans le contrôle de la source de ces ondes.

La première méthode employée pour générer une onde de cisaillement dans un tissu a été de placer un vibreur externe à la surface du corps d'un patient. Le déplacement créé à la surface de la peau d'un patient se transmet effectivement à la fois sous la forme d'une onde de compression et d'une onde de cisaillement en profondeur, et le tout peut être observé par ultrasons ou IRM. Ce principe a notamment été utilisé avec succès pour mesurer la dureté du foie et en déduire son état de fibrose \cite{sandrin2003transient} -- un appareil industriel a d'ailleurs été commercialisé sous le nom de Fibroscan \cite{foucher2006diagnosis}.

La pression de radiation ultrasonore peut également induire un déplacement dans un tissu \cite{torr1984acoustic}. Ce déplacement, qui peut être simplement utilisé pour en déduire la dureté du tissu dans la zone focale (méthode ARFI pour Acoustic Radiation Force Imaging \cite{nightingale2002}), crée également une onde de cisaillement \textit{in-situ}. Ainsi, la technique peut être appliquée avec un seul transducteur sans nécessiter d'appareil supplémentaire, ce qui facilite la tâche aux médecins habitués à réaliser des échographies. De plus, si l'on effectue des images ultrasonores à une cadence dite \textit{ultrarapide} avec plusieurs milliers d'images par seconde, on peut suivre la propagation des ondes de cisaillement dans le tissu, ce qui permet de reconstruire directement une image d'élasticité \cite{bercoff2004supersonic}. Cependant, si des obstacles comme des couches de graisses ou de l'os sont présents entre le transducteur ultrasonore et la zone focale des ultrasons, l'image peut présenter des aberrations voire être impossible à former.

\subsection{L'utilisation de la force de Lorentz}
\begin{figure}[!ht]
 \begin{center}
	 \includegraphics[width=0.6\columnwidth]{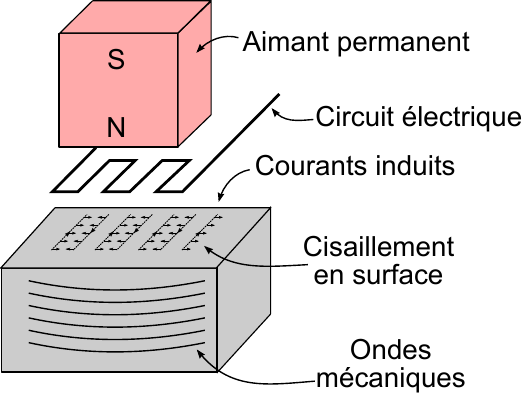}
	 \caption{Un circuit électrique crée par induction un courant électrique dans la surface d'un matériau ferromagnétique soumis à un champ magnétique. Des mouvements apparaissent par force de Lorentz, ce qui crée une combinaison d'ondes de cisaillement et de compression.}
	 \label{figElastoEMIntroSourceLorentzEMAT}
 \end{center}
\end{figure}

La force de Lorentz peut aussi être utilisée pour créer des ondes de cisaillement. En effet, lorsqu'un courant traverse un solide conducteur placé dans un champ magnétique, la force de Lorentz provoque un déplacement local du milieu. Selon le courant, ce déplacement peut se propager comme une onde de compression, de cisaillement ou une combinaison des deux. Ceci est utilisé dans les transducteurs électromagnétiques acoustiques (EMAT pour ElectroMagnetic Acoustic Transducer) \cite{thompson1981ultrasonic}. Ces transducteurs, dont l'un des designs est représenté figure \ref{figElastoEMIntroSourceLorentzEMAT}, sont utilisés dans l'industrie sur des matériaux ferromagnétiques. Ils sont composés d'un petit circuit électrique en forme de serpent ou de spirale et d'un aimant. Le circuit électrique induit un courant électrique dans le matériau. Le courant électrique combiné au champ magnétique de l'aimant est une source d'onde de compression et de cisaillement. Ces ondes sont alors détectées pour mesurer d'éventuels défauts dans le matériau.

L'application de ce principe aux tissus biologiques n'est pas évidente, principalement à cause de la faible conductivité électrique de ceux-ci -- typiquement un million de fois inférieure à celle d'un métal. Une différence importante concerne l'épaisseur de peau : à une fréquence donnée, l'épaisseur de peau des tissus biologiques est bien plus faible que celle des matériaux ferromagnétiques. Ainsi, à la différence de ces derniers, la source des ondes dans un tissu biologique ne sera pas localisée vers la surface et n'est pas ponctuelle. Cela peut être vu comme un avantage ou un inconvénient selon les applications.

Bien que la force de Lorentz ait déjà été utilisée pour produire des ondes mécaniques dans les tissus biologiques, avec la tomographie magnéto-acoustique par induction magnétique présentée dans la section \ref{secLFEITMethodesMATMI}, aucune méthode à l'heure actuelle ne s'est intéressée à l'utilisation de la force de Lorentz pour générer des ondes de cisaillement dans un but d'élastographie.

\section{Principe de la création d'ondes de cisaillement par force de Lorentz}
Dans l'élastographie par force de Lorentz, telle qu'étudiée dans cette thèse, un courant électrique est injecté dans un échantillon conducteur. La présence d'un champ magnétique perpendiculaire au courant électrique induit un déplacement par force de Lorentz. Ce déplacement est la source d'une onde de compression et d'une onde de cisaillement. Une barrette d'échographie observe alors la propagation des déplacements du milieu grâce à une technique de \textit{speckle-tracking} comme présenté par Ophir \textit{et al.} \cite{ophir1991}. Cette méthode compare des images échographiques à chaque instant et essaie d'en déduire les déplacements locaux à partir de corrélations entre les images. Comme pour les autres méthodes d'élastographie dynamique, la vitesse des ondes de cisaillement permet de retrouver l'élasticité des tissus.

Dans les expériences d'élastographie par force de Lorentz dans un échantillon de taille finie, les ondes de cisaillement peut être créé de différentes manières : soit par la force de Lorentz elle-même ; soit par conversion de mode à partir d'une onde de compression, aux bords d'un échantillon. Les deux derniers facteurs sont d'ailleurs communs aux autres techniques d'élastographie.

\subsection{La force volumique dans le milieu}
Comme précédemment, on suppose dans cette partie que le champ magnétique est dans la direction X et l'axe ultrasonore dans la direction Z. En élastographie par force de Lorentz, on applique une différence de potentiel $\Delta V$ entre deux électrodes. Cette différence de potentiel induit un champ électrique $\mathbf{E}=-\grad{V}$ dans tout l'espace. A l'intérieur du tissu, que l'on suppose isotrope de conductivité électrique $\underline{\sigma}$, la loi d'Ohm s'écrit $\mathbf{j}=\underline{\sigma} \mathbf{E}$ en suivant les notations introduites au chapitre \ref{secLFEIT}.

La force de Lorentz induite $\mathbf{F}$ peut par ailleurs s'écrire :
\begin{empheq}[box=\mybluebox]{align}
	\frac{d\mathbf{F}}{d\tau}=\mathbf{j}\wedge\mathbf{B}
	\label{eqElastoEMIntroForce1}
\end{empheq}
avec $d\tau$ un petit élémént de volume et $\mathbf{B}$ le champ magnétique.

En présence de cette force, l'équation de Navier s'écrit alors :
\begin{empheq}[box=\mybluebox]{align}
	\rho\frac{d^2\mathbf{u}}{dt^2} = (K + \frac{4}{3}\mu) \grad \Div \mathbf{u} + \mu \rot \rot \mathbf{u} + \mathbf{j}\wedge\mathbf{B}
	\label{eqElastoEMIntroForce2}
\end{empheq}

On supposera par la suite que la densité $\rho$ du milieu est homogène. Ce paramètre varie de moins de 10 \% dans les tissus mous du corps humain, donc cette hypothèse est généralement assez bien vérifiée.

En calculant la divergence de l'équation \ref{eqElastoEMIntroForce2}, on obtient avec $p=\Div \mathbf{u}$ :
\begin{empheq}[box=\mybluebox]{align}
		\frac{ d p^2}{dt^2} = v_p^2 \Delta p + \frac{1}{\rho}\Div (\mathbf{j}\wedge\mathbf{B})
	\label{eqElastoEMIntroOndeP1}
\end{empheq}

%En utilisant la formule d'analyse vectorielle $\Div (\mathbf{j}\wedge\mathbf{B}) = (\rot \mathbf{B}).\mathbf{j} - (\rot \mathbf{j}).\mathbf{B}$, et en remarquant que $\rot \mathbf{B}$ est nul si la source du champ magnétique se trouve en-dehors du tissu à observer (ce qui est une hypothèse raisonnable si le champ magnétique créé par le courant électrique dans le tissu est négligeable devant le champ magnétique extérieur), on obtient
%\begin{empheq}[box=\mybluebox]{align}
%		\frac{ d k^2}{dt^2} = v_k \Delta k - \frac{1}{\rho}(\rot \mathbf{j}).\mathbf{B} \\
%	\label{eqElastoEMIntroOndeP2}
%\end{empheq}
La source de l'onde de compression dépend donc simplement de la divergence du produit vectoriel de la densité de courant et du champ magnétique \cite{roth1994}.%, ou de manière équivalente, du rotationnel de la densité de courant.

En calculant le rotationnel de l'équation \ref{eqElastoEMIntroForce2}, on obtient avec $\mathbf{s}=\rot \mathbf{u}$ :
\begin{empheq}[box=\mybluebox]{align}
		\frac{ d \mathbf{s}^2}{dt^2} = v_s^2 \nabla \mathbf{s} + \frac{1}{\rho}\rot (\mathbf{j}\wedge\mathbf{B})
	\label{eqElastoEMIntroOndeS1}
\end{empheq}

On peut utiliser l'identité vectorielle suivante : $\rot (\mathbf{j}\wedge\mathbf{B}) = (\Div \mathbf{B})\mathbf{j} - (\Div \mathbf{j})\mathbf{B} + (\mathbf{B}.\grad)\mathbf{j} - (\mathbf{j}.\grad)\mathbf{B}$. En utilisant la deuxième équation de Maxwell $\Div \mathbf{B} = 0$ et en remarquant que $\Div \mathbf{j} = 0$ en l'absence d'accumulation de charges dans le milieu, on obtient l'équation :

\begin{empheq}[box=\mybluebox]{align}
		\frac{ d \mathbf{s}^2}{dt^2} = v_s^2 \Delta \mathbf{s} - \frac{1}{\rho}(\mathbf{B}.\grad)\mathbf{j} - \frac{1}{\rho}(\mathbf{j}.\grad)\mathbf{B}
	\label{eqElastoEMIntroOndeS2}
\end{empheq}

On peut alors distinguer deux cas notables. Si $\mathbf{B}$ est uniforme dans le milieu, que l'on suppose selon l'axe X, on a alors, en remplaçant $\mathbf{j}$ par $\sigma \mathbf{E}$, alors on a la relation :
\begin{empheq}[box=\mybluebox]{align}
		\frac{ d \mathbf{s}^2}{dt^2} = v_s^2 \Delta \mathbf{s} - \frac{\sigma B_x }{\rho}\frac{d\mathbf{E}}{dx} + \frac{B_x}{\rho} \frac{d\sigma}{dx} \mathbf{E}
	\label{eqElastoEMIntroOndeS3}
\end{empheq}
Si $\mathbf{E}$ est uniforme selon l'axe du champ magnétique, la source de l'onde de cisaillement dépend alors du gradient de $\sigma$ selon cet axe, et si $\sigma$ est uniforme selon l'axe du champ magnétique, la source de l'onde de cisaillement dépend alors du gradient de $\mathbf{E}$ selon cet axe.

Dans le deuxième cas, seul $\mathbf{j}$ est uniforme, que l'on supposera selon l'axe Y (le champ $\mathbf{B}$ n'étant plus uniforme), ce qui donne :
\begin{empheq}[box=\mybluebox]{align}
		\frac{ d \mathbf{s}^2}{dt^2} = v_s^2 \Delta \mathbf{s} - \frac{j_y}{\rho}\frac{d\mathbf{B}}{dy}
	\label{eqElastoEMIntroOndeS4}
\end{empheq}
La source de l'onde de cisaillement est alors due au gradient de $\mathbf{B}$ selon l'axe Y.

\subsection{La propagation d'une onde mécanique dans le milieu}
Il faut cependant bien comprendre que l'équation précédente indique la création locale d'une onde de cisaillement. Si l'on reprend l'équation de Navier avec une force volumique $\mathbf{f}(t)$ dépendant du temps, localisée en un point de coordonnées $\mathbf{r}$ et orientée dans une direction $i$, on obtient :
\begin{empheq}[box=\mybluebox]{align}
\rho\frac{d^2\mathbf{u}}{dt^2} - (K + \frac{1}{3}\mu) \grad \Div \mathbf{u} - \mu \mathbf{\Delta} \mathbf{u} = \mathbf{f}(t) \delta(\mathbf{r})
\label{eqElastoEMIntroVolumique}
\end{empheq}

La solution de cette équation est donnée par une fonction de Green, que l'on nommera $G(\mathbf{r},t)$. Connaissant la solution $G(\mathbf{r}, t)$ pour une source ponctuelle, la solution pour une source étendue spatialement $\mathbf{f}(t)S(\mathbf{r})$ est donnée par la convolution de la fonction de Green avec la fonction spatiale de la source soit $G(\mathbf{r}, t) \ast_\mathbf{r}S(\mathbf{r})$, où $\ast_\mathbf{r}$ indique la convolution sur la variable d'espace. La création de l'onde de cisaillement peut donc être calculée grâce à cette dernière équation.

Pour un volume isotrope, homogène et infini (sans conditions aux limites), le calcul dont le détail est donné en annexe donne :
\begin{empheq}[box=\mybluebox]{align}
	\begin{aligned}
	 u_i(\mathbf{r},t) & \left. =  \frac{1}{4 \pi \rho r^3}  \left( \frac{3 r_i r_j}{r^2} - \delta_{ij} \right) \int_{r/ v_p}^{r/ v_s} \tau \, f(t-\tau) \; \mathrm{d}\tau  \qquad \; \; \: \right\} \mathrm{champ~proche}
	\\
	& %\left.
	\begin{aligned}
	& \left. + \frac{1}{4 \pi \rho v_p^2 r} \; \frac{r_i r_j}{r^2} \; f \! \left( t-\frac{r}{v_p} \right) \qquad \qquad \quad \, \right\} \mathrm{onde~de~compression}
	\\
	& \left. + \frac{1}{4 \pi \rho v_s^2 r}  \left( \frac{r_i r_j}{r^2} - \delta_{ij} \right)  \; f \! \left( t-\frac{r}{v_s} \right) \quad \right\} \mathrm{onde~de~cisaillement}
	\end{aligned} %\right\} \mathrm{champ~lointain}
	\label{eqElastoEMIntroGreen}
	\end{aligned} 
\end{empheq}
avec $r = \|\mathbf{r}\|$ et $i,j$ les directions X, Y et Z. Ainsi, on voit que la solution est la somme de trois termes : deux termes de champs lointain, décroissant en $\frac{1}{r}$, et un terme de champ proche, avec une décroissance plus rapide. Le champ lointain se décompose lui-même en deux termes, celui correspondant à des ondes de compression polarisées longitudinalement se propageant à la vitesse $v_p$, et un deuxième correspondant à des ondes de cisaillement polarisées latéralement se propageant à la vitesse $v_s$. Ces deux termes correspondent d'ailleurs à la propagation d'ondes planes dans un milieu isotrope, homogène et infini comme vu au début du chapitre. Cependant, le terme de champ proche ne peut pas être décomposé de la même façon en deux termes distincts. On peut donc, proche de la source, observer des ondes de cisaillement polarisées longitudinalement. Ces dernières ont d'ailleurs été mises en évidence expérimentalement par Catheline \textit{et al.} \cite{catheline1998interferometrie}.

Le déplacement total du tissu par force de Lorentz nécessite donc la résolution des fonctions de Green.

\subsection{Conversion de mode}
Enfin, des ondes de cisaillement peuvent être produites par conversion de mode. Ceci arrive lorsqu'une onde mécanique rencontre un obstacle lors de sa propagation (paroi rigide par exemple). Dans ce cas, une onde de compression peut donner lieu à la création d'une onde de cisaillement. Les conditions aux bords d'un échantillon vont donc influer sur l'apparition d'ondes de cisaillement.

\section{Dispositif expérimental}
\begin{figure}[!ht]
   \begin{minipage}[c]{.5\linewidth}
		\begin{center}
	   		\includegraphics[width=1\linewidth]{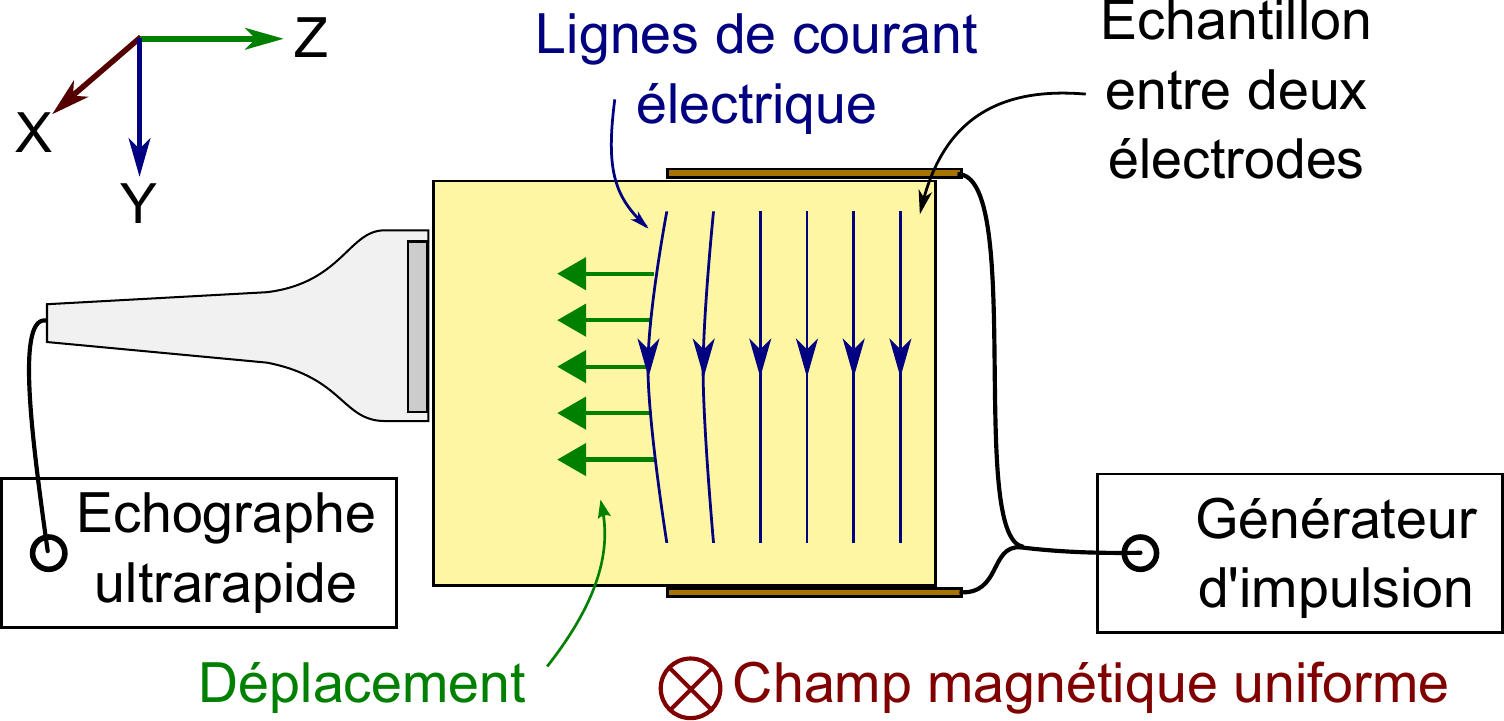}
		\end{center}
   \end{minipage} \hfill
   \begin{minipage}[c]{.5\linewidth}
		\begin{center}
	   		\includegraphics[width=1\linewidth]{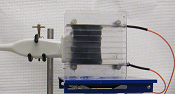}
		\end{center}
   \end{minipage}
	\caption{Un courant électrique est injecté par deux électrodes dans un échantillon placé dans un champ magnétique. La propagation du déplacement induit est observée par une sonde ultrasonore.}
	 \label{figElastoEMDispositif}
\end{figure}

Le dispositif est représenté sur la figure \ref{figElastoEMDispositif}.

Le principe de l'expérience est de faire circuler une impulsion électrique brève dans un échantillon placé dans un champ magnétique, et d'observer les mouvements induits avec une sonde ultrasonore.

Le courant électrique était composé d'une sinusoide de 10 ms à 80 V pic à pic et était émis par un générateur (AFG 3022B, Tektronix, Beaverton, OR, USA) couplé à un amplificateur (600W A500, Behringer, Willich, Germany). Une résistance de 1 Ohm placée en série permettait de connaitre l'intensité du courant, comprise en 0,2 et 1 A pic à pic selon les expériences. Ce courant était injecté par deux électrodes en cuivre 10x3x0,1 cm$^3$ dans l'échantillon. L'échantillon est un fantôme d'alcool polyvinylique (PVA) de taille 6x4x4 cm$^3$ composé de PVA à 50 g.L$^{-1}$ \cite{culjat2010review}, de 10 g/L de sel pour lui donner une conductivité électrique de 2 S.m$^{-1}$, un peu supérieure à celle des tissus biologiques, et de 1 g.L$^{-1}$  de poudre de graphite.

Cet échantillon était placé dans l'entrefer de l'aimant en U utilisé dans les parties précédentes (photographie figure \ref{figHydroEMProto1} et simulation de champ magnétique figure \ref{figHydroEMProto2}) pour avoir un champ magnétique global de 350 mT.

Le déplacement était visualisé %%de deux manières, soit avec une sonde ultrasonore monoélément à 8 MHz () reliée à un échographe monoélément (US-Key, LeCoeur Electronique, Chuelles, France) qui à chaque acquisition était déplacée verticalement afin de réaliser une image complète comme dans une échographie mode A, soit %%
avec une barrette multiéléments à 10 MHz fonctionnant en mode ultrarapide relié à un échographe Verasonics (Verasonics V-1, Redmond, WA, USA). Dans le mode ultrarapide, on émet une onde plane ultrasonore qui va être réfléchie par les interfaces d'impédance acoustique. On peut alors localiser celles-ci avec les temps d'arrivée des ondes réfléchies sur les différents éléments du transducteur. Avec une vitesse ultrasonore d'environ 1500 m.s$^{-1}$, on arrive donc théoriquement pour une profondeur d'exploration de 10 cm à une cadence d'image de 7500 images par seconde. La qualité de l'image peut alors être améliorée en faisant de l'imagerie composée, en utilisant plusieurs angles d'incidence de visualisation \cite{sandrin1999}). Dans notre dispositif, nous avions une cadence de %500 lignes par secondes pour la sonde monoélément et
1000 images par seconde avec un seul angle pour la barrette multiéléments.

La poudre de graphite présente dans l'échantillon faisait office de diffuseurs acoustiques. Ainsi, les échantillons présentaient un \textit{speckle} sur les images ultrasonores. Lors de la propagation de l'onde, le tissu était légèrement déformé, ce qui déplace les grains de \textit{speckle} de quelques micromètres. En cherchant un maximum de corrélation entre une ligne et la même ligne sur l'image suivante, il est possible de déduire le mouvement probable du tissu en cet endroit le long de la ligne ultrasonore, comme développé dans la thèse de S. Catheline \cite{catheline1998interferometrie}. Concrètement, il est possible d'observer une amplitude des déplacements inférieure au micromètre, alors que la longueur d'onde des ultrasons est approximativement de 150 \micro m (longueur d'onde à 10 MHz) : ceci est rendu possible par l'efficacité des algorithmes de \textit{speckle-tracking} qui, comme les méthodes interférométriques en optique, atteignent des résolution spatiales bien en-dessous de la longueur d'onde.

Les signaux représentant les déplacements en chaque point au cours du temps étaient enfin filtrés par le logiciel Matlab. Pour cela, la transformée de Fourier temporel du signal était calculée, la valeur nulle était donnée pour les fréquences supérieures à 200 Hz, et la transformée de Fourier inverse était calculée. Ceci permettait de diminuer le bruit présent sur les images de déplacements.

\section{Expériences de création d'ondes de cisaillement par force de Lorentz}

La première étape de ce travail a été d'obtenir une première observation expérimentale d'onde S par force de Lorentz dans de la matière molle. Un soin particulier a été porté à l'objectivation du phénomène, pour d'une part s'assurer que l'on crée effectivement des ondes de cisaillement, et d'autre part de vérifier que ces ondes ne sont pas induites par un effet indésirable comme le mouvement des électrodes ou une dilation par effet Joule.

\subsection{Création d'onde de cisaillement par conversion de mode}
La première expérience a consisté simplement à voir si l'injection d'un courant électrique dans un tissu plongé dans un champ magnétique donnait un déplacement du tissu suffisamment important pour être visible. De plus, le dispositif était fait de sorte que l'on puisse observer la création d'une onde de cisaillement par conversion de mode sur les bords de l'échantillon, grâce au contact avec des parois rigides.

\begin{figure}[!ht]
\begin{minipage}[c]{.5\linewidth}
 \begin{center}
	 \includegraphics[height=2.8cm]{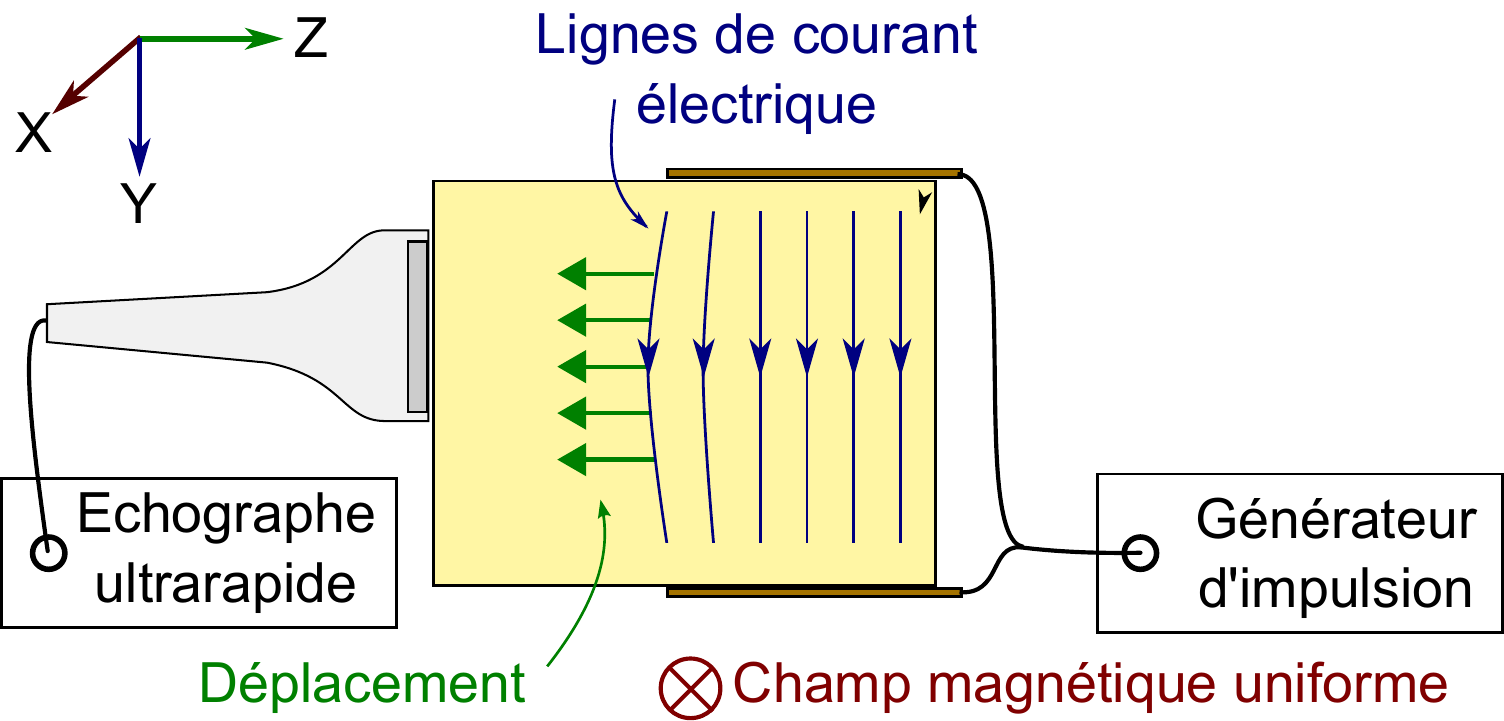}
 \end{center}
\end{minipage}
\begin{minipage}[c]{.5\linewidth}
	 \includegraphics[height=2.8cm]{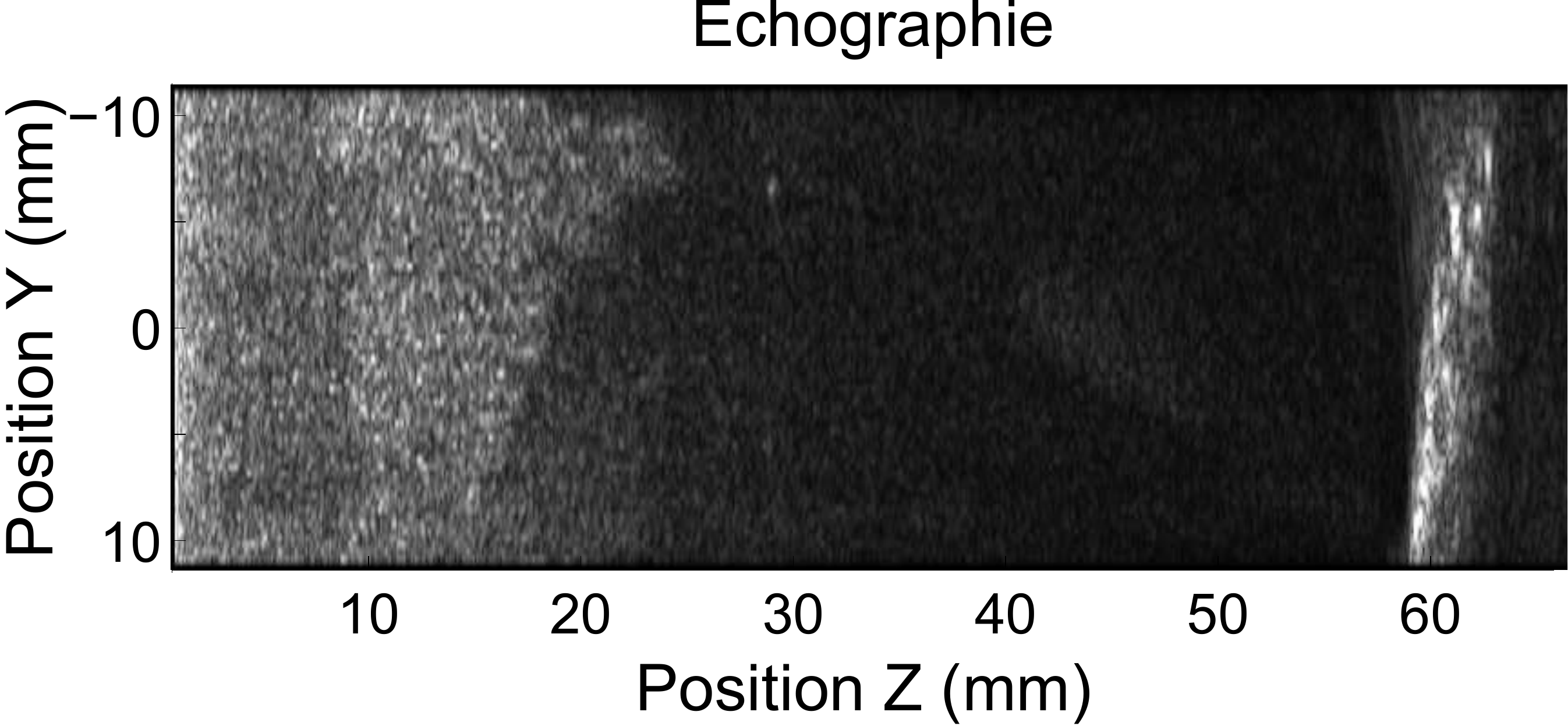}
\end{minipage}
	 \caption{(a) \textbf{Schéma de l'expérience de création d'onde de cisaillement par conversion de mode} -- Un courant électrique est injecté par deux électrodes planes dans un fantôme soumis à un champ magnétique. Cela entraine des déplacements observés par une sonde ultrasonore. (b) \textbf{Echographie obtenue par la sonde ultrasonore} -- Le fantôme a un aspect granulaire dû à l'ajout de graphite. L'interface du fond du fantôme est visible sur la droite.}
	\label{figElastoEMExpMatGlobal}
\end{figure}

\paragraph{Matériel et méthodes}
Une expérience a été faite avec deux électrodes en contact respectivement avec la partie supérieure et inférieure de l'échantillon, le long de la tranche, comme représenté sur le schéma \ref{figElastoEMExpMatGlobal}-(a). L'échographie du milieu est représentée figure \ref{figElastoEMExpMatGlobal}-(b). L'apparence granuleuse de l'image est due à l'ajout de graphite.

\begin{figure}[!ht]
   \begin{minipage}[c]{.5\linewidth}
		\begin{center}
	   		\includegraphics[width=1\linewidth]{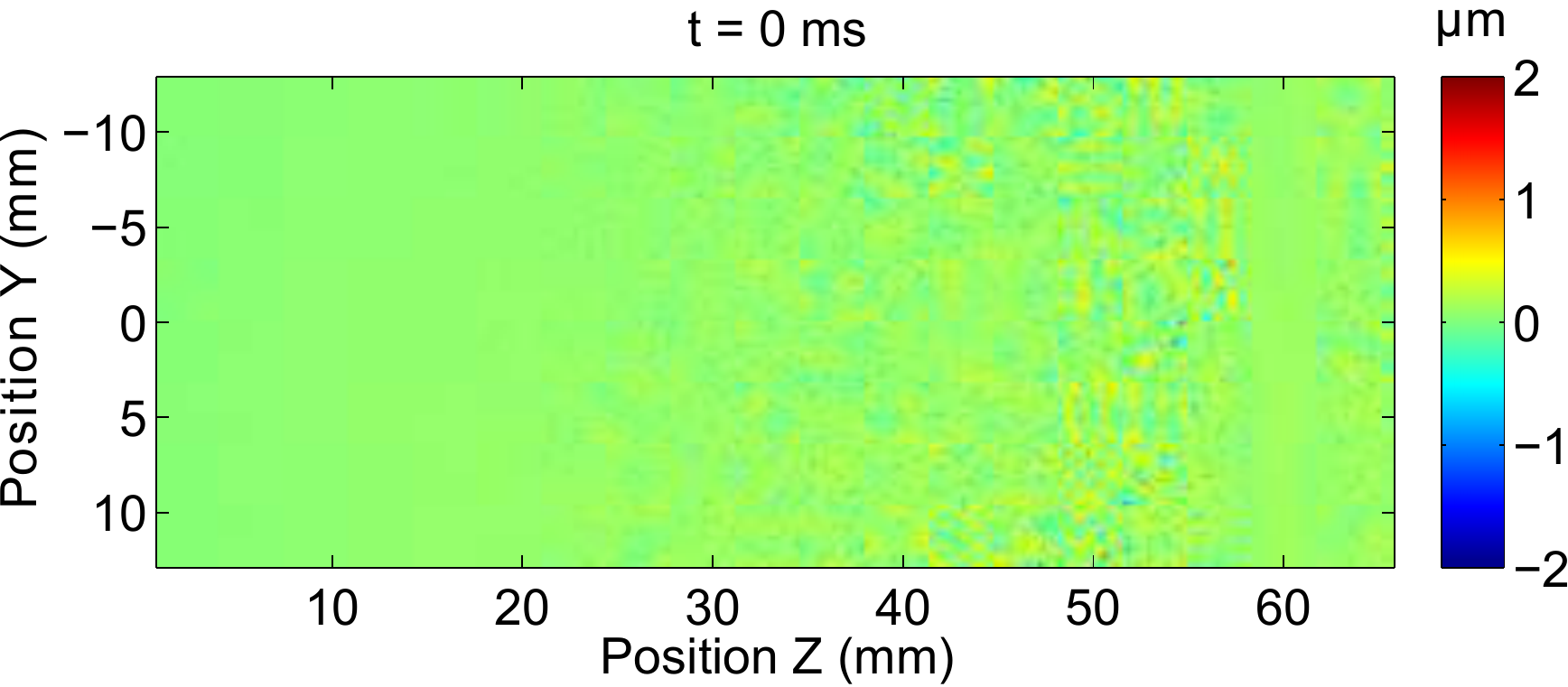}
		\end{center}
   \end{minipage} \hfill
   \begin{minipage}[c]{.5\linewidth}
		\begin{center}
	   		\includegraphics[width=1\linewidth]{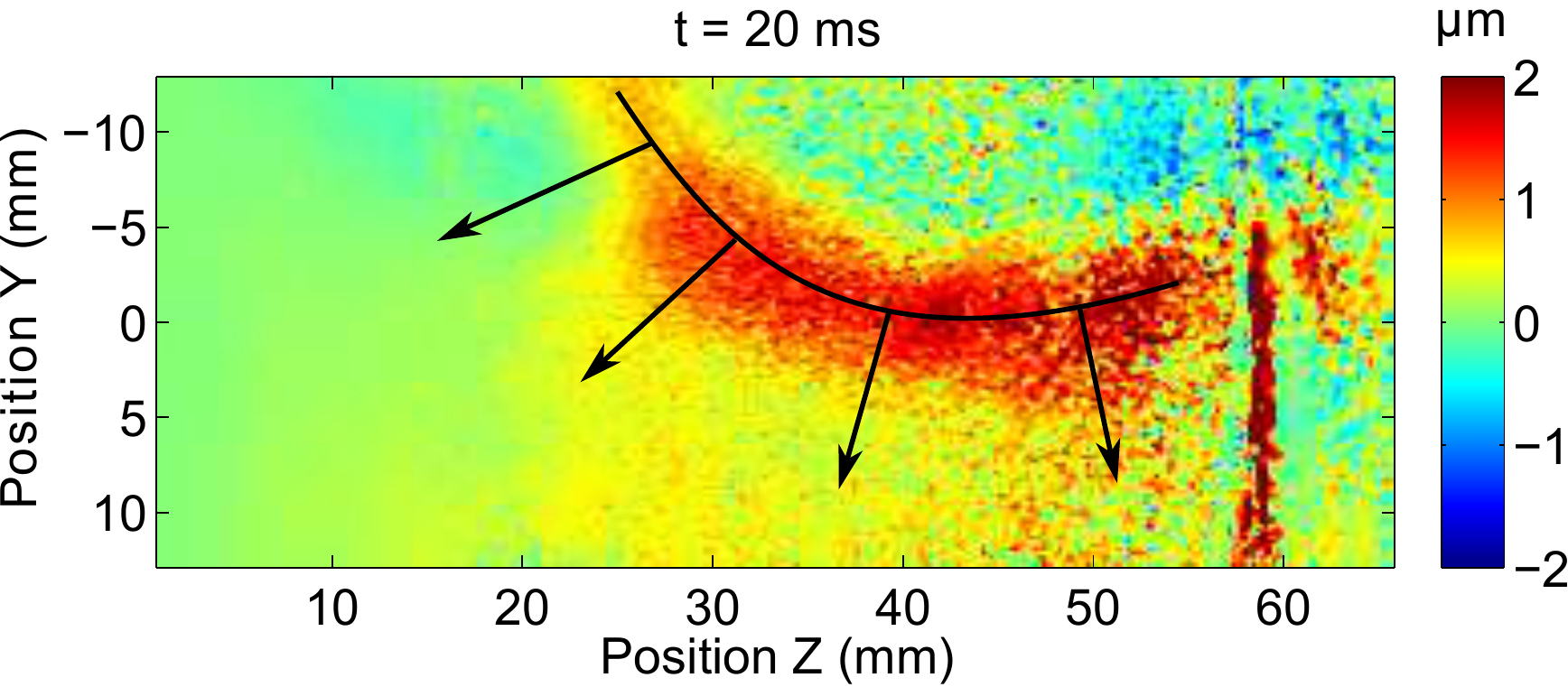}
		\end{center}
   \end{minipage}
   \begin{minipage}[c]{.5\linewidth}
		\begin{center}
	   		\includegraphics[width=1\linewidth]{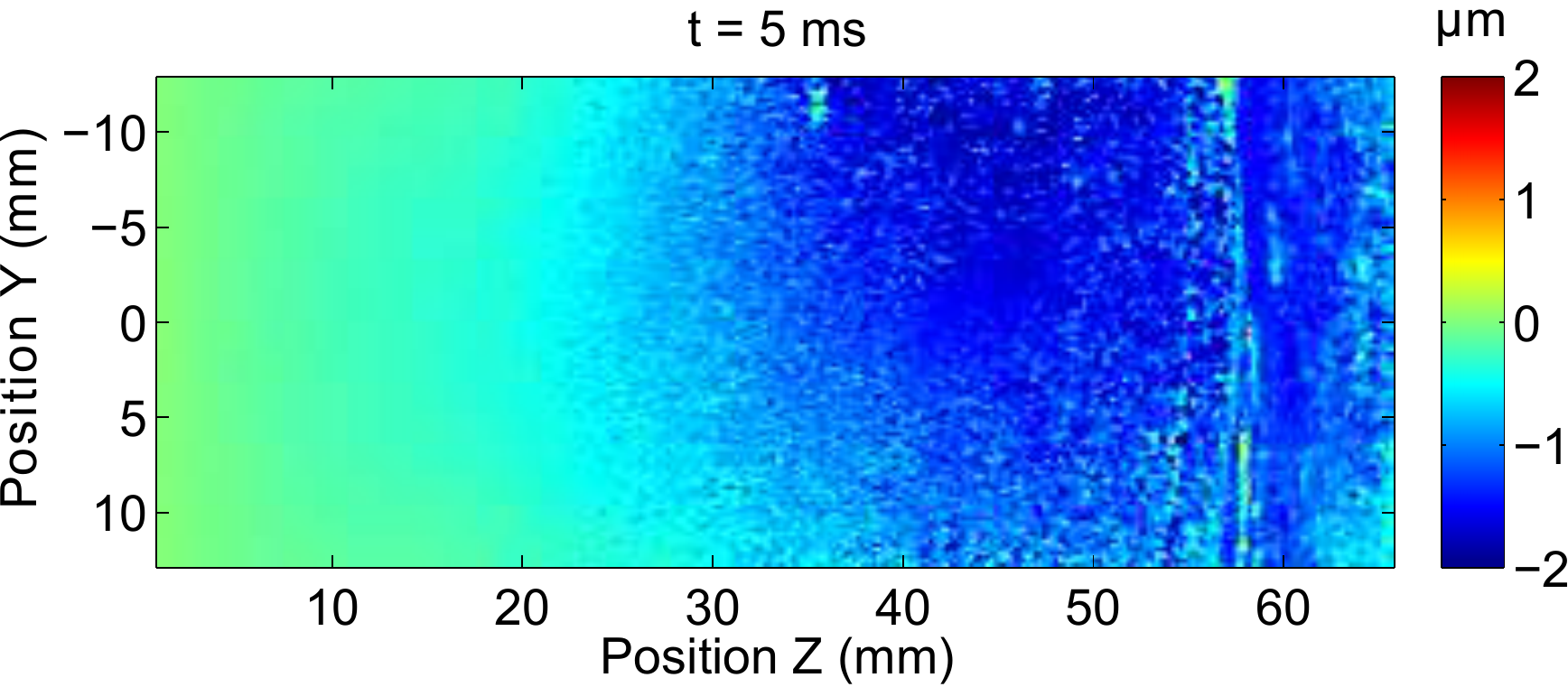}
		\end{center}
   \end{minipage} \hfill
   \begin{minipage}[c]{.5\linewidth}
		\begin{center}
	   		\includegraphics[width=1\linewidth]{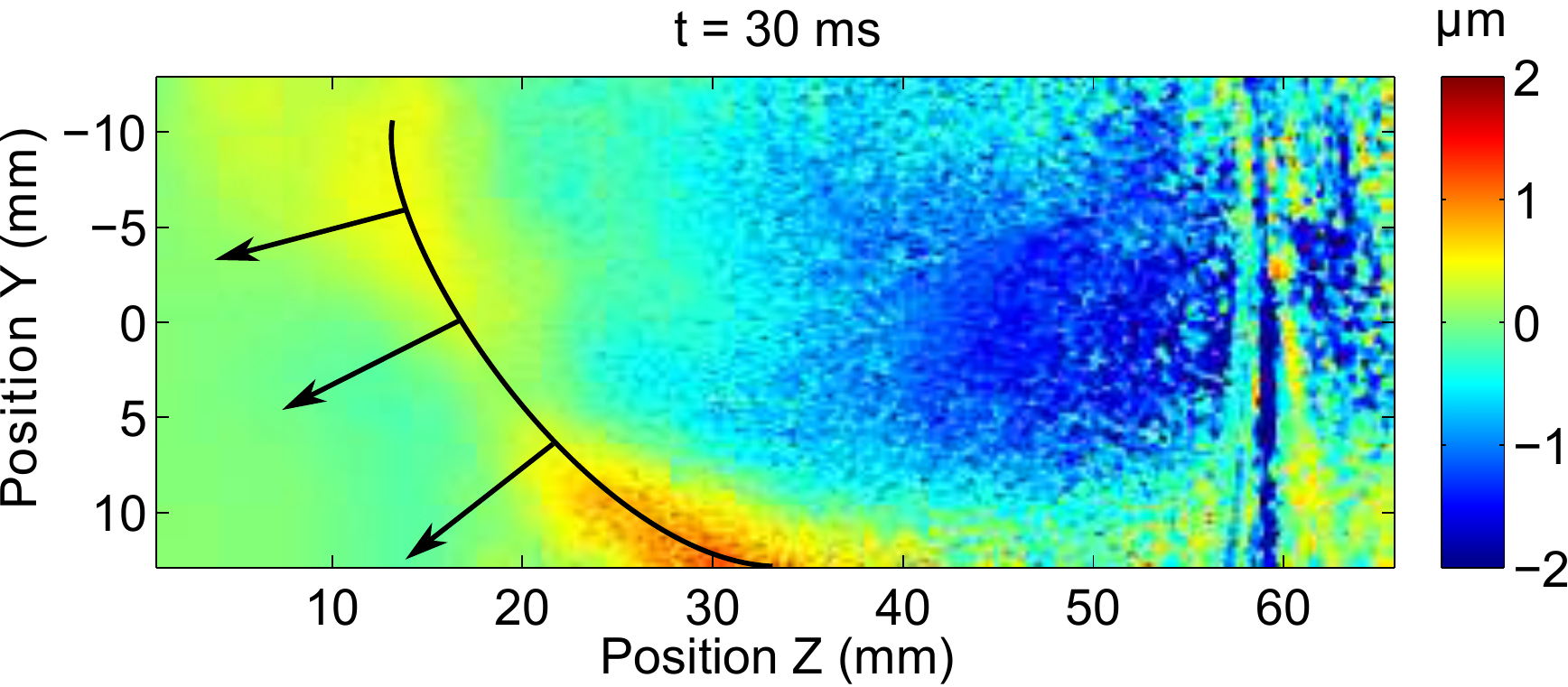}
		\end{center}
   \end{minipage}
   \begin{minipage}[c]{.5\linewidth}
		\begin{center}
	   		\includegraphics[width=1\linewidth]{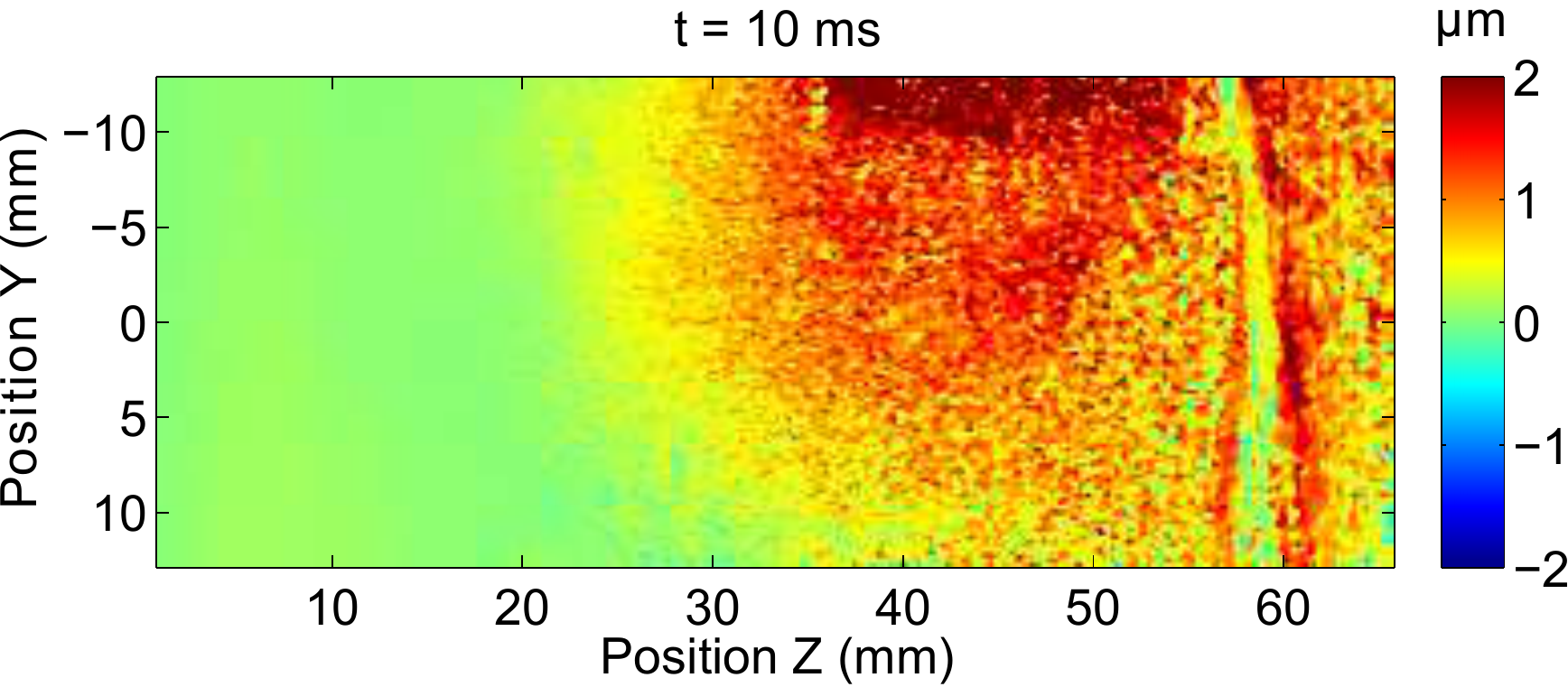}
		\end{center}
   \end{minipage} \hfill
   \begin{minipage}[c]{.5\linewidth}
		\begin{center}
	   		\includegraphics[width=1\linewidth]{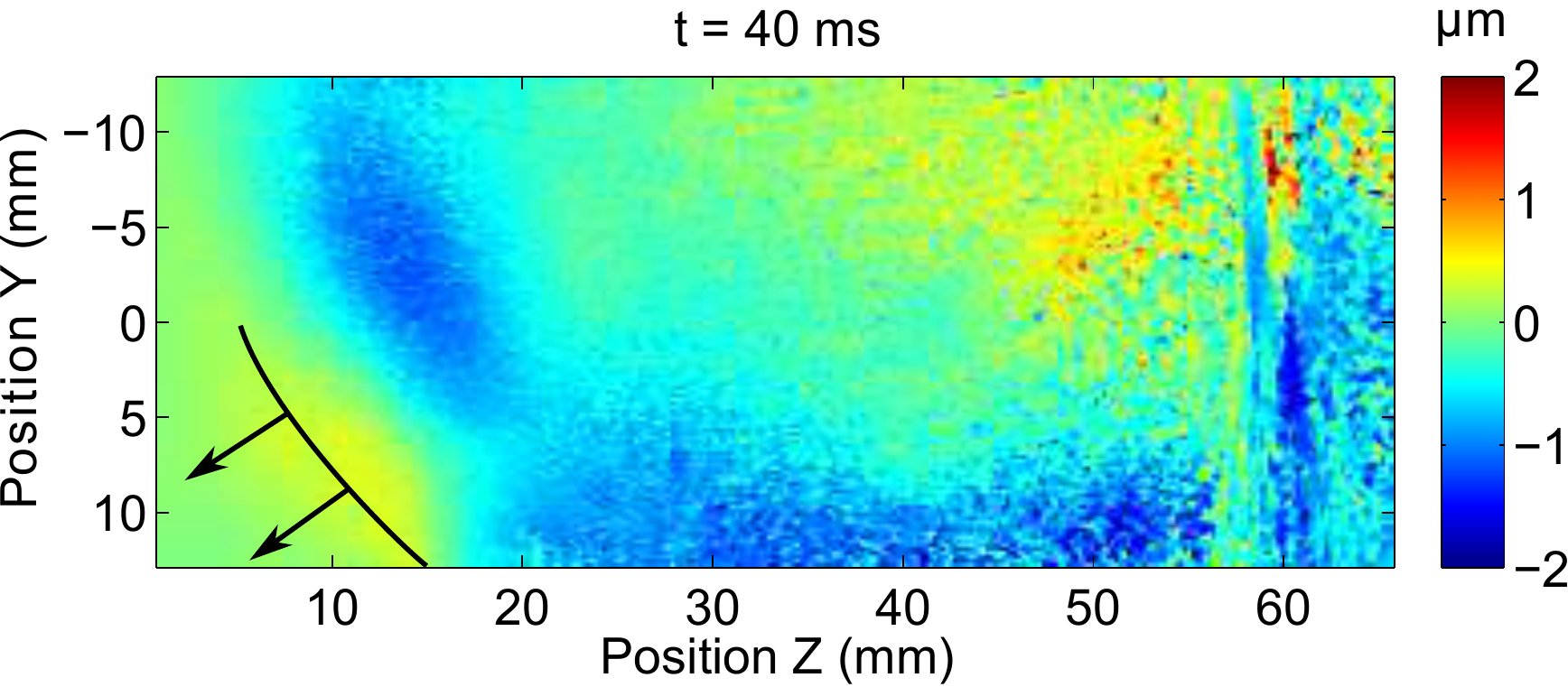}
		\end{center}
   \end{minipage}
   \begin{minipage}[c]{.5\linewidth}
		\begin{center}
	   		\includegraphics[width=1\linewidth]{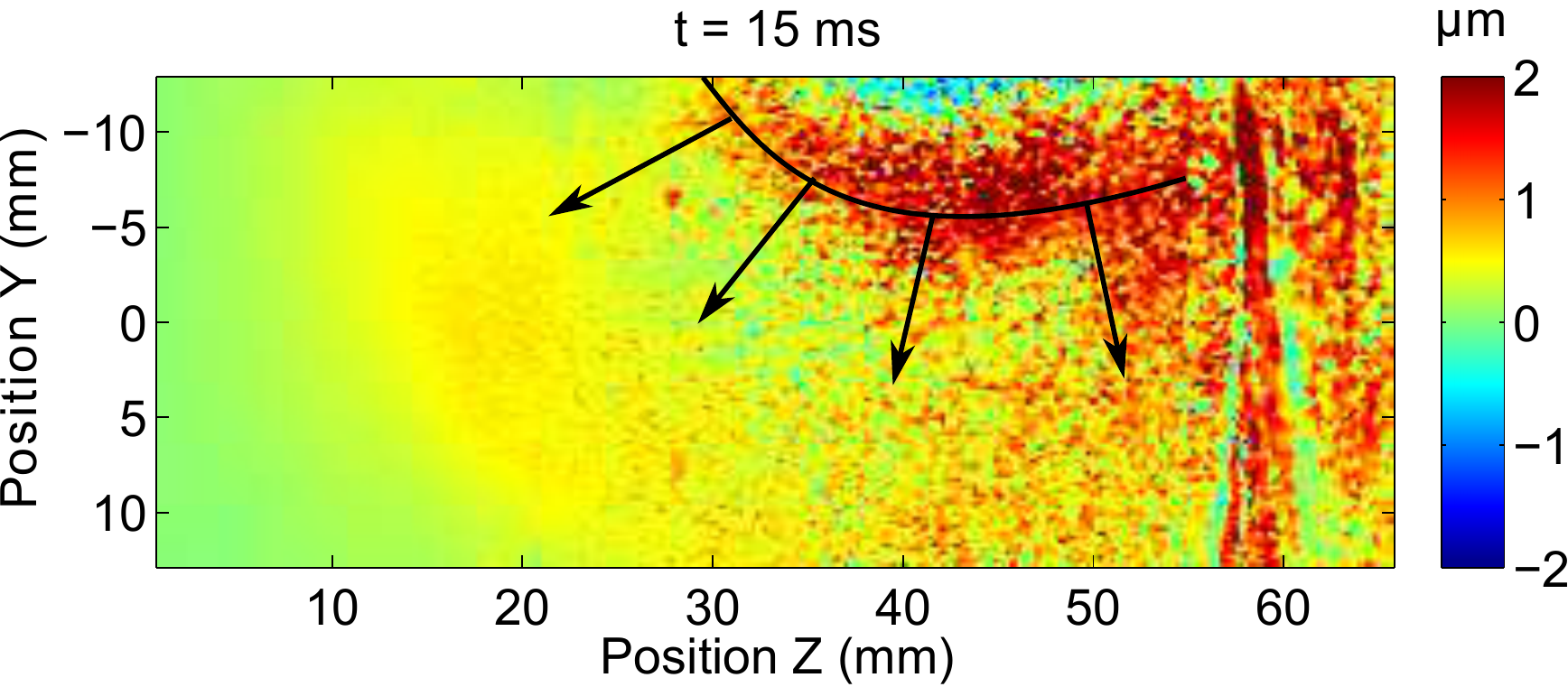}
		\end{center}
   \end{minipage} \hfill
   \begin{minipage}[c]{.5\linewidth}
		\begin{center}
	   		\includegraphics[width=1\linewidth]{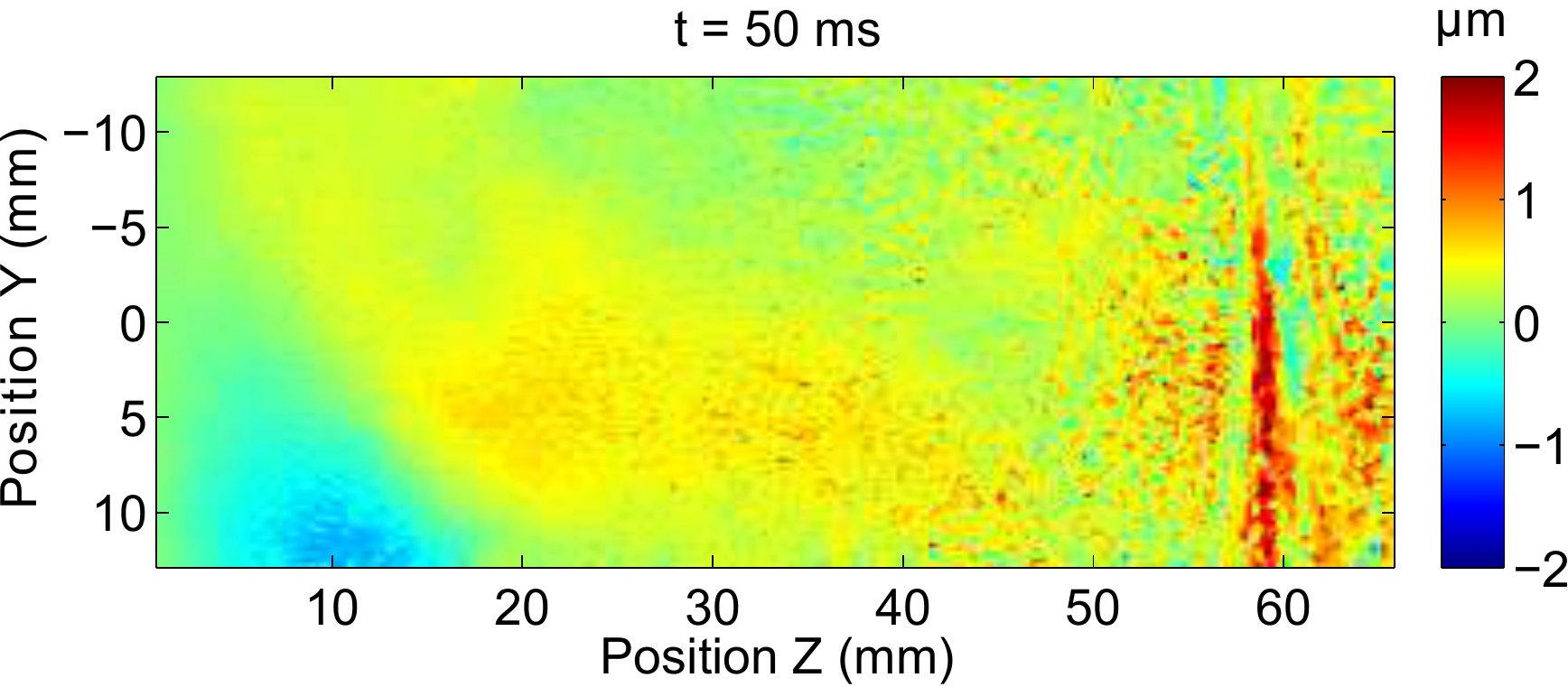}
		\end{center}
   \end{minipage}
   \caption{\textbf{Résultats de l'expérience de création d'ondes de cisaillement par conversion de mode} -- Champ des déplacements dus à la traversée d'un courant électrique par deux électrodes situées au-dessus et en-dessous d'un fantôme placé dans un champ magnétique, à 0, 5, 10, 15, 20, 30, 40, 50 ms après le début du passage du courant (plan YZ). On observe une onde de cisaillement provenant d'un bord de l'échantillon, produite par conversion de mode.}
   \label{figElastoEMExpResGlobal}
\end{figure}

\paragraph{Résultats et discussions}
 Les cartes des déplacements à 0, 5, 10, 15, 20, 30, 40 et 50 ms sont représentés sur la figure \ref{figElastoEMExpResGlobal}.

Au moment de déclencher le courant, aucun déplacement n'est visible.  Lorsqu'on injecte un courant électrique dans un sens (partie positive de la sinusoïde émise), à 5 ms, un premier déplacement du fantôme apparait, globablement dans le même sens. Puis, à 10 ms, correspondant à l'injection d'un courant dans le sens opposé (partie négative de la sinusoïde émise), on observe un déplacement globalement dans le sens contraire. A 15, 20, 30, 40 et 50 ms, on observe la propagation d'une onde avec une longueur d'onde centimétrique. Cette onde correspond à une onde de cisaillement dont la polarisation n'est pas transversale. Comme les champs $\mathbf{E}$ et $\mathbf{B}$ sont quasiment uniformes et que la conductivité électrique du fantôme est homogène, il ne s'agit pas de la force qui crée intrinsèquement l'onde de cisaillement. Cependant, une déformation importante a lieu au voisinage des plaques, ce qui donne lieu par conversion de mode à la création d'ondes de cisaillement.

\subsection{Création d'onde de cisaillement par force de Lorentz}
Si la force a une extension spatiale limitée, elle produit un déplacement qui se propage selon l'équation \ref{eqElastoEMIntroGreen}. Ainsi, même si la force ne produit pas de cisaillement intrinsèque, la taille de la source peut en produire. Les ondes de cisaillement ont alors une polarisation à la fois transverse et longitudinale, qui n'est transverse qu'en champ lointain. Nous avons réalisé ici une expérience avec une source ayant un rayonnement de type dipolaire pour induire une onde de cisaillement.

\begin{figure}[!ht]
\begin{minipage}[c]{.5\linewidth}
 \begin{center}
	 \includegraphics[height=2.8cm]{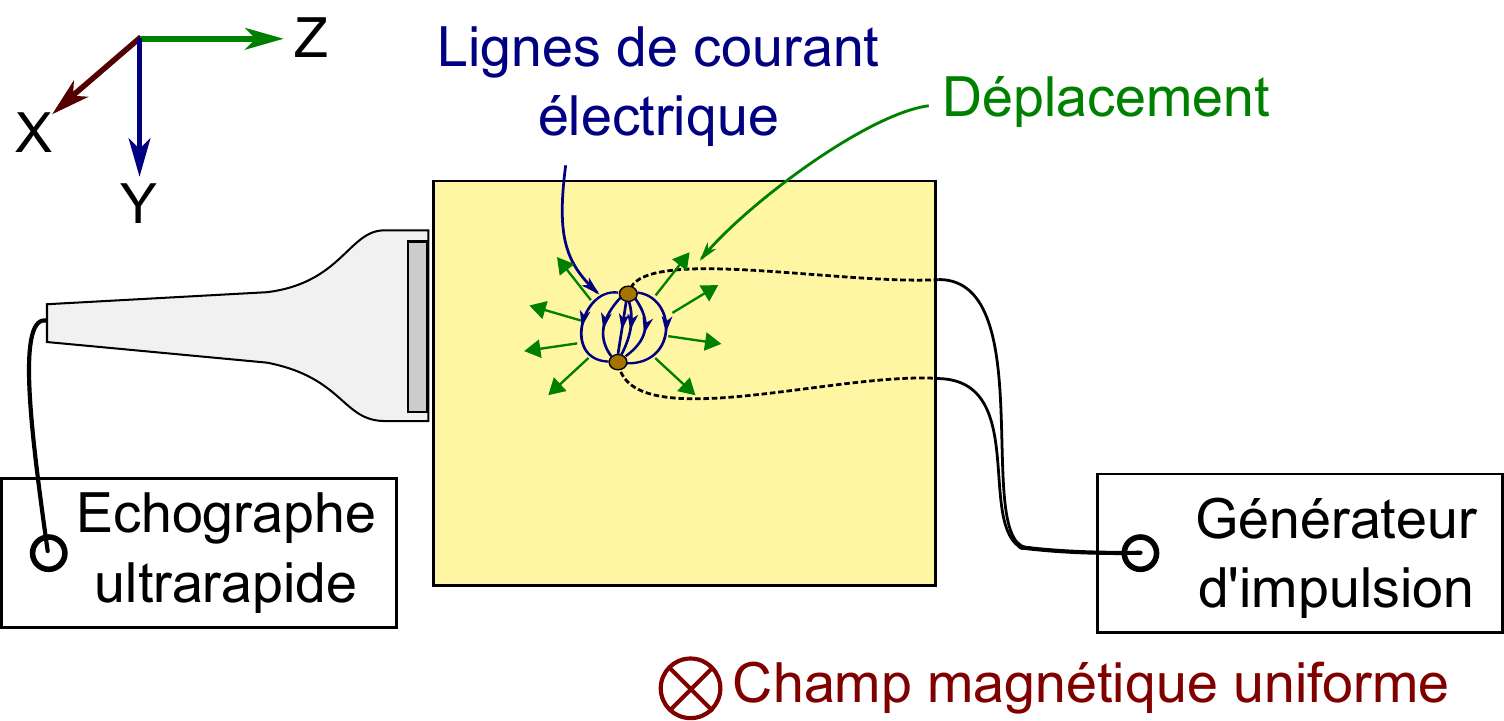}
 \end{center}
\end{minipage}
\begin{minipage}[c]{.5\linewidth}
	 \includegraphics[height=2.8cm]{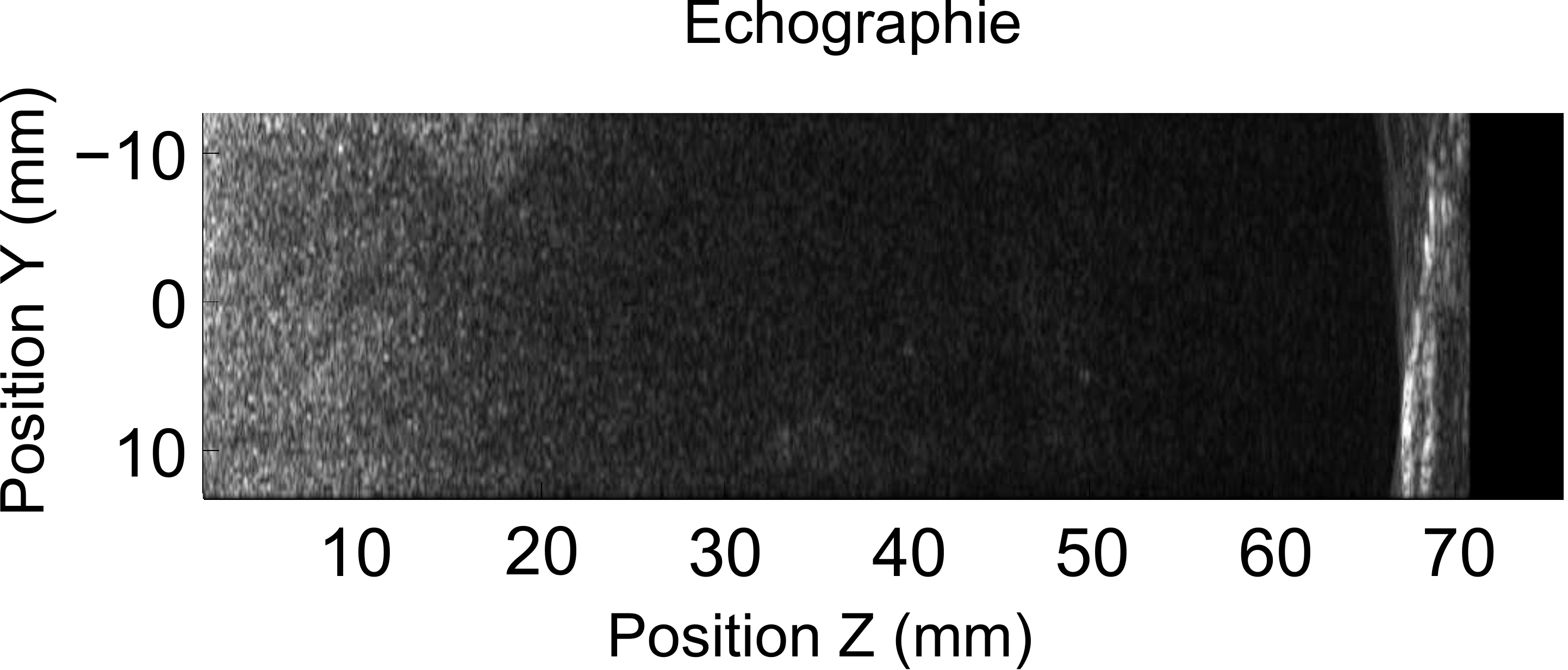}
\end{minipage}
	 \caption{(a) \textbf{Schéma de l'expérience sur la création d'onde de cisaillement par force de Lorentz} -- Deux électrodes filaires créent un courant électrique dans un fantôme soumis à un champ magnétique, et les déplacements sont observés par une sonde ultrasonore en mode ultrarapide. (b) \textbf{Echographie obtenue par la sonde ultrasonore} -- On distingue au milieu deux points blancs correspondant à la zone où se trouvent les deux électrodes filaires.}
	\label{figElastoEMExpMatElectrodeCentre}
\end{figure}

\paragraph{Matériel et méthodes}
Une expérience a été réalisée où deux électrodes filaires sont placées au milieu du fantôme, comme représenté sur la figure \ref{figElastoEMExpMatElectrodeCentre}-(a). Le courant électrique émis restait localisé entre celles-ci. L'échographie du milieu est représenté sur la figure \ref{figElastoEMExpMatElectrodeCentre}-(b). On distingue deux points blancs correspondants à la position des deux électrodes filaires au milieu du fantôme.

\begin{figure}[!ht]
   \begin{minipage}[c]{.5\linewidth}
		\begin{center}
	   		\includegraphics[width=.95\linewidth]{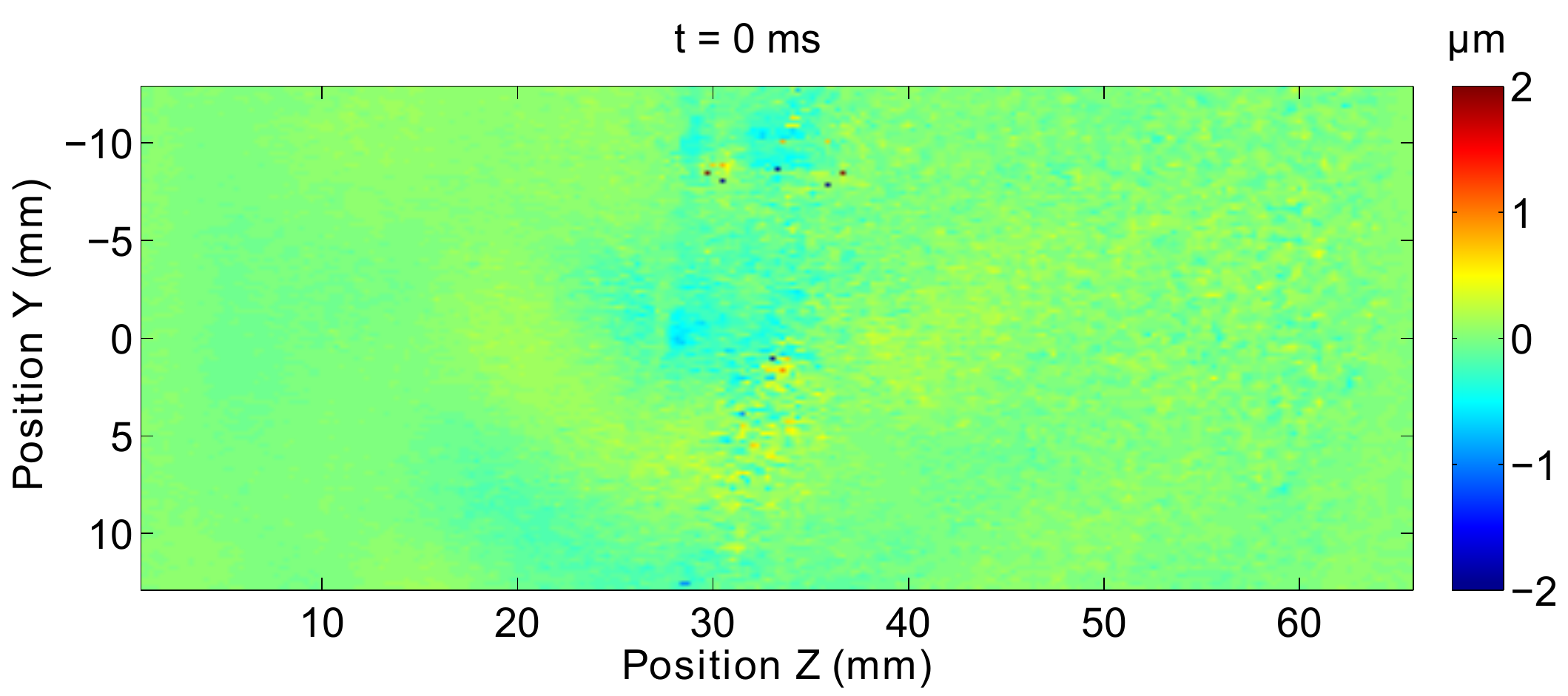}
		\end{center}
   \end{minipage} \hfill
   \begin{minipage}[c]{.5\linewidth}
		\begin{center}
	   		\includegraphics[width=.95\linewidth]{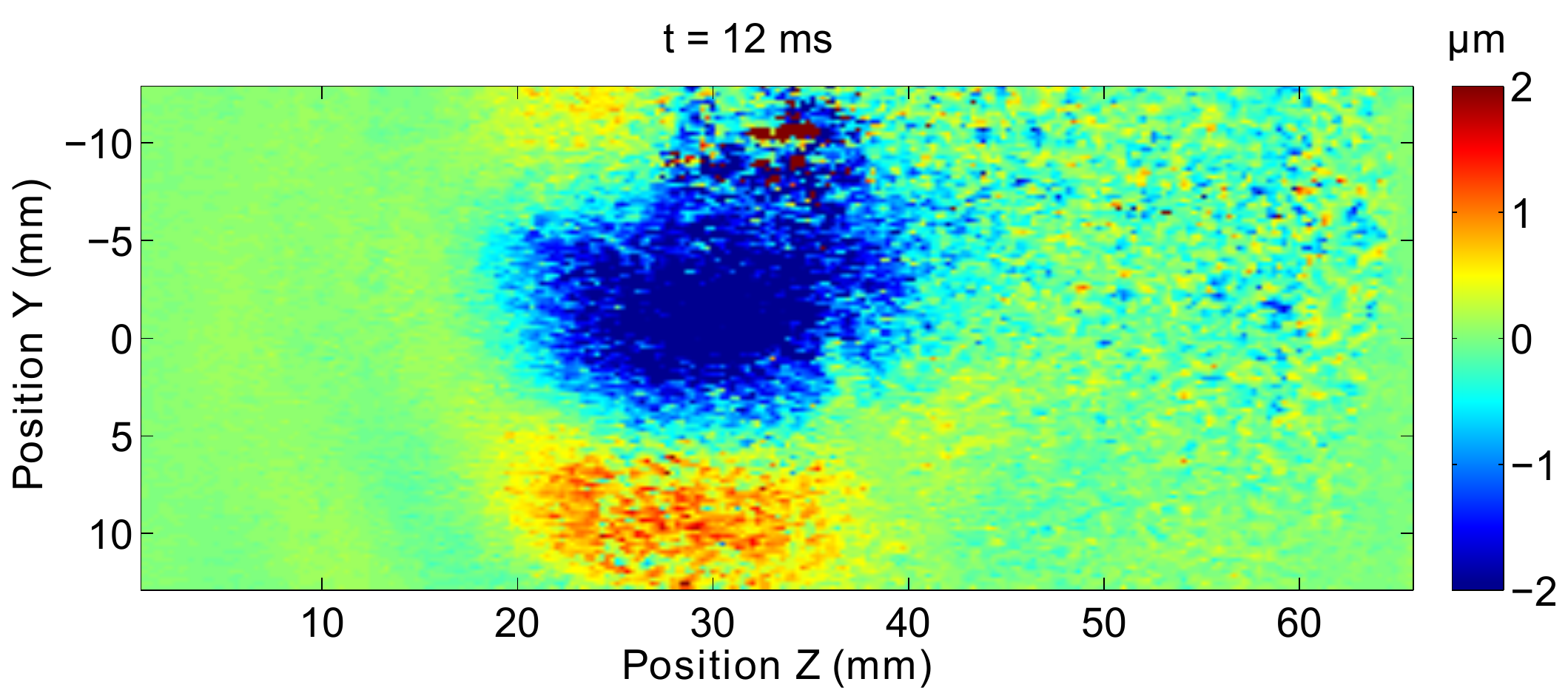}
		\end{center}
   \end{minipage}
   \begin{minipage}[c]{.5\linewidth}
		\begin{center}
	   		\includegraphics[width=.95\linewidth]{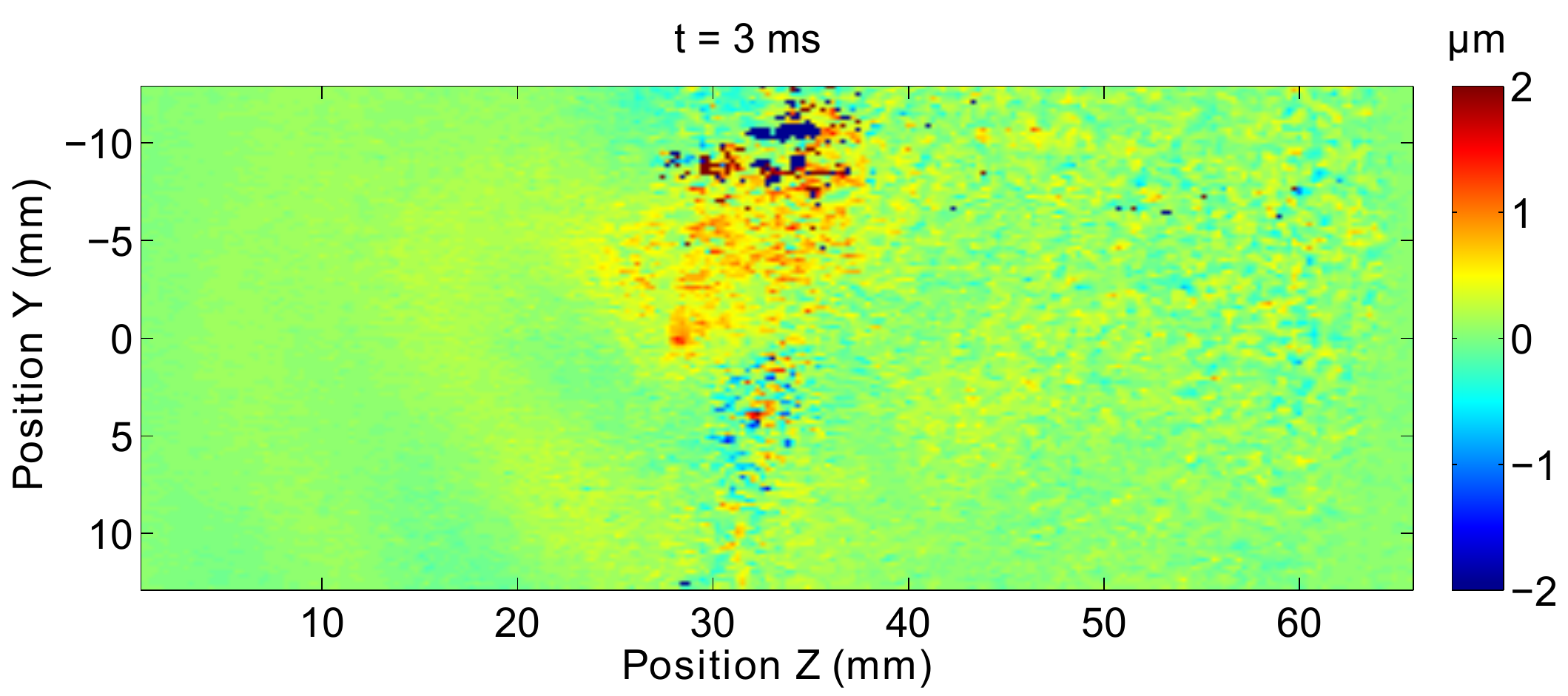}
		\end{center}
   \end{minipage} \hfill
   \begin{minipage}[c]{.5\linewidth}
		\begin{center}
	   		\includegraphics[width=.95\linewidth]{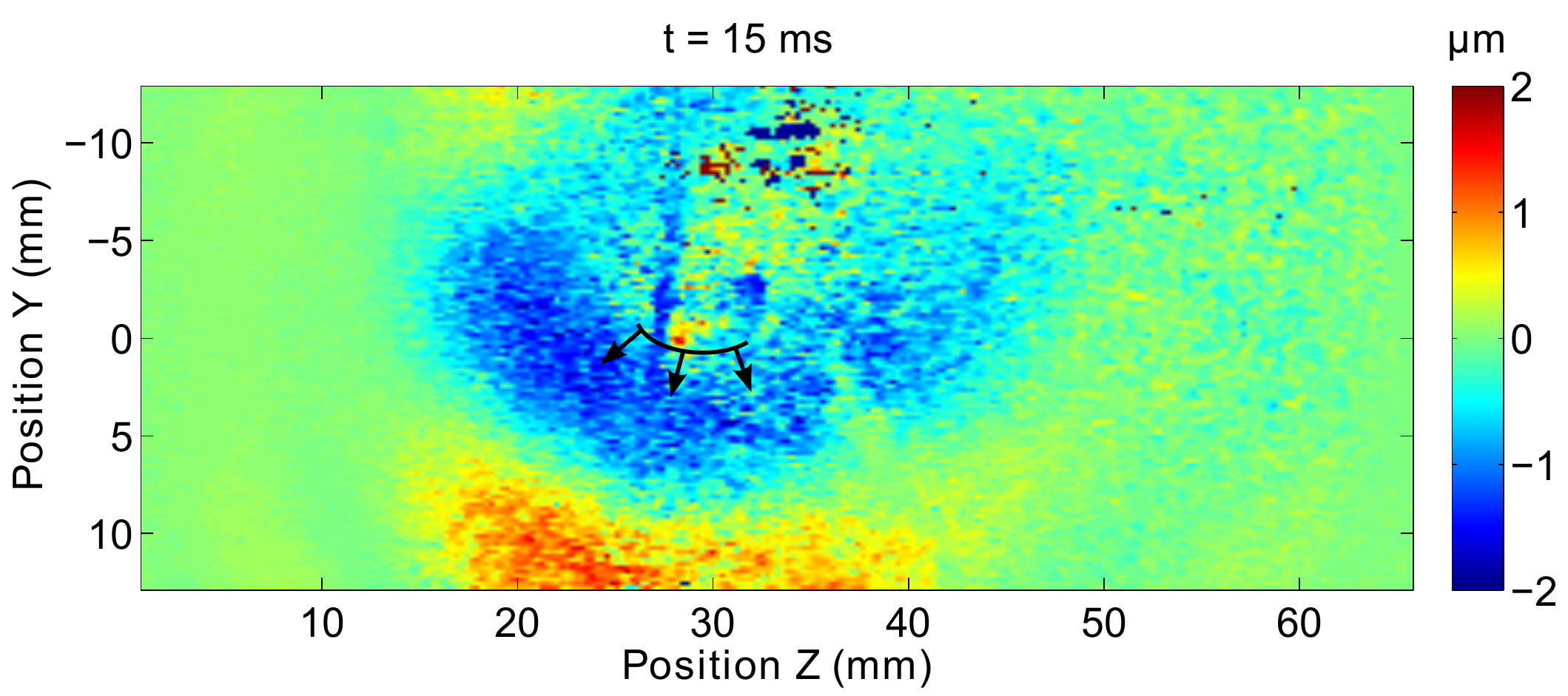}
		\end{center}
   \end{minipage}
   \begin{minipage}[c]{.5\linewidth}
		\begin{center}
	   		\includegraphics[width=.95\linewidth]{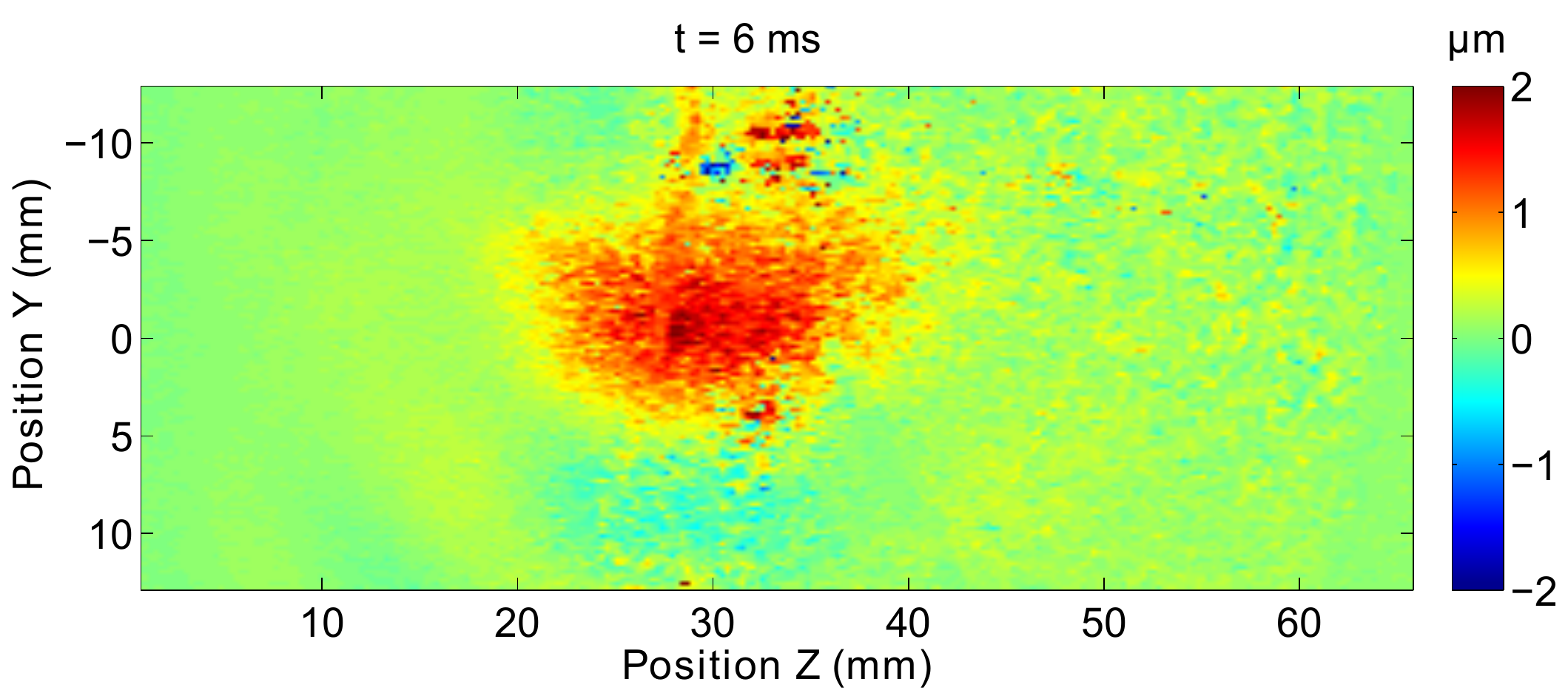}
		\end{center}
   \end{minipage} \hfill
   \begin{minipage}[c]{.5\linewidth}
		\begin{center}
	   		\includegraphics[width=.95\linewidth]{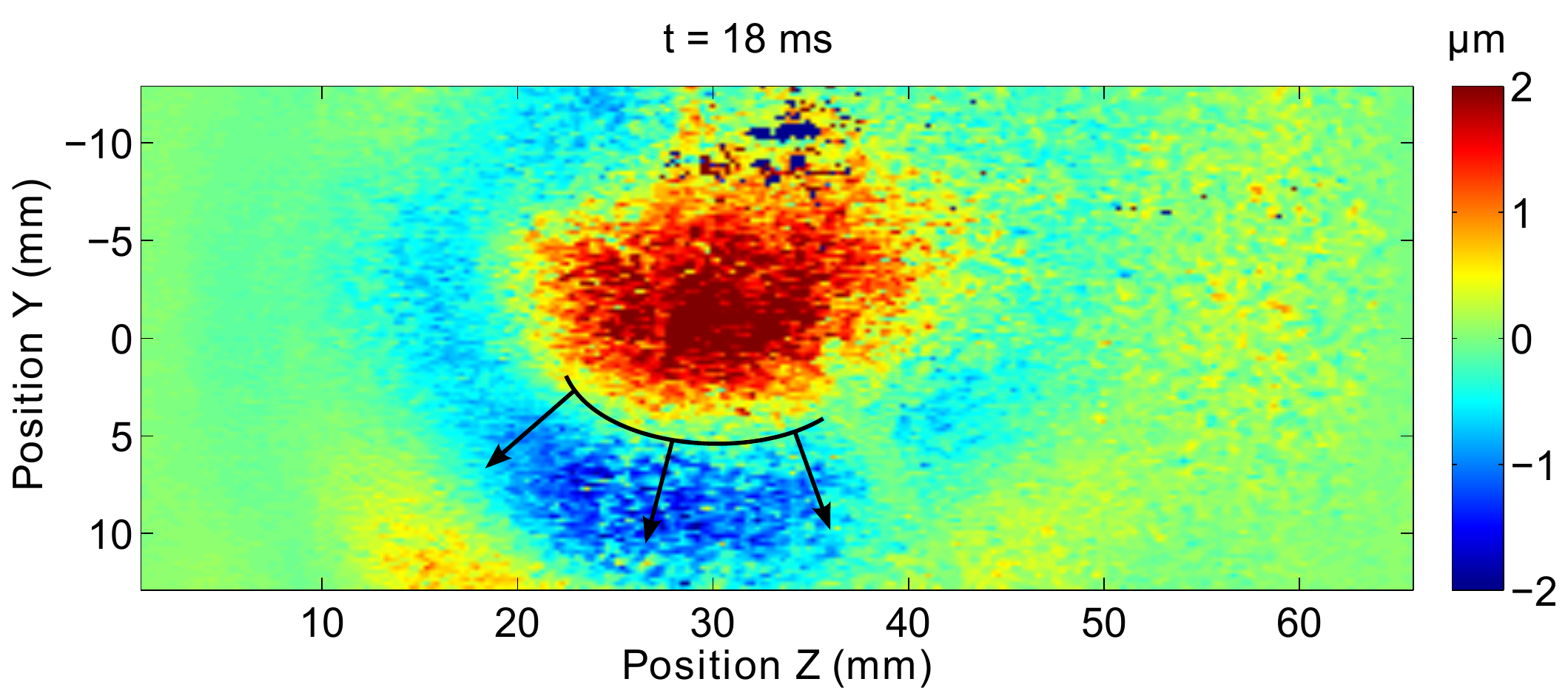}
		\end{center}
   \end{minipage}
   \begin{minipage}[c]{.5\linewidth}
		\begin{center}
	   		\includegraphics[width=.95\linewidth]{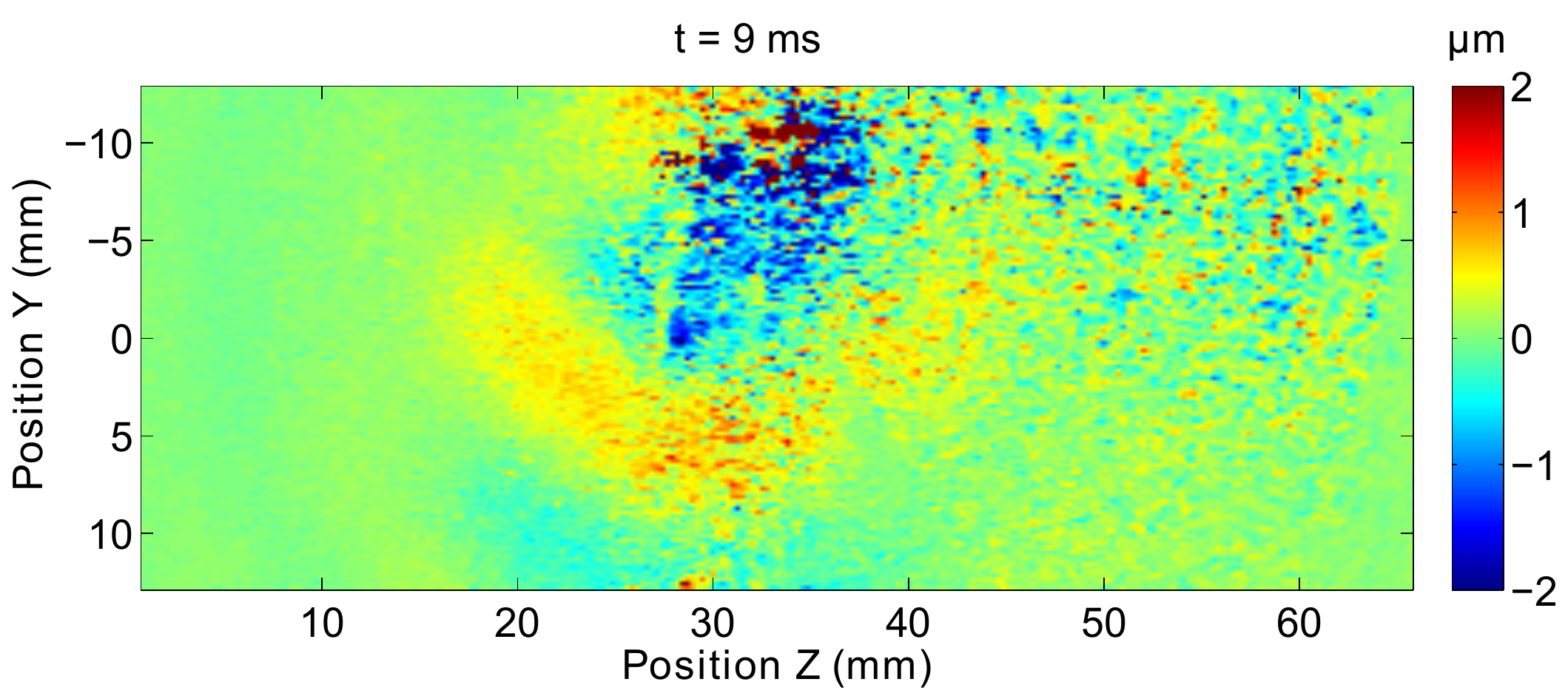}
		\end{center}
   \end{minipage} \hfill
   \begin{minipage}[c]{.5\linewidth}
		\begin{center}
	   		\includegraphics[width=.95\linewidth]{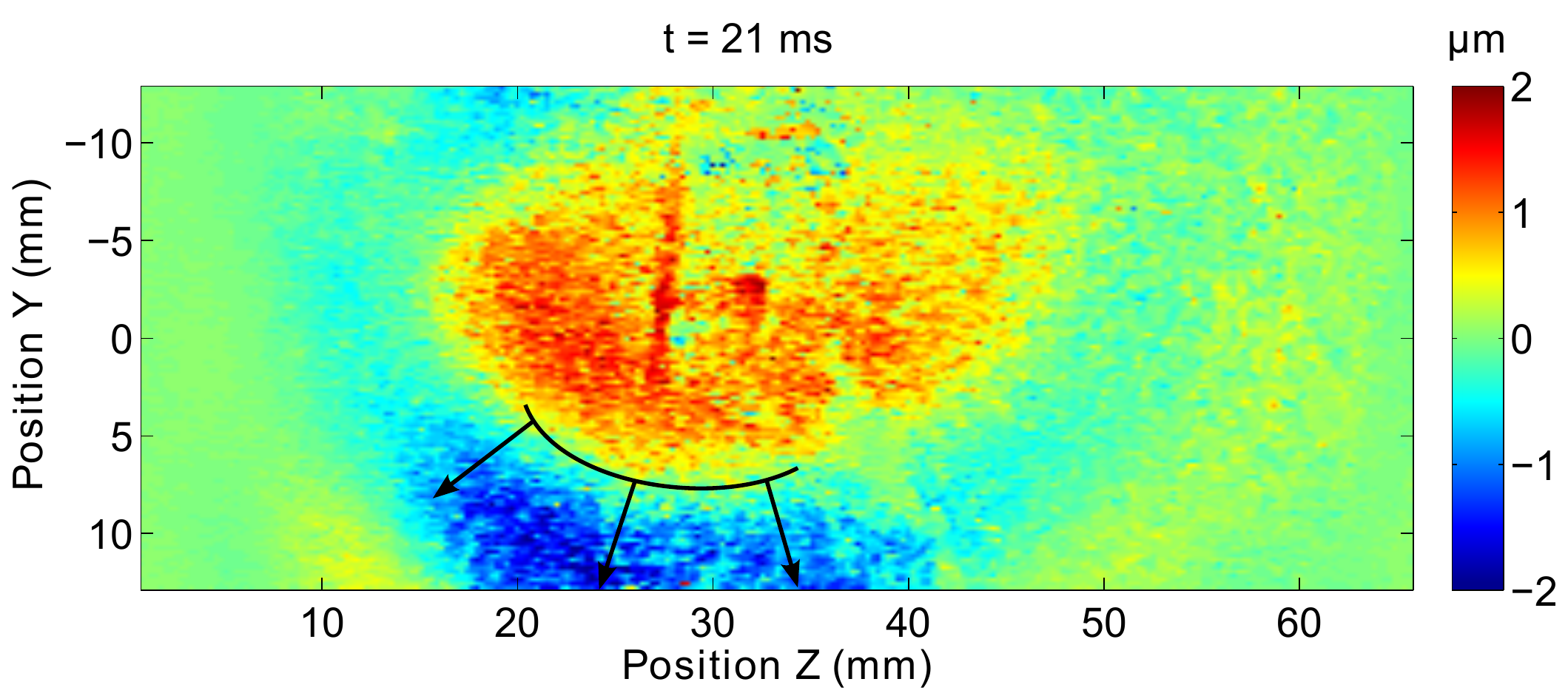}
		\end{center}
   \end{minipage}
   \begin{minipage}[c]{.5\linewidth}
		\begin{center}
	   		\includegraphics[width=.95\linewidth]{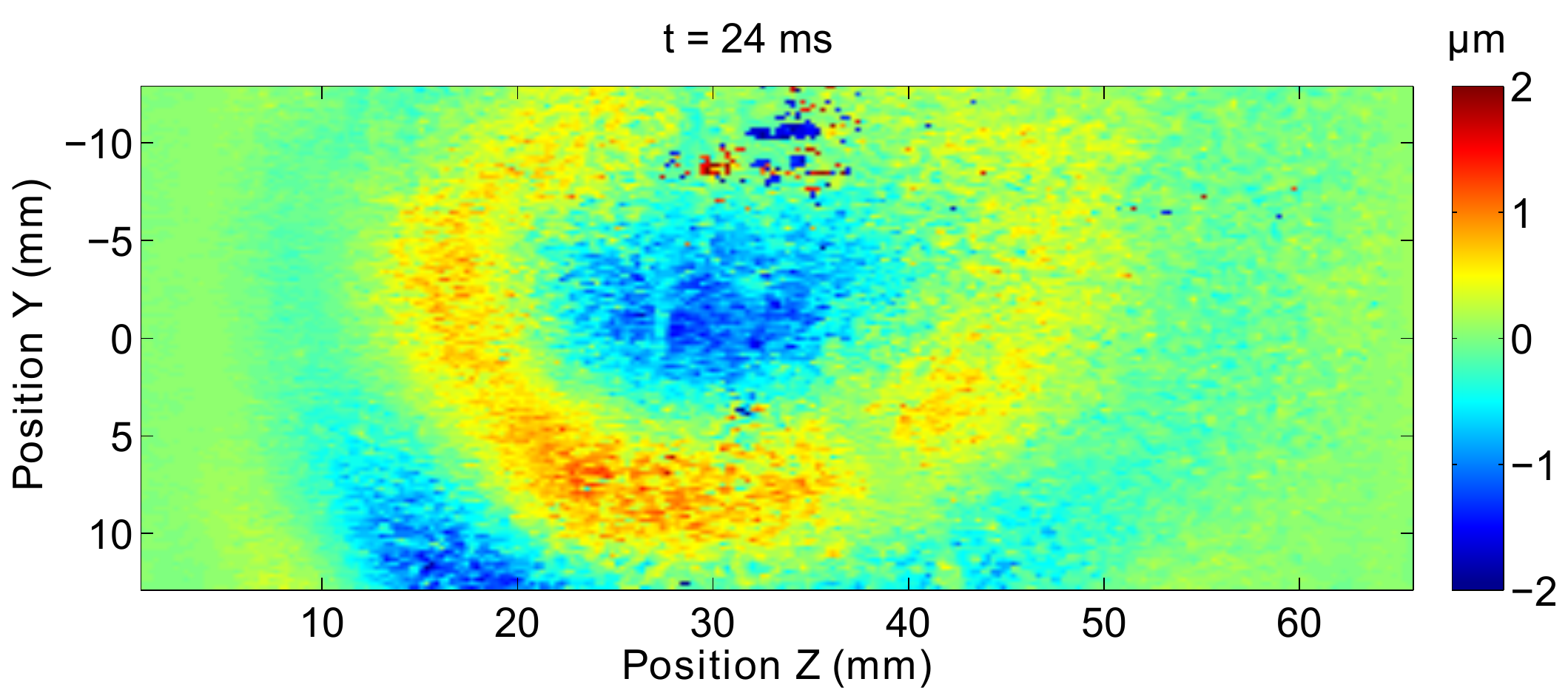}
		\end{center}
   \end{minipage}
   \begin{minipage}[c]{.5\linewidth}
		\begin{center}
	   		\includegraphics[width=.95\linewidth]{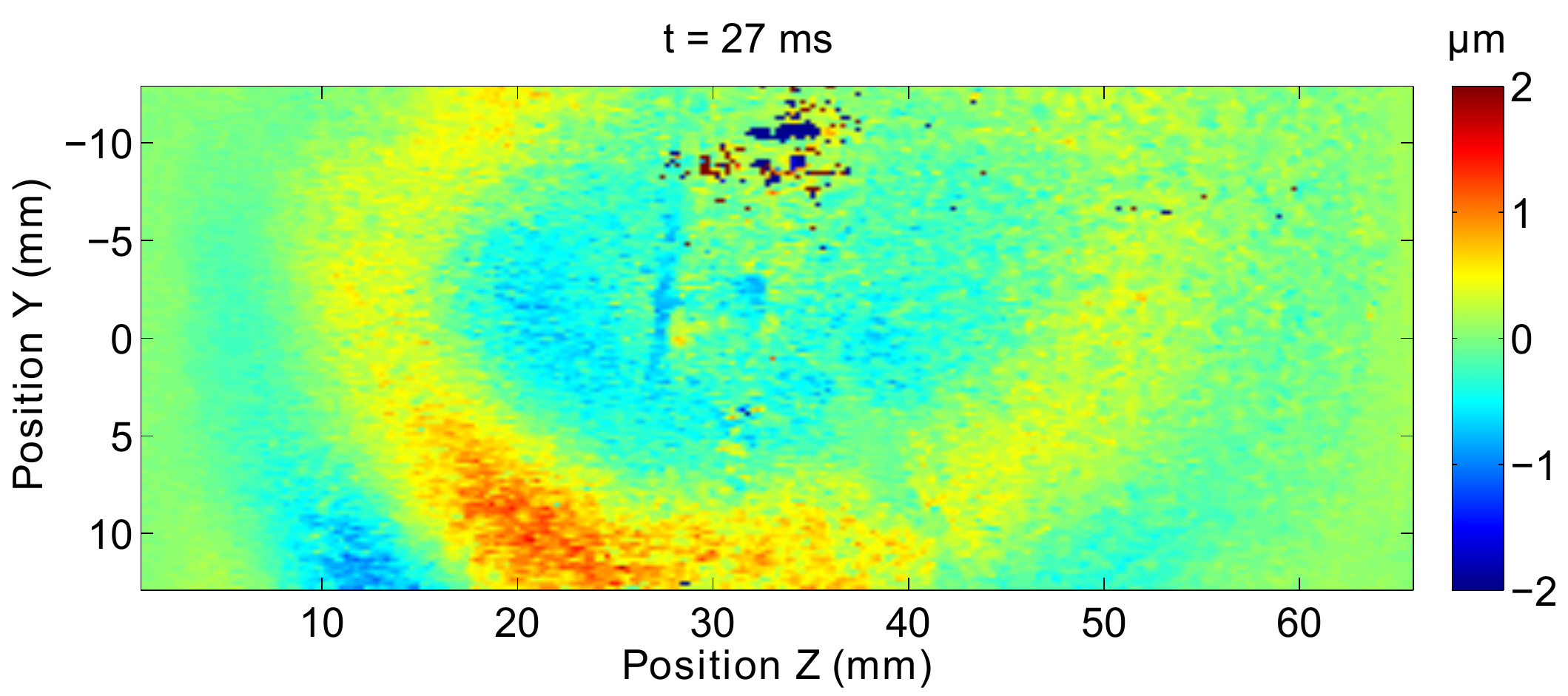}
		\end{center}
   \end{minipage}
 	\caption{\textbf{Schéma de l'expérience sur la création d'onde de cisaillement par force de Lorentz} -- Champ des déplacements dus à la traversée d'un courant électrique dans un fantôme par deux électrodes filaires au centre, le tout placé dans un champ magnétique, à 0, 3, 6, 9, 12, 15, 18, 21, 24 et 27 ms après le début du passage du courant (plan YZ). On observe une onde se propageant dans toutes les directions, avec une amplitude plus élevée dans la direction Y. Ceci est caractéristique de la directivité d'une onde de cisaillement en champ proche.}
 	\label{figElastoEMExpResElectrodeCentre} 
\end{figure}

\paragraph{Résultats et discussions}
La figure \ref{figElastoEMExpResElectrodeCentre} représente les champs des déplacements de l'échantillon à 0, 3, 6, 9, 12, 15, 18, 21, 24 et 27 millisecondes ms après le début du passage du courant. On observe, notamment sur les champs à 10, 13, 15 et 18 ms des déplacements correspondant à une onde de cisaillement. De plus, les déplacements sont d'amplitude plus importante au-dessus et en-dessous de la paire d'électrodes que sur les côtés, même si du même ordre de grandeur : ceci est caractéristique de la directivité d'une onde de cisaillement en champ proche, avec la présence d'ondes de cisaillement polarisées longitudinalement.

\subsection{Influence éventuelle du mouvement des électrodes}
Avec un champ magnétique global autour du fantôme et des électrodes, deux phénomènes peuvent se produire, sans être exclusifs : l'onde de cisaillement peut être créée par un déplacement dû (1) au courant électrique traversant le fantôme plongé dans le champ magnétique, ou bien (2) au mouvement des électrodes elles-mêmes. Même si les électrodes sont bien fixées, des mouvements de quelques micromètres d'amplitude restent possibles et peuvent se transmettre au reste du fantôme. Une expérience avec un champ magnétique localisé permet donc d'observer l'origine du mouvement.

\begin{figure}[!ht]
 \begin{center}
	 \includegraphics[width=0.4\columnwidth]{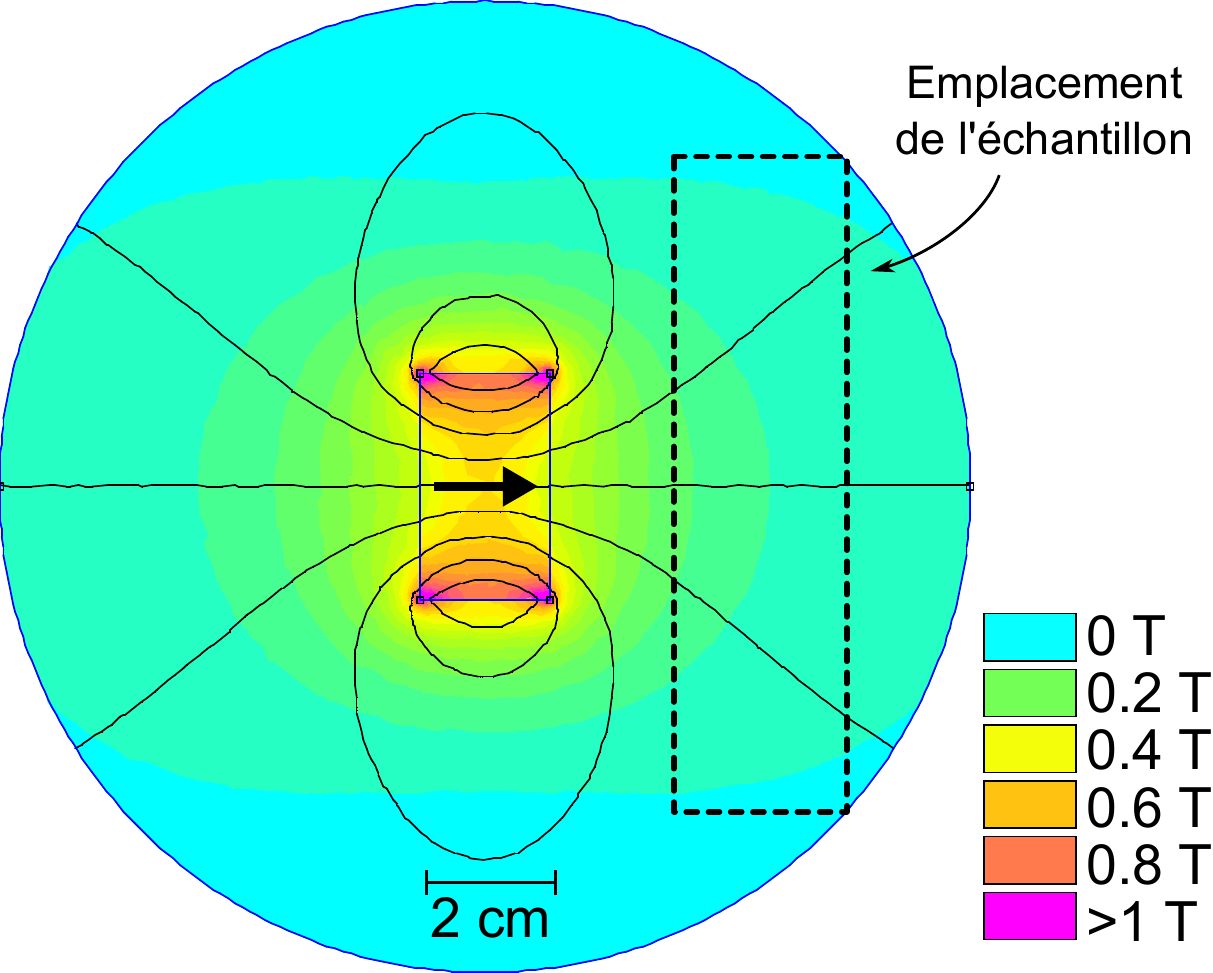}
	 \caption{Simulation du champ magnétique créé par un aimant de 35x20x20 mm$^3$ avec les lignes de champ en noir et la direction de l'aimant indiqué par une flèche (plan XY). Le champ magnétique est égal à 0,15 $\pm$ 0,05 T à 2 cm de l'aimant.}
	 \label{figElastoEMChampMagnetique}
 \end{center}
\end{figure}

\begin{figure}[!ht]
\begin{minipage}[c]{.5\linewidth}
 \begin{center}
	 \includegraphics[height=2.8cm]{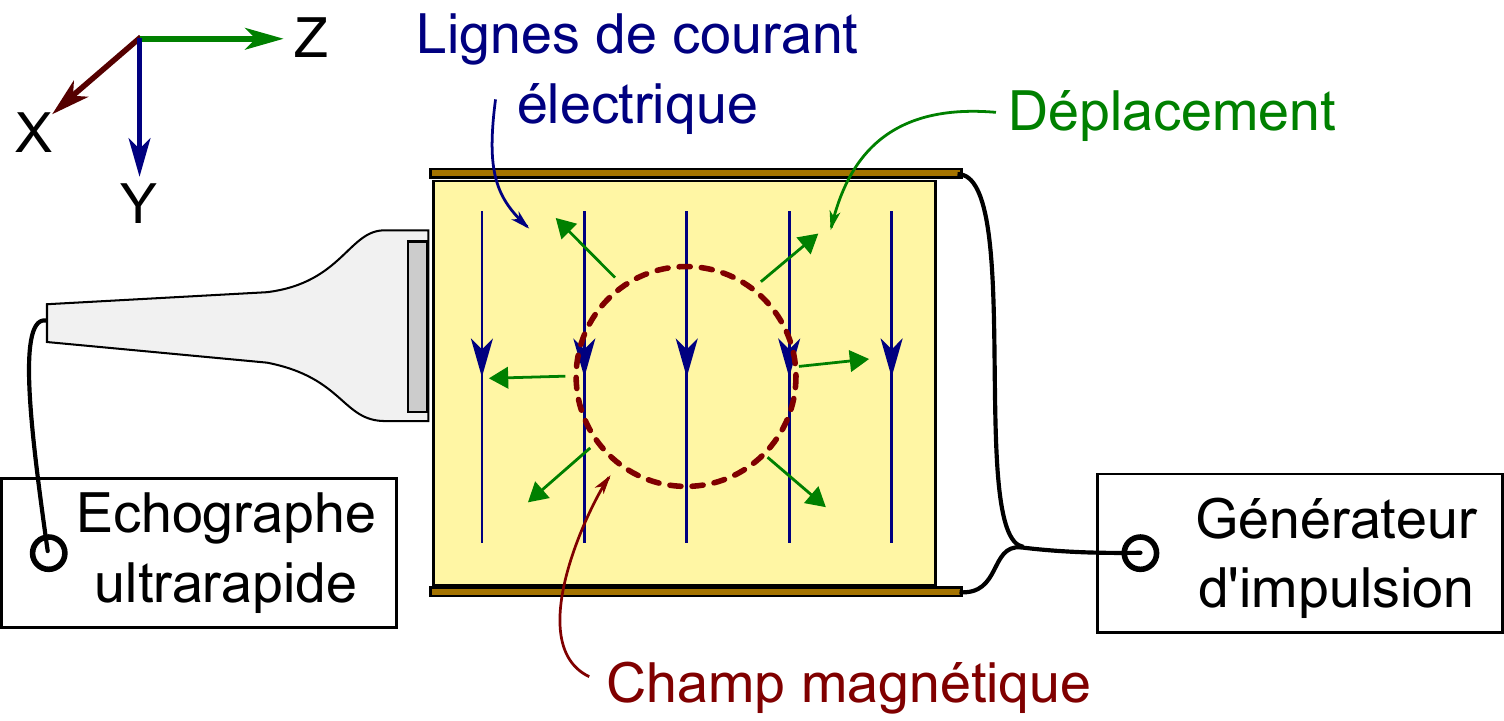}
 \end{center}
\end{minipage}
\begin{minipage}[c]{.5\linewidth}
	 \includegraphics[height=2.8cm]{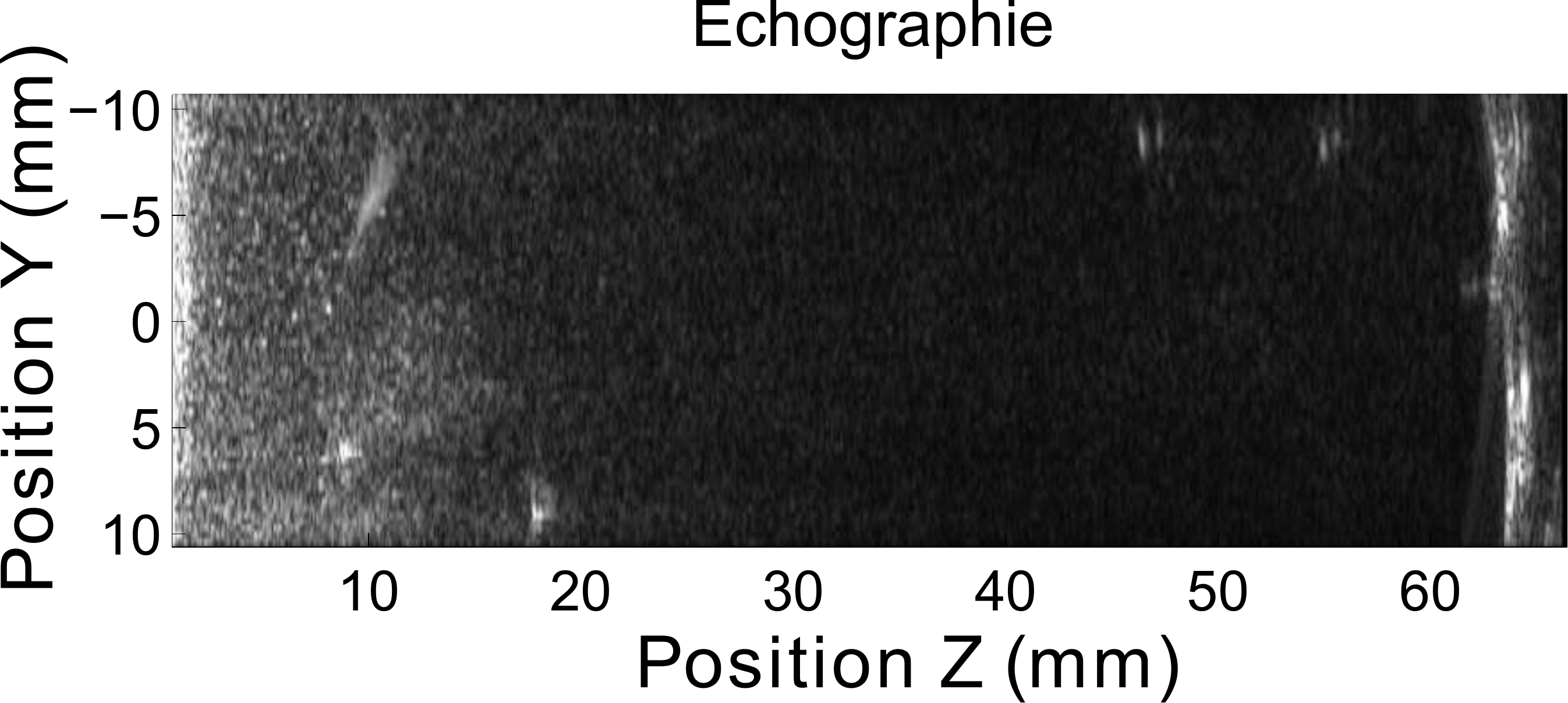}
\end{minipage}
	 \caption{(a) \textbf{Schéma de l'expérience sur l'influence éventuelle du mouvement des électrodes} -- Deux électrodes longues créent un courant électrique homogène dans un fantôme dont seulement une partie est soumise à un champ magnétique, et les déplacements sont observés par une sonde ultrasonore en mode ultrarapide. (b) \textbf{Echographie obtenue par la sonde ultrasonore} -- L'interface du fond du fantôme est visible sur la droite.}
	\label{figElastoEMExpMatGradientChamp}
\end{figure}

\paragraph{Matériel et méthodes}
Une expérience a été réalisée où deux électrodes rectangulaires en contact avec deux côtés du fantôme émettaient un courant électrique, tandis qu'un petit aimant de 35x20x20 mm$^3$ produisait un champ magnétique localisé sur le fantôme, comme représenté sur la figure \ref{figElastoEMExpMatGradientChamp}-(a). Cet aimant donnait un champ magnétique localisé de 150$\pm$50 mT à une distance de 2 cm, comme représenté sur la simulation de champ magnétique \ref{figElastoEMChampMagnetique} réalisée dans le plan XY. Une attention particulière a été portée pour que cet aimant ne soit pas en contact avec le fantôme et qu'il soit tenu par un support indépendant du fantôme -- mis à part la table de l'expérience -- pour éviter tout contact électrique ou mécanique. Tout mouvement ne peut donc provenir que de l'intérieur du fantôme et non pas par contact avec les électrodes ou l'aimant.

L'échographie du milieu est représenté sur la figure \ref{figElastoEMExpMatGradientChamp}-(b). On observe l'interface du fond du fantôme sur la droite.

\paragraph{Résultats et discussions}
La figure \ref{figElastoEMExpResElectrodeCentre} représente six champs de déplacements au cours du temps, à  0, 2, 5, 7, 10 et 12 millisecondes après le début du passage du courant électrique. On peut voir que les ondes sont issues de la zone à fort gradient de champ magnétique. Ici, le mouvement est créé au milieu du fantôme, donc il n'y a pas d'effets de bords ou de conversion de mode. Cela signifie que le mouvement a bien été produit au sein de l'échantillon par la force de Lorentz et non pas par contact avec les électrodes. L'amplitude des déplacements est environ dix fois plus faible que dans les expériences précédentes, mais le champ magnétique est également moins intense. Le phénomène (1) n'est donc pas négligeable devant le phénomène (2) lors des autres expériences.

\begin{figure}[!ht]
   \begin{minipage}[c]{.5\linewidth}
		\begin{center}
	   		\includegraphics[width=1\linewidth]{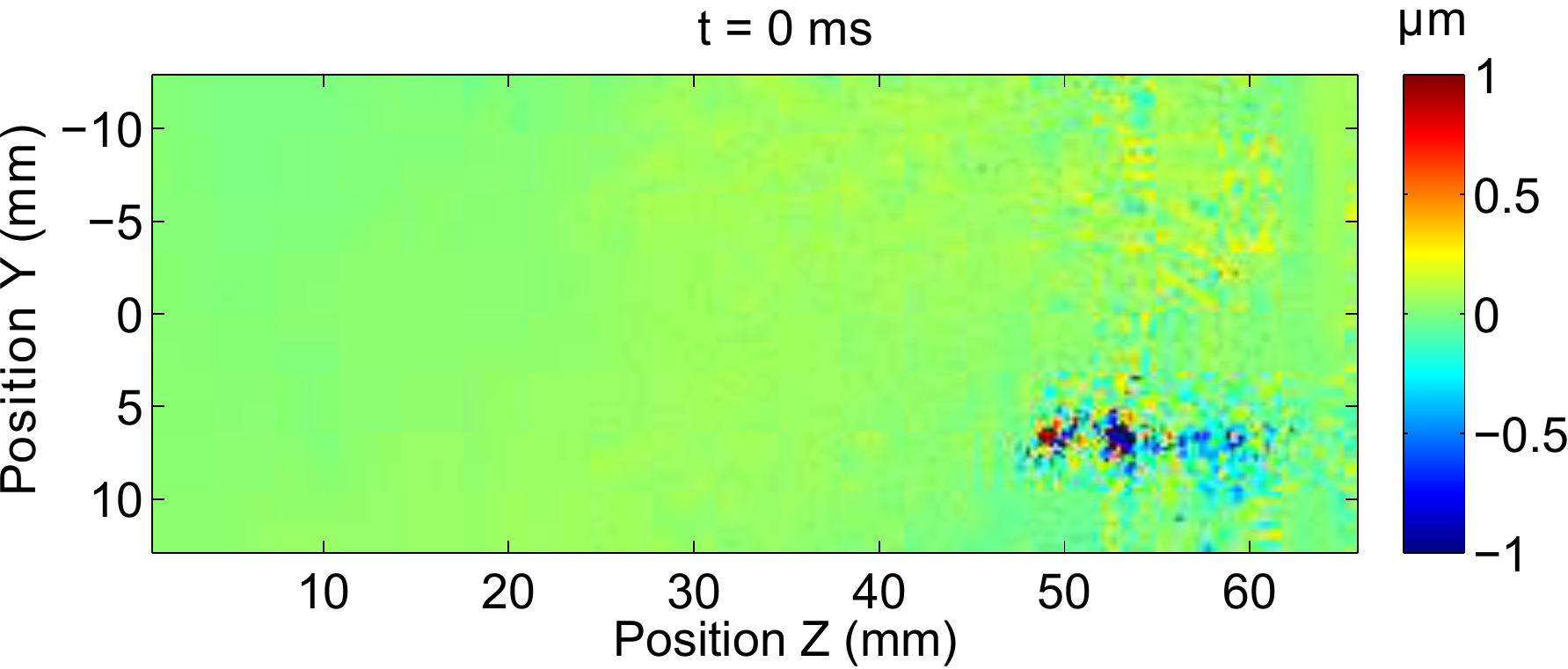}
		\end{center}
   \end{minipage}
   \begin{minipage}[c]{.5\linewidth}
		\begin{center}
	   		\includegraphics[width=1\linewidth]{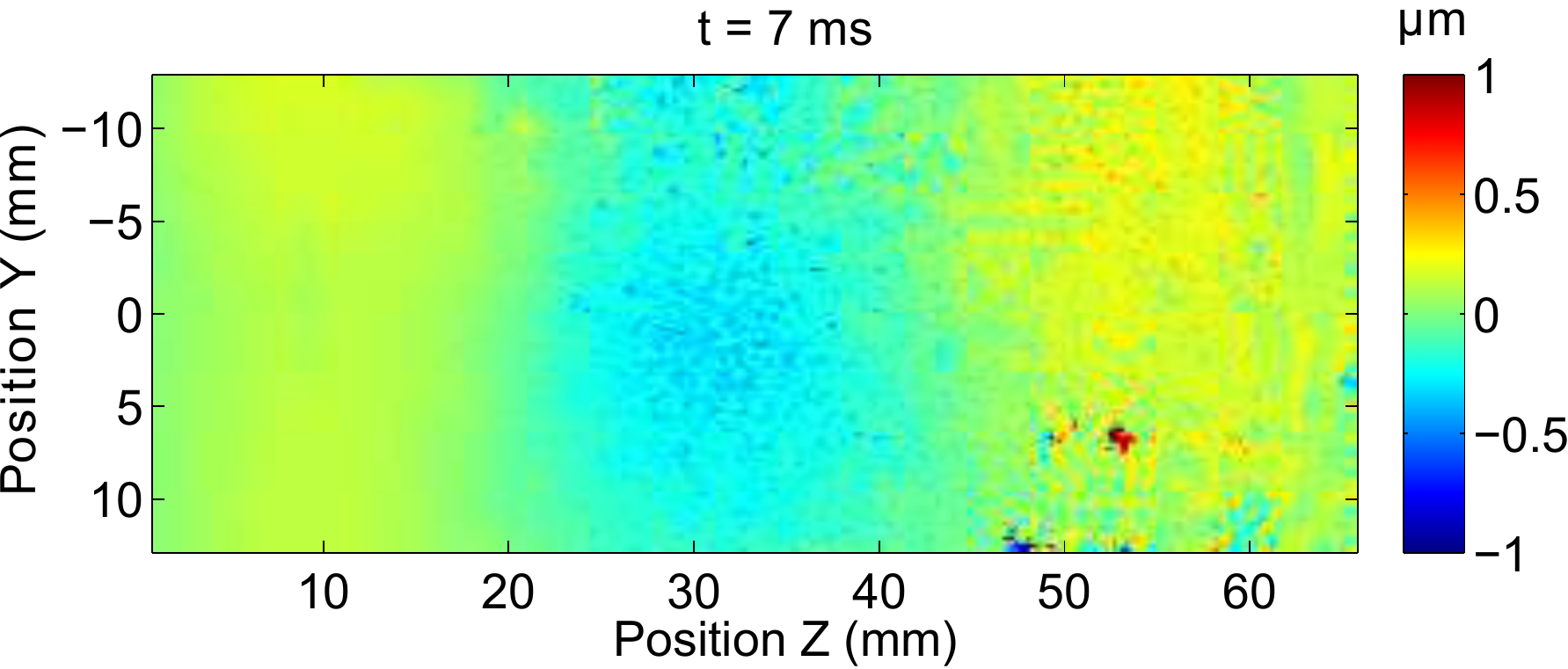}
		\end{center}
   \end{minipage}
   \begin{minipage}[c]{.5\linewidth}
		\begin{center}
	   		\includegraphics[width=1\linewidth]{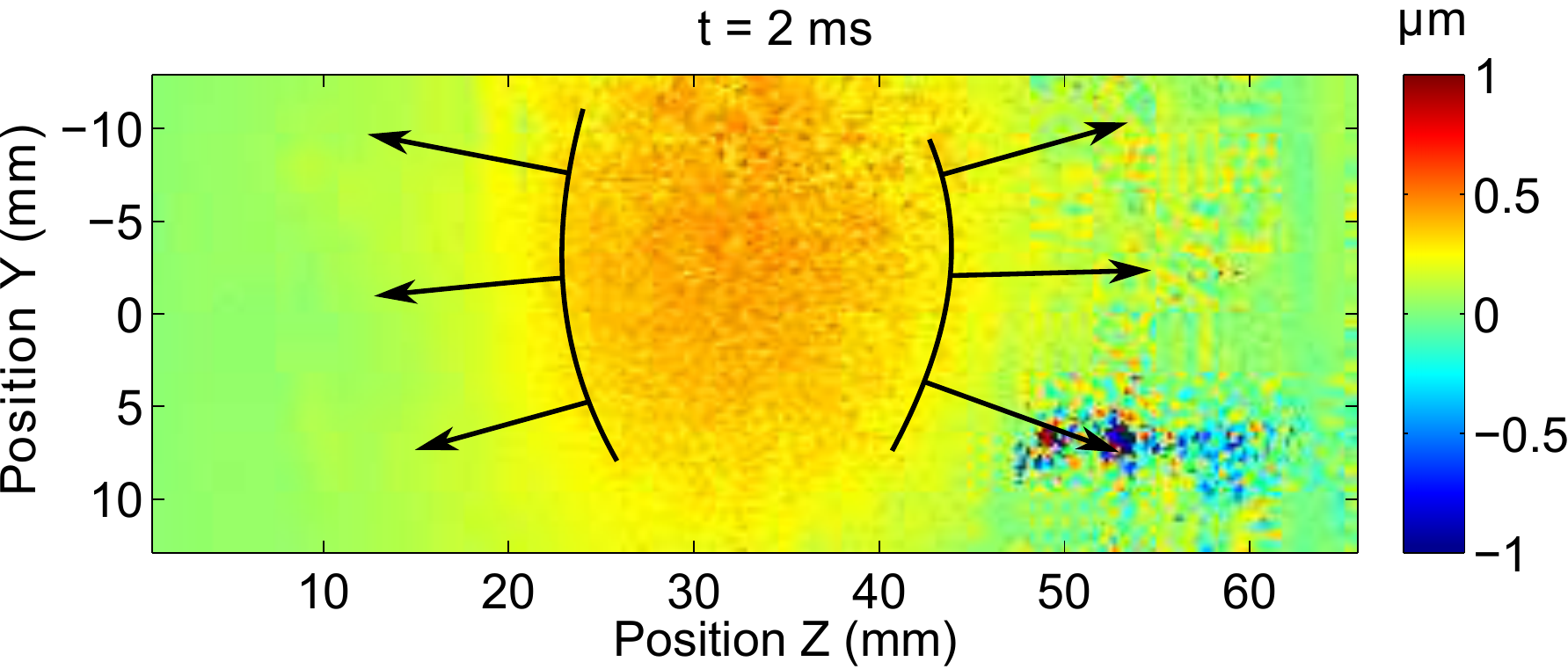}
		\end{center}
   \end{minipage} \hfill
   \begin{minipage}[c]{.5\linewidth}
		\begin{center}
	   		\includegraphics[width=1\linewidth]{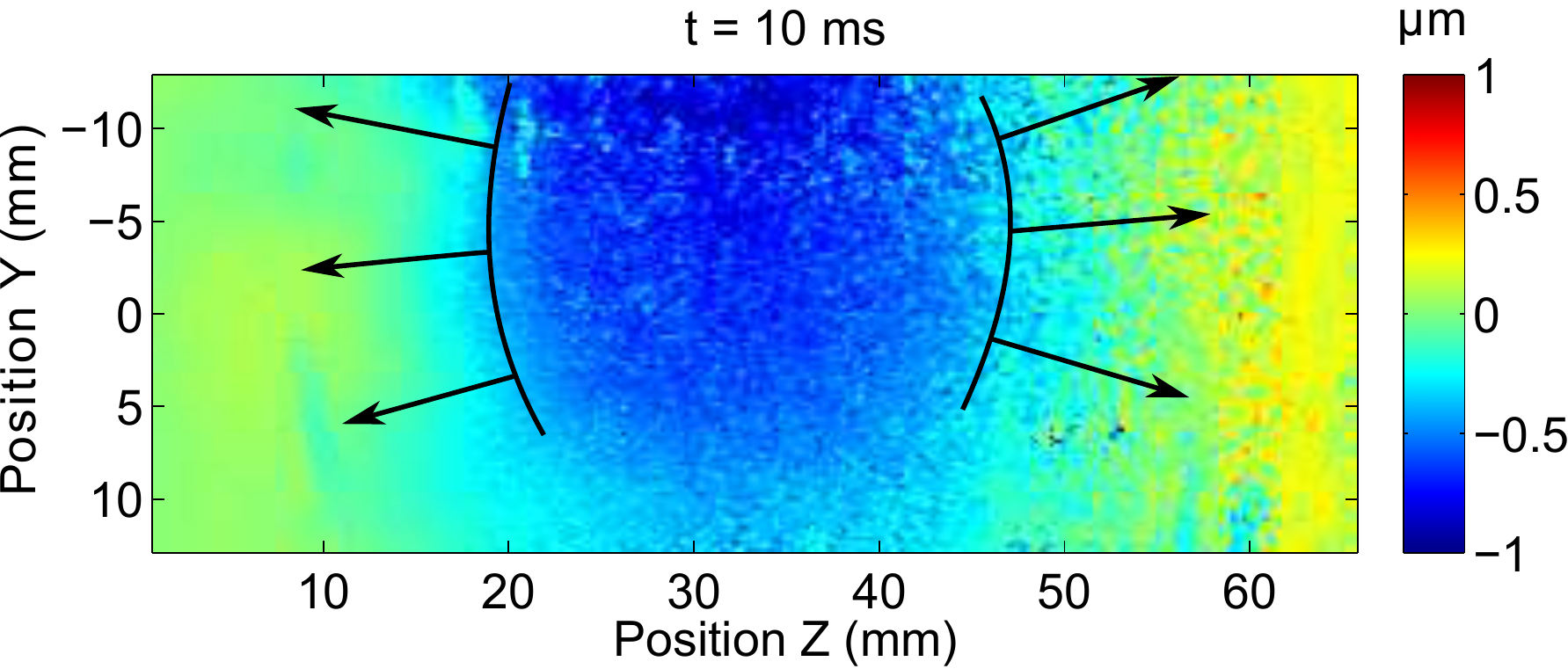}
		\end{center}
   \end{minipage}
   \begin{minipage}[c]{.5\linewidth}
		\begin{center}
	   		\includegraphics[width=1\linewidth]{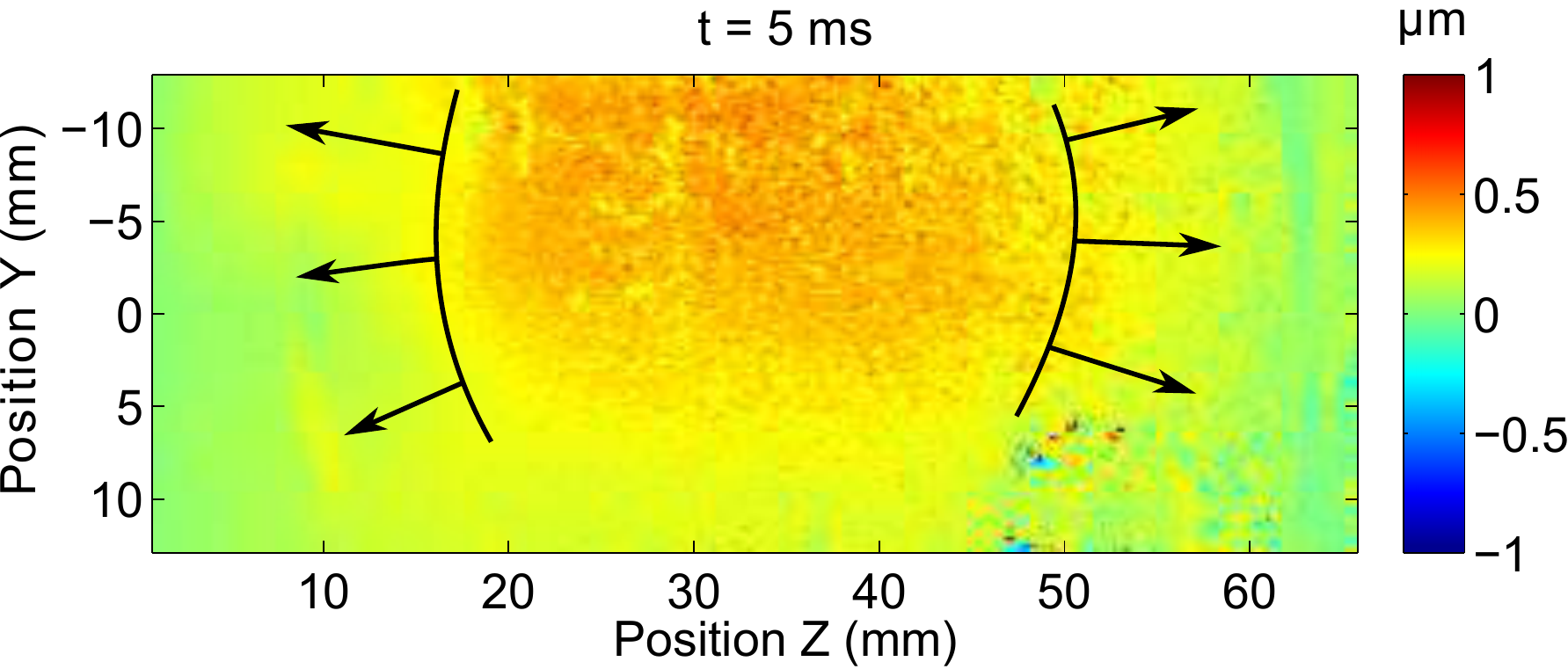}
		\end{center}
   \end{minipage} \hfill
   \begin{minipage}[c]{.5\linewidth}
		\begin{center}
	   		\includegraphics[width=1\linewidth]{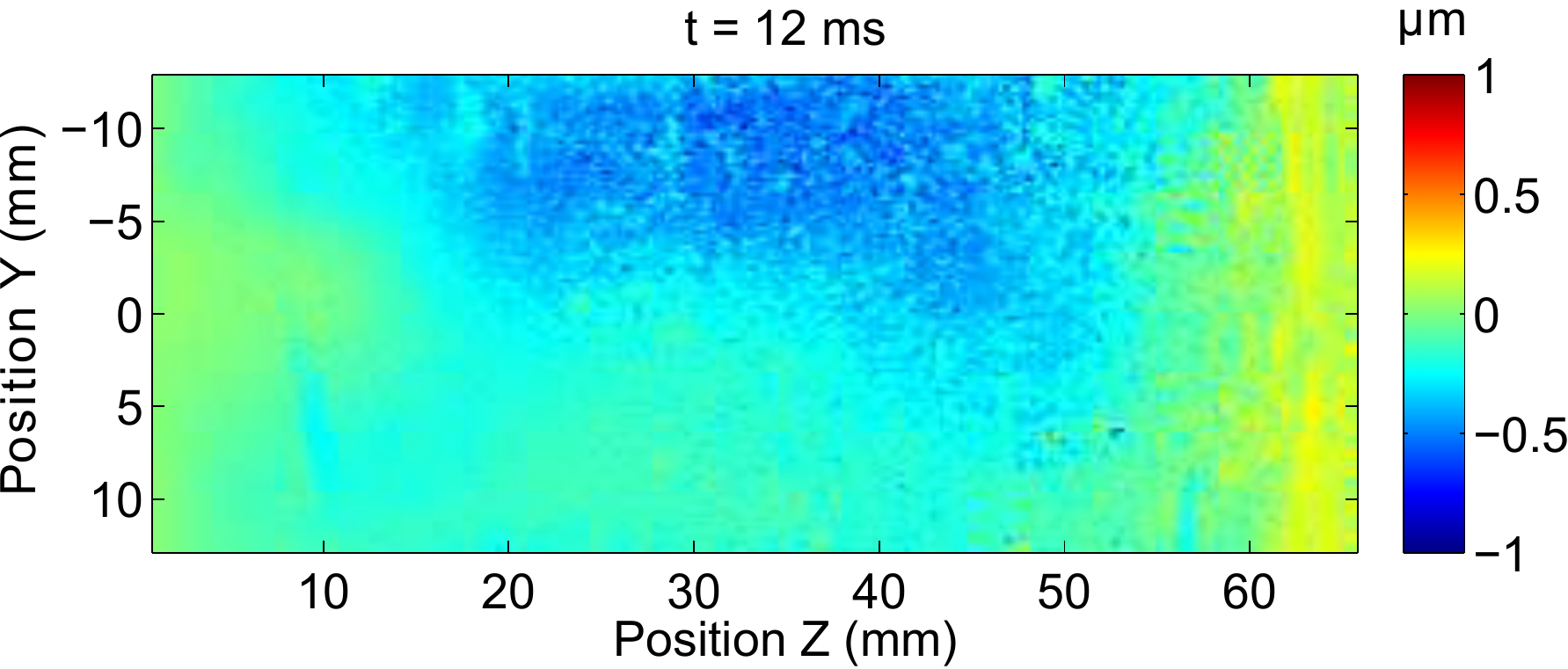}
		\end{center}
   \end{minipage}
 	\caption{\textbf{Résultats de l'expérience sur l'influence éventuelle du mouvement des électrodes} -- Champ des déplacements dus à la traversée d'un courant électrique dans le fantôme avec un champ magnétique localisé, à 0, 2, 5, 7, 10 et 12 ms après le début du passage du courant (plan YZ). On observe que le mouvement initial apparait au centre du fantôme, là où le champ magnétique est présent, et non pas des bords, où se trouvent les électrodes.}
 	\label{figElastoEMExpResGradientChamp} 
\end{figure}

\subsection{Influence éventuelle de l'effet Joule}
Tout passage d'un courant dans un conducteur échauffe celui-ci par effet Joule. Cet échauffement provoque une dilatation locale, ce qui pourrait être la source du déplacement observé, à l'instar de la technique d'imagerie photoacoustique.

\begin{figure}[!ht]
\begin{minipage}[c]{.5\linewidth}
 \begin{center}
	 \includegraphics[height=2.8cm]{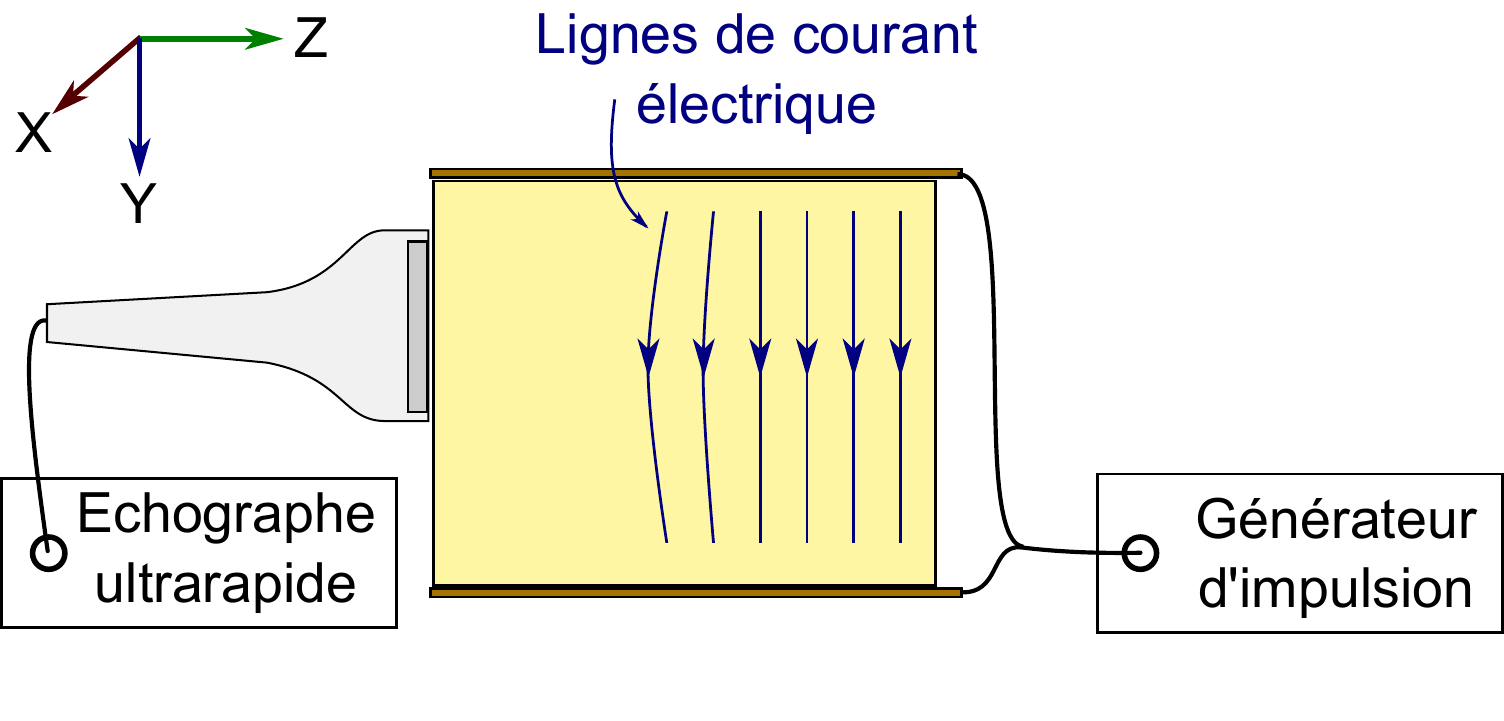}
 \end{center}
\end{minipage}
\begin{minipage}[c]{.5\linewidth}
	 \includegraphics[height=2.8cm]{figElastoEMExpEffetJouleEcho.pdf}
\end{minipage}
	 \caption{(a) \textbf{Schéma de l'expérience sur l'influence de l'effet Joule} -- Deux électrodes longues créent un courant électrique homogène dans un fantôme en l'absence de champ magnétique, et les déplacements sont observés par une sonde ultrasonore en mode ultrarapide. (b) \textbf{Echographie obtenue par la sonde ultrasonore} -- L'interface du fond du fantôme est visible sur la droite.}
	\label{figElastoEMExpMatEffetJoule}
\end{figure}

\paragraph{Matériel et Méthodes}
Une expérience a été mise en oeuvre sans champ magnétique, en retirant l'aimant, comme représenté sur la figure \ref{figElastoEMExpMatEffetJoule}-(a). Si l'effet Joule a un effet important, on devrait alors voir un déplacement lors du passage du courant. L'échographie du milieu est représentée figure \ref{figElastoEMExpMatEffetJoule}-(b), où l'on voit l'interface du fond du fantôme sur la droite.

\begin{figure}[!ht]
   \begin{minipage}[c]{.5\linewidth}
		\begin{center}
	   		\includegraphics[width=1\linewidth]{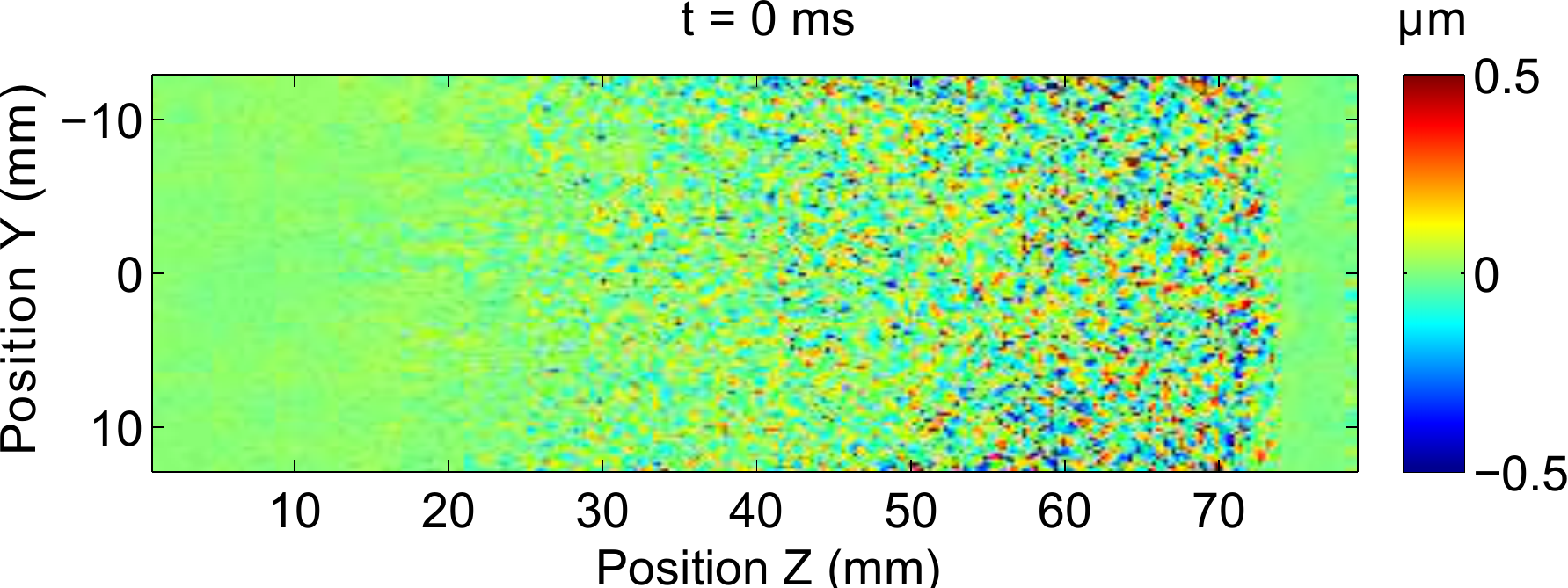}
		\end{center}
   \end{minipage} \hfill
   \begin{minipage}[c]{.5\linewidth}
		\begin{center}
	   		\includegraphics[width=1\linewidth]{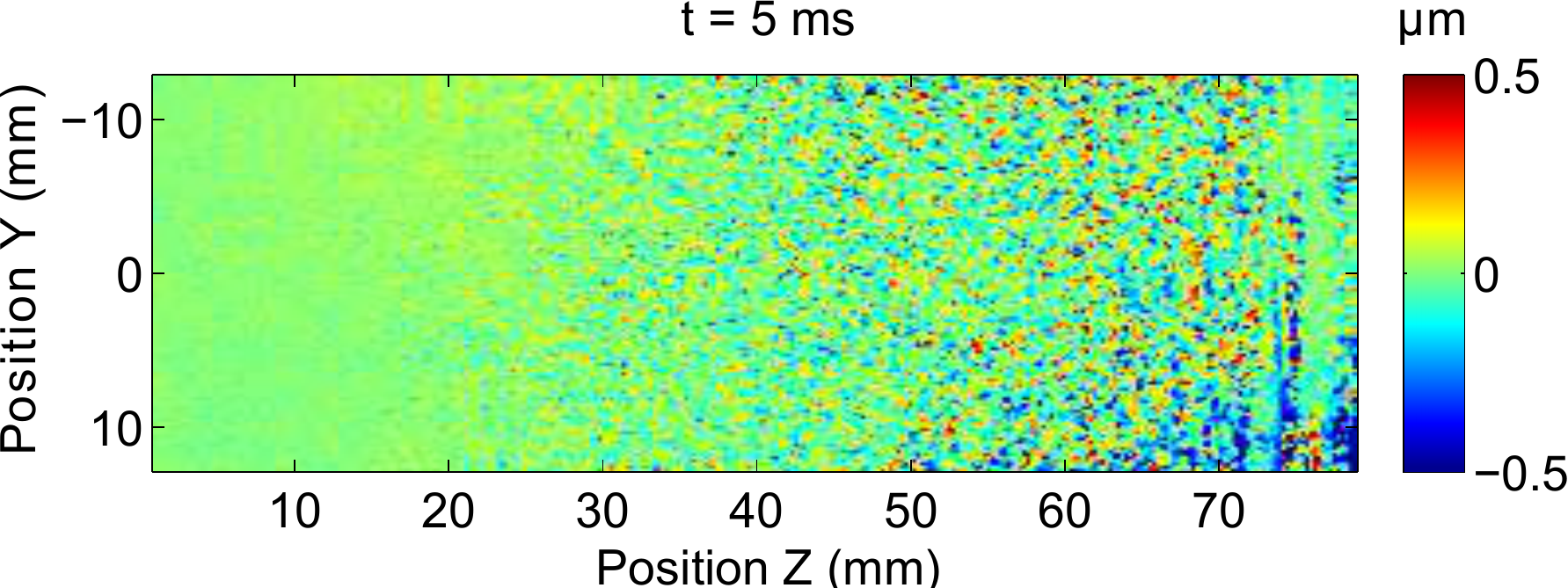}
		\end{center}
   \end{minipage}
   \begin{minipage}[c]{.5\linewidth}
		\begin{center}
	   		\includegraphics[width=1\linewidth]{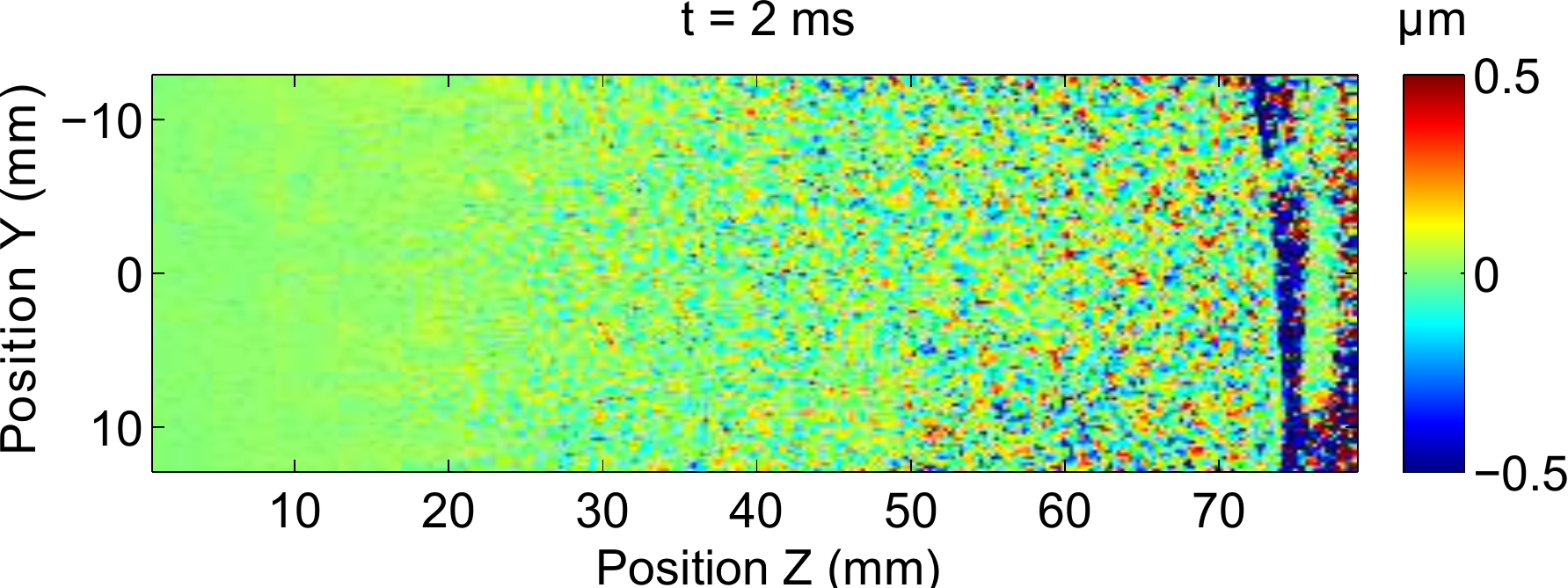}
		\end{center}
   \end{minipage} \hfill
   \begin{minipage}[c]{.5\linewidth}
		\begin{center}
	   		\includegraphics[width=1\linewidth]{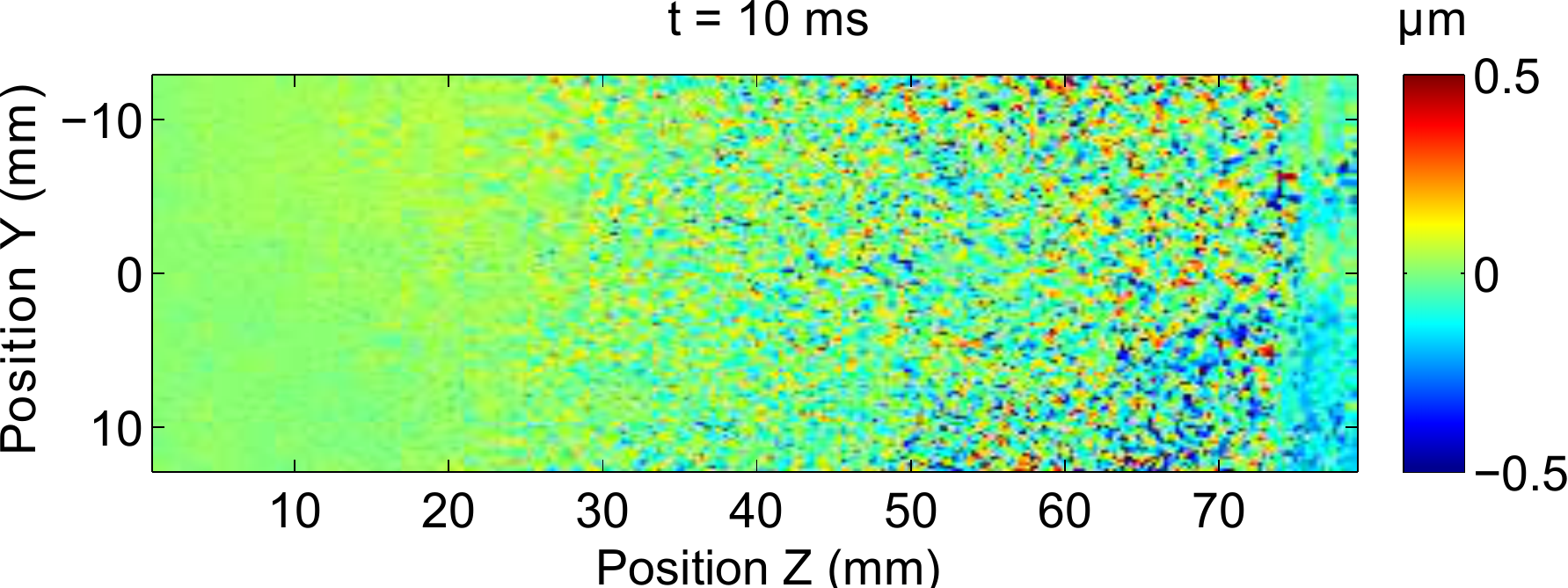}
		\end{center}
   \end{minipage}
 	\caption{\textbf{Résultats de l'expérience sur l'influence de l'effet Joule} -- Champ des déplacements dus à un courant électrique injecté par deux électrodes placé au-dessus et en-dessous du fantôme, sans champ magnétique, à 0, 2, 5 et 10 ms (plan YZ). On n'observe aucune propagation d'onde et quasiment aucun mouvement, ce qui indique que la dilatation due au chauffage par effet Joule n'explique pas les ondes observées.}
 	\label{figElastoEMExpResEffetJoule} 
\end{figure}

\paragraph{Résultats et discussions}
La figure \ref{figElastoEMExpResEffetJoule} représente quatre champs de déplacements au cours du temps à 0, 2, 5 et 10 millisecondes après le début du passage du courant. La barre de couleur va de -0,5 \micro m à 0,5 \micro m, donc d'une amplitude quatre fois plus faible que l'expérience initiale. Le courant mesuré est de 0,5 A pour une résistance du fantôme de 250 Ohm, ce qui est similaire aux expériences précédentes. On n'observe aucune propagation de mouvement dans le milieu, bien que l'interface du fond semble légèrement déplacée. Ce dernier phénomène est très probablement dû à des artefacts de l'algorithme de \textit{speckle-tracking}, qui essaie de trouver un déplacement même en l'absence de \textit{speckle}. La dilatation par effet Joule n'explique donc pas les phénomènes observés dans les parties précédentes.

\subsection{Bilan partiel}
Ces expériences ont montré que la force de Lorentz peuvent être utilisée pour induire des ondes de cisaillement dans les tissus mous. Les ondes de cisaillement peut apparaitre à cause de la force elle-même ou par conversion de mode lors de la propagation d'une onde mécanique.

Il a par ailleurs été montré que les déplacements observés n'étaient pas des artefacts dû au mouvement des électrodes elles-mêmes, qui peuvent se déplacer à cause de la force de Lorentz, ou à la dilatation par effet Joule.

\section{Expériences d'élastographie par force de Lorentz}

L'objectif de cette section est de montrer qu'il est possible de réaliser des expériences d'élastographie avec des ondes de cisaillement induites par force de Lorentz. Nous avons mesuré une vitesse d'onde de cisaillement dans un fantôme homogène, puis dans un fantôme composé de deux couches. Enfin, la méthode a été appliquée sur un échantillon \textit{ex-vivo} de foie.

\subsection{Elastographie sur un fantôme homogène}

\begin{figure}[!ht]
\begin{minipage}[c]{.5\linewidth}
 \begin{center}
	 \includegraphics[height=2.8cm]{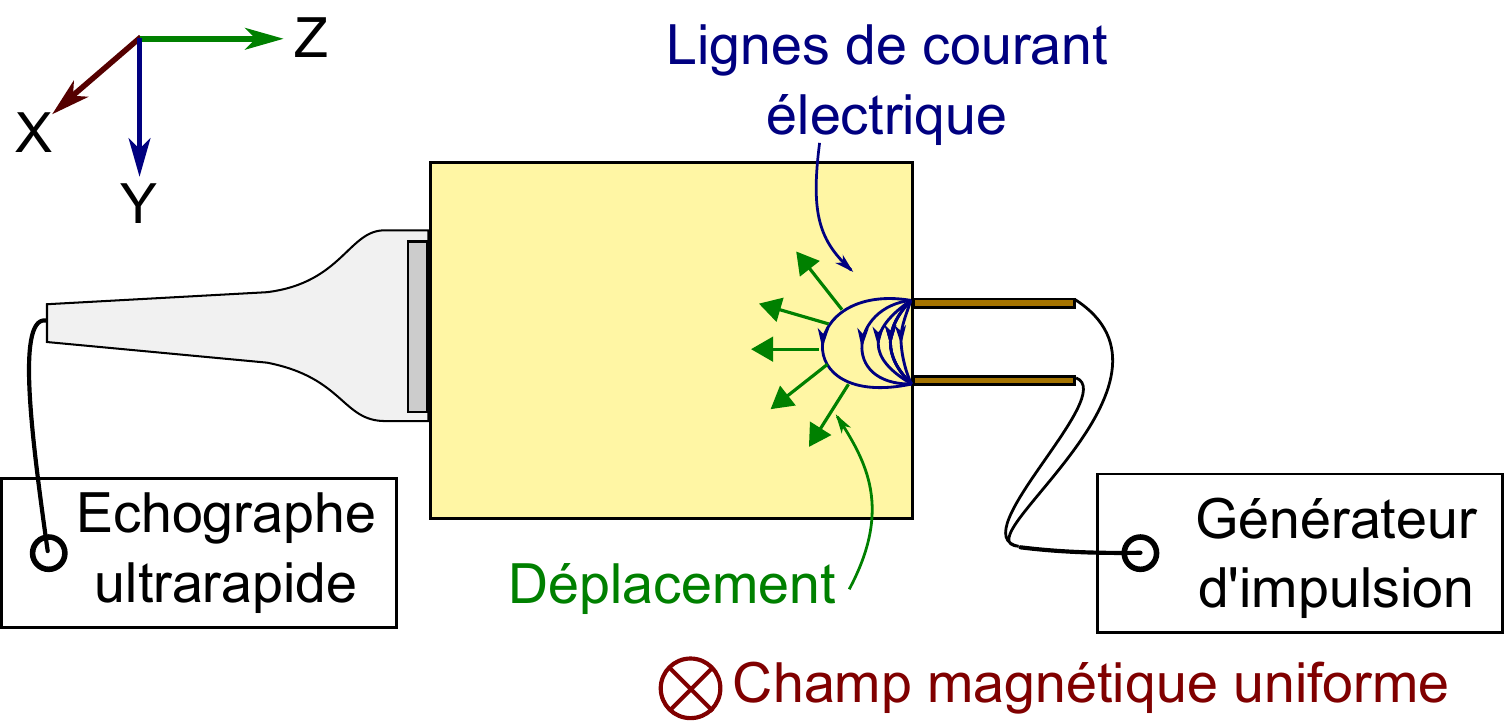}
 \end{center}
\end{minipage}
\begin{minipage}[c]{.5\linewidth}
	 \includegraphics[height=2.8cm]{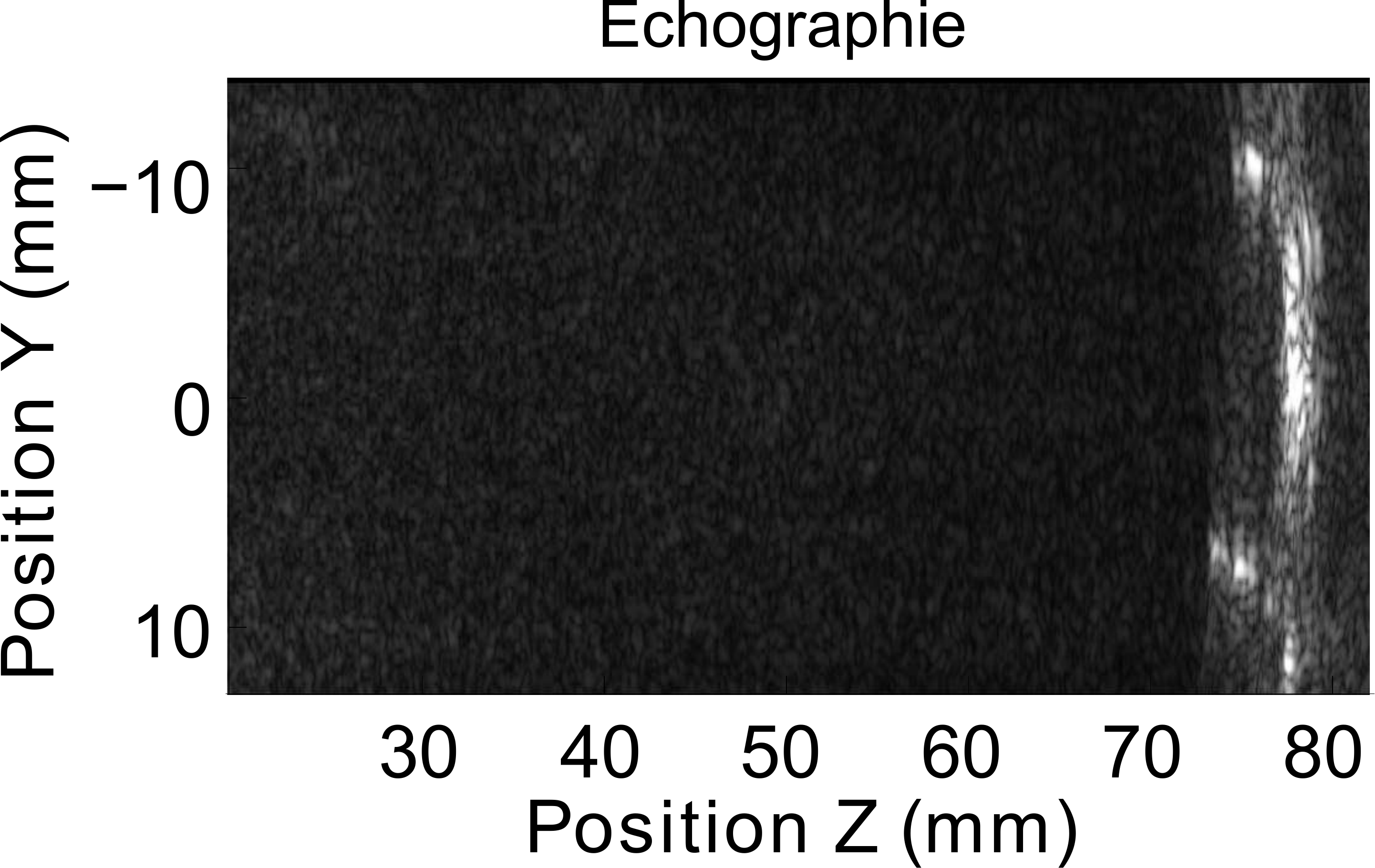}
\end{minipage}
	 \caption{(a) \textbf{Schéma de l'expérience d'élastographie sur un fantôme homogène} -- Deux électrodes proches créent un courant électrique dans un fantôme soumis à un champ magnétique, et les déplacements sont observés par une sonde ultrasonore en mode ultrarapide. (b) \textbf{Echographie obtenue par la sonde ultrasonore} -- On distingue au milieu deux points blancs correspondant à la zone de contact avec les deux électrodes.}
	\label{figElastoEMExpMatElectrodeFond}
\end{figure}

\paragraph{Matériel et méthodes}
Une expérience a été faite avec deux électrodes en contact avec le fond du fantôme, le long de la tranche. La barrette ultrasonore était placée de l'autre côté, comme représenté sur le schéma \ref{figElastoEMExpMatElectrodeFond}-(a).
L'échographie du milieu est représentée figure \ref{figElastoEMExpMatElectrodeFond}-(b), où l'on voit au fond deux points blancs correspondant à la zone de contact avec les deux électrodes.

\begin{figure}[!ht]
   \begin{minipage}[c]{.5\linewidth}
		\begin{center}
	   		\includegraphics[width=1\linewidth]{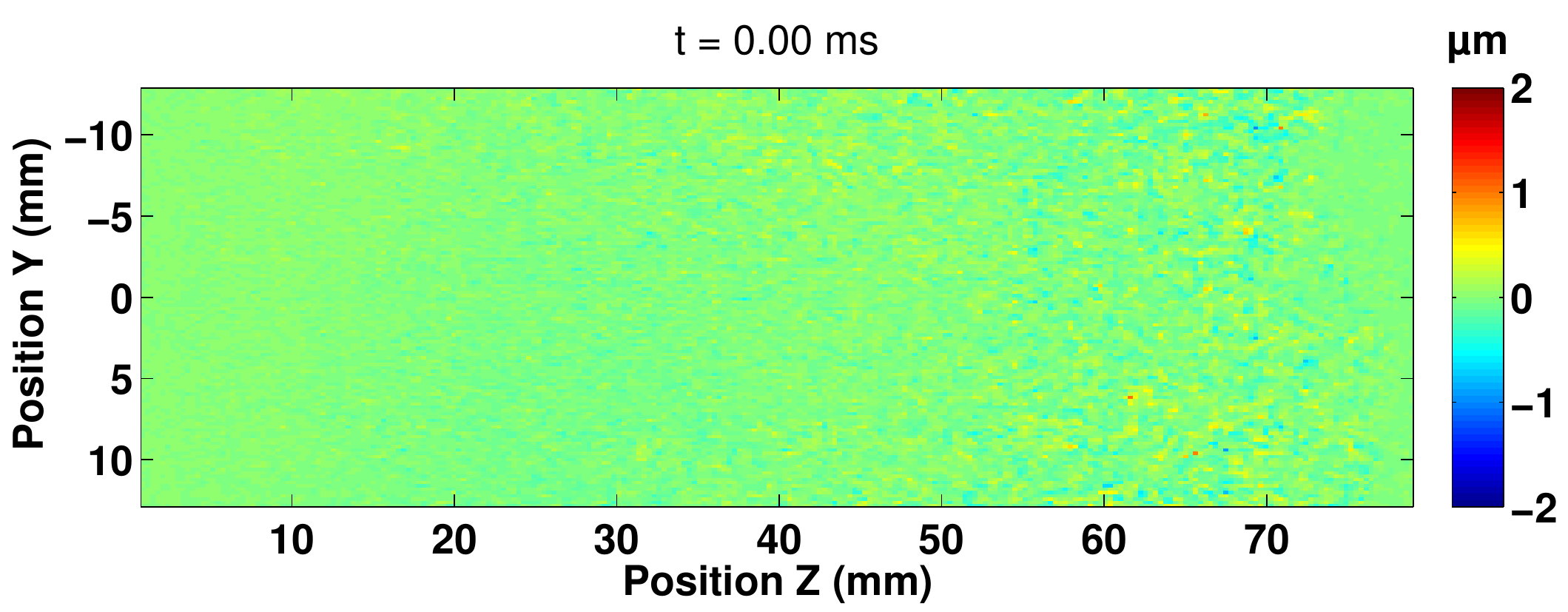}
		\end{center}
   \end{minipage} \hfill
   \begin{minipage}[c]{.5\linewidth}
		\begin{center}
	   		\includegraphics[width=1\linewidth]{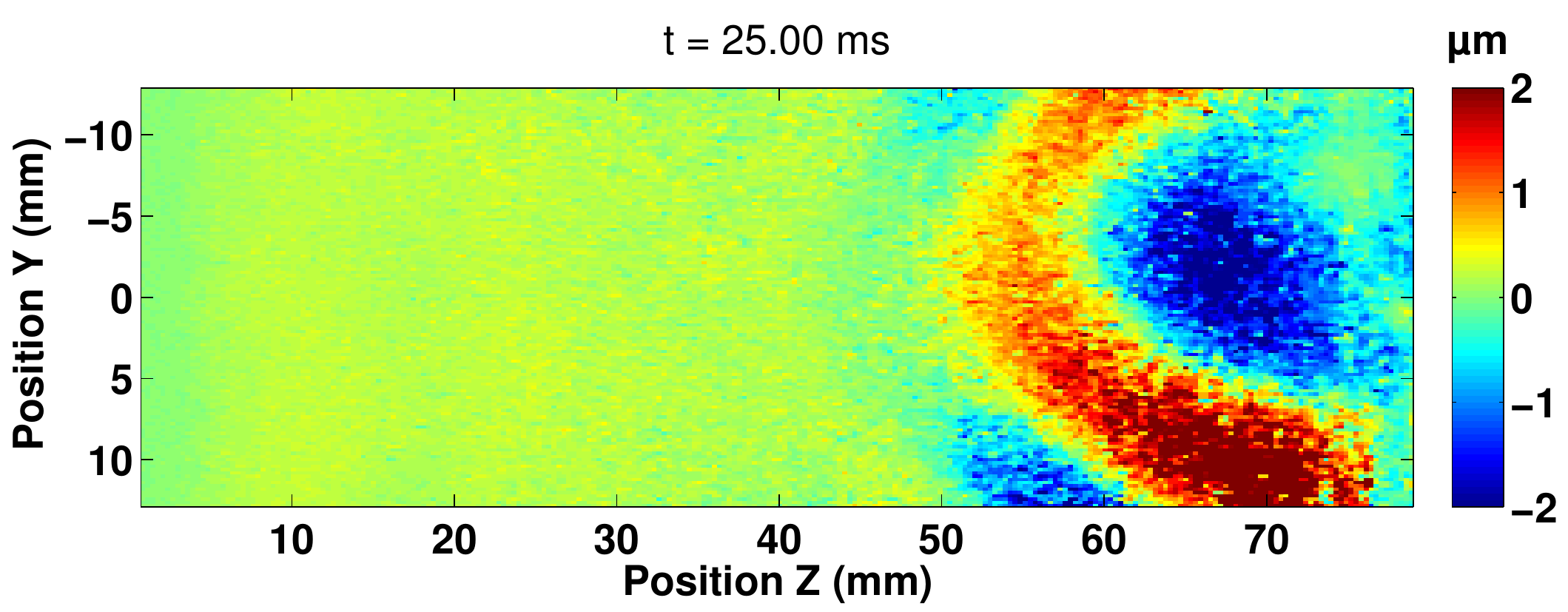}
		\end{center}
   \end{minipage}
   \begin{minipage}[c]{.5\linewidth}
		\begin{center}
	   		\includegraphics[width=1\linewidth]{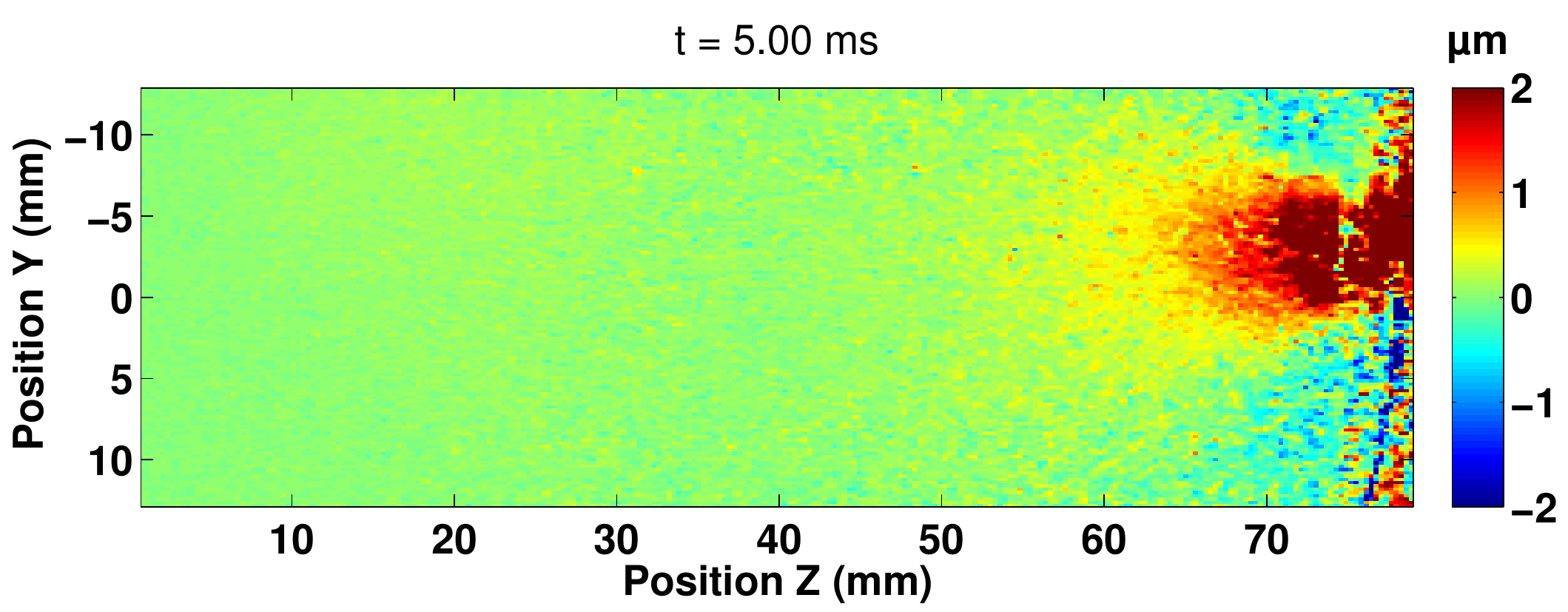}
		\end{center}
   \end{minipage}
   \begin{minipage}[c]{.5\linewidth}
		\begin{center}
	   		\includegraphics[width=1\linewidth]{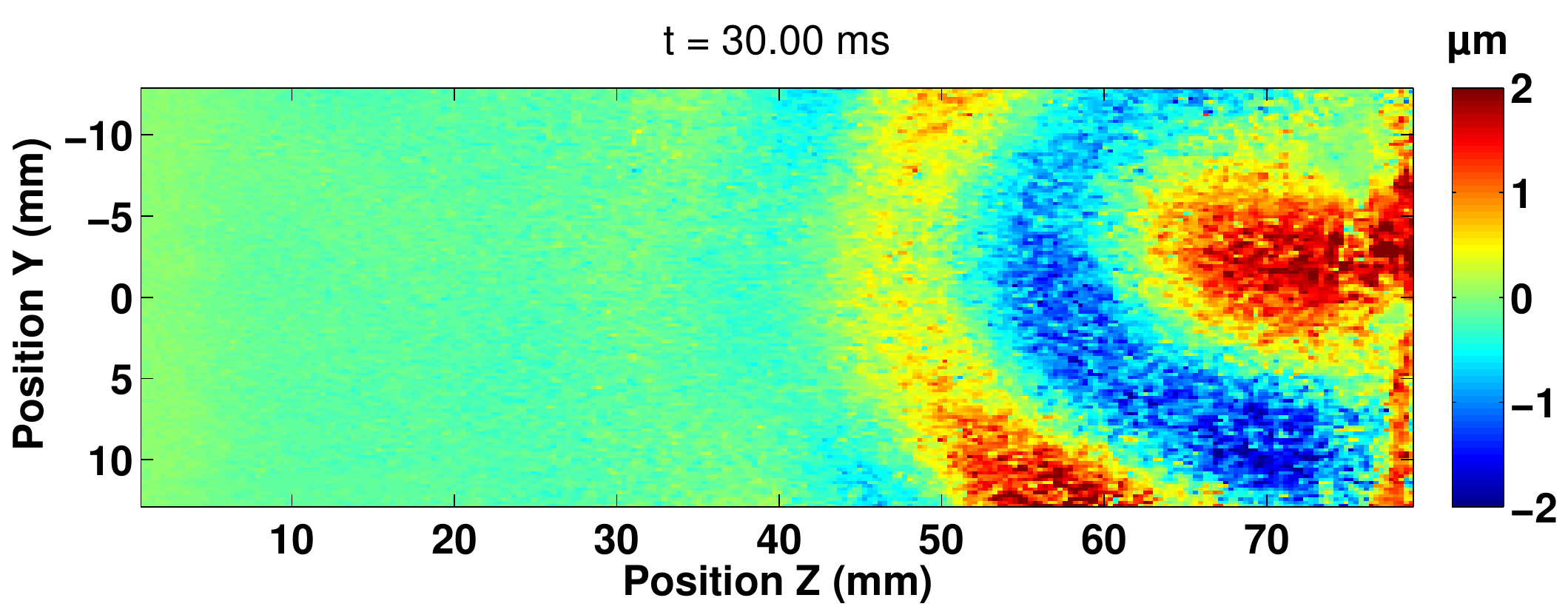}
		\end{center}
   \end{minipage} \hfill
   \begin{minipage}[c]{.5\linewidth}
		\begin{center}
	   		\includegraphics[width=1\linewidth]{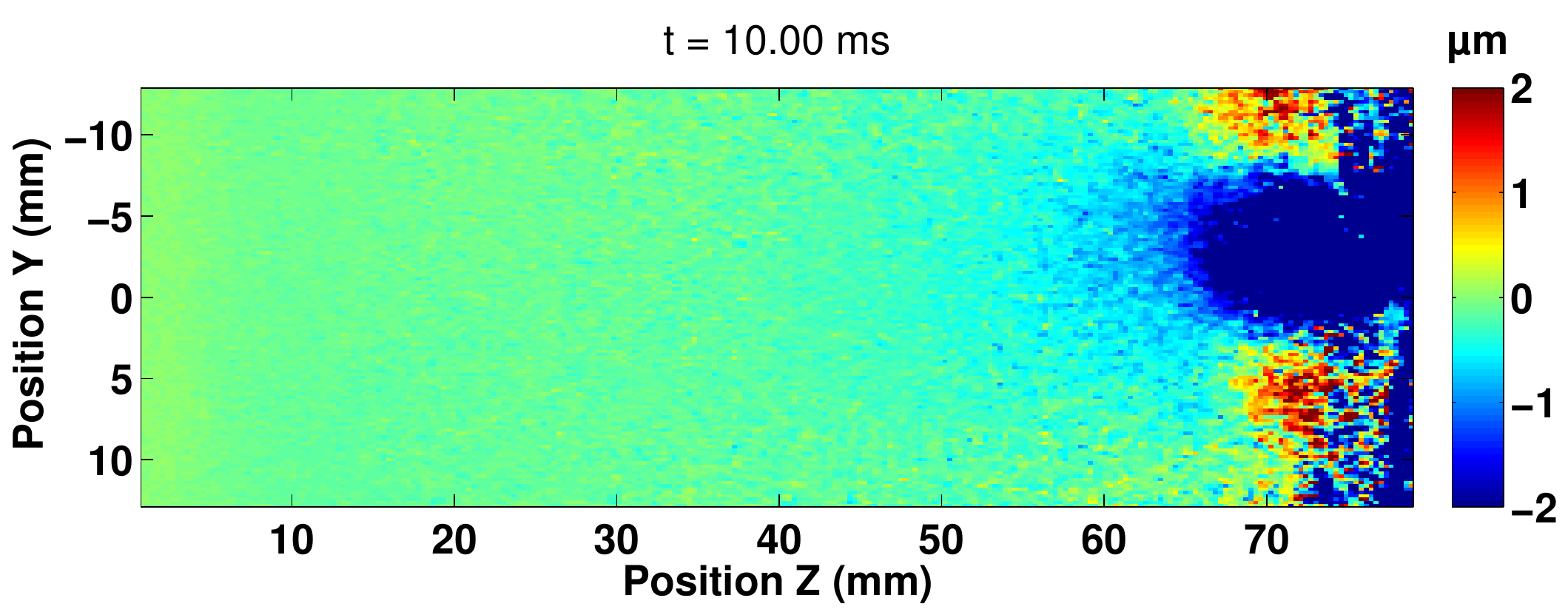}
		\end{center}
   \end{minipage}
   \begin{minipage}[c]{.5\linewidth}
		\begin{center}
	   		\includegraphics[width=1\linewidth]{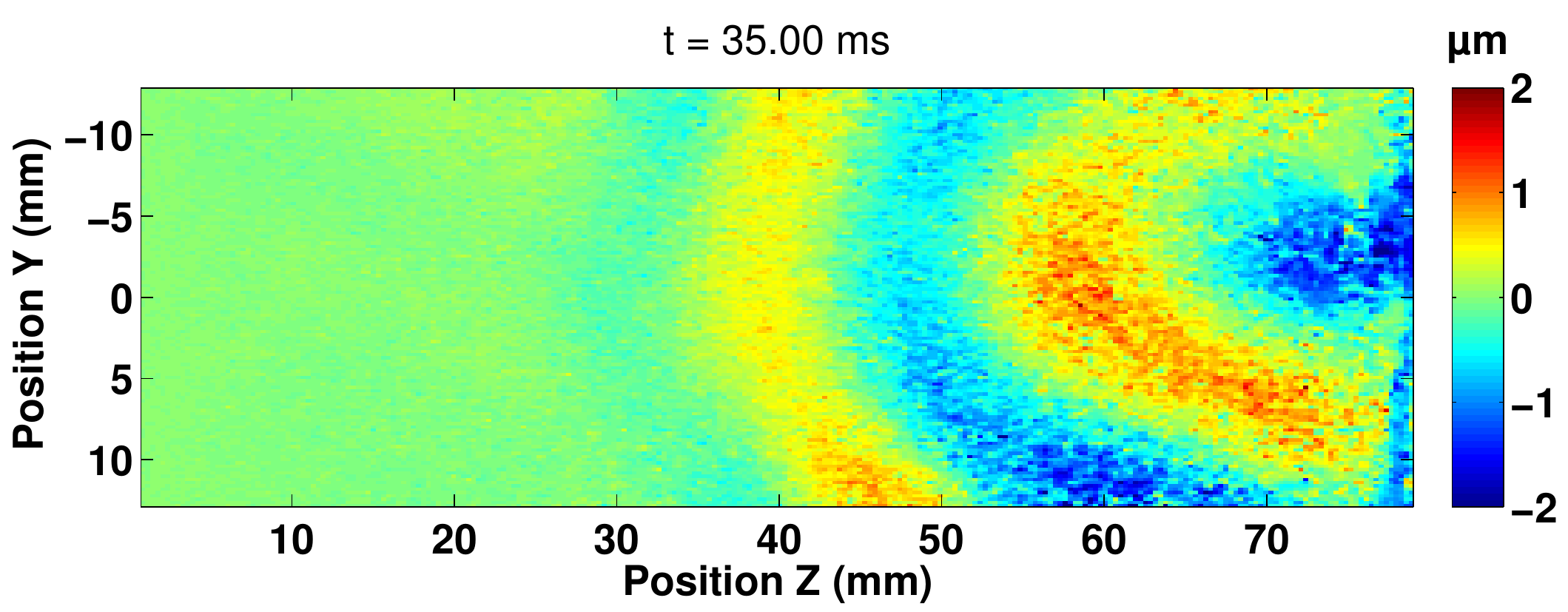}
		\end{center}
   \end{minipage} \hfill
   \begin{minipage}[c]{.5\linewidth}
		\begin{center}
	   		\includegraphics[width=1\linewidth]{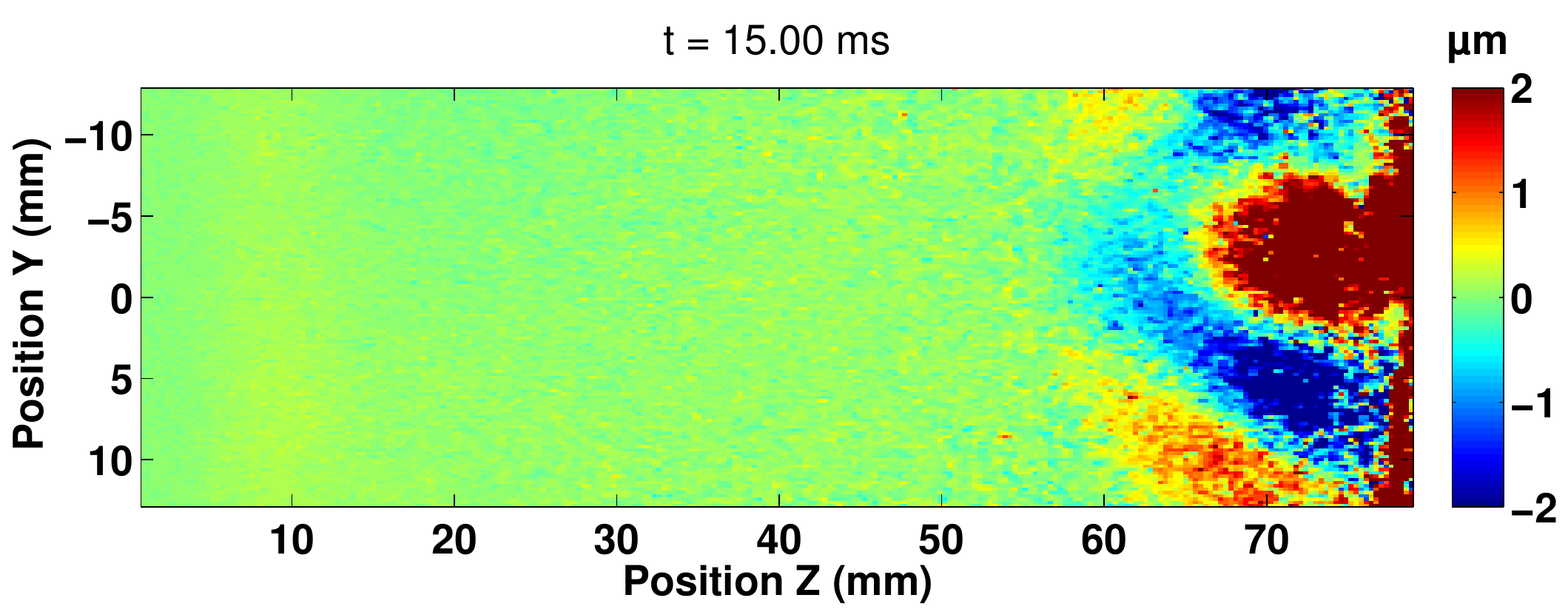}
		\end{center}
   \end{minipage}
   \begin{minipage}[c]{.5\linewidth}
		\begin{center}
	   		\includegraphics[width=1\linewidth]{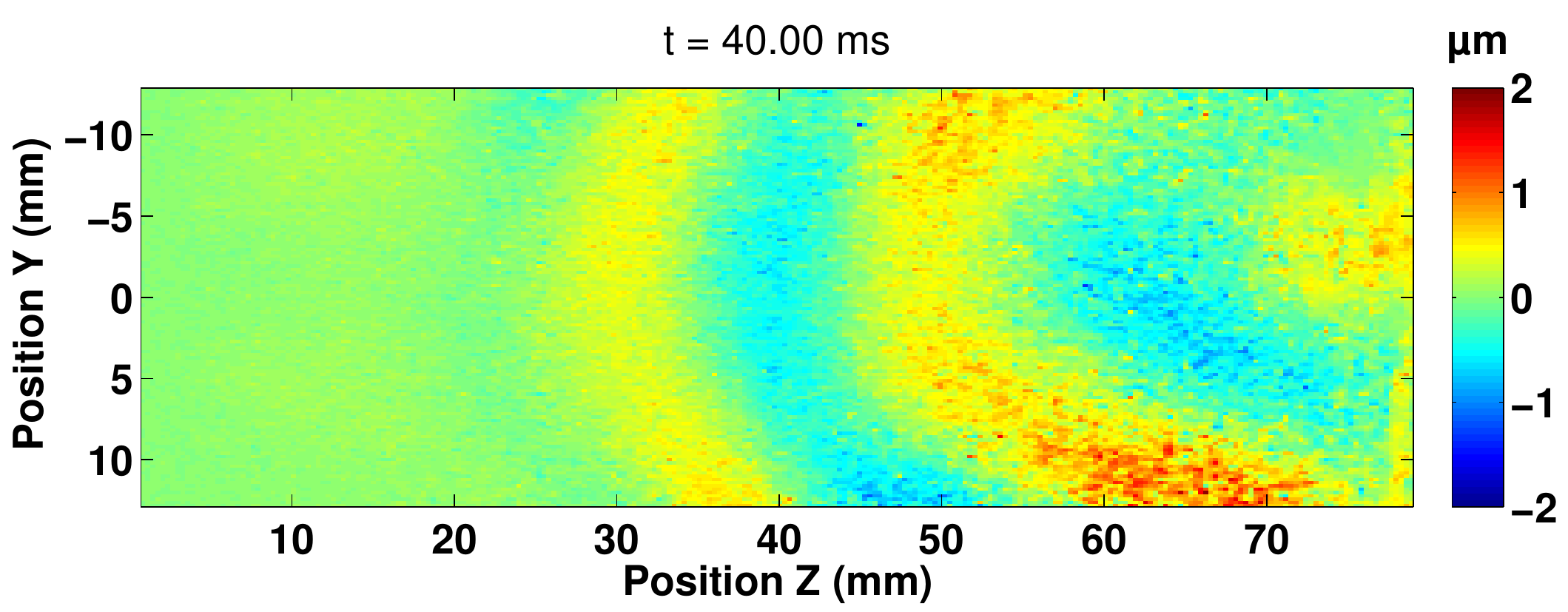}
		\end{center}
   \end{minipage} \hfill
   \begin{minipage}[c]{.5\linewidth}
		\begin{center}
	   		\includegraphics[width=1\linewidth]{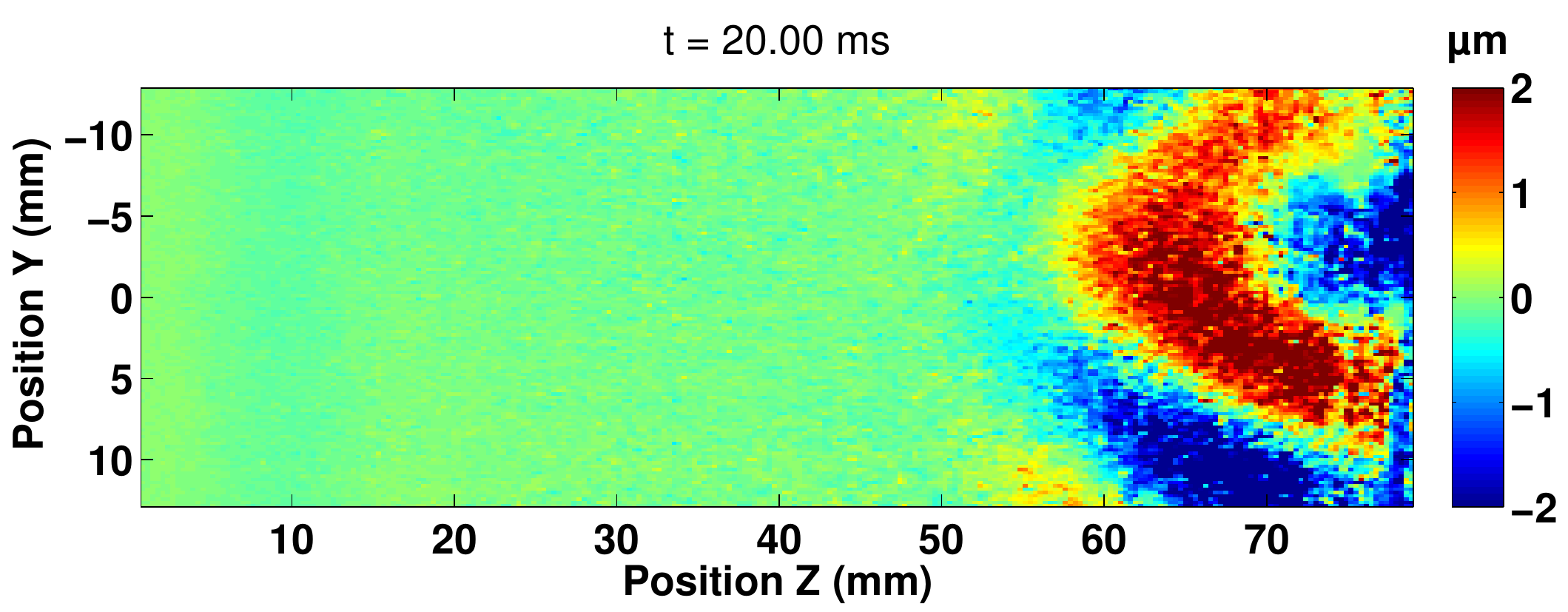}
		\end{center}
   \end{minipage}
   \begin{minipage}[c]{.5\linewidth}
		\begin{center}
	   		\includegraphics[width=1\linewidth]{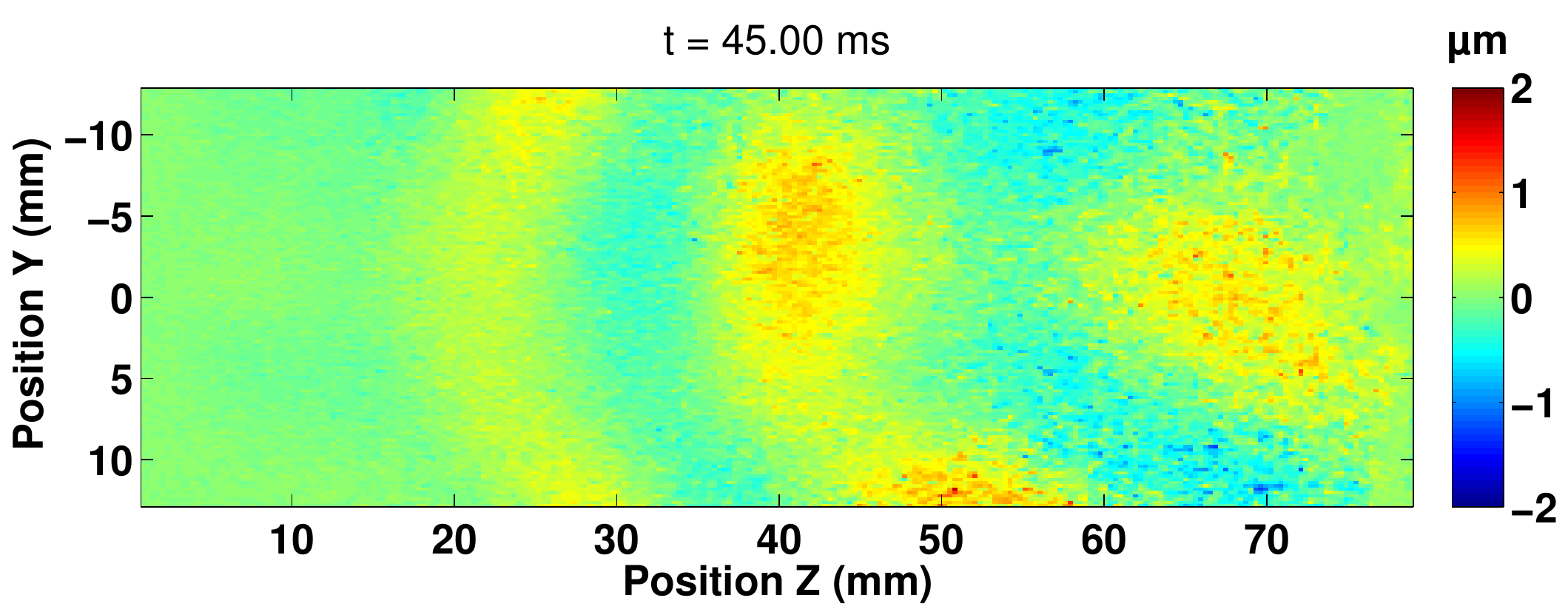}
		\end{center}
   \end{minipage}
 	\caption{\textbf{Résultats de l'expérience d'élastographie sur un fantôme homogène} -- Champ des déplacements dus à un courant électrique injecté par deux électrodes au fond du fantôme placé dans un champ magnétique, à 0, 5, 10, 15, 20, 25, 30, 35, 40 et 45 ms (plan YZ). On observe une propagation des ondes à une vitesse de 1,3 $\pm$ 0,1 m/s.}
 	\label{figElastoEMExpResElectrodeFond} 
\end{figure}

\begin{figure}[!ht]
 \begin{center}
	 \includegraphics[width=0.8\columnwidth]{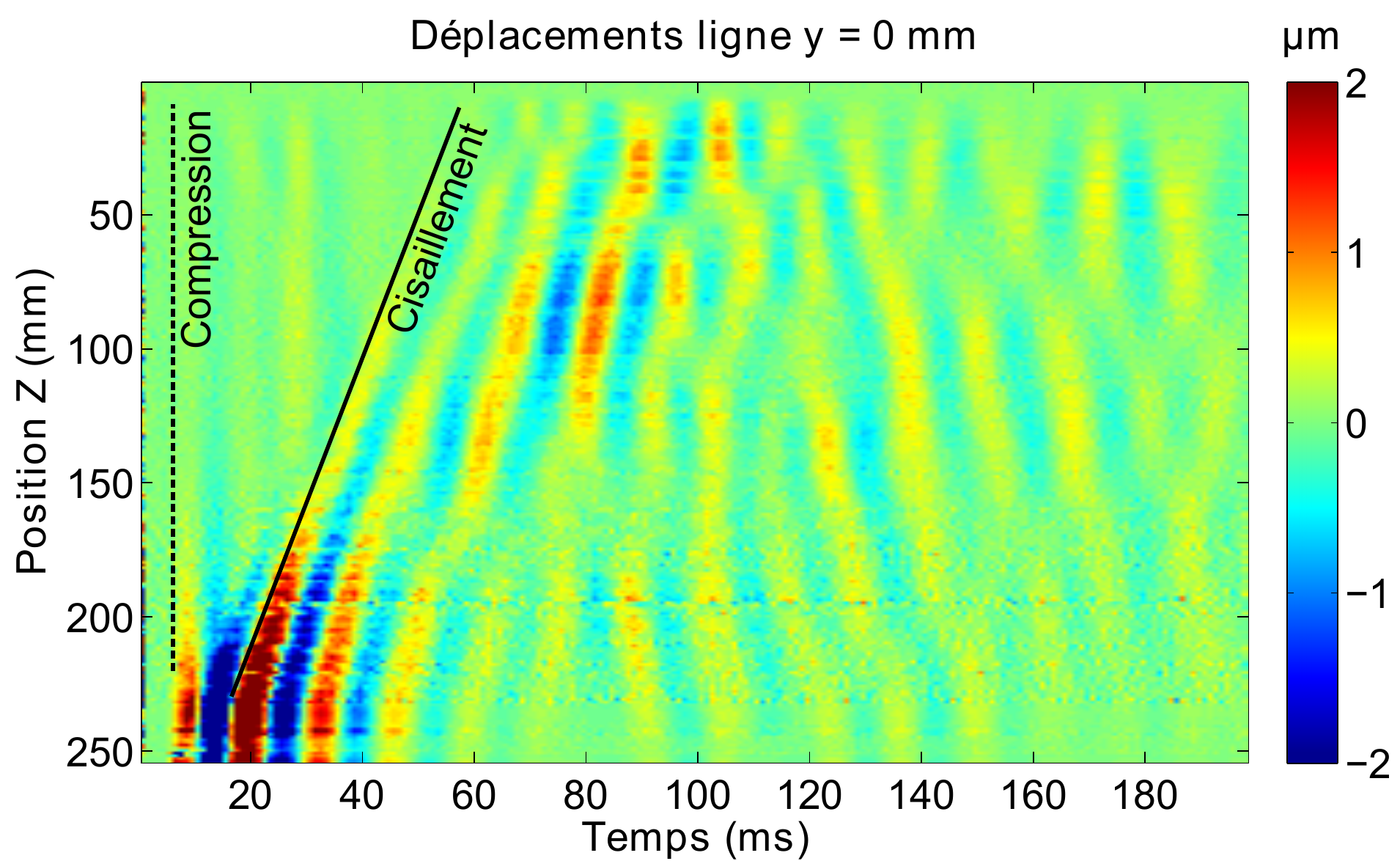}
	 \caption{\textbf{Résultats de l'expérience sur la vitesse des ondes} -- Déplacements locaux selon une ligne ultrasonore au cours du temps. On observe d'abord une vitesse de propagation très élevée (pointillés noirs), correspondant à la propagation d'une onde de compression. Puis une autre onde se propageant à 1,3 $\pm$ 0,1 m/s (ligne pleine noire), correspondant à une onde de cisaillement.}
	 \label{figElastoEMExpResElectrodeFondSismo}
 \end{center}
\end{figure}

\paragraph{Résultats et discussions}
La figure \ref{figElastoEMExpResElectrodeFond} représente des champs de déplacement au cours du temps à 0, 5, 10, 15, 20, 25, 30, 35, 40 et 45 millisecondes après le début du passage du courant. On observe un mouvement entre 0 et 10 ms qui correspond au courant électrique émis, puis à la propagation de ce mouvement au cours du temps. Une mesure le long de la ligne centrale en fonction du temps, représentée sur la figure \ref{figElastoEMExpResElectrodeFondSismo}, donne une vitesse de propagation des ondes de 1,3 $\pm$ 0,1 m/s. Cette vitesse est compatible avec la vitesse des ondes de cisaillement, entre 1 et 10 m.s$^{-1}$ pour un tissu mou. Avec la formule $v_s = \sqrt{\mu/\rho}$, cela correspond pour une densité $\rho$ de 1000 kg.m$^{-3}$ à une élasticité $E$ de 5,0 $\pm$ 0,8 kPa.

L'onde de cisaillement créée est similaire au rayonnement d'une source ponctuelle de surface en champ proche, d'où la présence d'ondes de cisaillement polarisées longitudinalement.

\subsection{Expérience sur un fantôme bicouche}

\begin{figure}[!ht]
\begin{minipage}[c]{.5\linewidth}
 \begin{center}
	 \includegraphics[height=2.8cm]{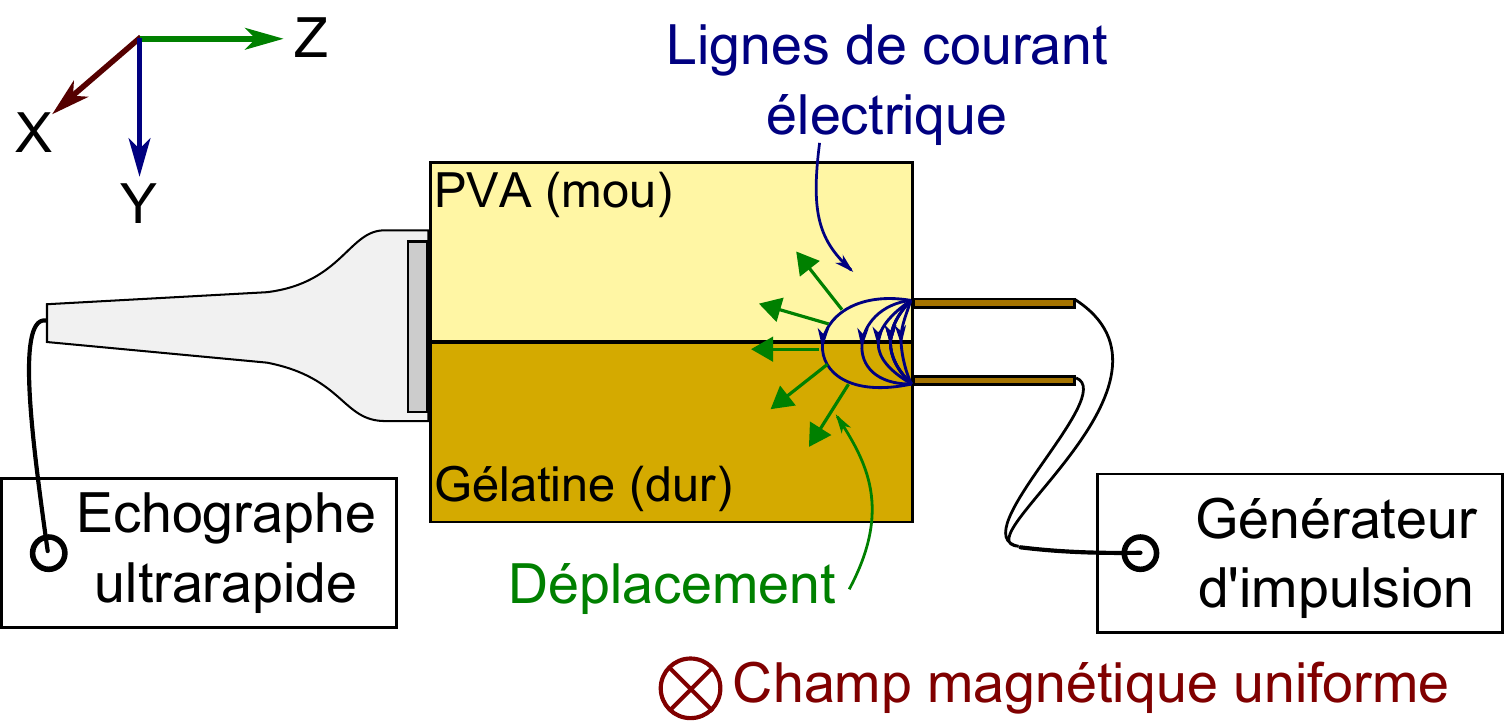}
 \end{center}
\end{minipage}
\begin{minipage}[c]{.5\linewidth}
	 \includegraphics[height=2.8cm]{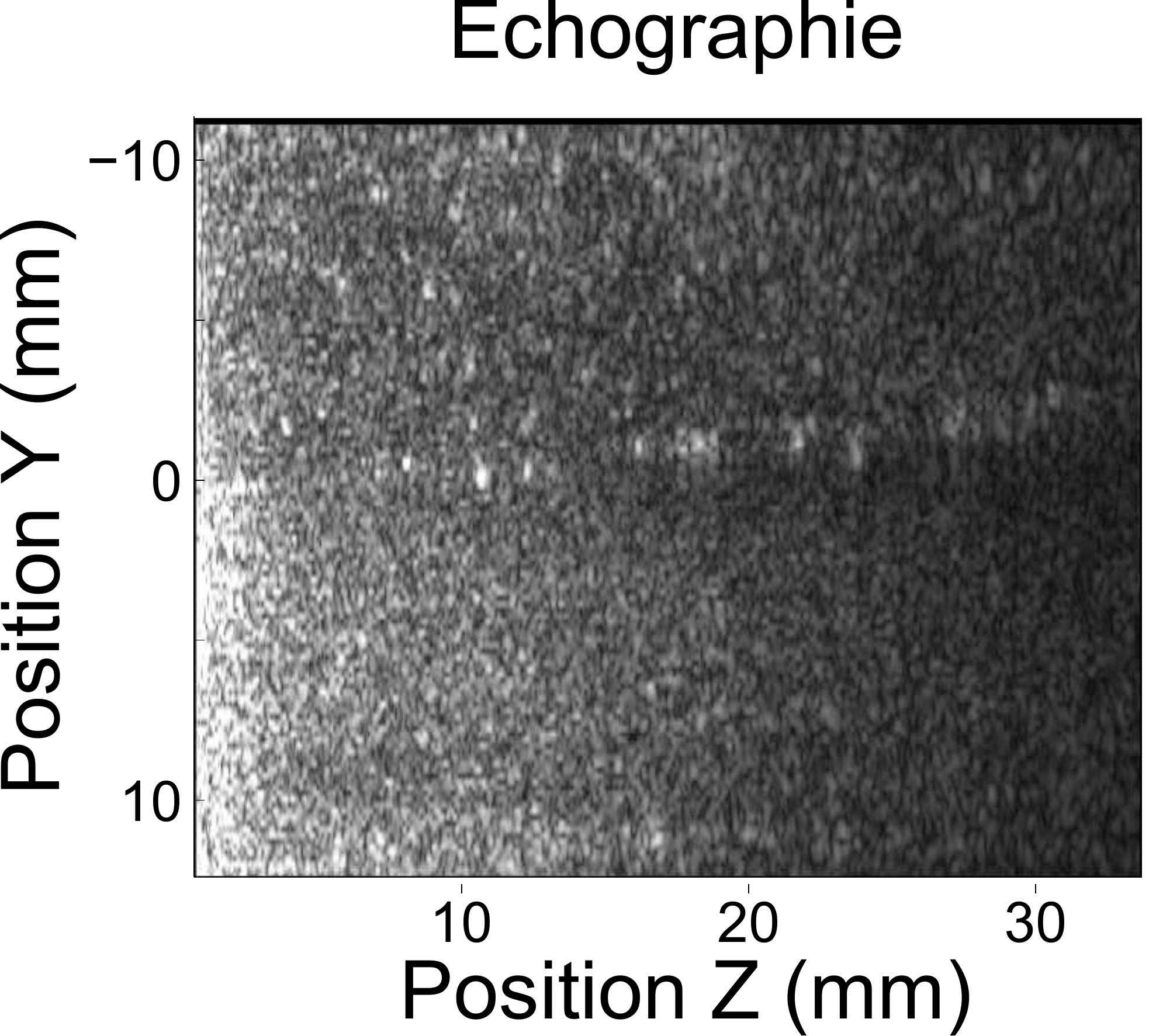}
\end{minipage}
	 \caption{(a) \textbf{Schéma de l'expérience d'élastographie sur un fantôme bicouche} -- Deux électrodes proches créent un courant électrique dans un fantôme soumis à un champ magnétique, et les déplacements sont observés par une sonde ultrasonore en mode ultrarapide. Le fantôme présente deux couches d'élasticité différentes, une plus dure sur le dessus et une plus molle en-dessous. (b) \textbf{Echographie obtenue par la sonde ultrasonore} -- La ligne plus claire horizontale au centre de l'échographie correspondant à la séparation entre les deux couches, mais est à peine visible.}
	\label{figElastoEMExpMatBicoucheElasto}
\end{figure}

\paragraph{Matériel et méthodes}
Une expérience a été faite avec deux électrodes en contact avec le fond du fantôme, le long de la tranche. La barrette ultrasonore était placée de l'autre côté, comme représenté sur le schéma \ref{figElastoEMExpMatBicoucheElasto}-(a). Cette fois-ci, le fantôme était composé de deux parties d'élasticité différente. La couche supérieure, plus dure, était faite en PVA, tandis que la couche inférieure, plus molle, était faite en gélatine. Le fantôme ne touchait pas les parois, car comme vu lors de la toute première expérience, le contact crée des ondes de cisaillement. La cadence d'acquisition était quatre fois plus élevée (4000 Hz au lieu de 1000 Hz) afin d'avoir une meilleure résolution temporelle, mais en contrepartie, on observe des déplacements quatre fois plus faibles entre chaque image.

L'échographie du milieu est représentée figure \ref{figElastoEMExpMatBicoucheElasto}-(b), où l'on distingue l'interface entre les deux couches par la ligne plus claire au centre. L'interface du fond n'est pas visible.

\begin{figure}[!ht]
   \begin{minipage}[c]{.5\linewidth}
		\begin{center}
	   		\includegraphics[width=.75\linewidth]{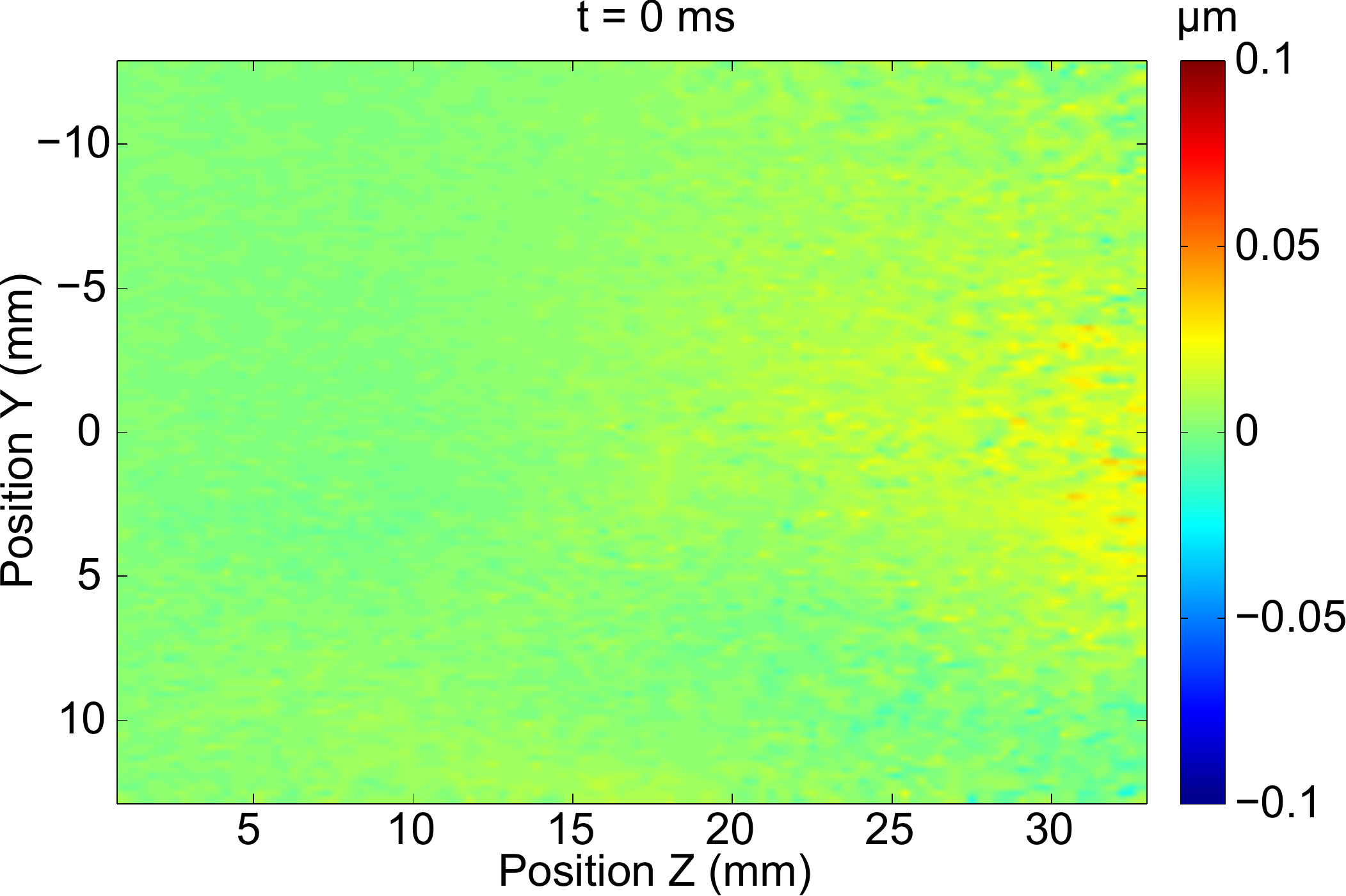}
		\end{center}
   \end{minipage} \hfill
   \begin{minipage}[c]{.5\linewidth}
		\begin{center}
	   		\includegraphics[width=.75\linewidth]{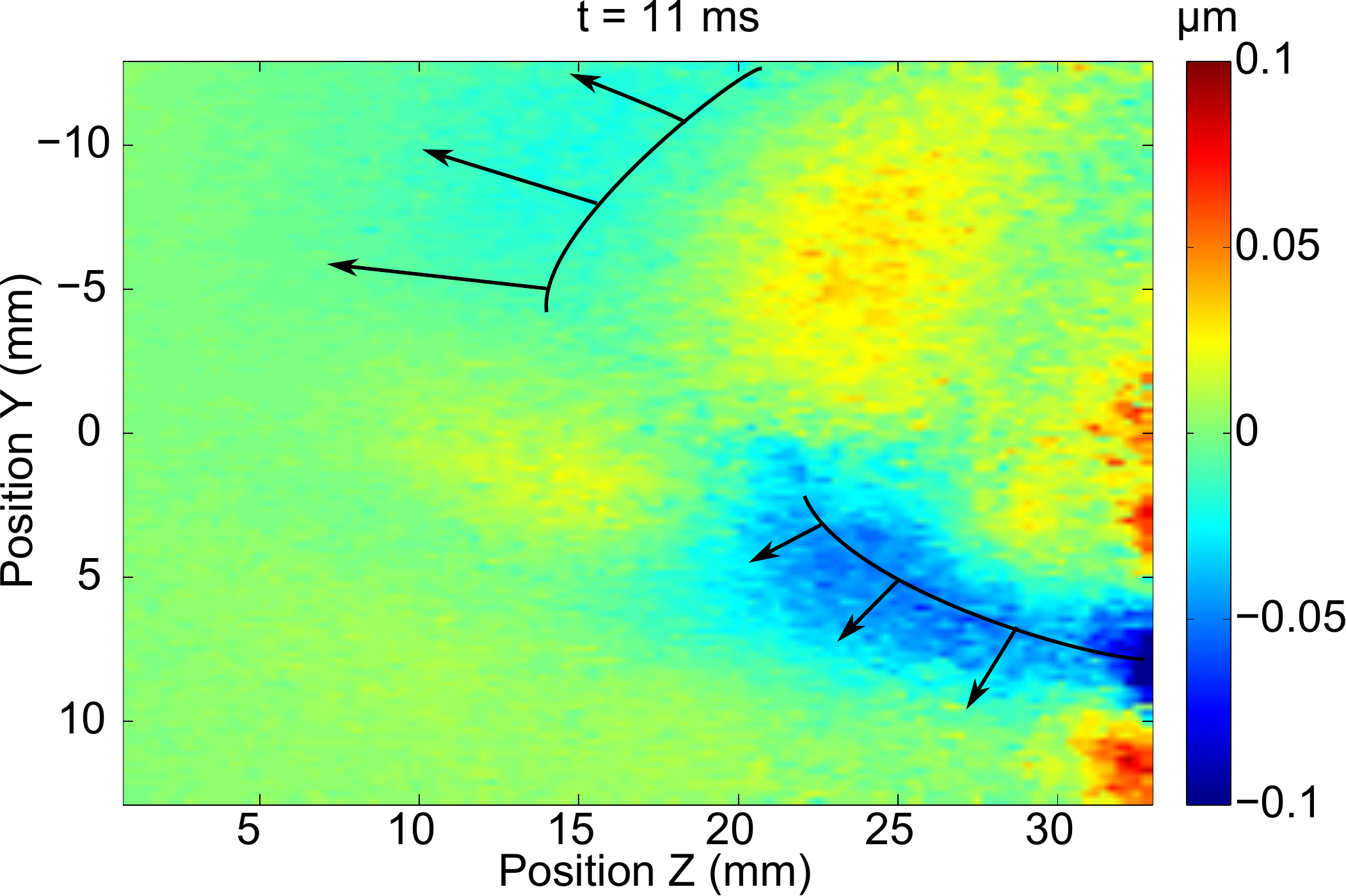}
		\end{center}
   \end{minipage}
   \begin{minipage}[c]{.5\linewidth}
		\begin{center}
	   		\includegraphics[width=.75\linewidth]{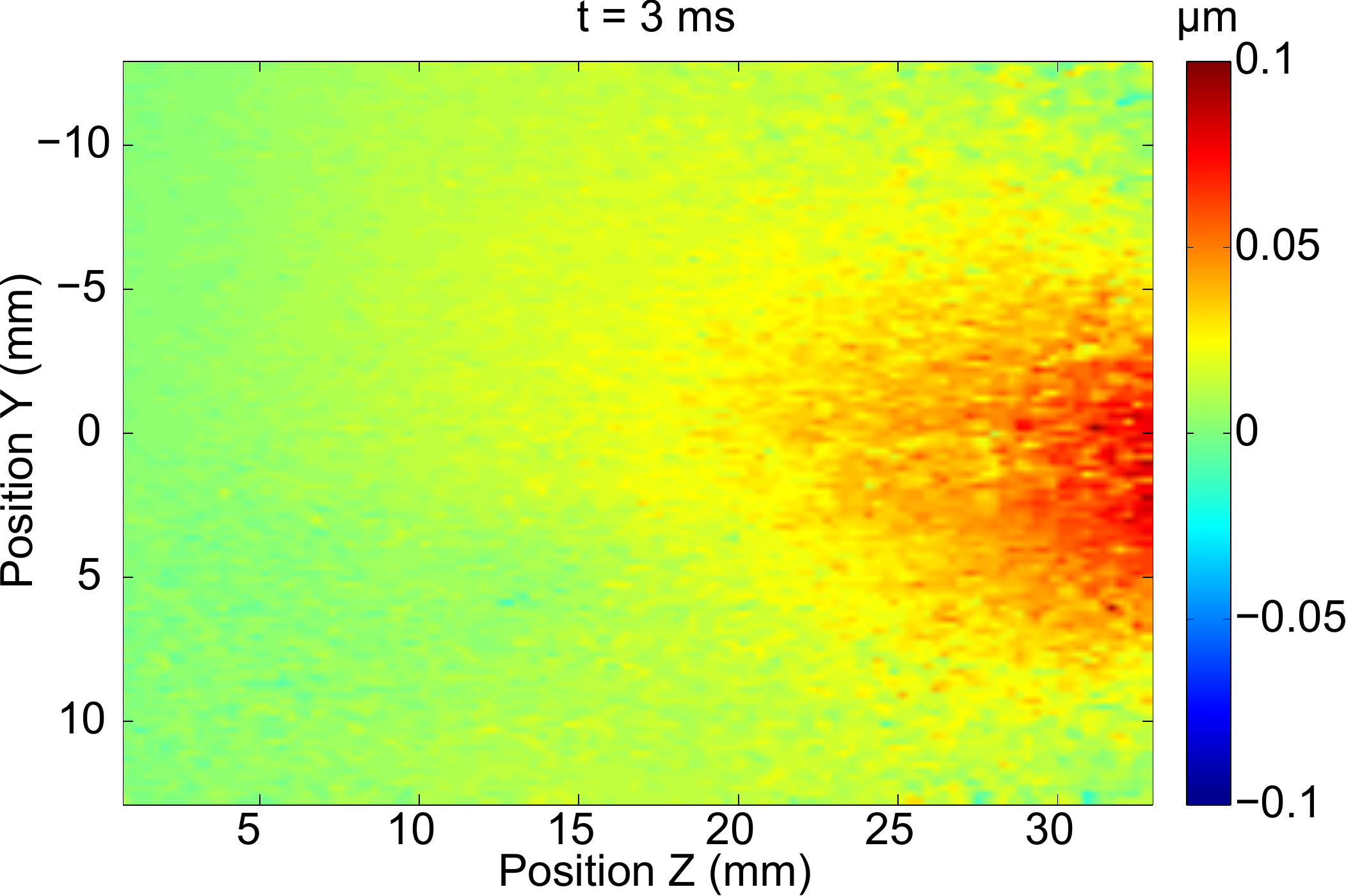}
		\end{center}
   \end{minipage} \hfill
   \begin{minipage}[c]{.5\linewidth}
		\begin{center}
	   		\includegraphics[width=.75\linewidth]{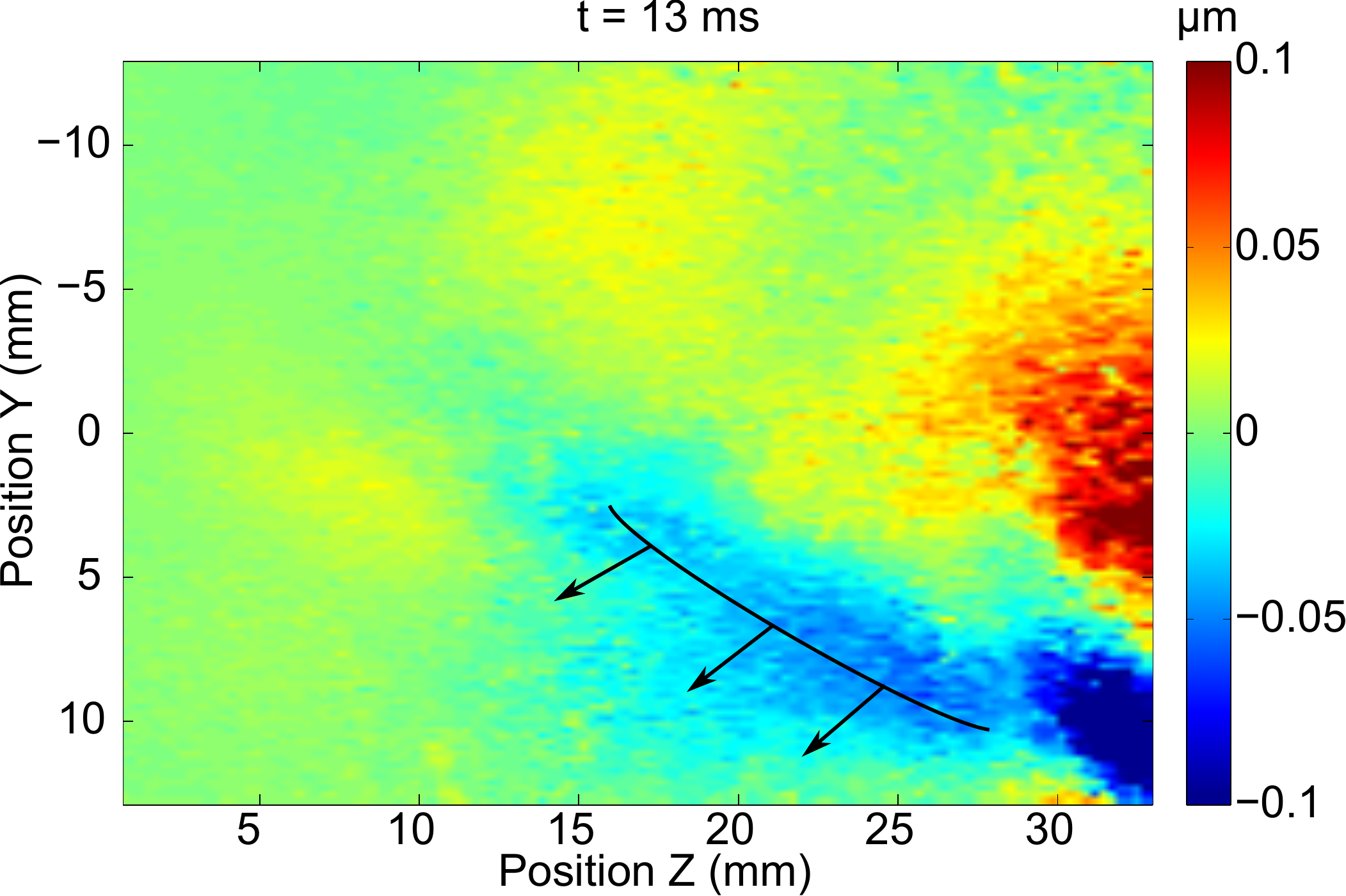}
		\end{center}
   \end{minipage}
   \begin{minipage}[c]{.5\linewidth}
		\begin{center}
	   		\includegraphics[width=.75\linewidth]{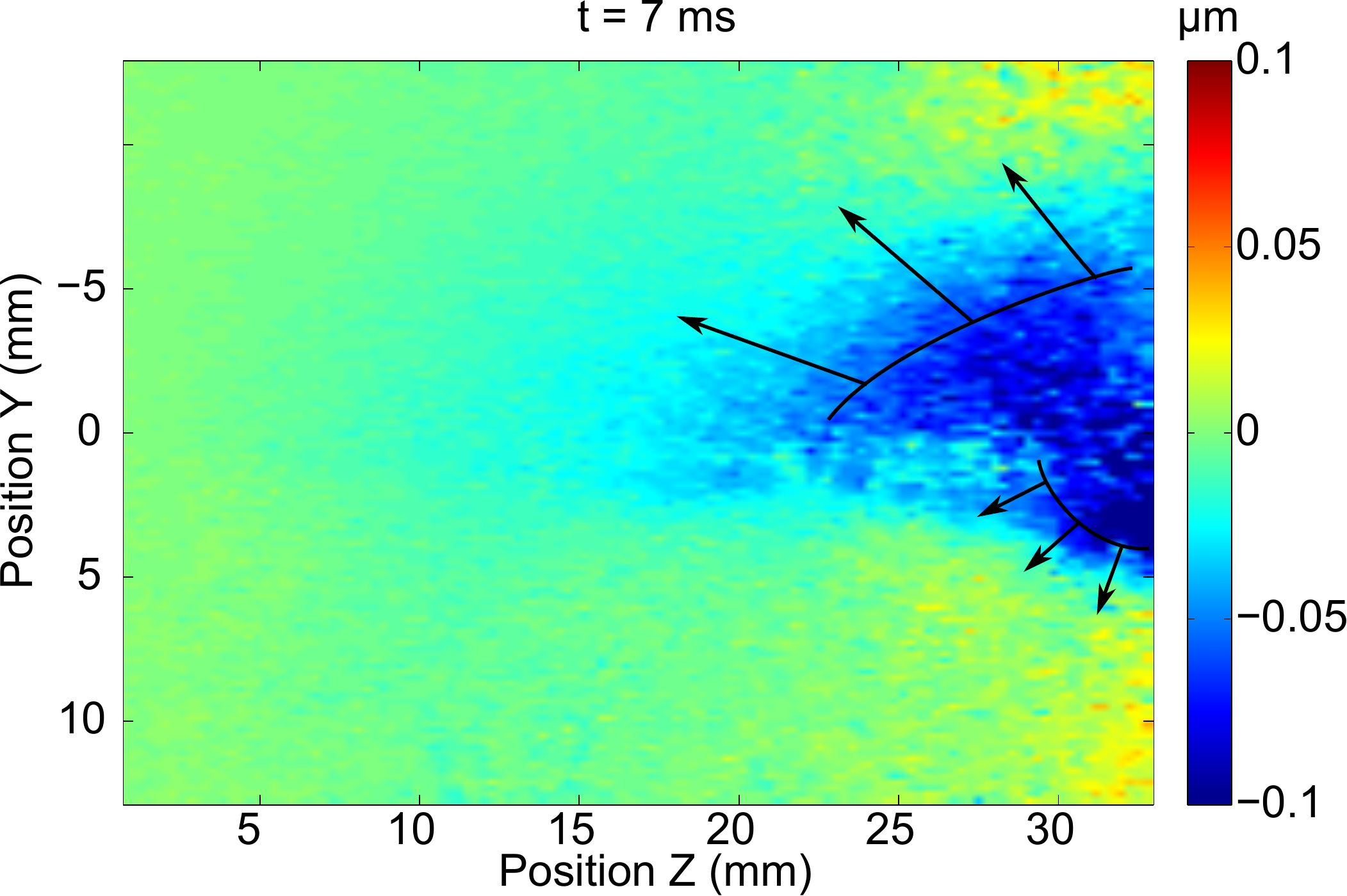}
		\end{center}
   \end{minipage} \hfill
   \begin{minipage}[c]{.5\linewidth}
		\begin{center}
	   		\includegraphics[width=.75\linewidth]{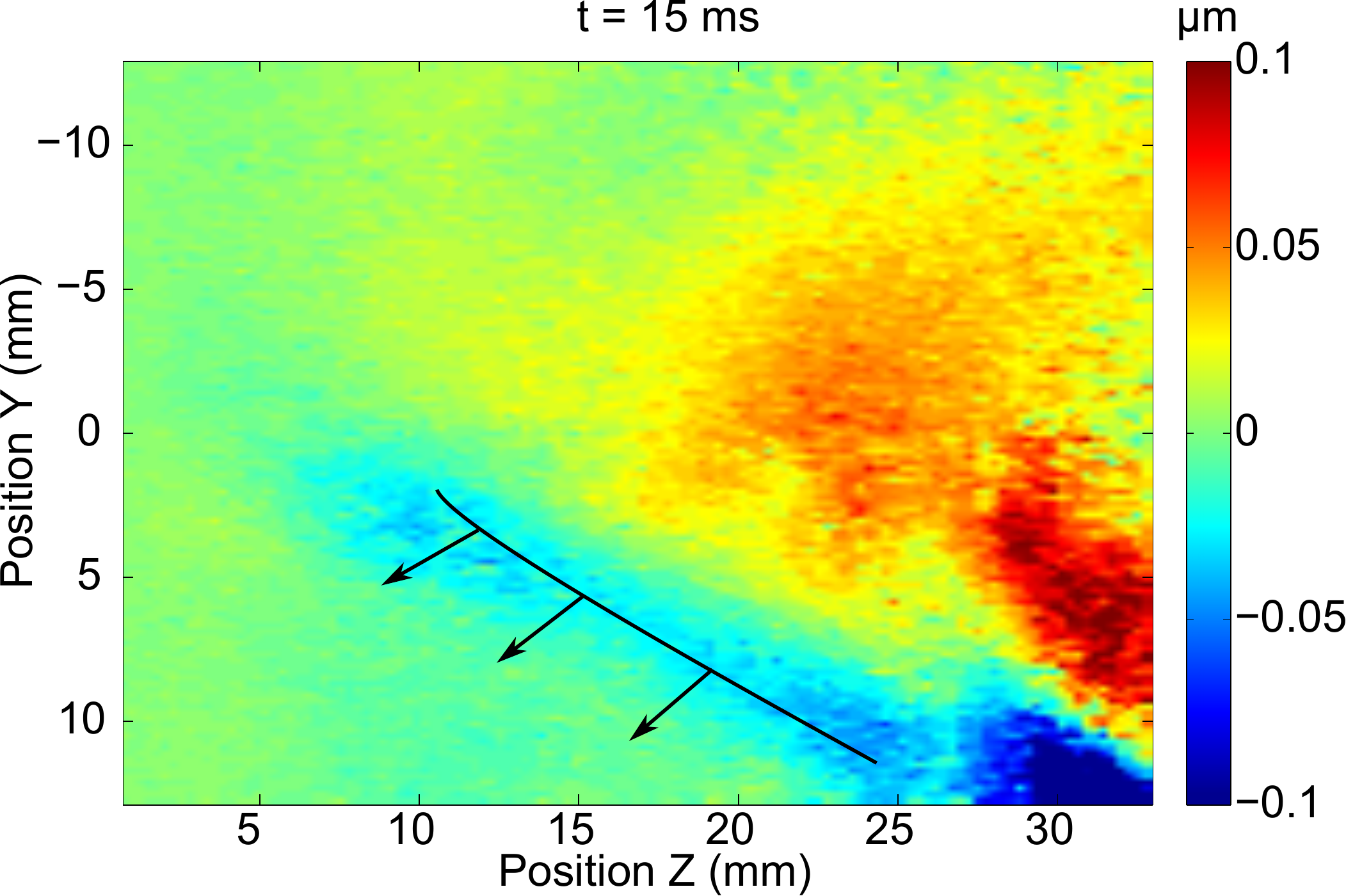}
		\end{center}
   \end{minipage}
   \begin{minipage}[c]{.5\linewidth}
		\begin{center}
	   		\includegraphics[width=.75\linewidth]{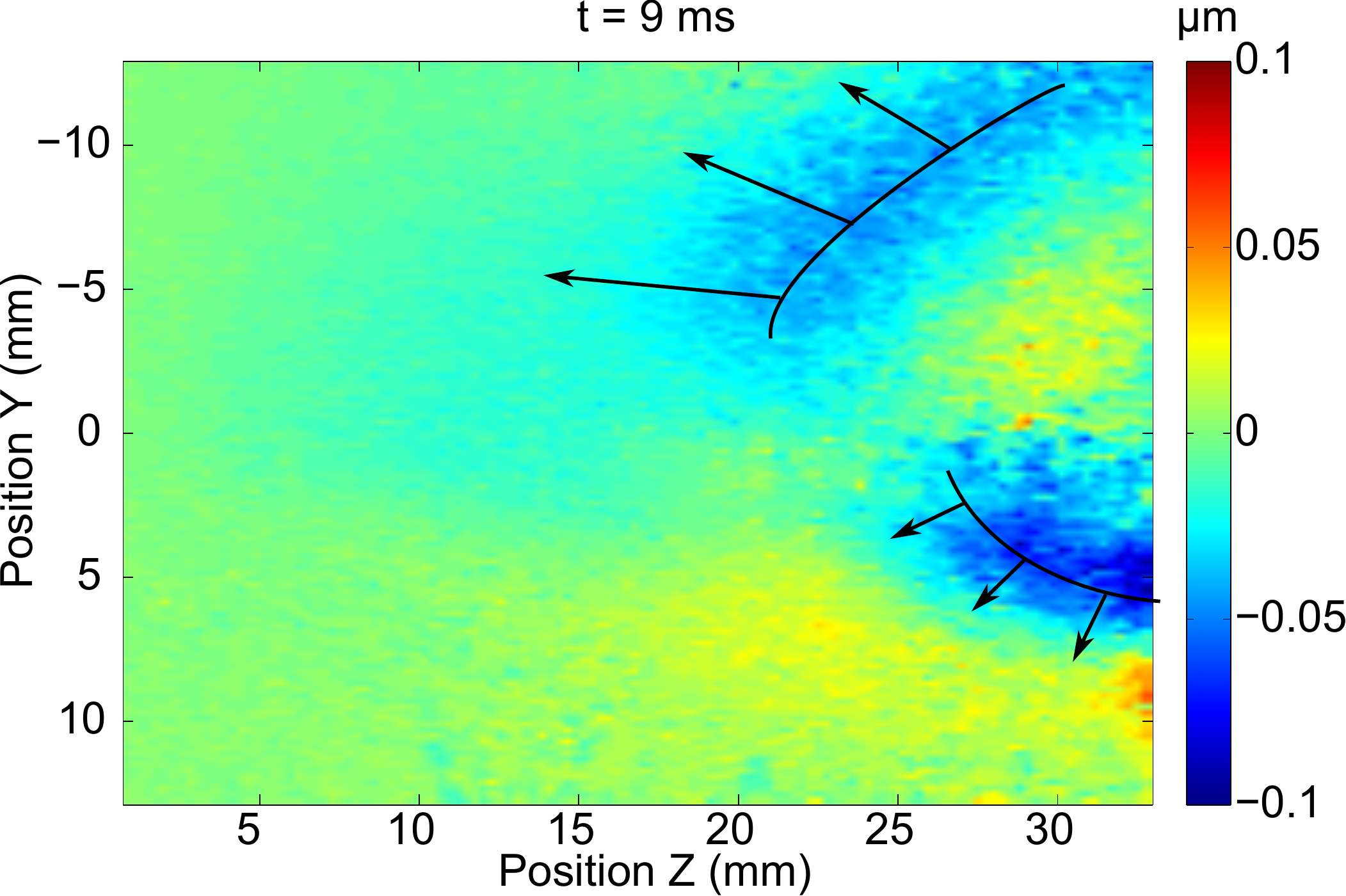}
		\end{center}
   \end{minipage} \hfill
   \begin{minipage}[c]{.5\linewidth}
		\begin{center}
	   		\includegraphics[width=.75\linewidth]{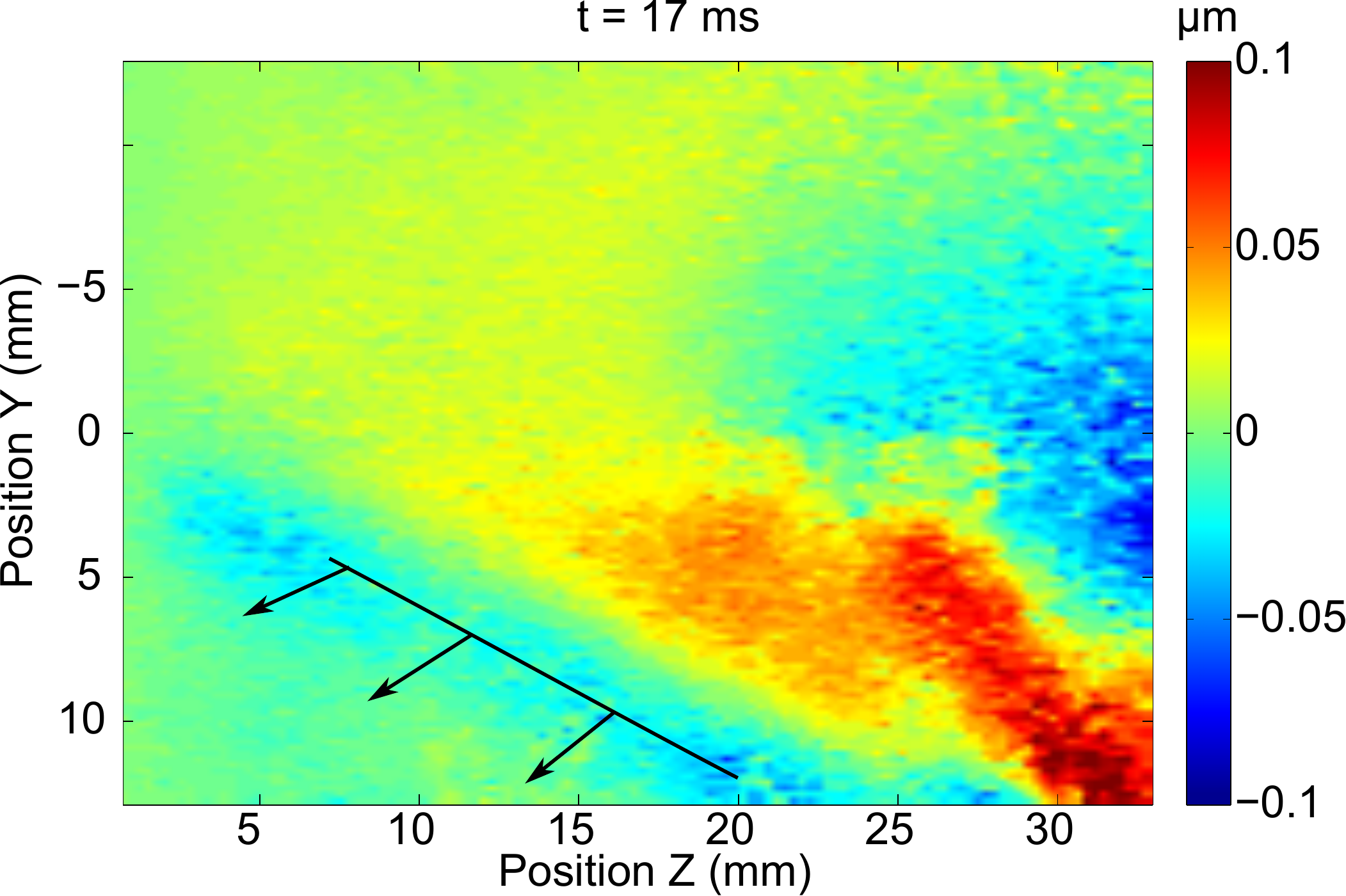}
		\end{center}
   \end{minipage}
 	\caption{\textbf{Résultats de l'expérience d'élastographie sur un fantôme bicouche} -- Champ des déplacements dus à un courant électrique injecté par deux électrodes au fond du fantôme placé dans un champ magnétique, 0, 3, 7, 9, 11, 13, 15 et 17 ms (plan YZ). Une onde de compression se propageant rapidement et sans différence entre les deux parties apparait à t = 3 ms. A 7, 9 et 11 ms, on voit l'onde en bleu se propager à une vitesse plus élevée dans la partie supérieure que dans la partie inférieure, de même avec l'onde rouge à 13, 15 et 17 ms. Les ondes de cisaillement se propagent donc à des vitesses différentes dans la partie supérieure et dans la partie inférieure, respectivement à 5,1 $\pm$ 0,1 et 2,5 $\pm$ 0,1 m.s$^{-1}$.}
 	\label{figElastoEMExpResBicoucheElasto} 
\end{figure}

\paragraph{Résultats et discussions}
La figure \ref{figElastoEMExpResBicoucheElasto} représente huit champs de déplacements au cours du temps à 0, 3, 7, 9, 11, 13, 15 et 17 millisecondes après le début du passage du courant. L'onde initiale visible à t = 3 ms est une onde de compression. A 7, 9 et 11 ms, une onde de cisaillement en bleu apparait et se propage à deux vitesses différentes dans la partie supérieure et dans la partie inférieure. De même, l'onde de cisaillement en rouge visible à 13, 15 et 17 ms se propage à des vitesses différentes entre les deux zones. La vitesse des ondes de cisaillement dans les parties supérieure et inférieure vaut respectivement 5,1 $\pm$ 0,1 et 2,5 $\pm$ 0,1 m.s$^{-1}$, ce qui correspond à des élasticités $E_1$ = 78 $\pm$ 3 kPa et $E_2$ = 19 $\pm$ 2 kPa.

\subsection{Expérience sur un morceau de foie}

\begin{figure}[!ht]
\begin{minipage}[c]{.5\linewidth}
 \begin{center}
	 \includegraphics[height=2.8cm]{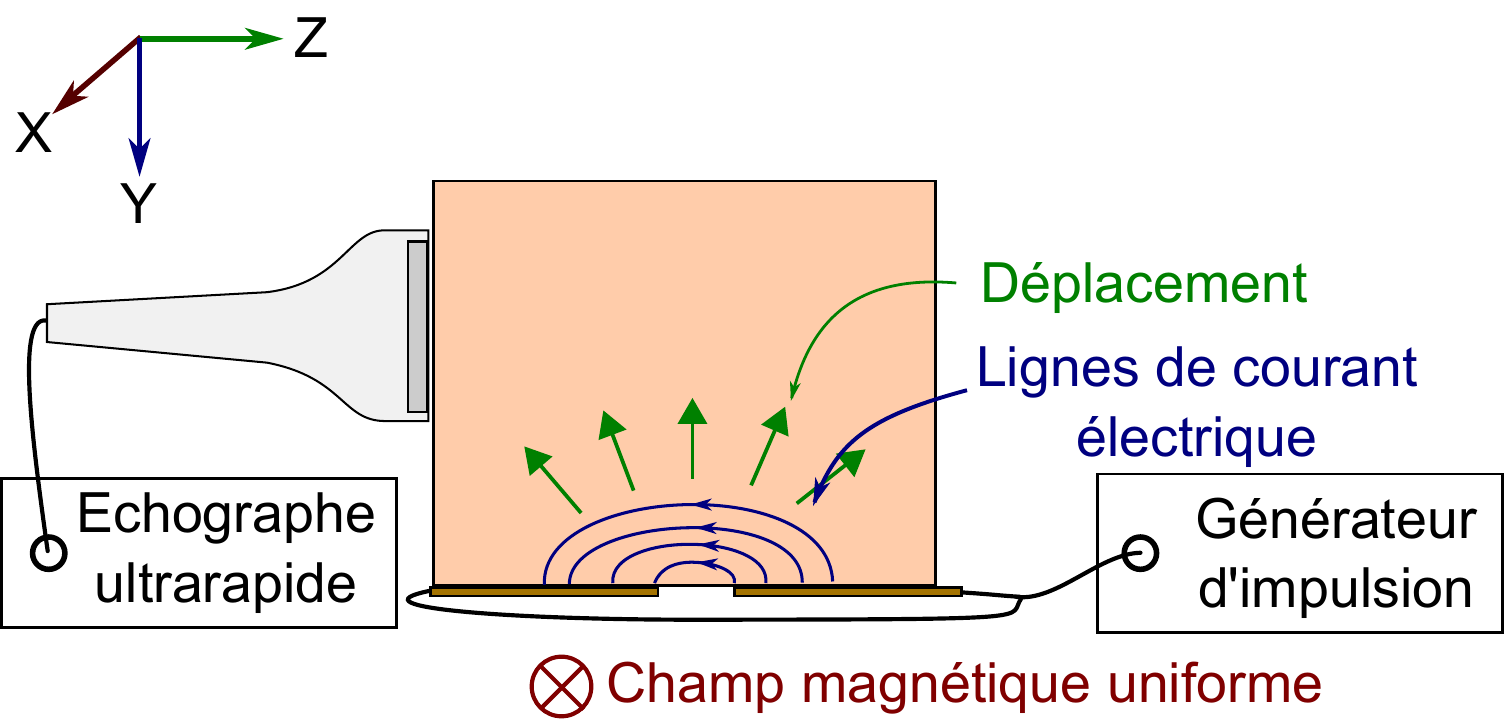}
 \end{center}
\end{minipage}
\begin{minipage}[c]{.5\linewidth}
	 \includegraphics[height=2.8cm]{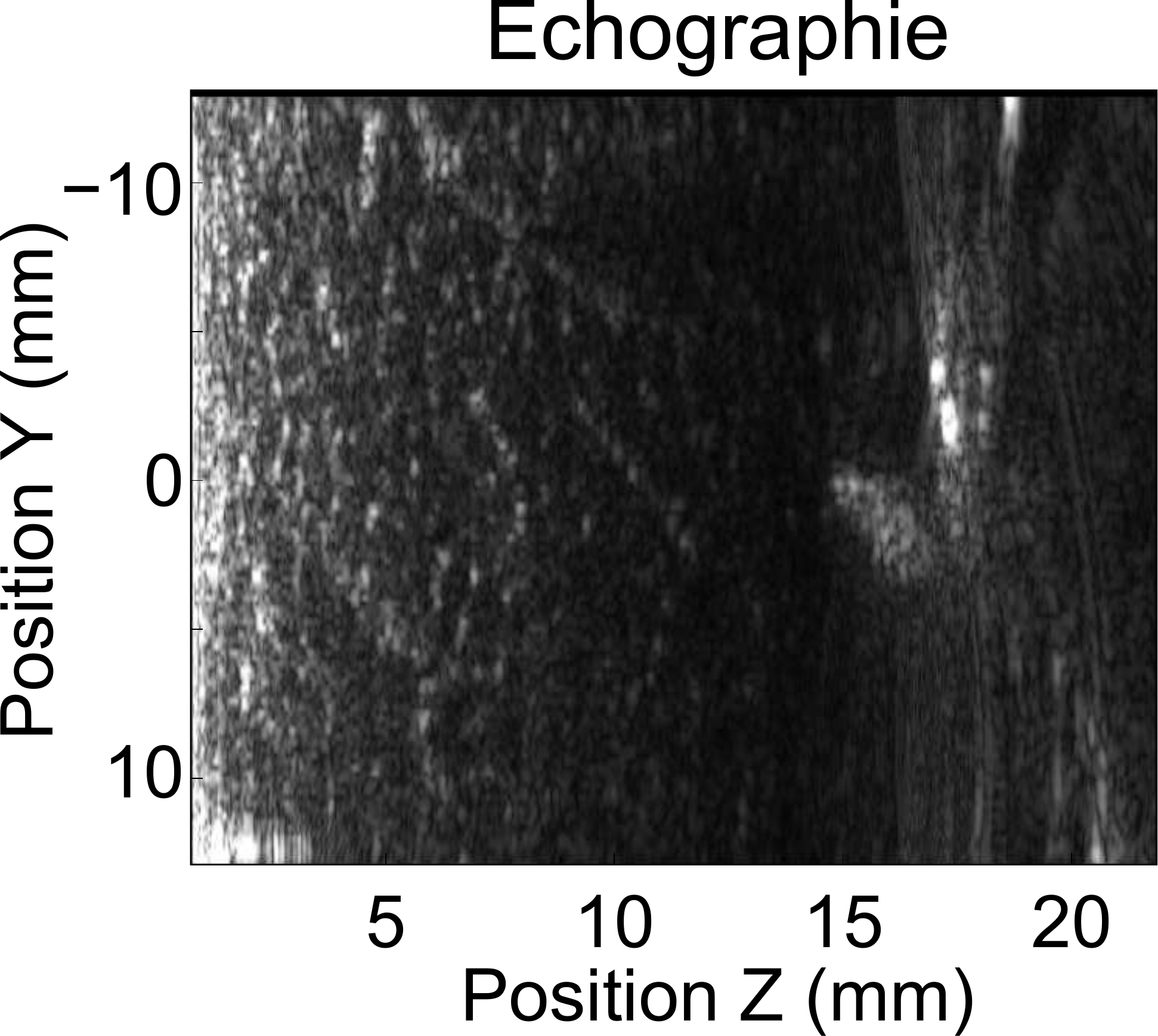}
\end{minipage}
	 \caption{(a) \textbf{Schéma de l'expérience d'élastographie sur un échantillon de foie} -- Deux électrodes proches créent un courant électrique dans un échantillon de foie soumis à un champ magnétique, et les déplacements sont observés par une sonde ultrasonore en mode ultrarapide placé sur le côté. (b) \textbf{Echographie obtenue par la sonde ultrasonore} -- On observe de nombreux points clairs et une atténuation importante.}
	\label{figElastoEMExpMatFoie}
\end{figure}

\paragraph{Matériel et méthodes}
Un foie congelé de porc abattu dans un bloc opératoire a été récupéré. Ce foie a été dégelé dans une solution physiologique, composée d'eau distillée et de chlorure de sodium à 9\% pour avoir une conductivité électrique proche de celle des tissus mous. Un morceau de foie de 3x3x3 cm$^3$ a ensuite été découpé au scalpel et posé au-dessus de deux électrodes, avec la barrette ultrasonore sur le côté, comme représenté sur le schéma \ref{figElastoEMExpMatFoie}-(a). Les configurations présentées dans les expériences précédentes ne pouvaient en effet pas être utilisées, car l'échantillon de foie avait une atténuation beaucoup plus forte, donc on ne pouvait pas vraiment faire d'images de profondeur supérieure à 2 cm, comme on peut le voir sur l'échographie figure \ref{figElastoEMExpMatFoie}-(b). Cette configuration permettait donc d'observer la propagation d'une onde de cisaillement dans une direction Y.

\begin{figure}[!ht]
   \begin{minipage}[c]{.32\linewidth}
		\begin{center}
	   		\includegraphics[width=1\linewidth]{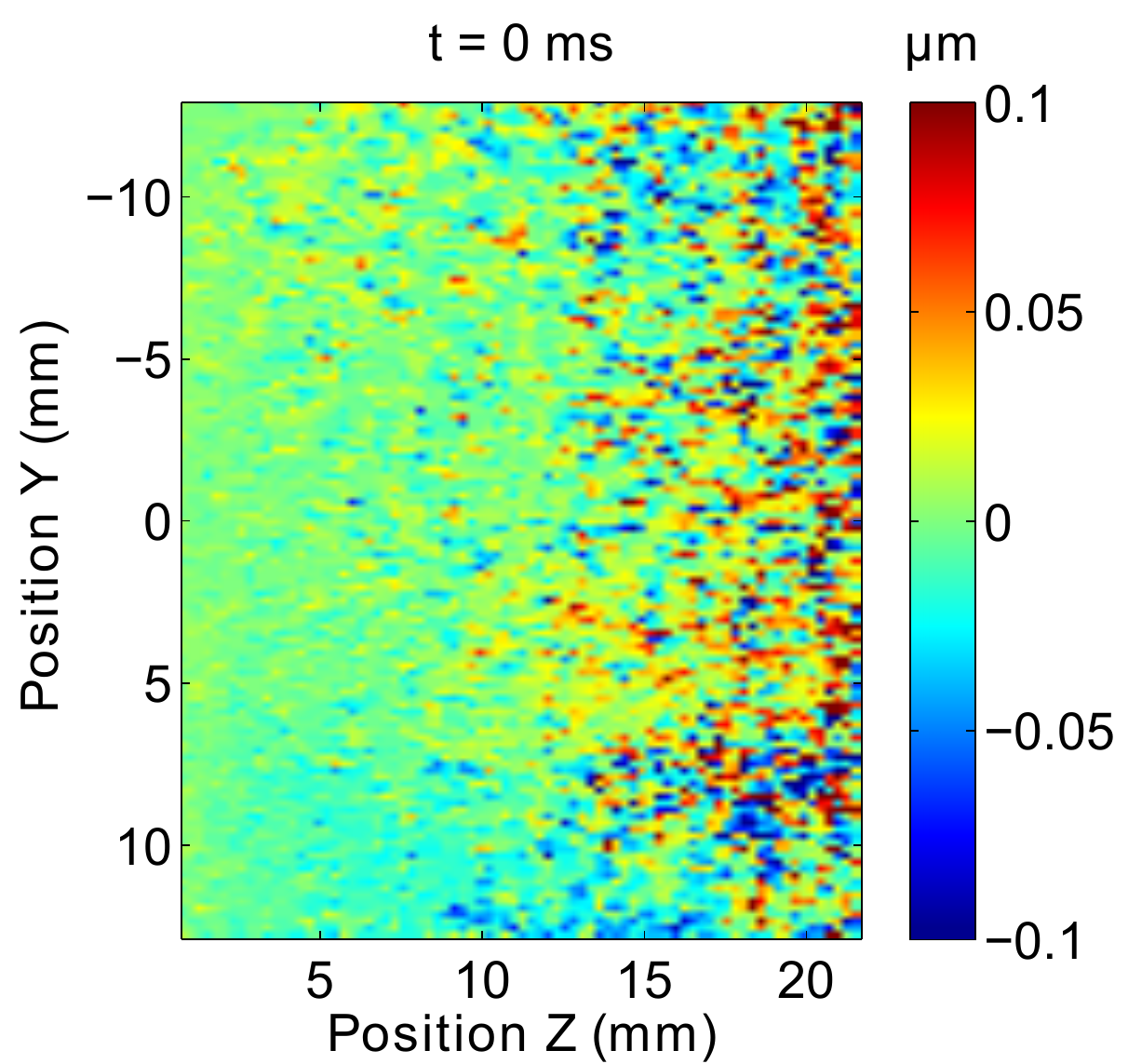}
		\end{center}
   \end{minipage} \hfill
   \begin{minipage}[c]{.32\linewidth}
		\begin{center}
	   		\includegraphics[width=1\linewidth]{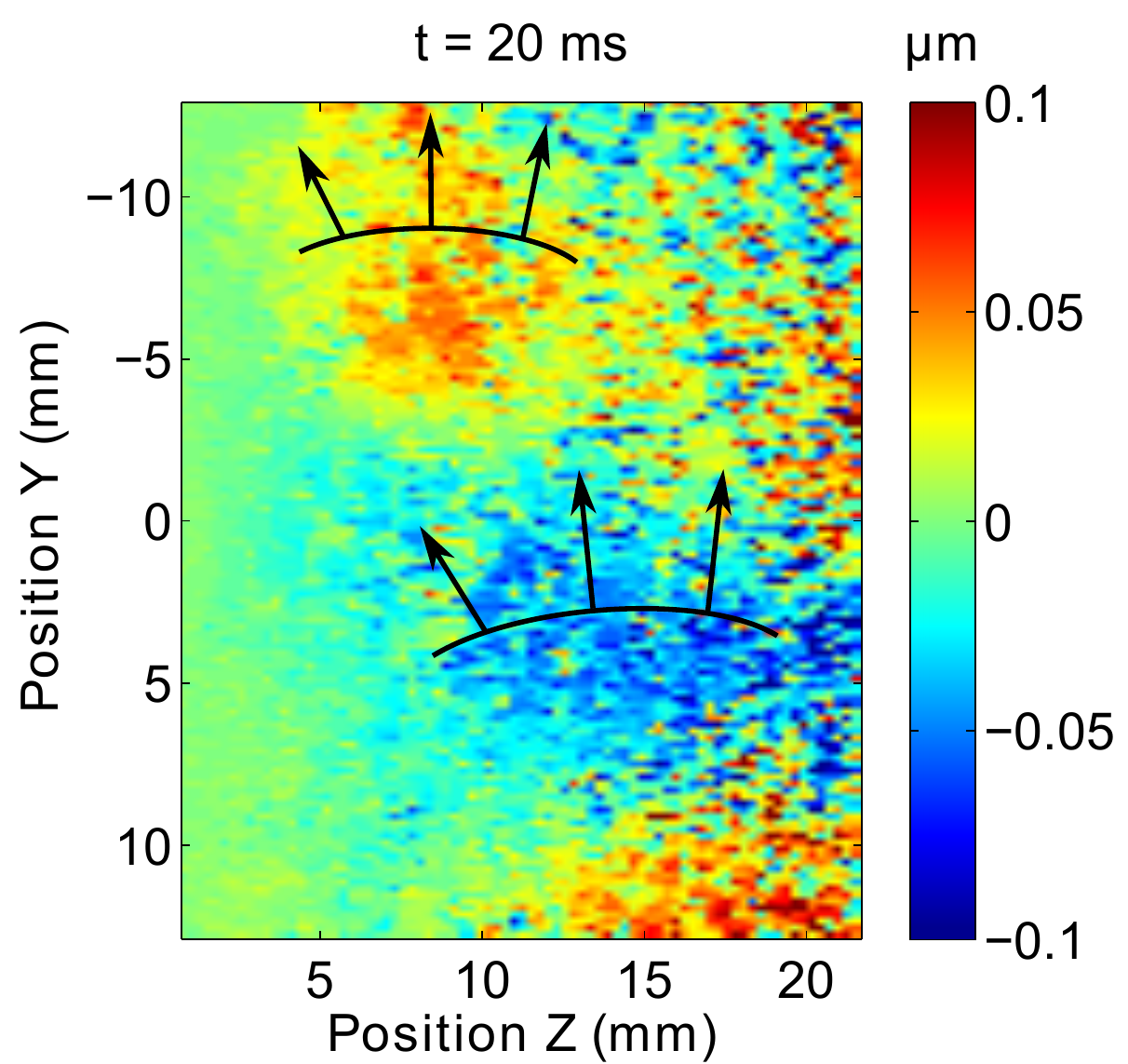}
		\end{center}
   \end{minipage}
   \begin{minipage}[c]{.32\linewidth}
		\begin{center}
	   		\includegraphics[width=1\linewidth]{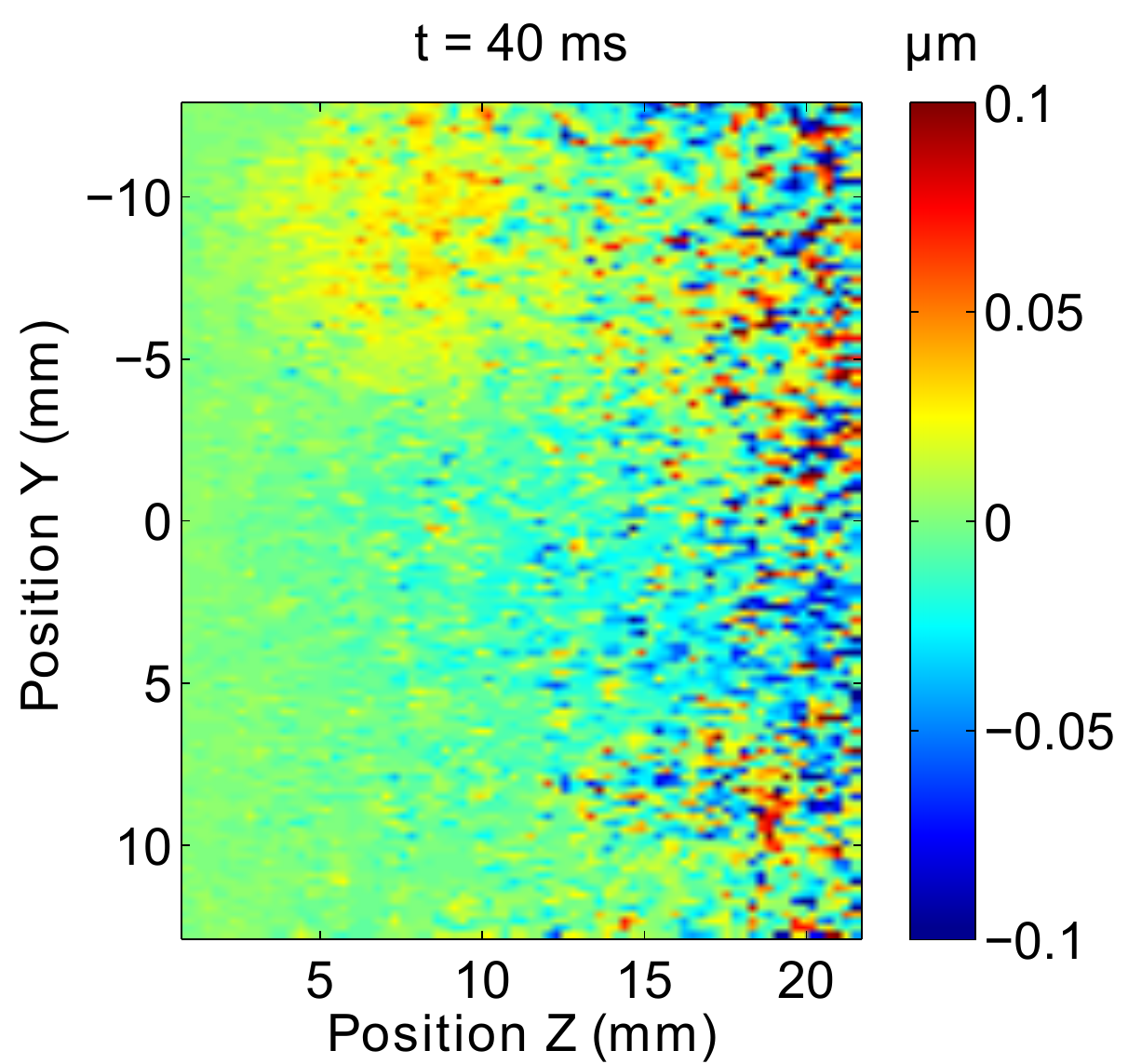}
		\end{center}
   \end{minipage} \hfill
   \begin{minipage}[c]{.32\linewidth}
		\begin{center}
	   		\includegraphics[width=1\linewidth]{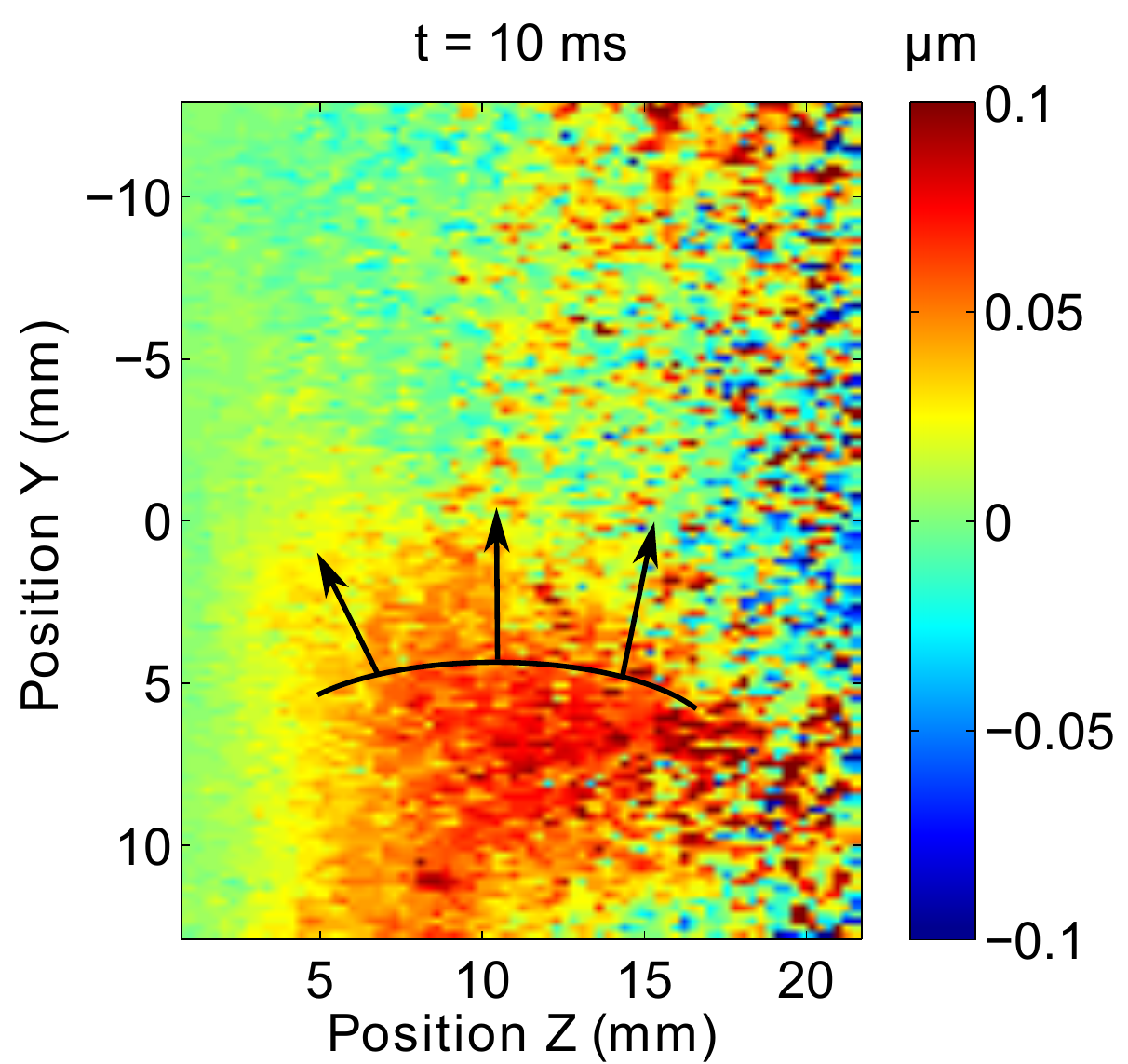}
		\end{center}
   \end{minipage}
   \begin{minipage}[c]{.32\linewidth}
		\begin{center}
	   		\includegraphics[width=1\linewidth]{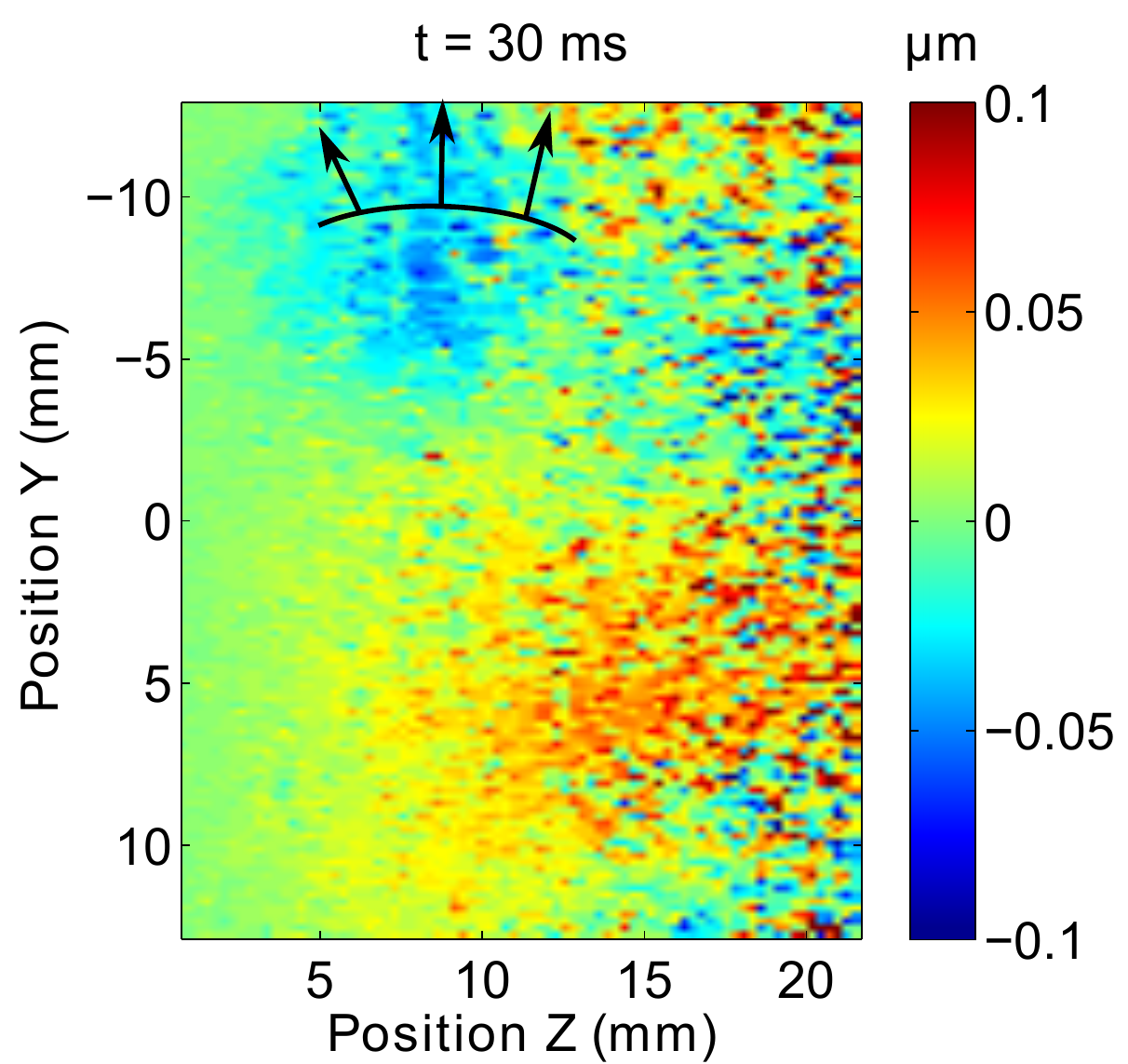}
		\end{center}
   \end{minipage} \hfill
   \begin{minipage}[c]{.32\linewidth}
		\begin{center}
	   		\includegraphics[width=1\linewidth]{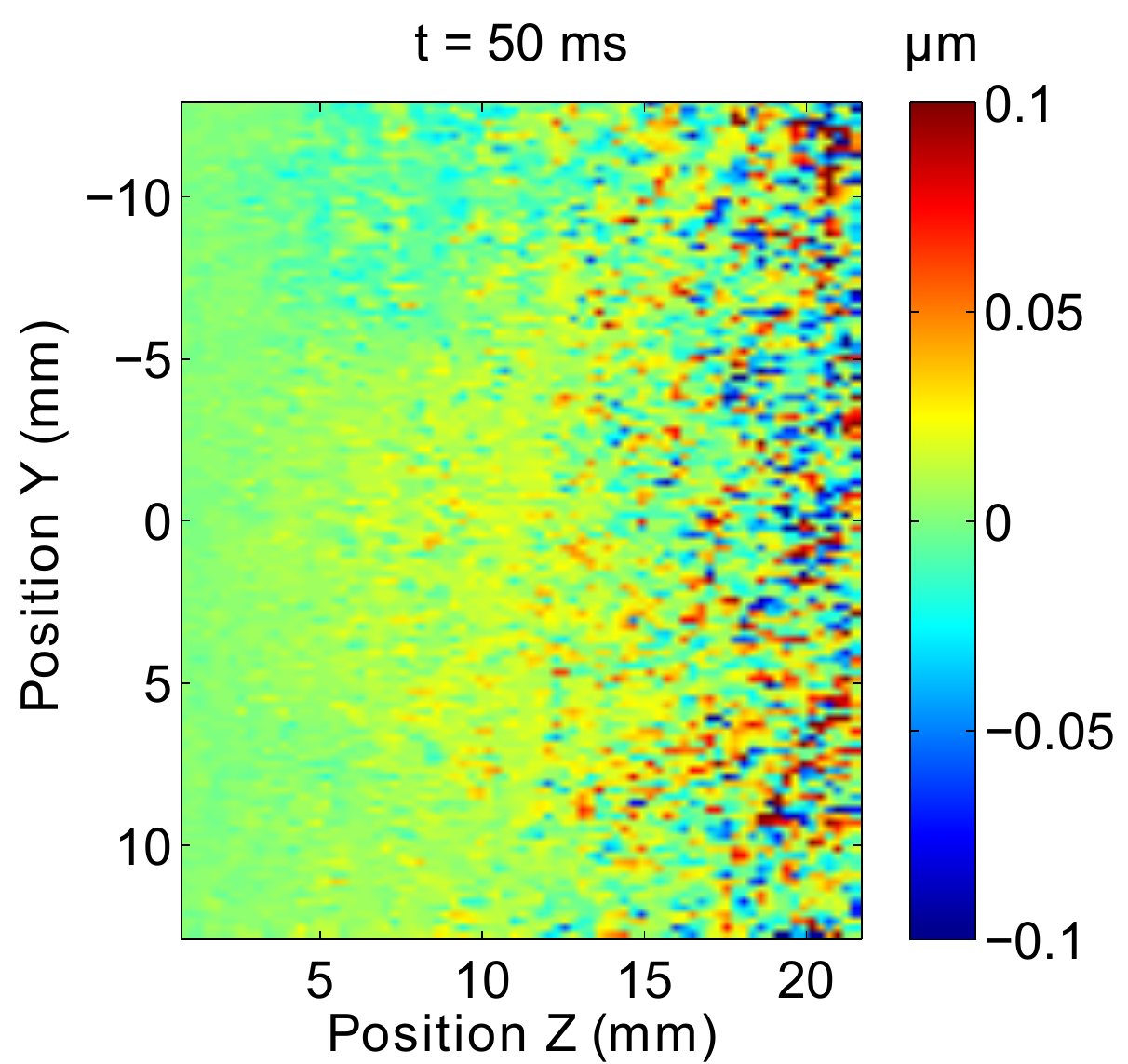}
		\end{center}
   \end{minipage}
 	\caption{\textbf{Résultats de l'expérience d'élastographie sur un échantillon de foie} -- Champ des déplacements dus à un courant électrique injecté par deux électrodes sous un échantillon de foie placé dans un champ magnétique à 0, 10, 20, 30, 40 et 50 ms (plan YZ). Le bruit est plus important que dans les fantômes. On observe une onde de cisaillement se propageant à 0,5 $\pm$ 0,1 m.s$^{-1}$, ce qui correspond à un module d'Young $E$ de 800 $\pm$ 300 kPa.}
 	\label{figElastoEMExpResFoie} 
\end{figure}

\paragraph{Résultats et discussions}
La figure \ref{figElastoEMExpResFoie} représente des champs de déplacements à 0, 10, 20, 30, 40 et 50 millisecondes après injection du courant. Un bruit plus important est visible dans cet échantillon par rapport aux fantômes. On observe néanmoins la propagation d'ondes de cisaillement vers le haut, à une vitesse de 0,5 $\pm$ 0,1 m.s$^{-1}$. Cela correspont à un module d'Young $E$ = 800 $\pm$ 300 Pa, ce qui confirme que l'échantillon était plutôt mou.

\subsection{Bilan partiel}
Ces expériences ont montré que les ondes de cisaillement induites par effet Joule peuvent être réalisées pour des expériences d'élastographie. Le courant injecté devra cependant être diminué d'un facteur cent environ avant de pouvoir tester l'expérience sur des animaux.

\section{Expériences d'imagerie de conductivité électrique}
Si l'on a un champ magnétique et un champ électrique uniformes, seules les inhomogénéités de conductivité électrique peuvent donner lieu à des ondes de cisaillement, si l'on ne prend pas en compte les conversions de modes par effets de bord. En s'intéressant à la source des déplacements plutôt qu'à leur propagation, on peut obtenir des cartes de conductivité électrique des tissus. Des expériences ont donc été réalisées pour voir comment l'on pourrait détecter des variations de conductivité électrique avec ce principe.

\subsection{Expérience de localisation d'inhomogénéité de conductivité électrique}
Nous avons d'abord voulu détecter une inhomogénéité de conductivité électrique forte, en plaçant un fil électrique au milieu du fantôme.

\begin{figure}[!htb]
\begin{minipage}[c]{.5\linewidth}
 \begin{center}
	 \includegraphics[height=2.8cm]{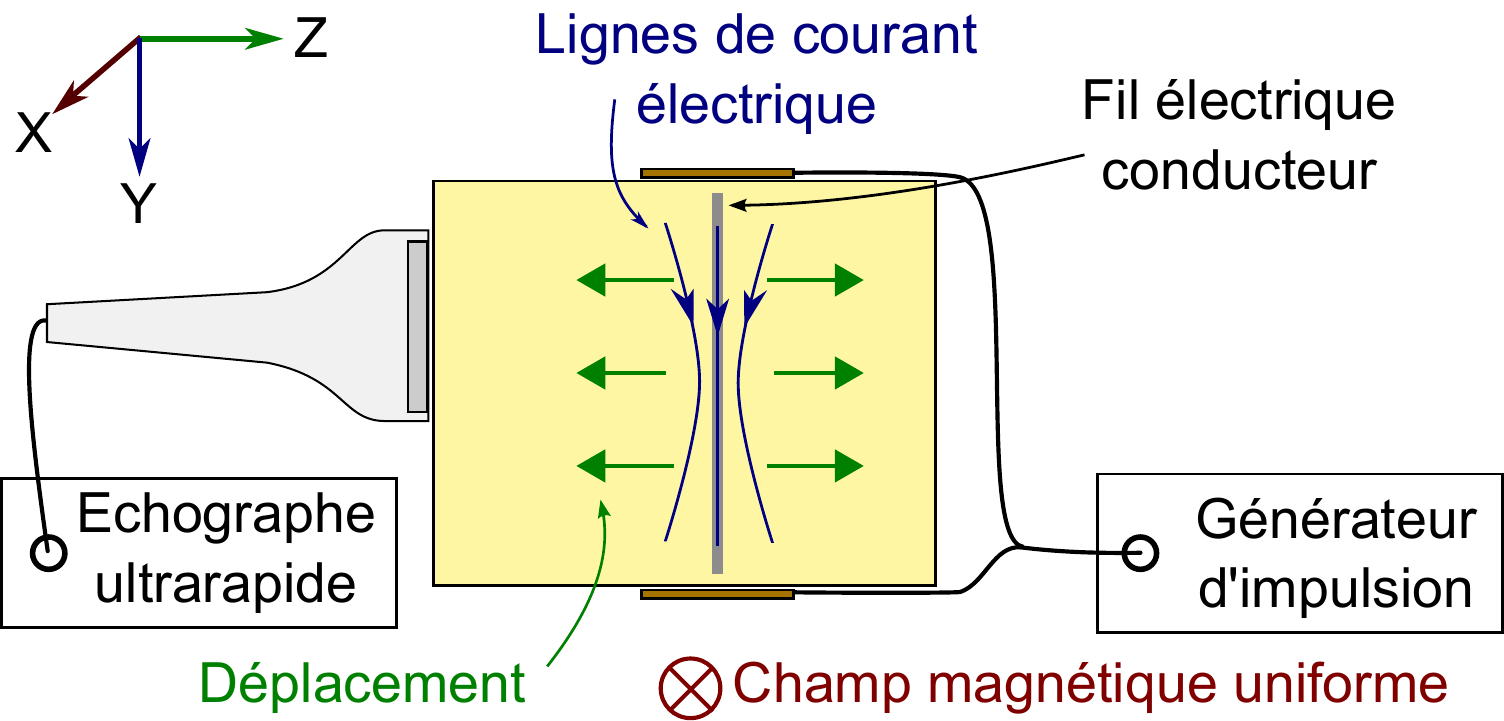}
 \end{center}
\end{minipage}
\begin{minipage}[c]{.5\linewidth}
	 \includegraphics[height=2.8cm]{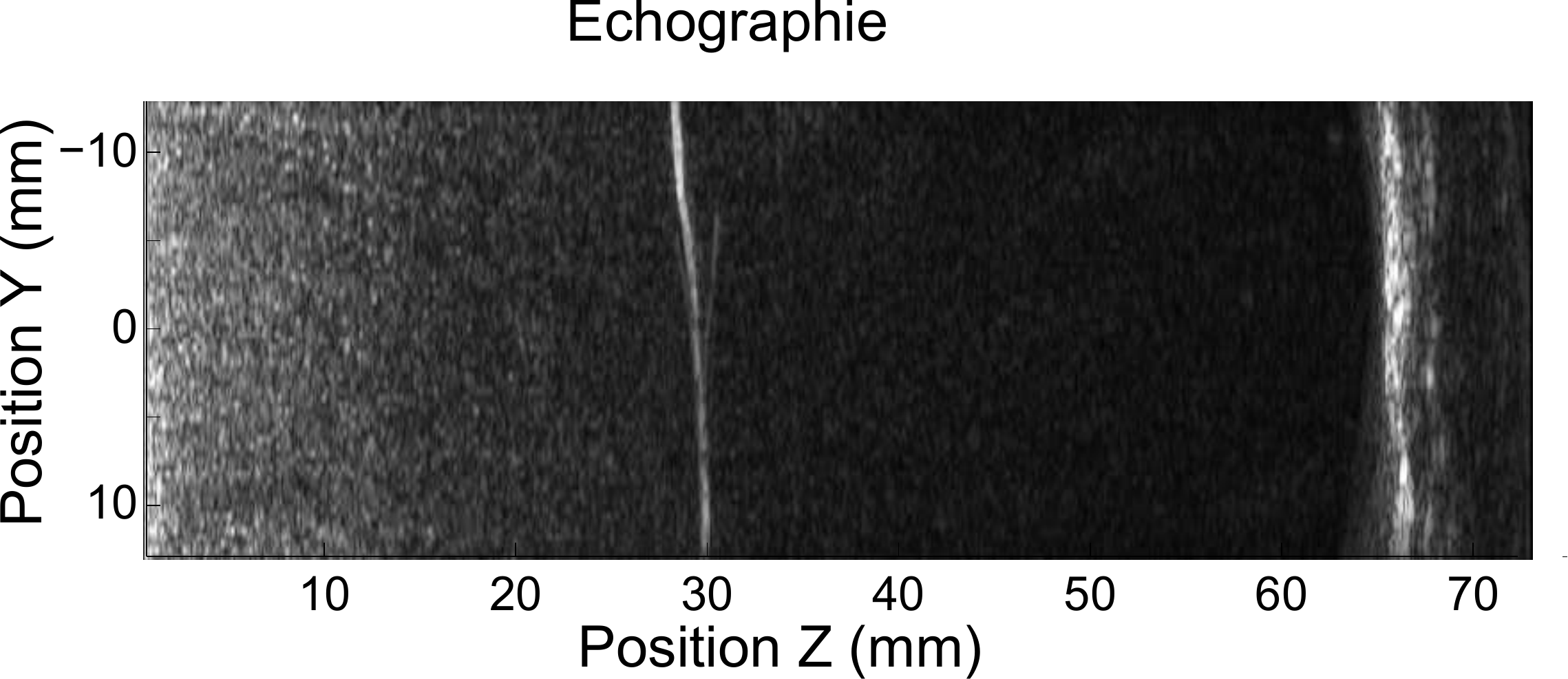}
\end{minipage}
	 \caption{(a) \textbf{Schéma de l'expérience sur la localisation d'inhomogénéité de conductivité électrique} -- Deux électrodes proches créent un courant électrique dans un fantôme soumis à un champ magnétique et contenant un fil électrique, et les déplacements sont observés par une sonde ultrasonore en mode ultrarapide. (b) \textbf{Echographie obtenue par la sonde ultrasonore} -- L'échographie a été prise dans le plan XZ. On distingue au milieu un point blanc correspondant au fil électrique et l'interface du fond sur la droite.}
	\label{figElastoEMExpMatInhomogeneite}
\end{figure}

\paragraph{Matériel et méthodes}
Une expérience, représentée figure \ref{figElastoEMExpMatInhomogeneite}-(a), a été faite avec deux électrodes en contact respectivement avec la partie supérieure et inférieure de l'échantillon. L'échantillon était un fantôme composé de 5\% de PVA, 0,1 \% de graphite et de 5\% de sel, ainsi qu'un fil électrique de 100 \micro m de diamètre dans la direction Y. Ce fil a été placé pendant la solidification du fantôme afin d'éviter de percer ce dernier par introduction du fil. Ce fil est environ cent mille fois plus conducteur que le reste du fantôme, afin d'avoir une inhomogénéité marquée. L'image acquise par la barrette ultrasonore était cette fois-ci dans le plan XZ, de façon à être perpendiculaire au fil électrique. L'échographie du milieu est représentée figure \ref{figElastoEMExpMatInhomogeneite}-(b).

\begin{figure}[!htb]
   \begin{minipage}[c]{.5\linewidth}
		\begin{center}
	   		\includegraphics[width=.75\linewidth]{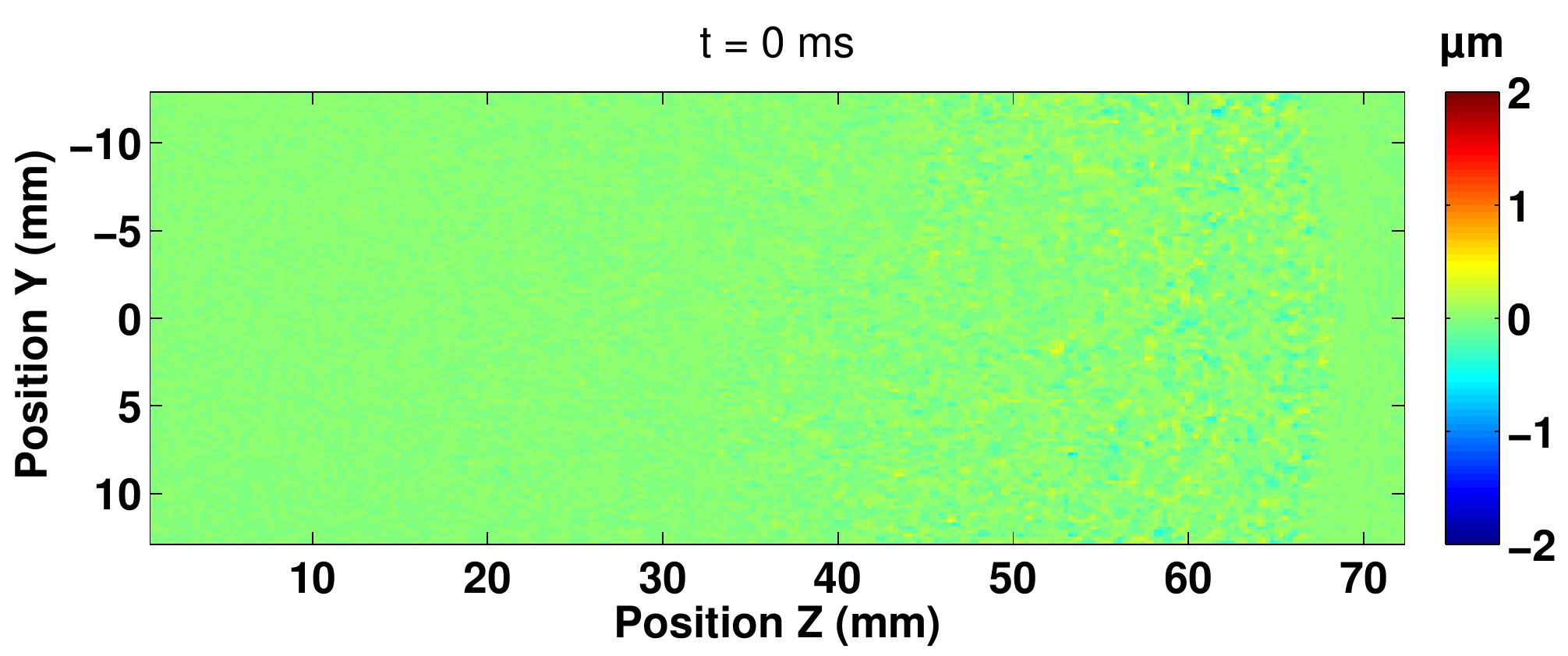}
		\end{center}
   \end{minipage} \hfill
   \begin{minipage}[c]{.5\linewidth}
		\begin{center}
	   		\includegraphics[width=.75\linewidth]{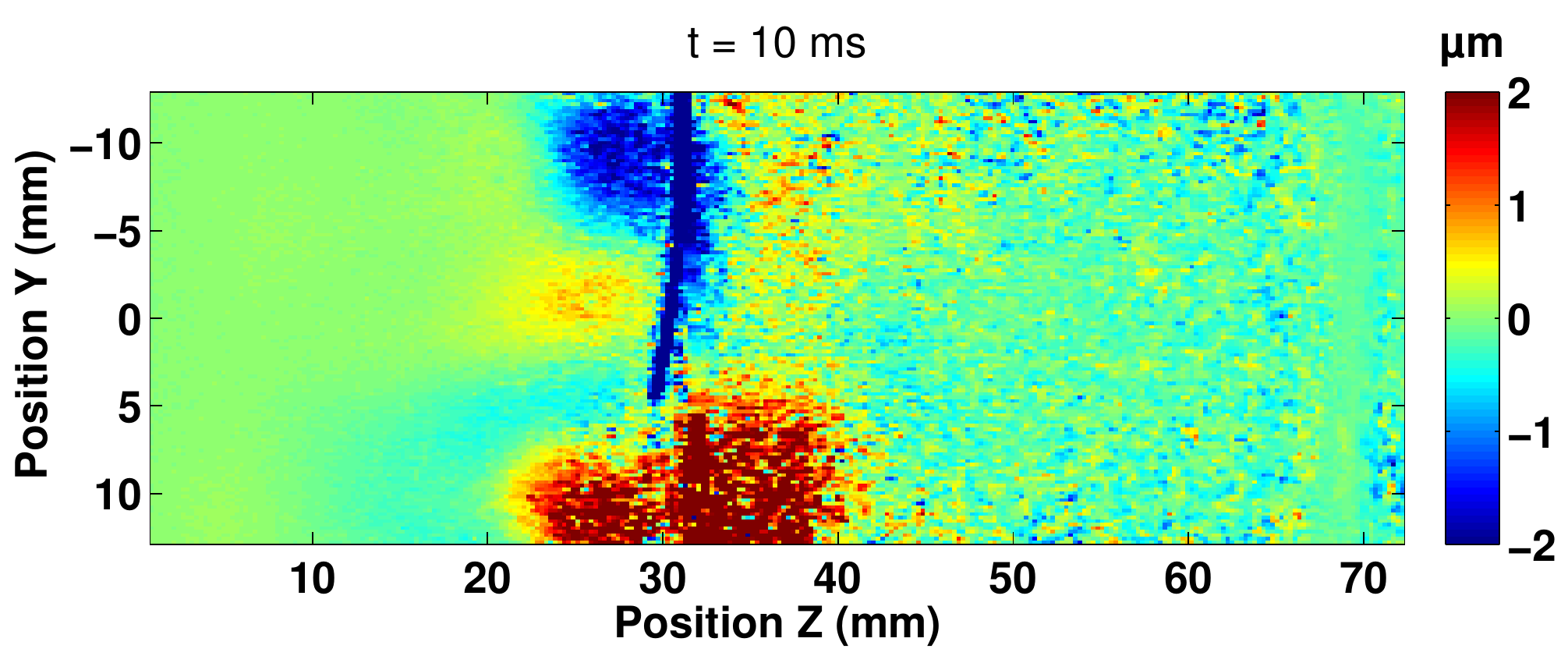}
		\end{center}
   \end{minipage}
   \begin{minipage}[c]{.5\linewidth}
		\begin{center}
	   		\includegraphics[width=.75\linewidth]{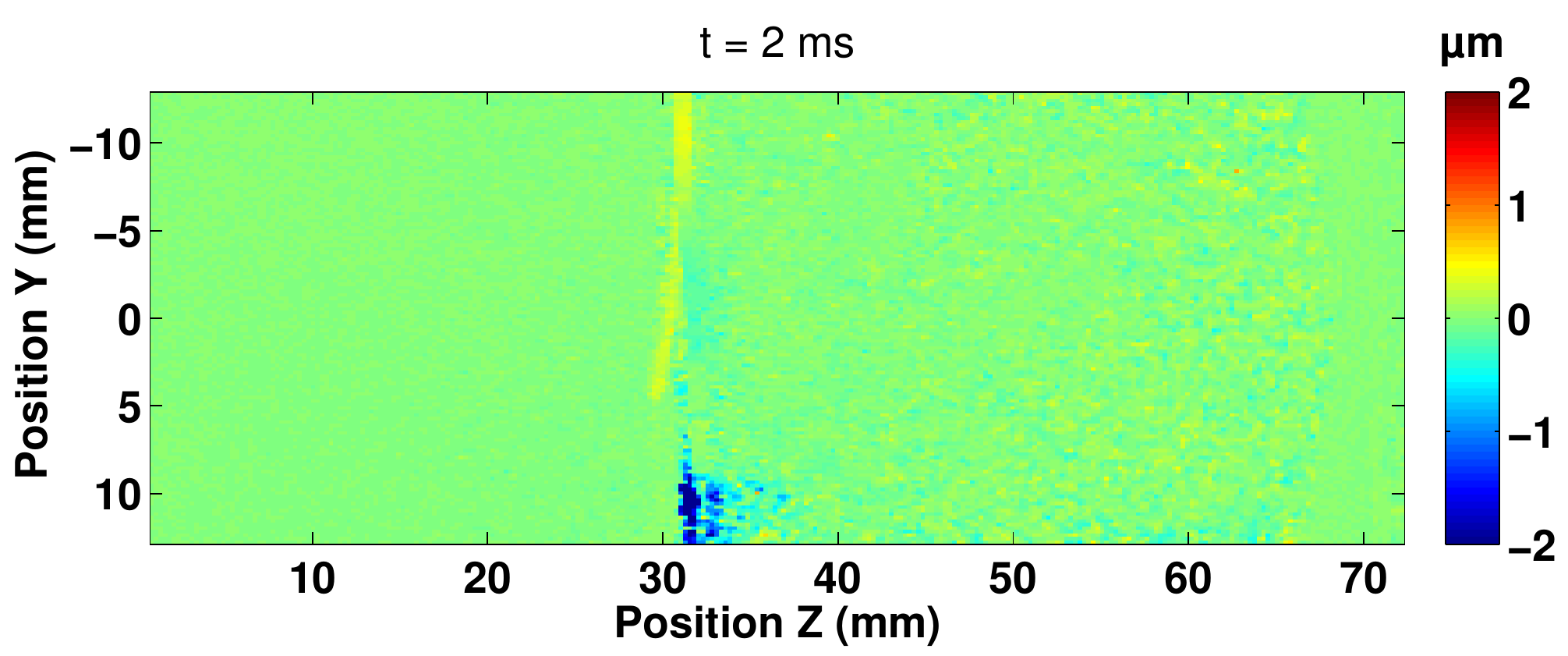}
		\end{center}
   \end{minipage} \hfill
   \begin{minipage}[c]{.5\linewidth}
		\begin{center}
	   		\includegraphics[width=.75\linewidth]{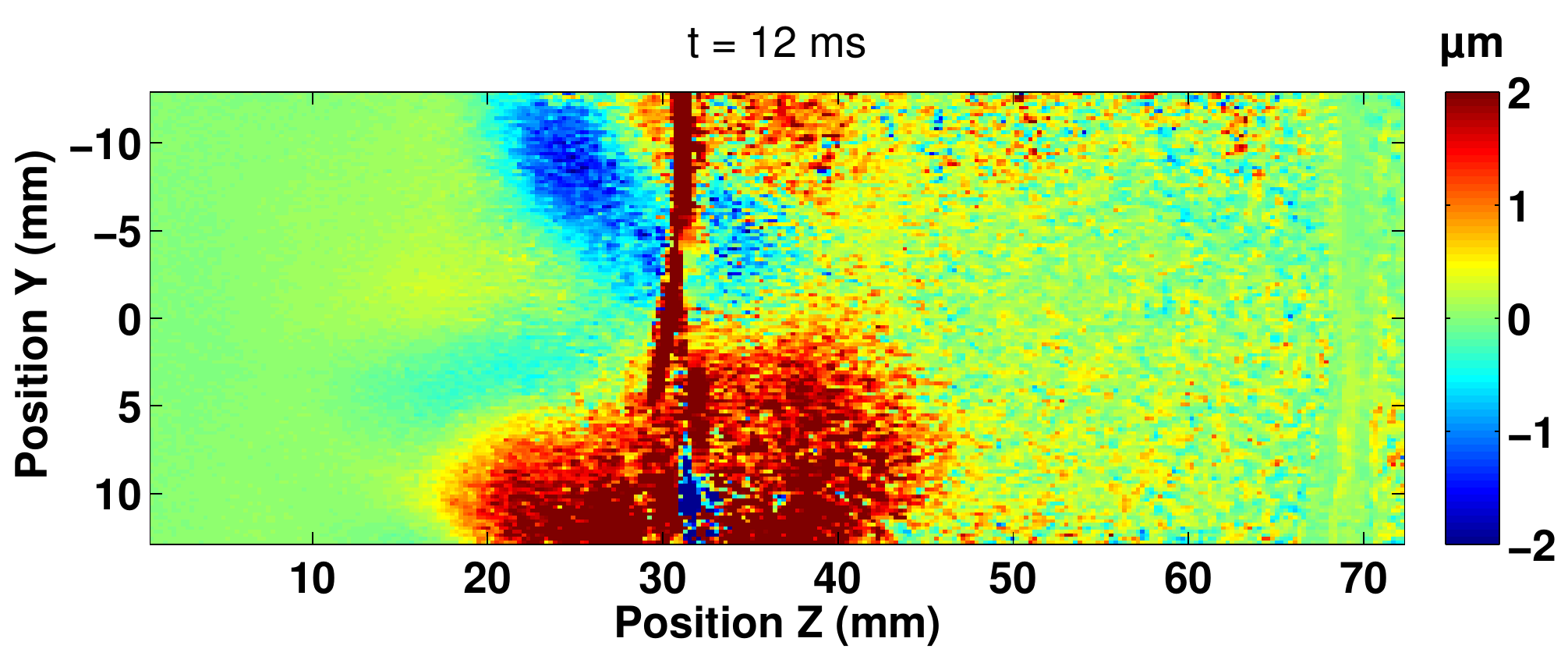}
		\end{center}
   \end{minipage}
   \begin{minipage}[c]{.5\linewidth}
		\begin{center}
	   		\includegraphics[width=.75\linewidth]{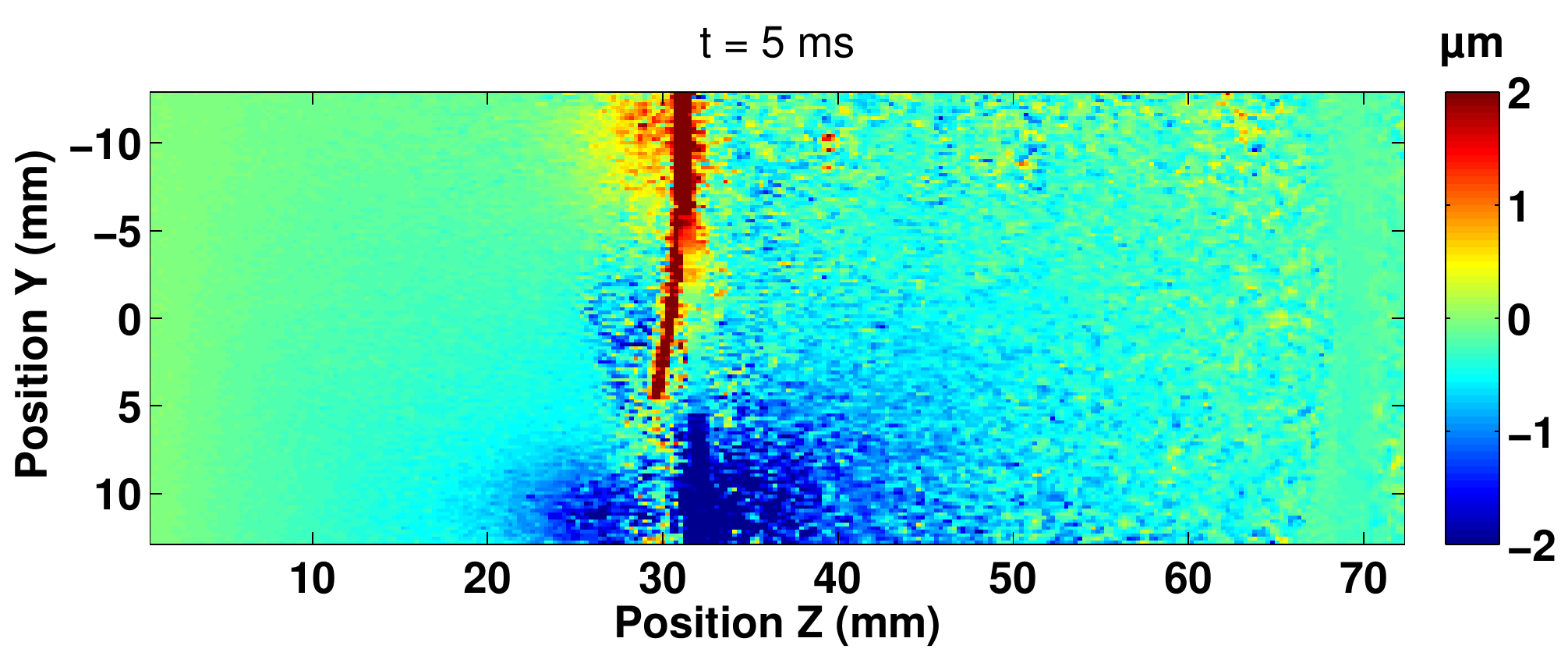}
		\end{center}
   \end{minipage} \hfill
   \begin{minipage}[c]{.5\linewidth}
		\begin{center}
	   		\includegraphics[width=.75\linewidth]{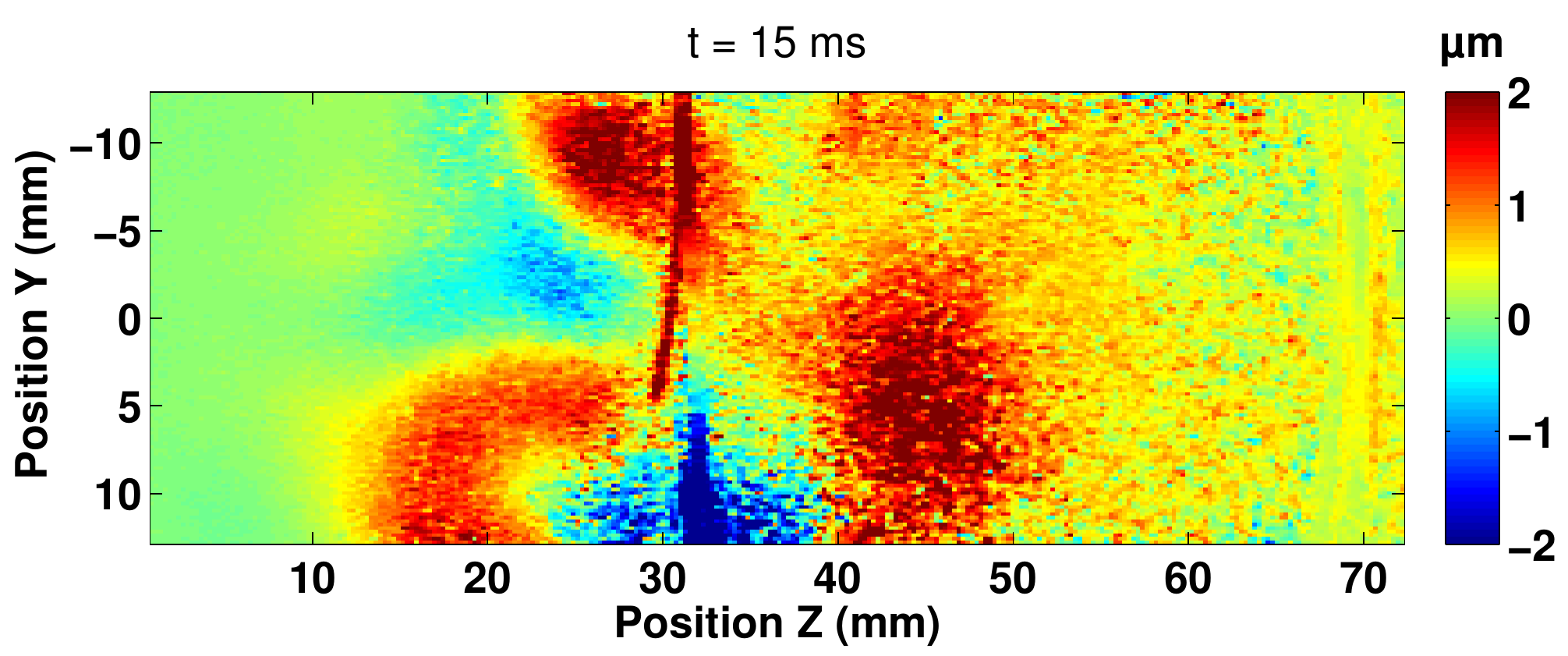}
		\end{center}
   \end{minipage}
   \begin{minipage}[c]{.5\linewidth}
		\begin{center}
	   		\includegraphics[width=.75\linewidth]{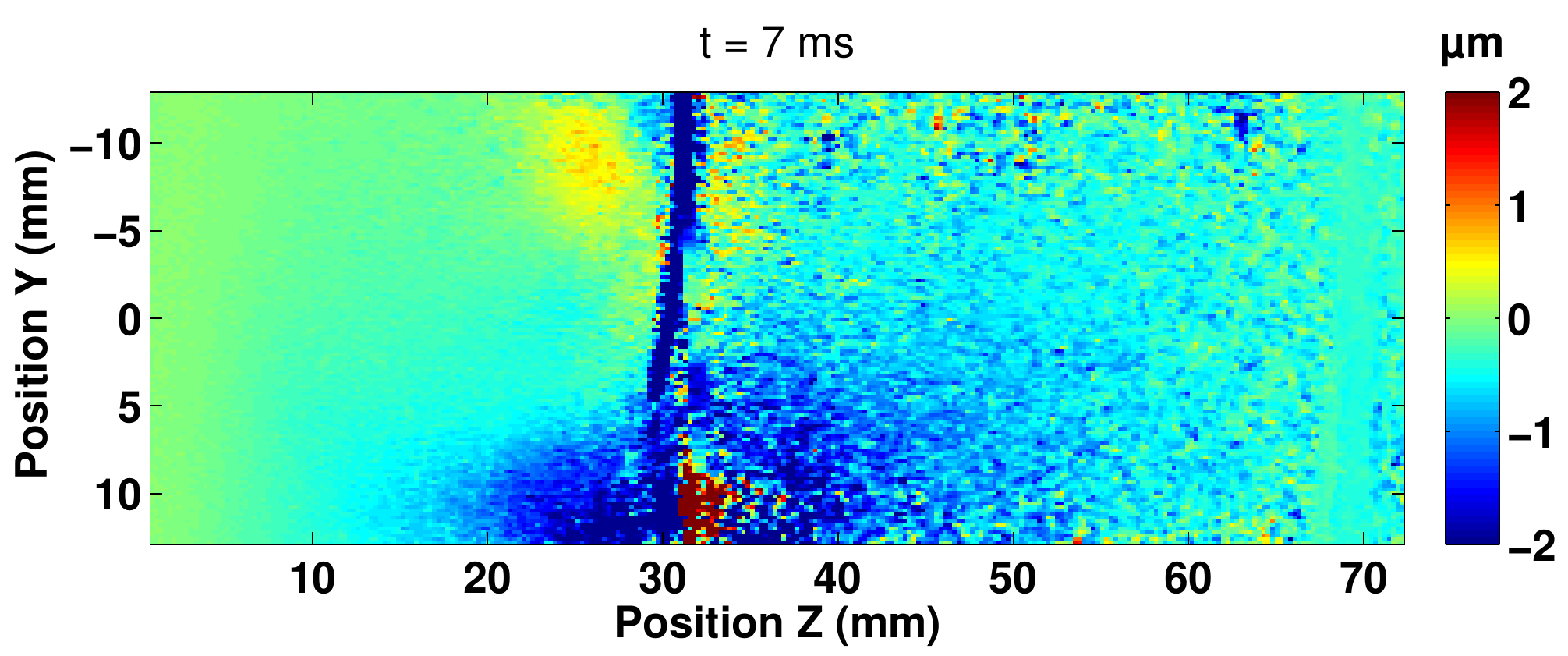}
		\end{center}
   \end{minipage} \hfill
   \begin{minipage}[c]{.5\linewidth}
		\begin{center}
	   		\includegraphics[width=.75\linewidth]{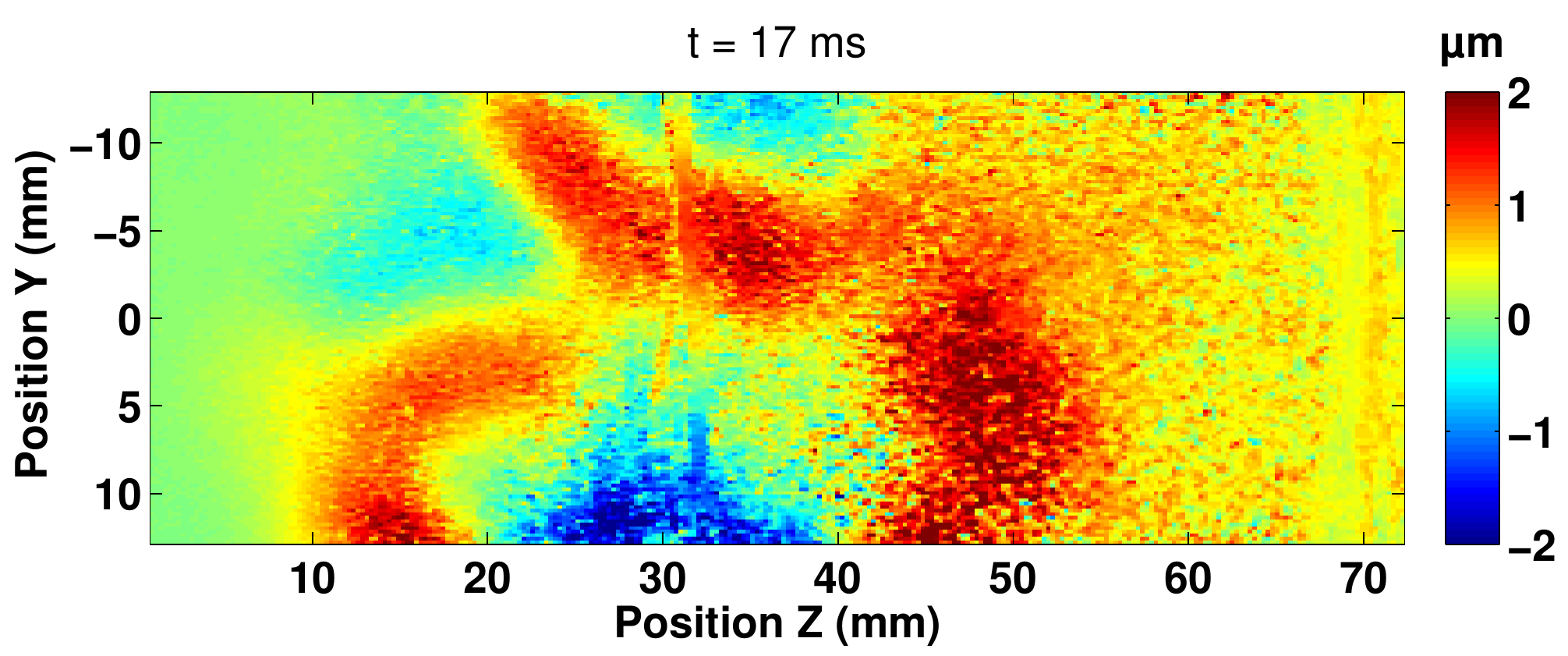}
		\end{center}
   \end{minipage}
 	\caption{\textbf{Résultats d'expérience sur la localisation d'inhomogénéité de conductivité électrique} -- Champ des déplacements due à la traversée d'un courant électrique dans un fantôme placé dans un champ magnétique à  0, 2, 5, 7, 10, 12, 15 et 17 ms après le début du passage du courant. On observe une onde de cisaillement créée au niveau du fil électrique se propageant dans toutes les directions.}
 	\label{figElastoEMExpResInhomogeneiteVert} 
\end{figure}

\begin{figure}[!htb]
   \begin{minipage}[c]{.5\linewidth}
		\begin{center}
	   		\includegraphics[width=1\linewidth]{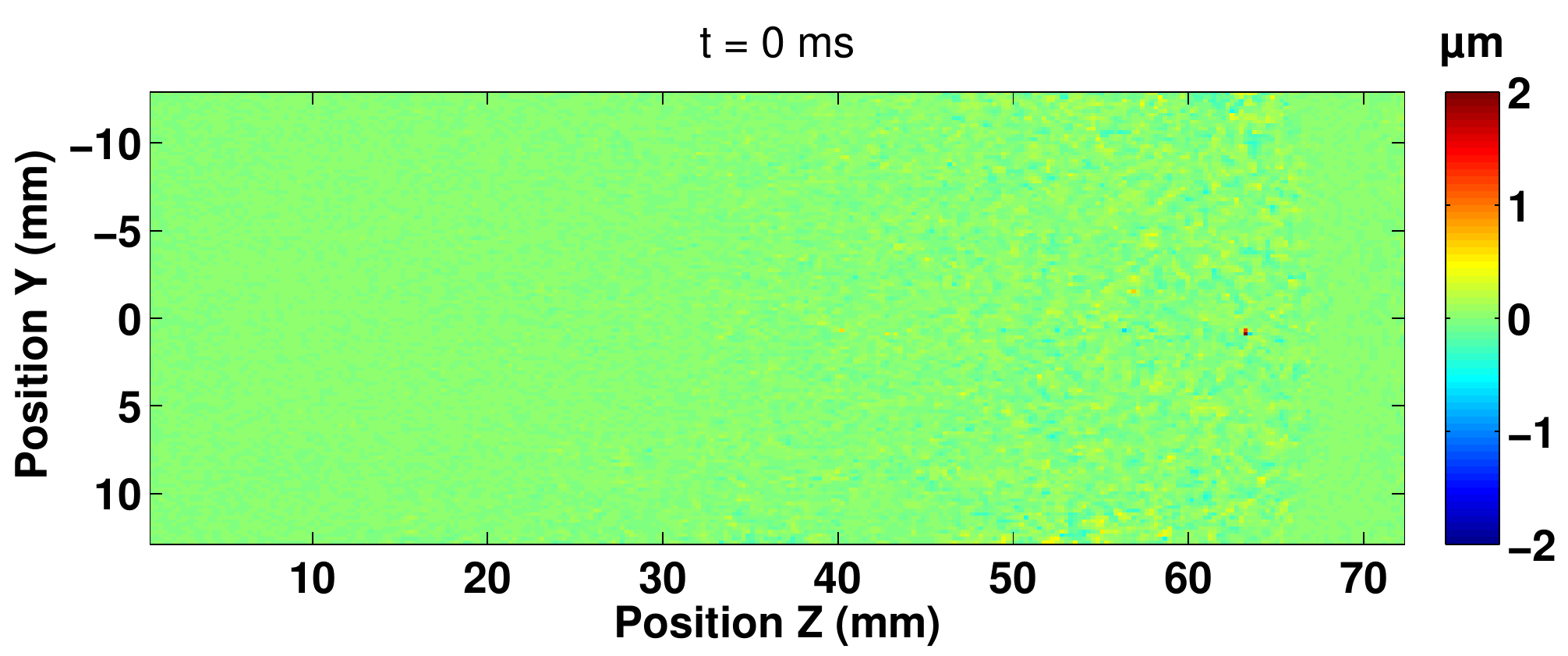}
		\end{center}
   \end{minipage} \hfill
   \begin{minipage}[c]{.5\linewidth}
		\begin{center}
	   		\includegraphics[width=1\linewidth]{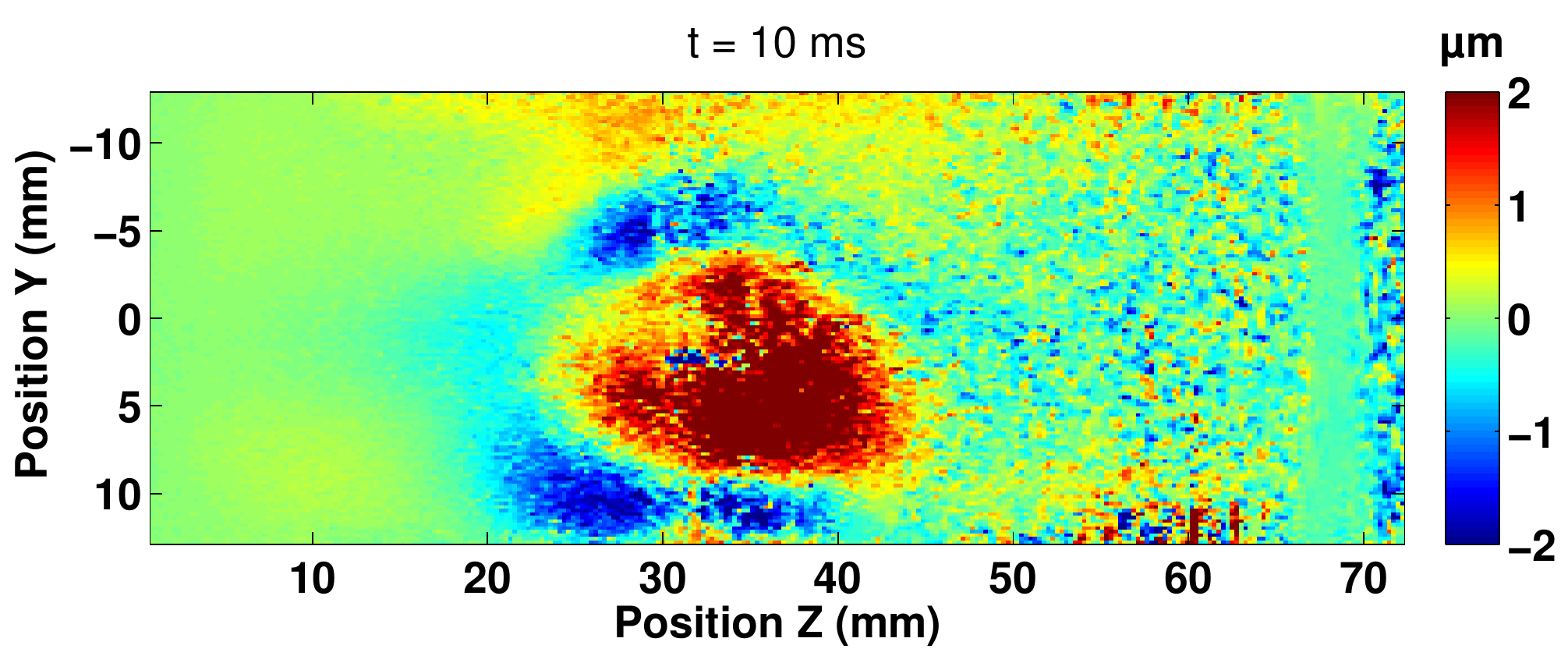}
		\end{center}
   \end{minipage}
   \begin{minipage}[c]{.5\linewidth}
		\begin{center}
	   		\includegraphics[width=1\linewidth]{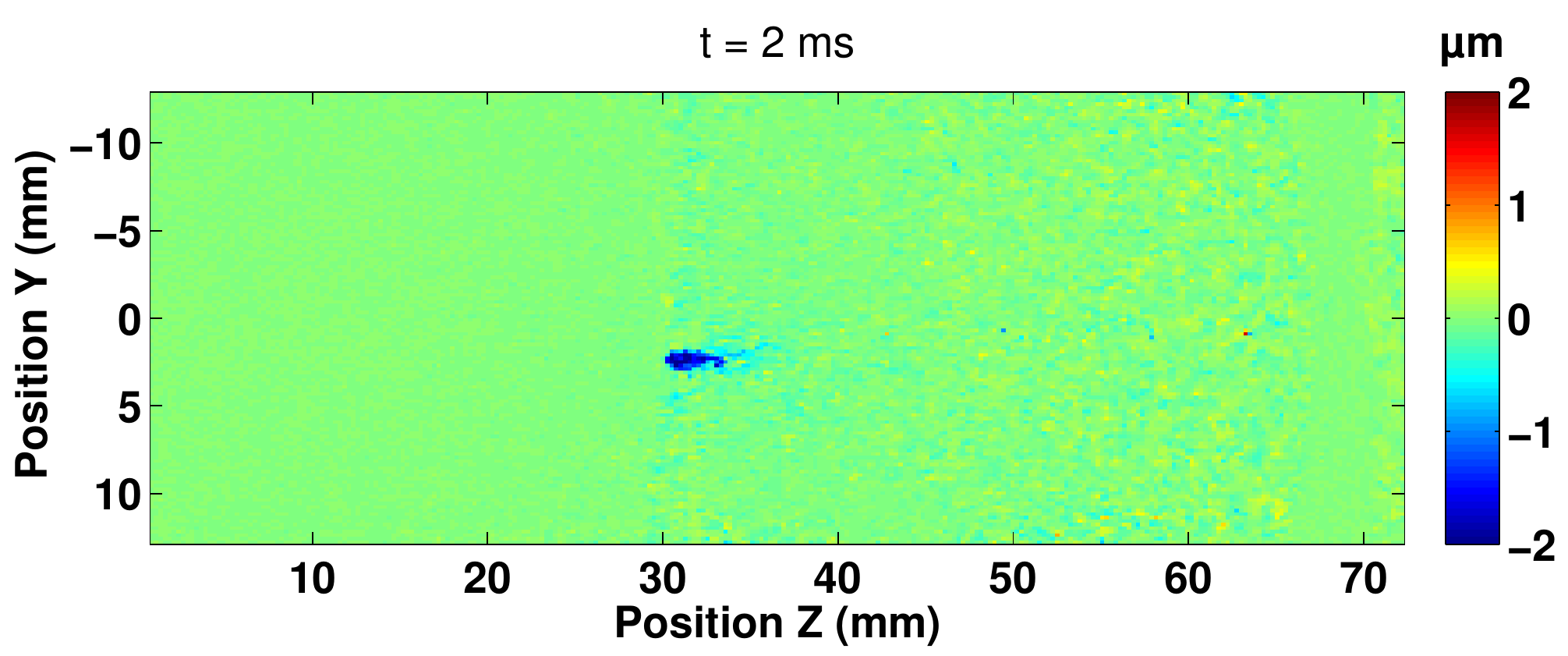}
		\end{center}
   \end{minipage} \hfill
   \begin{minipage}[c]{.5\linewidth}
		\begin{center}
	   		\includegraphics[width=1\linewidth]{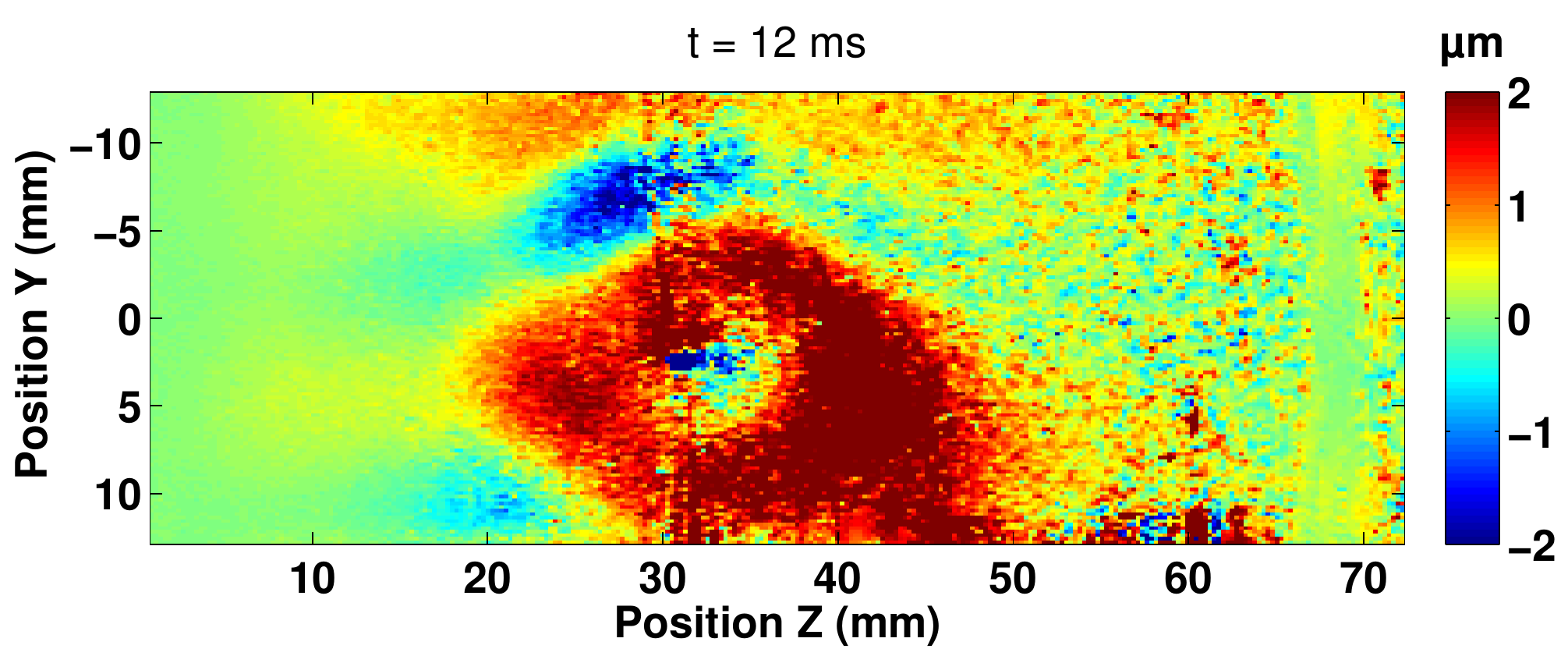}
		\end{center}
   \end{minipage}
   \begin{minipage}[c]{.5\linewidth}
		\begin{center}
	   		\includegraphics[width=1\linewidth]{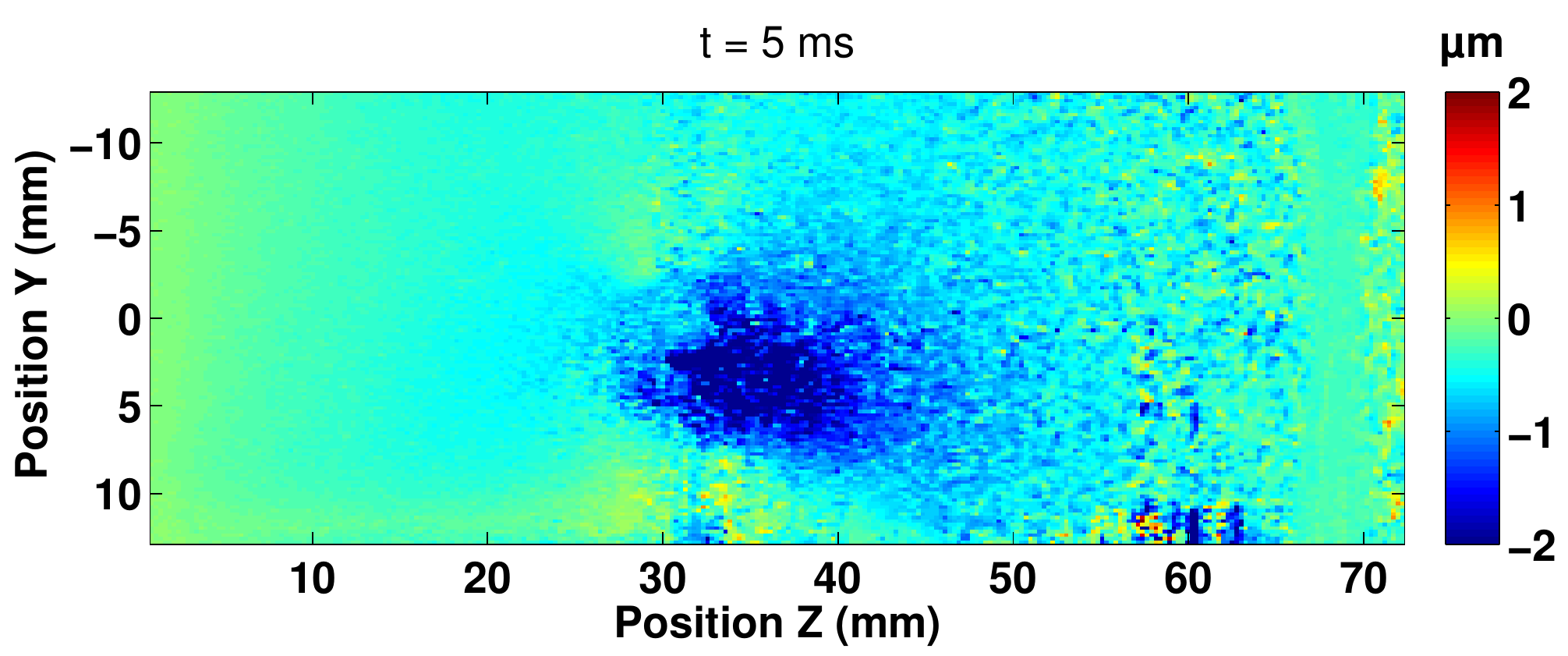}
		\end{center}
   \end{minipage} \hfill
   \begin{minipage}[c]{.5\linewidth}
		\begin{center}
	   		\includegraphics[width=1\linewidth]{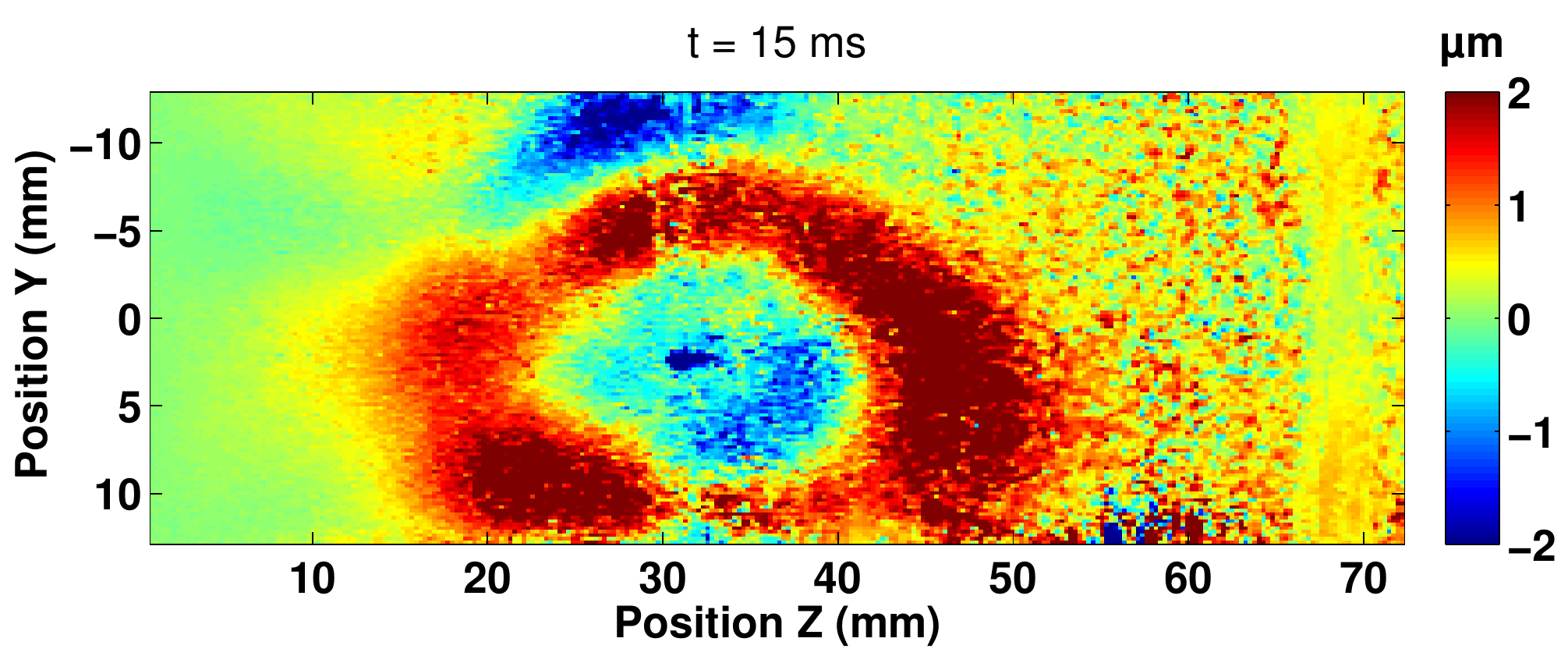}
		\end{center}
   \end{minipage}
   \begin{minipage}[c]{.5\linewidth}
		\begin{center}
	   		\includegraphics[width=1\linewidth]{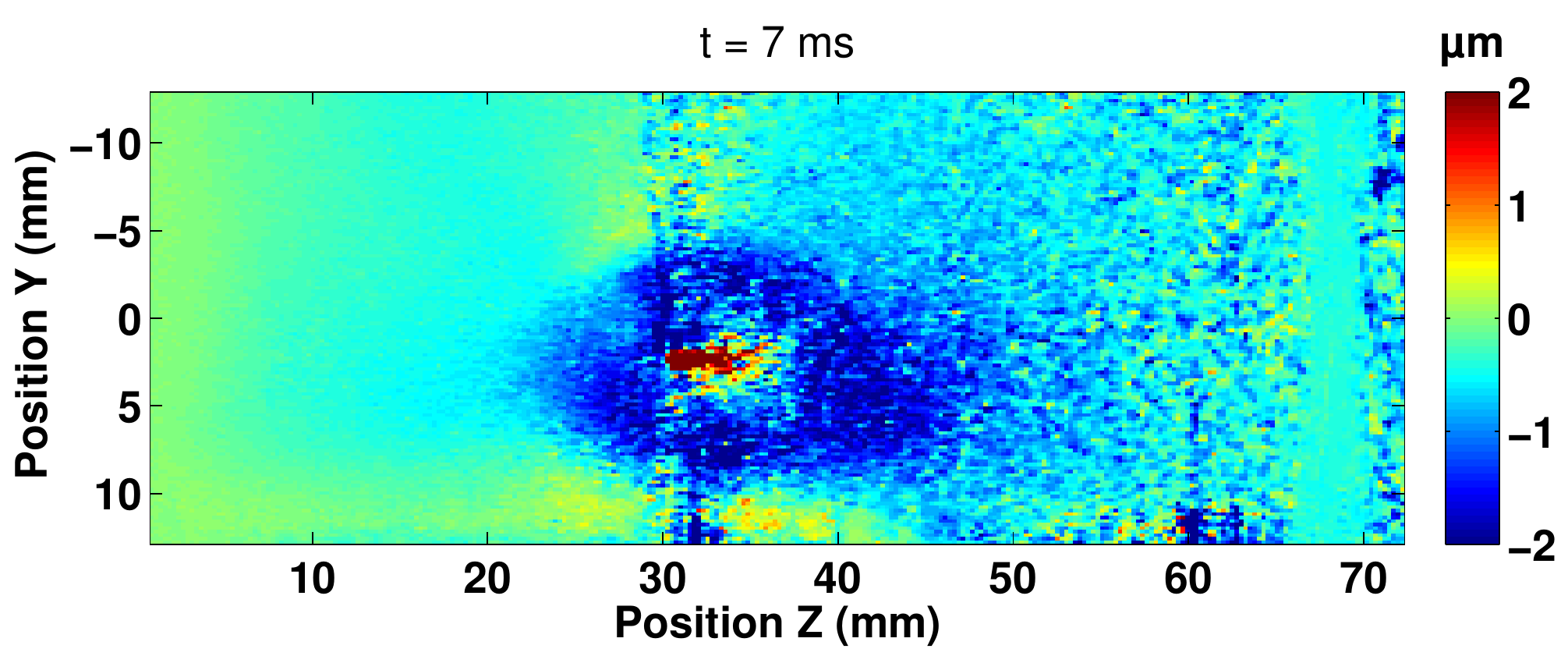}
		\end{center}
   \end{minipage} \hfill
   \begin{minipage}[c]{.5\linewidth}
		\begin{center}
	   		\includegraphics[width=1\linewidth]{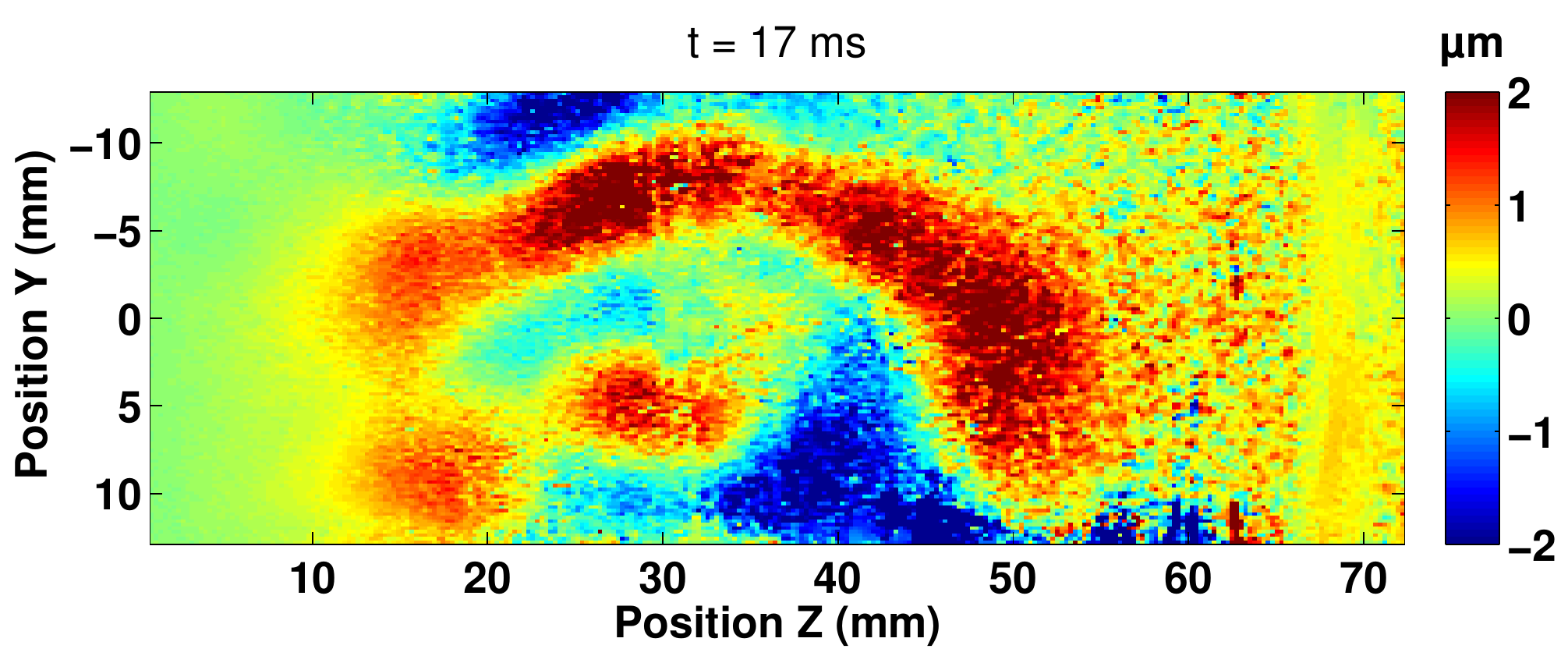}
		\end{center}
   \end{minipage}
 	\caption{\textbf{Résultats d'expérience sur la localisation d'inhomogénéité de conductivité électrique} -- Champ des déplacements due à la traversée d'un courant électrique dans un fantôme placé dans un champ magnétique à  0, 2, 5, 7, 10, 12, 15 et 17 ms après le début du passage du courant (plan XZ). On observe une onde de cisaillement créée au niveau du fil électrique se propageant dans toutes les directions.}
 	\label{figElastoEMExpResInhomogeneiteHori} 
\end{figure}

\paragraph{Résultats et discussions}
La figure \ref{figElastoEMExpResInhomogeneiteVert} représente les champs de déplacements à 0, 2, 5, 7, 10, 12, 15 et 17 ms. Les champs de déplacements dans un plan perpendiculaire (plan XZ) aux mêmes temps ont également été représentés, sur la figure \ref{figElastoEMExpResInhomogeneiteHori}. On peut voir dans les deux cas que des ondes de cisaillement sont créées au niveau du fil, donc que l'on peut trouver grâce à des ondes de cisaillement des inhomogénéités de conductivité électrique.Les inhomogénéités dans un tissu biologique seraient bien sûr moins contrastées que dans cette expérience, et des développements en terme de qualité d'image sont à prévoir. Nous voyons cependant que l'inhomogénéité est particulièrement détectable au début de l'injection du courant, avec un mouvement d'amplitude plus forte.

\subsection{Expérience dans un fantôme avec deux conductivités électriques différentes}
La force de Lorentz dépend de la densité de courant, donc le déplacement résultant aussi. La densité de courant électrique dépend, d'après la loi d'Ohm, de la conductivité électrique et du champ électrique. Une expérience a donc été menée pour évaluer l'influence de la conductivité électrique sur l'amplitude du déplacement.

\begin{figure}[!htb]
\begin{minipage}[c]{.5\linewidth}
 \begin{center}
	 \includegraphics[height=2.8cm]{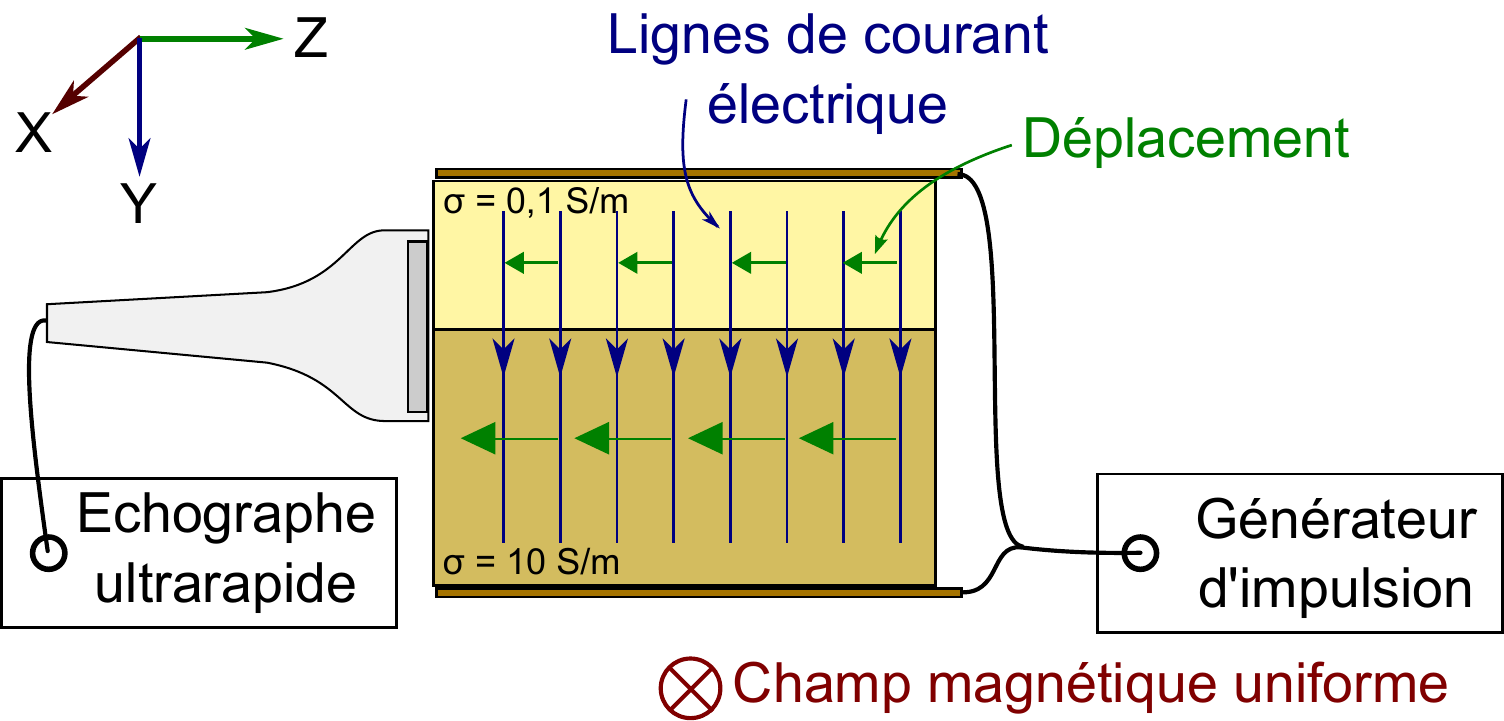}
 \end{center}
\end{minipage}
\begin{minipage}[c]{.5\linewidth}
	 \includegraphics[height=2.8cm]{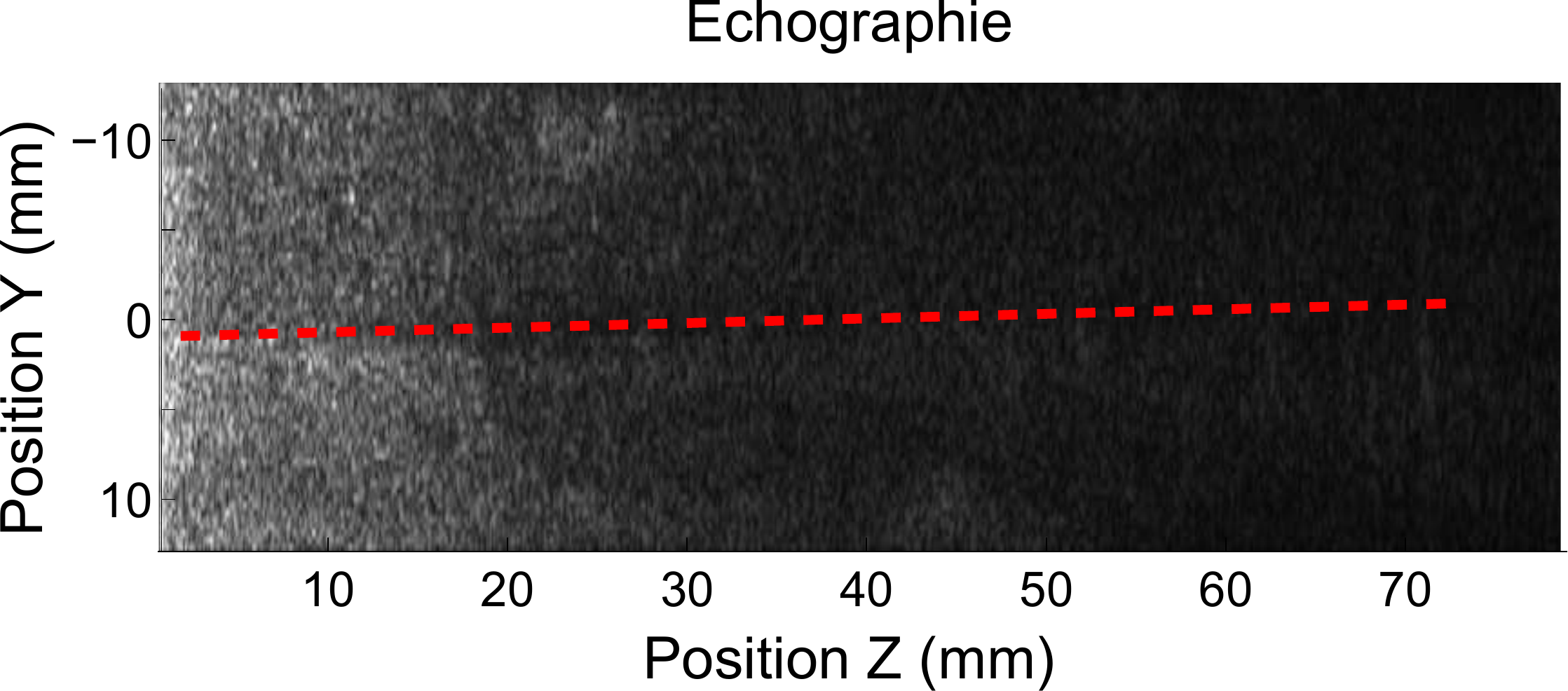}
\end{minipage}
	 \caption{(a) \textbf{Schéma de l'expérience dans un fantôme avec deux conductivités électriques différentes} -- Deux électrodes proches créent un courant électrique dans un fantôme soumis à un champ magnétique, et les déplacements sont observés par une sonde ultrasonore en mode ultrarapide. Le fantôme présente deux couches de conductivité électrique différentes, une plus isolante sur le dessus et une plus conductrice en-dessous. (b) \textbf{Echographie obtenue par la sonde ultrasonore} -- La ligne plus claire horizontale au centre de l'échographie correspondant à la séparation entre les deux couches, mais est à peine visible.}
	\label{figElastoEMExpMatBicoucheConduc}
\end{figure}

\paragraph{Matériel et méthodes}
Une expérience, représentée figure \ref{figElastoEMExpMatBicoucheConduc}-(a) a été faite avec deux électrodes en contact respectivement avec la partie supérieure et inférieure de l'échantillon, le long de la tranche. L'échantillon est composé d'une couche de deux morceaux de gélatine de 3 cm d'épaisseur, composée de 5\% de PVA, 0,1\% de graphite et l'un à 0,05\% de sel (salinité de l'eau du robinet), l'autre à 5\% de sel. Cela donne pour la couche supérieure une conductivité électrique de 0,1 S/m et celle inférieure de 10 S/m. Le fantôme a été fait en une fois, en commençant par couler la partie inférieure, et en attendant la gélification de celle-ci pour couler la partie supérieure. Cela permet d'éviter une interface d'air entre les deux, tout en ayant une élasticité identique des deux côtés. L'échographie du milieu est représentée figure \ref{figElastoEMExpMatBicoucheConduc}-(b). La séparation entre les deux couches est à peine visible (ligne horizontale au centre).

\begin{figure}[!htb]
   \begin{minipage}[c]{.5\linewidth}
		\begin{center}
	   		\includegraphics[width=1\linewidth]{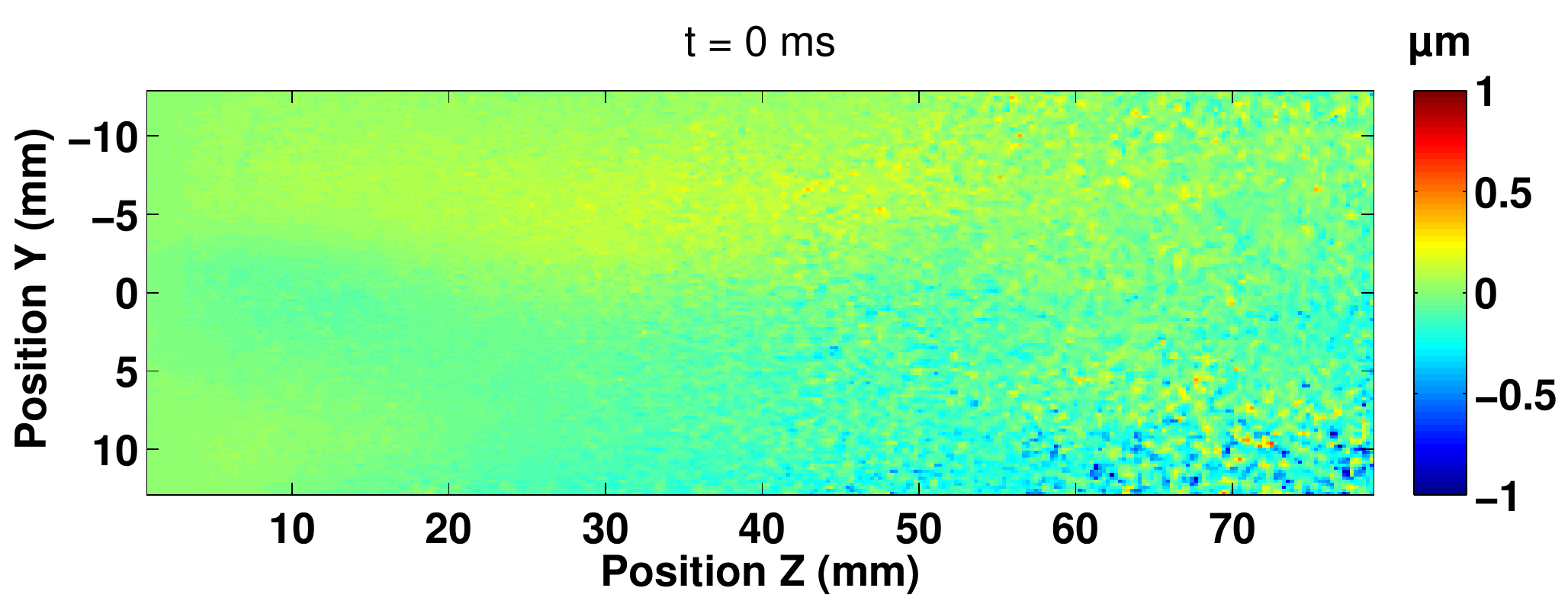}
		\end{center}
   \end{minipage} \hfill
   \begin{minipage}[c]{.5\linewidth}
		\begin{center}
	   		\includegraphics[width=1\linewidth]{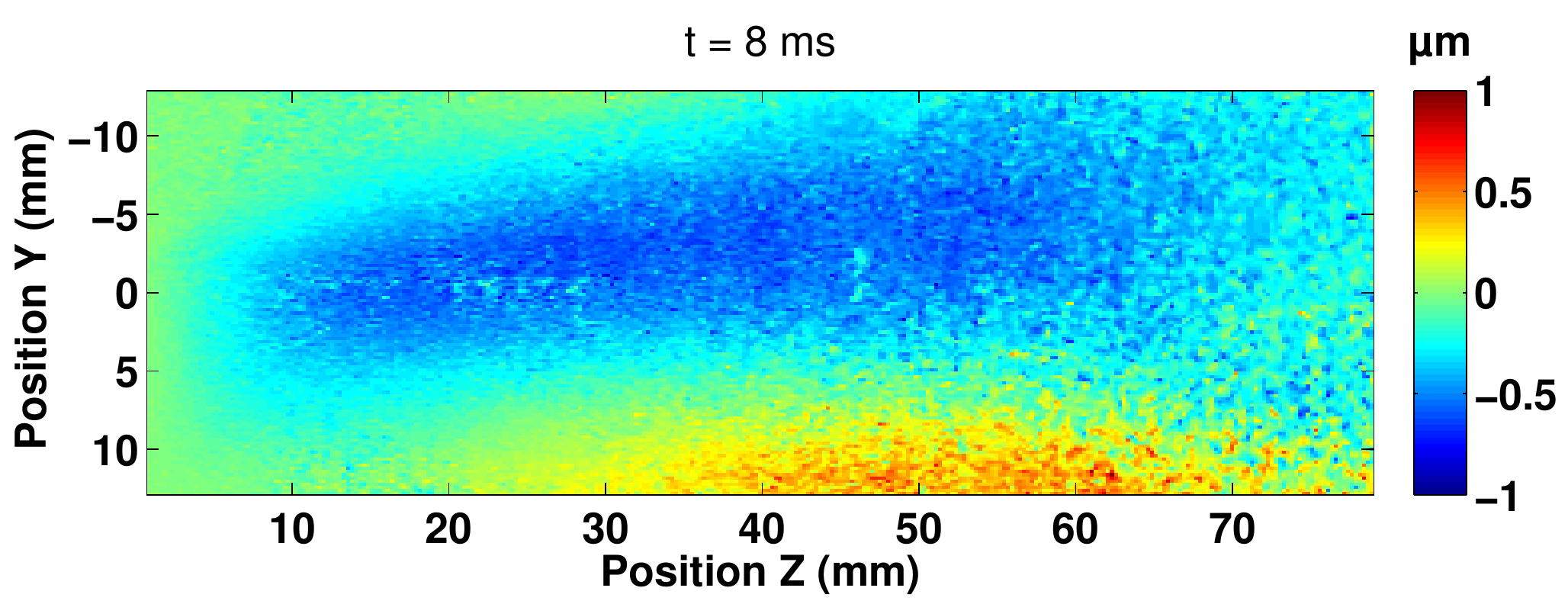}
		\end{center}
   \end{minipage}
   \begin{minipage}[c]{.5\linewidth}
		\begin{center}
	   		\includegraphics[width=1\linewidth]{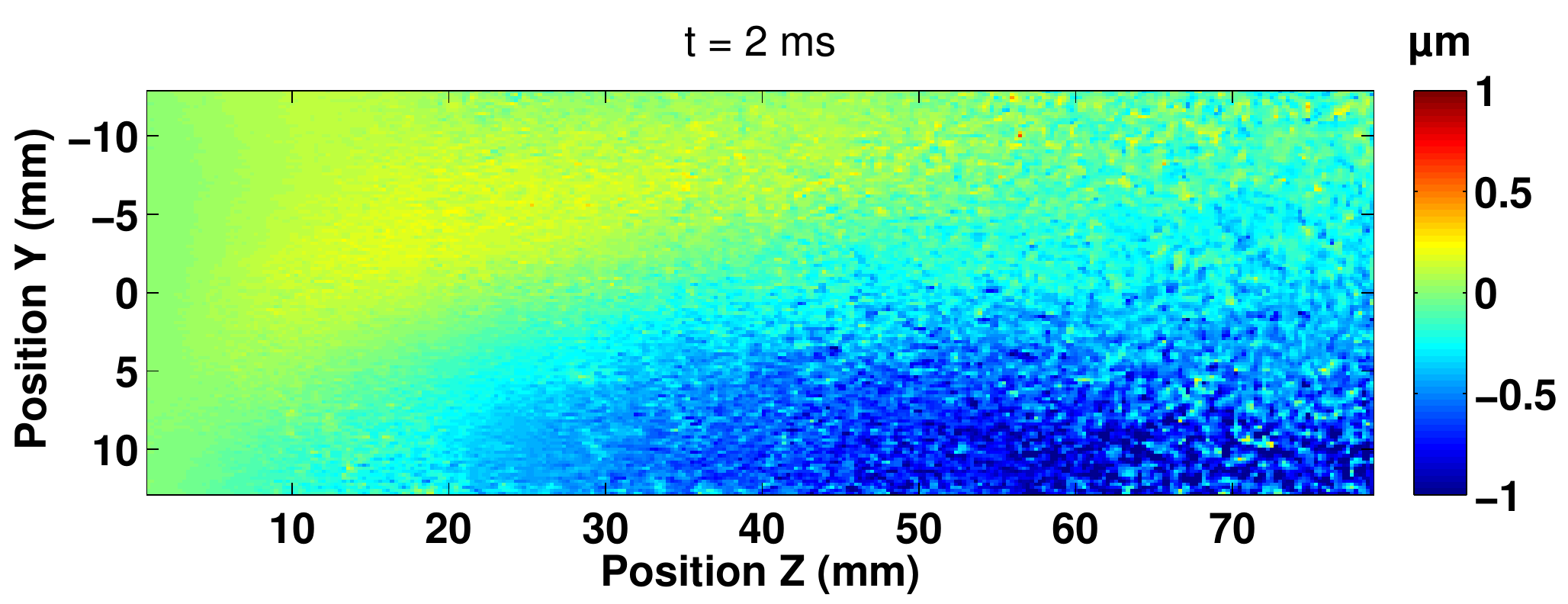}
		\end{center}
   \end{minipage} \hfill
   \begin{minipage}[c]{.5\linewidth}
		\begin{center}
	   		\includegraphics[width=1\linewidth]{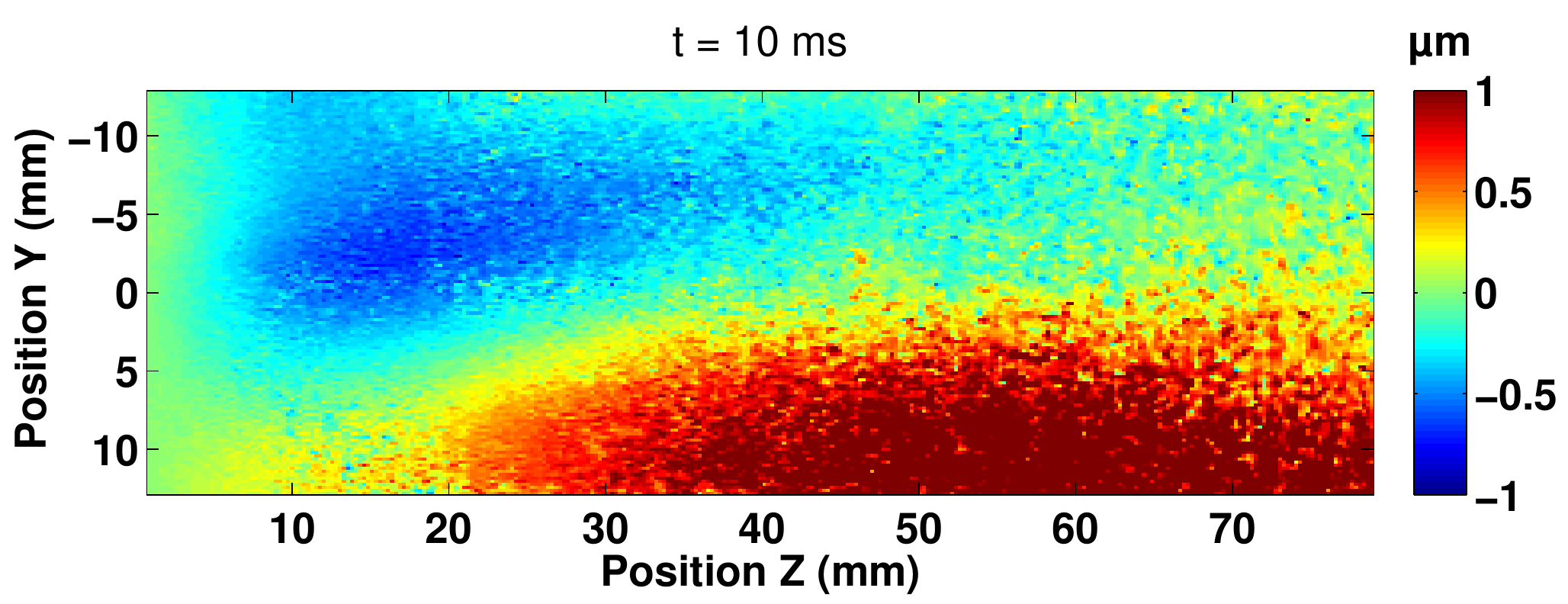}
		\end{center}
   \end{minipage}
   \begin{minipage}[c]{.5\linewidth}
		\begin{center}
	   		\includegraphics[width=1\linewidth]{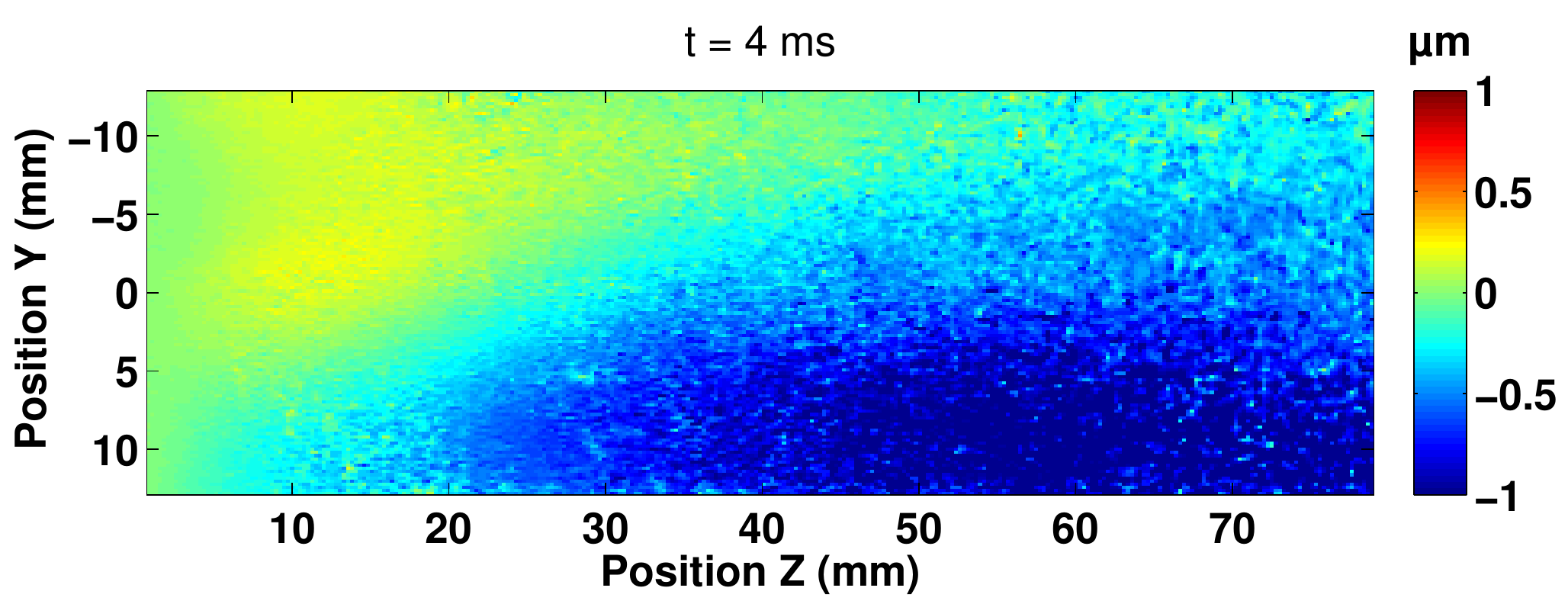}
		\end{center}
   \end{minipage} \hfill
   \begin{minipage}[c]{.5\linewidth}
		\begin{center}
	   		\includegraphics[width=1\linewidth]{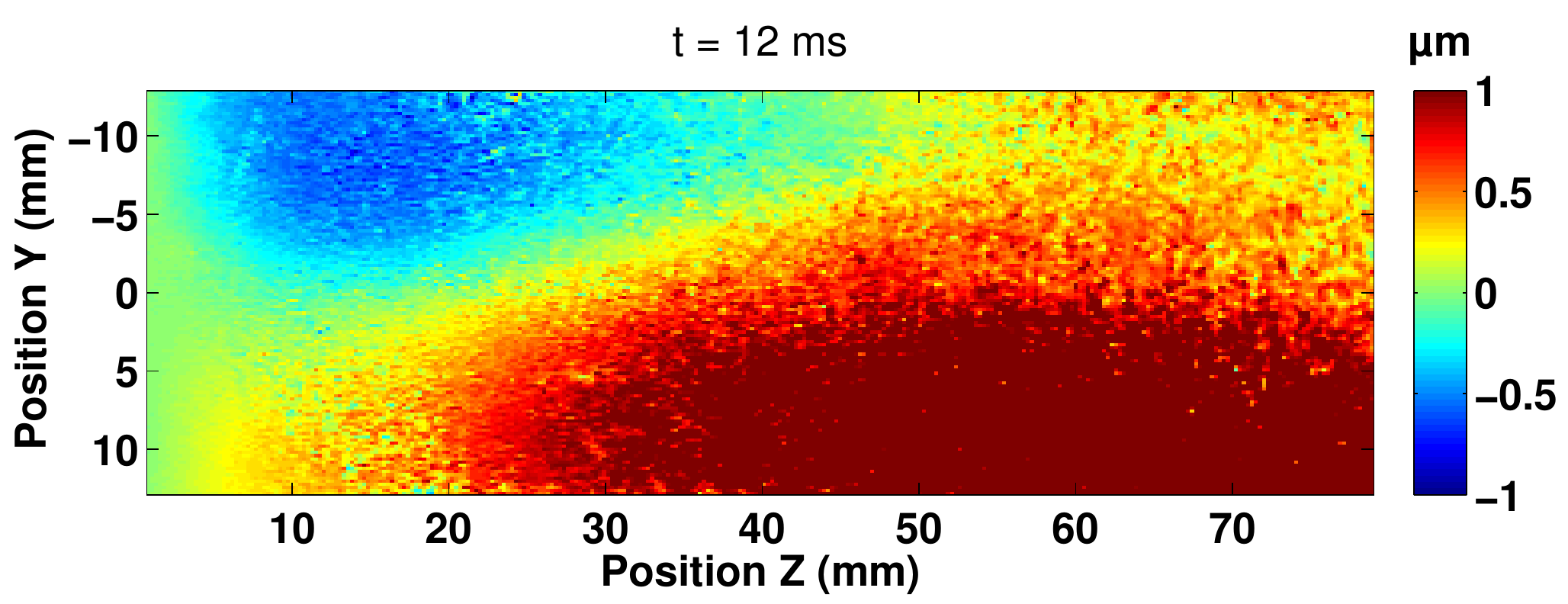}
		\end{center}
   \end{minipage}
   \begin{minipage}[c]{.5\linewidth}
		\begin{center}
	   		\includegraphics[width=1\linewidth]{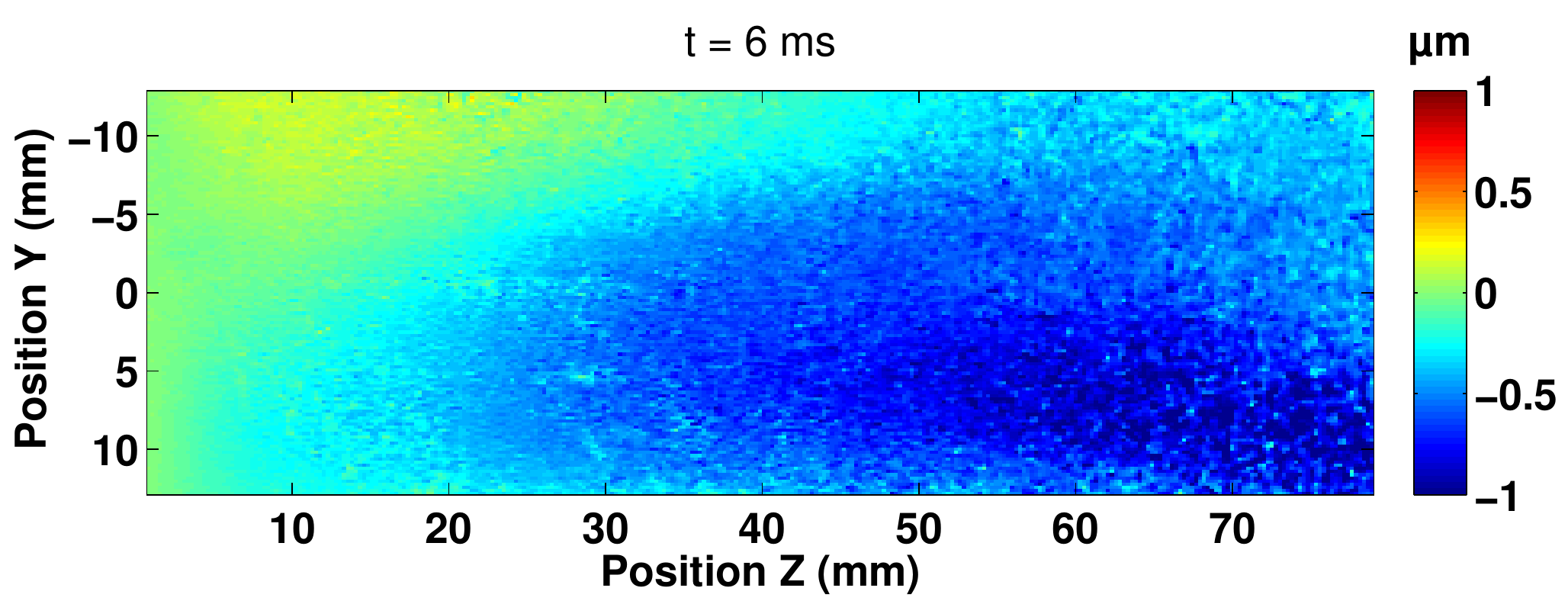}
		\end{center}
   \end{minipage}
   \begin{minipage}[c]{.5\linewidth}
		\begin{center}
	   		\includegraphics[width=1\linewidth]{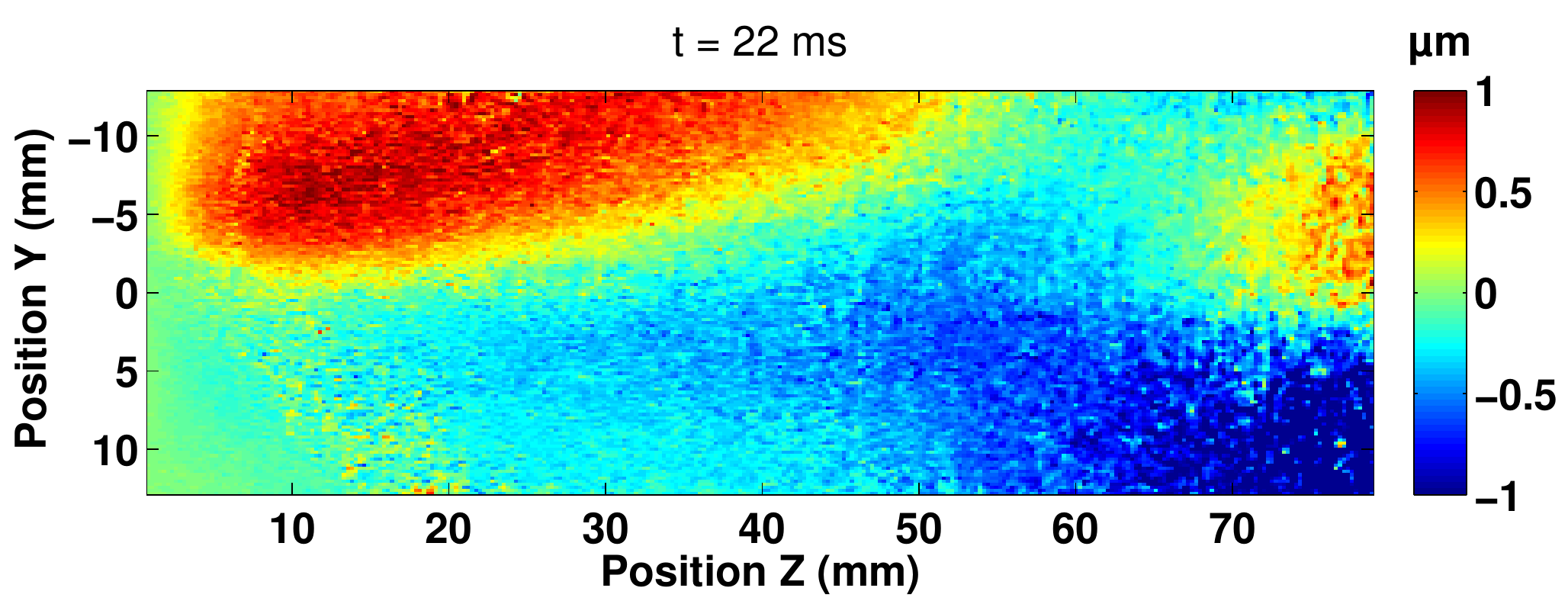}
		\end{center}
   \end{minipage}
 	\caption{\textbf{Expérience dans un fantôme avec deux conductivités électriques différentes} -- Champ des déplacements dus à la traversée d'un courant électrique dans un fantôme contenant un fil électrique, le tout placé dans un champ magnétique, à  0, 2, 4, 6, 8, 10, 12 et 22 ms après le début du passage du courant (plan YZ). On observe une amplitude de déplacement plus élevée dans la zone inférieure, de plus forte conductivité électrique.}
 	\label{figElastoEMExpResBicoucheConduc} 
\end{figure}

\paragraph{Résultats et discussions}
Les champs de déplacement à 0, 2, 4, 6, 8 et 10 millisecondes après injection du courant sont représentées sur la figure \ref{figElastoEMExpResBicoucheConduc}. La première image n'indique aucun déplacement. Celles à 2 et 4 ms montrent clairement une amplitude de déplacement plus importante dans la partie inférieure que dans la partie supérieure. Les images à 8 et 10 ms, prises au moment où le courant passe dans un sens inverse par rapport aux deux images précédentes, montrent également une amplitude de déplacement plus importante dans la partie inférieure. Les images suivantes, dont celle à 22 ms, sont faites de mouvements plus complexes à interpréter et ne montrent aucune différence d'amplitude entre les deux parties. D'autre part, on n'observe aucun déplacement dans la zone proche de la sonde ultrasonore, probablement à cause du contact entre celle-ci et le fantôme qui gêne tout déplacement.

\subsection{Bilan partiel}
A travers deux expériences, nous avons vu que la conductivité électrique avait une influence sur les déplacements induits force de Lorentz. On peut ainsi envisager des moyens de faire des images de conductivité électrique grâce à cette méthode.

\section{Bilan du chapitre}

\subsection{Principaux résultats}
Le but de ce chapitre était d'étudier la génération d'ondes de cisaillement en appliquant un courant électrique dans un tissu placé dans un champ magnétique.

La partie théorique a permis de voir que la force de Lorentz peut générer des ondes de compression et de cisaillement dans un milieu homogène. Pour un champ magnétique uniforme, il faut avoir une variation de la conductivité électrique ou du champ électrique dans la direction du champ magnétique pour générer des ondes. Il est cependant difficile de vérifier expérimentalement ces résultats à cause des effets de bords qui donnent lieu aussi à des ondes de cisaillement.

Les premières expériences se sont d'abord attachées à élaborer un dispositif de détection des ondes de cisaillement et de s'assurer que les ondes créées sont bien dues à la force de Lorentz. Après plusieurs dispositifs non décrits ici, nous avons choisi un échographe Verasonics fonctionnant en mode ultrarapide. La génération des ondes de cisaillement se sont faites par l'injection d'un courant électrique basse fréquence (50-100 Hz) d'amplitude de quelques centaines de milliampères dans un échantillon placé dans un champ magnétique créé par un aimant permanent. Cela a permis d'observer des déplacements qui se propagent sous forme d'ondes de cisaillement. Puis, nous avons vérifié que le déplacement était bien créé au sein de l'échantillon par la force de Lorentz et non pas par contact avec les électrodes ou le support, en utilisant un champ magnétique localisé. Enfin, nous avons retiré le champ magnétique pour vérifier que le phénomène observé n'est pas dû à l'effet Joule.

Par la suite, nous avons voulu voir si on peut utiliser ces ondes de cisaillement pour réaliser des expériences d'élastographie dynamique, technique basée sur la vitesse de propagation d'ondes de cisaillement. Nous avons pu calculer une vitesse de propagation dans un fantôme homogène qui donnait une élasticité crédible. Puis, nous avons pu voir dans un autre fantôme une vitesse de propagation nettement differente entre les deux couches d'élasticités différentes. Enfin, des manipulations sur un organe \textit{ex-vivo} ont été réalisées, avec toutefois une profondeur d'exploration moindre.

D'autre part, la théorie indique que les inhomogénéités de conductivité électrique donnent lieu à des ondes de cisaillement. Nous avons donc regardé s'il était possible de donner des informations sur la conductivité électrique des tissus à partir des déplacements générés par la force de Lorentz. Nous avons pu en particulier localiser de cette façon une inhomogénéité de conductivité électrique, à savoir un fil électrique conducteur dans un fantôme. D'autre part, nous avons observé deux amplitudes différentes dans un fantôme composé de deux couches de conductivités électriques différentes. 

\subsection{Avenir de la méthode}
Pour l'instant, les deux techniques présentées ne sont pas directement applicable au corps humain. En particulier, l'intensité du courant électrique, de quelques centaines de milliampères, est trop élevée par rapport au maximum acceptable dans le corps humain, de quelques milliampères. Cependant, la puissance du champ magnétique peut être facilement accrue d'un facteur dix en utilisant le champ magnétique d'un IRM. Une voie à explorer est donc de tester la méthode par IRM. En effet, un champ magnétique intense y est déjà présent ; et comme un champ magnétique variable induit un champ électrique, il serait possible d'utiliser les bobines de champs pour induire le courant électrique. De plus, de nombreuses techniques d'élastographie par IRM existent, ce qui permet d'avoir des algorithmes de visualisation de la propagation des ondes de cisaillement.

Une autre piste de développement serait d'induire un courant électrique grâce à un champ magnétique variable, comme cela est fait dans la technique de Tomographie Magnéto-Acoustique par Induction Magnétique. Le courant électrique risque cependant de ne pas être induit en un endroit précis mais dans tout un volume. Par exemple, on pourrait induire des ondes de cisaillement dans le cerveau de cette façon, d'autant plus si le courant électrique est créé par induction magnétique.

Il est difficile de prévoir aujourd'hui quel est l'aspect le plus intéressant de l'imagerie d'ondes de cisaillement générées par force de Lorentz. Nous avons proposé d'utiliser le phénomène pour des techniques d'élastographie dynamique, et ainsi observer des ondes de cisaillement en profondeur -- mais les variations de conductivité électrique dans les tissus peuvent perturber la localisation exacte de la source.

L'autre aspect est la détection d'inhomogénéités de conductivité électrique. Les dernières manipulations a montré que l'on peut détecter des inhomogénéités. Plusieurs pistes se présentent : essayer de localiser la source des ondes de cisaillement par retournement temporel ; ne regarder que les premières images de déplacements, juste après injection de courant ; augmenter la fréquence de l'impulsion ultrasonore, afin de garder un déplacement par force de Lorentz mais empêcher toute propagation par onde de cisaillement ; envoyer un signal blanc et à partir des mouvements d'amplitude statistiquement plus élevées, en déduire les zones les plus conductrices. Cette dernière approche profitera certainement des développements de l'élastographie passive au sein du laboratoire qui utilise des techniques de corrélation puissantes.

\chapter{Conclusion}
\label{secConclusion}

\section{Bilan général}
Cette thèse s'est intéressée à plusieurs applications de la force de Lorentz pour l'acoustique médicale. L'intérêt de la force de Lorentz est de relier un mouvement mécanique à un courant électrique via un champ magnétique. Cette force a été appliquée en acoustique médicale, qui regroupe l'utilisation d'ondes acoustiques pour l'imagerie médicale et la thérapie. Les trois principales applications étudiées sont liées par la même physique, mais ont chacune leurs spécificités propres : la métrologie et la mécanique des fluides pour la première partie, l'impédance électrique des tissus et les méthodes de reconstruction ultrasonore pour la deuxième, ou encore l'élastographie pour la troisième.

La première application envisagée dans cette thèse a été l'étude d'un hydrophone ultrasonore basé sur la force de Lorentz. Une étude théorique a donné une relation entre la vitesse d'un fluide due à une onde acoustique et la vitesse d'un fil placé dans ce fluide. A partir de cette relation, une preuve de concept expérimentale a été envisagée. Puis, un prototype d'hydrophone a été fabriqué et caractérisé en suivant notamment une norme internationale, sur des aspects de résolution spatiale, de réponse fréquentielle, de directivité, de sensibilité, de résistance à haute pression et d'influence de la taille des fils. Le principal avantage de l'hydrophone réside dans sa résistance à haute pression, tandis que le point principal à améliorer concerne la réponse fréquentielle haute, limitée pour l'instant à environ 1 MHz.

La deuxième application étudiée a été une technique d'imagerie médicale introduite sous le nom d' ``Imagerie par Effet Hall'' en 1998, renommée ``Tomographie d'Impédance Electrique par Force de Lorentz'' dans cette thèse car le phénomène étudié ne relève pas strictement de l'effet Hall. Cette technique permet d'obtenir une image de l'impédance électrique des tissus, qui offre potentiellement un bon contraste entre les différents types de tissus, les tissus sains ou pathologiques, les tissus chauffés ou non, etc. Elle utilise des ultrasons focalisés dans un tissu biologique lui-même placé dans un champ magnétique. La vibration due aux ultrasons dans un champ magnétique induit par force de Lorentz un courant électrique. La mesure de ce courant permet d'en déduire les variations de conductivité électrique le long d'une ligne ultrasonore. Dans cette thèse, un dispositif expérimental a été élaboré. Après avoir objectivé le signal, des images de conductivité électrique de morceau de gélatine et de muscle animal ont été réalisées. Par la suite, nous avons essayé d'observer la formation d'une lésion thermique grâce au dispositif. Les images obtenues sont de faible résolution spatiale à cause de la taille importante du faisceau ultrasonore et de la fréquence utilisée mais montrent néanmoins le potentiel offert par la méthode.

Enfin, une troisième application a été de générer des ondes de cisaillement par force de Lorentz. Dans cette technique, on injecte un courant électrique dans un tissu biologique placé dans un champ magnétique. La force de Lorentz induit alors un déplacement, qui se propage à basse fréquence sous la forme d'une onde de cisaillement. Cette onde de cisaillement peut alors être utilisée pour réaliser des expériences d'élastographie. Le travail ici a consisté à concevoir la technique, réaliser une étude théorique et mettre en place un dispositif expérimental. Nous nous sommes tout d'abord assuré que le signal que nous détections était bien créé par force de Lorentz. Puis, nous avons fait de l'élastographie ultrasonore par le principe, d'abord sur des échantillons de gélatine, puis sur des tissus biologiques animaux. Par ailleurs, nous avons remarqué lors des expériences la possibilité d'obtenir des images de conductivité électrique par la méthode. Des expériences sur des échantillons de gélatine avec inclusion métallique ont été réalisé comme preuves de concept.

\section{Chronologie}
Le début de cette thèse s'est penché sur la tomographie d'impédance électrique par force de Lorentz, en se basant sur les précédents travaux d'Amalric Montalibet et de Jacques Jossinet au sein du laboratoire. L'élaboration d'un dispositif fonctionnel a pratiquement pris une année, notamment avec l'aide précieuse de Jean-Martial Mari : le courant électrique induit est en effet très faible et facilement perturbé par des parasites extérieurs. Les premières mesures étaient faites sur des fantômes de gélatine salée. A cause de la faiblesse du courant électrique, ces fantômes étaient souvent remplacés par un fil électrique afin de vérifier les différents éléments de l'expérience (ultrasons, champ magnétique, amplification) grâce à sa conductivité électriqué élevée. Nous nous sommes alors aperçus que cela constituait un hydrophone rudimentaire. J'ai donc continué, en parallèle à la technique d'imagerie, à étudier ce dispositif. Un modèle hydrodynamique a alors été élaboré avec Bruno Gilles pour relier le courant mesuré à la pression ultrasonore. Nous avons finalement opté pour un design à fil long couplé à une technique de tomographie pour assurer une bonne résolution spatiale. Par la suite, nous avons voulu réaliser un prototype fonctionnel qui a été caractérisé en suivant les recommandations d'une norme IEC, avec l'aide d'un stagiaire, Benjamin Roussel. Enfin, deux stagiaires, Fayssel Hamimi puis Mickaël Parisi, ont participé à l'élaboration de simulations avec le logiciel Comsol pour mieux comprendre les phénomènes se produisant autour du fil électrique -- certaines expériences donnant des résultats non prévus par notre modèle hydrodynamique. Cette partie consacrée à l'hydrophone a représenté la majorité de mes communications scientifiques, avec des présentations à cinq congrès internationaux dont deux \textit{proceedings} et deux publications, dans les revues ``Applied Physics Letters'' et ``IEEE Transactions in Ultrasonics, Ferroelectrics and Frequency Controls'' (actuellement en cours de révision).

Les résultats sur l'imagerie d'impédance électrique ne donnaient cependant pas des résultats aussi bons qu'espérés, et il m'a été difficile de présenter des résultats particulièrement meilleurs que ceux déjà publiés. Les images résultent d'un compromis entre plusieurs facteurs : une hausse de la fréquence ultrasonore améliore la résolution spatiale mais diminue l'amplitude des signaux, et il a donc fallu se contenter d'une résolution spatiale de l'ordre du centimètre. Ces expériences ont néanmoins mené à un article publié dans ``Innovation and Research in BioMedical engineering''. Nous avons un peu étudié la partie théorique avec un stagiaire, Alexandre Petit, sans que les résultats soient confirmés par les expériences. En parallèle, un chercheur et deux doctorants du département de Mathématiques Appliqués de l'ENS, Habib Ammari, Laurent Seppecher et Pierre Millien, se sont intéressés à une modélisation mathématique du phénomène. Cela a permis de confronter un modèle théorique de la méthode sur des données expérimentales ``réelles''.

Ces résultats présentaient des débouchés moins prometteurs que la méthode en sens ``inverse'', où l'on crée un mouvement à partir d'un courant électrique et d'un champ magnétique. Cette dernière partie de ma thèse a été initiée avec l'arrivée de Stefan Catheline au laboratoire grâce à son expertise en élastographie. Ainsi, au lieu d'utiliser la force de Lorentz en sens inverse à des fréquences ultrasonores, comme le fait l'équipe de Bin He depuis plusieurs années, nous avons choisi de se placer à basse fréquence pour créer des ondes de cisaillement. Avec l'aide de Rémi Souchon et d'une stagiaire, Sandra Montalescot, nous avons réussi après quelques temps à détecter l'onde de cisaillement recherchée -- mais des effets stroboscopiques rendaient les résultats difficilement présentables tels quels. Nous avons donc poursuivi le travail avec un transducteur monoélément, notamment suite à l'arrivée de Florian Cartellier en stage. Nous pouvions alors obtenir des images en déplaçant le transducteur ligne par ligne. Cela prenait du temps, mais on obtenait des films de propagation d'ondes convaincants. L'objectivation du signal a été un obstacle, car nous n'étions pas sûrs que les mouvements étaient vraiment dus à un déplacement du fantôme et non pas des électrodes. Puis, avec l'arrivée de l'échographe ultrarapide Verasonics, sur lequel Ali Zorgani et Rémi Souchon ont beaucoup travaillé, nous avons pu faire des images en mode ultrarapide. Le dispositif était alors prêt pour faire une caractérisation avancée. Cela a convaincu l'INSERM de déposer un brevet sur ce sujet, et nous avons ensuite communiqué extensivement sur ce sujet à la fin de ma thèse : présentation à trois congrès internationaux dont un \textit{proceeding} et une publication (actuellement en cours d'écriture).

\section{Communications scientifiques}
\renewcommand{\labelitemi}{\textbullet}
\paragraph{Articles dans des journaux à comité de lecture}
\begin{itemize}
\item \textit{Imaging of Shear Waves Induced by Lorentz Force in Soft Tissues},\\{P Grasland-Mongrain}, S Catheline, R Souchon, F Cartellier, A Zorgani, JY Chapelon, C Lafon,\\{Physical Review Letters} (soumis)
\item \textit{Lorentz Force Hydrophone Characterisation},\\{P Grasland-Mongrain}, JM Mari, B Gilles, B Roussel, A Poizat, JY Chapelon, C Lafon,\\{IEEE Ultrasonics, Ferroelectrics and Frequency Control}, 2014 (in press)
\item \textit{Lorentz Force Electrical Impedance Tomography},\\{P Grasland-Mongrain}, JM Mari, JY Chapelon, C Lafon,\\{Innovation and Research in BioMedical engineering}, 34 (4), pp 357-360, 2013
\item \textit{Electromagnetic hydrophone with tomographic system for absolute velocity field mapping},\\{P Grasland-Mongrain}, JM Mari, B Gilles, JY Chapelon, C Lafon,\\{Applied Physics Letters}, 100 (24), 2012
\item \textit{Ordered packings of bubbles in columns of square cross-section},\\S Hutzler, J Barry, {P Grasland-Mongrain}, D Smyth, D Weaire,\\{Colloids and Surfaces A: Physicochemical and Engineering Aspects}, 344 (1), 2009 (travaux précédant la thèse)
\end{itemize}
 
\paragraph{Brevets} \footnote{Inventeurs par ordre alphabétique}
\begin{itemize}
\item \textit{Shear Wave Imaging Method and Installation for Collecting Information on a Soft Solid},\\S Catheline, JY Chapelon, P Grasland-Mongrain, C Lafon, R Souchon, \\Europe n° 13305987.3, 11 juillet 2013
\end{itemize}

%\newpage
\paragraph{Présentations à des congrès internationaux} \footnote{En gras : personne ayant fait la présentation}
\begin{itemize}
\item \textit{Ultrasound velocity mapping with Lorentz Force Hydrophone},\\\textbf{P Grasland-Mongrain}, JM Mari, B Gilles, B Roussel, A Poizat, JY Chapelon, C Lafon,\\{Acoustical Society of America Symposium}, San Francisco, 1-4 décembre 2013
\item \textit{Thermal lesion imaging using Lorentz force : Proofs of concept},\\\textbf{P Grasland-Mongrain}, S Catheline, R Souchon, F Cartellier, A Zorgani, JY Chapelon, C Lafon,\\{Acoustical Society of America Symposium}, San Francisco, 1-4 décembre 2013
\item \textit{Imaging of Shear Waves Induced by Lorentz Force in Soft Tissues},\\\textbf{P Grasland-Mongrain}, S Catheline, R Souchon, F Cartellier, A Zorgani, JY Chapelon, C Lafon,\\{International Tissue Elasticity Conference}, Lingfield (UK), 1-4 octobre 2013
\item \textit{Imaging of Shear Waves Induced by Lorentz Force} - avec proceeding,\\\textbf{P Grasland-Mongrain}, S Catheline, R Souchon, F Cartellier, A Zorgani, S Montalescot, JY Chapelon, C Lafon,\\{IEEE Ultrasonics, Ferroelectrics and Frequency Control}, Prague, 21-25 juillet 2013
\item \textit{Lorentz Force Hydrophone Prototype Characterization},\\\textbf{P Grasland-Mongrain}, JM Mari, B Gilles, B Roussel, A Poizat, JY Chapelon, C Lafon,\\{IEEE Ultrasonics, Ferroelectrics and Frequency Control}, Prague, 21-25 juillet 2013
\item \textit{Electromagnetic hydrophone for high-intensity focused ultrasound (HIFU) measurement} - avec proceeding,\\\textbf{P Grasland-Mongrain}, JM Mari, B Gilles, JY Chapelon, C Lafon,\\{Acoustical Society of America Symposium, Montréal, 2-7 juin 2013}
\item \textit{Electromagnetic Hydrophone for High Intensity Focused Ultrasound Measurement},\\P Grasland-Mongrain, JM Mari, B Gilles, JY Chapelon, \textbf{C Lafon},\\{International Symposium on Thérapeutic Ultrasound}, Shanghai, 12-15 mai 2013 (poster)
\item \textit{Electromagnetic Hydrophone Characterization},\\P Grasland-Mongrain, JM Mari, B Gilles, JY Chapelon, \textbf{C Lafon},\\{International Congress on Ultrasonics}, Singapour, 2-5 mai 2013
\item \textit{Electromagnetic tomographic ultrasonic sensor} - avec proceeding,\\\textbf{P Grasland-Mongrain}, JM Mari, B Gilles, JY Chapelon, C Lafon,\\{Acoustical Society of America Symposium}, Hong-Kong, 13-18 mai 2012
\item \textit{Detection of fat tissues with ultrasonically-induced Lorentz force},\\\textbf{P Grasland-Mongrain}, JM Mari, C Lafon,\\{Acoustics}, Nantes, 23-27 avril 2012
\item \textit{Combination of shape-constrained and inflation deformable models with application to the segmentation of the left atrial appendage} - avec proceeding,\\\textbf{P Grasland-Mongrain}, J Peters, O Ecabert,\\{IEEE International Symposium on Biomedical Imaging}, Rotterdam, 14-17 avril 2010 (poster) (travaux précédant la thèse)
\end{itemize}

\paragraph{Présentations à des congrès nationaux} \footnote{En gras : personne ayant fait la présentation}
\begin{itemize}
\item \textit{Lorentz force prototype to characterize HIFU pressure fields},\\\textbf{P Grasland-Mongrain}, JM Mari, JY Chapelon, C Lafon,\\{Colloque Recherche en Imagerie et Technologies pour la Santé}, Bordeaux, 8-11 avril 2013 (poster)
\item \textit{Détection de gradients de conductivité électrique par ultrasons},\\\textbf{P Grasland-Mongrain}, JM Mari, JY Chapelon, C Lafon,\\{Colloque Recherche en Imagerie et Technologies pour la Santé}, Bordeaux, 8-11 avril 2013 (poster)
\item \textit{Détection de gradients de conductivité électrique par ultrasons},\\\textbf{P Grasland-Mongrain}, JM Mari, C Lafon,\\Journées Imag'in vivo, Lyon, 11-13 décembre 2012 (poster)
\item \textit{Détection de gradients de conductivité électrique par ultrasons},\\\textbf{P Grasland-Mongrain}, JM Mari, C Lafon,\\Journées Campus Santé, Lyon, 25-26 juin 2012 (poster)
\item \textit{Détection de tissus adipeux par force de Lorentz induite par ultrasons},\\\textbf{P Grasland-Mongrain}, JM Mari, JY Chapelon, C Lafon,\\Journées Signal et Image en Acoustique Médicale, Paris, 1-2 décembre 2011
\item \textit{Mesure d'anisotropie de conductivité électrique par interaction acousto-magnétique},\\\textbf{P Grasland-Mongrain}, JM Mari, C Lafon,\\Journées d'Acoustique Physique, Sous-Marine et Ultrasonore, Lille, 8-10 juin 2011
\end{itemize}

\paragraph{Séminaires grand public}
\begin{itemize}
\item \textit{L'imagerie médicale aujourd'hui... et demain ?},\\Ecole Normale Supérieure de Cachan, 2 décembre 2013 \\Ecole Normale Supérieure de Lyon, 30 janvier 2013

\item \textit{La recherche en imagerie médicale},\\Ecole Normale Supérieure de Cachan, 10 mars 2011, 16 février 2012, 14 février 2013 (3 séminaires)
\end{itemize}

\paragraph{Prix et distinctions}
\begin{itemize}
\item \textit{2e prix meilleure présentation étudiante}, Acoustical Society of America Symposium, Montréal, 2-7 juin 2013
\item \textit{Prix jeune chercheur}, Colloque Recherche en Imagerie et Technologies pour la Santé, Bordeaux, 8-11 avril 2013
\item \textit{Prix du meilleur poster}, Journées Imag'in vivo, Lyon, 11-13 décembre 2012
\item \textit{Prix du meilleur poster}, Journées Campus Santé, Lyon, 25-26 juin 2012
\end{itemize}

\appendix
\clearpage
\chapter*{Annexe : Calcul du déplacement pour une force ponctuelle}
\addcontentsline{toc}{chapter}{Annexe : Calcul du déplacement pour une force ponctuelle}
\label{secAnnexeForcePonctuelle}

Il est intéressant de connaître la solution de l'équation de Navier pour une force ponctuelle, avec~$\mathbf{f}(t)$ une fonction temporelle quelconque. On suppose pour le moment qu'elle est orientée une direction X.

\begin{empheq}[box=\mybluebox]{align}
\rho\frac{d^2\mathbf{u}}{dt^2} - (K + \frac{4}{3}\mu) \grad \Div \mathbf{u} - \mu \rot \rot \mathbf{u} = \mathbf{f}(t) \delta(\mathbf{r})
\label{eqAnnexeSource}
\end{empheq}

Le calcul de cette équation est donné dans un livre de Aki et Richards \cite{aki1980quantitative}. Le principe consiste décomposer le vecteur déplacement comme $\mathbf{u} = \grad(\phi) + \rot(\mathbf{\psi})$, et on cherche les potentiels~$\phi$ et~$\mathbf{\psi}$ associés. On commence par effectuer la même décomposition pour le terme source, afin d'obtenir les potentiels dits de Helmholtz correspondants. On peut montrer que l'on a la décomposition suivante où l'on note $r = \|\mathbf{r}\|$ la norme du vecteur $\mathbf{r}$ :

\begin{empheq}[box=\mybluebox]{align}
 \mathbf{f}(t)\delta(\mathbf{r}) =\grad(\Phi_f) + \rot(\mathbf{\Psi_f})
\label{eqAnnexeDecompositionF}
\end{empheq}

avec :

\begin{empheq}[box=\mybluebox]{align}
\begin{cases}
			\Phi_f (\mathbf{r},t) = - \frac{\|f(t)\|}{4 \, \pi} \; \frac{\partial}{\partial x} \left( \frac{1}{r} \right) \\			
			\mathbf{\Psi_f} (\mathbf{r},t) =   \frac{\|f(t)\|}{4 \, \pi}	\left( 0, \; \frac{\partial}{\partial  z} \left( \frac{1}{r} \right) , \; -\frac{\partial}{\partial  y}  \left( \frac{1}{r} \right) \right)
\end{cases}
\label{eqAnnexePotentielsHelmholtz}
\end{empheq}

On peut alors réécrire~\eqref{eqAnnexeSource} comme deux équations portant sur les potentiels~$\phi$ et~$\mathbf{\psi}$.

\begin{empheq}[box=\mybluebox]{align}
\begin{cases}
			\rho \frac{d^2\phi}{dt^2}  - (K + \frac{4}{3}\mu) \; \Delta \phi  = - \frac{\|f(t)\|}{4 \, \pi} \; \frac{\partial}{\partial x} \left( \frac{1}{r} \right) \\								
			\rho \frac{d^2\mathbf{\psi}}{dt^2}  -  \mu \; \Delta \mathbf{\psi} = \frac{\|f(t)\|}{4 \, \pi}	\left( 0, \; \frac{\partial}{\partial  z} \left( \frac{1}{r} \right) , \; -\frac{\partial}{\partial  y}  \left( \frac{1}{r} \right) \right)
\end{cases}
\label{eqAnnexePotentielsU}
\end{empheq}

L'étape suivante consiste à résoudre~\eqref{eqAnnexePotentielsU}. On obtient alors les potentiels suivants, où l'on a posé~$v_p~=~\sqrt{\frac{K + \frac{4}{3}\mu}{\rho}}$ et~$v_s~=~\sqrt{\frac{\mu}{\rho}}$.

\begin{empheq}[box=\mybluebox]{align}
\left\{
\begin{aligned}
			&\phi(\mathbf{r},t)  = - \frac{1}{4  \pi  \rho} \frac{\partial}{\partial x} \left( \frac{1}{r} \right) \int_{0}^{\frac{r}{\alpha}} \tau \, f(t-\tau) \; \mathrm{d}\tau 
			\\									
			&\mathbf{\psi}(\mathbf{r},t)  =  - \frac{1}{4  \pi  \rho}   \left( 0, \; \frac{\partial}{\partial  z} \left( \frac{1}{r} \right) , \; -\frac{\partial}{\partial  y}  \left( \frac{1}{r} \right) \right) \int_{0}^{\frac{r}{\beta}} \tau \, F(t-\tau) \; \mathrm{d}\tau 
\end{aligned}
\right.
\end{empheq}

On passe ensuite à des directions quelconques $i,j$ à la place de X, Y et Z. La dernière étape du calcul consiste à reconstruire~$\mathbf{u} = \grad(\phi) + \rot(\mathbf{\psi})$, et avec $\frac{r_i}{r} =  \frac{\partial r}{\partial r_i}$ et $\frac{\partial^2}{\partial r_i \, \partial r_j} \; \frac{1}{r} = \left( \frac{3 r_i r_j}{r^2} - \delta_{ij} \right)$ :
\begin{empheq}[box=\mybluebox]{align}
\begin{aligned}
 u_i(\mathbf{r},t) & \left. =  \frac{1}{4 \pi \rho r^3}  \left( \frac{3 r_i r_j}{r^2} - \delta_{ij} \right) \int_{r/ v_p}^{r/ v_s} \tau \, f(t-\tau) \; \mathrm{d}\tau  \qquad \; \; \: \right\} \mathrm{champ~proche}
\\
& %\left.
\begin{aligned}
& \left. + \frac{1}{4 \pi \rho v_p^2 r} \; \frac{r_i r_j}{r^2} \; f \! \left( t-\frac{r}{v_p} \right) \qquad \qquad \quad \, \right\} \mathrm{onde~de~compression}
\\
& \left. + \frac{1}{4 \pi \rho v_s^2 r}  \left( \frac{r_i r_j}{r^2} - \delta_{ij} \right)  \; f \! \left( t-\frac{r}{v_s} \right) \quad \right\} \mathrm{onde~de~cisaillement}
\end{aligned} %\right\} \mathrm{champ~lointain}
\label{eqAnnexeGreen}
\end{aligned} 
\end{empheq}

\bibliographystyle{plain3}
\bibliography{biblio}

\begin{thebibliography}{100}

\bibitem{fda1985}
{Guide for measuring and reporting acoustic output of diagnostic ultrasound
  medical devices}.
\newblock {\em Food and Drug Administration}, 510(K), 1985.

\bibitem{medical1993}
{EC Medical Devices Directive (93/42/EEC)}.
\newblock {\em Official Journal of the European Communities}, 36(L169), 1993.

\bibitem{aberg2004skin}
P.~Aberg, I.~Nicander, J.~Hansson, P.~Geladi, U.~Holmgren, and S.~Ollmar.
\newblock {Skin cancer identification using multifrequency electrical
  impedance-a potential screening tool}.
\newblock {\em Biomedical Engineering, IEEE Transactions on},
  51(12):2097--2102, 2004.

\bibitem{ahmed2009minimally}
H.~U. Ahmed, C.~Moore, and M.~Emberton.
\newblock {Minimally-invasive technologies in uro-oncology: the role of
  cryotherapy, HIFU and photodynamic therapy in whole gland and focal therapy
  of localised prostate cancer}.
\newblock {\em Surgical oncology}, 18(3):219--232, 2009.

\bibitem{aki1980quantitative}
K.~Aki and P.~G. Richards.
\newblock {\em {Quantitative seismology}}.
\newblock Freeman San Francisco, 1980.

\bibitem{ammari2008}
H.~Ammari.
\newblock {\em {An introduction to mathematics of emerging biomedical
  imaging}}.
\newblock Springer, 2008.

\bibitem{ammari2012b}
H.~Ammari, J.~Garnier, and W.~Jing.
\newblock {Resolution and stability analysis in acousto-electric imaging}.
\newblock {\em Inverse Problems}, 28(8):084005, 2012.

\bibitem{ammari2004b}
H.~Ammari, O.~Kwon, J.~K. Seo, and E.~J. Woo.
\newblock {T-Scan electrical impedance imaging system for anomaly detection}.
\newblock {\em SIAM Journal on Applied Mathematics}, 65(1):252--266, 2004.

\bibitem{anand2004monitoring}
A.~Anand and P.~J. Kaczkowski.
\newblock {Monitoring formation of high intensity focused ultrasound (HIFU)
  induced lesions using backscattered ultrasound}.
\newblock {\em Acoustics Research Letters Online}, 5:88, 2004.

\bibitem{bacon1982characteristics}
D.~R. Bacon.
\newblock {Characteristics of a PVDF membrane hydrophone for use in the range
  1-100 MHz}.
\newblock {\em Sonics and Ultrasonics, IEEE Transactions on}, 29(1):18--25,
  1982.

\bibitem{barnett2000international}
S.~B. Barnett, G.~R. Ter~Haar, M.~C. Ziskin, H.-D. Rott, F.~A. Duck, and
  K.~Maeda.
\newblock {International recommendations and guidelines for the safe use of
  diagnostic ultrasound in medicine}.
\newblock {\em Ultrasound in medicine \& biology}, 26(3):355--366, 2000.

\bibitem{bercoff2004supersonic}
J.~Bercoff, M.~Tanter, and M.~Fink.
\newblock {Supersonic shear imaging: a new technique for soft tissue elasticity
  mapping}.
\newblock {\em Ultrasonics, Ferroelectrics and Frequency Control, IEEE
  Transactions on}, 51(4):396--409, 2004.

\bibitem{bernard2007these}
L.~Bernard.
\newblock {\em {Caract{\'e}risation {\'e}lectrique des tissus biologiques et
  calcul des ph{\'e}nom\`{e}nes induits dans le corps humain par des champs
  {\'e}lectromagn{\'e}tiques de fr{\'e}quence inf{\'e}rieure au GHz.}}
\newblock PhD thesis, \'{E}cole Centrale de Lyon, 2007.

\bibitem{bernier1995practical}
C.~A. Bernier, L.~Huntsman, and R.~Martin.
\newblock {A practical approach to measuring an intravascular ultrasonographic
  imaging system beam pattern}.
\newblock {\em Journal of ultrasound in medicine}, 14(5):367--373, 1995.

\bibitem{bossy2005}
E.~Bossy, L.~Sui, T.~W. Murray, and R.~A. Roy.
\newblock {Fusion of conventional ultrasound imaging and acousto-optic sensing
  by use of a standard pulsed-ultrasound scanner}.
\newblock {\em Optics letters}, 30(7):744--746, 2005.

\bibitem{brown2003electrical}
B.~Brown.
\newblock {Electrical impedance tomography (EIT): a review}.
\newblock {\em Journal of medical engineering \& technology}, 27(3):97--108,
  2003.

\bibitem{bucaro1977}
J.~Bucaro, H.~Dardy, and E.~Carome.
\newblock {Fiber-optic hydrophone}.
\newblock {\em Journal of Acoustical Society of America}, 62(5):1302--1304,
  1977.

\bibitem{catheline1998interferometrie}
S.~Catheline.
\newblock {\em {Interf{\'e}rom{\'e}trie-Speckle ultrasonore: Application {\`a}
  la mesure d'{\'e}lasticit{\'e}}}.
\newblock PhD thesis, Universit{\'e} Paris-Diderot-Paris VII, 1998.

\bibitem{cathignol1997}
D.~Cathignol, P.~Faure, and F.~Chavrier.
\newblock {Acoustic field of plane or spherical transducers}.
\newblock {\em Acustica}, 83:410--418, 1997.

\bibitem{cheney1999}
M.~Cheney, D.~Isaacson, and J.~C. Newell.
\newblock {Electrical impedance tomography}.
\newblock {\em SIAM review}, 41(1):85--101, 1999.

\bibitem{cole1941dispersion}
K.~S. Cole and R.~H. Cole.
\newblock {Dispersion and absorption in dielectrics I. Alternating current
  characteristics}.
\newblock {\em The Journal of Chemical Physics}, 9:341--351, 1941.

\bibitem{cox2007}
B.~Cox and P.~Beard.
\newblock {The frequency-dependent directivity of a planar Fabry-Perot polymer
  film ultrasound sensor}.
\newblock {\em IEEE Transactions on Ultrasonics, Ferroelectrics and Frequency
  Control}, 54:394--404, 2007.

\bibitem{culjat2010review}
M.~O. Culjat, D.~Goldenberg, P.~Tewari, and R.~S. Singh.
\newblock {A review of tissue substitutes for ultrasound imaging}.
\newblock {\em Ultrasound in medicine \& biology}, 36(6):861--873, 2010.

\bibitem{darrigol2000electrodynamics}
O.~Darrigol.
\newblock {\em {Electrodynamics from Ampère to Einstein}}.
\newblock Oxford University Press, 2000.

\bibitem{davidson1991ultrasonic}
F.~Davidson.
\newblock {Ultrasonic power balances}.
\newblock In {\em Output measurements for medical ultrasound}, pages 75--90.
  Springer, 1991.

\bibitem{debye1929polar}
P.~J.~W. Debye.
\newblock {\em {Polar molecules}}, volume 172.
\newblock Dover New York, 1929.

\bibitem{dereggi1977}
A.~Dereggi, S.~Edelman, S.~Roth, H.~Warner, and J.~Wynn.
\newblock {Piezoelectric polymer receiving arrays for ultrasonic applications}.
\newblock {\em Journal of Acoustical Society America}, 61:17--18, 1977.

\bibitem{desjouy2013}
C.~Desjouy, A.~Poizat, B.~Gilles, C.~Inserra, and J.-C. Bera.
\newblock {Control of inertial acoustic cavitation in pulsed sonication using a
  real-time feedback loop system}.
\newblock {\em Journal of Acoustical Society of America}, 134(2), 2013.

\bibitem{dirac1931quantised}
P.~Dirac.
\newblock {Quantised Singularities in the Electromagnetic Field}.
\newblock {\em Proceedings of the Royal Society A}, 133(60), 1931.

\bibitem{duck1990physical}
F.~A. Duck.
\newblock {\em {Physical Properties of Tissue: A Comprehensive Reference
  Network}}.
\newblock Academic Press, London, 1990.

\bibitem{einstein1905elektrodynamik}
A.~Einstein.
\newblock {Zur Elektrodynamik Bewegter K{\"o}rper}.
\newblock {\em Annalen der physik}, 322(10), 1905.

\bibitem{etienne1997}
J.~Etienne, L.~Filipczynski, T.~Kujawska, and B.~Zienkiewicz.
\newblock {Electromagnetic hydrophone for pressure determination of shock wave
  pulses}.
\newblock {\em Ultrasound in Medicine and Biology}, 23(5):747--754, 1997.

\bibitem{filipczynski1969}
L.~Filipczynski.
\newblock {Absolute measurements of particle velocity, displacement or
  intensity of ultrasonic pulses in liquids and solids}.
\newblock {\em Acustica}, 21:173--180, 1969.

\bibitem{flax1982}
L.~Flax, J.~Cole, R.~De~Paula, and J.~Bucaro.
\newblock {Acoustically induced birefringence in optical fibers}.
\newblock {\em Optical Society of America, Journal}, 72:1159--1162, 1982.

\bibitem{foucher2006diagnosis}
J.~Foucher, E.~Chanteloup, J.~Vergniol, L.~Castera, B.~Le~Bail, X.~Adhoute,
  J.~Bertet, P.~Couzigou, and de~V.~Ledinghen.
\newblock {Diagnosis of cirrhosis by transient elastography (FibroScan): a
  prospective study}.
\newblock {\em Gut}, 55(3):403--408, 2006.

\bibitem{frerichs2002detection}
I.~Frerichs, J.~Hinz, P.~Herrmann, G.~Weisser, G.~Hahn, T.~Dudykevych,
  M.~Quintel, and G.~Hellige.
\newblock {Detection of local lung air content by electrical impedance
  tomography compared with electron beam CT}.
\newblock {\em Journal of applied physiology}, 93(2):660--666, 2002.

\bibitem{fricke1932}
H.~Fricke.
\newblock {The theory of electrolytic polarization}.
\newblock {\em Dublin Philosophical Magazine}, 14(90):310--318, 1932.

\bibitem{fry1979biological}
F.~J. Fry.
\newblock {Biological effects of ultrasound -- A review}.
\newblock {\em Proceedings of the IEEE}, 67(4):604--619, 1979.

\bibitem{fry1954determination2}
W.~J. Fry and R.~B. Fry.
\newblock {Determination of absolute sound levels and acoustic absorption
  coefficients by thermocouple probes - experiment}.
\newblock {\em The Journal of the Acoustical Society of America}, 26:311, 1954.

\bibitem{fry1954determination1}
W.~J. Fry and R.~B. Fry.
\newblock {Determination of absolute sound levels and acoustic absorption
  coefficients by thermocouple probes - theory}.
\newblock {\em The Journal of the Acoustical Society of America}, 26:294, 1954.

\bibitem{gabriel1996dielectric1}
C.~Gabriel, S.~Gabriel, and E.~Corthout.
\newblock {The dielectric properties of biological tissues: I. Literature
  survey}.
\newblock {\em Physics in medicine and biology}, 41(11):2231, 1996.

\bibitem{gabriel1996dielectric2}
S.~Gabriel, R.~Lau, and C.~Gabriel.
\newblock {The dielectric properties of biological tissues: II. Measurements in
  the frequency range 10 Hz to 20 GHz}.
\newblock {\em Physics in medicine and biology}, 41(11):2251, 1996.

\bibitem{gelet1996treatment}
A.~Gelet, J.~Chapelon, R.~Bouvier, R.~Souchon, C.~Pangaud, A.~Abdelrahim,
  D.~Cathignol, and J.~Dubernard.
\newblock {Treatment of prostate cancer with transrectal focused ultrasound:
  early clinical experience}.
\newblock {\em European urology}, 29(2):174--183, 1996.

\bibitem{gianfelice2003}
D.~Gianfelice, A.~Khiat, M.~Amara, A.~Belblidia, and Y.~Boulanger.
\newblock {MR Imaging--guided Focused US Ablation of Breast Cancer:
  Histopathologic Assessment of Effectiveness - Initial Experience1}.
\newblock {\em Radiology}, 227(3):849--855, 2003.

\bibitem{grant1991electromagnetism}
I.~Grant and W.~Phillipse.
\newblock {\em {Electromagnetism}}.
\newblock 2nd edition, 1991.

\bibitem{grasland2013LFEIT}
P.~Grasland-Mongrain, J.-M. Mari, J.-Y. Chapelon, and C.~Lafon.
\newblock {Lorentz Force Electrical Impedance Tomography}.
\newblock {\em Innovation and Research in BioMedical engineering}, 2013 (in
  press).

\bibitem{grimnes2002interface}
S.~Grimnes, O.~Rikshospitalet, and H.~P. Schwan.
\newblock {Interface phenomena and dielectric properties of biological tissue}.
\newblock {\em Encyclopedia of Surface and Colloid Science}, 2002.

\bibitem{haemmerich2003}
D.~Haemmerich, S.~Staelin, J.~Tsai, S.~Tungjitkusolmun, D.~Mahvi, and
  J.~Webster.
\newblock {In vivo electrical conductivity of hepatic tumours}.
\newblock {\em Physiological measurement}, 24(2):251, 2003.

\bibitem{halbach1980}
K.~Halbach.
\newblock {Design of permanent multipole magnets with oriented rare earth
  cobalt material}.
\newblock {\em Nuclear Instruments and Methods}, 169:1--10, February 1980.

\bibitem{hall1977progress}
D.~C. Hall, M.~K. Goldstein, and G.~H. Stein.
\newblock {Progress in manual breast examination}.
\newblock {\em Cancer}, 40(1):364--370, 1977.

\bibitem{halter2008electrical}
R.~J. Halter, A.~Schned, J.~Heaney, A.~Hartov, S.~Schutz, and K.~D. Paulsen.
\newblock {Electrical impedance spectroscopy of benign and malignant prostatic
  tissues}.
\newblock {\em The Journal of urology}, 179(4):1580--1586, 2008.

\bibitem{hanafy1991}
A.~Hanafy and C.~Zanelli.
\newblock {Quantitative real-time pulsed Schlieren imaging of ultrasonic
  waves}.
\newblock In {\em Ultrasonics Symposium}, pages 1223--1227. IEEE, 1991.

\bibitem{harris1988hydrophone}
G.~R. Harris.
\newblock {Hydrophone measurements in diagnostic ultrasound fields}.
\newblock {\em Ultrasonics, Ferroelectrics and Frequency Control, IEEE
  Transactions on}, 35(2):87--101, 1988.

\bibitem{harris2000}
G.~Harris, R.~Preston, and A.~DeReggi.
\newblock {The impact of piezoelectric PVDF on medical ultrasound exposure
  measurements, standards, and regulations}.
\newblock {\em Ultrasonics, Ferroelectrics and Frequency Control, IEEE
  Transactions on}, 47(6):1321--1335, 2000.

\bibitem{hodnett1997strategy}
M.~Hodnett and B.~Zeqiri.
\newblock {A strategy for the development and standardisation of measurement
  methods for high power/cavitating ultrasonic fields: review of high power
  field measurement techniques}.
\newblock {\em Ultrasonics sonochemistry}, 4(4):273--288, 1997.

\bibitem{hu2011}
G.~Hu, E.~Cressman, and B.~He.
\newblock {Magnetoacoustic imaging of human liver tumor with magnetic
  induction}.
\newblock {\em Applied physics letters}, 98:023703, 2011.

\bibitem{hu2011magnetoacoustic}
G.~Hu and B.~He.
\newblock {Magnetoacoustic imaging of electrical conductivity of biological
  tissues at a spatial resolution better than 2 mm}.
\newblock {\em PloS one}, 6(8):e23421, 2011.

\bibitem{jackson1998classical}
J.~D. Jackson.
\newblock {\em {Classical Electrodynamics}}.
\newblock John Wiley and Sons, 3d edition, 1998.

\bibitem{jespersen1998}
S.~K. Jespersen, J.~E. Wilhjelm, and H.~Sillesen.
\newblock {Multi-angle compound imaging}.
\newblock {\em Ultrasonic imaging}, 20(2):81--102, 1998.

\bibitem{jossinet1998phenomenology}
J.~Jossinet, B.~Lavandier, and D.~Cathignol.
\newblock {The phenomenology of acousto-electric interaction signals in aqueous
  solutions of electrolytes}.
\newblock {\em Ultrasonics}, 36(1):607--613, 1998.

\bibitem{jossinet2002electrical}
J.~Jossinet, E.~Marry, and A.~Montalibet.
\newblock {Electrical impedance endo-tomography: imaging tissue from inside}.
\newblock {\em Medical Imaging, IEEE Transactions on}, 21(6):560--565, 2002.

\bibitem{jossinet1999review}
J.~Jossinet and M.~Schmitt.
\newblock {A review of parameters for the bioelectrical characterization of
  breast tissue}.
\newblock {\em Annals of the New York Academy of Sciences}, 873(1):30--41,
  1999.

\bibitem{jossinet2005impedance}
J.~Jossinet, C.~Trillaud, and S.~Chesnais.
\newblock {Impedance changes in liver tissue exposed in vitro to high-energy
  ultrasound}.
\newblock {\em Physiological measurement}, 26(2):S49, 2005.

\bibitem{kennedy2005high}
J.~E. Kennedy.
\newblock {High-intensity focused ultrasound in the treatment of solid
  tumours}.
\newblock {\em Nature reviews cancer}, 5(4):321--327, 2005.

\bibitem{kennedy2004high}
J.~Kennedy, F.~Wu, G.~Ter~Haar, F.~Gleeson, R.~Phillips, M.~Middleton, and
  D.~Cranston.
\newblock {High-intensity focused ultrasound for the treatment of liver
  tumours}.
\newblock {\em Ultrasonics}, 42(1):931--935, 2004.

\bibitem{klann2005}
M.~Klann and C.~Koch.
\newblock {Measurement of spatial cross sections of ultrasound pressure fields
  by optical scanning means}.
\newblock {\em Ultrasonics, Ferroelectrics and Frequency Control, IEEE
  Transactions on}, 52(9):1546--1554, 2005.

\bibitem{korpel1987}
A.~Korpel, D.~Mehrl, and H.~Lin.
\newblock {Schlieren imaging of sound fields}.
\newblock In {\em IEEE Ultrasonics Symposium}, 1987.

\bibitem{kwon2002}
O.~Kwon, E.~J. Woo, J.-R. Yoon, and J.~K. Seo.
\newblock {Magnetic resonance electrical impedance tomography (MREIT):
  simulation study of J-substitution algorithm}.
\newblock {\em Biomedical Engineering, IEEE Transactions on}, 49(2):160--167,
  2002.

\bibitem{landau1987}
L.~Landau and E.~Lifshitz.
\newblock {\em {Fluid Mechanics}}, volume~6 of {\em Course of Theoretical
  Physics}.
\newblock Butterworth-Heinemann, 2nd edition, 1987.

\bibitem{lavandier2009experimental}
B.~Lavandier, A.~Gleizal, and J.-C. B{\'e}ra.
\newblock {Experimental assessment of calvarial bone defect re-ossification
  stimulation using low-intensity pulsed ultrasound}.
\newblock {\em Ultrasound in medicine \& biology}, 35(4):585--594, 2009.

\bibitem{lavandier2000}
B.~Lavandier, J.~Jossinet, and D.~Cathignol.
\newblock {Experimental measurement of the acousto-electric interaction signal
  in saline solution}.
\newblock {\em Ultrasonics}, 38(9):929--936, 2000.

\bibitem{lionheart1997electrical}
W.~R. Lionheart, F.~J. Lidgey, C.~N. McLeod, K.~S. Paulson, M.~K. Pidcock, and
  Y.~Shi.
\newblock {Electrical Impedance Tomography for high speed chest imaging}.
\newblock {\em Physica Medica}, 13(1):247--249, 1997.

\bibitem{martin1980use}
C.~Martin and A.~Law.
\newblock {The use of thermistor probes to measure energy distribution in
  ultrasound fields}.
\newblock {\em Ultrasonics}, 18(3):127--133, 1980.

\bibitem{martin1986ultrasound}
X.~Martin, J.~Mestas, D.~Cathignol, J.~Margonari, and J.~Dubefnard.
\newblock {Ultrasound stone localisation for extracorporeal shock wave
  lithotripsy}.
\newblock {\em British journal of urology}, 58(2-4):349--352, 1986.

\bibitem{mcadams1995tissue}
E.~McAdams and J.~Jossinet.
\newblock {Tissue impedance: a historical overview}.
\newblock {\em Physiological measurement}, 16(3A):A1, 1995.

\bibitem{mcadams1996problems}
E.~McAdams and J.~Jossinet.
\newblock {Problems in equivalent circuit modelling of the electrical
  properties of biological tissues}.
\newblock {\em Bioelectrochemistry and bioenergetics}, 40(2):147--152, 1996.

\bibitem{melodelima2009thermal}
D.~Melodelima, W.~A. N'Djin, H.~Parmentier, S.~Chesnais, M.~Rivoire, and J.-Y.
  Chapelon.
\newblock {Thermal ablation by high-intensity-focused ultrasound using a toroid
  transducer increases the coagulated volume. Results of animal experiments}.
\newblock {\em Ultrasound in medicine \& biology}, 35(3):425--435, 2009.

\bibitem{montalibet2002these}
A.~Montalibet.
\newblock {\em {Etude du couplage acousto-magn{\'e}tique: d{\'e}tection des
  gradients de conductivit{\'e} {\'e}lectrique en vue de la caract{\'e}risation
  tissulaire}}.
\newblock PhD thesis, Institut Nationale des Sciences Appliqu{\'e}es de Lyon,
  2002.

\bibitem{montalibet2002scanning}
A.~Montalibet, J.~Jossinet, and A.~Matias.
\newblock {Scanning electric conductivity gradients with ultrasonically-induced
  Lorentz force}.
\newblock {\em Ultrasonic imaging}, 23(2):117--132, 2001.

\bibitem{montalibet2001electric}
A.~Montalibet, J.~Jossinet, A.~Matias, and D.~Cathignol.
\newblock {Electric current generated by ultrasonically induced Lorentz force
  in biological media}.
\newblock {\em Medical and Biological Engineering and Computing}, 39(1):15--20,
  2001.

\bibitem{muthupillai1995magnetic}
R.~Muthupillai, D.~Lomas, P.~Rossman, J.~Greenleaf, A.~Manduca, and R.~Ehman.
\newblock {Magnetic resonance elastography by direct visualization of
  propagating acoustic strain waves}.
\newblock {\em Science}, 269(5232):1854--1857, 1995.

\bibitem{newman1973observations}
D.~R. Newman.
\newblock {Observations of cylindrical waves reflected from a plane interface}.
\newblock {\em The Journal of the Acoustical Society of America}, 53:1174,
  1973.

\bibitem{newman1998history}
P.~G. Newman and G.~S. Rozycki.
\newblock {The history of ultrasound}.
\newblock {\em Surgical clinics of north America}, 78(2):179--195, 1998.

\bibitem{nightingale2002}
K.~Nightingale, M.~S. Soo, R.~Nightingale, and G.~Trahey.
\newblock {Acoustic radiation force impulse imaging: in vivo demonstration of
  clinical feasibility}.
\newblock {\em Ultrasound in medicine \& biology}, 28(2):227--235, 2002.

\bibitem{oh2003}
S.~H. Oh, B.~I. Lee, E.~J. Woo, S.~Y. Lee, M.~H. Cho, O.~Kwon, and J.~K. Seo.
\newblock {Conductivity and current density image reconstruction using harmonic
  Bz algorithm in magnetic resonance electrical impedance tomography}.
\newblock {\em Physics in Medicine and Biology}, 48(19):3101, 2003.

\bibitem{olafsson2006}
R.~Olafsson, R.~S. Witte, K.~Kim, S.~Ashkenazi, and M.~O'Donnell.
\newblock {Electric current mapping using the acousto-electric effect}.
\newblock In {\em Medical Imaging}, pages 61470O--61470O. International Society
  for Optics and Photonics, 2006.

\bibitem{olson1931}
H.~Olson.
\newblock {The Ribbon Microphone}.
\newblock {\em Journal of Society of Motion Picture and Television Engineers},
  16(6):695--708, 1931.

\bibitem{ophir1991}
J.~Ophir, I.~Cespedes, H.~Ponnekanti, Y.~Yazdi, and X.~Li.
\newblock {Elastography: a quantitative method for imaging the elasticity of
  biological tissues}.
\newblock {\em Ultrasonic imaging}, 13(2):111--134, 1991.

\bibitem{pitts2000three}
T.~A. Pitts and J.~F. Greenleaf.
\newblock {Three-dimensional optical measurement of instantaneous pressure}.
\newblock {\em The Journal of the Acoustical Society of America}, 108:2873,
  2000.

\bibitem{pitts2002optical}
T.~A. Pitts and J.~F. Greenleaf.
\newblock {Optical measurement of ultrasonic Poynting and velocity vector
  fields}.
\newblock {\em Ultrasonics, Ferroelectrics and Frequency Control, IEEE
  Transactions on}, 49(2):193--203, 2002.

\bibitem{pitts1994tomographic}
T.~A. Pitts, J.~F. Greenleaf, J.-y. Lu, and R.~R. Kinnick.
\newblock {Tomographic Schlieren imaging for measurement of beam pressure and
  intensity}.
\newblock In {\em Ultrasonics Symposium, 1994. Proceedings., 1994 IEEE},
  volume~3, pages 1665--1668. IEEE, 1994.

\bibitem{pitts2001optical}
T.~A. Pitts, A.~Sagers, and J.~F. Greenleaf.
\newblock {Optical phase contrast measurement of ultrasonic fields}.
\newblock {\em Ultrasonics, Ferroelectrics and Frequency Control, IEEE
  Transactions on}, 48(6):1686--1694, 2001.

\bibitem{pop2003}
M.~Pop, A.~Molckovsky, L.~Chin, M.~C. Kolios, M.~A. Jewett, and M.~D. Sherar.
\newblock {Changes in dielectric properties at 460 kHz of kidney and fat during
  heating: importance for radio-frequency thermal therapy}.
\newblock {\em Physics in Medicine and Biology}, 48(15):2509, 2003.

\bibitem{pourcelot1974applications}
L.~Pourcelot.
\newblock {Applications cliniques de l'examen Doppler transcutane}.
\newblock {\em Velocimetrie Ultrasonore Doppler}, 34:780--785, 1974.

\bibitem{preston1991}
R.~Preston.
\newblock {\em Output measurements for medical ultrasound}.
\newblock Springer Verlag, Berlin, 1991.

\bibitem{provost2013ultrafast}
J.~Provost, W.~Kwiecinski, M.~Fink, M.~Tanter, and M.~Pernot.
\newblock {Ultrafast Acoustoelectric Tomography}.
\newblock In {\em Ultrasonics, Ferroelectrics and Frequency Control, IEEE
  symposium}, pages 55--55, 2013.

\bibitem{raum1997}
K.~Raum and W.~O'Brien~Jr.
\newblock {Pulse-echo field distribution measurement technique for
  high-frequency ultrasound sources}.
\newblock {\em Ultrasonics, Ferroelectrics and Frequency Control, IEEE
  Transactions on}, 44(4):810--815, 1997.

\bibitem{retat2011}
L.~Retat.
\newblock {\em {Characterisation of the Acoustic, Thermal and Histological
  Properties of Tissue Required for High Intensity Focused Ultrasound ({HIFU})
  Treatment Planning}}.
\newblock PhD thesis, University of London, 2011.

\bibitem{rewcastle2006high}
J.~C. Rewcastle.
\newblock {High intensity focused ultrasound for prostate cancer: A review of
  the scientific foundation, technology and clinical outcomes}.
\newblock {\em Technology in cancer research \& treatment}, 5(6):619--625,
  2006.

\bibitem{roth2011}
B.~J. Roth.
\newblock {The role of magnetic forces in biology and medicine}.
\newblock {\em Experimental Biology and Medicine}, 236(2):132--137, 2011.

\bibitem{roth1994}
B.~J. Roth, P.~J. Basser, and J.~P. Wikswo.
\newblock {A theoretical model for magneto-acoustic imaging of bioelectric
  currents}.
\newblock {\em IEEE transactions on biomedical engineering}, 41(8):723--728,
  1994.

\bibitem{roth1998comments}
B.~J. Roth, J.~P. Wikswo~Jr, H.~Wen, and R.~Balaban.
\newblock {Comments on Hall Effect Imaging}.
\newblock {\em Biomedical Engineering, IEEE Transactions on},
  45(10):1294--1296, 1998.

\bibitem{routh1891}
E.~Routh.
\newblock {\em {A Treatise on Analytical Statics: The parallelogram of forces.
  Forces acting at a point. Parallel forces. Forces in two dimensions. On
  friction. The principle of work. Forces in three dimensions. Graphical
  statics. Centre of gravity. On strings. The machines}}.
\newblock A Treatise on Analytical Statics: With Numerous Examples. University
  Press, 1891.

\bibitem{sandrin1999}
L.~Sandrin, S.~Catheline, M.~Tanter, X.~Hennequin, and M.~Fink.
\newblock {Time-resolved pulsed elastography with ultrafast ultrasonic
  imaging}.
\newblock {\em Ultrasonic imaging}, 21(4):259--272, 1999.

\bibitem{sandrin2003transient}
L.~Sandrin, B.~Fourquet, J.-M. Hasquenoph, S.~Yon, C.~Fournier, F.~Mal,
  C.~Christidis, M.~Ziol, B.~Poulet, F.~Kazemi, et~al.
\newblock {Transient elastography: a new noninvasive method for assessment of
  hepatic fibrosis}.
\newblock {\em Ultrasound in medicine \& biology}, 29(12):1705--1713, 2003.

\bibitem{sarvazyan1998}
A.~P. Sarvazyan, O.~V. Rudenko, S.~D. Swanson, J.~B. Fowlkes, and S.~Y.
  Emelianov.
\newblock {Shear wave elasticity imaging: a new ultrasonic technology of
  medical diagnostics}.
\newblock {\em Ultrasound in medicine \& biology}, 24(9):1419--1435, 1998.

\bibitem{schneider1996quantitative}
B.~Schneider and K.~K. Shung.
\newblock {Quantitative analysis of pulsed ultrasonic beam patterns using a
  Schlieren system}.
\newblock {\em Ultrasonics, Ferroelectrics and Frequency Control, IEEE
  Transactions on}, 43(6):1181--1186, 1996.

\bibitem{schwan1957conductivity}
H.~Schwan and C.~Kay.
\newblock {The conductivity of living tissues}.
\newblock {\em Annals of the New York Academy of Sciences}, 65(6):1007--1013,
  1957.

\bibitem{seo2011magnetic}
J.~K. Seo and E.~J. Woo.
\newblock {Magnetic resonance electrical impedance tomography (MREIT)}.
\newblock {\em SIAM review}, 53(1):40--68, 2011.

\bibitem{sharf1999}
Y.~Sharf, G.~Clement, and K.~Hynynen.
\newblock {Absolute measurement of ultrasound pressure using high magnetic
  fields}.
\newblock {\em IEEE Transactions on Ultrasonics, Ferroelectrics and Frequency
  Control}, 46(6):1504--1511, 1999.

\bibitem{shaw2008buoyancy}
A.~Shaw.
\newblock {A buoyancy method for the measurement of total ultrasound power
  generated by HIFU transducers}.
\newblock {\em Ultrasound in medicine \& biology}, 34(8):1327--1342, 2008.

\bibitem{shaw2008calibration}
A.~Shaw and M.~Hodnett.
\newblock {Calibration and measurement issues for therapeutic ultrasound}.
\newblock {\em Ultrasonics}, 48:234--252, 2008.

\bibitem{smith1989hydrophones}
R.~A. Smith.
\newblock {Are hydrophones of diameter 0.5 mm small enough to characterise
  diagnostic ultrasound equipment?}
\newblock {\em Physics in Medicine and Biology}, 34(11):1593, 1989.

\bibitem{somaglino2011delivrance}
L.~Somaglino.
\newblock {\em {D{\'e}livrance par ultrasons de chimioth{\'e}rapie
  encapsul{\'e}e dans des liposomes sono-sensibles: contr{\^o}le et dosage de
  la cavitation inertielle ultrasonore}}.
\newblock PhD thesis, Universit{\'e} Claude Bernard-Lyon I, 2011.

\bibitem{staudenraus1993fibre}
J.~Staudenraus and W.~Eisenmenger.
\newblock {Fibre-optic probe hydrophone for ultrasonic and shock-wave
  measurements in water}.
\newblock {\em Ultrasonics}, 31(4):267--273, 1993.

\bibitem{ter2000intervention}
G.~{Ter Haar}.
\newblock {Intervention and therapy}.
\newblock {\em Ultrasound in medicine \& biology}, 26:S51--S54, 2000.

\bibitem{thompson1981ultrasonic}
R.~B. Thompson.
\newblock {Ultrasonic shear wave transducer}, October~13 1981.
\newblock US Patent 4,295,214.

\bibitem{torr1984acoustic}
G.~Torr.
\newblock {The acoustic radiation force}.
\newblock {\em Am. J. Phys}, 52(5):402--408, 1984.

\bibitem{towe1988}
B.~C. Towe and M.~R. Islam.
\newblock {A magneto-acoustic method for the noninvasive measurement of
  bioelectric currents}.
\newblock {\em Biomedical Engineering, IEEE Transactions on}, 35(10):892--894,
  1988.

\bibitem{vershuur1993hidden}
G.~L. Verschuur.
\newblock {\em {Hidden Attraction : The History And Mystery Of Magnetism}}.
\newblock New York: Oxford University Press, 1993.

\bibitem{wagner1983}
R.~F. Wagner, S.~W. Smith, J.~M. Sandrik, and H.~Lopez.
\newblock {Statistics of speckle in ultrasound B-scans.}
\newblock {\em IEEE TRANS. SONICS ULTRASONICS.}, 30(3):156--163, 1983.

\bibitem{wen1998}
H.~Wen, J.~Shah, and R.~S. Balaban.
\newblock {Hall effect imaging}.
\newblock {\em Biomedical Engineering, IEEE Transactions on}, 45(1):119--124,
  1998.

\bibitem{wild1951use}
J.~WILD and D.~NEAL.
\newblock {Use of high-frequency ultrasonic waves for detecting changes of
  texture in living tissues.}
\newblock {\em Lancet}, 1(6656):655, 1951.

\bibitem{wildi2000}
T.~Wildi and G.~Sybille.
\newblock {\em {Electrotechnique. 3{\`e}me {\'e}dition}}.
\newblock De Boeck Universit{\'e}, 2000.

\bibitem{woo2008}
E.~J. Woo and J.~K. Seo.
\newblock {Magnetic resonance electrical impedance tomography (MREIT) for
  high-resolution conductivity imaging}.
\newblock {\em Physiological measurement}, 29(10):R1, 2008.

\bibitem{xu2007}
Y.~Xu, S.~Haider, and A.~Hrbek.
\newblock {Magneto-Acousto-Electrical Tomography: A New Imaging Modality for
  Electrical Impedance}.
\newblock In {\em 13th International Conference on Electrical Bioimpedance and
  the 8th Conference on Electrical Impedance Tomography}, pages 292--295.
  Springer, 2007.

\bibitem{xu2005}
Y.~Xu and B.~He.
\newblock {Magnetoacoustic tomography with magnetic induction (MAT-MI)}.
\newblock {\em Physics in medicine and biology}, 50(21):5175, 2005.

\bibitem{zhang2004}
H.~Zhang and L.~V. Wang.
\newblock {Acousto-electric tomography}.
\newblock In {\em Biomedical Optics}, pages 145--149. International Society for
  Optics and Photonics, 2004.

\end{thebibliography}

\addcontentsline{toc}{chapter}{Credits images}
%\section*{Credits images}
\chapter*{Credits images}
\thispagestyle{vide}

\begin{itemize}
	\item \label{source11} Image \ref{figIntroAcoustiqueMedicaleHIFU} : Sonde Ablatherm {www.edap-tms.com} - reproduction autorisée par les auteurs
	\item \label{source10} Image \ref{figIntroAcoustiqueMedicaleEchographie} : {commons.wikimedia.org/wiki/File:Medecine\_Echographie.jpg} - licence libre
	\item \label{source9} Image \ref{figHydroEMIntroFibreOptique} : manuel Fiber Optic Probe Hydrophone 2000, RP Acoustics - reproduction autorisée par l'auteur
	\item \label{source8} Image \ref{figHydroEMIntroSchlieren} : {en.wikipedia.org/wiki/File:Schlieren\_HIFU\_1MHz.JPG} - domaine public
	\item \label{source3} Image \ref{figHydroEMIntroHydroEM}-(a) : {runeberg.org/nfcr/0059.html} - domaine public
	\item \label{source4} Image \ref{figHydroEMIntroHydroEM}-(b) : {en.wikipedia.org/wiki/File:Rca44.png}, LuckyLouie 23:22, 4 November 2007 (UTC) - licence Creative Commons (CC-BY-SA)
	\item \label{source5} Image \ref{figHydroEMIntroHydroEM2} : reproduction autorisée par les auteurs
	\item \label{source13} Image \ref{figLFEITIntroConducElecTissus1} : reproduction personnelle
	\item \label{source6} Image \ref{figLFEITIntroConducElecTissus2} : reproduction personnelle
	\item \label{source12} Image \ref{figLFEITMethodesEIT}-(b) : {eprints.ma.man.ac.uk/1053/01/trieste.pdf} - reproduction autorisée par les auteurs
	\item \label{source1} Image \ref{figLFEITMethodesMREIT} : reproduction autorisée par les auteurs
	\item \label{source2} Image \ref{figLFEITMethodesMATMI} : reproduction autorisée par les auteurs
	\item \label{source7} Image \ref{figElastoEMIntroElasticiteTissus} : reproduction personnelle
\end{itemize}

\cleardoublepage
\addcontentsline{toc}{chapter}{Abstract}
\chapter*{Abstract}
\thispagestyle{vide}
{The ability of the Lorentz force to link a mechanical displacement to an electrical current presents a strong interest for medical acoustics, and three applications were studied in this thesis.

In the first part of this work, a hydrophone was developed for mapping the particle velocity of an acoustic field. This hydrophone was constructed using a thin copper wire and an external magnetic field. A model was elaborated to determine the relationship between the acoustic pressure and the measured electrical current, which is induced by Lorentz force when the wire vibrates in the acoustic field of an ultrasound transducer. The built prototype was characterized and its spatial resolution, frequency response, sensitivity, robustness and directivity response were investigated.

An imaging method called Lorentz Force Electrical Impedance Tomography was also studied. In this method, a biological tissue is vibrated by ultrasound in a magnetic field, which induces an electrical current by Lorentz force. The electrical impedance of the tissue can be deduced from the measurement of the current. This technique was applied for imaging a gelatin phantom, a beef muscle sample, and a thermal lesion in a chicken breast sample. This showed the method may be useful for providing additional contrast to conventional ultrasound imaging.

Finally, this thesis demonstrated that shear waves can be generated in soft tissues using Lorentz force. This work was performed by applying an electrical current with two electrodes in a soft solid placed in a magnetic field. Shear waves were observed in gelatin phantom and liver sample. The speed of the shear waves were used to compute elasticity and their source to map the electrical conductivity of the samples.}

\newpage

\chapter*{Résumé}
\addcontentsline{toc}{chapter}{Résumé}
\thispagestyle{vide}
{La capacité de la force de Lorentz à relier un déplacement mécanique à un courant électrique présente un intérêt certain pour l'acoustique médicale, et trois applications ont été étudiées dans cette thèse.

Dans la première partie de ce travail, un hydrophone a été développé pour effectuer des champs de vitesse acoustique. Cet hydrophone était constitué d'un fil de cuivre vibrant dans un champ magnétique. Un modèle a été élaboré pour déterminer une relation entre la pression acoustique et le courant électrique mesuré, qui est induit par force de Lorentz lorsque le fil vibre dans un champ acoustique. Un prototype a ensuite été conçu et sa résolution spatiale, sa réponse fréquentielle, sa sensibilité, sa résistance et sa réponse directionnelle ont été examinées.

Une méthode d'imagerie appelée Tomographie d'Impédance Electrique par Force de Lorentz a aussi été étudiée. Dans cette méthode, un tissu biologique est déplacé par ultrasons dans un champ magnétique, ce qui induit un courant électrique par force de Lorentz. L'impédance électrique du tissu peut ensuite être déduite de la mesure du courant. Cette technique a été appliquée pour réaliser des images d'un fantôme de gélatine, d'un muscle de b\oe uf, et d'une lésion thermique dans un échantillon de poulet. Cela a montré que la méthode est potentiellement utile pour fournir un contraste supplémentaire à des images ultrasonores classiques.

Enfin, cette thèse a démontré que des ondes de cisaillement peuvent être générées dans des tissus mous par force de Lorentz. Cela a été réalisé en appliquant un courant électrique par deux électrodes dans un solide mou placé dans un champ magnétique. Des ondes de cisaillement ont été observées dans des échantillons de gélatine et de foie. La vitesse des ondes de cisaillement a été utilisée pour calculer l'élasticité et leur source pour cartographier la conductivité électrique des échantillons.}

\end{document}